\documentclass{article}
\usepackage{times}
\usepackage[scaled=.90]{helvet}
\usepackage{courier}
\usepackage[dvips]{color}
\usepackage{amssymb}
\usepackage{amsmath}
\usepackage{psfrag}
\usepackage{subfigure}
\usepackage{wrapfig}
\usepackage[latin1]{inputenc}
\usepackage{icrctc07}
\begin{document}
\begin{center}
\begin{figure}

\includegraphics[width=0.15\textwidth]{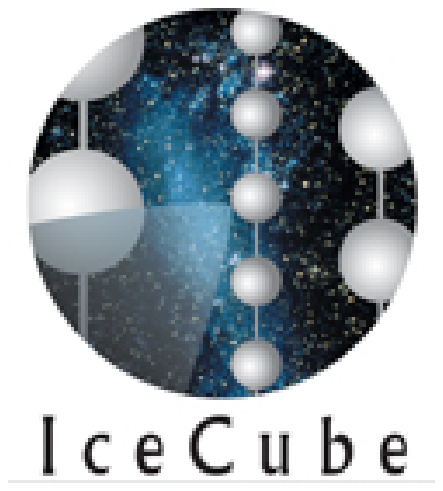} \hfill  \includegraphics[width=0.35\textwidth]{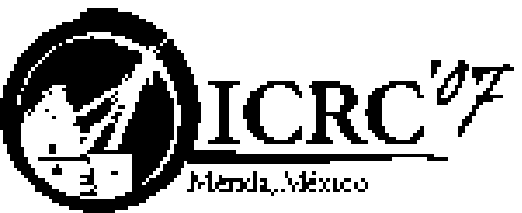}
\end{figure}
{\Huge \bf The IceCube Collaboration:  \\
\vskip 0.5cm
 \Large contributions to the\\
30  $^{th}$  International Cosmic Ray Conference (ICRC 2007), \\
Merida, Yucatan Mexico, \\
August  2007}
\end {center}
\vspace{2cm}
\begin{center}
{\large \bf Abstract}
\end{center}
This paper bundles 40 contributions by the IceCube collaboration that were submitted to the 30$^{th}$ International Cosmic Ray Conference ICRC 2007.  The  articles cover studies on cosmic rays and atmospheric neutrinos, searches for non-localized,  extraterrestrial $\nu_{e}$, $\nu_{\mu}$ and $\nu_{\tau}$ signals, scans for steady and intermittent neutrino point sources,  searches for dark matter candidates, magnetic monopoles and other exotic particles, improvements in analysis techniques, as well as future detector extensions. 

The IceCube observatory will be finalized in 2011 to form a cubic-kilometer ice-Cherenkov detector at the location of the geographic South Pole.  At the present state of construction, IceCube consists of  52 paired IceTop surface tanks  and 22 IceCube strings with a total of 1426 Digital Optical Modules deployed at depths up to 2350 m. The observatory also integrates the 19 string AMANDA subdetector, that was completed in 2000 and extends IceCube's reach to lower energies.  Before the deployment of IceTop, cosmic air showers were registered with the 30 station SPASE-2 surface array.

IceCube's low noise Digital Optical Modules are very reliable,  show a uniform response and record waveforms of arriving photons that are resolvable with nanosecond precision over a large dynamic range.  Data acquisition, reconstruction and simulation software are running in production mode and the analyses, profiting from the improved data quality and increased overall sensitivity, are well under way. 
%
%\newpage
%\title{The IceCube Collaboration}
%\shorttitle{The IceCube Collaboration}
 %\authors{%
 \newpage
 %\begin{onecolumn}
 \begin{center}
{\large \bf The IceCube Collaboration}
\end{center}
 M.~Ackermann$^{34}$,
 J.~Adams$^{11}$,
 J.~Ahrens$^{22}$,
 K.~Andeen$^{21}$,
 J.~Auffenberg$^{33}$,
 X.~Bai$^{24}$,
 B.~Baret$^{9}$,
 S.~W.~Barwick$^{16}$,
 R.~Bay$^{5}$,
 K.~Beattie$^{7}$
% ...
 \\
 %}
 %
 %\afiliations{
% ...
$^{5}$  Dept.~of Physics, University of California, Berkeley, CA 94720, USA \\
% ...
$^{7}$  Lawrence Berkeley National Laboratory, Berkeley, CA 94720, USA \\
% ...
$^{9}$  Vrije Universiteit Brussel, Dienst ELEM, B-1050 Brussels, Belgium \\
% ...
$^{11}$  Dept.~of Physics and Astronomy, University of Canterbury, Private Bag 4800, Christchurch, New Zealand \\
% ...
$^{16}$  Dept.~of Physics and Astronomy, University of California, Irvine, CA 92697, USA \\
% ...
$^{21}$  Dept.~of Physics, University of Wisconsin, Madison, WI 53706, USA \\
 $^{22}$  Institute of Physics, University of Mainz, Staudinger Weg 7, D-55099 Mainz, Germany \\
% ...
$^{24}$  Bartol Research Institute and Dept.~of Physics and Astronomy, University of Delaware, Newark, DE 19716, USA \\
% ...
$^{33}$  Dept.~of Physics, University of Wuppertal, D-42119 Wuppertal, Germany \\
 $^{34}$  DESY, D-15735 Zeuthen, Germany \\
% ...
% }
%\maketitle
% \begin{document}
%  \begin{onecolumn}
\begin{center}
{\large \bf Acknowledgments}
\end{center}
 %{\bf Acknowledgments:}
 We acknowledge the support from the following agencies:
 National Science Foundation-Office of Polar Program,
 National Science Foundation-Physics Division,
 University of Wisconsin Alumni Research Foundation,
 Department of Energy, and National Energy Research Scientific Computing Center
 (supported by the Office of Energy Research of the Department of Energy),
 the NSF-supported TeraGrid system at the San Diego Supercomputer Center (SDSC),
 and the National Center for Supercomputing Applications (NCSA);
 Swedish Research Council,
 Swedish Polar Research Secretariat,
 and Knut and Alice Wallenberg Foundation, Sweden;
 German Ministry for Education and Research,
 Deutsche Forschungsgemeinschaft (DFG), Germany;
 Fund for Scientific Research (FNRS-FWO),
 Flanders Institute to encourage scientific and technological research in industry (IWT),
 Belgian Federal Office for Scientific, Technical and Cultural affairs (OSTC);
 the Netherlands Organisation for Scientific Research (NWO);
 M.~Ribordy acknowledges the support of the SNF (Switzerland);
 A.~Kappes and J.~D.~Zornoza acknowledge support by the EU Marie Curie OIF Program.
 %\end{document}
\newpage
\begin{center}
{\large \bf Table of Contents}
\end{center}
\begin{enumerate}
\item A.~Karle for the IceCube Collaboration, {\it IceCube - construction status, performance results of the 22 string detector}, pages 7-10.
\item A. ~Gross, C. ~Ha, C. ~Rott, M. Tluczykont, E. Resconi, T. Deyoung, and G. Wikstr\"om for the IceCube Collaboration, {\it The combined AMANDA and IceCube neutrino telescope}, pages 11-14.
\item T.K.~ Gaisser for the IceCube Collaboration, {\it Performance of the IceTop array}, pages 15-18.
\item T.~Kuwabara, J.W.~Bieber and R.~Pyle, {\it Heliospheric physics with IceTop}, page 19-22.
\item K.G.~Andeen, C.~Song and K.~Rawlins for the SPASE2 and IceCube Collaborations, {\it Measuring cosmic ray composition at the 
knee with SPASE-2 and AMANDA-II}, page 23-26.
\item C.~Song, P.~Niessen and K.~Rawlins for the IceCube Collaboration, {\it Cosmic Rays in IceCube: Composition-sensitive observables}, page 27-30.
\item K.~James, X.~Bai, T.K.~Gaisser, J.~Hinton, P.~Niessen, T.~Stanev, S.~Tilav, and A.~Watson for the SPASE-2 and IceCube Collaborations,
{\it Search for TeV gamma-rays from point sources with SPASE-2}, pages 31-34. 
\item S.~Klein and D.~Chirkin for the IceCube Collaboration, {\it Study of high pt muons in air showers with IceCube}, pages 35-38.
\item X.~Bai, T.K.~Gaisser, T.~Stanev, and T.~Waldenmaier for the IceCube Collaboration, {\it IceTop/IceCube coincidences}, pages 39-42. 
\item S.~Klepser, F.~Kislat, H.~Kolanoski, P.~Niessen, and A.~ Van Overloop for the IceCube Collaboration, {\it Lateral distribution of 
air shower signals and initial energy spectrum above 1 PeV from IceTop}, pages 43-46. 
\item L.~ Demir\"ors, M.~Beimforde, J.~Eisch, J.~Madsen, P.~Niessen, G.M.~Spiczak, S.~Stoyanov, and S.~Tilav for the IceCube 
Collaboration, {\it IceTop tank response to muons}, pages 47-50. 
\item J.M.~Clem, P.~Niessen and S.~Stoyanov for the IceCube Collaboration, {\it Response of IceTop tanks to low-energy particles}, pages 51-54.
\item J.~Ahrens and J.L.~Kelley for the IceCube Collaboration, {\it Testing alternative oscillation scenarios with atmospheric neutrinos using AMANDA-II data from 2000 to 2003}, pages 55-58.
\item J.~Pretz for the IceCube Collaboration, {\it Atmospheric muon neutrino analysis with IceCube}, pages 59-62. 
\item J.D.~Zornoza and D.~Chirkin for the IceCube Collaboration, {\it Muon energy reconstruction and atmospheric neutrino spectrum unfolding with the IceCube detector}, pages 63-66.
\item K.~Hoshina, J.~Hodges and G.C.~Hill for the IceCube Collaboration, {\it Searches for a diffuse flux of extra-terrestrial muon 
neutrinos with AMANDA-II and IceCube}, pages 67-70.
\item K. M\"unich and J.~L\"unemann for the IceCube Collaboration, {\it Measurement of the atmospheric lepton energy spectra with 
AMANDA-II}, pages 71-74.
\item L.~Gerhardt for the IceCube Collaboration, {\it Multi-year search for UHE diffuse neutrino flux with AMANDA-II}, pages 75-78.
\item G.C.~Hill for the IceCube Collaboration, {\it Likelihood deconvolution of diffuse prompt and extra-terrestrial neutrino fluxes 
in the AMANDA-II detector}, pages 79-82.
\item O.~Tarasova, M.~Kowalski and M.~Walter, {\it Search for neutrino-induced cascades with AMANDA data taken in 2000-2004}, pages 83-86.
\item A.~Ishihara for the IceCube Collaboration, {\it EHE neutrino search with the IceCube 9 string array}, pages 87-90.
\item J.~Bolmont, B.~ Voigt and R. Nahnhauer for the IceCube Collaboration, {\it Very high energy electromagnetic cascades in the LPM regime with IceCube}, pages 91-94.
\item J.~Kiryluk, M.V.~D'Agostino, S.R.~Klein, C.~ Song, and D.R.~Williams for the IceCube Collaboration, 
{\it IceCube performance with artificial light sources: the road to cascade analyses}, pages 95-98.
\item J.~Braun, A.~Karle and T.~Montaruli for the IceCube Collaboration, {\it Neutrino point source search strategies for AMANDA-II 
and results from 2005}, pages 99-102.
\item R.~Franke, R.~Lauer, M.~Ackermann, and E.~Bernardini for the IceCube Collaboration, {\it Point source analysis for cosmic 
neutrinos beyond PeV energies with AMANDA and IceCube}, pages 103-106.
\item C.~Finley, J.~Dumm and T.~Montaruli for the IceCube Collaboration, {\it  Nine-string IceCube point source analysis}, pages 107-110.
\item J.P.~H\"ulss and C.~Wiebusch for the IceCube Collaboration, {\it Search for signatures of extra-terrestrial neutrinos
 with a multipole analysis of the AMANDA-II sky-map}, pages 111-114.
\item K.~Satalecka, E.~Bernardini, M.~Ackermann, and M.~Tluczykont for the IceCube Collaboration, {\it Cluster Search for neutrino flares from pre-defined directions}, pages 115-118.
\item R.~Porrata for the IceCube Collaboration, {\it All-sky search for transient sources of neutrinos using give years of AMANDA-II data}, pages 119-122.
\item M.~Ackermann, E.~Bernardini, N.~Gallante, F.~Goebel, M.~Hayashida, K.~Satalecka, M.~Tluczykont and R.M.~Wagner for the IceCube and MAGIC Collaboration, {\it Neutrino triggered Target of Opportunity (NToO) test run with AMANDA-II and MAGIC}, pages 123-126.
\item A.~Kappes, M.~Kowalski, E.~Strahler, and I.~Tabaoda for the IceCube Collaboration, {\it Detecting GRBs with IceCube and optical follow-up observations}, pages 127-130.
\item D.~Hubert and A.~Davour for the IceCube Collaboration, {\it Search for neutralino dark matter with the AMANDA neutrino telescope}, pages 131-134. 
\item G. Wikstr\"om for the IceCube Collaboration, {\it Prospects of dark matter detection in IceCube}, pages 135-138.
\item H.~Wissing for the IceCube Collaboration, {\it Search for relativistic magnetic monopoles with the AMANDA-II detector}, pages 139-142.
\item A.~Pohl and D.~ Hardtke for the IceCube Collaboration, {\it Subrelativistic Particle Searches with the AMANDA-II detector}, pages 143-146.
\item B. ~Christy, A.~Olivas and D.~Hardtke for the IceCube Collaboration, {\it Exotic particles searches with IceCube}, pages 147-150.
\item D.~Chirkin for the IceCube Collaboration, {\it Effect of the improved data acquisition system of IceCube on its neutrino-detection 
capabilities}, pages 151-154.
\item J.~Lundberg for the IceCube Collaboration, {\it Improved Cherenkov light propagation methods for the IceCube neutrino telescope}, pages 155-158. 
\item S.~Grullon, D.J.~Boersma, G.~Hill, K.~Hoshina, and K.~Mase for the IceCube Collaboration, {\it Reconstruction of high energy muon events in IceCube using waveforms}, pages 159-162.
\item H.~Landsman for the IceCube Collaboration, L.~Ruckmann, G.S.~Varner, {\it Radio detection of GZK neutrinos - AURA status and plans}, pages 163-166.
\end{enumerate}
%
%\end{onecolumn}
%Overview papers
%
%Karle_IC_status_icrc07v2.7.pdf
%icrc_AMANDA_integration.pdf
%gaisser-HE-1-5.pdf
%
%%
% International Cosmic Ray Conference 2007 Merida Yucatan Mexico
% In this file you will find detailed instructions to correctly
% typeset your document.
%
% By: Victor De la Luz
% vdelaluz@inaoep.mx
% Mexico City

%Class Required
%%% for classical LaTeX
%\documentclass[dvips]{article}
%%% for PDFLaTeX
%\documentclass[pdftex]{article}
%The ICRC Style
%(This package is the last package in the usepackage list)
%If you need import other package you need write it first.
%\usepackage{icrctc07}

%The paper title
\title{IceCube - construction status and performance results of the 22 string detector
}
%Short title to print in the headers to the final publication (Not showed in this print).
%\shorttitle{Short title}

%All paper authors
\authors{Albrecht Karle$^{1}$,  for the IceCube Collaboration$^{2}$ }
%Short title to print in the headers to the final publication (Not shown in this print).
%\shortauthors{Author and et al.}
%All the affiliations.
\afiliations{$^1$University of Wisconsin-Madison\\ $^2$See special section of these proceedings }
\email{karle@icecube.wisc.edu}

%The abstract.
\abstract{The IceCube neutrino observatory is a cubic-kilometer ice-Cherenkov detector being constructed in the deep ice at the geographic South Pole. After a successful construction season ending in February 2007, IceCube consists of 22 strings and 26 IceTop stations with a total of 1424 Digital Optical Modules (DOMs) deployed at depths up to 2450m. Together with the commissioning of the central laboratory building and central DAQ electronics, this allowed IceCube to begin early operations and data analysis.  The goal is to complete construction of the final configuration of 80 strings and IceTop stations in 2011.  First results from the 22-string configuration and an overview of the project will be presented.}

%%%%%%%%%%%%%%%%%%%% B E G I N   D O C U M E N T%%%%%%%%%%%%%%%%%%%%%%%
%\begin{document}
\maketitle
%Begin the section.
\section{Overview}

\vskip -0.4cm

The IceCube neutrino observatory is a kilometer-scale neutrino
telescope currently under construction at the South Pole. 
The existing
AMANDA-II array, the precursor of IceCube,  will be surrounded by and integrated into the IceCube array \cite{gross-icrc}.  IceCube is
designed to detect astrophysical neutrino fluxes at energies from a
few 100\,GeV up to the highest energies of $10^{9}$ GeV \cite{karle_performance},
\cite{karle_pdd}.

\begin{tabular}{|c|c|c|c|}
\hline
Project    &  Strings &   IceTop    &     \#  of\\
Year        &   deployed   &  stations   &  Sensors  \\
\hline
2004/05    &   1     &   4    &    76     \\
2005/06   &   8     &   12   &    528   \\
2006/07   &   13     &   10   &   820   \\
\hline
2007 total   &   22 &   26   &   1424   \\
\hline
\end{tabular}
\\

\begin{figure}
\begin{center}
\noindent
%\fbox{\hbox{\vbox{\hsize=60mm \hfill \vspace{80mm} }}}
%uncomment next line to include real image
%\includegraphics [width=0.2\textwidth]{figura.ps}
\includegraphics[width=0.48\textwidth]{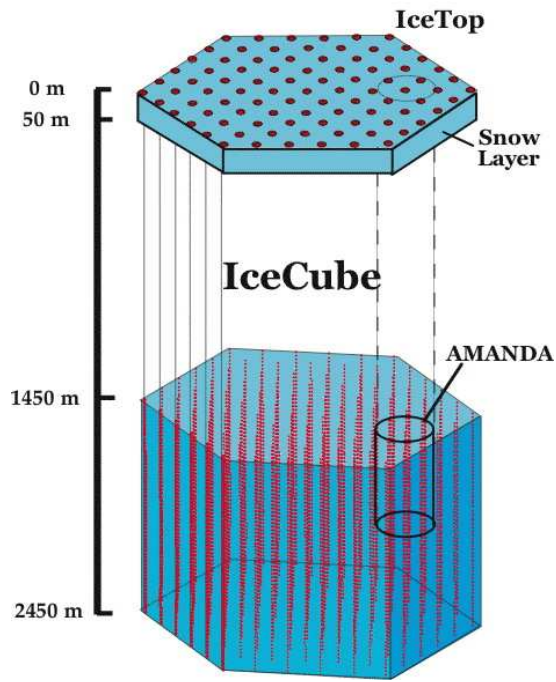}
\end{center}
\vskip -0.4cm
\caption{Schematic view of the IceCube array consisting of 80 strings 
with 60 sensors on each string.  The surface array 
IceTop consists of 160 detectors,  two of which 
are associated with each string. }
\label{karle_fig1}
\end{figure}

\begin{figure}[htb]
\centering
 \includegraphics[width=0.48\textwidth]{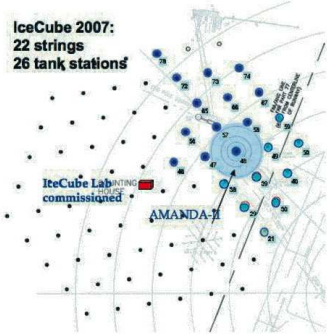}
%\hspace*{\fill}
\vskip -0.4cm
\caption{Schematic view of current geometry}
\label{karle_layout}
\end{figure}

The IceCube neutrino observatory at the South Pole will consist of
4800 optical sensors - digital optical modules (DOMs) - installed on 80
strings at depths of 1450\,m to 2450\,m in the Antarctic
Ice, and 320 sensors deployed in 160 IceTop \cite{gaisser-icrc} detectors in pairs on the ice surface 
directly above the strings.  
Each sensor consists of a 
photomultiplier tube connected to a waveform-recording data
acquisition circuit capable of resolving pulses with nanosecond
precision and having a dynamic range of at least 250 photoelectrons
per 10\,ns.  
Construction started at the South Pole in November 2004. 
A total of 1424 sensors have been installed to date 
on 22 strings and in 26 IceTop surface detector stations.
The table below summarizes the construction status as of February 2007.

\vskip -0.3cm

\subsection{Electrical and mechanical structure}

It was a design goal to avoid single
point failures in the ice, as the sensors are not accessible once the ice refreezes.  
High reliability and ease of maintenance were other design goals.
A string consists of the following major configuration items: a cable from the counting
house to the string location, a cable from the surface to 2450\,m depth,
and 60 optical sensors.  
30 twisted-pair copper cables packaged in 15 twisted quads are used to
provide power and communication to 60 sensors.  
%The shielded main cable is about 2500\,m long and
%42\,mm in diameter with a weight of 6\,t. 
%Three additional quads provide communication to pressure sensors 
%and local communication between optical sensors.
To reduce the amount of cable, two sensors are 
operated on the same wire pair, one terminated and 
one unterminated.  
Neighboring sensors are 
connected to enable fast local coincidence triggering
in the ice.  
%Each sensor has a direct connection to the 
%data acquisition computers in the central counting house. 

\begin{figure}[htb]
\centering
 \includegraphics[width=0.48\textwidth]{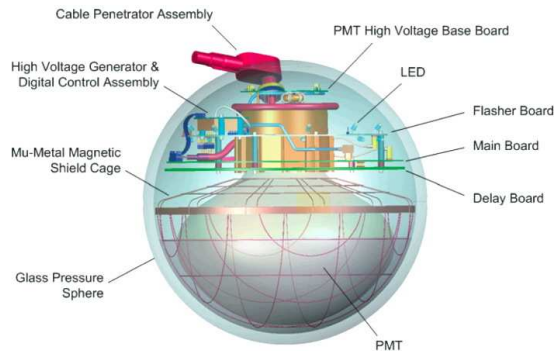}
%\hspace*{\fill}
\vskip -0.4cm
\caption{Schematic view of a Digital Optical Module.}
\label{karle_DOM}
\end{figure}

A schematic view of an optical sensor is shown in Fig. \ref{karle_DOM}.  
An optical sensor consists of a 25-cm-diameter 
photomultiplier tube (PMT) embedded in a glass pressure vessel of 
32.5-cm diameter. 
The HAMAMATSU R7081-02 PMT has ten dynodes, 
allowing operation at a gain of at least $5 \cdot 10^{7}$.  
The average gain  
is set to $1.0\cdot 10^{7}$, providing a single photoelectron 
amplitude of about 5\,mV.   
The signals are digitized by a fast analog transient waveform recorder 
(ATWD, 300 MSPS)  and by a fADC (40 MSPS).  
%The ATWD uses a 128-sample, switched- 
%capacitor array which is operated at a sampling rate of 3.3 ns/sample.
The PMT signal is amplified by 3 different gains ($\times$0.25, $\times$2, $\times$16)  
to extend the dynamic range of the ATWD to 16 bits.
The linear dynamic range of the sensor is 400 photoelectrons in
15\,ns; the integrated dynamic range is of more than 
5,000 photoelectrons in 2 $\mu$s.
%The PMT analog pulse is delayed by 75\,ns on a separate circuit board
%to account for the time needed to make the trigger decision
%and initiate the ATWD for readout.
The digital electronics on the main board are based on a 
%400k-gate, 
field-programmable gate array (FPGA) which contains a 
32-bit CPU, 8\,MB of flash storage, and 32\,MB of RAM. 
A small communications program stored in ROM 
allows communication to be established with 
the surface computer system and new programs to be downloaded to the DOM.

The flasher board is an optical calibration device which is integrated in each DOM. 
The amplitude of the LED pulses can be adjusted over a wide range
 up to a brightness of $9\cdot10^{10}$ photons at a wavelength of about 
405\,nm. 
%The flasherboard allows for a variety of calibration functions, 
%e.g. timing and geometry verification. 

%The high voltage is generated by an  HV generator, 
%which is located on a separate board.  The base is 
%a simple resistor chain with appropriate capacitors.  
%%Part selection and electronic circuit bords were designed for 
%%high reliability.  
%%All electronics of the DOM were designed for high reliability 
%%and stress testing was performed to screen for high quality. 
%% For example all
%% parts exposed to high voltage are derated by at least a factor of 2.5.
%Low power consumption was a design goal, the power of a single DOM being 
%3\,W in normal operation.

\vskip -0.3cm

\section{Data acquisition and online data processing}
\vskip -0.3cm

\begin{figure}
\begin{center}
\includegraphics[width=0.45\textwidth]{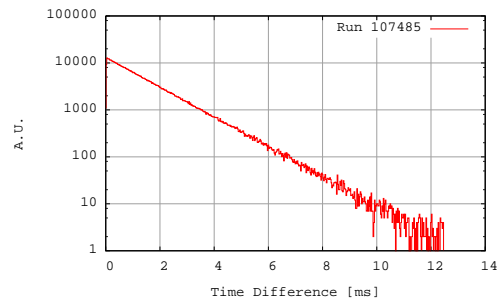}
\end{center}
\vskip -0.4cm
\caption{The time difference between subsequent events is shown for one run.}
\label{deadtime}
\end{figure}

All digitized photomultiplier pulses are to be sent to the surface.  
A local coincidence (LC) trigger scheme is used 
to apply data compression for isolated hits, which are mostly noise pulses. 
%For the LC trigger, they are defined as pulses for 
%which no signal was recorded 
%in the neighboring sensors within $\approx\pm800$\,ns. 
% In 2005 isolated hits were suppressed entirely. 
% In the future the isolated
%All other hits are being transmitted with full waveform information. 
%In the system to be used in 2007, the isolated hits are not transmitted.  
Every string is connected to one server called a stringhub, which includes 8 custom PCI cards. They provide power, communication and time calibration to the sensors.  
The stringhub sorts the hits in time and buffers them until the trigger and eventbuilding process is complete.  The digital architecture allows deadtime-free data acquisition
(Fig. \ref{deadtime})
with the exception of runstop and start times and maintenance times. 
A joint eventbuilder combines signals from the AMANDA-II array with IceCube data. 
The raw data rate is on the order of 100 GB/day, which are written to tapes. 
An online processing and filtering cluster extracts interesting phenomena, such as all upgoing muons,  
high-energy events, IceTop-In-ice coincidences,  cascade events, 
events from the direction of the moon, 
events that are interesting for dark matter search and events in coincidence  with GRB.
%In addition, 
% some prescaled minimum-bias events, calibration and monitoring and 
%the supernova data streams are separated.  
The filtered data stream (of order 20 GB/day) is then transmitted by satellite 
to the Northern Hemisphere to be stored and 
archived in the data center.
% at UW-Madison, the host institution of IceCube project.  
The data will then be prepared for physics data analysis by the working groups 
in the collaboration.

\vskip -0.3cm

\section{Drilling and detector installation}
\vskip -0.3cm

The strings are installed in holes which are drilled using the 
enhanced hot-water drill (EHWD). 
The drill consists of numerous pump and heating systems, hoses, a drill 
tower and a complex control system.  It delivers a thermal power of 5 
MW.   
The average time required for drilling a hole 60 cm in diameter to a depth of 2450\,m was 

$\sim$34 hours in the most recent construction season. 
The subsequent installation of a string with 60 DOMs required typically
12 hours.  
Overall, the construction cycle time between two strings was ~3 days,
which allowed the installation of 2 strings per week.  
With some optimizations in set-up time and an improved technique
for drilling through the firn layer, we expect to 
install up to 18 strings 
between December 2007 and January 2008. 
Based on the past season, the long-term construction schedule remains unchanged 
with completion expected in January 2011. 

%i define "serious issues" (like dead or not communicating) and "slight issues" (like LC probs)
%let's see:
%1424 DOMs deployed.
%# with serious issues:  16  (0 from 04/05, 9 from 05/06, 7 from 06/07)
%# with slight or serious issues:   34  (0 from 04/05, 16 from 05/06, 18 from 06/07)
%
%ie 1424 DOMs deployed, 1.1% have serious issues, 1.3% have slight issues.

All sensors undergo a final acceptance test at their production sites 
before being shipped to the South Pole.
They are again tested briefly on the ice prior to deployment. 
The installation and the subsequent freeze-in process (with temporary pressures up 
to more than 400 bar) places unusual demands on the string hardware. 
Yet, the survival rate of optical sensors is very high.  
For 1424 optical sensors deployed to date, only 16 (1.1\%) are not usable; 
another 18 (1.3\%) have developed minor issues, some of which are expected to be resolved. 
97.6\% of all sensors have been commissioned with full functionality and 
are in operation to date.  Only two sensors failed after they were frozen in and commissioned.  A total of 1000 DOMyears of integrated operation has been accumulated as of May 2007.  
%As a comparison, 85 (13\%) of the 677 deployed OMs in AMANDA-II could not be 
%used for analysis in 2005 for technical reasons. 

\begin{figure}
\begin{center}
\includegraphics[width=0.45\textwidth]{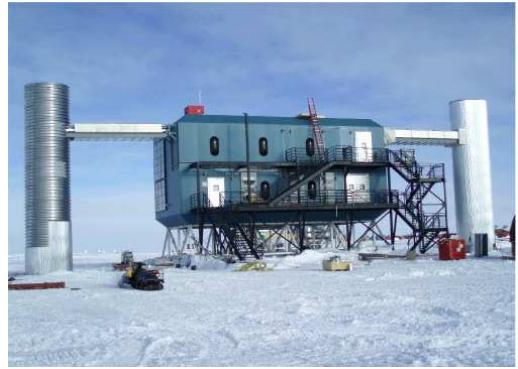}
\end{center}
\vskip -0.4cm
\caption{The IceCube Laboratory contains all surface electronics and 
server farms for data acquisition and online data processing.}
\label{ICL}
\end{figure}

\vskip -0.3cm

\section{Operation and performance characteristics}
\vskip -0.3cm

The detector electronics and software are designed 
to require minimal maintenance at the remote location.
For example, the time calibration system, a critical 
part of any neutrino telescope, is designed to be a
self-calibrating, integral part of the readout system 
(in contrast to the AMANDA detector, which required 
manual calibration of all analog detector channels).
The strings are calibrated as soon as they are frozen in, 
allowing for gradual commissioning of the instrument. 

%
%\begin{figure}[htb]
%\centering
% \includegraphics[width=0.5\textwidth]{lcrates-icrc-2.eps}
%%\hspace*{\fill}
%\caption{Observed number of hits/DOM are plotted versus depth for all strings. }
%\label{noise}
%\end{figure}

%\begin{enumerate}

All sensors have precise quartz oscillators to provide local clocks, which are synchronized every  few seconds to the central  GPS clock.  
Using LED flashers, it was possible to verify the time resolution 
to a precision of less than 2\,ns on average.
Studies with muons and flashers have shown that the
timing is stable over periods of months \cite{joanna-icrc}.
Another important performance parameter is the
%single-photoelectron, 
dark-noise rate of the sensors.  
There is no known natural background of light in the deep ice other than 
light generated by cosmic particles.  
The noise rates for DOMs in the deep ice are $\approx$700\,Hz.
The rate is $\sim$320\,Hz with an applied dead time 
of 50\,$\mu$s.  
The very low noise rates of the sensors are critical for the detection
 of the low energy neutrino emission associated with supernova core 
collapse. 

\vskip -0.3cm
\begin{figure}[htb]
\centering
\includegraphics[width=0.42\textwidth]{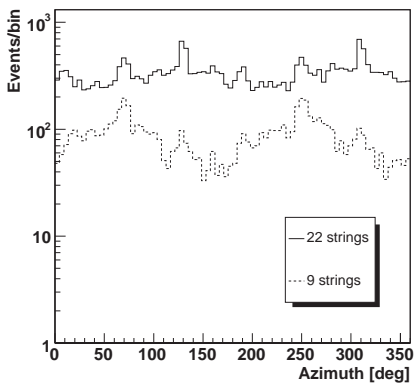}
%\hspace*{\fill}
\vskip -0.5cm
\caption{Azimuth distribution of atmospheric muons observed in the 9-string array in 2006  and the 22-string array in 2007.  }
\label{zenazi}
\end{figure}

The 9- and 22-string arrays trigger on atmospheric muons at a rate of  140Hz and 520Hz, respectively.  
The 22 string trigger condition requires an 8-fold coincidence within 5\,$\mu$sec.  Several characteristic figures of AMANDA-II, IC9, IC22 and IC80 are compared in table\,2.
The 22-string configuration has a significantly higher effective area and overall sensitivity.  
Fig. \ref{zenazi}  shows the azimuth distribution of cosmic-ray muons for one hour of livetime of the 9- string array and the 22-string array for events with at least 20 DOMs and 3 strings hit.  The azimuth distribution for IC22 is more even, and the overall rate is visibly higher as the detector is now sensitive in all directions.  

%Fig. \ref{zenazi}  shows the azimuth distribution of cosmic-ray muons
%% for one hour of livetime
% in the 9- string array 
% %and the 22-string array.  
% The azimuth distribution reflects geometric acceptance of the array.  
%% and reconstruction effects at low energies. It also shows that the overall rate is visibly higher.   
First physics analyses have already been performed using data of the IceCube 9 string array\cite{pretz-icrc, finley-icrc, hoshina-icrc}.  The start of regular science operations with IC22
is scheduled for May 2007 and will continue in this configuration until March 2008.

\begin{table}[t]
\begin{center}
\begin{tabular}{l|ccccl}
%\footnotesize
\hline
-                                                    &     A-II     &     IC9       & IC22     & IC80   \\
\hline
Instr. Volume/km$^3$              &  .016       &     0.044        &    0.18    &  0.9       \\
%\# of OMs                                 & 677         &   604     & 1424   &  5120     \\
\# of sensors (in ice)                       & 677         &   540      & 1320     &  4800            \\
$\mu_{Atm.}$/Hz                        & 80            &   140      &    550    & 1650         \\
%$\nu_{Atm.}$/day        &  22            &    57    &             & 2246         \\
%$\nu_{Atm.}^{cuts}$/day    & 5.2    &    3.6           &     17        & 250                \\
%$E_{med}^{\nu_{Atm.}}$/TeV    & 0.62           &    1.3       &         &        \\
Ang. res./$^{\circ}$ (10TeV)                  & 2.0           &    2.0           &         &  0.7               \\
%$A^\nu_{eff}/m^2$ after cuts     & 0.3            &     .7       &    ~7      & 1       \\
%$\Phi^{Pointsource}_{sensit.}$*         & 1.1           &     1.2          &    0.5        & 0.12       \\
\hline
\end{tabular}
 \caption{Some performance parameters for the AMANDA-II and IceCube 9-,  22- 
and  80-string detector configurations. Rates are given for cosmic ray muons at trigger level.
The rate for the 80-string array is based on simulations \cite{performance}.  
%*Livetimes: 1000days \cite{amanda-point}, 130d\cite{pretz-icrc}, 1 year for IC22 and IC80\cite{performance}, units: $10^{-7}GeV cm^{-2} s^{-1}$. 
}
\end{center}
%\normalsize
\label{table2}
\end{table}
\vskip -0.5cm

\section{Acknowledgments}
\vskip -0.3cm

We acknowledge the support from the following agencies: National Science Foundation-Office of Polar Program, National Science Foundation-Physics Division, University of Wisconsin Alumni Research Foundation; Swedish Research Council, Swedish Polar Research Secretariat, and Knut and Alice Wallenberg Foundation, Sweden; German Ministry for Education and Research, Deutsche Forschungsgemeinschaft (DFG), Germany; Fund for Scientific Research (FNRS-FWO), Flanders Institute to encourage scientific and technological research in industry (IWT), Belgian Federal Office for Scientific, Technical and Cultural affairs (OSTC); the Netherlands Organisation for Scientific Research (NWO).

%\subsection{Bibtex references}
%This is the reference to .bib file (Without .bib!)
%\bibliography{libros}

%This in the bibtex style, is ok.
%\bibliographystyle{plain}

%\end{document}

\setcounter{figure}{0}
%Class Requeried
%\documentclass[dvips]{article}
%The ICRC Style
%\usepackage{icrctc07}
\title{The combined AMANDA and IceCube Neutrino Telescope}
\shorttitle{The combined AMANDA and IceCube Neutrino Telescope}
\authors{ A.~Gross$^1$, C.~Ha$^2$, C.~Rott$^2$,
        M.~Tluczykont$^3$ , E.~Resconi$^1$,T.~DeYoung$^2$, G.~Wikstr\"om$^4$
        for the IceCube Collaboration$^5$} 
\shortauthors{A.~Gross and et al}

\afiliations{ $^1$ MPI f\"ur Kernphysik, Saupfercheckweg 1, D-69117 Heidelberg,
        Germany\\
        $^2$ Pennsylvania State University, Department of Physics, 
        University Park PA 16803, USA \\
        $^3$ DESY, Platanenallee 6, D-15738 Zeuthen, Germany\\
        $^4$ Department of physics, Stockholm University, AlbaNova, S-10691
        Stockholm, Sweden\\
        $^5$ see special section of these proceedings
}
\email{gross@mpi-hd.mpg.de}
\abstract{The IceCube Neutrino Telescope is currently under construction at the
geographic South Pole and will eventually  
instrument a volume of one cubic kilometer by 2011. It currently consists of
22 strings with 60 Digital Optical Modules each.  
Additionally the AMANDA detector has been fully integrated into IceCube
operation. This includes hardware synchronisation, combined  
triggering, common event building and a combined data analysis strategy. Monte
Carlo simulations of a combined AMANDA + IceCube  
detector will be presented. The results of the simulations were used to
implement an online filtering on data provided by the Joint
Event Builder collecting data from both detectors. Data taken
synchronously from both detectors serve for Monte Carlo verification. 
We discuss the impact of the AMANDA integration  
on the effective area, track reconstruction and event selection for the muon
neutrino detection channel.
% and the sensitivity to astrophysical neutrino sources.
In particular, we study fully and
partially contained events at low energy.  
%The identification of starting muon tracks allows for a separation of neutrino
%induced from air shower induced downgoing muons and thus for an analysis of the
%southern sky. 
An online filter marks candidates for contained events using peripheral
optical modules as a veto against atmospheric muons.  
The effective interaction volume for this filter
is presented. 
%Furthermore, contained events are especially interesting for
%atmospheric neutrino studies. The sensitivity of the combined detector 
%lies  above the energy range accessible by other atmospheric neutrino detectors.
%These events are
%especially interesting for atmospheric neutrino studies, as  
%IceCube's sensitivity lies above the range accessible by other atmospheric
%neutrino detectors.  
 }

%\setcounter{page}{1}
%
%\begin{document}
\maketitle
\section{Introduction}
In its 2007 configuration, IceCube consists of
22 strings in operation with 60 digital optical modules each. For
details on its performance see \cite{IceCube}. With the 
deployment of 13 additional strings in the 2006/07 polar summer, the detector
surrounds now its predecessor AMANDA.  
%The primary design goal of the IceCube detector was to maximize the collection
%volume for muon tracks with energies above 1-10 TeV.  
%The relatively wide spacing between strings of 125 m in IceCube is a result of
%optimizing its design to the collection volume for muon tracks with energies
%above 1-10 TeV.  
%With a spacing of typically 30-40 m, the AMANDA detector
%has a higher efficiency to lower energy muons and cascades.
Since IceCube has a wide string spacing of $125$ m, optimized for muon tracks
above a few TeV, the integration of AMANDA with its denser array adds an
important part to the low energy reach of the combined detector.

\begin{figure}[ht]
\begin{center}
\includegraphics*[width=0.45\textwidth,angle=0,clip]{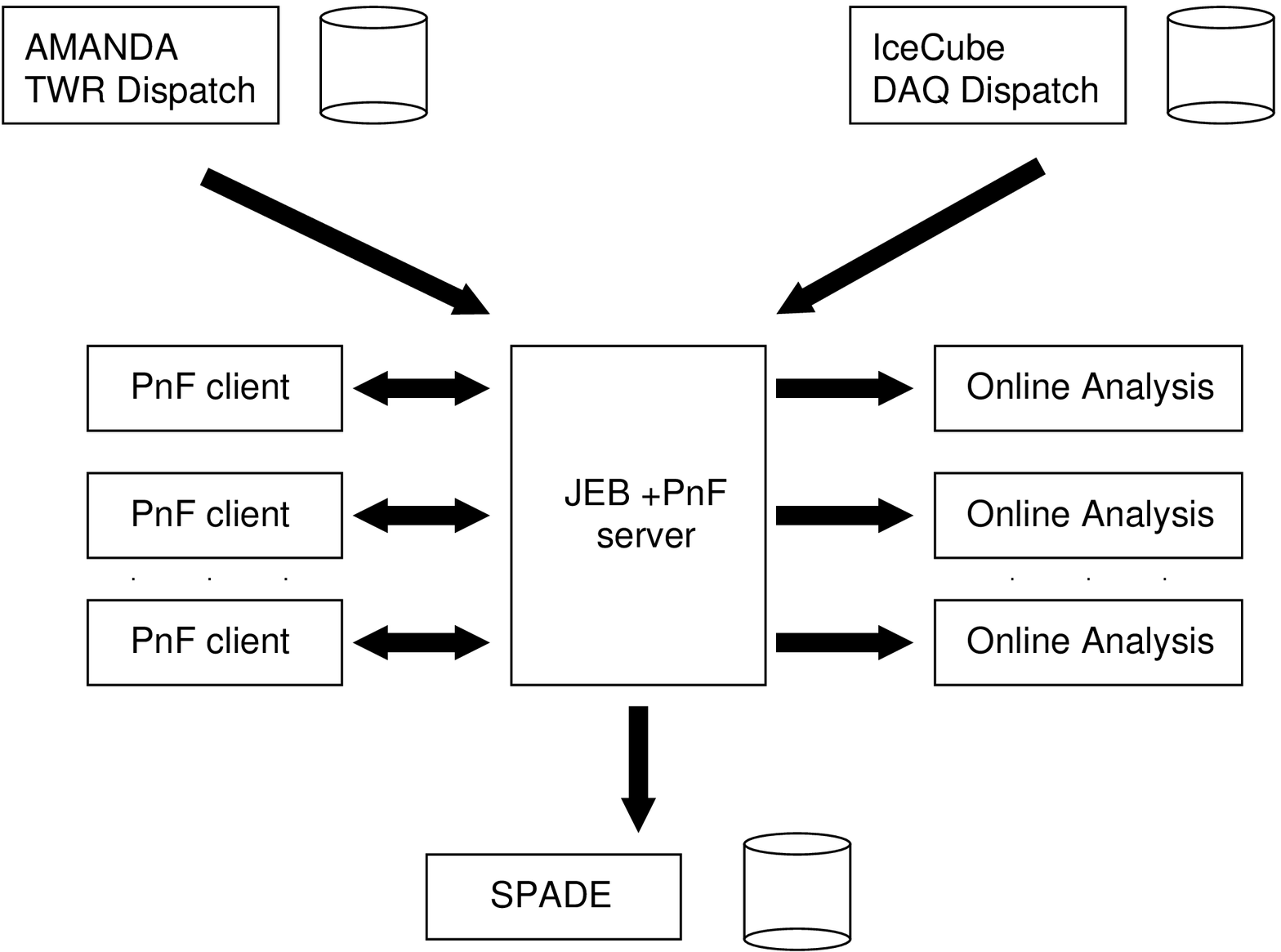}
\caption{\label{AIIJEB}Data flow in the combined AMANDA and IceCube neutrino
  telescopes using the JEB. The
processing and filtering (PnF) clients reconstruct the combined events
within a few seconds of data acquisition.  Online analysis is then
performed before the data is transferred to the SPADE system for satellite
transmission and tape archiving.}
\end{center}
\end{figure}

%The implementation of the the FADC based TWRDaq to the AMANDA
%detector \cite{TWR} in the years 2003-2005 allowed for a lower trigger 
%threshold 
%of $13$ instead of $24$ hit modules 
%and significantly improved its
%capabilities at low energies. 
The implementation of a new DAQ system to the AMANDA
detector \cite{TWR} in the years 2003-2005 allowed for a reduction of the
multiplicity trigger threshold. By this the energy threshold of AMANDA
has been lowered below 50 GeV. Hence it is capable to complement IceCube at
low energies and 
consequently, the AMANDA
detector has been fully integrated into the IceCube detector. 
This includes a
common run control,
triggering, event building and online filtering.  
Every time the AMANDA
detector is triggered, a readout request is sent to the IceCube
detector. Since the energy threshold of AMANDA is lower, no triggering
requests from IceCube to AMANDA are needed. As shown in Fig.~\ref{AIIJEB}, the
Joint Event Builder  (JEB) receives 
data from both detectors, merges events on a time coincidence base and
provides the data to the online filtering. The online filtering selects events
of interest for physics analyses and transfers the selected data
to the Northern Hemisphere. With this filtering the relevant physics data can
be quickly analyzed despite the constraints of  limited satellite bandwidth
available for data transfer from the South Pole.   

Monte Carlo (MC) studies of the performance of the combined detector 
in muon neutrino channel are
presented in this paper. 
%, for
%the cascade channel see \cite{cascade_integrated}.
The combined detector provides an improved performance at low
 energies: the IceCube strings directly adjacent to AMANDA enlarge the densely
instrumented region, provide a longer lever arm
and thus improve the angular resolution.
This reduces the background for low
  energy neutrinos from point-like sources compared to previous AMANDA
  analyses \cite{5yr_PS}. 
%The dense instrumentation provides reconstruction algorithms
%with sufficient information also for low energy events. 
Since AMANDA is now
completely embedded into the IceCube array, the identification of
starting and contained tracks becomes possible using IceCube as a veto. The
identification of contained events allows a better measurement of the energy.
Additionally, with this
technique, the 
rejection of down-going atmospheric muons is possible and thus,
the detector is sensitive to sources in the southern sky.
Furthermore, analyses for different neutrino flavors
will use the combined detector as well to improve the low-energy performance.
\section{Low energy physics with the combined detector}
With its enhanced performance at low energies the combined detector will
  have an improved sensitivity to WIMPs (see \cite{WIMP_ICRC}) and sources with
  steep energy 
  spectra or cut-offs below 10 TeV like the Crab nebula \cite{HESS_Crab}. 
  In  particular, the search for time-variable sources will profit from this
  enhancement since 
  their localization in space and time significantly reduces the number of
  background events. An example for such a source is LS I+61 303 emitting
  TeV photons periodically with a power law index of -2.6 \cite{MAGIC_LSI}.
  Another region of high interest is the Galactic Center which contains a
  TeV gamma-ray source \cite{HESS}. 
  As it lies in the southern sky it
  was not accessible for AMANDA up to now.
  But also the analysis of atmospheric neutrinos will benefit and might
  even allow the detection of neutrino oscillation effects in the energy
  range 10 -- 100 GeV and test for non-standard oscillation scenarios.

%The physics goals of the AMANDA integration are the improvement of the
%sensitivity to WIMPs, the
%analysis of atmospheric neutrinos and point sources with a steep spectrum or a
%cut-off 
%energy below 10 TeV. 
%A detection of neutrino oscillation effects in the energy range 10 GeV - 100
%GeV would provide a check of
%Lorentz invariance and allow to check non-standard oscillation scenarios. 
%For point sources, many Galactic candidate sources
%including the Crab Nebula show a spectrum steeper than $E^{-2}$ and a
%high-energy cut-off \cite{HESS_Crab}. 

%Time-variable sources are of particular interest for low energy
%point-source analysis, since the localisation in space and time reduces
%significantly the background and thus compensate the increased flux of
%atmospheric neutrinos at these energies. An example for a low energy time
%variable source is LS +61 303, emitting TeV photons periodically witha power
%law spectral index of $-2.6$ \cite{MAGIC_LSI}

% A region of special
%interest accesible with the combined detectors is the galactic center,
%including the H.E.S.S.\ source associated 
%with the central black hole of the Milky Way \cite{HESS}. As most of the
%energy of contained events is deposited  
%inside the array, their energy can be reliably reconstructed. 

%This white paper examines the physics topics that are more effectively
%addressed with augmented sensitivity to low energy (> 1 TeV) events. 

\section{Online filtering and data analysis}
Two filtering strategies make use of the combined detectors.
The first strategy aims for an improved performance for up-going muon tracks,
by adding a low energy online filter for combined
data to the standard IceCube filter for up-going muons. 
Additionally, a filter using the veto strategy identifies events contained in
 the AMANDA
array and opens a sensitivity window to the southern sky. 
%Furthermore the analysis of  
%starting tracks is not limited to muon neutrinos from the northern
%hemisphere.
%The background rejection and the effective area of the contained event filter
%using the veto method specific to the combined detectors 
%are discussed. 
%The application of veto techniques to improve the low energy performance of
%IceCube only addresses the same physics. 
In addition to the integration of AMANDA, the implementation of a string
trigger improves the detection of vertical low energy tracks with IceCube.

{\bf The up-going muon filter} \\ [1mm]
The low energy up-going muon track filter uses all hits from both detectors to
reach a decision. It is complementary to an up-going muon
track filter defined on IceCube hits only. The JAMS
reconstruction\footnote{JAMS is based on a cluster search in the abstract
  space spanned by the distance of the hit to the track and the time residual. The time residual is the difference of the measured hit time
  and the
  passing time of the Cherenkov cone for an assumed track. }
was chosen for the low energy filter. Events with a reconstructed zenith
angle larger than $75^\circ$ are selected. The combination with the IceCube 
only filter allows to constrain the use of this relatively slow algorithm to
events with hits 
in the AMANDA detector not passing the IceCube filter and having less than
$20$ hits in IceCube. For events with more hits, the additional information
from AMANDA does not result in a significantly better filtering efficiency.

\begin{figure}[ht]
\begin{center}
\includegraphics*[width=0.4\textwidth,angle=0,clip]{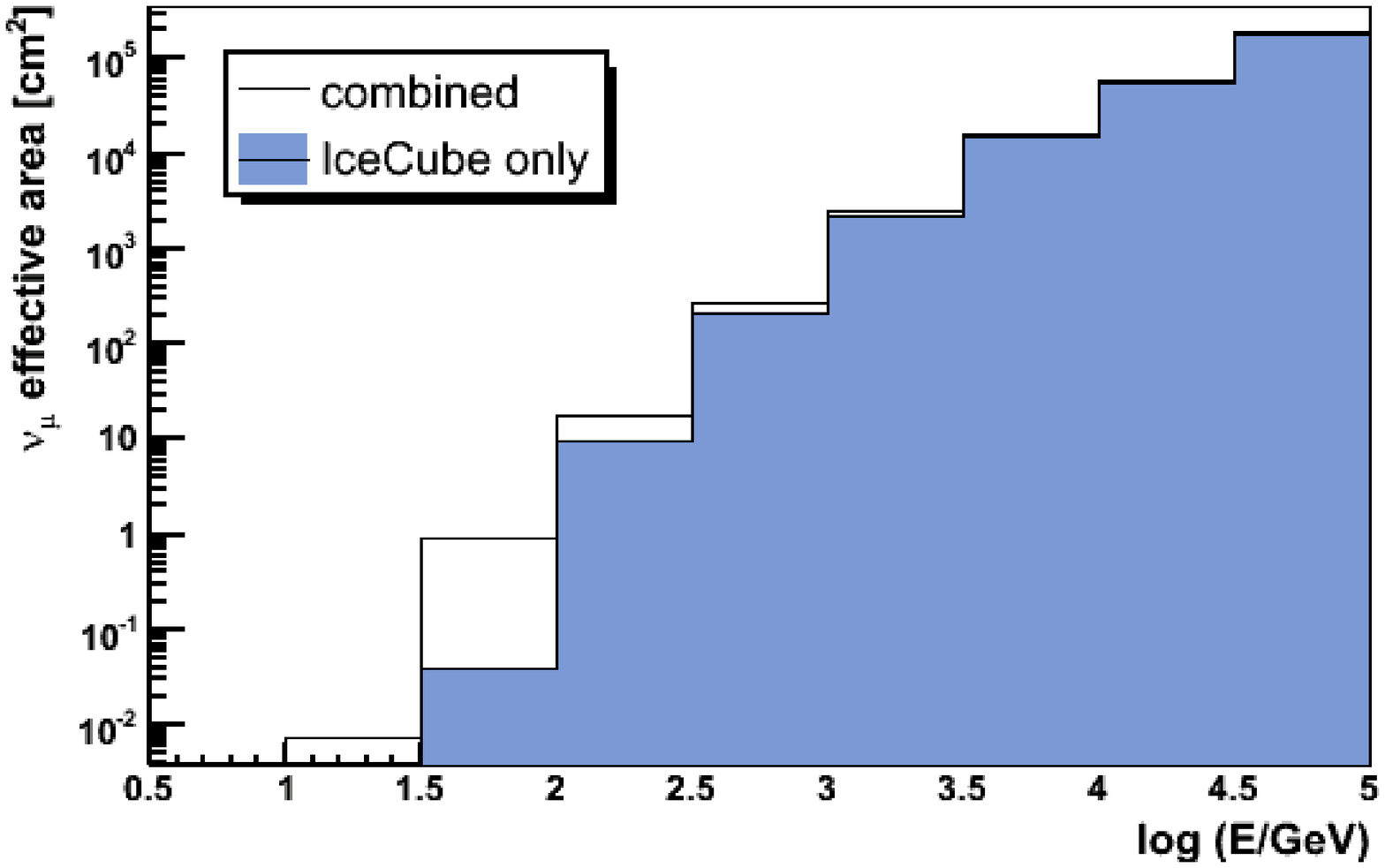}
\caption{\label {eff_area_upmu_L1}Effective area of the combined detectors  in
  comparison to IceCube only at
  online filter level.}
\end{center}
\end{figure}

\begin{figure}[ht]
\begin{center}
\includegraphics*[width=0.4\textwidth,angle=0,clip]{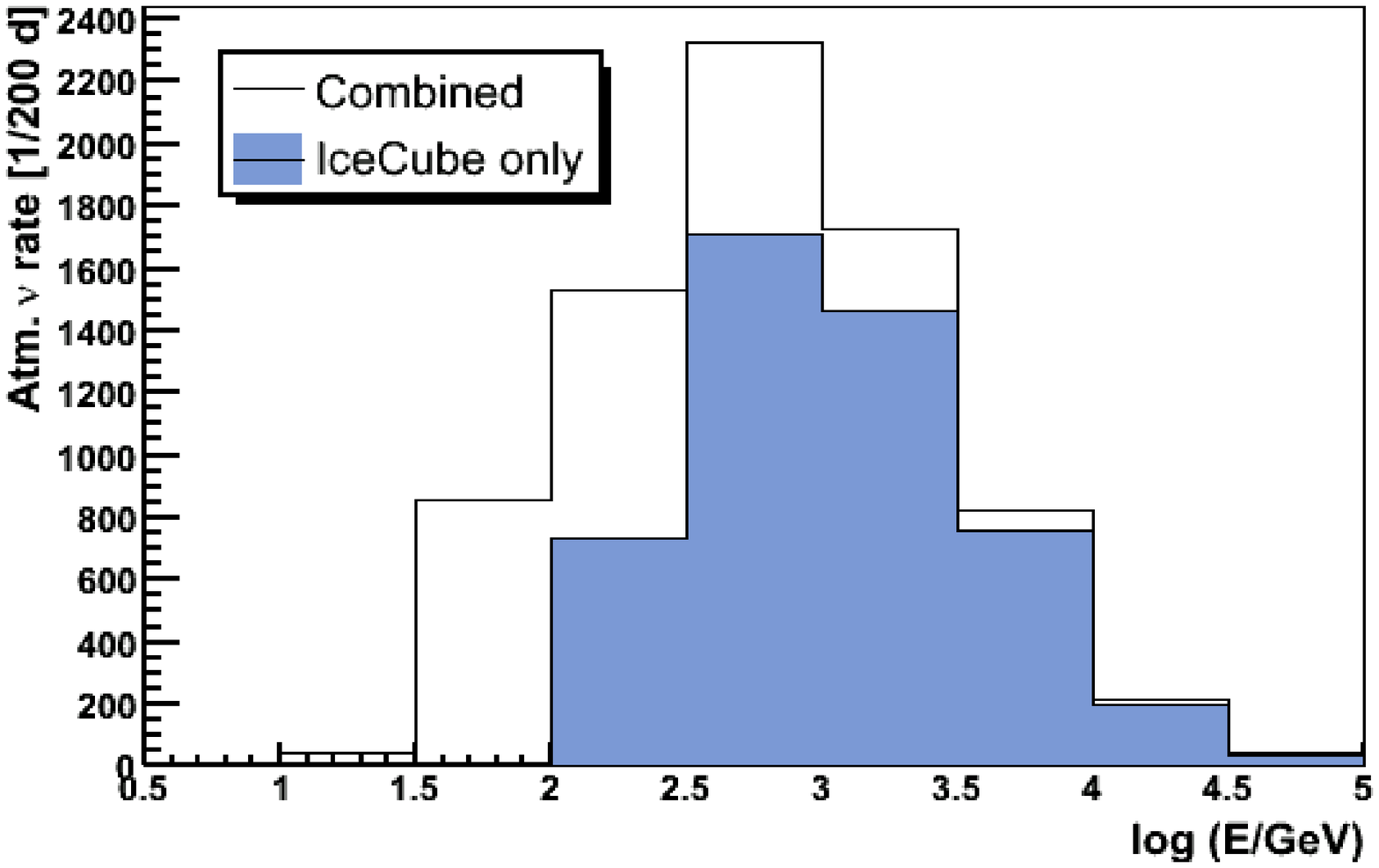}
\caption{\label {atm_rate_upmu_L1}Atmospheric neutrino rate at online filter
  level for a generic run period of 200 days.}
\end{center}
\end{figure}

The effective area for muon neutrinos of the combined detectors using the
combined online filter is shown in 
Fig.~\ref{eff_area_upmu_L1} in comparison to the IceCube only filter. 
Figure~\ref{atm_rate_upmu_L1} shows the resulting
expected rate of atmospheric neutrinos \cite{bartol}. It is worth noting that
the combined 
detector detects atmospheric neutrinos over four orders of magnitude in energy
between 10 GeV and 100 TeV.

The neutrino signal efficiency of the combined filter is above $90\%$ over the
wide energy range from 10 GeV to 100 PeV. The rejection of the
atmospheric muon background is above $95\%$, where 
less than $0.5\%$ of all events 
%(i.e.\ $1.5\%$ of events evaluated by this
%filter) 
are passing the JAMS filter on combined events. That demonstrates that the
background  
of atmospheric muons is not significantly increased by the AMANDA integration. 

\begin{figure}[ht]
\begin{center}
\includegraphics*[width=0.4\textwidth,angle=0,clip]{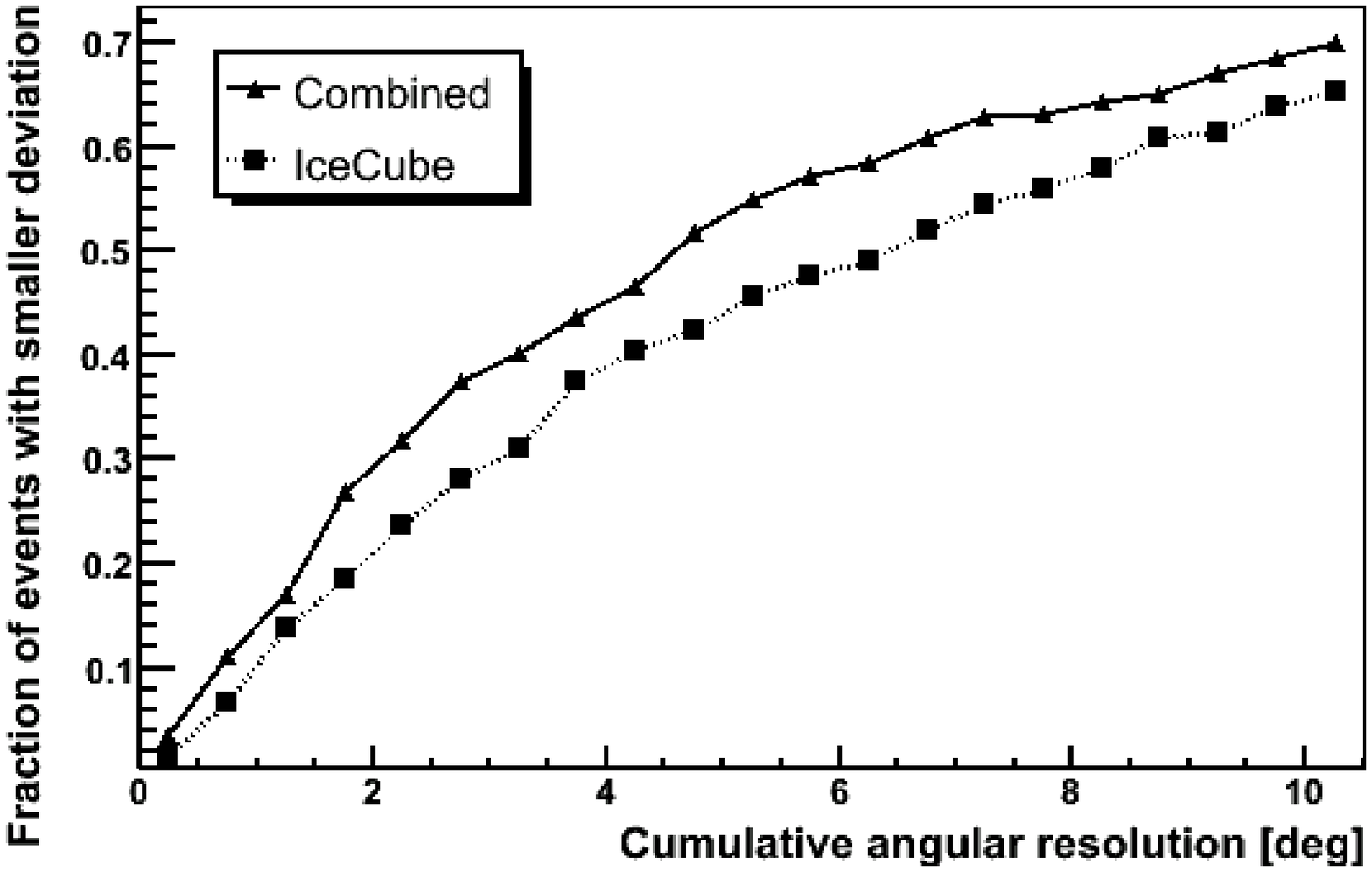}
\caption{\label {AIIangres}Cumulative distribution of the
  angle between simulated and reconstructed track for the
  combined detectors and IceCube only.}
\end{center}
\end{figure}

A first study of the angular resolution in the low energy regime ($E<10$ TeV)
was conducted. For this study, events triggering both detectors separately have
been selected and a full likelihood reconstruction \cite{recopaper} has been
applied. As
shown in Fig.~\ref{AIIangres},
a slight improvement was found.

\begin{figure}[ht]
\begin{center}
\includegraphics*[width=0.4\textwidth,angle=0,clip]{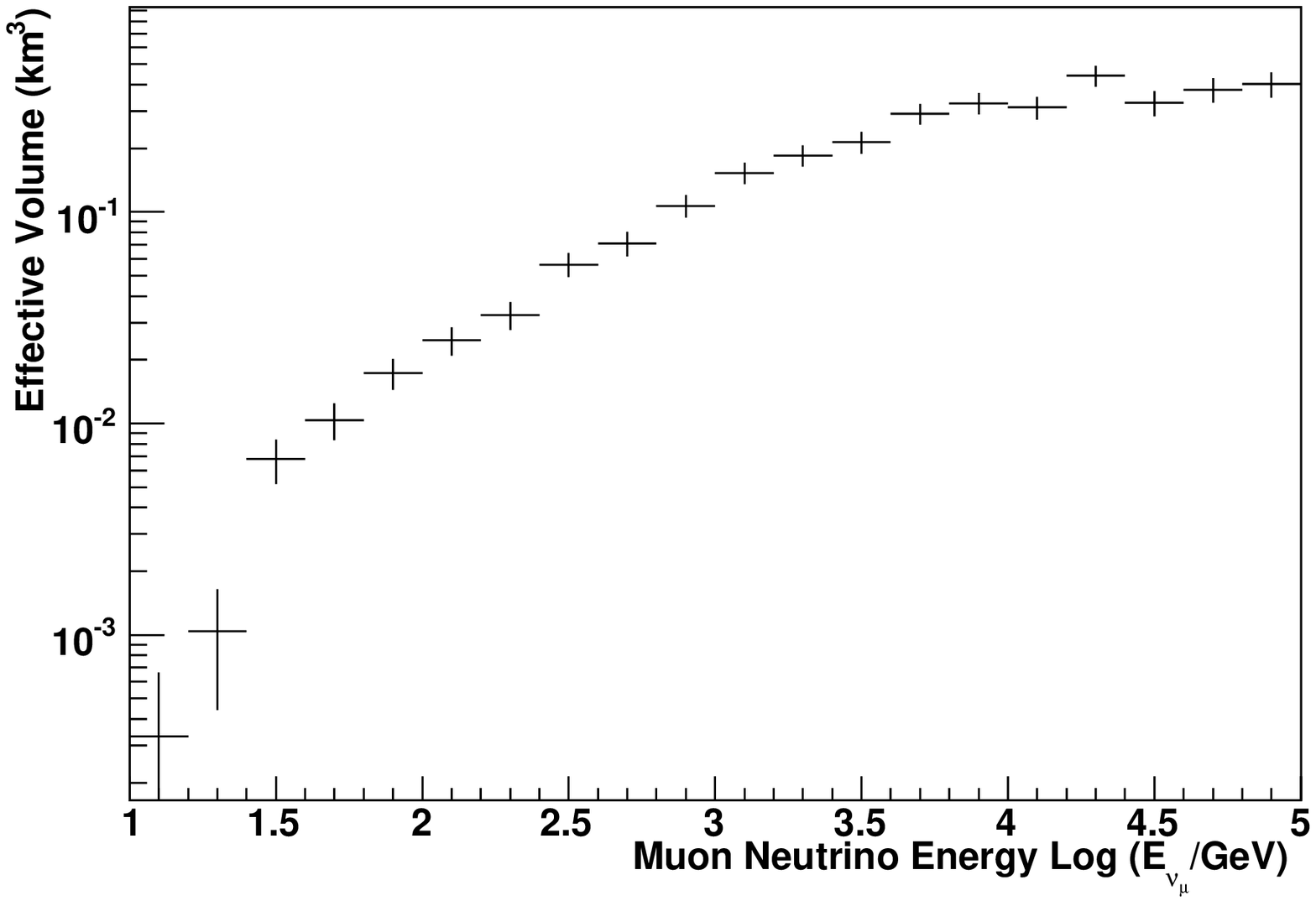}
\caption{\label{eff_volume_contained} Effective interaction volume of the
  contained event filter.}
\end{center}
\end{figure}

{\bf A filter for low-energy contained events} \\[1mm]
As IceCube surrounds the AMANDA detector its outer strings and top-layers
can be used to veto through-going tracks and especially study low
energy (100~GeV -- 1~TeV) fully or partially contained tracks with $4\pi$
sensitivity. Figure \ref{eff_volume_contained} shows
the effective 
volume for these events at filter level. 
%In this analysis we use the
The combined AMANDA-IceCube detector is used to reconstruct tracks and point them back
to their origin. Reconstructed tracks that deposite no light in one or more
peripheral strings despite of a high probability to do so assuming a through-going
track, are more likely to be due to muon neutrino
interactions rather than atmospheric muon background.
Furthermore, the charged current interaction of the muon neutrino in the
detector produces a cascade with a track attached to it. This topologically
differs from a through-going muon track and can be studied in the recorded
waveforms and leading edge times. A dedicated reconstruction algorithm is
currently under development.

%\begin{figure}[ht]
%\begin{center}
%\includegraphics*[width=0.35\textwidth,angle=0,clip]{fig_c1.eps}
%\caption{\label{fig_c1} Comparison of the trigger efficiency for muon
%neutrinos that produce at least one hit in the detector for IceCube's
%Multiplicity 8 trigger and string trigger.}
%\end{center}
%\end{figure}

\begin{figure}[ht]
\begin{center}
\includegraphics*[width=0.4\textwidth,angle=0,clip]{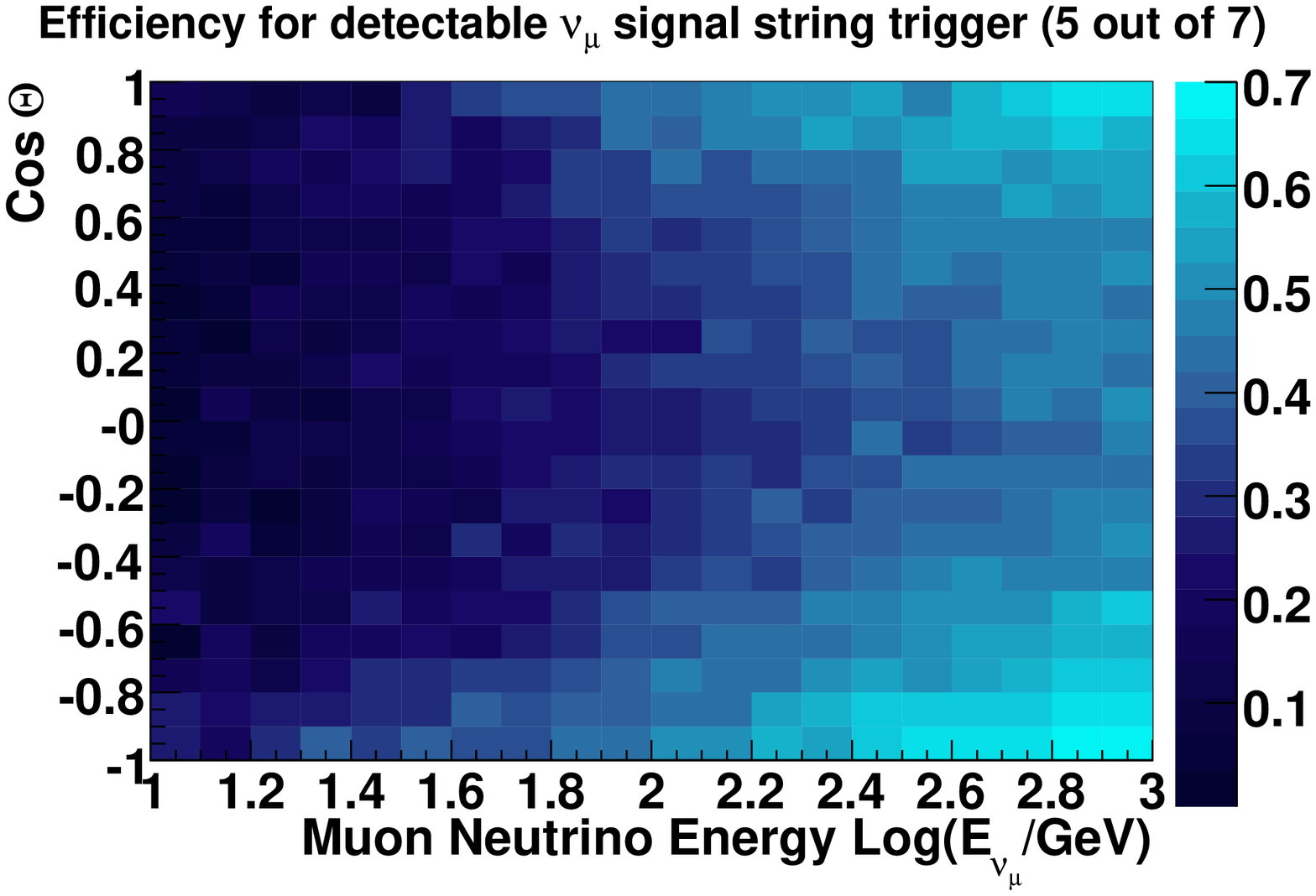}
\caption{\label{fig_c2} String trigger efficiency for muon
neutrinos that produce at least one hit in the detector as a function of muon
neutrino energy and zenith angle.}
\end{center}
\end{figure}

{\bf A string trigger for vertical low-energy events in IceCube} \\[1mm]
We are currently implementing a new string trigger for IceCube that requires
5 DOMs to be hit out of a sequence of 7 DOMs on a single string. 
%Furthermore it
%is required that these hits do not occur near the top of the string to
%reduce the trigger rate on down-going muons. 
The upper most part of the string isd excluded to
reduce the trigger rate on down-going muons. 
%Figure \ref{fig_c1} compares
%the trigger efficiency for muon neutrinos that produce at least one hit in
%the detector to IceCubes Multiplicity 8 trigger and shows that it
%dramatically increases IceCubes low energy reach. 
In comparison to the standard IceCube trigger requiring 8 DOMs to be hit, for
energies below 100 GeV an
improvement by more than a factor of 10 is obtained.
Figure \ref{fig_c2} shows
the string trigger efficiency as a function of the muon neutrino energy and
zenith angle. The good performance for vertical tracks allows to compare the
fluxes of up- and down-going
 atmospheric neutrinos and the analysis of WIMP annihilations at
the center of the Earth.

%Given IceCube's geometry, it can be seen that especially
%vertical events are collected, which can be used to search for earth WIMPs
%or study low energy atmospheric neutrinos at an energy range of 10-100~GeV.
%In the later analysis we compare the rate of vertically up-going with
%vertically down-going neutrinos as a function of their energy. Some of the
%systematic uncertainties cancel out in this way. 
%Vertical events can be more
%reliably distinguished from background events as the muon neutrino events at
%the energy range studied are fully or partially contained and hence have a
%starting point inside the detector.

\section{Verification of MC simulations}
\begin{figure}[ht]
\begin{center}
\includegraphics*[width=0.35\textwidth,angle=0,clip]{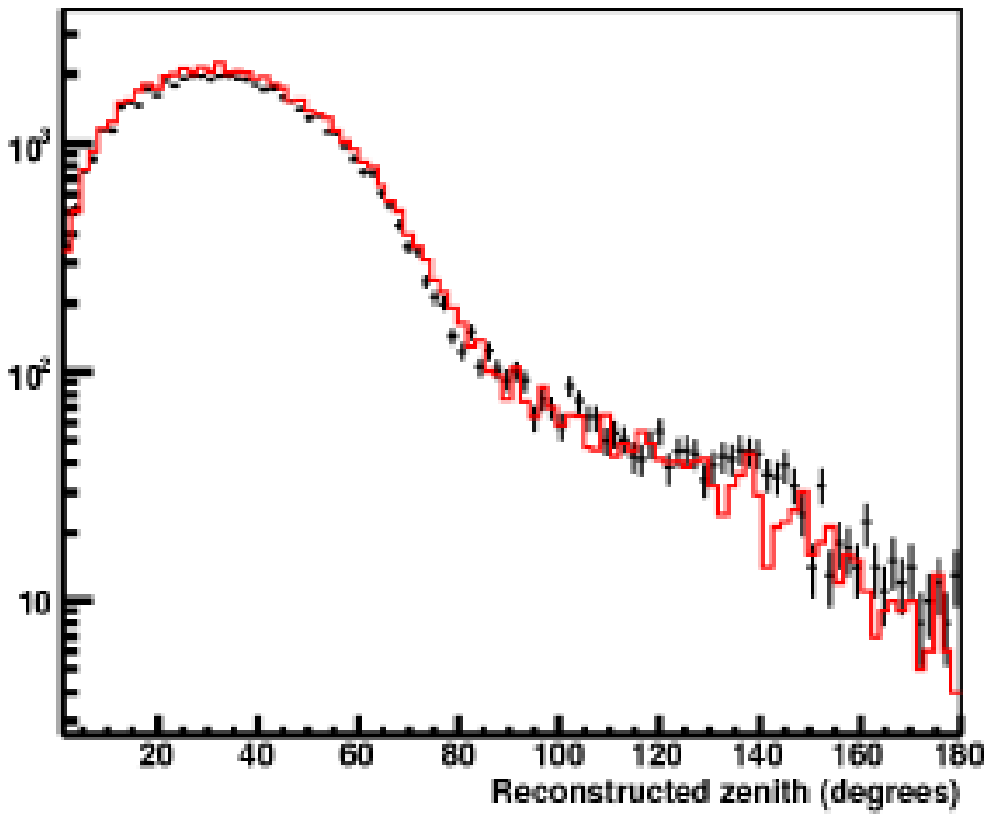}
\caption{\label{simu} 
AMANDA-IceCube (9 strings) JAMS zenith spectrum from integrated test run 2006.} 
\end{center}
\end{figure}
In order to check the viability of the MC simulation for the combined
detector, we have compared the distributions of various quantities between
data and simulation. The data for this comparison has been acquired in a
special integrated mode test run in 2006. As an example Fig.~\ref{simu}
shows the 
comparison of the reconstructed zenith angle spectrum for data and
MC. Other distributions, including that of the trigger rate and the number
of hit channel were also found to 
be in good agreement.

\section{Conclusions}
According to the preliminary results presented here, the combined IceCube and
AMANDA detector in its current configuration provides 
a significantly improved performance in the low energy regime.
The effective area for 
up-going muon neutrinos and the effective interaction volume for contained
down-going events at online filter level provide improved  possibilities to
investigate  atmospheric neutrinos as well as possible
astrophysical sources emitting neutrinos with energies below 10 TeV.
For the first time, the Galactic Center can be examined with a 
neutrino telescope on the Southern Hemisphere. 

\setcounter{figure}{0}
%%
% International Cosmic Ray Conference 2007 Merida Yucatan Mexico
% In this file you will find detailed instructions to correctly
% typeset your document.
%
% By: Victor De la Luz
% vdelaluz@inaoep.mx
% Mexico City

%Class Required
%\documentclass{article}
%The ICRC Style
%(This package is the last package in the usepackage list)
%If you need import other package you need write it first.
%\usepackage{icrctc07}

%The paper title
\title{Performance of the IceTop Array}
%Short title to print in the headers to the final publication (Not showed in this print).
\shorttitle{IceTop}

%All paper authors
\authors{Thomas Gaisser
  for the IceCube Collaboration(*)} 
%Short title to print in the headers to the final publication (Not shown in this print).
\shortauthors{T. Gaisser et al.}
%All the affiliations.
\afiliations{Bartol Research Institute, Department of Physics and Astronomy,
 University of Delaware, Newark, DE 19716, U.S.A.}
\email{gaisser@bartol.udel.edu~(*)see collaboration lists at end of these Proceedings}

%The abstract.
\abstract{We present an overview of the
status of IceTop, which now consists of 52 tanks
at 26 stations above the 22 deep strings of
IceCube.  Six months of good data were taken
with the previous 16 station-9 string version
of IceCube during 2006.}
%
%\begin{document}
\maketitle

%\begin{linenumbers}
\section{Introduction}
\begin{figure}
\vspace{-.5cm}
\begin{center}
\noindent
\includegraphics[width=0.5\textwidth]{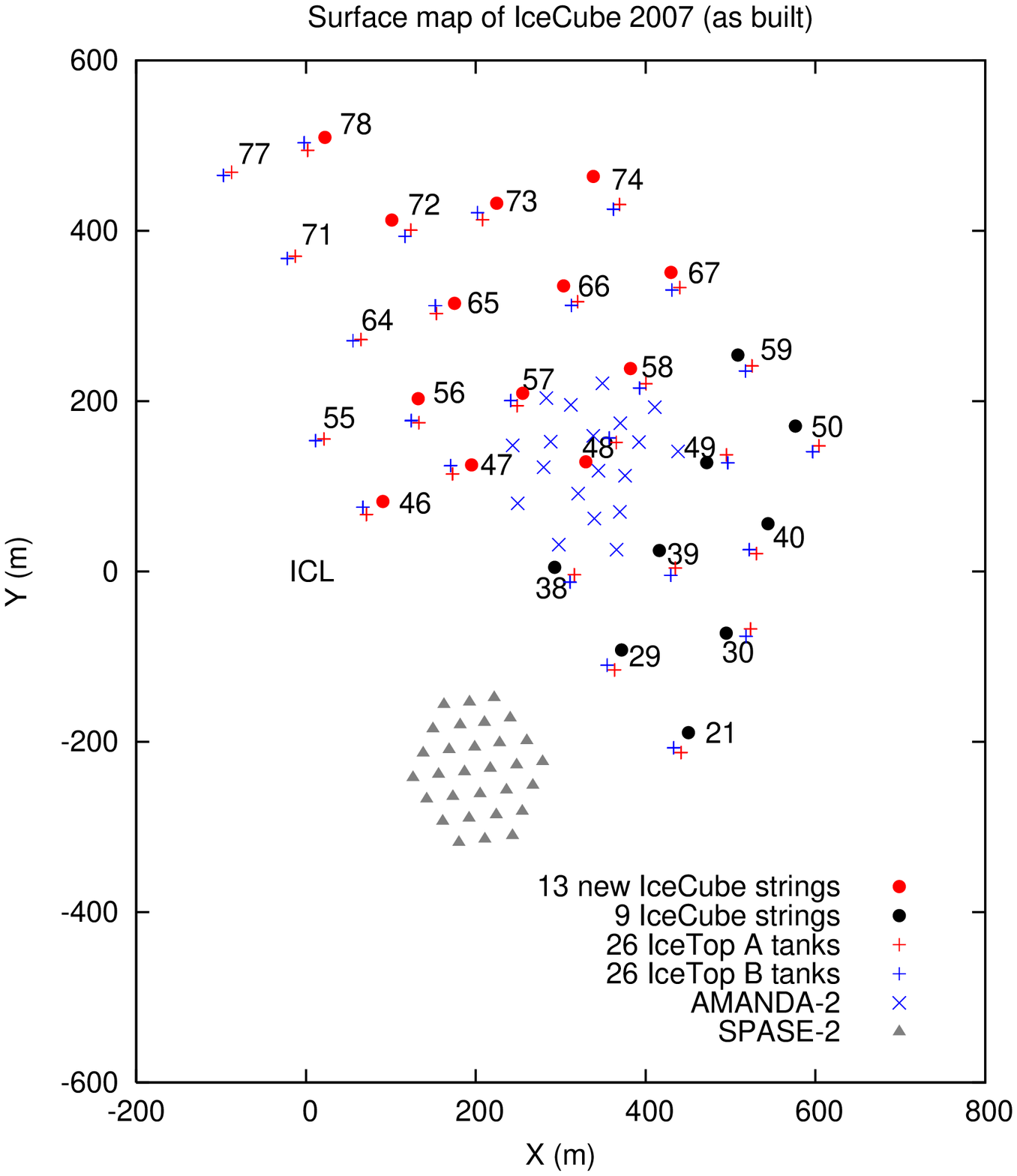}
\end{center}
\caption{Surface map of IceCube/IceTop in 2007.  When completed the array will
be symmetric around the IceCube Laboratory (ICL).
}
\label{ICRC0758_fig1}
\end{figure}

During 2006, IceCube ran with 
sixteen IceTop stations and nine IceCube strings. 
Ten more stations and thirteen more strings were 
deployed in the 2006-2007 austral summer, as shown in
Fig~\ref{ICRC0758_fig1}.  When complete, there will be 80 surface stations
and a similar number of deep strings in IceCube.

The IceTop air shower array consists of pairs of tanks (A and B) separated from
each other by 10 m.  Each IceTop station with its pair of tanks is associated
with an IceCube string.  Each tank is instrumented with two standard IceCube
digital optical modules (DOMs) operating at different gains to extend
the dynamic range of the tank.  This configuration has several advantages:
\begin{itemize}
\item Local coincidence between two tanks at a station is used to
select potential air shower signals from the high (typically 2 kHz) event
rate generated in each tank by uncorrelated photons, electrons and muons.
\item Comparison of signals seen by two DOMs within a tank can
be used to demonstrate that fluctuations in tank response are much
smaller than intrinsic fluctuations in air showers as measured by
comparing signals from the same shower in the two tanks at a 
station~\cite{performance05}.
\item Two identical sub-arrays (A-tanks and B-tanks) can be used
to measure shower front curvature, lateral distribution, timing and density fluctuations,
core location accuracy, angular resolution and other properties of showers.
\item By selecting coincident events in which an event in deep IceCube is
accompanied by exactly one hit station in the inner part of IceTop,
we can identify and tag a set of events that consist almost entirely of
single muons in the deep detector (as compared to the multi-muon events
typical of showers big enough to trigger several stations of IceTop).  Such
events are useful for calibration.
\item With 52 tanks we already have a total detector area of $140$~m$^2$, which will
grow to over $400$~m$^2$ when the detector is complete.  The monitoring
stream includes scalar rates of IceTop DOMs that can be used to observe
solar and heliospheric cosmic-ray activity.
\end{itemize}  
\section{Calibration of IceTop DOMs}
IceTop DOMs are calibrated and monitored with the continuous flux of
muons through the tanks~\cite{LeventICRC}.  Through-going muons give a
broad peak in the distribution of signals from the inclusive
flux of all particles that hit the tank.  For a vertical muon
the signal corresponds to a track length of 90 cm in ice.
The peak is calibrated with a muon telescope and with simulations.
Air shower signals are then expressed in terms of vertical equivalent
muons (VEMs) by comparing the integrated charge of the signal
to that of a vertical, through-going muon.  Regular calibration runs
provide monitoring information and a data base of calibration constants,
which is updated weekly.  The first 8 tanks deployed in December 2004
provide a 2.5 year timeline for studying stability of the response,
which generally varies slowly within a range of $\pm 5$\%.  In half the
cases (8/16) DOMs showed a sudden decrease in response ranging from 10\%
in two cases to 33\% in one.  The shifts occurred in mid-winter of 2006,
which was the first season that the tanks experienced operation
at the ambient winter temperature.  (In 2005 winter the freeze-control units
were still in operation.)

\section{Air showers in IceTop}
\begin{figure}[t]
    \includegraphics[width=0.5\textwidth]{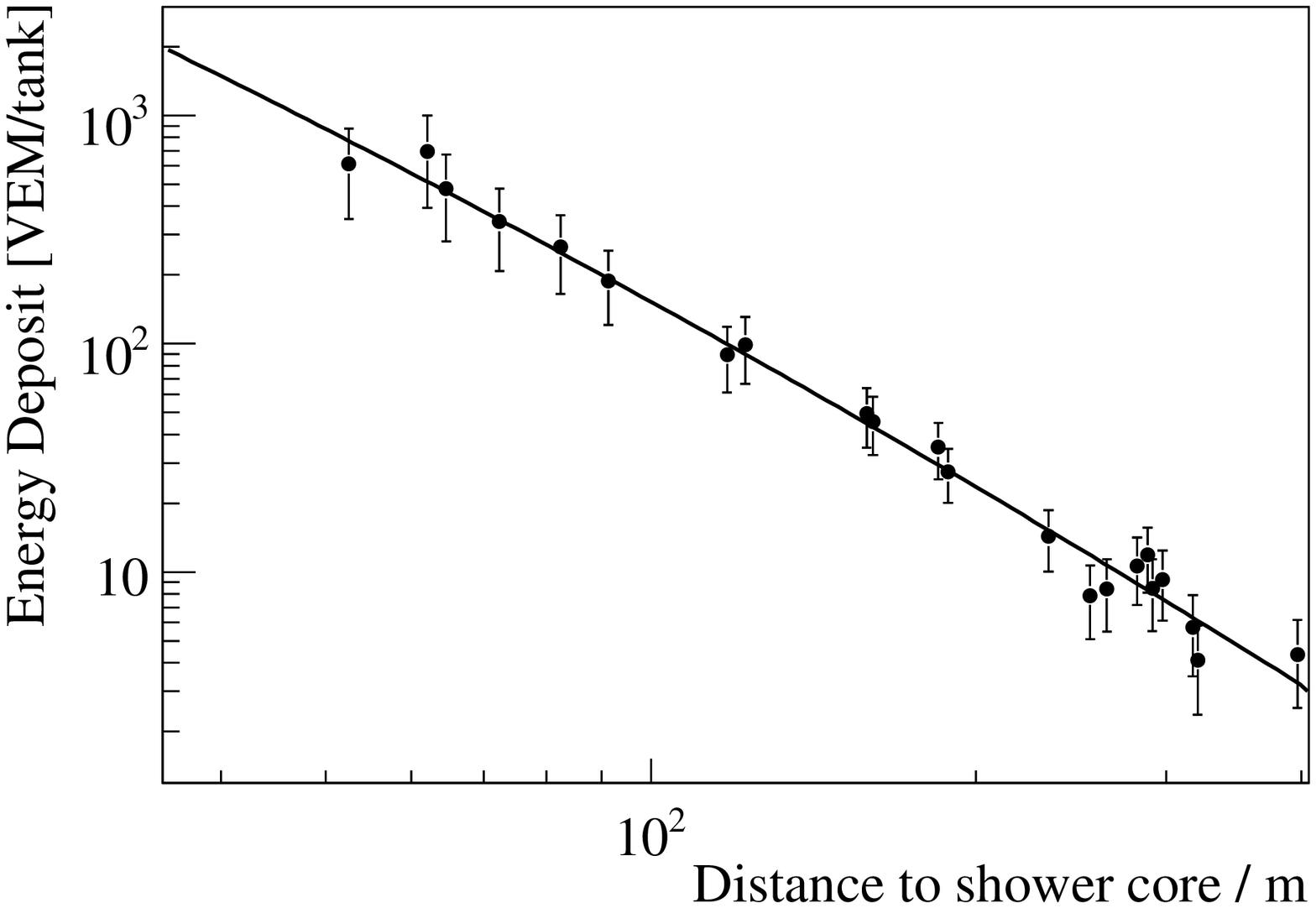}
    \caption{Signal vs distance from shower core with fitted lateral distribution
for an event with an estimated energy of $100$~PeV.}
    \label{ICRC0758_fig2}
\end{figure}

  With a spacing between stations of approximately $125$~m and a 
surface area per tank of $2.7$~m$^2$, the effective threshold
for IceTop is about $500$~TeV for a trigger requirement of five 
or more stations,~\cite{KlepserICRC}  
somewhat higher than the nominal threshold of $300$~TeV for showers
near the vertical that hit four or more stations.  (Here 
``effective threshold" is defined as the energy
above which the previously measured cosmic-ray flux through a defined area-solid angle
inside the array equals the observed rate of events.)
Figure~\ref{ICRC0758_fig2} shows an example of the lateral
distribution of signals in a large shower in units of VEMs.  
The line is the fitted lateral
distribution of energy deposition, which has a shape different from the
standard NKG function.~\cite{KlepserICRC}
The NKG function is appropriate for a scintillator array that is
relatively insensitive to the photonic part of the signal ($\gamma\rightarrow e^+ + e^-$).
Conversion of photons inside the tanks makes an important contribution to signals.

A convenient measure of primary cosmic-ray energy for showers observed in
IceTop is the fitted signal density in VEMs at $100$~m from the shower core ($S_{100}$).  
The mean energy for $S_{100} = 20$ is approximately 10 PeV for showers
with zenith angle less than $30^o$ and $\sim 100$ PeV for $S_{100} = 200$.
A functional relation for $S_{100}$ as a function of primary cosmic-ray energy 
and zenith angle for
protons is given in Ref.~\cite{KlepserICRC}, which
shows a preliminary energy spectrum extending from
$1$ to $100$ TeV based on this relation.
Because of fluctuations on the steep cosmic-ray spectrum, 
the mean primary energy for a measured
$S_{100}$ is smaller than the energy which gives the same average
$S_{100}$.  There are also systematic differences in the relation
for different primary masses (up to $25$~\% for Fe).  
The full energy spectrum analysis will
require an unfolding procedure to account for fluctuations.

The same events can be reconstructed independently by the sub-array
of A tanks and that of B tanks.  From the comparison one can obtain
an experimental measure of the accuracy of reconstruction.  Such a 
sub-array analysis indicates that core location can be determined to 
an accuracy of $13$~m and the reconstructed direction to about $2^\circ$. 

\section{Primary composition from coincident events}

An important physics goal 
is to use the downward moving events observed in coincidence by
IceTop and the deep IceCube 
detectors to study primary composition in the knee region and above.
The idea is to measure the distribution of energy deposition by muons in
the deep detector as a function of primary cosmic-ray energy
and hence to measure the fraction of heavy nuclei, which produce more
muons.
Previous studies of this type have been done by SPASE2-AMANDA-B10~\cite{SPAM}
and by EASTOP-MACRO~\cite{Aglietta} in the knee region.
Status of this analysis with 2006 IceCube data is presented in
Ref.~\cite{Song}.  

The full IceCube detector can cover the energy
range from $<10^{15}$~eV below the knee to $10^{18}$~eV.
Showers generated by primary cosmic rays in this
energy range produce multiple muons
with energy sufficient to reach the depth of
IceCube.  For primary energy of $10^{15}$~eV, for example,
proton-induced showers near the vertical
produce on average about 10 muons with $E_\mu\,>\,500$~GeV
and iron nuclei about $20$.  For higher primary energies,
the number of muons increases, and 
the multiplicity in showers generated by nuclei 
approaches asymptotically a factor of $A^{0.34}$ times
the muon multiplicity of a proton shower,
or $\approx 2.7$ for $A=56$.  

As a consequence of the high
altitude of IceTop, showers are observed near
maximum so the detector has good energy resolution,
which is important when the goal is to measure
changes in composition as a function of energy.  In
some currently favored models~\cite{Berezinsky,Volk}
the transition from galactic to extra-galactic cosmic rays
occurs in the decade between $10^{17}$ and $10^{18}$~eV.
In the model of Ref.~\cite{Berezinsky} the transition would
be characterized by a transition from heavy nuclei at the 
end of the galactic population to nearly all protons at
higher energy as the extra-galactic population dominates.
The details of the transition may in principle give
information about the cosmology of the extragalactic cosmic
ray sources if the change in composition can be measured with
sufficient precision and energy resolution~\cite{Allard}.

\section{Calibration of IceCube with IceTop}

Events reconstructed by IceTop that are also seen
in the deep strings can be used in a straightforward
way to calibrate event reconstruction in IceCube.
One can, for example, compare the directions reconstructed
by IceTop with the direction of the muon core reconstructed
by one of the algorithms used for muon reconstruction in
the neutrino telescope.  Examples of verification of timing
and direction with IceTop are given in Ref.~\cite{performance05}.
As noted above, however, showers that trigger IceTop normally
produce bundles of several (at 1 PeV) or many muons in the
deep detectors.

\begin{figure}[t]
    \includegraphics[width=0.4\textwidth]{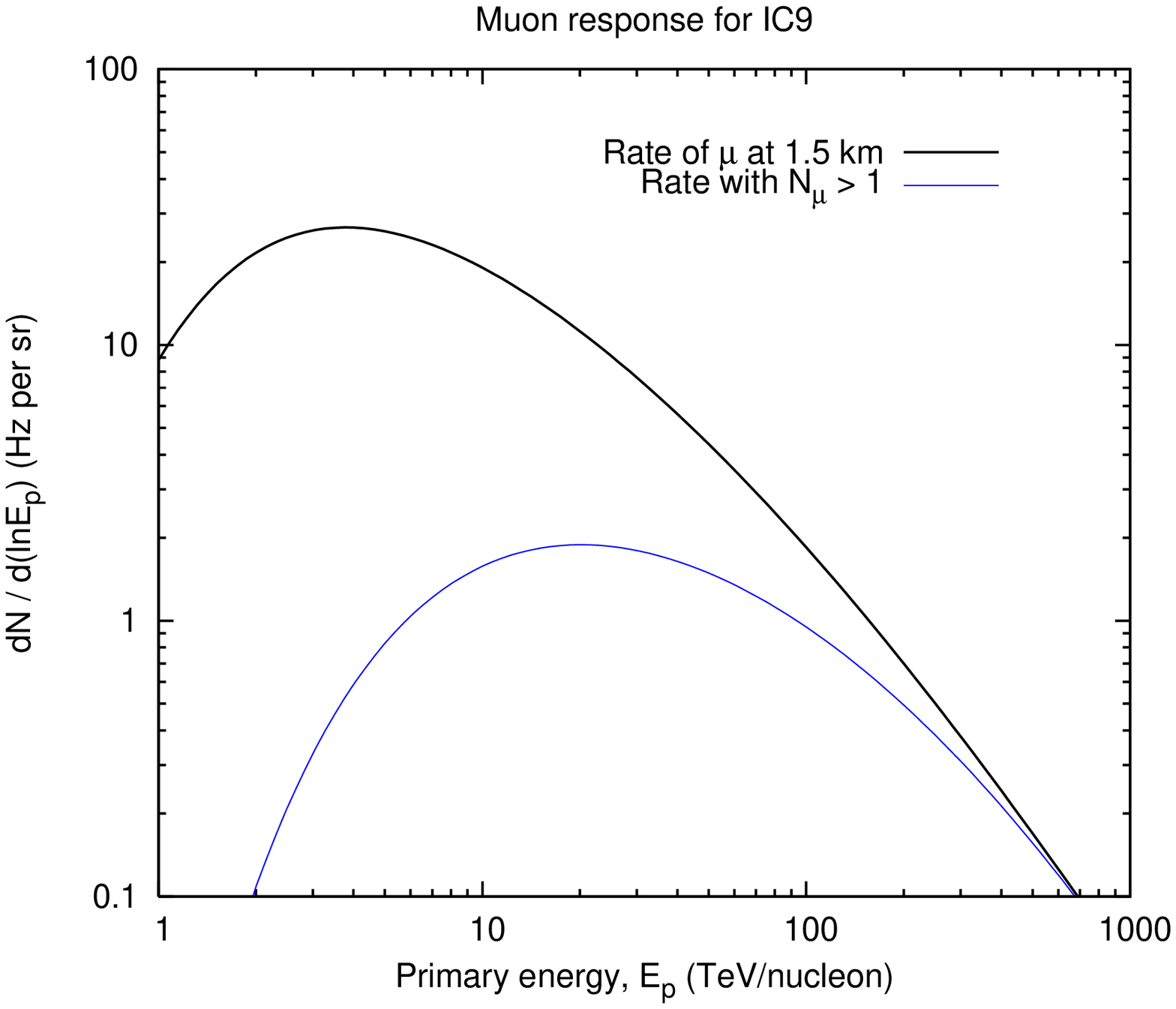}
    \caption{Distribution of primary cosmic-ray nucleons
that give rise to muons near the vertical at $1.5$~km in IceCube
(estimated from Ref.~\cite{crpp}).  The lower curve shows the
contribution of events with more than one muon entering the deep detector.
}
    \label{ICRC0758_fig3}
\end{figure}

In contrast, much of the atmospheric muon background in deep
IceCube consists of single muons, as does the target population
of neutrino-induced muons.
Figure~\ref{ICRC0758_fig3} shows the response function for atmospheric
muons at the top of the deep IceCube detector, $1.5$~km below
the surface.  About 90\% of downward events consist of
a single muon entering the deep detector. 
Most of these events are from cosmic-rays with primary energy $<$10~TeV.
  The region under
the lower curve shows the contribution of events with multiple
muons.  By selecting a sample of coincident events in which
both tanks at one and only one IceTop station are hit, it is possible to discriminate
against high-energy events and find a sample enriched in single muons.  
Coincidences involving only an interior IceTop station provide a
sample in which about 75\% are single muons in the deep detector.~\cite{Bai}  
The line from the hit station to the center of gravity
of hits in the deep detector can be compared with the direction
obtained from the muon reconstruction algorithm in the deep detector
alone to check the reconstruction algorithm on single muon tracks.
The analysis confirms that the same reconstruction algorithm used for $\nu_\mu$-induced
upward muons reconstructs most events with an accuracy of better than 2$^\circ$.~\cite{Bai}

\section{Heliospheric physics with IceTop}
The monitoring stream of IceCube includes the scalar
rates of both discriminators in each DOM.  Response
of IceTop DOMs to secondary cosmic rays at the surface
is discussed in~\cite{PeterJohn}.
Signals at the rate of $\sim$2~kHz are produced
by a combination of photons converting in the tanks,
and electrons and muons that enter the tanks.  Most of these
particles come from primary cosmic rays with energies
in the few GeV range.  Large heliospheric
events can produce sudden changes in the counting rate.
Depending on the nature and orientation of the event
(e.g. a coronal mass ejection associated with a large 
solar flare), one can detect either a decrease in
the flux of galactic cosmic rays as the magnetic activity
excludes the lower energy cosmic rays from the inner
heliosphere or an increase due to solar energetic particles
accelerated in the event.

As the setting of the discriminator is increased, the average signal
rate decreases as the contribution from the lower energy cosmic-rays
falls below threshold.  The response of a DOM
to the primary cosmic-ray spectrum can therefore be tuned
significantly by changing the discriminator threshold--
even within the constraint that the threshold must
remain below a fraction of the VEM peak.  This gives the
possibility of studying heliospheric phenomena with
unprecedented timing resolution and with significant energy resolution,
as discussed in~\cite{Takao}.

{\bf Acknowledgments} This work is supported by the U.S. National Science Foundation,
Grants No. OPP-0236449 and OPP-0602679.

%\end{linenumbers}

%\end{document}

%
%cosmic radiation
%
%icrc0729.pdf (helio)
%icrc1285.pdf  (Spase 2 composition)
%icrc2007_v45.pdf (Spase 2 composition analysis)
%icrc1294_v2 (composition sensitive variables)
%icrc0678.pdf (SPase-2 point sources)
%newmuonsv1.1.pdf
%icrc0328.pdf (IceCube IceTop coincidences)
%icrc0858.pdf (lateral)
%icrc1059.pdf (response to muons)
%icrc0408.pdf  (low energy response)
%
\setcounter{figure}{0}
\setcounter{table}{0}
%%
% International Cosmic Ray Conference 2007 Merida Yucatan Mexico
% In this file you will find detailed instructions to correctly
% typeset your document.
%
% By: Victor De la Luz
% vdelaluz@inaoep.mx
% Mexico City

%Class Required
%\documentclass{article}
%The ICRC Style
%(This package is the last package in the usepackage list)
%If you need import other package you need write it first.
%\usepackage{icrctc07}

%The paper title
\title{Heliospheric Physics with IceTop}
%Short title to print in the headers to the final publication (Not showed in this print).
\shorttitle{Heliospheric Physics with IceTop}

%All paper authors
\authors{
T. Kuwabara$^{1,2}$, J. W. Bieber$^{1}$, and R. Pyle$^{1}$}
%Short title to print in the headers to the final publication (Not shown in this print).
\shortauthors{Author and et al.}
%All the affiliations.
\afiliations{ $^1$Bartol Research Institute and Department of
Physics and Astronomy, University of Delaware\\ $^2$For the
IceCube Collaboration, described in a special section of these
proceedings}

\email{takao@bartol.udel.edu}

%The abstract.
\abstract{ IceTop is an air shower array now under construction at
the South Pole. It is the surface component of IceCube, an
observatory primarily focused on cosmic neutrinos. When completed,
IceTop will have approximately 500 square meters of collecting
area in the form of 160 separate ice Cherenkov detectors. These
detectors are sensitive to electrons, photons, muons and neutrons.
With the high altitude and low geomagnetic cutoff at the South
Pole, IceTop promises to have unprecedented statistical precision,
coupled with spectral sensitivity that can be used to observe
solar energetic particles and transient phenomena in the flux of
galactic cosmic rays. We discuss the potential of IceCube to
contribute to heliospheric physics in general, and present a
preliminary analysis of a complex interplanetary disturbance that
occurred in August of 2006. }

%%%%%%%%%%%%%%%%%%%% B E G I N   D O C U M E N T%%%%%%%%%%%%%%%%%%%%%%%
%\begin{document}
\maketitle

%Begin the section.
\section{Introduction}
IceTop is an air shower array now under construction at the South
Pole as the surface component of the IceCube neutrino telescope.
When completed, IceTop will have approximately 500 square meters
of ice Cherenkov collecting area arranged in an array of 80
stations on a 125 m triangular grid. Each station consists of two,
two meter diameter tanks filled with ice to a depth of 90 cm.
Tanks are instrumented with two Digital Optical Modules (DOM)
operated at different gain settings to provide appropriate dynamic
range to cover both large and small air showers. Each DOM contains
a 10 inch photomultiplier and an advanced readout system capable
of returning the full waveform of more complex events. For the
present analysis we use two discriminator counting rates recorded
in each DOM. For historical reasons, the two discriminators are
termed SPE (Single Photo Electron), and MPE (Multi Photo
Electron). As used in IceTop the SPE threshold corresponds
typically to 10 photoelectrons, and the MPE threshold to 20
photoelectrons.

\begin{figure} [t]
\begin{center}
\includegraphics [width=0.48\textwidth]{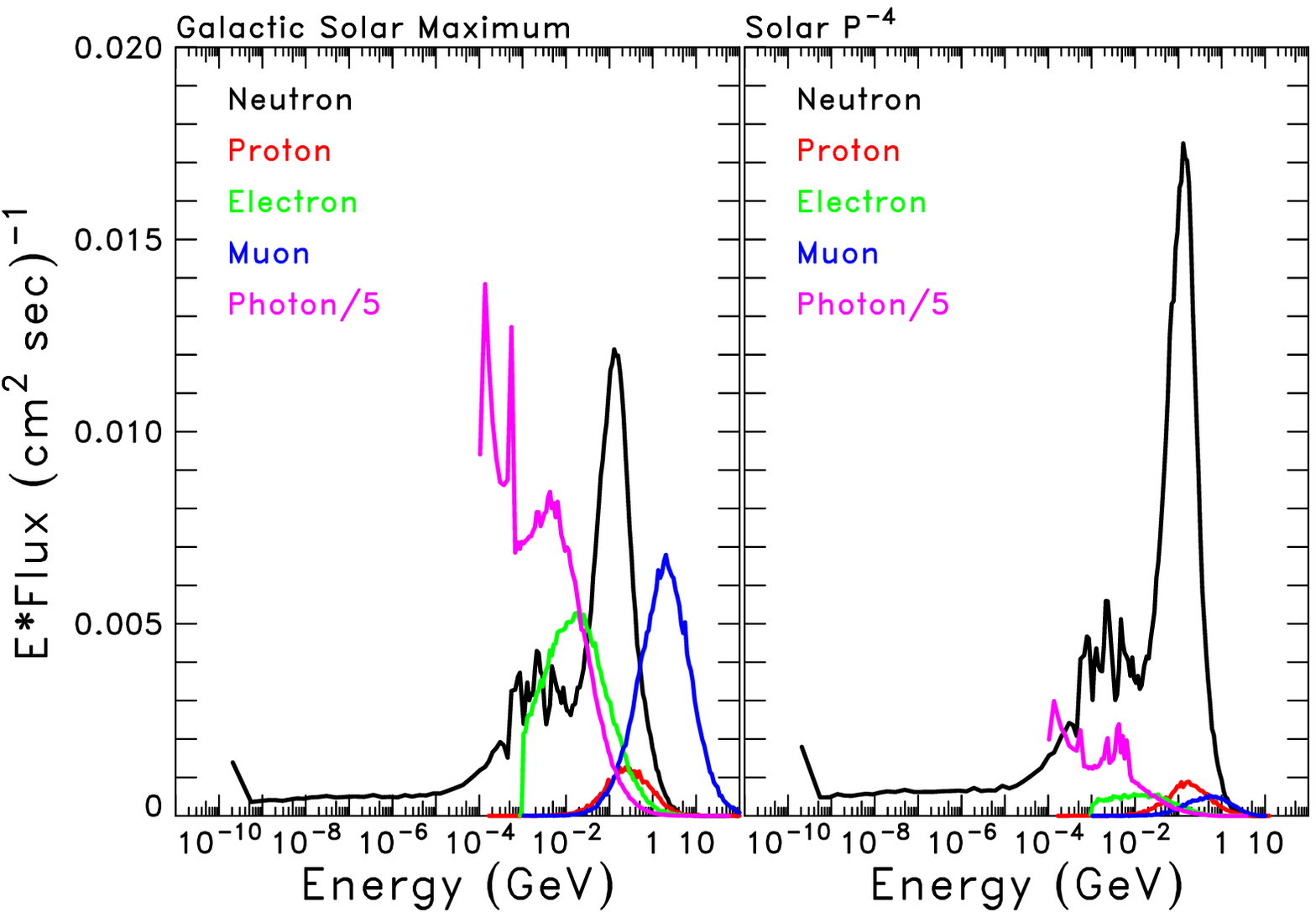}
\end{center}
\caption{Calculated secondary particle spectra at the South Pole.
Left: Galactic solar maximum. Right: Solar flare particle event
normalized to produce a doubling of the count rate of a standard
(NM64) neutron monitor.}\label{fig1}
\end{figure}

Due to the high altitude (2835m) and the nearly zero geomagnetic
cutoff at the South Pole, secondary particle spectra at ``ground''
level retain a significant amount of information on the spectra of
the primary particles. This is illustrated in Figure~1, which
summarizes the result of a FLUKA \cite{Fasso:1993} calculation of
the secondary spectra due to galactic cosmic rays at solar maximum
(left panel) and a typical solar flare particle event (right
panel). Of course the solar spectrum would be superimposed on the
galactic background. It is beyond the scope of this brief paper to
show this in detail, but because the IceTop tanks are thick enough
to totally absorb many of the incident particles the signal
distribution in the tank contains information on the primary
spectrum. More details are provided in a companion paper
\cite{Clem:2007}.

\subsection{Barometer Correction}
As with a neutron monitor, the counting rate of an IceTop detector
shows a strong dependence on barometric pressure. From simulation
and observation, it has been shown that barometric correction
coefficients vary with the threshold energy of secondary cosmic
rays \cite{Shamos:1966} \cite{Dorman:2004}. The energy sensitivity
of IceTop detectors is nicely illustrated by the barometric
coefficients we derive for them. By considering time periods in
which there appears to be little variation in the primary particle
intensity, it is possible to make a phenomenological estimate of
the appropriate pressure correction by means of a simple
correlation between detector counting rate and barometric
pressure. Figure~2 shows this correlation for the two thresholds
of an individual DOM. Note in particular the small but significant
difference in the slope of the correlation, and hence the derived
barometric correction.

\begin{figure}
\begin{center}
\includegraphics [width=0.48\textwidth]{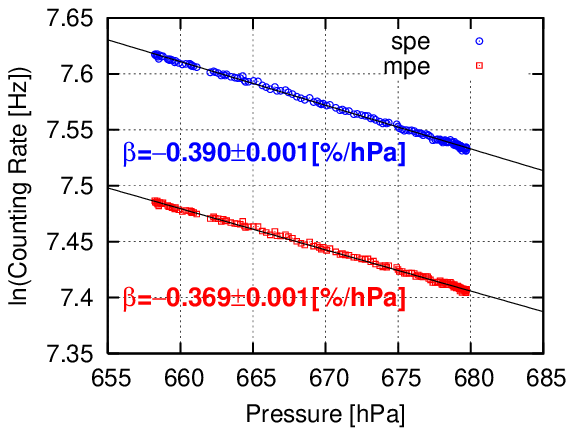}
\end{center}
\caption{Correlation of scaler rate with pressure for one DOM on
October 8-9, 2006.}\label{fig2}
\end{figure}

In 2006 a total of 32 tanks were operational. Figure~3 shows the
derived correction for each DOM (red squares for the MPE
discriminators and blue circles for the SPE discriminators) plotted as a
function of the counting rate of the discriminator. At that time
the tanks were all operating at the same nominal setting, but they
had not been calibrated, so in fact the discriminators were
triggering over a range of physical light levels. The correlation
of correction with light level is nearly perfect. Those
discriminators with lower counting rates, corresponding to higher
light thresholds, have markedly lower barometric corrections. This
is just what is expected since these signals should result
preferentially from higher energy primaries. We are in the process
of trying to use this information, plus simulations and
calculations, to establish an energy response function for the
tanks. For the remainder of this paper we rely on the approximate
response functions derived from the FLUKA calculation that
produced the plots shown in Figure~1.

\begin{figure}
\begin{center}
\includegraphics [width=0.48\textwidth]{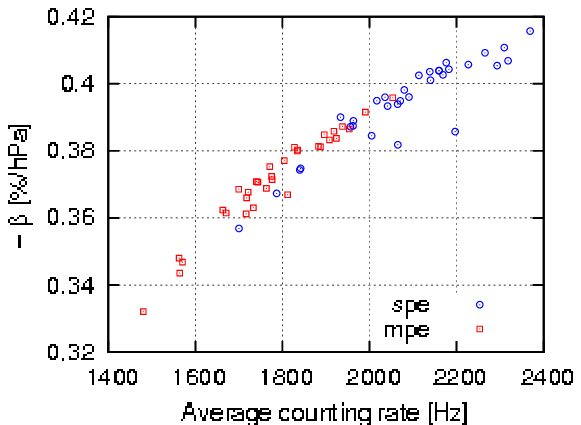}
\end{center}
\caption{Pressure correction coefficient for all DOM as a function
of scaler rate.}\label{fig3}
\end{figure}

\begin{figure*} [t]
\begin{center}
\includegraphics [width=0.82\textwidth]{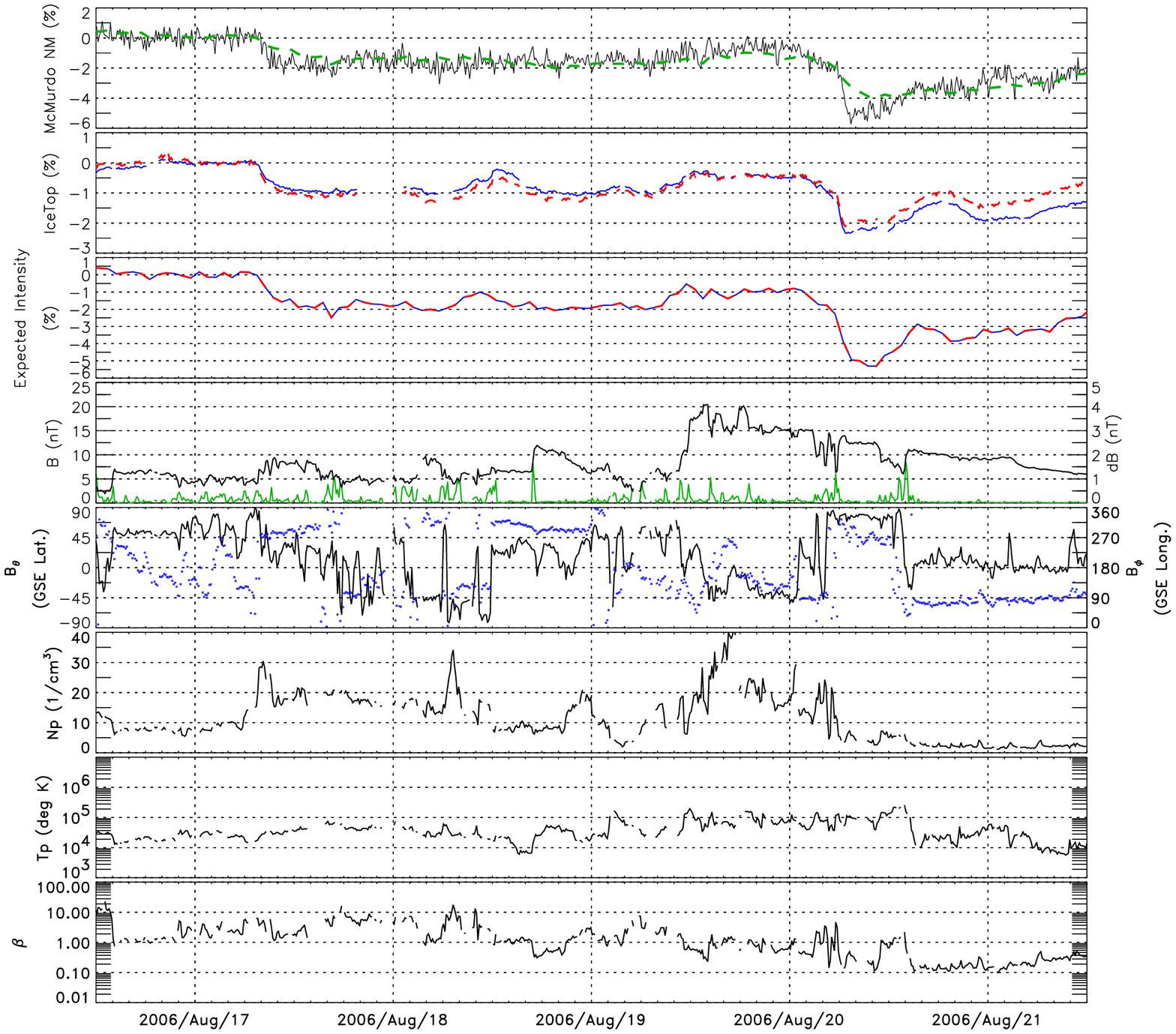}
\end{center}
\caption{August 17-21, 2006. From top: (1) McMurdo monitor (black)
and Spaceship Earth isotropic component (green dashed). (2) IceTop SPE
(blue) and MPE (red dashed) scaler rate, 32 DOM average. (3) IceTop model
prediction. (4) Interplanetary magnetic field magnitude (black)
and derivative (green). (5) Field direction latitude (black) and
longitude (blue dots). (6) Plasma density. (7) Plasma temperature. (8)
Plasma $\beta$.}\label{fig4}
\end{figure*}

\subsection{Heliospheric Event}
In Figure~4 we show several data sets characterizing a
heliospheric event in August 2006. The IceTop measurements are
shown in the second panel. We have averaged the SPE (blue) and MPE
(red dashed) counting rates for all 32 DOM, after individually applying
the barometric corrections described in the previous section. Ten
minute averages are shown, all expressed as percent changes
relative to the normalization interval on August 17 prior to the
first decrease. For comparison the top panel shows the similarly
treated counting rate of the McMurdo neutron monitor. While the
event is generally similar in the two detectors, the remarkably
better counting statistics of IceTop stand out. In IceTop the
total counting rate for the SPE channel is $\sim$64 kHz  (2 kHz
from each DOM), while the total counting rate of the 18NM64
McMurdo neutron monitor is $\sim$0.3 kHz {\cite{Bieber:1995}.

From McMurdo alone, one might characterize this event as a double
Forbush decrease \cite{Forbush:1938}. Both decreases are
associated with structures in what is evidently an interplanetary
coronal mass ejection (ICME) containing at least one shock and
multiple regions with different magnetic field and plasma
parameters. However in IceTop the two decreases appear quite
different. The second fits the conventional pattern in which the
magnitude tends to scale inversely with primary rigidity
\cite{Morishita:1990}. Note that the decrease is consistently
larger in low threshold SPE channel than high threshold MPE, and
also that the higher rigidity particles tend to recover more
rapidly.

In contrast, during the first decrease the higher energy channel
shows a (slightly) larger deviation. There is also an intriguing
feature in the IceTop data on August 18, near the time of a large
change in the interplanetary magnetic field direction, that is not
observed in the McMurdo neutron monitor. The {\it Spaceship Earth}
neutron monitor network \cite{Bieber:1995} measures a significant
anisotropy during the event, which we can model as a dipole
anisotropy with a time variable magnitude and direction
superimposed on a time varying isotropic cosmic ray flux. The
isotropic component of our model fit is shown as the green dashed curve
superimposed on the McMurdo data in Figure~4. The deviations of
the McMurdo data from this line can only result from anisotropy
since the Spaceship Earth stations have well matched energy
response. Because IceTop has inherent spectral resolution it is
possible for anisotropy to produce an apparent spectral feature.
Even though the low and high rigidity channels of IceTop are
derived from the same physical detector, the low and high rigidity
particles will come from somewhat different asymptotic directions.

Using calculated response functions appropriate to the different
discriminator levels, and asymptotic directions calculated as a
function of rigidity, it is straightforward to convolute the two
to make a specific prediction for IceTop. The third panel of
Figure~4 gives the result of such a calculation under the simplest
possible assumption, anisotropy independent of energy. We have
used response functions that predict the observed counting rate
corresponding to thresholds of ten photoelectrons (blue curve) and
fifty photoelectrons (red dashed curve). On the scale at which the figure
is reproduced it is not possible to see the small difference in
the curves. Our conclusion is that the observed splitting of the
red dashed and blue solid curves in the second panel results from spectral
variation. We note that the overall time structure of IceTop data,
and in particular the marked difference from McMurdo, is
consistent with the dipole model derived from {\it Spaceship
Earth}. The amplitude predicted for IceTop is understandably too
large, particularly in the second decrease, because at these
discriminator thresholds IceTop is observing at a higher average
energy. Although IceTop is geographically further south than
McMurdo, it is magnetically further north. Thus McMurdo looks
nearly perpendicular to the ecliptic, whereas Pole has a mid
latitude viewing direction.

The high statistical precision of IceTop may translate even small
anisotropy into apparent spectral variation, and this must be
taken into account in the analysis of interplanetary events.
However there is no indication that the feature on August 18
results from such an effect. It is not clear at this time just
what aspect of the complicated plasma and magnetic field structure
at the time is responsible for the unusual spectral variation of
the high energy cosmic rays. What is clear is that the high time
resolution and energy resolution provided by IceTop will usher in
a new era in the study of the propagation of GeV particles in the
heliosphere.

\section{Acknowledgements}
This work is supported in part by U.S. NSF awards OPP-0236449,
ATM-0527878 and OPP-0602679.

%\bibliography{ICRC0729/icrc0729}
%\bibliographystyle{plain}
%\end{document}

\setcounter{figure}{0}
\setcounter{table}{0}
% International Cosmic Ray Conference 2007 Merida Yucatan Mexico
%
%\documentclass{article}
%\usepackage{icrctc07}
%
\title{Measuring Cosmic Ray Composition at the Knee with SPASE-2 and AMANDA-II}
\shorttitle{Measuring the Cosmic Ray Composition}
\authors{K. G. Andeen$^{1}$, C. Song$^{1}$ and K. Rawlins$^{2}$ for
the IceCube Collaboration $^A$.}
\shortauthors{K. G. Andeen and et al}
\afiliations{$^1$IceCube Collaboration, University of
Wisconsin-Madison, 1150 University Ave, Madison, WI\\ $^2$ University
of Alaska Anchorage, 3211 Providence Dr, Anchorage, AK }
\email{kandeen@icecube.wisc.edu ; $^A$ See special section of these proceedings}
 
\abstract{Important information pertaining to the origin of
high-energy cosmic rays can be gained by studying their mass
composition in the region of the knee ($\sim$~3 PeV).  Thus, air
showers have been observed at the South Pole using the SPASE-2
detector, which measures the electronic component at the surface, and
the AMANDA-II neutrino telescope, which measures the coincident muonic
component in deep ice.  These two components, together with a Monte
Carlo simulation and a well-understood analysis method, yield the
relative cosmic ray composition in the knee region.  We report on the
efficacy of a new neural network technique for obtaining a composition
result with the SPASE-2/AMANDA-II detectors.}

%%%%%%%%%%%%%%%%%%%% B E G I N   D O C U M E N T%%%%%%%%%%%%%%%%%%%%%%%
%\begin{document}
\maketitle

\section{Introduction}
Cosmic ray composition studies can provide a greater understanding of
the origin of cosmic rays, and thus lead to an increased understanding
of the physical processes which accelerate these particles to Earth.
At energies up to 10$^{14}$~eV, the mass composition of cosmic rays
can be measured directly; however, due to the low flux, the mass
composition above 10$^{14}$~eV must currently be gleaned from indirect
measurements, involving the examination of the extensive air shower
produced by the primary particle in the atmosphere.  By utilizing more
than one component of the air shower, such as the electronic and
muonic components, an analysis technique can be developed that leads
to a composition measurement.

\section{Detectors and Reconstruction}
The detectors used for this analysis the South Pole Air Shower
Experiment (SPASE-2) and the Antarctic Muon And Neutrino Detector
Array (AMANDA-II).  The SPASE-2 detector is situated on the surface of
the South Pole and is composed of 30 stations in a 30~m triangular
grid.  Each station contains four 0.2~m$^2$ scintillators.  The
AMANDA-II detector lies beneath the surface of the ice, located such
that the center-to-center separation between AMANDA-II and SPASE-2 is
about 1730~m, with an angular offset of 12$^\circ$.  AMANDA-II
consists of 677 optical modules (OMs) deployed on 19 detector strings
at depths between 1500 and 2000~m.  Each OM contains a photomultiplier
tube which can detect the Cherenkov light emitted by particles--namely
muon bundles--passing through the ice.  Besides a composition
analysis, this coincident detector configuration allows for
calibration as well as measurement of the angular resolution of the
AMANDA-II detector \cite{Ahrens}.

For this preliminary analysis, coincident data from the years
2003-2005 are used, with a total livetime of 369 days.  For comparison
with the data, Monte Carlo simulated proton and iron showers with
energies between 100~TeV and 100~PeV have been produced using the
MOCCA air shower generator \cite{MOCCA} and the SIBYLL v1.7
interaction model \cite{SIBYLL}.  These events are then propagated
through the ice, and the detector response of AMANDA-II is simulated
using AMASIM.  An E$^{-1}$ spectrum is used for generation, but for
analysis the events are re-weighted to the cosmic ray energy spectrum
of E$^{-2.7}$ at energies below the knee at 3~PeV, and E$^{-3.0}$
above it. Both the data and Monte Carlo are then put through the same
reconstruction chain.

The first step in the reconstruction is to find the incoming direction
of the air shower, as well as the core position and shower size.  The
direction can be computed from the arrival times of the charged
particles in the SPASE-2 scintillators, while the shower core position
and shower size are acquired by fitting the lateral distribution of
particle density to the Nishimura-Kamata-Greisen (NKG) function and
then evaluating the fit at a fixed distance from the center of the
shower (in this case 30~m) \cite{Dickinson}.  This parameter, called
S30, has units of particles/m$^2$ and will be referred to throughout
this paper as a measure of the electronic part of the air shower .

The next step in the reconstruction provides a measure of the muon
component of the air shower, which is carried out using the
combination of the two detectors.  The core position of the shower
measured at SPASE-2 is kept fixed as a vertex from which $\theta$ and
$\phi$ are varied in the ice to obtain a good fit of the track
direction in AMANDA-II.  The expected lateral distribution function
(LDF) of the photons from the muon bundle in AMANDA-II is then
computed, fit to the OM hits, and evaluated at a perpendicular
distance of 50~m from the center of the shower \cite{Kathsthesis}.
This parameter, called K50, has units of photoelectrons/OM and will be
used throughout the rest of this paper as the measure of the muon
component of the air shower.

\section{Analysis Details}
Once the reconstruction has been completed, it is important to find
and eliminate poorly reconstructed events.  Thus, as in the previous
analysis \cite{Kathsthesis}, events have been discarded which:

\begin{itemize}
\item have cores outside either the area of SPASE-2 or the
volume of AMANDA-II,
\item have too low an energy to be well-reconstructed in both
detectors,
\item have an unphysical reconstructed attenuation length of
light in the ice.
\end{itemize}

After these cuts have been made, it can be seen in Figure
\ref{k50s30plot} that our two main observables, S30 and K50, form a
parameter space in which primary energy and primary mass separate.
This is expected, as the showers associated with the heavier primaries
develop earlier in the atmosphere and hence have more muons per
electron by the time they reach the surface than the showers
associated with lighter primaries \cite{Gaisser}.  This means that
K50, which is proportional to the number of muons in the ice, will be
higher for heavier primaries than for lighter primaries of the same
S30, as is observed.

In the three-year data set used for this analysis, 105,216 events
survive all quality cuts.  It is interesting to notice that in the
previous analysis, using the SPASE-2/AMANDA-B10 detector, the final
number of events for one year was 5,655.  Furthermore, the larger
detector used here is sensitive to higher energy events.  The
significant increases in both statistics and sensitivity are the basis
for performing a new analysis.

\begin{figure}
  \begin{center}
    \noindent
    \includegraphics[width=0.5\textwidth]{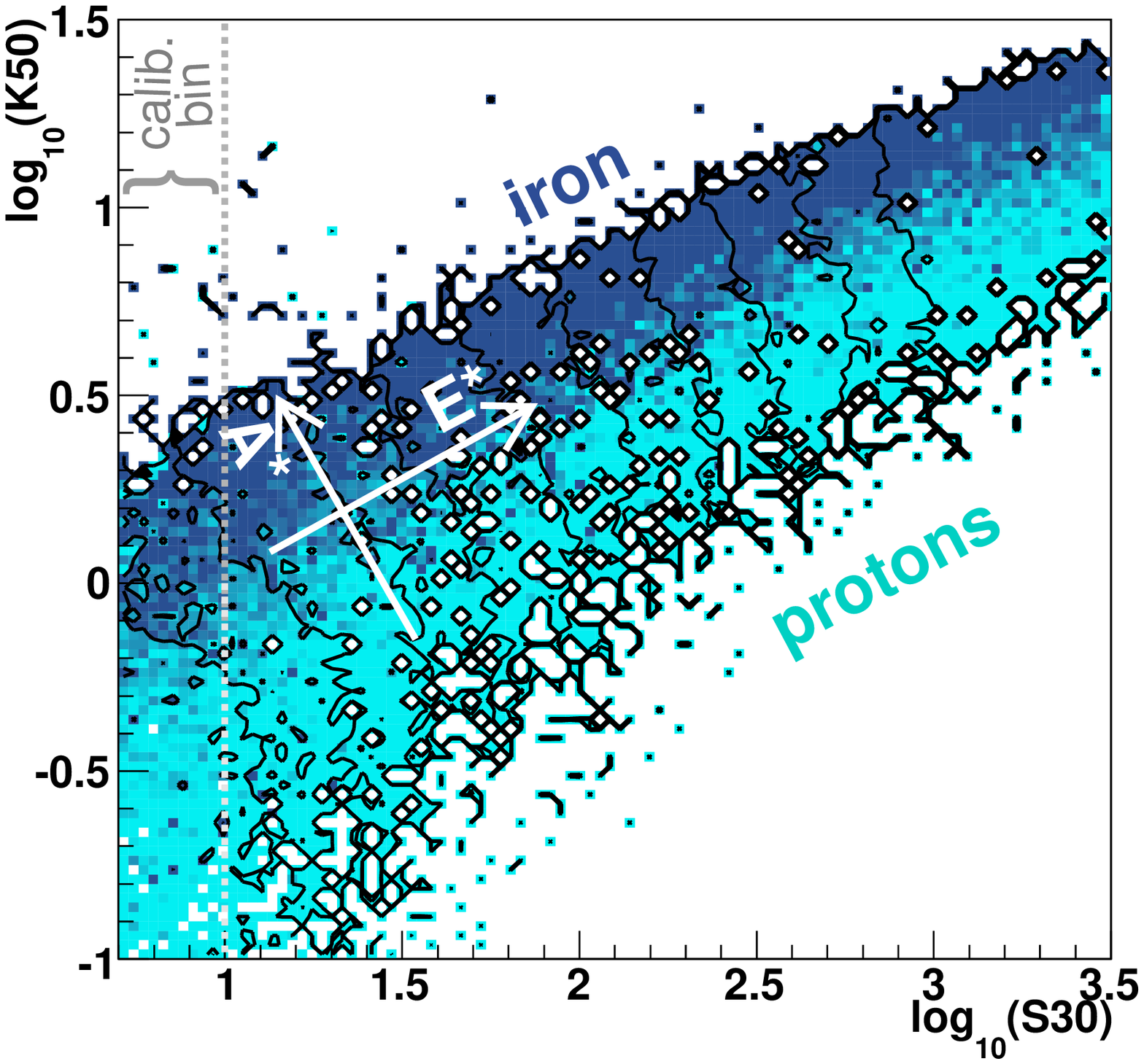}
  \end{center}
  \caption{\label{k50s30plot}The two main observables, log$_{10}$(K50)
    vs log$_{10}$(S30), in the Monte Carlo simulation.  The black contour
    lines depict gradients in energy.  The axes along which mass (A*) and
    energy (E*) change in a roughly linear way are drawn in white, and the
    low-energy calibration bin is also labeled.}
\end{figure}

\begin{figure}[ht]
  \begin{center}
    \noindent
    \includegraphics*[width=0.49\textwidth]{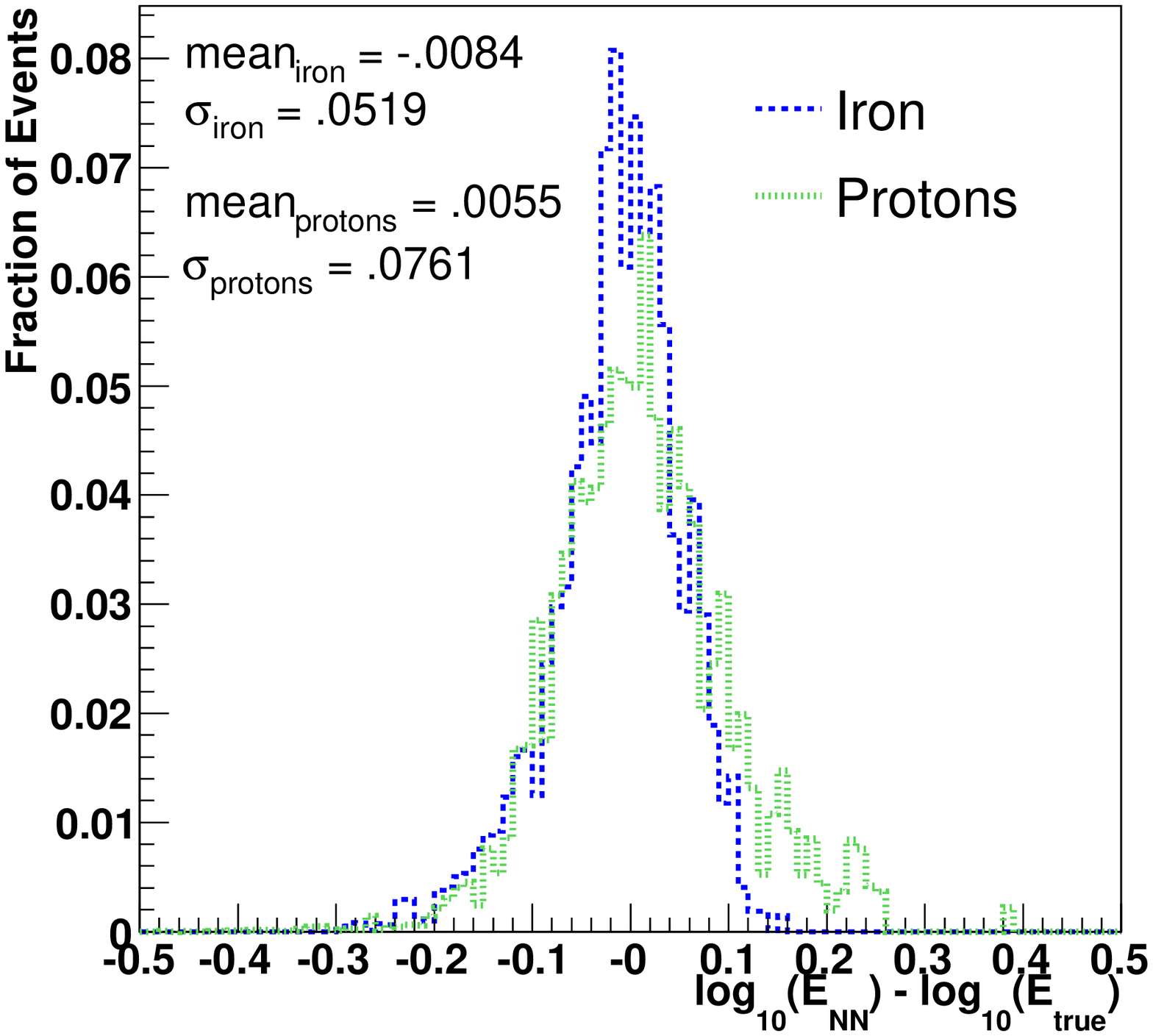}
  \end{center}
  \caption{\label{energyresolution}The energy resolution of the neural
    network for output energies between 1 and 10~PeV for proton and
    iron showers.}
\end{figure}

\subsubsection {Calibration}
To accurately measure the composition using both electron and muon
information reconstructed as described above, the Monte Carlo
simulations must represent the overall amplitude of light in the ice
very well.  However, the overall light amplitude is subject to
systematic errors in the simulation.  Therefore, it is important to
calibrate the composition measurement at low energies where direct
measurements of cosmic ray composition are available from balloon
experiments.  A vertical ``slice'' of events from Figure
\ref{k50s30plot}, corresponding to S30 between 5 and 10~m$^{-2}$, is
used to perform this calibration.  The K50 values of the data adjusted
by an offset, chosen such that the distribution of K50 best matches a
50$\%$-50$\%$ mixture of protons and iron \cite{Kathsjournal,
Kathsthesis}.  This mixture corresponds to $<$lnA$>$~=~2, which is an
approximation to the value indicated by direct measurements
\cite{Hoerandel}.

\begin{figure}[ht]
  \begin{center}
    \noindent
    \includegraphics[width=0.49\textwidth]{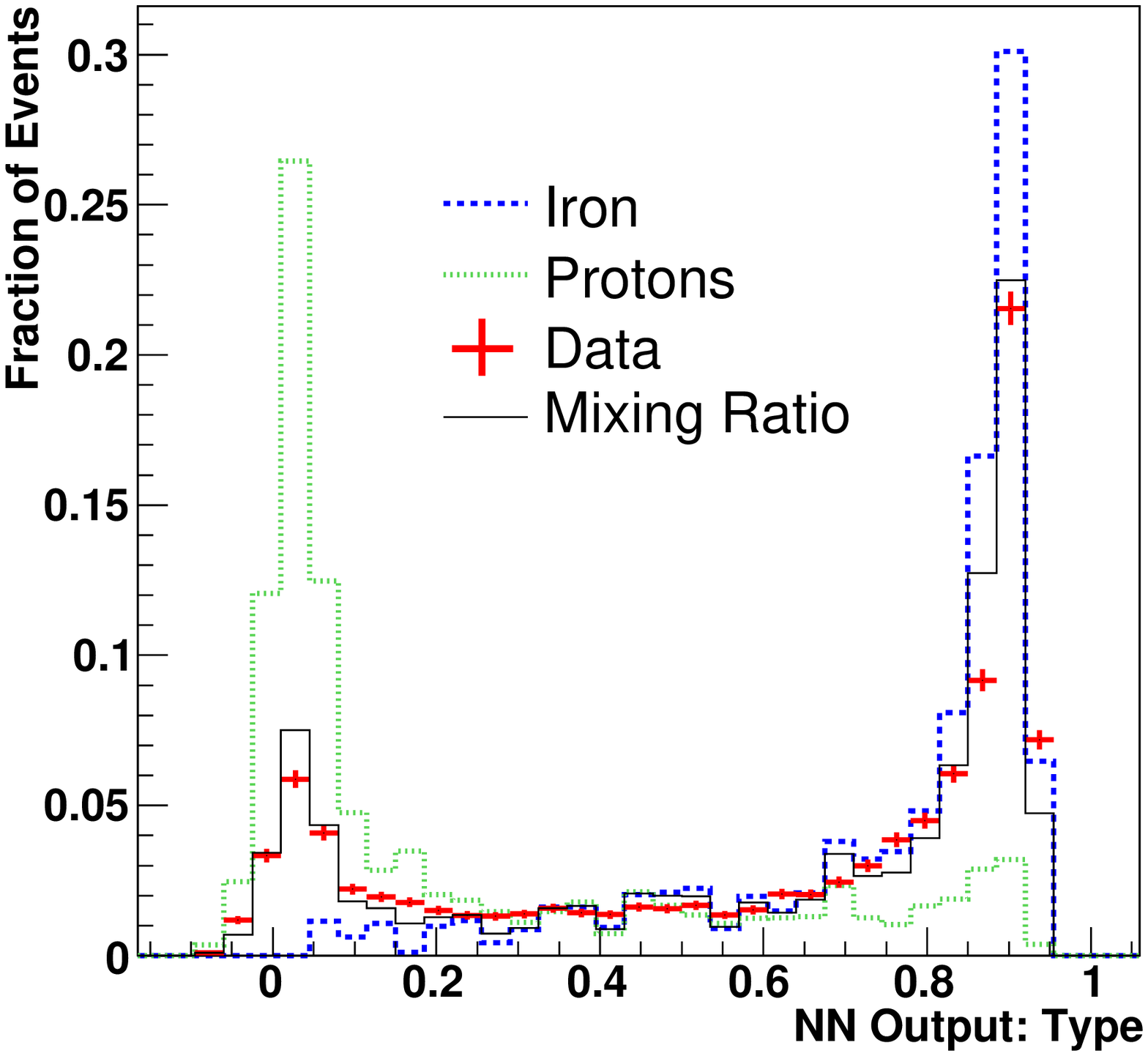}
  \end{center}
  \caption{\label{nnmassplot} The neural network output for particle
    type with log$_{10}$(E$_{NN}$/GeV) between 6.0 and 6.2.  The
    three-year data set is compared to the Monte Carlo generated
    proton and iron showers, and a mixing ratio is found which
    represents the data. }
\end{figure}

\subsubsection {The Neural Network}

Similar past analyses \cite{Kathsjournal} exploited the fact that the
relationship between K50/S30 and mass/energy is approximately linear.
One can then rotate to the mass/energy coordinate plane, labeled as
A*/E* in Figure \ref{k50s30plot}, and utilize further analysis
techniques to extract the energy and mean log mass after the rotation.
However, the relationship is not perfectly linear, nor should exact
linearity necessarily be expected.  In fact, as seen in Figure
\ref{k50s30plot}, the non-linear effects become more pronounced at
higher energies.  As the data set for this new analysis has more
statistics at high energies than previous analyses, it has become
important to find a technique that can resolve these events with
accuracy.  A neural network should be able to take these non-linear
effects into account.

The neural network chosen for this analysis was the
TMultiLayerPerceptron class from ROOT, which is a simple, feed-forward
network, although other neural networks were also tested with similar
results.  The network configuration which best separates the pure
proton from the pure iron scenarios and yields the best energy
resolution in the Monte Carlo was a very simple 2:5:2 network, meaning
there are two input variables, five hidden nodes, and two output
variables.  In this case, the two input variables are log$_{10}$(K50)
and log$_{10}$(S30), and the two outputs are energy and particle type
(0 for protons, 1 for iron).  The network is trained on half of the
Monte Carlo and tested on the other half (to evaluate its
effectiveness) before being applied to the data.  Figure
\ref{energyresolution} shows the energy resolution of the neural
network for proton and iron showers.  The ``type'' output of the
neural network for one energy bin is plotted in Figure
\ref{nnmassplot}.  Notice that, since it was trained on pure proton
and iron samples, the neural network tends to classify every event
strictly as one or the other, resulting in the strong peaks in the
data at 0 and 1.  It is expected that the simulation of more primary
nuclei would yield a more accurate result.

It is assumed that the data can be described by some mixture of proton
and iron showers, and a technique is developed to find the mixing
ratio in each energy bin which best fits the data.  In order to find
this proportion, the proton, iron, and data outputs are normalized and
a minimization technique is applied.  The result is one mixing ratio
for each ``slice'' in energy; an example of this is shown by the solid
black line in Figure \ref{nnmassplot} This method was verified using
various mixtures of proton and iron simulations as input ``data'' and
comparing with the non-mixed monte-carlo results.  The ratio of heavy
particles in each energy bin can also be expressed as the mean log
mass.  The difference between $<$lnA$>$ for the neural network
technique described herein and $<$lnA$>$ for a rotation method similar
to that used for the previous SPASE-2/AMANDA-B10 analysis is reported
in Figure \ref{resultplot}.  (Note that the same data set was used for
both methods.)

\section{Discussion}

It is clear from Figure \ref{resultplot} that the percent difference
in $<$lnA$>$ between the two types of analysis methods is generally
quite small, especially below log$_{10}$(E/GeV)~=~6.8, which is the
highest energy measured in the previous analysis.  Furthermore, it
seems promising that the percent difference increases at higher
energies where the neural network is expected to be more reliable.
The systematic errors for this data sample have yet to be fully
examined, and a new Monte Carlo simulation with a variety of primary
nuclei--including helium, carbon and oxygen in addition to protons and
iron--is currently being generated.  Nevertheless, there is a clear
indication that the neural network technique is a valid method for
understanding SPASE-2/AMANDA-II data, and it is hoped that, together
with the new simulation and new data from the IceCube/IceTop
coincident detectors, this new technique will allow us to probe
energies up to 10$^{18}$~eV.

\begin{figure}[ht]
  \begin{center}
    \noindent
    \includegraphics[width=0.49\textwidth]{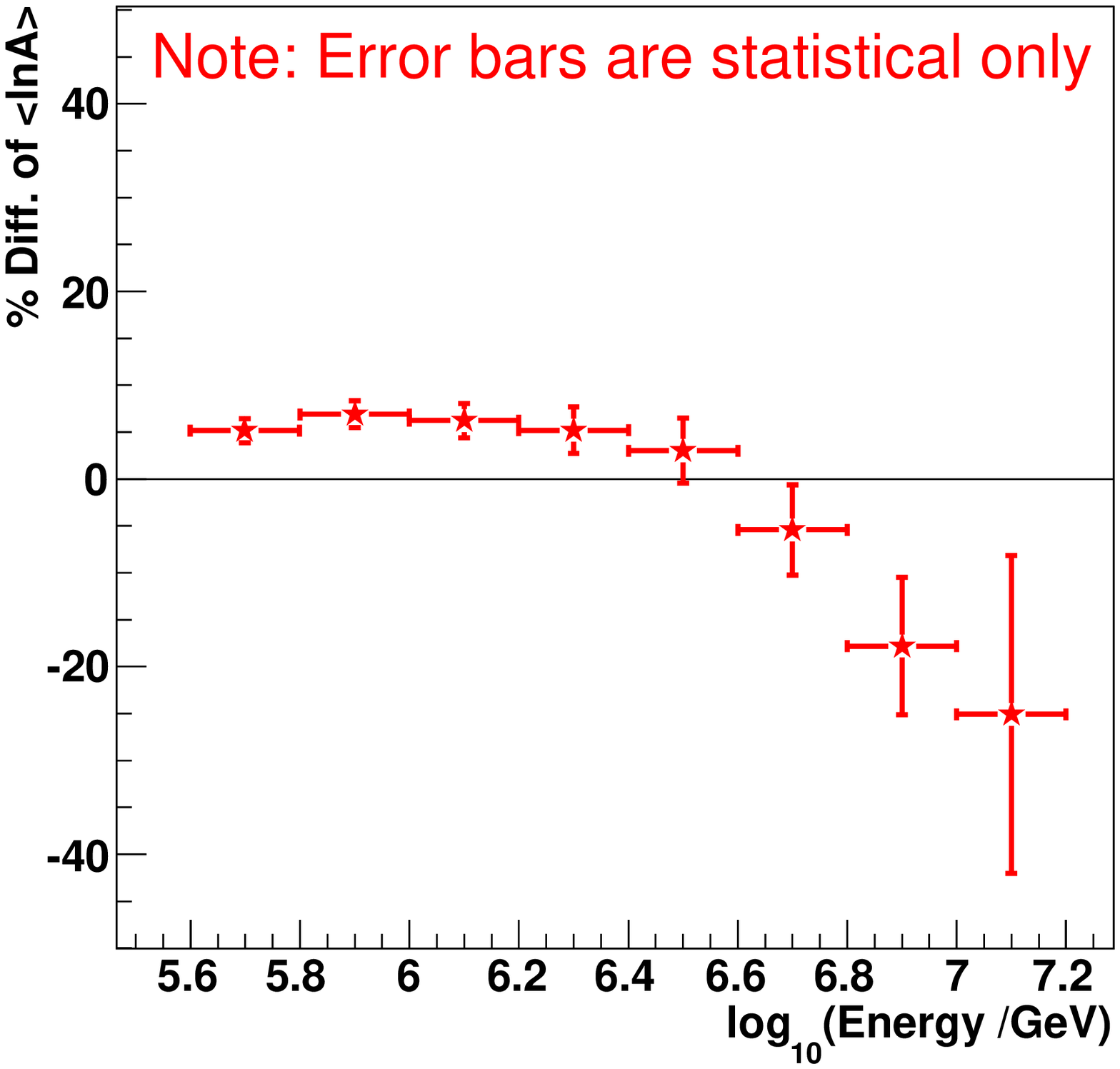}
  \end{center}
  \caption{\label{resultplot} The percent difference in $<$lnA$>$
    between two analysis techniques applied to the same three-years
    of SPASE-2/AMANDA-II data.}
\end{figure}

\section{Acknowledgements}
The authors would like to acknowledge support from the Office of Polar
Programs of the United States National Science Foundation.
%
%\bibliography{ICRC1285/libros}
%
%\bibliographystyle{plain}

%\end{document}
 %Kareen
\setcounter{figure}{0}
\setcounter{table}{0}
%%
% International Cosmic Ray Conference 2007 Merida Yucatan Mexico
% In This file you will find detailed instructions to correctly
% typeset your document.
%
%
%

%Class Requeried
%\documentclass{article}
%The ICRC Style
%\usepackage{icrctc07}

%The paper title
\title{Cosmic Rays in IceCube: Composition-Sensitive Observables}
%Short title to print in the headers to the final publication (Not showed in this print).
\shorttitle{Cosmic Rays in IceCube: Composition-Sensitive Observables}
%All paper authors
\authors{C. Song$^{1}$, P. Niessen$^{2}$ and K. Rawlins$^{3}$ 
         for the IceCube collaboration$^{*}$}
%Short title to print in the headers to the final puplication (Not showed in this print).
\shortauthors{}
%All the affiliations.
\afiliations{$^1$ University of Wisconsin, 5th Floor Suite, 222 W.
		  Washington Ave. Madison, WI 53717, USA \\
	     $^2$ Bartol Research Institute, University of Delaware, 
                  104 Center Mall, Newark, DE 19716, USA \\ 
             $^3$ University of Alaska, 3211 Providence Drive,
                  Anchorage, AK 99508, USA \\
	     $^*$ See a special section of the proceedings} 
\email{csong@icecube.wisc.edu}

%The abstract.
\abstract{Cosmic ray showers that trigger the IceTop surface array 
generate high energy muons that are measured by the IceCube  
detector. The large surface and underground area of this 3-dimensional 
instrument at completion guarantees significant statistics for shower 
energies up to about 1 EeV. Since the number of muons is sensitive to the 
type of the primary cosmic ray nucleus, these events can be used for 
the measurement of cosmic ray composition. Using the data taken in the 
existing array, we measure the observables sensitive to the primary 
mass as a function of shower energy estimated by the surface array. 
The result is compared to simulations of the coincident events of 
different primary nuclei.}

%\email{aastex-help@aas.org}

%%%%%%%%%%%%%%%%%%%% B E G I N   D O C U M E N T%%%%%%%%%%%%%%%%%%%%%%%
%\begin{document}
\maketitle
%Begin the section.

\section{Introduction}

Cosmic rays follow a steep power-law spectrum which spans a wide energy range
up to a few 10$^{20}$ eV. One of the interesting features in the all-particle
energy spectrum is that the cosmic ray spectrum steepens around 3 PeV,
which is called the `knee'. The origin of the knee is generally understood
to be due to the limiting energy attained during the acceleration process 
and/or leakage of charged particles from the galaxy. The mass composition
of cosmic rays at the knee region provides important clues to their origin.

The IceCube Observatory located at the South Pole, a 3-dimensional instrument
which consists of the IceTop surface detector and IceCube optical sensor arrays, 
is uniquely configured to measure cosmic ray composition. The IceTop surface 
array will consist of 80 pairs of frozen water tanks which measure the energy 
deposition 
at the surface, and 80 strings of 60 digital optical modules (DOMs) in ice will
measure Cherenkov photons from muon bundles. The DOMs are attached to a
cable every 17 m, between depths of 1,450 and 2,450 m. A pair of the IceTop
tanks separated by 10 m is located above each IceCube string and a tank
employs two DOMs which are identical to in-ice DOMs but with different PMT 
gains, which results in a wide dynamic range.

\section{Data and simulation}

IceTop/IceCube coincident data taken in 2006 were used for this analysis.
In 2006, 16 pairs of IceTop tanks and 9 IceCube strings were operational. 
Events were recorded when the following trigger conditions were satisfied:
6 hits within 2~$\mu$s for IceTop DOMs, and 8 hits within 5~$\mu$s
for in-ice DOMs. The coincident rate is about 0.2~Hz. A threshold of 300 
TeV allows us to measure cosmic rays below the knee.

Air shower events were simulated with CORSIKA\cite{Corsika}, and
GHEISHA\cite{Gheisha} and SIBYLL-2.1\cite{Sibyll} were selected as the low 
and high energy hadronic interaction models, respectively. Proton and iron
showers were generated over an area of 4.5 km$^{2}$ covering the IceTop array,
from energies of 50 TeV to 5 PeV, using the South Pole atmospheric 
model\cite{Atmos}. The events were generated according to $E^{-1}$ spectrum 
and re-weighted to the cosmic ray energy spectrum with spectral index of 
-2.7 below the knee at 3 PeV, and -3.0 above it.

As a first guess, the shower core is determined by calculating the center of
gravity of tank positions by weighting with the square root of pulse 
amplitude. The shower direction is determined on the basis of shower front 
arrival times measured by the IceTop tanks. The energy deposition at the surface 
as a function of distance from the shower core is fitted to the function given 
by\cite{Stefan}:

\begin{figure}
\begin{center}
\includegraphics[width=0.48\textwidth]{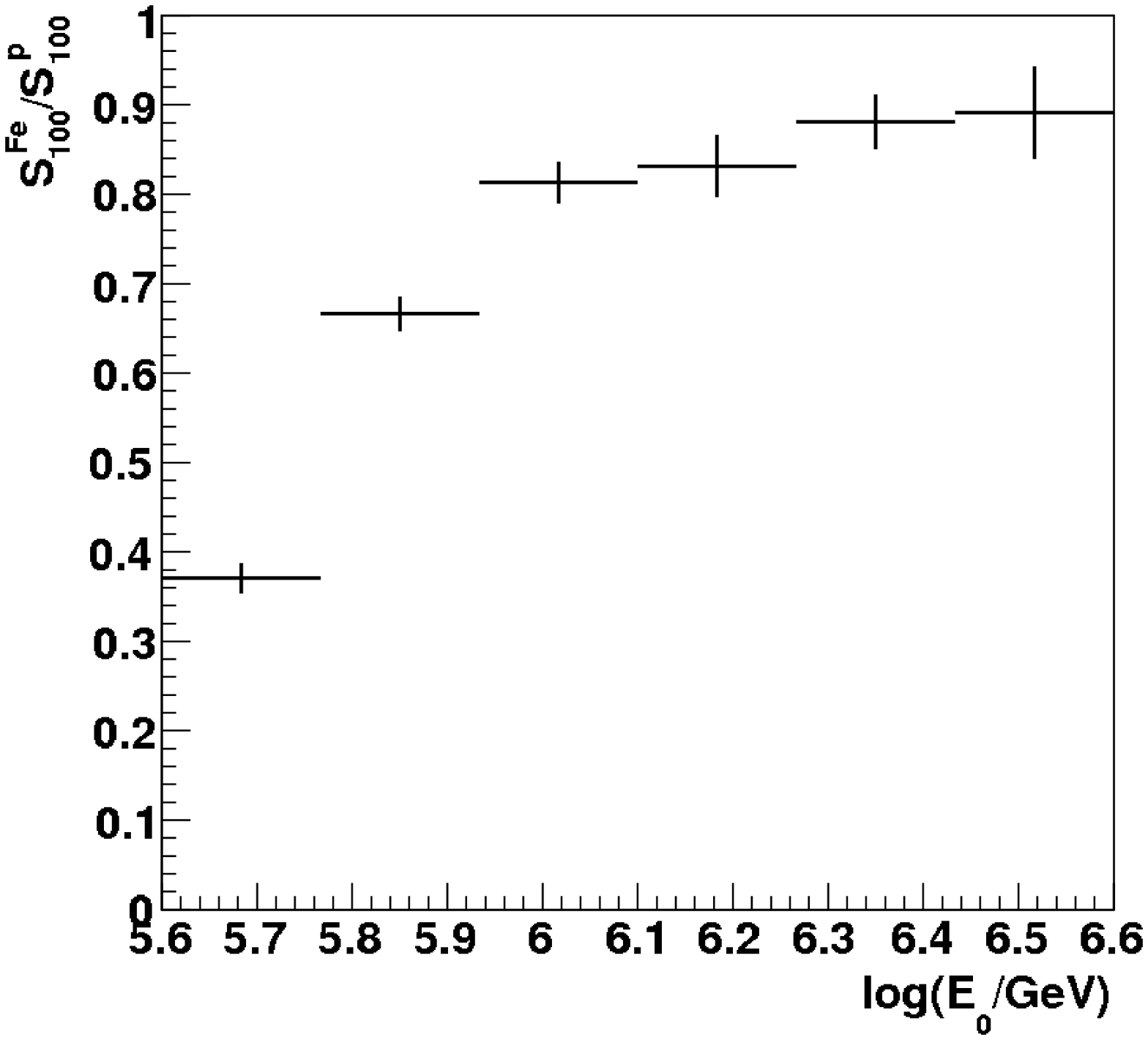}
\caption{Ratio of $S_{100}^{Fe}$ to $S_{100}^{p}$ as a function of the total energy per nucleus ($E_{0}$).}
\label{energy}
\end{center}
\end{figure}

\begin{equation}
f(r) = S_{100} \left( \frac{r}{100\mathrm{m}} \right) ^{-\beta -\kappa
\log(r/100\mathrm{m}) }
\end{equation}
where $r$ is a distance from shower core, $\kappa$ is 0.303 for hadronic 
showers, and $S_{100}$ is the signal in vertical equivalent muon (VEM) per 
tank at 100 m from the shower core. The parameter $\beta$ is roughly 
correlated with shower age via $s = -0.94 \beta  + 3.4$. $S_{100}$ is an 
energy estimator and depends on primary mass, as shown in Figure 
\ref{energy}. 

%Primary energy of 1 PeV corresponds roughly to log($S_{100}$) = 0.6 
%for IceTop-IceCube coincident events. However, $S_{100}$ also depends 
%on primary mass to some extent. 

The events which passed the following quality cuts are used in this study:

\begin{itemize}
\item Reconstructed shower core lands 60 m inside of IceTop array.
\item $\beta$ in Eq. (1) is less than 6.
\item Reconstructed zenith angle is less than 20$^\circ$.
\item The number of hit strings is greater than 1.
%\item Lateral distribution fit does not fail.
\end{itemize}

The number of hit strings is required to be equal to or greater than 2
since the lateral distribution fit in ice which will be described in the 
next session fails if a reconstructed track is vertical.

\section{Cosmic ray composition}

\begin{figure}
\begin{center}
\includegraphics[width=0.48\textwidth]{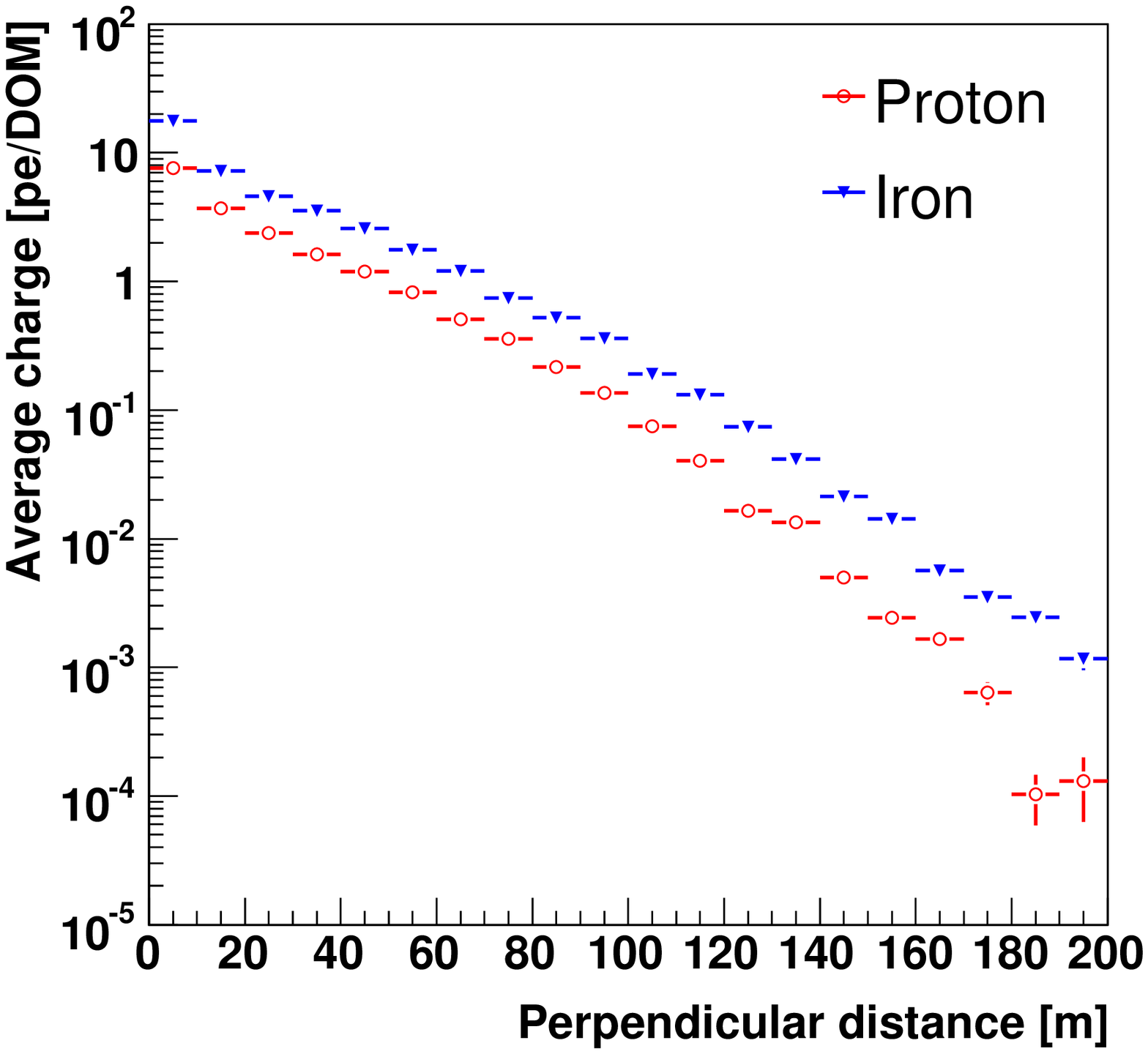} \caption{Average charge per in-ice DOM is shown as a function of a perpendicular distance from a primary track for proton and iron showers [0.5 $<$ log($S_{100}$) $<$ 1.3].}
\label{ldf}
\end{center}
\end{figure}

The IceCube detector is located deep in ice, so only muons can reach the 
detector, and useful information about primary cosmic rays can be inferred
from muon bundles with the 3-dimensional instrument. The total number 
of muons in a bundle is dependent on the type of primary nucleus. 
Cherenkov photons from the muon bundle are detected by optical sensors in 
ice, and the photon intensity is measured as a function of perpendicular
distance from a primary muon track and fitted by an exponential function. 
The primary muon track is the shower axis determined by the IceTop array.
Figure \ref{ldf} shows the average charge per in-ice DOM as a function of the 
distance from a primary track to each hit DOM in a range of $S_{100}$ 
between 0.5 and 1.3 showing separation between proton and iron showers.
It was found, for the SPASE/AMANDA detectors, that the photon intensity at 50 m 
($K_{50}$) is most sensitive to the mass of primary cosmic rays\cite{Kath}. 
Ranging-out of muons and depth dependence of light scattering in the 
ice are taken into account in the lateral distribution fit. However, these 
corrections are not made in Figure \ref{ldf}. 
%For IceCube, $K_{60}$ is %chosen simply because 60 m is half of string 
%spacing, but further studies are needed to choose the distance at which 
%the $K$ parameter is most stable. Figure \ref{s-k} shows that 
%$\log(K_{60})$ increases roughly linearly with $\log(S_{100})$ between 
%0.5 and 1.3. 
Once we find all observables sensitive to primary mass, we will feed them 
into a neural network (see \cite{Karen} for detailed description) for 
composition analysis. 

%\begin{figure}
%\begin{center}
%\includegraphics[width=0.48\textwidth]{icrc1294_fig03.eps} \caption{Log($K_{60}$) is shown as a function of log($S_{100}$) for proton and iron showers.}
%\label{s-k}
%\end{center}
%\end{figure}

Figure \ref{charge} shows the average charge as a function of DOM number
for proton and iron showers. Overall the average charge decreases with depth, 
featuring changes in the optical properties of ice. For instance, a thick 
dust layer observed by a dust logger during string deployment is seen around  
DOM 36. Figure \ref{chdist} shows the same as Figure \ref{charge} but 
with three different distance ranges only for proton showers, and 
indicates that using the hits close to muon bundles gives measurement
less dependent on ice properties. An appropriate correction for the dust 
layer needs to be made, or those DOMs around the dust layer can be removed 
in the analysis.

\begin{figure}
\begin{center} 
\includegraphics[width=0.48\textwidth]{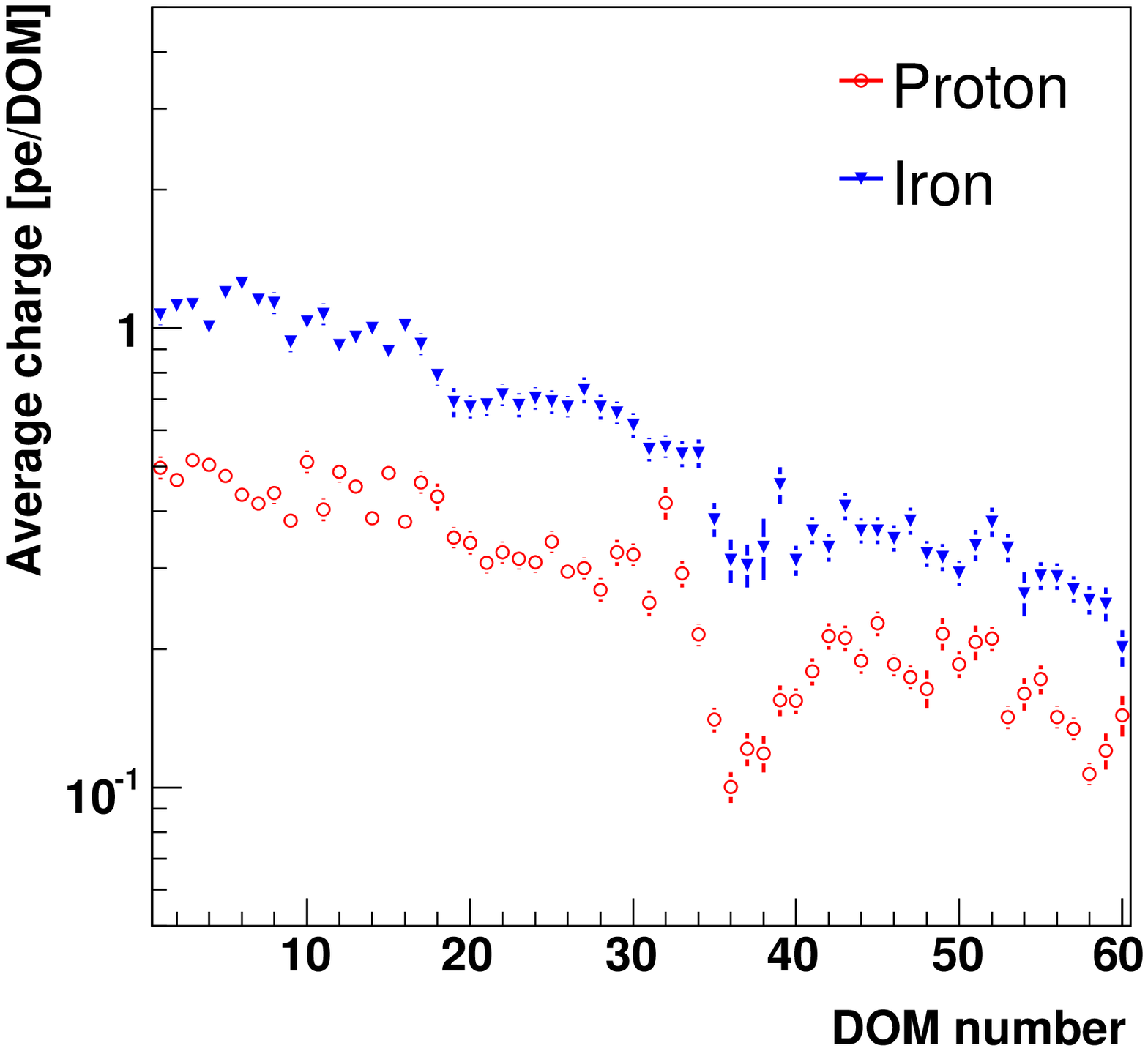}
\caption{Average charge vs. DOM number for proton and iron showers [0.5 $<$ log($S_{100}$) $<$ 1.3].}
\label{charge}
\end{center}
\end{figure}

\begin{figure}
\begin{center} 
\includegraphics[width=0.48\textwidth]{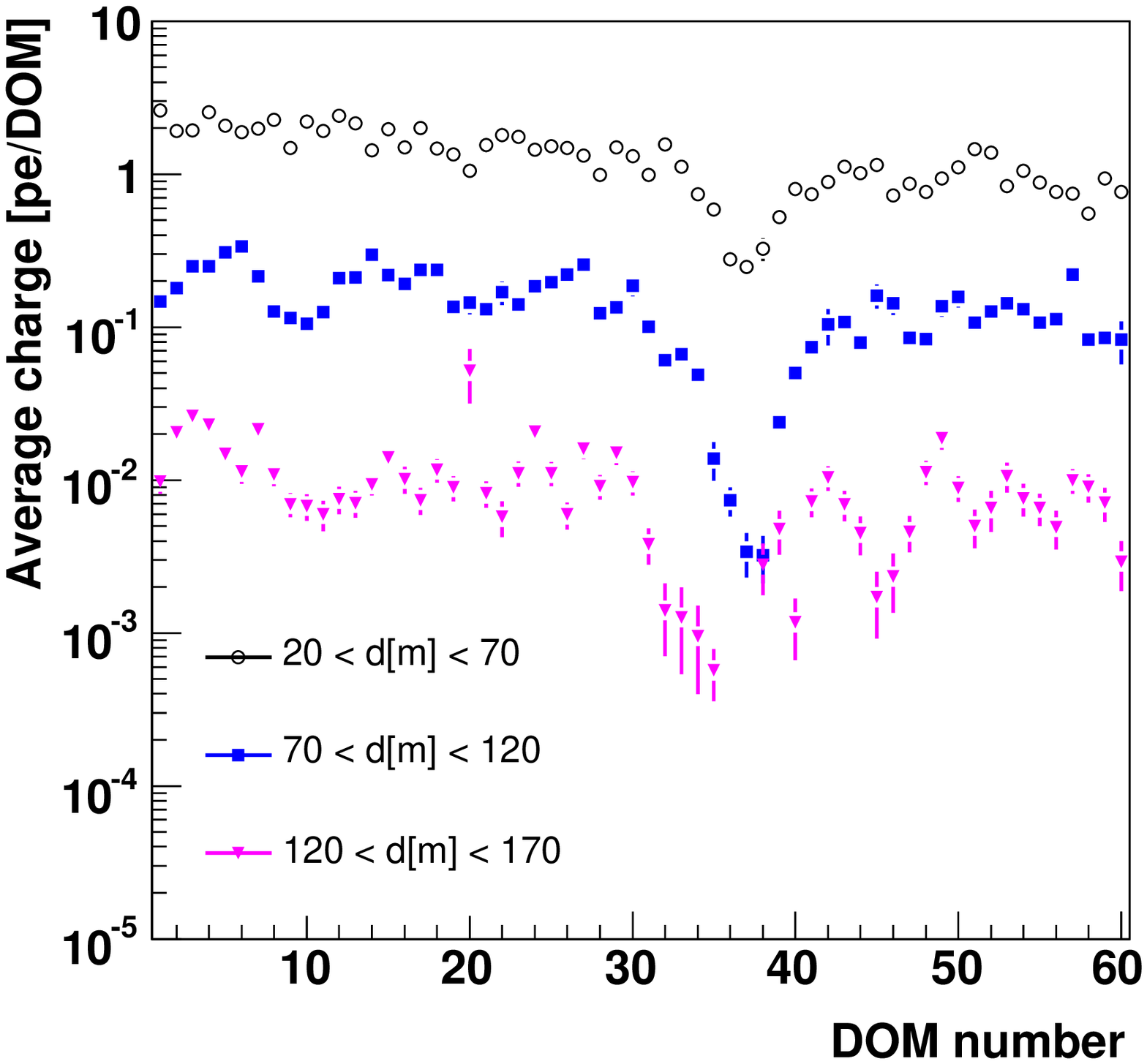}
\caption{Average charge vs. DOM number for proton showers only at different distance ranges [0.5 $<$ log($S_{100}$) $<$ 1.3].}
\label{chdist}
\end{center}
\end{figure}

\begin{figure*}
\begin{center}
\includegraphics[width=0.90\textwidth]{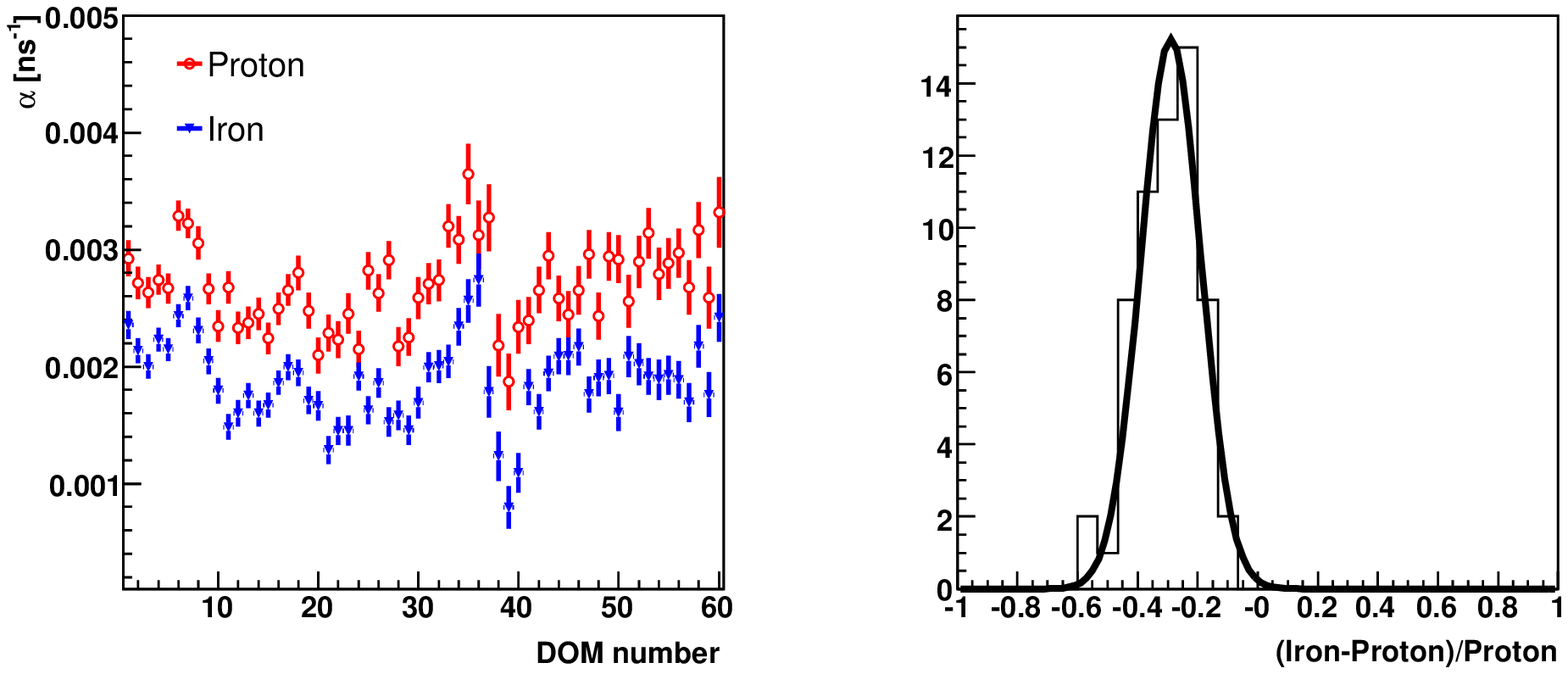}
\caption{Slope ($\alpha$) of the time residual distribution as a function of DOM number (left) and distribution of $(\alpha^{Fe}-\alpha^{p})/\alpha^{p}$ (right) are shown.}
\label{residual}
\end{center}
\end{figure*}

In addition to charge, we looked into timing information to see whether or 
not it is sensitive to primary mass. The size of the muon bundle 
depends on the type of the primary nucleus at a given energy and can affect 
the time residual (observed minus expected times from the primary muon 
track). The expected time is the travel time of a direct Cherenkov photon 
from the primary muon track to each hit DOM. The time residual distribution 
is fitted by $\exp(-\alpha t)$ 
from 50 to 400 ns where the tail of the distribution is straight in log 
scale, and the slope, $\alpha$, of the distribution as a function of DOM number 
is shown in Figure \ref{residual}.
Separation between proton and iron showers is seen, and the slope varies 
depending on depth of DOM and rises at dusty layers.

\section{Discussion}

We investigated observables sensitive to primary mass. In addition 
to charge information from the DOMs in ice, the slope of the time residual
distribution seems to be sensitive to the type of the primary cosmic ray, 
though it has dependence of optical properties of ice. However the 
dependence of ice properties can be reduced by making an appropriate 
correction for dusty layers or by excluding the DOMs in the thick dust 
layer around DOM 36. Moreover, DOMs close to a muon bundle appear to
be best suited for such an analysis. Once we have all observables 
sensitive to primary mass, the neural network can be employed for cosmic 
ray composition studies.

\section{Acknowledgments} 

This work is supported by the U.S. National Science Foundation, Grants 
No. OPP-0236449 and OPP-0602679 (University of Delaware).

%\end{document}

\setcounter{figure}{0}
\setcounter{table}{0}
%%
% International Cosmic Ray Conference 2007 Merida Yucatan Mexico
% In this file you will find detailed instructions to correctly
% typeset your document.
%
% By: Victor De la Luz
% vdelaluz@inaoep.mx
% Mexico City

%Class Required
%\documentclass{article}
%The ICRC Style
%(This package is the last package in the usepackage list)
%If you need import other package you need write it first.
%\usepackage{icrctc07,lineno}

%The paper title
\title{Search for TeV gamma-rays from point sources with SPASE2}
%Short title to print in the headers to the final publication (Not showed in this print).
\shorttitle{SPASE2 point source search}

%All paper authors
\authors{Kory James$^1$, X. Bai$^{1}$, T.K. Gaisser$^{1}$, Jim Hinton$^{2}$,
 Peter Niessen$^{1}$, 
 Todor Stanev$^{1}$, Serap Tilav$^{1}$, and Alan Watson$^{2}$ for the SPASE2
 and IceCube Collaborations$^*$} 
%Short title to print in the headers to the final publication (Not shown in this print).
\shortauthors{X.Bai and et al.}
%All the affiliations.
\afiliations{$^1$Bartol Research Institute, Department of Physics and Astronomy, University of Delaware, Newark, DE 19716, U.S.A.\\ 
$^2$School of Physics \& Astronomy, University of Leeds, LS2 9JT UK } 
\email{stanev@bartol.udel.edu, $^*$ see special section of these proceedings}

%The abstract.
\abstract{The South Pole Air Shower Experiment (SPASE2) began operation 
 in 1996 and took data until it was decommissioned in December
 2006. We are currently analyzing those of the 205 million reconstructed 
 events that were taken during the last five years. In this paper 
 we report on a search for 100 TeV gamma-rays from three 
 specific Southern hemisphere point sources discovered by 
 HESS. that may have gamma-ray spectra extending to energies
 higher than 50 TeV.} 

%\begin{document}
\maketitle

%%\begin{linenumbers}
\section{Introduction}

The SPASE2 scintillator array at the Amundsen-Scott South-Pole 
station is at an altitude of 2835 m.a.s.l., corresponding to a 
year-round average atmospheric overburden of $695\, {\rm g cm^{-2}}$. 
The total area within the perimeter of the array is 
$16,000\, {\rm m^{2}}$~\cite{spase2}.
For this search we use data taken during the last five years 
with livetime of 171+167+204+307+322=1171 days = 3.21 years. 

In this work, we focus on the following three HESS sources:\\
\ {\em a)} The shell-type supernova remnant RX J0852.0-4622~\cite{hess1}.
It has a spectrum observed 
in the energy range between 500 GeV and 15 TeV, which can be 
well described by a power law with a spectral index of 
2.1$\pm$0.1$_{stat}\pm$0.2$_{syst}$ and a differential flux at 
1~TeV of (2.1$\pm$0.2$_{stat}\pm$0.6$_{syst}$)$\times$10$^{-11}$ cm$^{-2}$s$^{-1}$
TeV$^{-1}$. 
The corresponding integral flux above 1 TeV was measured to be 
(1.9$\pm$0.3$_{stat}\pm$0.6$_{syst}$)$\times$10$^{-11}$ cm$^{-2}$s$^{-1}$. 
\\
\ {\em b)} The Supernova Remnant MSH 15-52. 
Its image~\cite{hess2} reveals an elliptically shaped emission 
region around the pulsar PSR B1509-58. The overall 
energy spectrum from 280 GeV up to 40 TeV 
can be fitted by a power law with spectral index 
$\alpha$=2.27$\pm$0.03$_{stat}\pm$0.20$_{syst}$ and a 
differential flux at 1 TeV of (5.7$\pm$0.2$_{stat}\pm$1.4$_{syst}$)
$\times$10$^{-12}$ TeV$^{-1}$cm$^{-2}$s$^{-1}$.
\\
\ {\em c)} The unidentified TeV $\gamma$-ray source close 
to the galactic plane named HESS J1303-631~\cite{hess3} is  
an extended source with a width of 
an assumed intrinsic Gaussian emission profile of 
$\sigma$ = (0.16$\pm$0.02)$^{o}$. The measured 
energy spectrum can be described by a power-law 
$\mathrm{d}N/\mathrm{d}E = N_{0}\cdot (E/TeV)^{-\alpha}$ with a 
spectral index of 
$\alpha$=2.44$\pm$0.05$_{stat}\pm$0.2$_{syst}$ 
and a normalization of N$_{0}$=(4.3$\pm$0.3$_{stat}$)$\times$10$^{-12}$
TeV$^{-1}$cm$^{-2}$s$^{-1}$.

\section{Energy estimate}

 The particle density at 30 meters from the shower
 core, $S_{30}$, is used by the SPASE2 experiment to
 estimate the primary particle energy.
 Monte Carlo simulations tell us that the $S_{30}$
 for 100 TeV $\gamma$-rays is higher than for 100 TeV
 proton. The Monte Carlo simulates cascades as well as
 the response of the air shower array using Corsika
 with the 2.1 version of the Sibyll~\cite{ICRC0678_Sibyll}
 interaction model.

 Currently a Monte Carlo estimate is available for all 
 showers with zenith angles between 20$^o$ and 50$^o$.
 For example, at $S_{30}$
 of 3 m$^{-2}$, $E_\gamma$ is about 120 TeV, while
 $E_p$ is 180 TeV. We will perform more simulations 
 to determine the energy dependence as a function of the
 zenith angle. 

\section{Angular resolution}

 The angular resolution of an air shower array is much worse
 that that of an air Cherenkov telescope. We have estimated
 the SPASE2 angular resolution in two different ways - 
 using the experimental data with sub-array comparison and
 with Mote Carlo calculations.
  
 In the sub-array approach the SPASE2 array is divided into two 
 parts. For each one the shower angle is estimated separately.
 The space angle between the two sub-arrays
 is used to study the angular resolution.

 Monte Carlo events after the standard shower reconstruction
 were also used to determine the angular resolution.
 The results from both methods fully agree with each other
 at higher energy. At threshold the sub-array approach 
 suffers from statistical fluctuations because there are
 not enough detectors that respond to the showers.

 Fig.~\ref{angle} shows the integral distribution of the square
 of the space angle difference between the true direction of the
 simulated shower and the reconstructed direction $\Psi^2$ for
 $\gamma$-ray showers with $S_{30}$$>$ 3 m$^{-2}$.
 The $\Psi^2$ value that contains 68\% of all events is (2.1$^o)^2$.
 For showers of $S_{30}$ $<$ 3 m$^{-2}$
 this number is (3.3$^o)^2$. Proton showers in both energy ranges
 show slightly worse angular resolution.

 \begin{figure}[thb]
    \includegraphics[width=0.45\textwidth]{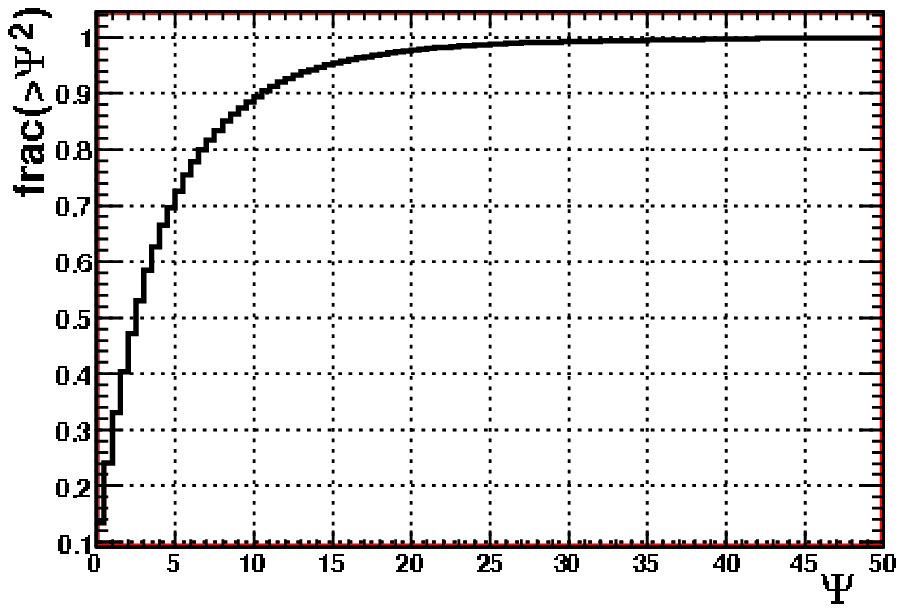}
    \caption{Integral distribution of the $\Psi^2$ values
     (in square degrees) derived
     from simulation of $\gamma$-induced showers.}
    \label{angle}
\end{figure}

\section {Systematic errors}

 There are several possible sources of systematic errors in
 the  data set.
 One is that at the beginning of 2002 the electronics of the
 shower ray was updated with a consequent increase of its
 threshold. For this reason we will first
 use the five years data taken after 2001.

 A second source is that the response of SPASE2 has 2\%
 variation with azimuth. Since the array typically has a
 lower duty cycle in the antarctic summer this could lead
 to a background that is not completely uniform in
 right ascension.

\section{The background}

 We studied the possible anisotropies by looking at the 
 scrambled RA distribution in different declination bins.
 Initially our data set was {\em blinded}. Scrambling was performed
 by shifting the real RA by a random amount. Figure~\ref{rasd}
 shows the rms value over the Gaussian expectation in Gaussian
 standard deviations $\sigma$ for zenith 
 angles from 20$^o$ to 50$^o$.
 In this case the average number of entries per bin is
 1.37 million and the standard deviation of Fig.~\ref{rasd}
 is 1.17$\times$10$^3$ showers.
 Out of 60 bins 38 bins show deviation by less
 than 1$\sigma$
 \begin{figure}[thb]
    \includegraphics[width=0.23\textwidth]{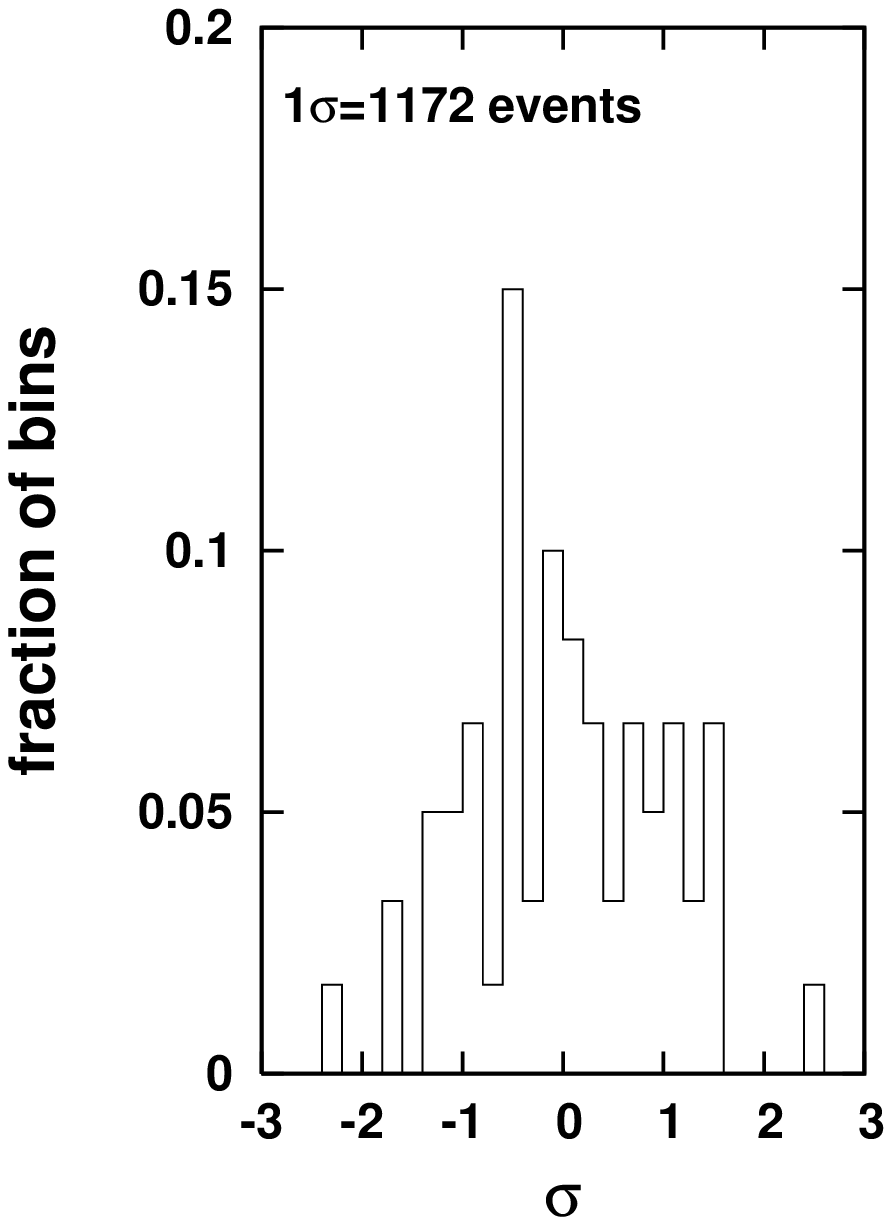}
    \includegraphics[width=0.23\textwidth]{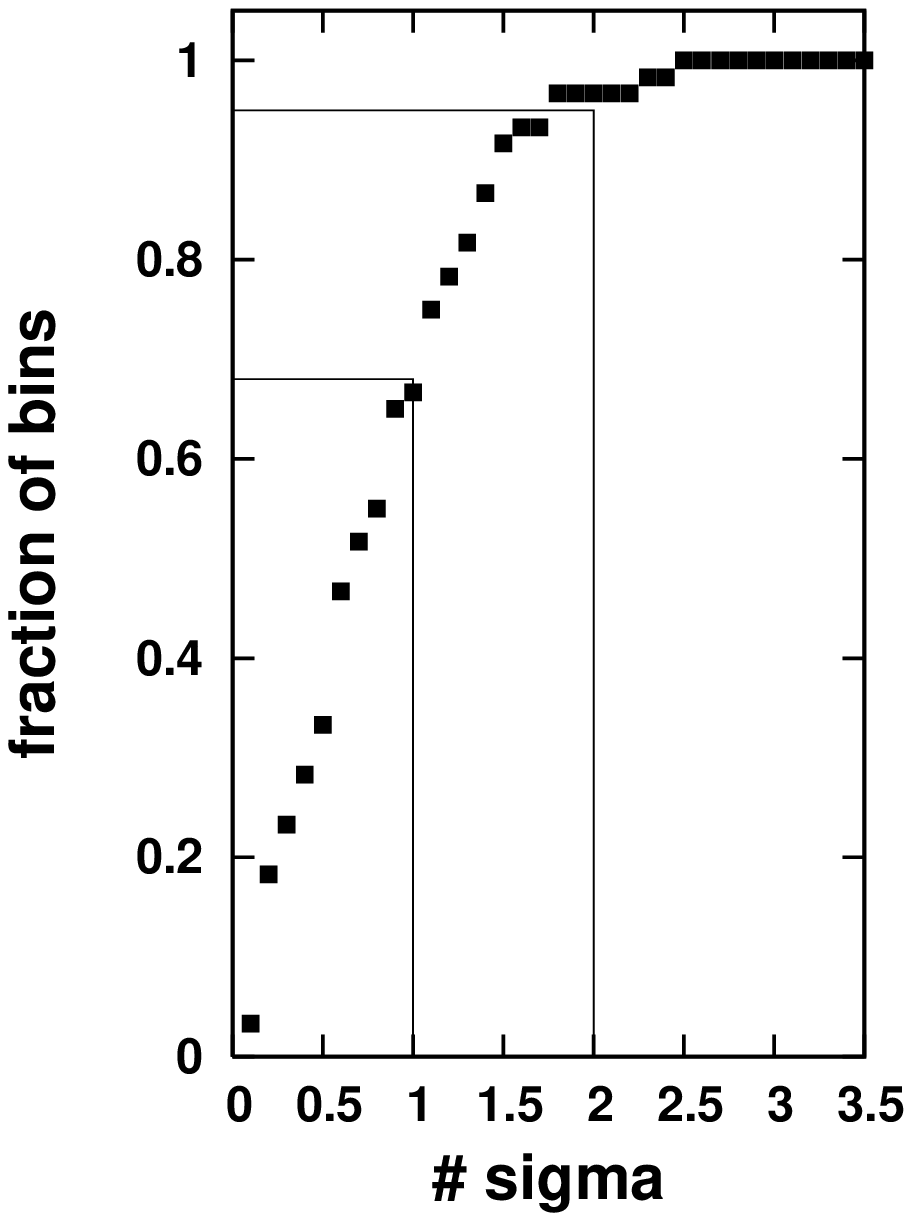}
    \caption{Left-hand panel: Distribution of the deviation
     from the average for 60 6$^o$ RA bins. Right-hand panel:
     Integral distribution in number of $\sigma$.}
    \label{rasd}
\end{figure}
 and 3 bins have deviations of more than 2$\sigma$ which 
 fully agrees with a Gaussian distribution.

 We also looked at these distributions for smaller zenith 
 angle bins similar to those that we will use in the source
 search. Fig.~\ref{nev} shows the scrambled RA distribution
 in 6$^o \times$6$^o$ bins for the zenith angle band of 
 41$^o$ to 47$^o$, which almost coincides with one of the 
 sources. The results are very similar to those for the 
 wider zenith angle band.
 \begin{figure}[thb]
    \includegraphics[width=0.23\textwidth]{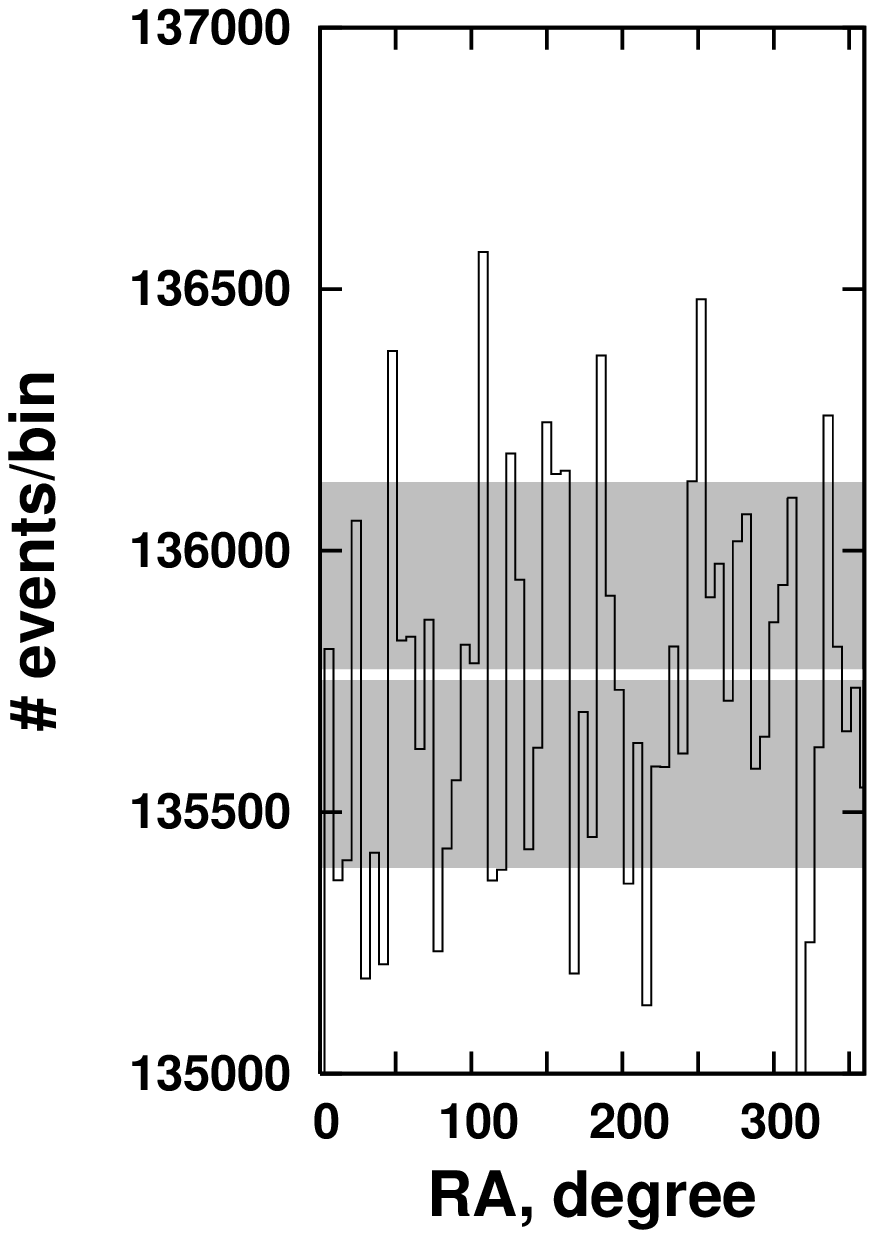}
    \includegraphics[width=0.23\textwidth]{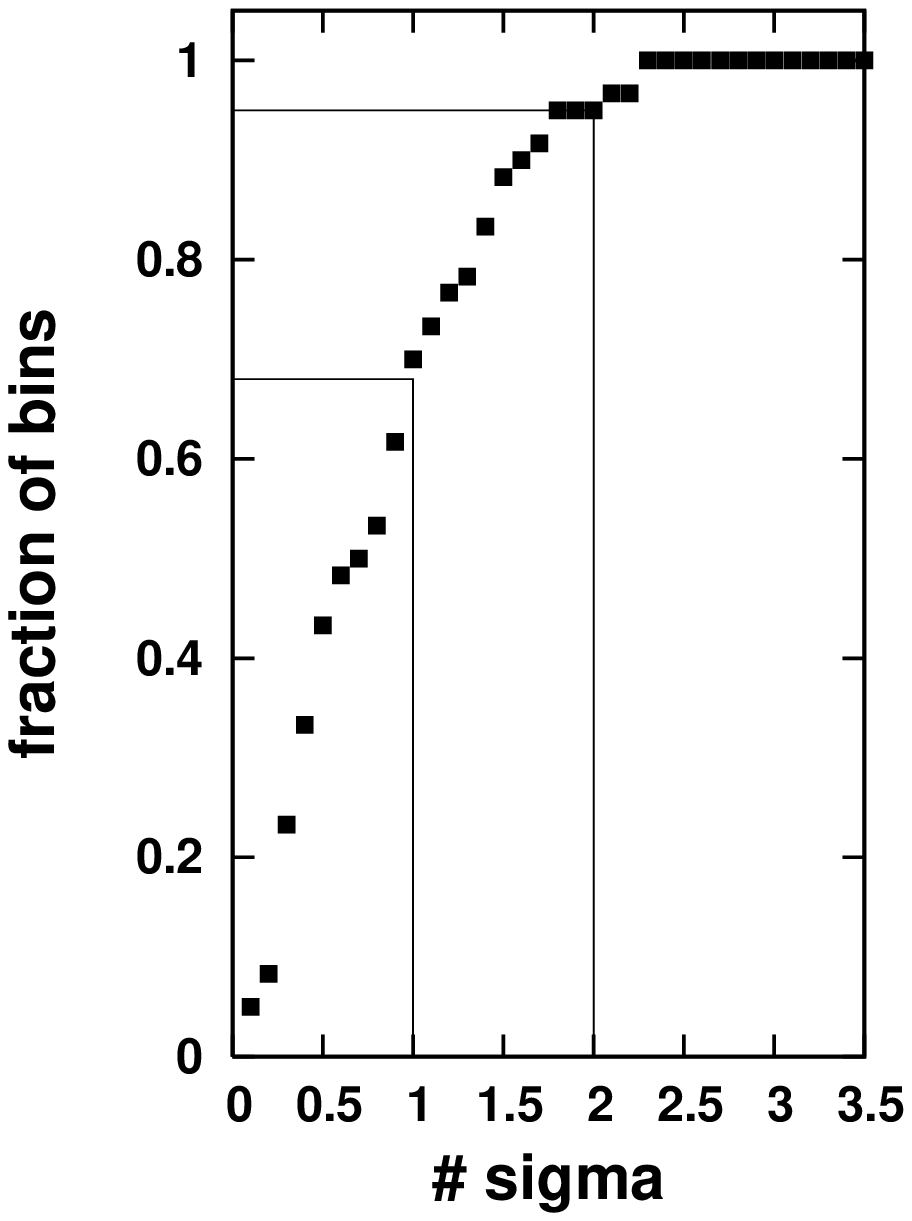}
    \caption{Left-hand panel: Number of events per bin in
     the declination band 41$^o$- 47$^o$. The average is shown
     with a white line and the shaded area represents
     $\pm$1$\sigma$. Right-hand panel:
     Integral distribution in number of $\sigma$ for the
     declination band.}
    \label{nev}
\end{figure}

\subsection{Angular bins} 

 The angular bins recommended for source search with air shower
 arrays~\cite{ngp} correspond to an elliptical region with 
 axes equal to 1.59$\sigma_0$ where $\sigma_0$ is the angular
 resolution of the detector. We decided to use equal solid angle
 which means that the major axis of the ellipses are bigger
 at low zenith angles. We will search separately for showers
 with $S_{30}$ higher and lower than 3 m$^{-2}$.
 The angular resolution for $S_{30}>$3 m$^{-2}$ is 2.1$^o$ and 
 is about 3.3$^o$ for lower energy showers.
 The search ellipses would be correspondingly wider for lower
 energy showers. The search ellipses for the three sources 
 and the two $S_{30}$ values are plotted in relative RA units 
 in Fig.~\ref{search}.
 \begin{figure}[thb]
    \includegraphics[width=0.45\textwidth]{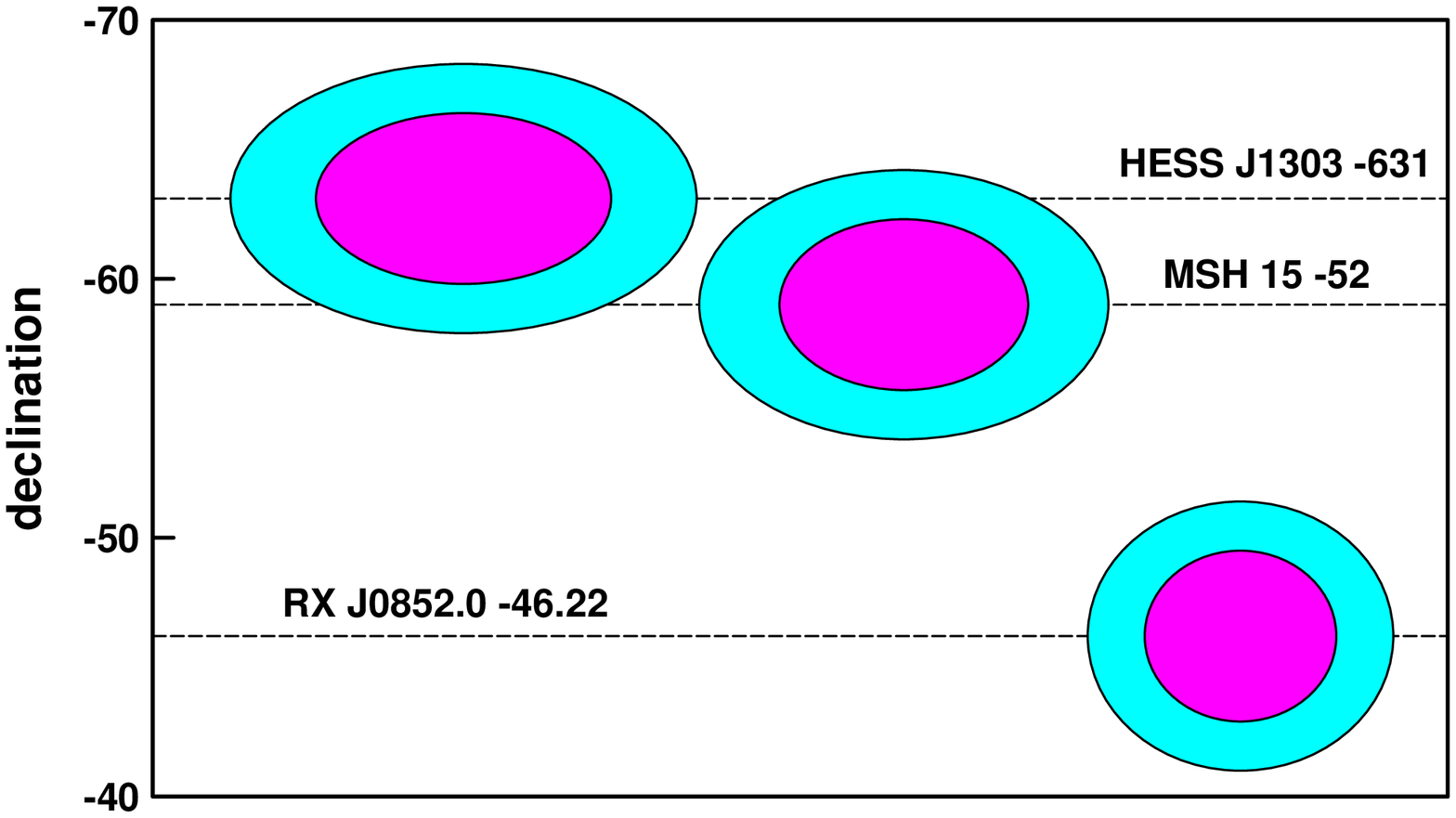}
    \caption{Relative sizes of the search ellipses for the
     three sources and the two \protect$S_{30}$ values -
     light shading is for \protect$S_{30}<$3 and the dark
     shading is for \protect$S_{30}>$3.}
    \label{search}
\end{figure}
 Since the angular area of these bins (and correspondingly
 the number of background events in them) is higher than those
 used in the previous section the expected detection probability 
 is slightly different.

\section{Signal expectations} 
 
 Because of its flat energy spectrum the source  RX J0852.0-4622
 offers the highest chance for detection if its spectrum does
 not cut off. It is, however at the highest zenith angle of the
 3 sources studied. We will first look at the 2005 data set.
 Assuming conservatively the area of SPASE2 to be 10$^8$ cm$^2$
 and  livetime in 2005 of 2.65.10$^7$ s, we expect to have
 321 (149) events above $E^{thr}_\gamma$ 100 (200) TeV.
 At zenith angle of 43.8$^o$ this would roughly correspond to 
 $S_{30}$ values of 1 and 3 m$^{-2}$. There may be some contribution
 from lower energy gamma ray showers but the array efficiency
 below 100 TeV is less than one and we need further Monte Carlo
 studies to estimate it.

 The backgrounds estimated from the two search ellipses for
 RX J0852.0-46.22 (excluding the source bins) are respectively
 38656 (13739) per bin for $S_{30} <$3 ($S_{30} >$3). The background
 for the lower energy showers is higher because of the much steeper
 cosmic ray spectrum compared to the $\gamma$=1.1 for the source.
 The expected number of gamma showers thus corresponds to 
 0.88$\sigma$ for  $S_{30} <$3 and 1.27$\sigma$ for $S_{30} >$3.
 SPASE2 is not, by far, the best detector for $\gamma$-ray 
 astronomy, but the chance of detection is reasonable for a flat
 source spectrum and no cut off.

 The other two $\gamma$-ray sources are less intense and can
 produce not more than several tens of events even if their spectra
 do not cut off. For this reason we will present only the results 
 for RX J0852.0-46.22.

\section{Results from the 2005 search}
   
 Figure~\ref{RXJ} shows the observed number of showers
 from the direction of  RX J0852.0-46.22 in the
 2005 data set (which we unblinded first) for the two $S_{30}$
 values. Note that the bins do not cover the whole 24 hours of RA
 in the zenith angle band because of the requirement for equal
 space angle bins. The missing phase space is always less than
 one bin width. 
 \begin{figure}[thb]
    \includegraphics[width=0.45\textwidth]{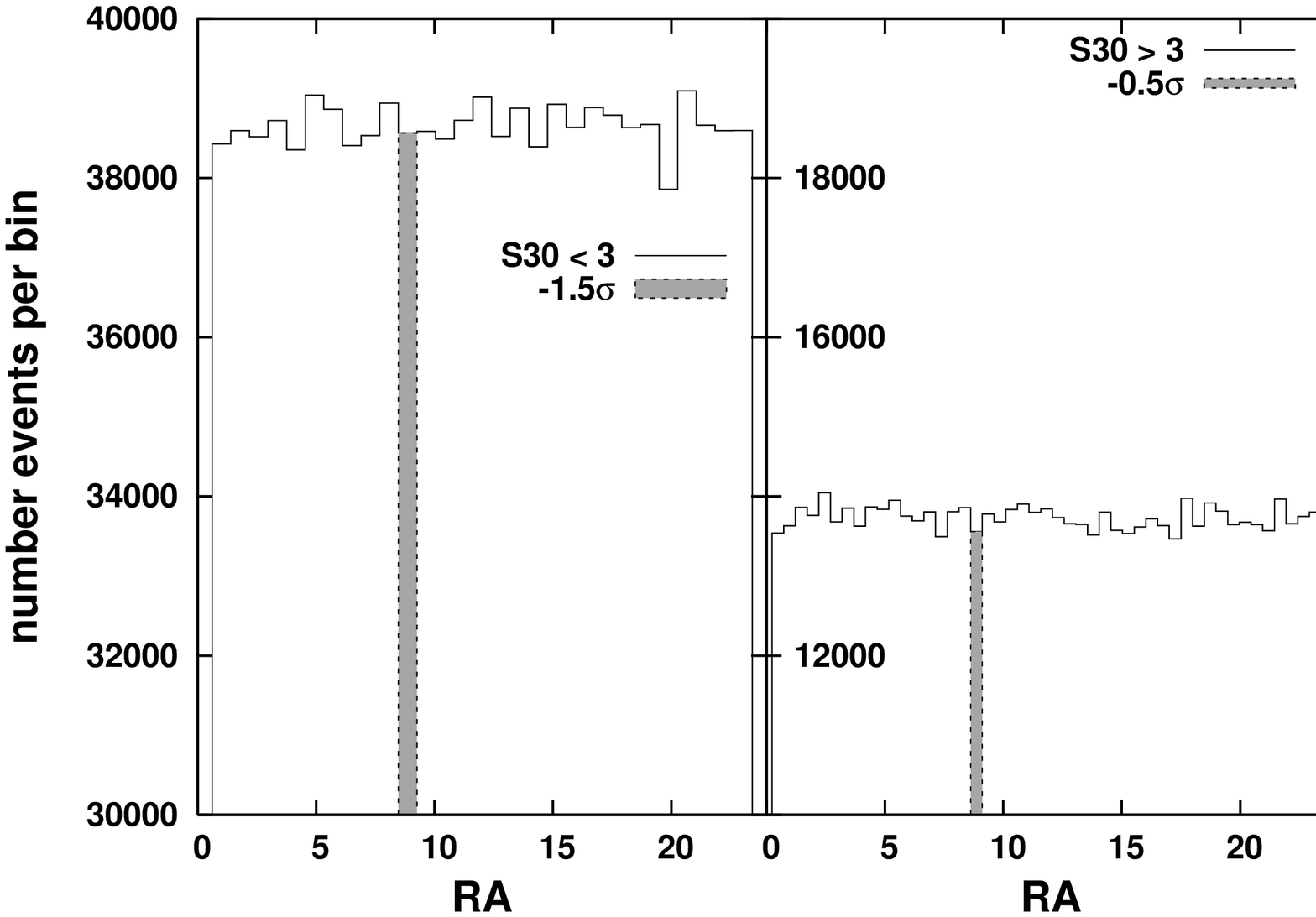}
    \caption{Observed number of showers from the position of the
     source RX J0852.0-4622 for the two energy bins.}
    \label{RXJ}
\end{figure}
 Both searches give negative results. In the $S_{30} <$3 sample 
 we see -1.5$\sigma$ from the average expected background.
 In the higher energy range the lack of events is smaller 
 (-0.5$\sigma$). 

\section{Conclusion}
  The search for 100 TeV $\gamma$-ray signal from   RX J0852.0-46.22
 in the SPASE2 data set for 2005 gave negative results - we did not
 observe any showers above the expected cosmic ray background.
  However, based on the preliminary simulation used here to 
 relate $S_{30}$ to primary energy, we find a limit based on one year
 data that is nearly inconsistent with the continuation of the
 spectrum of RX J0852.0-46.22 to 100 TeV without a steepening of its
 spectrum. We therefore plan to pursue this analysis and to 
 search separately in all five years data and then combine the results,
 possibly using a more sensitive unbinned search.
 We will use a detailed simulation of $\gamma$-ray and cosmic ray
 showers appropriate for the declination of this source which
 corresponds to a zenith angle of 43.8$^o$.
 
\section{Acknowledgments} 
The work is supported by the US National Science Foundation under 
Grant Nos. OPP-9601950, ANT-0602679 and OPP-0236449, University of 
Wisconsin-Madison, and from the U.K. Particle Physics and Astronomy 
Research Council. The authors gratefully acknowledge the support 
from the U.S. Amundsen-Scott South Pole station.

%%\end{linenumbers}

%\end{document}

\setcounter{figure}{0}
\setcounter{table}{0}
%%
% International Cosmic Ray Conference 2007 Merida Yucatan Mexico
% In this file you will find detailed instructions to correctly
% typeset your document.
%
% By: Victor De la Luz
% vdelaluz@inaoep.mx
% Mexico City

%Class Required
%\documentclass{article}
%The ICRC Style
%(This package is the last package in the usepackage list)
%If you need import other package you need write it first.
%\usepackage{icrctc07}

%The paper title
\title{Study of High $p_T$ Muons in Air Showers with IceCube}
%Short title to print in the headers to the final publication (Not showed in this print).
\shorttitle{High $p_T$ muons}

%All paper authors
\authors{Spencer R. Klein$^{1}$ and Dmitry Chirkin$^{1}$ for the IceCube Collaboration$^2$ }
%Short title to print in the headers to the final publication (Not shown in this print).
\shortauthors{S. Klein et al.}
%All the affiliations.
\afiliations{$^1$Lawrence Berkeley National Laboratory, CA, 94720 USA\\
$^2$ see the special section of these proceedings\\}
\email{srklein@lbl.gov}

%The abstract.
\abstract{With it's 1 km$^2$ area, IceCube and the associated IceTop
surface detector array are large enough to study high $p_T$ muon production in
air showers.  The muon $p_T$ will be determined from the muon energy
and it's distance from the core.  A few thousand high $p_T$ muons are
expected to be observable each year in the full array.  The flux of
high $p_T$ muons may be computed using perturbative QCD calculations;
the cross-section is sensitive to the composition of the incident particles.}

%%%%%%%%%%%%%%%%%%%% B E G I N   D O C U M E N T%%%%%%%%%%%%%%%%%%%%%%%
%\begin{document}
\maketitle
%Begin the section.
\section{Introduction}

The number of muons produced in cosmic-ray air showers is sensitive to
the nuclear composition of the incident particles.  Previous studies
of the cosmic-ray composition have used relatively low ($\approx
1$ GeV) or high ($\approx 1$ TeV) energy muons.  These studies relied on 
muon counting.  Relating the muon count to the composition requires a model
for the hadronic interactions; most of the muons come from $\pi/K$ decay;
the bulk of the mesons are produced at low transverse momentum ($p_T$) with
respect to the direction of the incident particle.  The production of these
low $p_T$ particles cannot be described in perturbative QCD (pQCD), so 
phenomenological models must be used.

In contrast, the production of particles with $p_T >\sim 2$ GeV/c is calculable in pQCD.
We label these tracks high $p_T$ particles, and consider their
production in cosmic-ray air showers.
High $p_T$ muons come from the decay of charm and bottom quarks, and from
$\pi/K$ produced in jets.  Both of these processes can be described by
pQCD, allowing for calculations of the energy and $p_T$ spectra
for different incident nuclei.  The predictions depend sensitively on the
composition of the incident nuclei - neglecting shadowing, a nucleus with
energy $E$ and atomic number $A$ has the same parton distribution as $A$ nucleons,
each with energy $E/A$.  Nuclei with $A=1$ and $A=10$ have very different parton 
energy spectra.  

\section{High $p_T$ muons in Air Showers}

Previous studies of high-energy muons associated with air showers have involved 
relatively small detectors. AMANDA has measured muon bundles near the shower core,
but did not study the muon lateral distribution \cite{AMANDA}.  MACRO measured
the muon decoherence function for separations up to 65 m \cite{MACRO}.  The most
likely pair separation was  4m; only 1\% of the pairs have a separation greater
than 20 m. MACRO simulated air
showers and studied the pair separation as a function of the $p_T$ of the mesons
that produced the muons. The MACRO analysis established a
clear linear relationship between muon separation and $p_T$;  the mean
$p_T$ rose roughly linearly with separation, from 400 MeV/c at zero separation up
to 1.2 GeV/c at 50 meter separation.

IceCube will observe both high-energy muons and the associated surface air showers that 
accompany them. For muons with energy $E_\mu$ above 1 TeV, the muon energy is proportional
to the specific energy loss ($dE/dx$) that is measured by the deep detectors; 
the muon energy resolution is about 30\% in $\log_{10}(E_\mu)$ \cite{mmc,mmc2}.

The muon energy and distance from the shower core can be used to find
the $p_T$ of a muon \cite{ISVHECRI}:
\begin{equation}
d = \frac{h p_T}{E_\mu}.
\end{equation} 
Here, $h$ is the height of the primary cosmic-ray interaction in the
atmosphere.  $h$ follows an exponential distribution and depends somewhat on the cosmic-ray
composition. A full analysis would include these effects.  Here, we take $h=30$ km.

Secondary interactions (of particles produced by the first interaction)
are expected be only a small contribution to the high-energy flux, contributing 
at most $15\%$ of the muons \cite{prompt2}.  For muons
far from the core, multiple scattering is expected to be a small contribution $d$.

Here, we consider showers where the core is inside the 1 km$^2$ area of IceTop, and muons
following the core trajectory are inside the IceCube physical volume.  This corresponds
to about 0.3 km$^2$ sr total acceptance.  

IceCube will detect air showers above an energy threshold of about 300 TeV; for vertical showers,
the minimum muon energy is about 500 GeV.  The  rate for triggered IceTop-InIce coincidences for
the 9-string IceCube array is about 0.2 Hz \cite{bai3d}, or
about 6 million events/year.  The full 80-strings $+$ stations should produce a rate more than
an order of magnitude higher.  

For vertical showers with energies above 1 PeV, the core location is found with a resolution of about
13 meters, and the shower direction is measured to about 2 degrees
\cite{gaisser}.   This allows the core position to be extrapolated to 1500 m in depth with 
an accuracy of about 55 meters, corresponding to a $p_T$ uncertainty of 1.6 GeV/c for a 1 TeV muon.

Most of these air showers are accompanied by a muon bundle.  A high $p_T$ analysis
will select events with a muon (or bundle) near the core, and another muon 
at a large distance from it.  The near-core muon(s) can be used to refine
the core position, avoiding the extrapolation error.  The muon
positions at a given depth can be determined within a 10-20 meters, allowing for better
$p_T$ resolution.  
Figure \ref{fig:example} shows an example of an IceCube 22-string event that contains an
air shower that struck  IceTop stations, plus muon bundle.  Although the bulk of
the bundle follows the shower direction, as projected from IceTop, there is
a well-separated light source, consistent with a high $p_T$ muon, about 400 meters
from the bulk of the muon bundle.  This secondary track hits 12 DOMs on a single string. 

\begin{figure}
\begin{center}
\includegraphics[height=0.5\textheight]{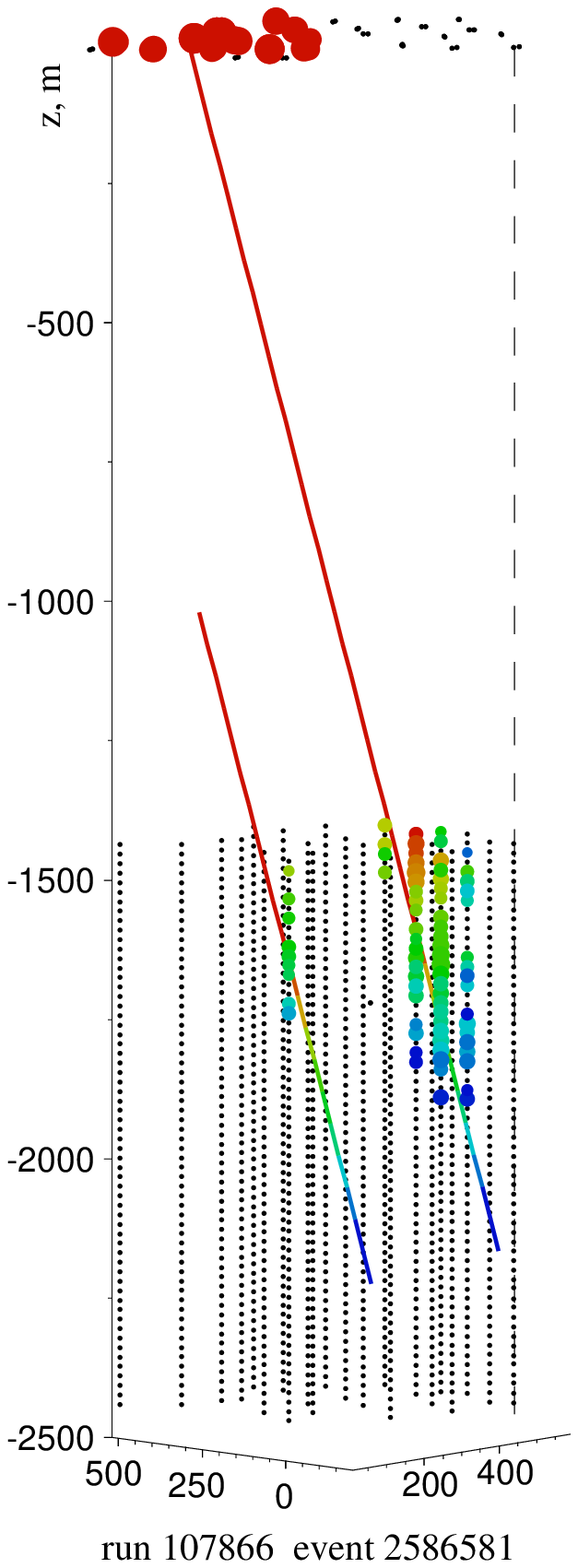}
\end{center}
\caption{An IceTop air shower accompanied by a muon bundle including an
apparent well-separated track.  The air shower hits 11 surface stations
(top of diagram).  A total of 96 IceCube DOMs are hit;
84 DOMs on four strings near the extrapolated air shower direction, plus 12
DOMs on another string, about 400 m from the projection.
\label{fig:example}
}
\end{figure}

For this analysis, the key performance issue is two-track resolution.  This 
remains to be determined.   However, the 125 m string spacing and the comparable (depth-dependent)
light absorption length set the scale for two-track resolution. Two muons 100 meters apart in IceCube
will largely deposit light in different strings; for a DOM near one muon, the first light 
from the farther muon will arrive about 500 nsec after the first light from the nearby muon.  If
the second muon (or muon bundle) is bright enough to illuminate a DOM 100 meters away, this late 
light will be temporally separate from that from the nearby muon.  A minimum ionizing
muon is not bright enough to be visible 100 meters away, but muon bundles may be.  
Here, we estimate that IceCube can reconstruct muon pairs that are separated by 100 meters; 
smaller separations may be possible with optimized tracking. 

For a muon with energy of 1 TeV, 100 meters separation corresponds to a $p_T$ of
3 GeV/c.  For a fixed separation, the
minimum $p_T$ rises linearly with muon energy, reaching $p_T > 150$ GeV/c for a 
50 TeV muon.  The highest energy muons are likely to come from high energy showers; 
the additional light will improve position reconstruction, and may allow for reconstruction at
smaller separation distances.
Still, there are unlikely to be useful events at higher energy/$p_T$.

\section{Rates}

High $p_T$ muons come from two sources: prompt muons from charm/bottom decays, and 
non-prompt muons from decays of high $p_T$ pions and kaons.  The charm  rates
have been discussed previously \cite{ISVHECRI,prompt}, about 600,000 muons per year with 
energy above 1 TeV are expected in the 0.3 km$^2$ acceptance.    Only 1-2\% of
these muons will have $p_T > 3$ GeV/c.  Still, this is a useful signal.  

Bottom quark production in air showers has received much less attention.  Although
$b\overline b$ production in air showers is only about 3\% of $c\overline c$
\cite{MRS}, the higher quark mass changes the kinematics, increasing the
importance of $b\overline b$ production at high $p_T$.   
At LHC energies, about 10\% of the muons from $b\overline b$ should satisfy
the $p_T > 3$ GeV/c cut, and, at high enough $p_T$, they should be the
dominant prompt contribution \cite{LHC}.

Although they are far more numerous than prompt muons, non-prompt muons have a much softer
$p_T$ spectrum. 
Non-prompt production may be estimated by using the measured $p_T$ spectrum
from $\pi$ produced in high-energy collisions.  
The PHENIX collaboration has parameterized their $\pi^0$ spectrum at mid-rapidity
with a power law: $dN/dp_T \approx 1/(1+p_T/p_0)^n$, where $p_0=1.219$ GeV/c, and
$n=9.99$ \cite{PHENIX}; about 1 in 200,000 $\pi^0$ has $p_T>3$ GeV/c.   This data is at 
mid-rapidity, while most muons seen in air
showers come from far forward production.  LHC will provide good data on forward
particle production at the relevant energies.  Here, we neglect this
difference and ignore the minor differences
between $\pi^0$ and $\pi^\pm\rightarrow\mu^\pm$ and 
$K^\pm\rightarrow\mu^\pm$ spectra.  With the acceptance discussed above,
IceCube expects to see more than 100 million muons/year associated with air showers, including
at least 500 of them with $p_T > 3$ GeV/c.

Overall, based on the standard cosmic-ray models, we expect $\approx 1,000-3,000$ muons 
with $p_T > 3$ GeV/c year.

\section{Muon spectral analysis \& Composition Analysis}

The 'cocktail' of charm, bottom and non-prompt muons is not so different
from that studied at RHIC \cite{RHICe}\cite{RHICmu}; the prompt fraction is
also not too different.  There, the muon $p_T$ spectrum is fitted to a mixture 
of prompt and non-prompt sources. In air showers, the accelerator beam is
unknown; it constitutes the initial object of study. 

The rate of high $p_T$ muons is sensitive to the cosmic-ray composition.  
High $p_T$ particles are produced in parton-parton collisions, and, as
Fig. \ref{fig:parton} shows, the
parton densities of a $10^{17}$ eV proton and of a $10^{17}$ eV $A=10$
nucleus are quite different.  In contrast to the usual presentation, these
are normalized to the parton energies, although the per-nucleon energies
are different for the two cases.  The nuclear distribution
cuts off at an energy of $10^{17}/A$ eV, limiting the maximum parton-parton
center of mass energy, and thereby constraining the possible muon kinematics.
Because of this, the yield of high-energy, high $p_T$
particles is much higher for protons than for heavier nuclei.  

Most of the muons seen by IceCube are produced in the forward region, where
a high$-x$ parton from the incident nucleus interacts with a low$-x$ parton from
a nitrogen or oxygen atom in the atmosphere.   The maximum muon energy is the
incident parton energy $E_p=x_pE_c$ where $x_p$ is the parton energy fraction and
$E_c$ is the cosmic-ray energy.  

In the far-forward limit, the
incident parton energy $x_{inc} = E_p/E_{incident} \approx E_\mu/E_{shower}$.
So, these muons are quite dependent on the high$-x$ partons that are
sensitive to nuclear composition.
\begin{figure}[t]
\begin{center}
\includegraphics [height=0.52\textheight,clip]{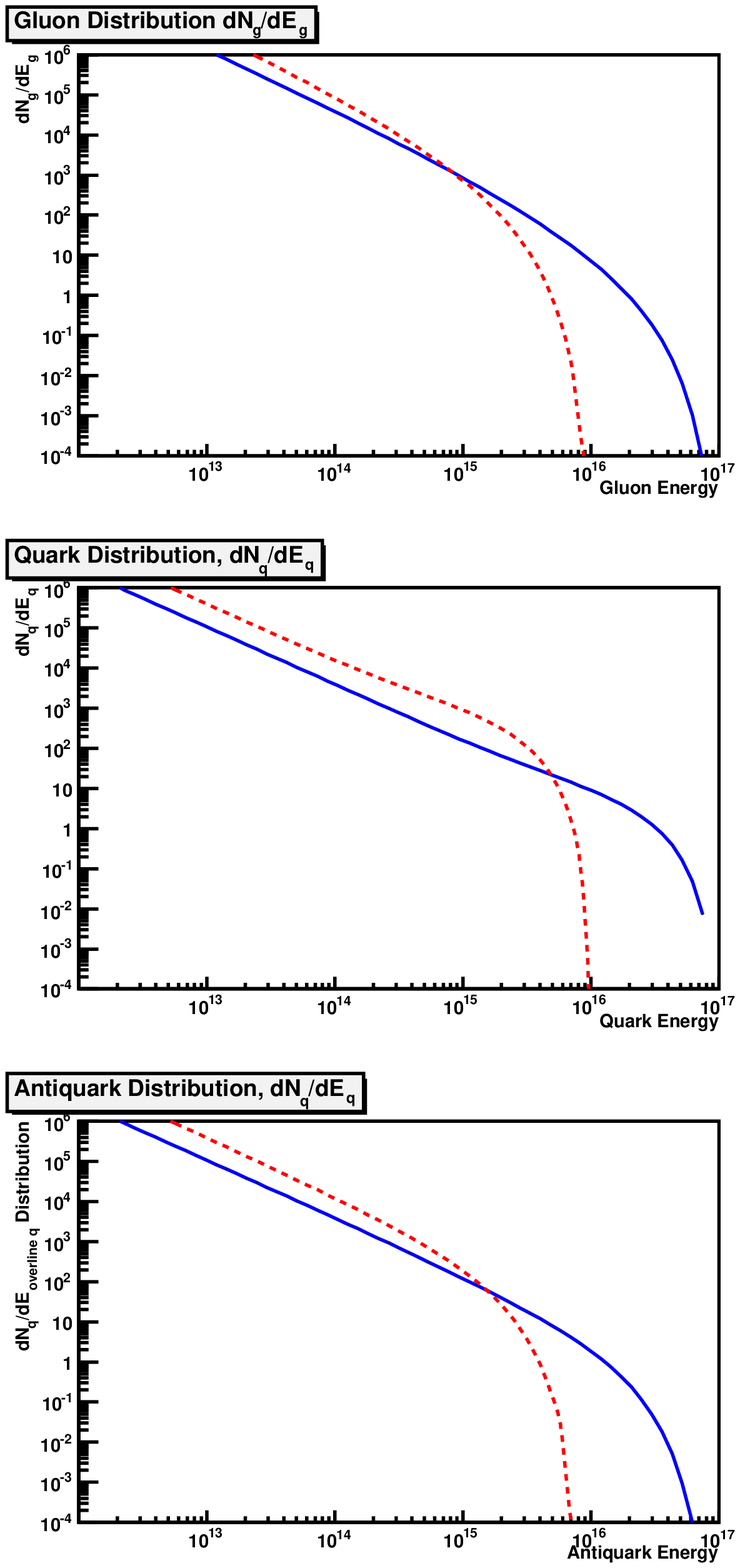}
\end{center}
\caption{Quark (top), gluon (middle) and antiquark (bottom) densities ($dN_{parton}/dE_{parton}$) for
a proton (solid line) and an $A=10$ nucleus (dashed lines).  The curves are based
on the MRST99 parton distributions \cite{partons} evaluated at $Q^2=1000$ GeV$^2$.
Nuclear shadowing is neglected.}
\label{fig:parton}
\end{figure}

\section{Conclusions}

IceCube is the first detector large enough to study high $p_T$ muon production
in cosmic-ray air showers.  A 100 meter minimum muon-shower core separation
would allow the study of muons with $p_T>3$ GeV/c; a few thousand of these muons
are expected each year.

By measuring the energy and core separation of muons
associated with air showers, the muon $p_T$ can be inferred.  The cross-sections
for high-$p_T$ muon production can be related to perturbative QCD calculations of
cosmic-ray interactions.  The rate of high $p_T$ muon production is very sensitive to
the cosmic-ray composition; pQCD based composition measurements offer an alternative to
existing cosmic-ray composition studies. 

We thank the U.S. National Science Foundation and Department of Energy, Office
of Nuclear Physics, and the agencies listed in Ref. \cite{albrecht}.

%\end{document}

\setcounter{figure}{0}
\setcounter{table}{0}
%%
% International Cosmic Ray Conference 2007 Merida Yucatan Mexico
% In this file you will find detailed instructions to correctly
% typeset your document.
%
% By: Victor De la Luz
% vdelaluz@inaoep.mx
% Mexico City

%Class Required
%\documentclass{article}
%The ICRC Style
%(This package is the last package in the usepackage list)
%If you need import other package you need write it first.
%\usepackage{icrctc07}

%The paper title
\title{IceTop/IceCube coincidences}
%Short title to print in the headers to the final publication (Not showed in this print).
\shorttitle{IceTop/IceCube}

%All paper authors
\authors{Xinhua Bai, Thomas Gaisser, Todor Stanev,
 \& Tilo Waldenmaier for the IceCube Collaboration $^*$} 
%Short title to print in the headers to the final publication (Not shown in this print).
\shortauthors{X.Bai et al.}
%All the affiliations.
\afiliations{Bartol research Institute, Department of Physics and Astronomy,
 University of Delaware, Newark, DE 19716, U.S.A.}
\email{bai@bartol.udel.edu; $^*$ see special section of these proceedings}

%The abstract.
\abstract{Atmospheric muons in IceCube are often accompanied
by air showers seen in IceTop when their trajectories
pass near the surface detectors.  By selecting events in
which only a single IceTop station on the surface is hit, we
can identify a class of events with high probability of having
a single muon in the deep detector.
In this work we use this tagged sample of atmospheric
muons as a calibration beam for IceCube.
}
%\begin{document}
\maketitle
%\begin{linenumbers}

\section{1. Introduction}
\vspace{-0.2cm}
In 2006 IceCube collected data with 
sixteen IceTop stations and nine in-ice strings,
as shown in Fig.~\ref{fig1T}. 
Ten more stations and thirteen more strings were 
deployed in 2006-07 austral summer~\cite{tomicrc07}. 
IceTop runs with a simple multiplicity trigger
that requires 6 or more digital optical modules (DOMs)
to have signals above threshold.  The configuration of
gain settings and DOMs in tanks is such that
IceTop triggers normally involve three or more
stations separated from each other by 125~m. 
Such showers typically have energies of several hundred TeV
and higher.  
The deep IceCube strings also have a simple multiplicity trigger
of 8 or more DOMs within 5~$\mu$sec.  The 8 DOMs need not be on the same
IceCube string.  Whenever there is an in-ice trigger, all IceTop DOMs
are read out for the previous 8~$\mu$sec.  This allows the possibility
of identifying small, sub-threshold showers on the surface in coincidence with muons
in deep IceCube.

\begin{figure}[t]
    \includegraphics[height=6.5cm,width=0.45\textwidth]{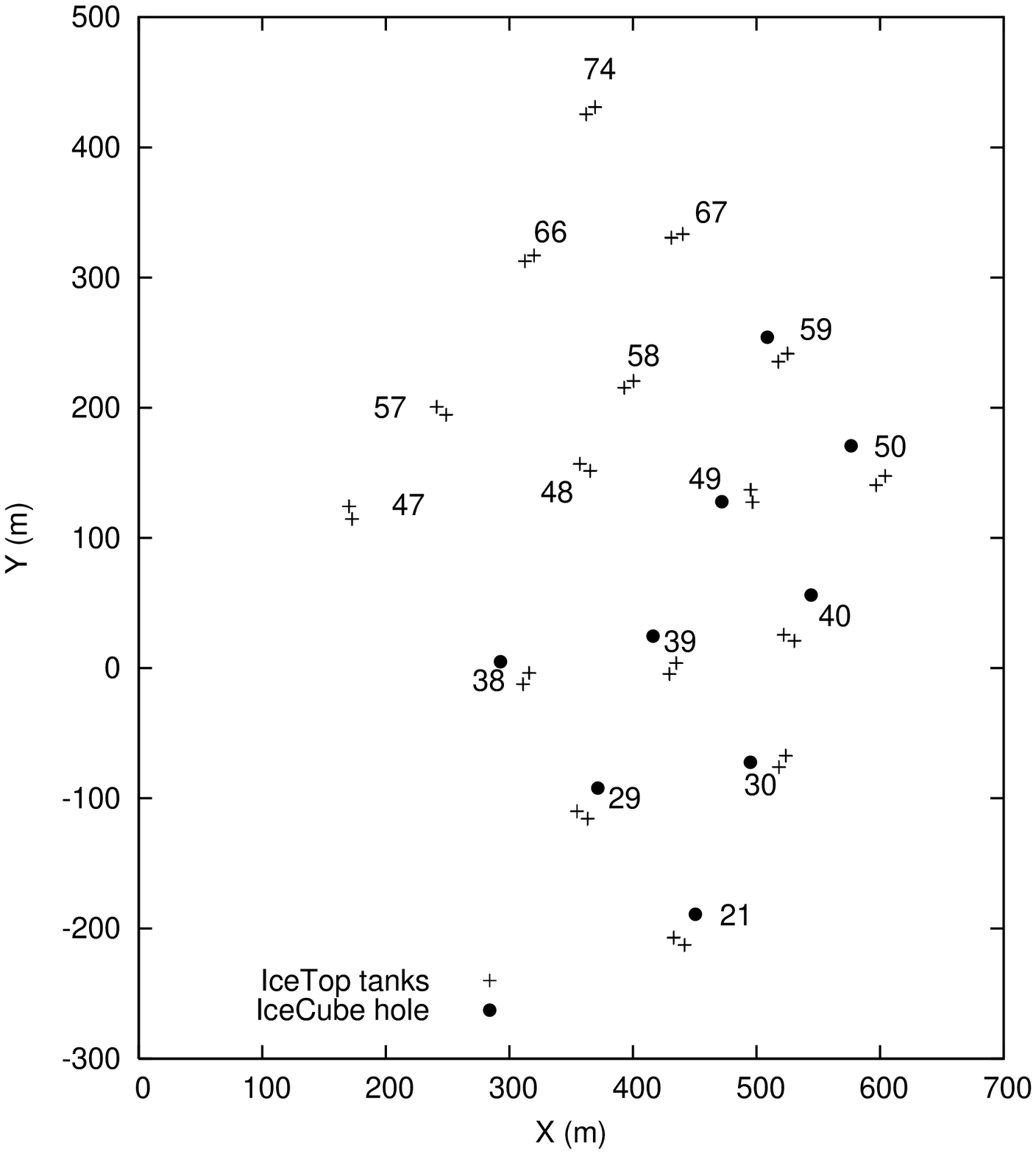}
    \caption{{\small Surface map of IceCube in 2006. Two tanks ($+$) 
are separated from each other by 10~m at each station. Each tank has 
one high-gain and one low-gain DOM.}}
    \label{fig1T}
\end{figure}

Events that trigger both the surface array
and deep IceCube can be reconstructed independently by the
air shower array on the surface and by the in-ice detector.
Such events can be used to verify the system timing and to survey 
the relative position of all active detection units, 
i.e. IceTop tanks or in-ice DOMs. The concept has been 
demonstrated in the SPASE2-AMANDA experiment~\cite{survey}.
Verification of timing with coincident events is now a routine
component of IceCube monitoring.  One can also 
compare the two independently
determined directions for the same events.  Showers big enough
to trigger IceTop, however, typically have several muons in the deep
detector.  One would also like to be able to tag
single muons in IceCube to have a set of events similar
to the $\nu_\mu$-induced muons that are the principal target of IceCube.
In this paper we describe how a sample enriched in single 
muons can be tagged with IceTop, and we illustrate the use
of this sample for verification of IceCube.
\begin{figure}[ht]
\begin{center}
\noindent
\includegraphics[height=6.8cm,width=0.47\textwidth]{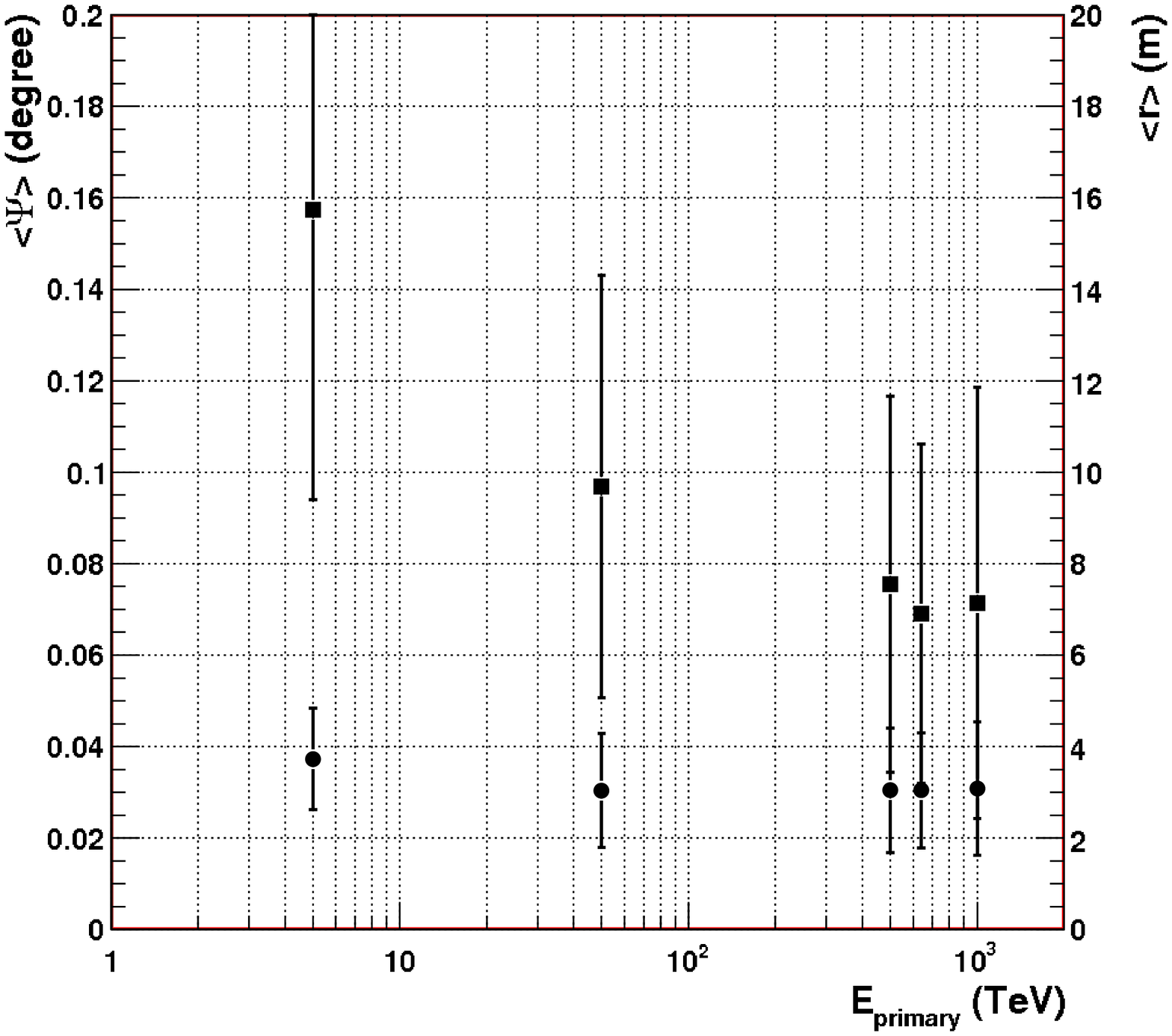}
\end{center}
\caption{{\small The average space angle $\Psi$ between muons
and air shower axis (solid circle, left vertical scale), the mean distance 
$r$ of muons from air shower core (solid square, right vertical scale) 
as function of primary proton energy. The error bars 
represent the $rms$ of $\Psi$ and $r$. Only muons with energy 
above 460 GeV on the surface are counted. Proton showers were produced 
at the South-Pole altitude by CORSIKA~\cite{corsika} with QGSJET as 
the high energy hadronic model.}}
\label{ICRC0328_fig1}
\end{figure}

\vspace{-0.2cm}
\section{2. Muons in air showers and their energy loss in the ice}
\vspace{-0.2cm}
The average number of high energy muons in an air shower can be 
parameterized as ~\cite{ICRC0328_crpp} 
%%\begin{equation}\label{eq:nmuon}
\vspace{-0.2cm}
\begin{displaymath}
N_{\mu,>E_{\mu}} = A\frac{0.0145TeV}{E_{\mu}cos(\theta)} 
(\frac{E_{0}}{AE_{\mu}})^{0.757}(1-\frac{AE_{\mu}}{E_{0}})^{5.25} 
%%\end{equation}
\end{displaymath}
in which $A$, $E_{0}$ and $\theta$ are the mass, total energy and 
zenith angle of the primary nucleus. Muons with energy high 
enough to trigger the in-ice detector are also nearly parallel with 
the air shower axis as shown in Fig.~\ref{ICRC0328_fig1}. 

The mean muon energy loss in matter is customarily 
expressed as 
%%\begin{equation}\label{eq:eloss}
%%\begin{displaymath}
$$\frac{dE}{dx} = -a(E)-b(E)\cdot E , $$
%%\end{displaymath}
%%\end{equation}
where $a(E)$ stands for ionization loss and $b(E)$ for 
stochastic energy loss due to pair production, 
photo-nuclear interactions and bremsstrahlung. 
As an approximation, $a(E)$ and $b(E)$ can be treated as
constants. For ice at the South-Pole, 
$a=0.26~GeV\, mwe^{-1}$ and $b=3.57\cdot 10^{-4}\, mwe^{-1}$, 
which are claimed with the systematic error of $\sim 3.7\%$. 
~\cite{pred01}. The least mean energy required for a muon to 
reach the top (1450~m) and the bottom (2450~m) of the in-ice detector is about 
460~GeV and 930~GeV.  For cosmic-ray protons of 500~TeV, typical of
showers that trigger IceTop, 
$\langle N_\mu\rangle\approx$~6 at 1450~m and $\approx$~2 at 2450~m.

\begin{figure}[ht]
    \includegraphics[height=5.5cm,width=0.47\textwidth]{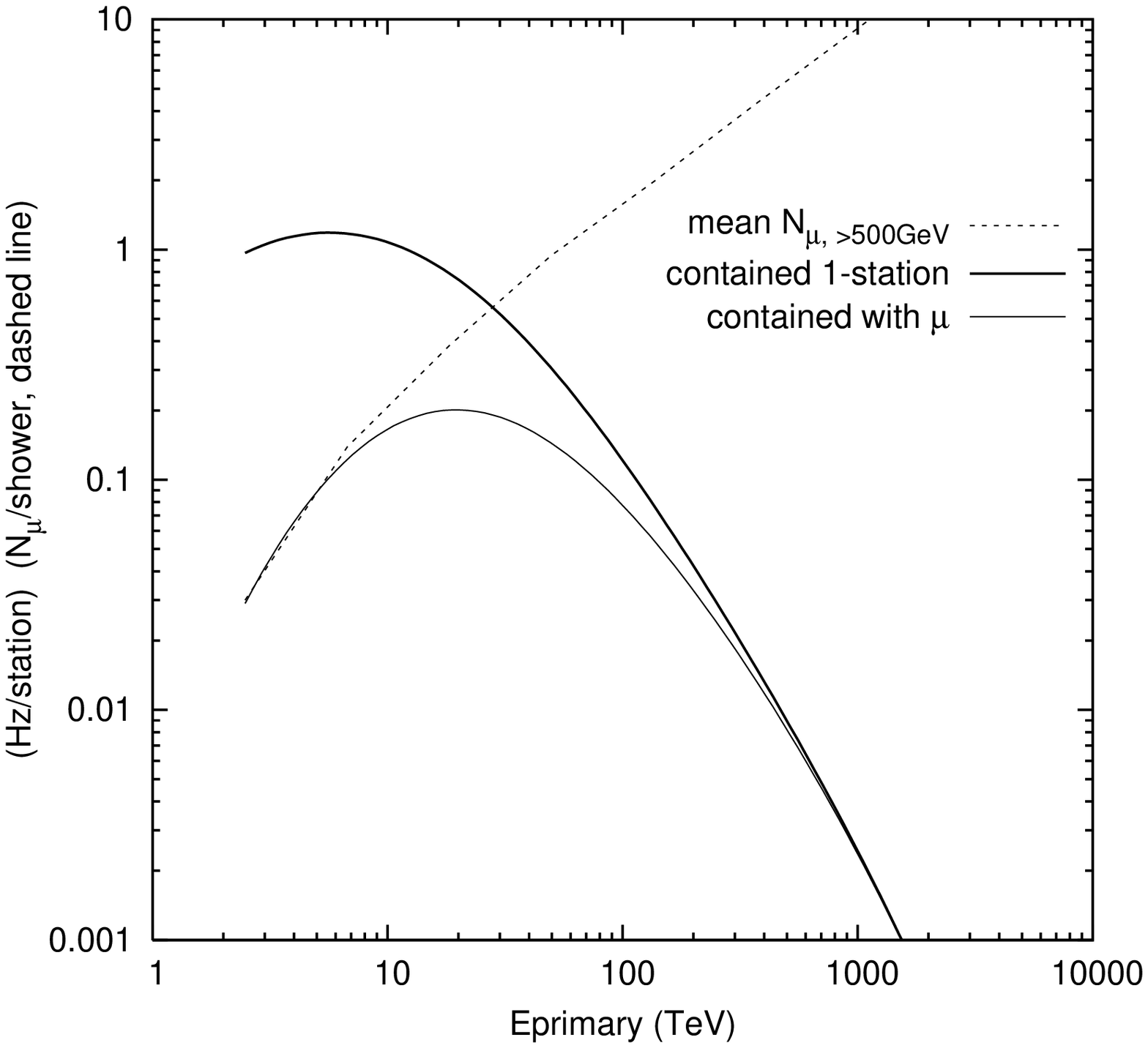}
    \caption{{\small Response function for single station events in IceTop. 
Only four contained stations (39, 48, 49 and 58) were considered. The dashed 
line represents the number of muons above 500~GeV at production in a 
proton shower. The lower curve shows the response function for events 
with one muon in the deep detector.}} 
    \label{fig2TG}
\end{figure}

We can select a sample of lower energy events by choosing in-ice
triggers with both tanks hit at exactly one IceTop station. We also 
require the single station is not on the periphery so that events
with energy high enough to hit both tanks at two or more IceTop 
stations are excluded from the sample. The concept is illustrated 
in Fig.~\ref{fig2TG} where we show an estimate of the distribution 
of primary cosmic-ray proton energies that give single station hits 
above 30MeV threshold in each tank. 
The lower curve shows the convolution of this response function with 
the probability of producing a muon with $E_\mu >$ 500~GeV. This 
corresponds to the distribution of primary energy that gives rise 
to the single station coincident event sample. About ninety percent
of this sample are generated by primaries with $E <$~100~TeV, and
about three quarters have only a single muon with $E_\mu >$~500~GeV at
production. 

\vspace{-0.2cm}
\section{3. Verification of time synchronization and depth of the DOMs} 
\vspace{-0.2cm}

A critical requirement for doing physics with IceCube is good time
synchronization among the individual DOMs in IceCube, including IceTop
together with accurate positions for the DOMs. Calibration with flashers
and survey by hole logging during deployment shows that timing synchronization
is at the level of 3~ns for a whole In-Ice string while the depth of 
individual DOMs are known with an accuracy of 50~cm~\cite{performance}. 
By using tagged, vertical muons
we can make a global check on the combination of time synchronization
and depth of the DOMs over a 2.5~km
baseline, from the surface to the deepest module on an IceCube string.
 To ensure that
the single station events are not caused by tails of big air showers outside
the array, only the inner stations of the IceTop array are used
together with the in-ice strings directly below
them.  With the 16 IceTop station and 9 in-ice string array in 2006, 
only stations 39 and 49 fulfill this requirement. 

For these two strings the muon speed has been
individually calculated for each DOM relative to the time $t_{0}$ at the
surface according to $v_i=d_i/(t_{i}-t_{0})$ where $d_i$ is the distance between the
station and the $i^{th}$ in-ice DOM.  Because of scattering in the ice, there 
is a distribution of arrival times of photons at each DOM relative to
the arrival time in the ideal case with no scattering. We represent
the distribution of delays by an exponential with a characteristic
delay $\tau$.  We then convolve this exponential distribution with 
a Gaussian resolution function to represent other uncertainties 
in the system.  The result is a Gaussian-convoluted exponential 
function as shown bellow. By fitting the distribution of arrival times 
at each DOM, we extract a fitted value of the arrival time $t_i$ at 
the $i$th DOM in the absence of scattering. 
\vspace{-0.2cm}
\begin{displaymath}
\frac{dN}{dt} = \frac{1}{2}\frac{N}{\tau} e^{-\frac{t-t_{i}}{\tau}}
e^{\frac{\sigma^{2}}{2\tau^2}}\cdot \mathrm{erfc}\left(\frac{t_{i} - t +
\frac{\sigma^{2}}{\tau}}{\sqrt{2}\sigma}\right)
\end{displaymath}
Other parameters here are the effective time resolution, $\sigma$, and
the mean number of hits $N$. The expression $\mathrm{erfc}$ represents 
the complementary error function~\cite{recipes}. 

The distribution of the relative muon speed to
the speed of light, $v_i/c$, is shown in Fig.~\ref{ICRC0328_fig2}, where we use
the surveyed values of $d_i$ to calculate the velocity.  
There are 60 DOMs on each string, 10 of which are not fitted
because of insufficient data, so there are 110 entries in
Fig.~\ref{ICRC0328_fig2}.  The $rms$ of 0.0015 of the distribution
of $v_i/c$ in Fig.~\ref{ICRC0328_fig2} reflects the uncertainties in 
the system timing, the location of DOMs and the true muon position 
on the surface. This corresponds to
upper limits on the uncertainty of 12~ns or 4~m over 2.5 km. 
Thus, although this method at present is not
as precise as the standard survey and calibration techniques,
it is useful to show by a complementary and independent method
that there are no significant deviation from expectation.
\begin{figure}[t]
    \includegraphics[height=5.5cm,width=0.5\textwidth]{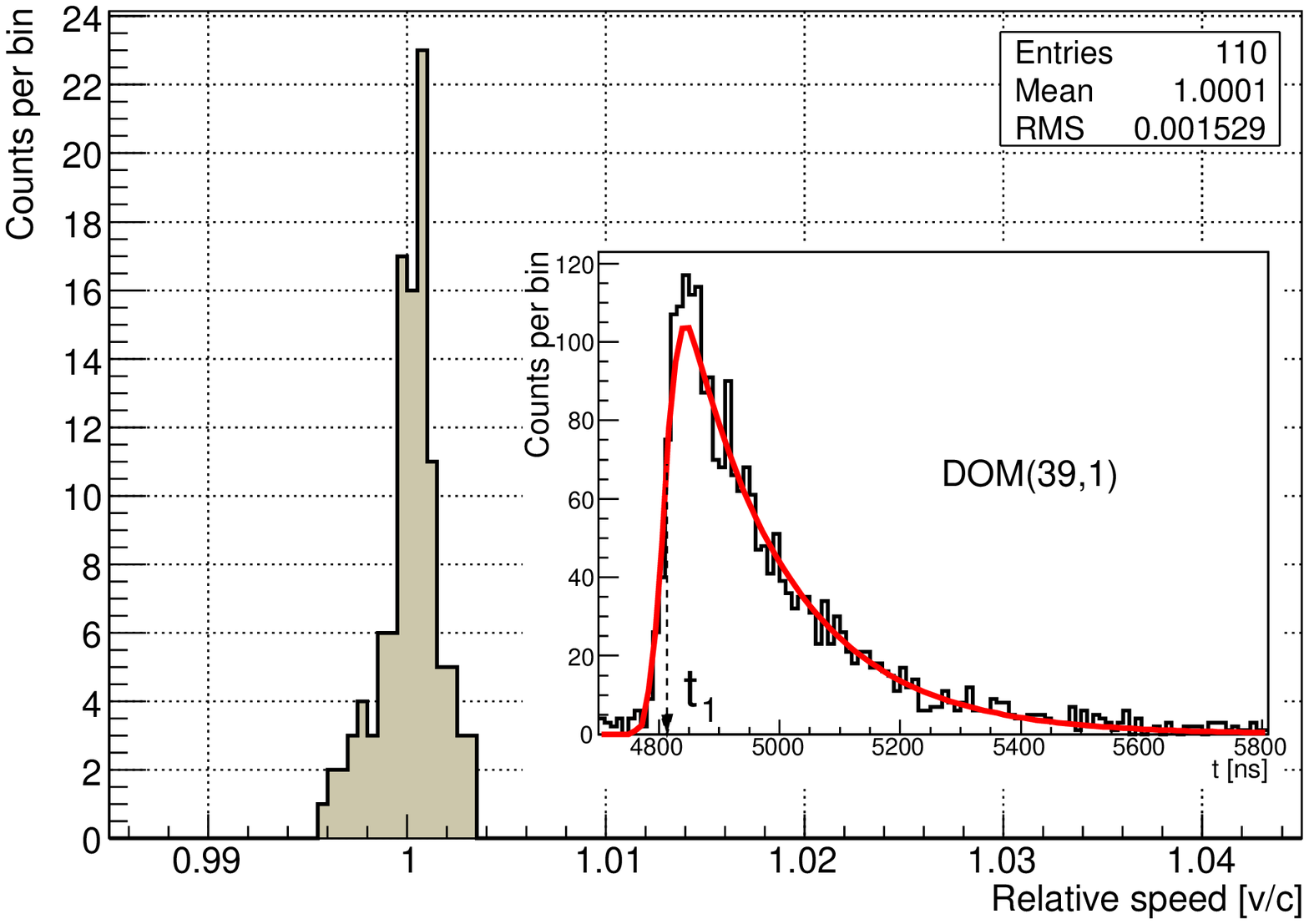}
    \caption{{\small The distribution of muon speed ($v$) relative
     to the speed of light ($c$). The cut-in
     entry shows the time delay on one in-ice DOM and the fit.
     See text for details.}}
    \label{ICRC0328_fig2}
\end{figure}

\vspace{-0.2cm}
\section{4. Muons in the in-ice detector}
\vspace{-0.2cm}
\subsection{4.1 Muon direction} 
\vspace{-0.2cm}
Small air showers trigger a single IceTop station 
efficiently only when the shower core is close to the 
station. Since high energy muons are nearly parallel to 
the shower axis, the line connecting the station on the surface 
and the center of gravity (COG) of triggered in-ice 
DOMs approximates the muon trajectory closely.  If we use half
the string spacing to estimate the accuracy of the location
of the track at the surface and at 1500~m, we find
that the direction of the track should be determined
to an accuracy of about 3$^\circ$.  
Fig.~\ref{ICRC0328_fig4} shows a comparison between the zenith angle 
defined by this line and by an independent in-ice reconstruction. 
The events in the solid-circle histogram have charge more 
than 5 photo-electrons in the triggered IceTop station. 
Those under the solid-triangle histogram have charge more 
than 400 photo-electrons, indicating a core closer to the 
station and/or slightly higher primary energy. The mean 
of $\delta(\theta)$ decreased from ~0.37 degree in the low-density
sample to 
~0.13 degrees in the high density sample.% The agreement is reasonably good. 
The $rms$ are about 3.7 degrees and 3.4 degrees respectively for 
the two groups.  Given the estimated 3~$^\circ$ uncertainty in the
estimation of the direction by this method, the good agreement
indicates that the in-ice reconstruction algorithm has an accuracy
of 2$^\circ$ or better for events near the vertical. Nevertheless, 
further investigation is needed to understand these events with 
zenith offset larger than 6 degrees. 

%The width is consistent with the geometric 
%uncertainties on the surface and the possible offset of 
%the COG vertex in depth from the true muon track. Nevertheless, 
%further investigation is needed to understand these events with 
%zenith offset larger than 6 degrees. 
\begin{figure}[t]
    \includegraphics[height=6.3cm,width=0.5\textwidth]{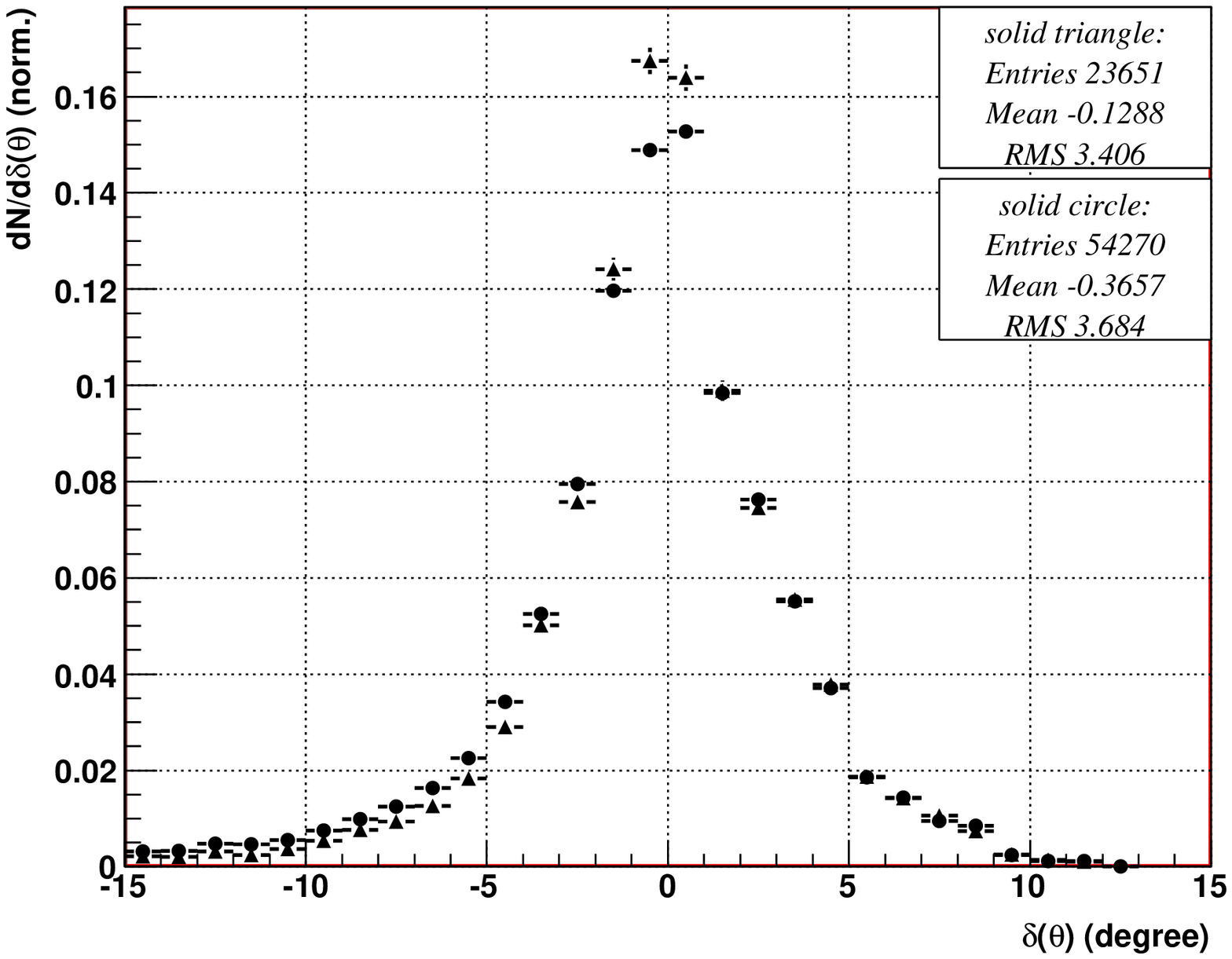}
    \caption{{\small The difference between the zenith angle defined
by the line connecting triggered IceTop station and the 
COG of triggered in-ice DOMs and that by the in-ice 
reconstruction. See text for details.}}
    \label{ICRC0328_fig4}
\end{figure}

\vspace{-0.2cm}
\subsection{4.2 Uncorrelated, coincident atmospheric muons in IceCube}
\vspace{-0.2cm}

An important source of background for upward-moving neutrino-induced
muons in IceCube is the subset of events in which two uncorrelated atmospheric
muons pass through the detector in the same trigger window.
Such events, which are estimated to constitute about 3\%
of the trigger rate in the full cubic kilometer IceCube~\cite{Henrike}, 
are of concern because the time sequence of hits
in the combined event can easily have an upward component.
It will be useful to tag a subset of such events with
IceTop for study and to check that they are efficiently
filtered.  At present, however, with the smaller detector
the fraction of accidental coincidences is much smaller, and
IceTop can only tag a very small fraction of them.
The rate of identified single station coincidences in 2006
was about 0.075 Hz per station, so 1.2~Hz over the sixteen
station array.  An estimate of the rate of tagged double
uncorrelated events is therefore $\sim$10$^{-{\rm 5}}$~Hz, somewhat
about one per day.  For comparison, 
the trigger rate of the 9-string IceCube
in 2006 was 146 Hz.  

{\bf {\Large Acknowledgments}} The work is supported by the 
US National Science Foundation under Grant No. OPP-0236449 
(IceCube), University of Wisconsin-Madison and NSF Grant 
OPP-0602679 at the University of Delaware. The authors 
gratefully acknowledge the support from the U.S. 
Amundsen-Scott South Pole station.

%\end{linenumbers}

%\end{document}

\setcounter{figure}{0}
\setcounter{table}{0}
%%
% International Cosmic Ray Conference 2007 Merida Yucatan Mexico
% In This file you will find detailed instructions to correctly
% typeset your document.
%
%
%

%Class Requeried
%\documentclass[dvips]{article}
%The ICRC Style
%\usepackage{icrctc07}

\newcommand{\eh}[1]{\,\mathrm{#1}}
\newcommand{\todo}[1]{(\textit{TODO:#1})}

%The paper title
\title{Lateral Distribution of Air Shower Signals and Initial Energy Spectrum above 1 PeV from IceTop}
%Short title to print in the headers to the final publication (Not showed in this print).
\shorttitle{IceTop LDF and Energy Reconstruction}
%All paper authors
\authors{S. Klepser$^{1}$, F. Kislat$^{2}$, H. Kolanoski$^{2}$, P. Niessen$^{3}$, A. Van Overloop$^{4}$
for the IceCube Collaboration$^{5}$
}
%Short title to print in the headers to the final puplication (Not showed in this print).
\shortauthors{S. Klepser et al}
%All the affiliations.
\afiliations{$^1$ DESY, D-15735 Zeuthen, Germany\\
             $^2$ Institut f\"ur Physik, Humboldt-Universit\"at zu Berlin, D-12489 Berlin, Germany\\
             $^3$ Bartol Research Institute, University of Delaware, Newark, DE 19716, U.S.A.\\
             $^4$ Dept. of Subatomic and Radiation Physics, University of Gent, B-9000 Gent, Belgium\\
             $^5$ see special section of these proceedings
}
\email{stefan.klepser@desy.de%, $^5$ see special section of these proceedings
}

%\linenumbers

%The abstract.
\abstract{The IceTop surface detector array is part of the IceCube Neutrino Observatory that is presently
being built at the South Pole. In a triangular grid with a spacing of $125\eh{m}$, up  to 80 pairs of ice
Cherenkov tanks will be set up, 16 of which were already in operation in 2006.
The data from this array allows the reconstruction of a first preliminary energy spectrum
%With this array, it is possible to measure an energy spectrum
in the range of about $1\eh{PeV}$ to $100\eh{PeV}$. 
To reconstruct the primary energy of a cosmic ray particle, a fit to the lateral distribution of the air shower signals
has to be performed. We have developed a functional description of expected lateral distributions and of the corresponding
fluctuations of the measured signals. The function and its parameters have been tuned
in a CORSIKA simulation study with parametrised particle responses.
% and a description of signal fluctuations has been found
%by taking advantage of
%having two tanks separated by $10\eh{m}$ at each detector station.
%in a detailed detector simulation.
%Having two tanks separated by $10\eh{m}$ at each detector station,
From a detailed detector simulation, the fluctuations could be extracted and qualitatively compared
with experimental data.
%The fluctuations could be extracted from a detailed detector simulation and qualitatively compared
%with experimental data.
Some performance tests and an initial energy spectrum,
uncorrected for efficiency near threshold, are presented.
%Some performance tests and an initial, not acceptance-corrected energy spectrum is presented.
}

%%%%%%%%%%%%%%%%%%%% B E G I N   D O C U M E N T%%%%%%%%%%%%%%%%%%%%%%%
%\begin{document}
\maketitle
%Begin the section.

\section{Introduction} \label{sec:introduction}
When a high energy cosmic ray hits the earth's atmosphere, it induces an extensive air shower (EAS)
whose axis and energy can be reconstructed by detector arrays at ground level. In general,
the arrival times of the particles deliver the direction information while the signal strength distribution
is used to reconstruct the core and size of the shower. The shower size is usually represented by the signal $S_{R}$ at
a certain perpendicular distance $R$ from the shower axis (``core radius''). With the spacing of IceTop, $S_{100}$ at $R = 100\eh{m}$ proved to be
a stable and reliable quantity in the fit procedure.

The signal $S$ of an IceTop tank is derived from the charge of two photomultipliers that
are operated at different gains ($5 \cdot 10^4$ and $5 \cdot 10^6$ in 2006)
to enhance the dynamic range of the detector well above $10^5$. They
collect the Cherenkov photons produced by the shower particles
%when they traverse
in the $2.45\eh{m}^3$ of ice in each tank. The total signal
is proportional to the deposited energy
in the tank since the Cherenkov light and the deposited energy are both approximately
proportional to the track
lengths of the charged particles. Using atmospheric muons for calibration, the signals can
thus be converted to the detector-independent unit VEM (vertical
equivalent muon), which is equivalent to about $200\eh{MeV}$ of 
deposited energy \cite{vemcallevent}.

To estimate the energy of the primary particle and determine the shower core, a log-likelihood fit
is applied to the measured signals. This requires a lateral
distribution function (LDF) $S(r)$ at a given core radius, and a
parametrisation of the signal fluctuations.
%\begin{equation}
%  -\ln(\mathcal{L}) = \sum_{i_{station}}\left(\frac{\log_{10}S_i-\log_{10}S_{fit}}{2\sigma_S^2}\right)^2+\ln(\sigma_S)\quad-\quad\sum_{i_{silent}}\ln P_{silent}\quad+\quad const.
%\end{equation}
The likelihood also includes a term for stations without trigger.
%Since the readout of the signals of a station requires a coincidence between both tanks, the likelihood
%also has to account for the probability that the signals of both tanks are above threshold.
%In the next
%section, a suitable LDF and fluctuation parametrisation
%are described that is used in the following sections.
\nopagebreak
\section{LDF and Fluctuation Parametrisation}
\begin{figure*}
\begin{center}
\includegraphics[width=0.49\textwidth]{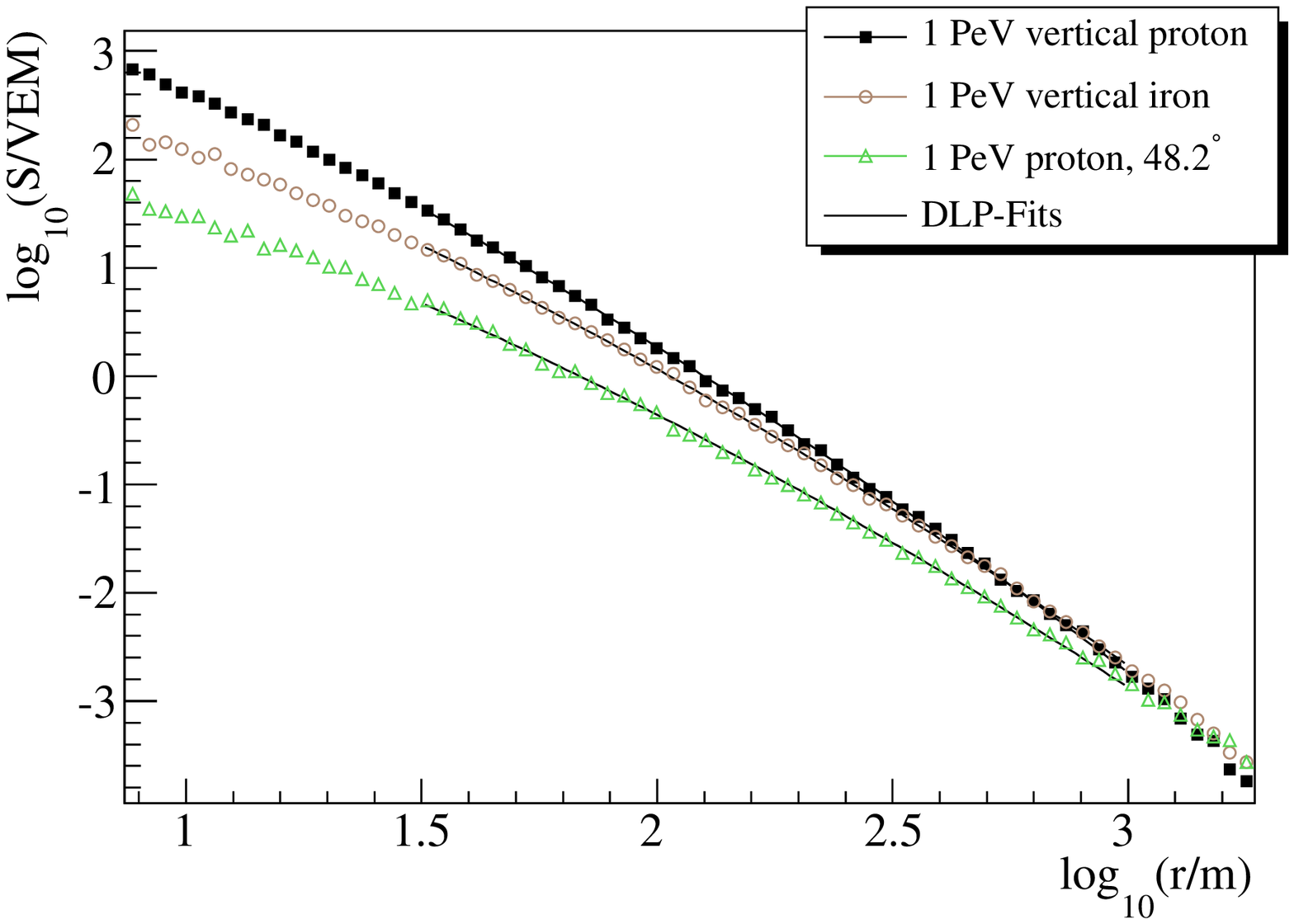}
\hfill
\includegraphics[width=0.49\textwidth]{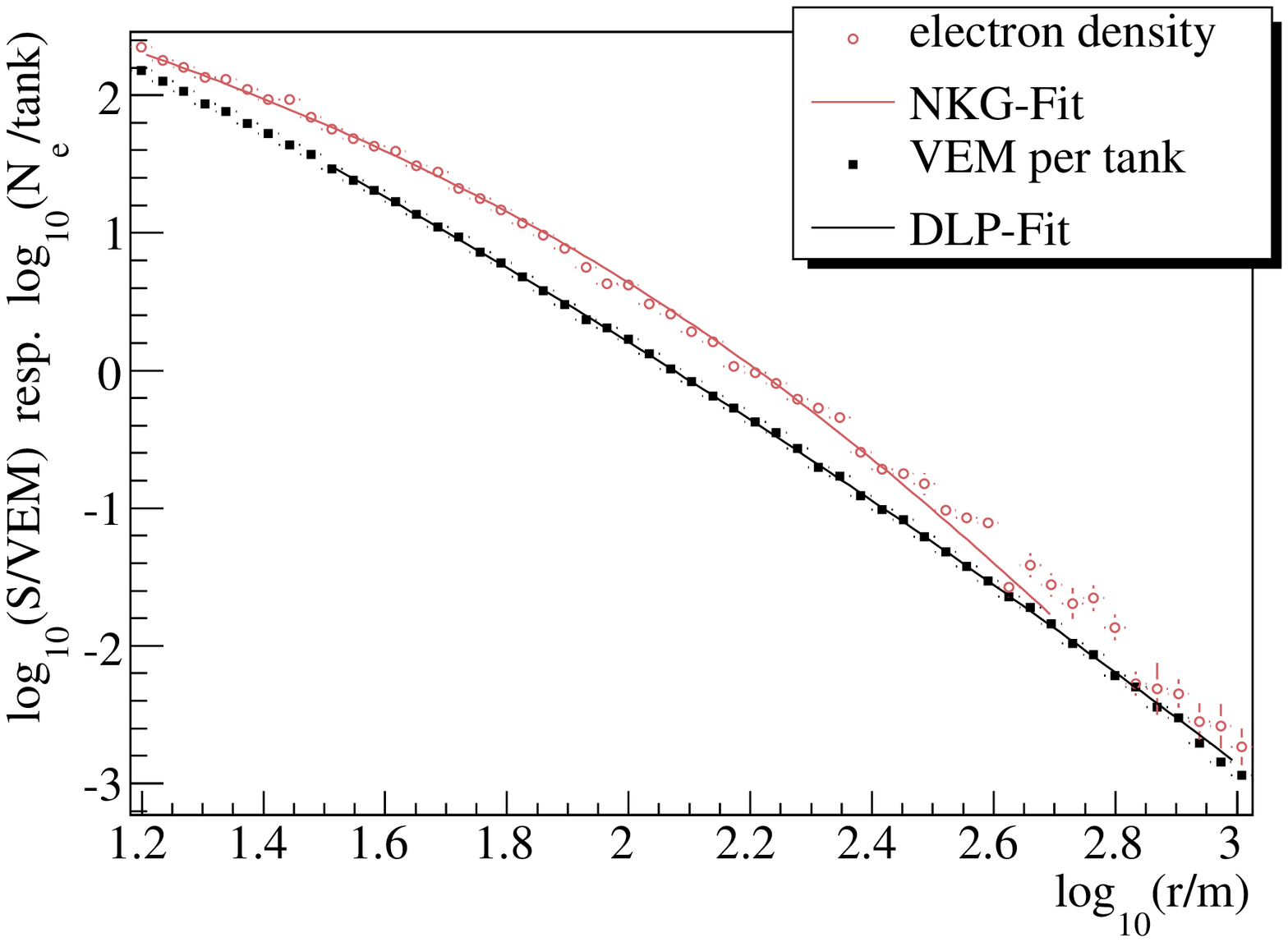}
\caption{\label{fig:vem_ld} 
Left: Derived lateral signal distributions of IceTop tanks for three different simulated showers,
% a vertical proton shower, an iron shower and an inclined 
%proton shower
fitted with the DLP function described in the text. Right: Comparison between lateral electron
density and tank signal distribution, fitted with NKG and DLP respectively.}
\end{center}
\end{figure*}
To find an appropriate LDF for IceTop, lateral distributions of CORSIKA
%\footnote{The 
%hadronic interaction models used in all simulations are Sibyll 2.1 for high energies and
%Fluka for low energies.} \cite{corsika}
shower simulations \cite{ICRC0858_Corsika} were analysed. The
hadronic interaction models used in all simulations are Sibyll 2.1 \cite{ICRC0858_Sibyll} for energies
above $80\eh{GeV}$ and
Fluka \cite{fluka} below that. Each shower particle was weighted with
an average response function $S_{j}(E)$ derived from single particle simulations that
were carried out with a Geant4-based detector simulation \cite{geant4}. The particle types
considered are $j=\{\gamma, e^{\pm},
\mu^{\pm}, p, \bar{p}, n, \bar{n}, \pi^{\pm}, K^{\pm,0}\}$, which are the most
abundant in air showers. Three examples of the distributions that
were found, and a comparison to the electron density distribution described by the NKG function \cite{nkg}
are given in Fig.\,\ref{fig:vem_ld}. It is remarkable that the main feature of the NKG function in double logarithmic
representation,
which is a bend with a maximal curvature approximately at the Moli\`ere Radius ($128\eh{m}$ at the South Pole \cite{cosmicrayreview}), cannot be seen in the
tank signal lateral distributions. This is presumably a consequence of the fact that the energy
deposition is not proportional to the particle number.

%The logarithmic signal difference between iron and proton showers for the shown $1\eh{PeV}$ showers is around 0.1 at $100\eh{m}$.
% and thus in the order of the
%fluctuations of $\log_{10}(S_{100})$.
%Since there are no detailed
%studies being done at present about the dependence of that number on primary energy and zenith angle,
%and it is probable that the chemical compositon changes at the observed energy range,
%this leads to a systematic error on the spectral index of $\sigma_{\gamma} \approx 0.1$.   

The function found to fit these distributions well in a range between 30 and $1000\eh{m}$ is a parabola in a double logarithmic representation (DLP),
which can be written as
\begin{equation}\label{eq:dlp}
          S(R) = S_{R_{0}}\left(\frac{R}{R_{0}}\right)^{-\beta - \kappa\, \log_{10}\left(\frac{R}{R_{0}}\right)}
\end{equation}
with $R_{0} = 100\eh{m}$ being the reference core radius, $\beta$ the slope at $R_{0}$, and $\kappa \approx 0.303$
%264(5)$
the
curvature of the parabola. This curvature is approximately a
constant for all hadronic showers and thus a fixed parameter for all fits on real data. The parameter $\beta$ is roughly
linearly connected to the shower age parameter of the NKG function via $s_{NKG} = -0.94\, \beta + 3.4$ 
%
%\begin{equation}\label{eq:beta_age}
%          s_{NKG} = -0.94\, \beta + 3.4
%\end{equation}
for all simulated angles, energies and nuclei.

To study the fluctuations $\sigma_S$ of the approximately log-normally distributed tank signals,
two analyses were done. Figure \ref{fig:fluc}
shows the comparison of the dependencies of $\sigma_S$ on $S$ that were found. The points designated with
``tank-to-tank''
indicate the outcome of a study of signal differences between the two tanks separated by $10\eh{m}$
at each detector
station. Shower fluctuations were thus measured directly in data and the result is compared to simulated
data that was produced with CORSIKA 
showers processed with a Geant4 detector simulation of the array. The lower points
are taken from a similar simulation with tanks set up in a ring-like structure. 
%, comparing the tank signals to the average signal in a whole ring.
Since the former is biased by uncertainties in reconstruction and shower intrinsic
correlations, and the latter depends on the quality of the detector simulation, the two
methods are not fully comparable but should yield results in the same order of magnitude. This could roughly
be verified, although the tank-to-tank
fluctuations have some features at higher amplitudes that are most likely an artefact from
misreconstructed cores that are very close to one of the tanks.
%close enough to a station to produce different signal expectation values in both tanks.
In the full array simulations described below, the parametrisation taken from the ring-like
simulation delivers a better core and energy resolution
and is therefore used in the fit. The dependence of $\sigma_S$ on the core radius was found to be in the order of
$15\eh{\%}$ for radii above $30\eh{m}$ and is therefore negligible.

With the parametrised CORSIKA simulations described above, it was found that for zenith angles $\theta < 50^{\circ}$, the 
dependence of $S_{R}$ on $x = \sec \theta$ can be described by parabolas (Fig.\,\ref{fig:ls100_theta}). Assuming that
the maximum of $\log_{10}S_{R}$ and its position $x_{max}$ linearly depend on $\log_{10}E$, a function
$S_{R}(\theta, E)$ was found that fits all data points and can be inverted analytically
to $E(S_{R}, \theta)$. For several $R$ between 
50 and $1000\eh{m}$, the parameters of $E(S_{R}, \theta)$ were interpolated such that
the conversion from $S_R$ to the primary energy can be done at any radius $R_{opt}$ that 
might be regarded optimal for physical or numerical reasons. Presently, to be
as independent as possible from the quality of the LDF, $R_{opt}$ is chosen event by event
in a way that $\log_{10}R_{opt}$ is the mean logarithmic core radius of all tanks that were actually used in the fit.
%With this method, the average $R_{opt}$ chosen is $103\eh{m}$.
\begin{figure}
\begin{center}
\includegraphics[width=0.48\textwidth]{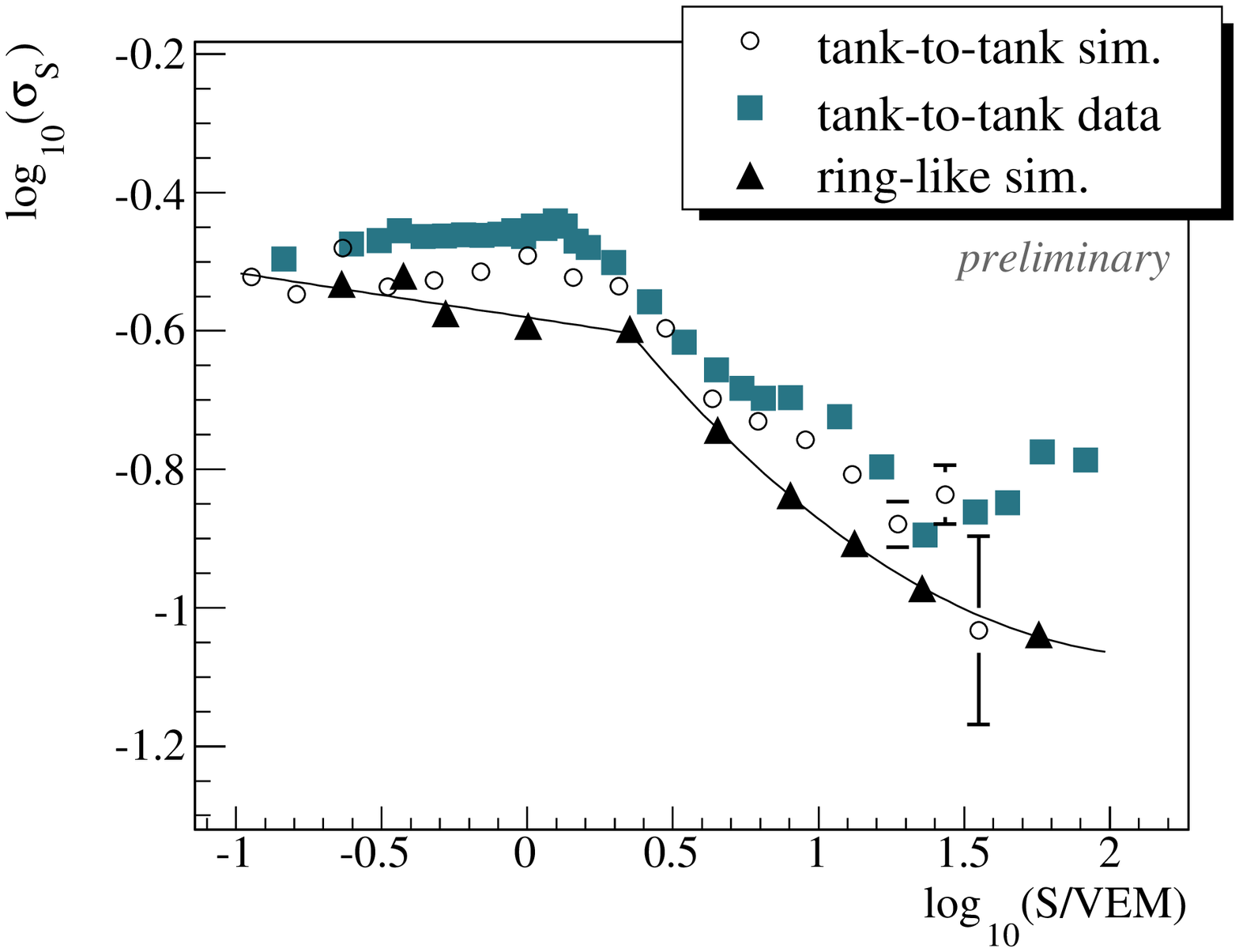}
\caption{\label{fig:fluc} Dependency of
the signal fluctuation $\sigma_S$ on the signal $S$ in data and different simulations (the error bars are partly smaller than
the markers). $\sigma_S$ designates the standard deviation of $\log_{10}(S)$.
%Apparently, the statistical
%behaviour of the pulse distributions changes around $1\eh{VEM}$.
The differences between the methods are discussed in the text.
The solid line indicates the parametrisation that was extracted
for the lateral fit.}
\end{center}
\end{figure}

This energy conversion does not yet take into account the influence of the primary mass. From the
shower size differences observed between proton and iron showers in the simulations ($\Delta \log_{10}S_{R} \approx 0.1$),
the systematic uncertainty on the spectral index of the following spectrum
can be estimated to be $\sigma_{\gamma} \approx 0.1$.
%logarithmic signal difference between iron and proton showers shown in fig.~\ref{fig:vem_ld},
%which is around 0.1 at $100\eh{m}$, the systematic uncertainty on the spectral index of the following spectrum
%can be estimated to $\sigma_{\gamma} \approx 0.1$.
% and thus in the order of the
%fluctuations of $\log_{10}(S_{100})$.
%Since there are no detailed
%studies being done at present about the dependence of that number on primary energy and zenith angle,
%and it is probable that the chemical compositon changes at the observed energy range,
%this leads to a systematic error on the spectral index of the following spectrum of $\sigma_{\gamma} \approx 0.1$.   

\begin{figure}
\begin{center}
\includegraphics[width=0.42\textwidth]{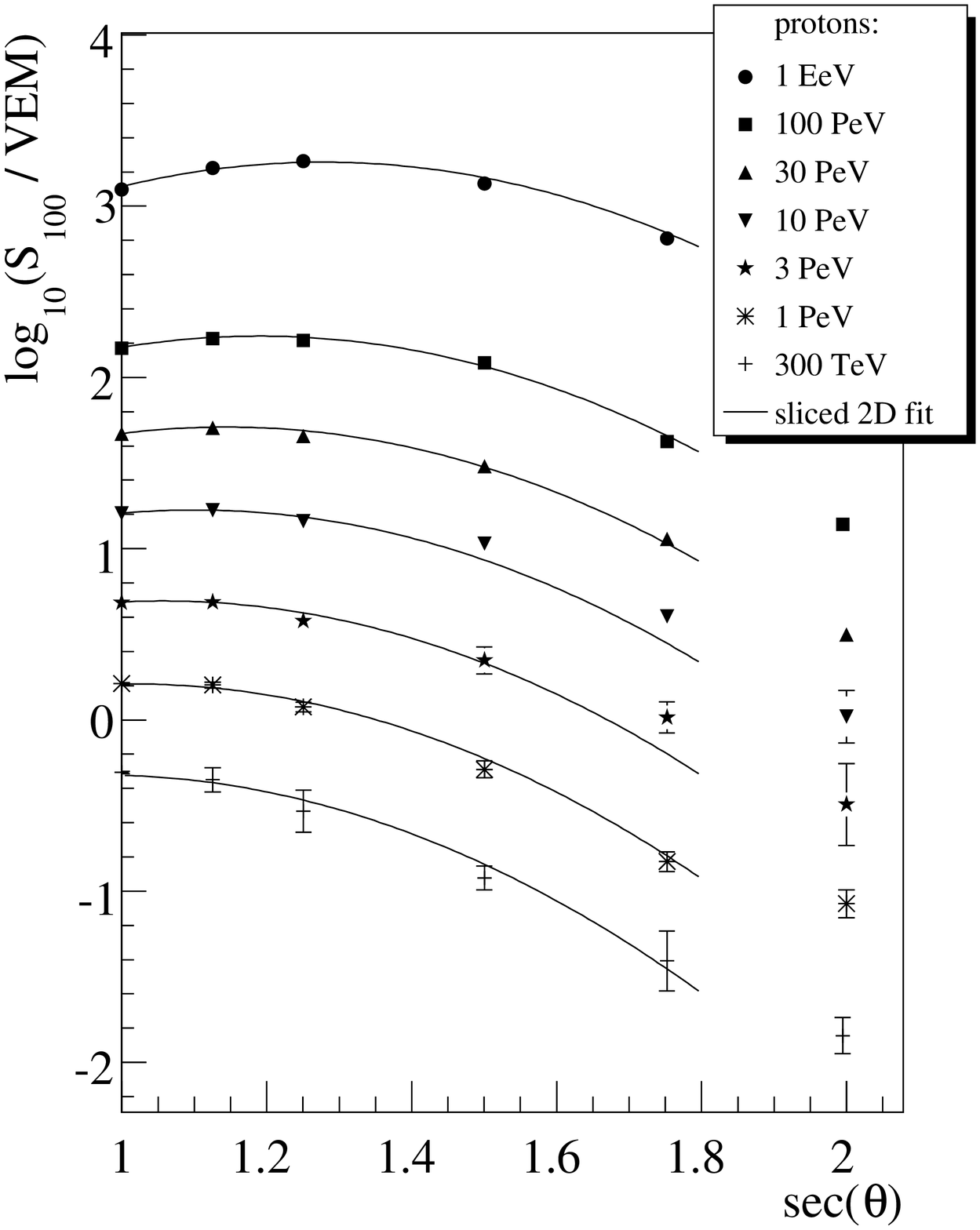}
\caption{\label{fig:ls100_theta}
CORSIKA simulations of $\log_{10}(S_{100})$ as a function of $\sec \theta$ for various energies. The lines are projections of the fit that was
performed on all data points simultaneously ($\chi^2/\mathrm{ndf} = 41.2 / 32$).}
\end{center}
\end{figure}

\section{Performance and Results}
To benchmark the performance of the LDF, CORSIKA simulations of $1\eh{PeV}$ vertical showers were
carried out on the 2006 array configuration, using the tank intersects of the shower particles and the above $S_{j}(E)$ tank response parametrisations
to scale the responses of the particles. The simulation also
includes the generation of PMT responses, digitisation and the behaviour of the IceCube trigger devices. Thus the simulated
raw data completely resembles the level and format of experimental raw data.
The quantities that serve to estimate the quality of the LDF are the core position resolution $\sigma_{core}$,
the energy resolution $\sigma_{\log_{10}E}$, the reconstruction efficiency $\epsilon$
and the mean of the $\chi^2$ distribution.
%Since the minimised likelihood function is not 
%the $\chi^2$ function, values somewhat above 1 are expected for this quantity. All of these numbers are estimated for
%events with a reconstructed core lying within the outer borders of the array. 
\begin{figure}
\begin{center}
\includegraphics[width=\columnwidth]{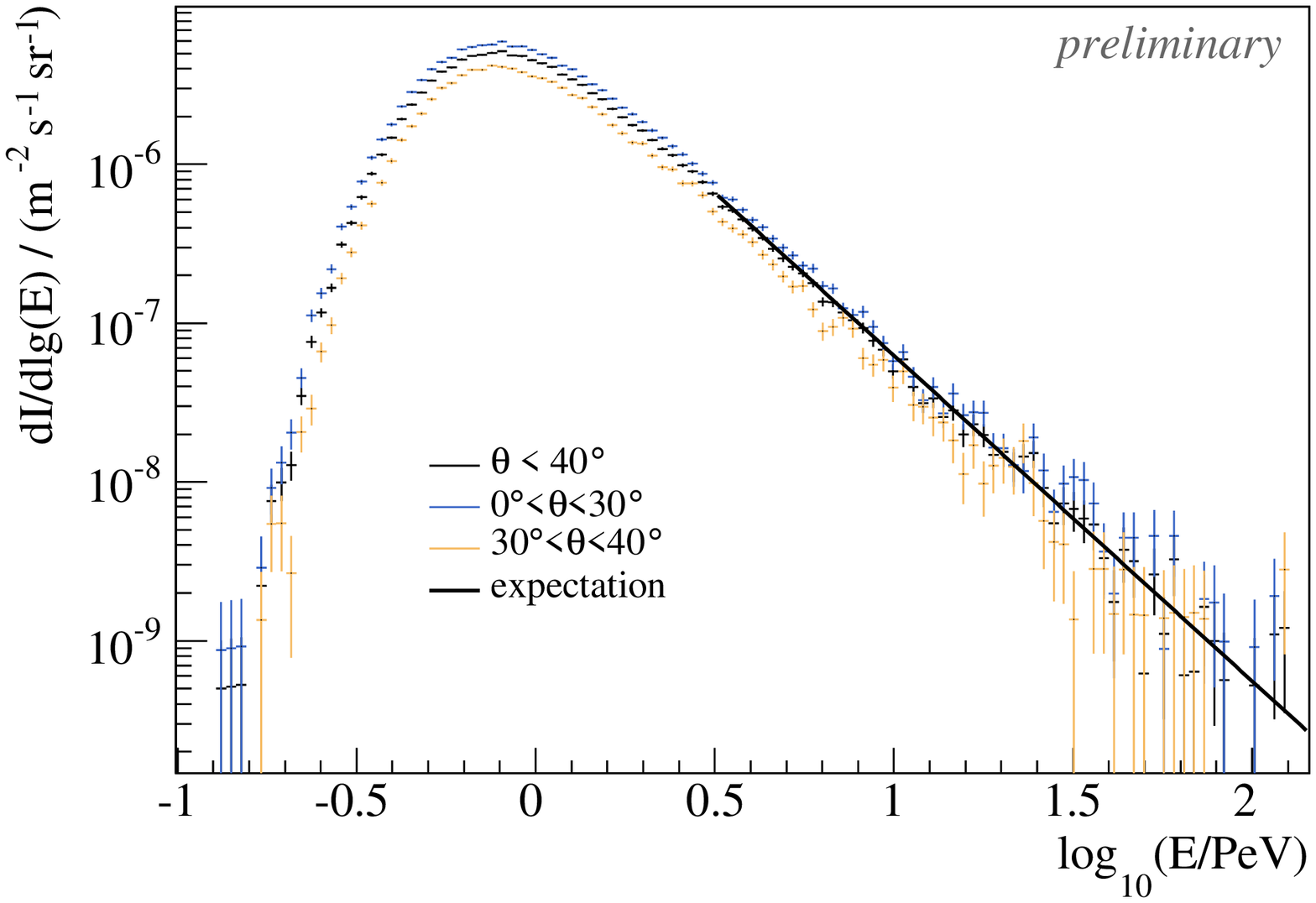}
\caption{\label{fig:espec} Preliminary, raw energy spectrum without acceptance correction. The difference between
high and low zenith range indicates the systematic uncertainty. Though not deconvoluted yet, the high-energetic part is compared to the
expected spectrum and agrees well with it (solid line, \cite{hoerandel}).}
\end{center}
\end{figure}

Compared to a simple power law and the NKG function,
the numbers found indicate a slight preference for the DLP function, especially concerning the reconstruction efficiency.
For vertical $1\eh{PeV}$ showers, the core and energy resolution are
$\sigma_{core} = 12.8\eh{m}$ and $\sigma_{\log_{10}E} = 0.094$. However, once a bigger array is available in
the coming years, this has to be reevaluated.

%\begin{table*}[t]
%    \centering
%    \begin{tabular}{|l||r@{.}l|r@{.}l|r@{.}l|r@{.}l|r@{.}l|}
%    \hline
%    LDF & \multicolumn{2}{c|}{$\sigma_{core}$ / m} & \multicolumn{2}{c|}{$\sigma_{(\log_{10}E)}$} & \multicolumn{2}{c|}{$\epsilon$}
%        & \multicolumn{2}{c|}{$(\chi^2/\mathrm{ndf})_{mean}$} \\ \hline\hline
%    power law    & 12    & 853(13)  & 0    & 091(3) & 0    & 74(6)    & 1    & 47(5)     \\
%    NKG          & 12    & 902(12)  & 0    & 097(4) & 0    & 85(7)    & 1    & 34(3)     \\
%    DLP          & 12    & 759(12)  & 0    & 094(3) & 0    & 87(7)    & 1    & 30(3)     \\
%    \hline
%    \end{tabular}
%    \caption{Results of a $1\eh{PeV}$ vertical shower simulation to compare the performance of
%             different LDFs. The compared quantities are core position resolution, energy resolution,
%             reconstruction efficiency and the mean of the achieved $\chi^2/\mathrm{n.d.f.}$-distribution.}
%    \label{tab:comparisons}
%\end{table*}

With the energy extracted as described above, a dataset with an effective lifetime of $0.692 \cdot 10^6\eh{s}$ was
analysed. Requiring 5 triggered stations, the reconstructed core to be $50\eh{m}$ inside the array and the zenith angle to be
$\theta < 40^{\circ}$, an exposure of $0.67 \cdot 10^{11}\eh{m}^2\mathrm{sr\,s}$ is achieved. 
%To reduce reconstruction uncertainties, only shower cores reconstructed to be $50\eh{m}$ inside the array
%were accepted, which is equivalent to a detection area of $71.3 \cdot 10^3\eh{m}^2$. With an additional
%requirement of $\theta < 40^{\circ}$, this leads
%to an exposure of $2.00 \cdot 10^{11}\eh{m}^2\mathrm{sr\,s}$.
In this dataset, 192507 shower
events were detected. From the known energy spectrum \cite{hoerandel} of charged cosmic
rays, one can estimate an effective reconstruction threshold
of $\sim 500\eh{TeV}$ and expect approximately 1000 events above $10\eh{PeV}$ and
10 events above $100\eh{PeV}$. In the dataset, 800 and 5 events were found respectively. 

The raw distribution of energy estimates without acceptance
correction is shown in Fig.\,\ref{fig:espec}. The high 
energy part, where the efficiency can be assumed to be constant and close to 1, the slope of the spectrum agrees well
with the slope of $\gamma \approx 3.05$ that is expected from other experiments, drawn as a solid line for comparison.
The absolute scale of the raw spectrum is lower than the 
expectation, which indicates the need for more simulations to tune the energy extraction and correct for efficiencies.

\section{Conclusion}
With the 2006 array configuration, we will be able to measure the cosmic ray energy
spectrum from $0.5$ to $100\eh{PeV}$. The signal
distributions are well understood, and applying advanced log-likelihood fits we are able to reconstruct the cores and sizes
of the measured showers with good precision.
Since February 2007, already 26 stations are in operation, which covers a third of the total planned area. 
This and the development of an unfolding procedure will enable IceTop to measure an energy spectrum well above $100\eh{PeV}$
at the end of 2007.

%{\bf Acknowledgments} This work is supported by the U.S. National Science Foundation, Grant No. OPP-0236449.

%This is the reference to .bib file (Whitout .bib!)

%\bibliography{ICRC0858/icrc0858}
%This in the bibtex style, is ok.
%\bibliographystyle{plain}

%\end{document}

\setcounter{figure}{0}
\setcounter{table}{0}
% Template article for preprint document class `elsart'
% SP 2006/04/26
%Class Requeried
%\documentclass[dvipdf]{article}
%The ICRC Style
%\usepackage[latin1]{inputenc}
%\usepackage{icrctc07}

%The paper title
\title{IceTop tank response to muons}
%Short title to print in the headers to the final publication (Not showed in this print).
\shorttitle{IceTop tank response to muons}
%All paper authors
\authors{L.~Demirörs$^{3}$, M.~Beimforde$^{1}$, J.~Eisch$^{2}$, J.~Madsen$^{4}$,
  P.~Nießen$^{3}$, G.~M.~Spiczak$^{4}$, S.~Stoyanov$^{3}$, S.~Tilav$^{3}$ for the IceCube Collaboration$^{5}$}

%Short title to print in the headers to the final puplication (Not showed in this print).
\shortauthors{L.~Demirörs and et al}

%All the affiliations.
\afiliations{$^{1}$Institut für Physik, Humboldt-Universität zu Berlin, D-12489 Berlin, Germany\\
  $^{2}$Dept.\ of Physics, University of Wisconsin, Madison, WI 53706, USA\\
  $^{3}$Bartol Research Inst., Dept.\ of Physics \&
  Astronomy, University of Delaware, Newark, DE 19716, USA\\
  $^{4}$Dept.\ of Physics, University of Wisconsin, River Falls, WI
  54022, USA\\
  $^{5}$see special section of these proceedings
}

\email{levent@udel.edu}

%%%%%%%%%%%%%%%%%%%%%%%%%%%%%%%%%%%%%%%%%%%%%%%%%%%%%%%%%%%%%%%%%%%%%%%%
%The abstract.
\abstract{%%
  Each digital optical module (DOM) of the IceTop air shower array is
  calibrated by identifying and understanding its muon response, which is
  measured in vertical equivalent muon (VEM). Special calibration runs
  and austral season measurements with a tagging telescope provide the
  basis for determining the VEM and monitoring its variation with time
  and temperature. We also study muons that stop and decay in the
  tank. The energy spectrum of the electrons from muon decay is well
  known (Michel spectrum) and can also be used as a calibration
  tool. Both spectra are compared to a GEANT4 based Monte Carlo
  simulation to gain a better understanding of the tank properties.
}

%%%%%%%%%%%%%%%%%%%%%%%%%%%%%%%%%%%%%%%%%%%%%%%%%%%%%%%%%%%%%%%%%%%%%%%%

\DeclareGraphicsExtensions{.eps,.ps}
\DeclareGraphicsRule{.ps}{eps}{.ps}{}
\DeclareGraphicsRule{.eps}{eps}{.eps}{}

\newcommand{\unit}[2]{#1\ensuremath{\,}#2}

%%%%%%%%%%%%%%%%%%%%%%%%%%%%%%%%%%%%%%%%%%%%%%%%%%%%%%%%%%%%%%%%%%%%%%%%
%\begin{document}

\maketitle

\abovecaptionskip1truemm

%%%%%%%%%%%%%%%%%%%%%%%%%%%%%%%%%%%%%%%%%%%%%%%%%%%%%%%%%%%%%%%%%%%%%%%%
\section{Introduction}
\label{sec:overview}

IceTop is an air shower array of ice--Cherenkov counters
\cite{Gaisser:2007mm, Stanev:2005nt}.
Each of its current 26 stations is made up of two IceTop tanks. The
tank shell is black, cross--linked polyethelyne, \unit{6}{mm} thick,
\unit{1.1}{m} high, and \unit{1.9}{m} in diameter. A second layer of
\unit{4}{mm} thickness, made out of zirconium fused polyethylene, is
molded on the inner surface to act as a diffusely reflective
liner (eight tanks deployed in 2005 have Tyvek linings). Each tank is
filled with \unit{90}{cm} of frozen water and then covered with
\unit{47}{g/cm$^2$} of perlite to provide insulation and a barrier to light
leaks around the fitted wooden tank cover.

The tank ice is viewed by two standard IceCube digital optical
modules (DOMs). They consist of a 10" Hamamatsu R7081--02
photo multiplier tube (PMT) and processing and readout
electronics. Two different types of digitizers are
used to process the PMT signal: a fast pipelined ADC (FADC) with 255 samples of
\unit{25}{ns} each, and two Analog Transient Wave Digitizer (ATWD)
chips, with three channels of up to 128 samples of about \unit{3.6}{ns}
each. The three channels are configured with different
pre--amplification factors to extend the DOM's dynamic range (for
details, cf.\ \cite{Achterberg:2006md}).

%%%%%%%%%%%%%%%%%%%%%%%%%%%%%%%%%%%%%%%%%%%%%%%%%%%%%%%%%%%%%%%%%%%%%%%%
\section{IceTop  setup for calibration runs}
\label{sec:calibration}

Periodic special IceTop calibration runs are carried out to serve two
purposes: one, to calibrate the conversion from integrated waveform to
vertical equivalent muon (VEM) for each DOM in a tank, and two, to monitor the
DOMs response's time dependence.

\begin{figure}[t]
  \includegraphics[width=\columnwidth]{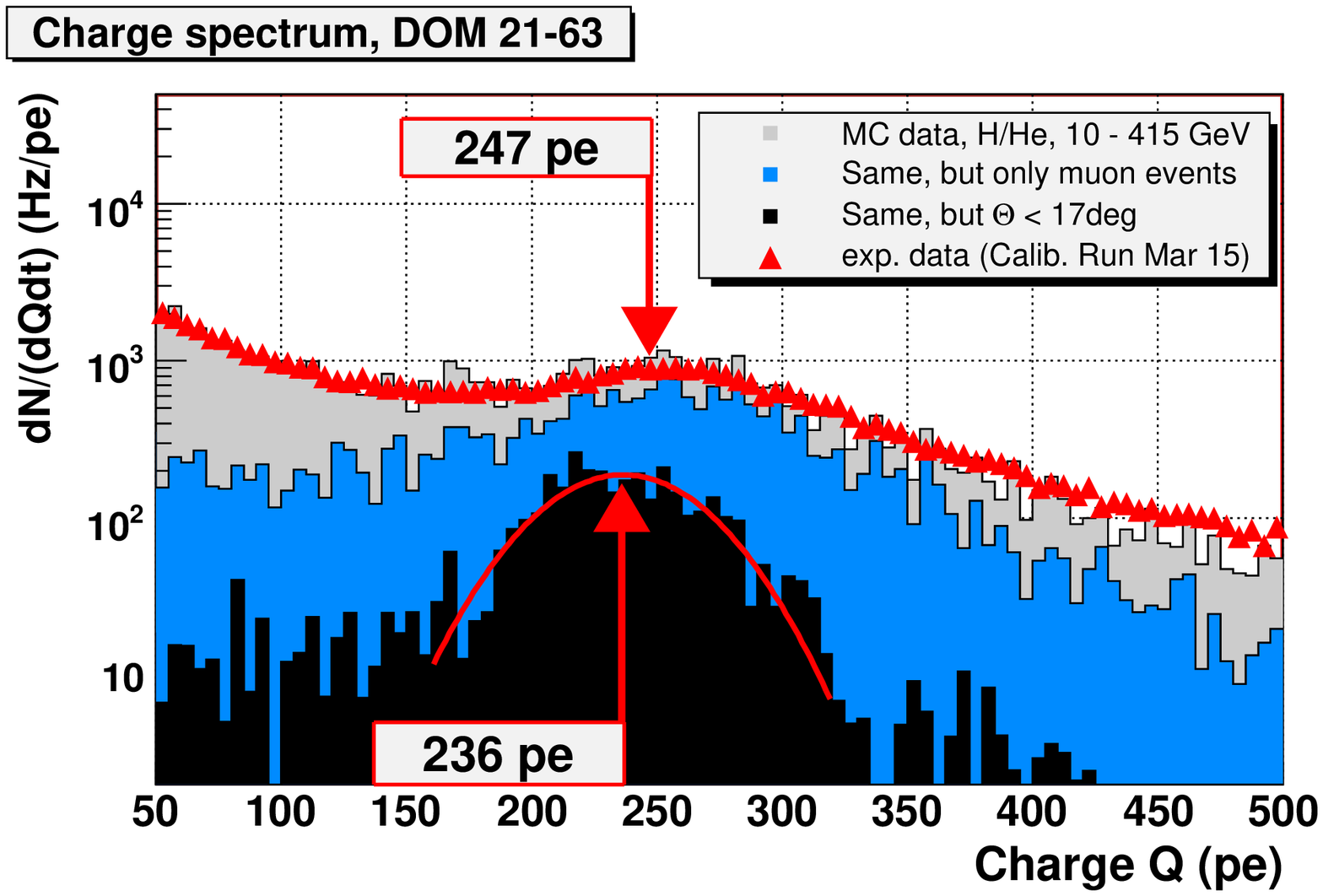}    
  \caption{MC simulated charge spectrum for DOM 21--63. See text for
    further explanations.}
  \label{fig:icrc1059_fig01}    
\end{figure}

The calibration run configuration differs from the regular one used for
air shower data runs. In this so--called singles mode, the local
coincidence between DOMs and the simple majority trigger are
disabled. All DOMs are
set to the same nominal gain of $5\cdot10{^6}$, while in the air
shower mode, the two DOMs in the same tank are set to different gains (in
2006, $5\cdot10{^6}$ and $5\cdot10{^4}$, resp.) to extend the dynamic
range of a tank. For the DOMs that are operated at the lower gain, the
VEM might differ due to changes in the collection efficiency of the PMT. 
Currently, that effect is not taken into account.

The data files are analyzed with an IceTop specific waveform
processing module written for the official offline software
suite. Each raw waveform, given in ATWD channel counts, is corrected
for the specific, ATWD chip--dependent pedestal pattern, and calibrated to
give charge. Further corrections include
the (optional) adjustment of any residual baseline offset and a
droop correction. Finally, the charge, given in
units of photo electrons (pe), is calculated by summing up all the
waveform bins. 

%%%%%%%%%%%%%%%%%%%%%%%%%%%%%%%%%%%%%%%%%%%%%%%%%%%%%%%%%%%%%%%%%%%%%%%%
\section{Calibration using through-going muons}
\label{sec:through_going_muons}

A DOM's response to a vertical muon passing an IceTop tank is defined
to be one VEM. The energy deposit of such a
muon is around \unit{200}{MeV} in the tank \cite{Beimforde:2007mb}. By
finding the vertical muon signal in the measured total charge
spectrum, the DOM--dependent charge--to--VEM conversion factor is
determined. However, single IceTop tanks cannot discriminate between
different particles or incident angles. Therefore, the relation between the
measured peak position of the total charge spectrum and the VEM must
be determined with simulations and the tagging telescope.

This is illustrated in Fig.~\ref{fig:icrc1059_fig01}. The measured
total charge spectrum is shown in triangles. The simulated total
charge spectrum (light grey) is obtained with GEANT4 based simulations. Using
Corsika \cite{Heck:1998vt} generated hydrogen
and helium air showers with primary energies between 10 and
\unit{415}{GeV} and angles up to \unit{70}{deg} as input, the DOM
response is simulated by generating and tracking the Cherenkov light
in a tank. Several tank and DOM properties, e.g.\ the reflectivities of the
sides and top, ice quality, PMT quantum efficiency, are taken into
account \cite{Clem:2007mm}.

\begin{figure}[t]
  \includegraphics[width=\columnwidth]{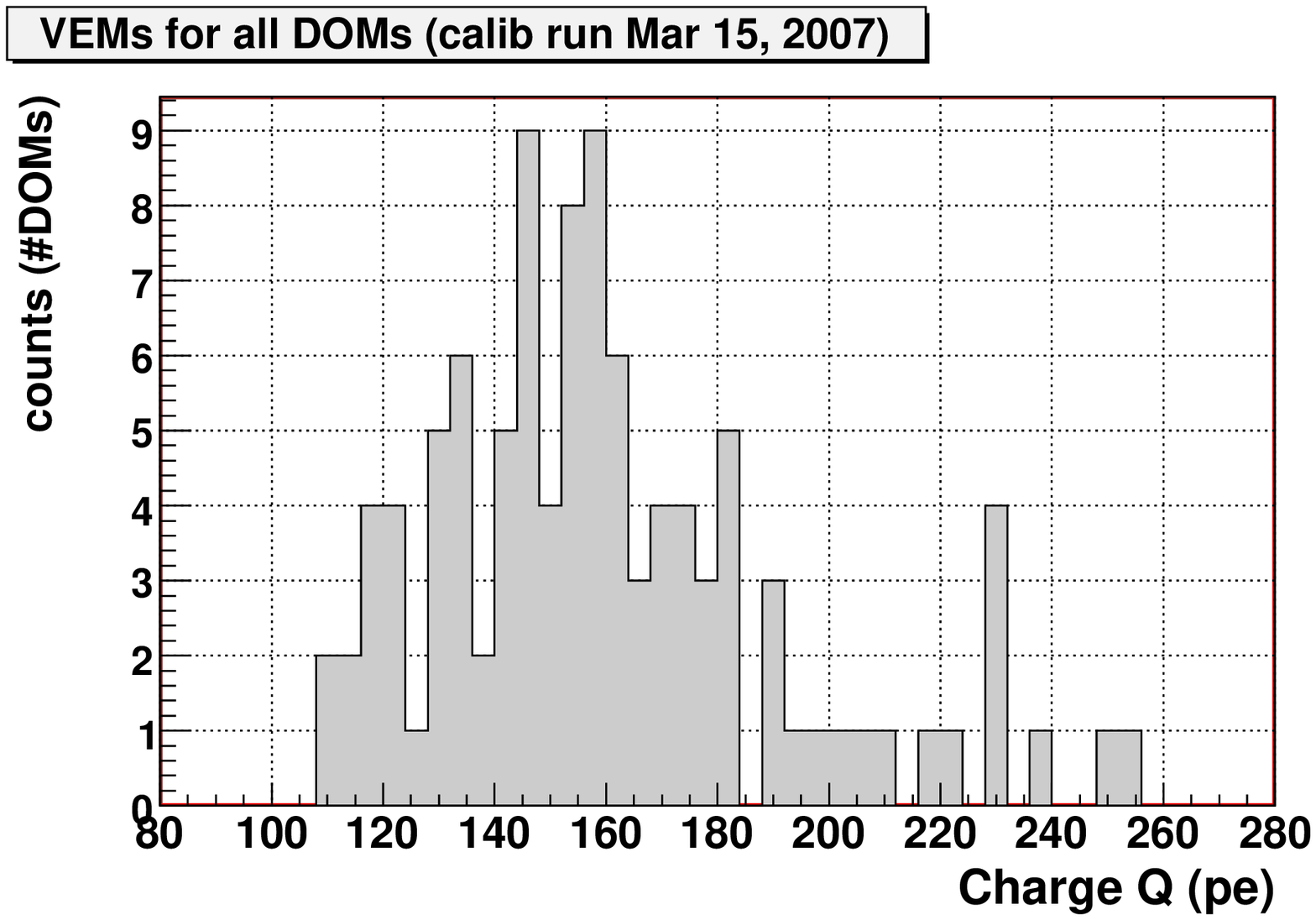}
  \caption{Distribution of VEMs for all DOMs.}
  \label{fig:icrc1059_fig02}
\end{figure}

Superimposed on the simulated total charge spectrum is the
contribution from only muons. Choosing a cut on the muons' incident
zenith angle that correponds to the angular acceptance of the
tagging telescope ($<$ \unit{17}{deg}), the black histogram is
obtained. It gives the best estimate for the VEM, which is  determined
as the mean of a Gaussian fit, \unit{236}{pe} for this particular DOM.

Comparing this to the peak position of the simulated total charge
spectrum, \unit{247}{pe}, gives a correction factor of about five
percent. This is the amount by which the measured total charge spectra's peak
positions have to be corrected to determine the VEM. Currently, it is
assumed that this correction factor is the same for all IceTop tanks.

The spread in VEM is shown in Fig.~\ref{fig:icrc1059_fig02} for a run
taken on March 15, 2007. The fluctuations in the response, even
between DOMs in the same tank, are the main reason to introduce the
VEM as a uniform, array--wide unit. 

The VEM response per DOM is tracked with regular
calibration runs. In Fig.~\ref{fig:icrc1059_fig03}, the VEM response over
time is shown for both DOMs in Tank 21b. Both DOMs exhibit a rather
stable VEM response, except for a sharp drop in DOM 21-64 around July
2006. In total, about half of all DOMs of the oldest tanks, deployed
in 2005, show a significant drop in their VEM response in mid--2006.
Though the specific cause of these changes in the DOM response is unknown,
evidence points to seasonal effects, i.e.\ the change in temperature
during the Antarctic winter.

\begin{figure}[t]
  \includegraphics[width=\columnwidth]{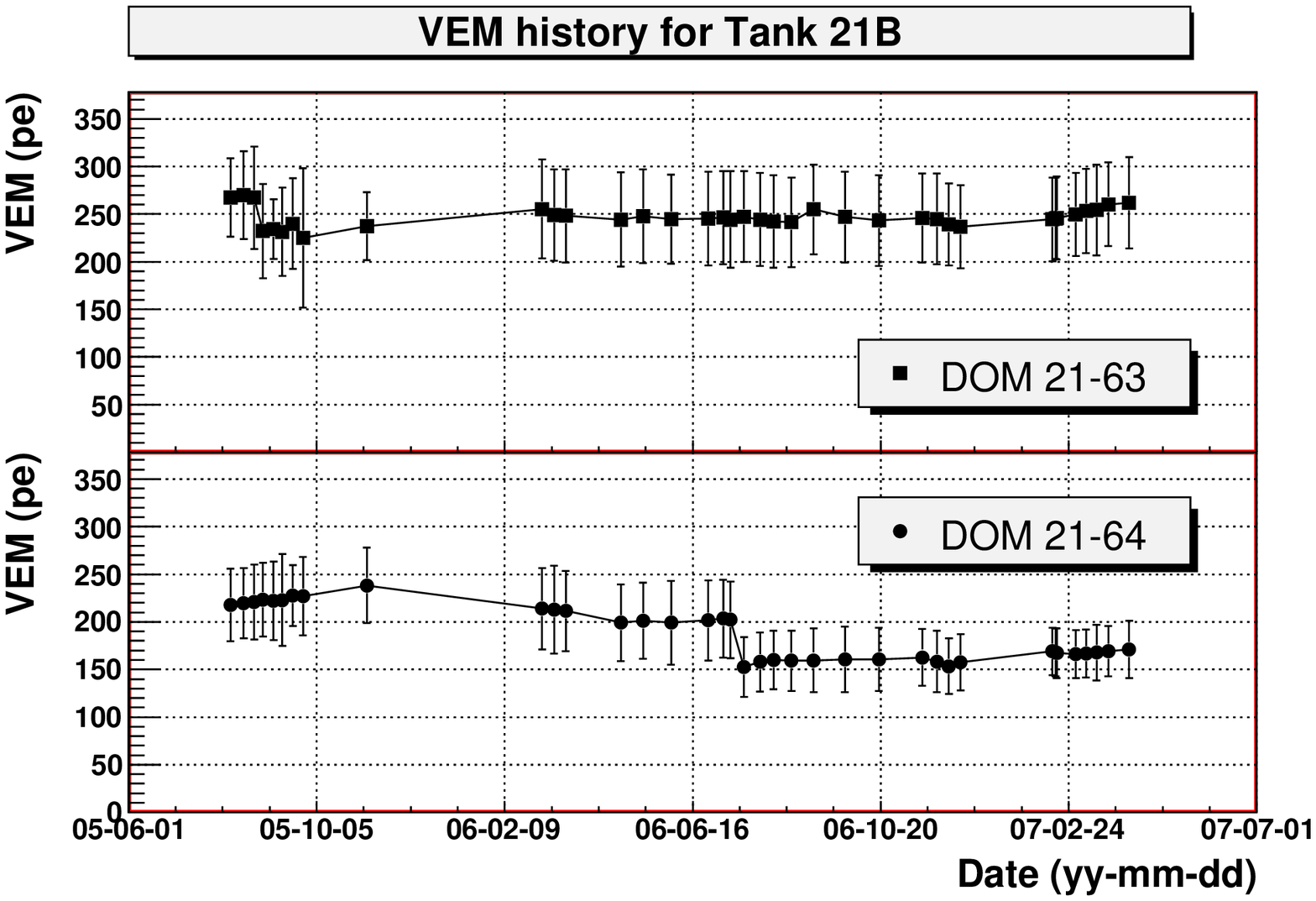}
  \caption{History of charge to VEM conversion for DOMs 21-63/64}
  \label{fig:icrc1059_fig03}
\end{figure}

%%%%%%%%%%%%%%%%%%%%%%%%%%%%%%%%%%%%%%%%%%%%%%%%%%%%%%%%%%%%%%%%%%%%%%%%
\section{Muon Telescope Measurements}
\label{sec:muon_telescope}

\begin{figure}[b]
  \begin{minipage}{0.5\columnwidth}
    \includegraphics[width=\textwidth]{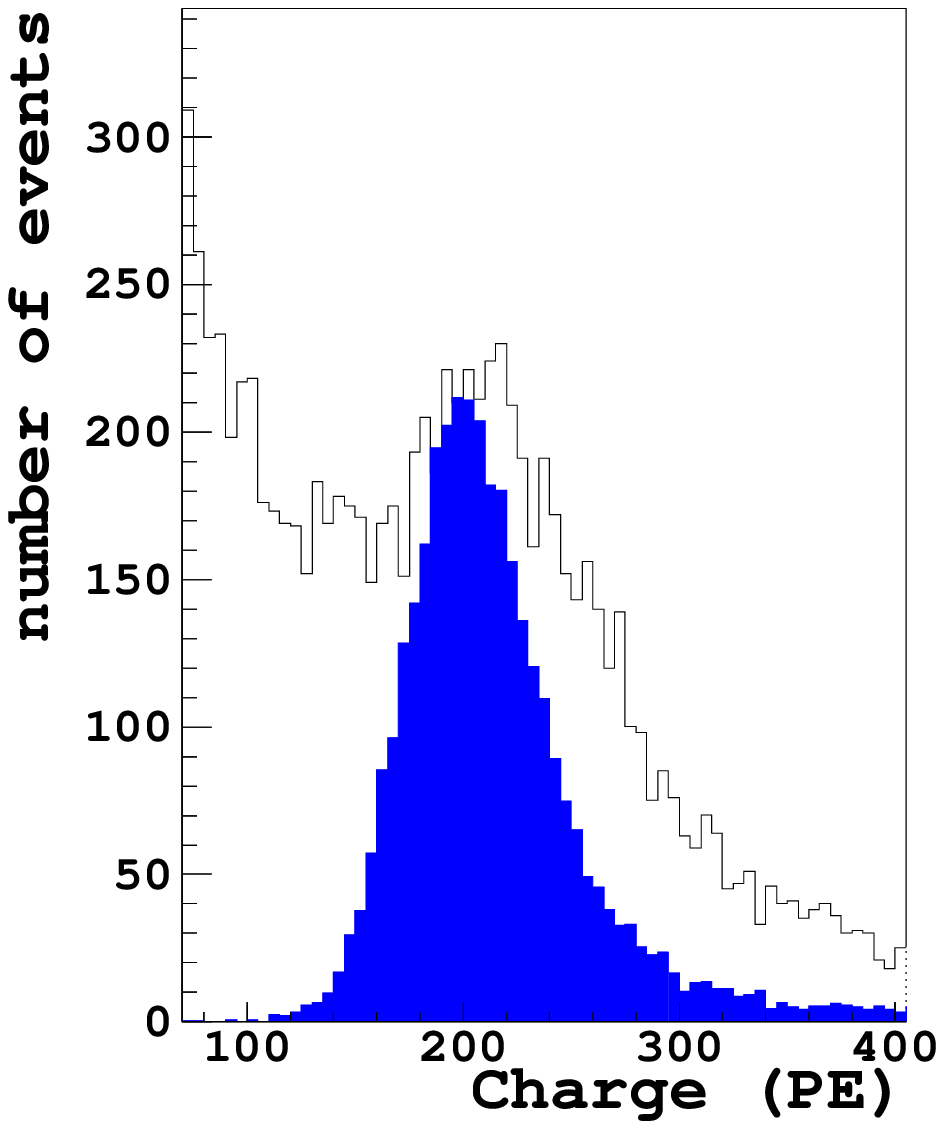}    
  \end{minipage}\hfill\begin{minipage}{0.5\columnwidth}
    \includegraphics[width=\textwidth]{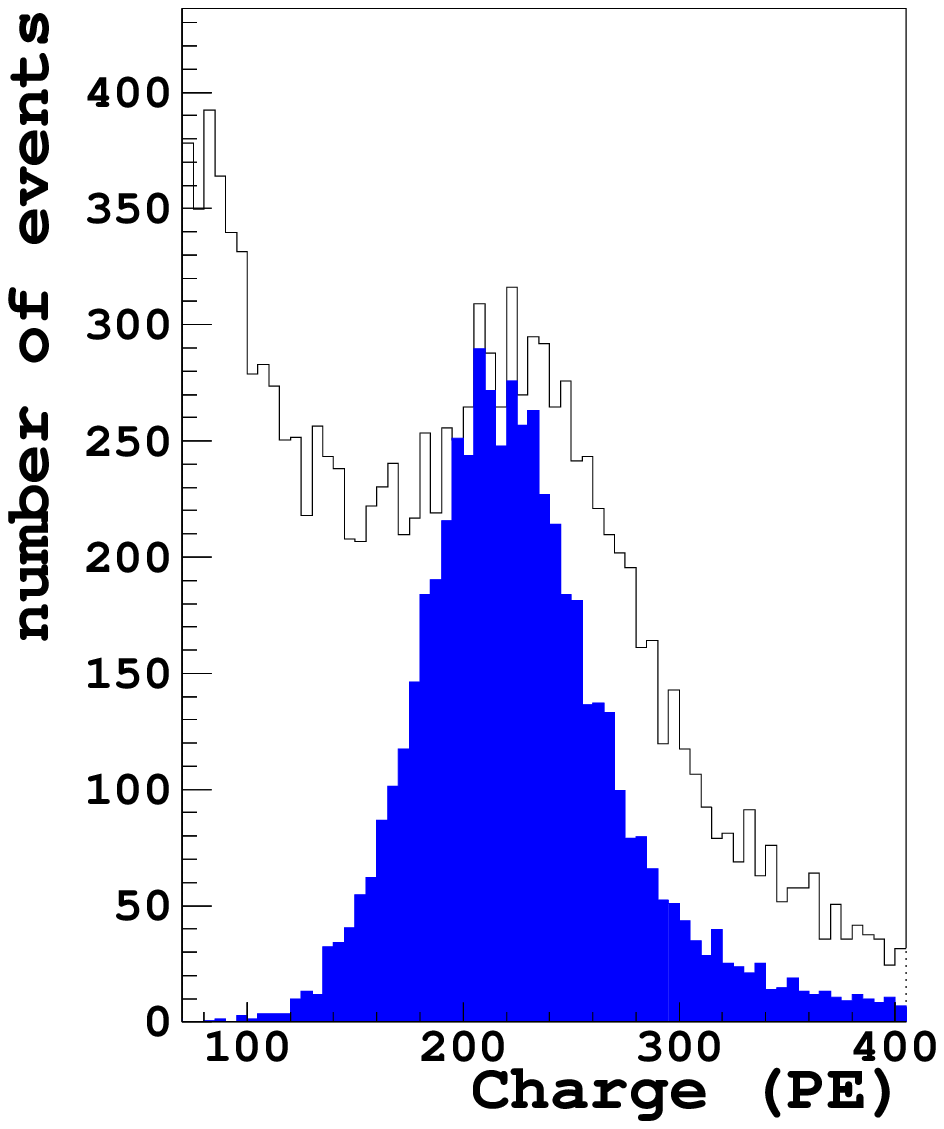}    
  \end{minipage}
  \caption{Total charge spectra (black) for tank 39b with tagged
    muon spectrum (blue) superimposed. See text for further explanation.}
  \label{fig:icrc1059_fig0405}
\end{figure}

A portable, solar--powered muon telescope was developed to tag muons that
have angles close to vertical ($<$ \unit{17}{deg}) and pass through
the center of the tank. With this device the VEM charge can be determined
independently from simulation.

The muon telescope is a completely autonomous device, having its own data
acquisition system and power supply. It measures signals in coincidence
between two scintillator slabs \unit{70}{cm}
apart and records the GPS clock time stamp on a Flash Media
drive. 

Measurements were taken during the polar season 2005/2006 on
tanks deployed one year earlier. Configuring the DOMs in a tank to
singles mode, data were taken for six hours. Matching the GPS time
stamps from both the muon telescope 
and the DOMs was done using a $[-2, 2]\,\mu$s time window. Thus, a
tagged quasi--vertical muon data set is obtained. Figure
\ref{fig:icrc1059_fig0405} shows the charge spectra for DOMs 63 and 64 in
Tank 39b and superimposed the tagged muon charge
spectra. If compared to Fig.~\ref{fig:icrc1059_fig01}, the tagged
spectra show some differences. This is mainly due to the fact that
in the simulation muons over the whole tank surface are accepted,
while the tagging telescope is positioned in the tank center. When the
statistics in simulation are improved, more realistic cuts can be
applied. Still, the qualitative difference between the tagged and the
full spectrum is well reproduced in the simulated spectrum.

%%%%%%%%%%%%%%%%%%%%%%%%%%%%%%%%%%%%%%%%%%%%%%%%%%%%%%%%%%%%%%%%%%%%%%%%
\section{Calibration using stopping muons}
\label{sec:stopping_muons}

\begin{figure}[t]
  \includegraphics[width=\columnwidth]{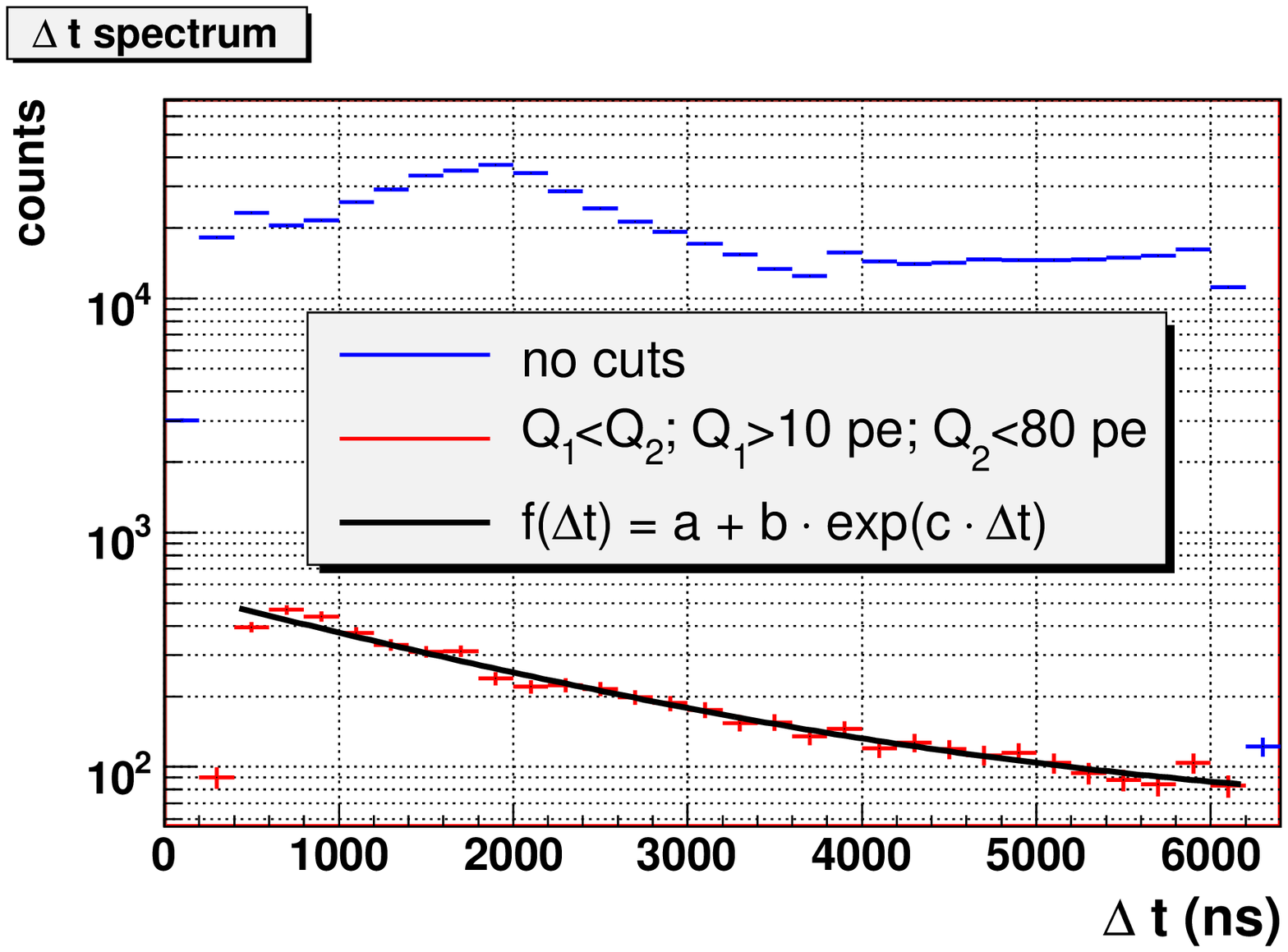}
  \caption{Time difference distributions between the two signals in
    a FADC trace. The exponential fit yields a lifetime of $\tau =
    $\unit{2.06$\pm$0.16}{$\mu$s}.}
  \label{fig:icrc1059_fig06}
\end{figure}

An IceTop tank stops muons of kinetic energies up to \unit{210}{MeV}
(vertical muons) and \unit{430}{MeV} (muon crossing through the tank
diagonally from an upper to a lower corner). After stopping, the muon
decays with its characteristic mean lifetime of \unit{2.19703}{$\mu$s}
into an
electron and an antineutrino--neutrino pair (neglecting muon
capture). The resulting energy distribution of the electron is the
well--known Michel spectrum. The maximum electron energy is
\unit{53}{MeV}, which corresponds to a range of less than
\unit{25}{cm} in the tank ice. Thus, most of the decay electrons are
well contained within the tank volume, making them a suitable
calibration sample.

A feasibility study was carried out by applying the method outlined in
\cite{Allison:2005ge} to the IceTop configuration. First, calibration
data from 2005 were analyzed to find FADC traces with two distinct 
signals. The time difference of those two signals is shown in
Fig.~\ref{fig:icrc1059_fig06} as the upper histogram. To suppress
background, stringent cuts were applied on the integrated charges
$Q_1$ and $Q_2$ of the primary and secondary signal, respectively. The cuts
were adjusted by using the GEANT4 based simulation from \cite{Clem:2007mm}.

Fitting the remaining time difference spectrum yields a lifetime of
$\tau =\, $\unit{2.06$\pm$0.16}{$\mu$s}, which is comparable to the muon
mean lifetime of \unit{2.2}{$\mu$s}.

To extract the Michel spectrum from the background, a difference
method is chosen that does not require the cuts imposed above. First,
two time windows are chosen, a ``decay'' window between 1 and
\unit{2}{$\mu$s}, and a ``crossing'' window between 5 and \unit{6}{$\mu$s}.
For both time windows, the integrated charge of the
second signal is calculated. By subtracting them from
each other, the Michel spectrum is obtained, which is compared to a
simulated spectrum in Fig.~\ref{fig:icrc1059_fig07}. Though the
simulation lacks statistics, it qualitatively describes the measured
spectrum rather well.

\begin{figure}[t]
  \includegraphics[width=\columnwidth]{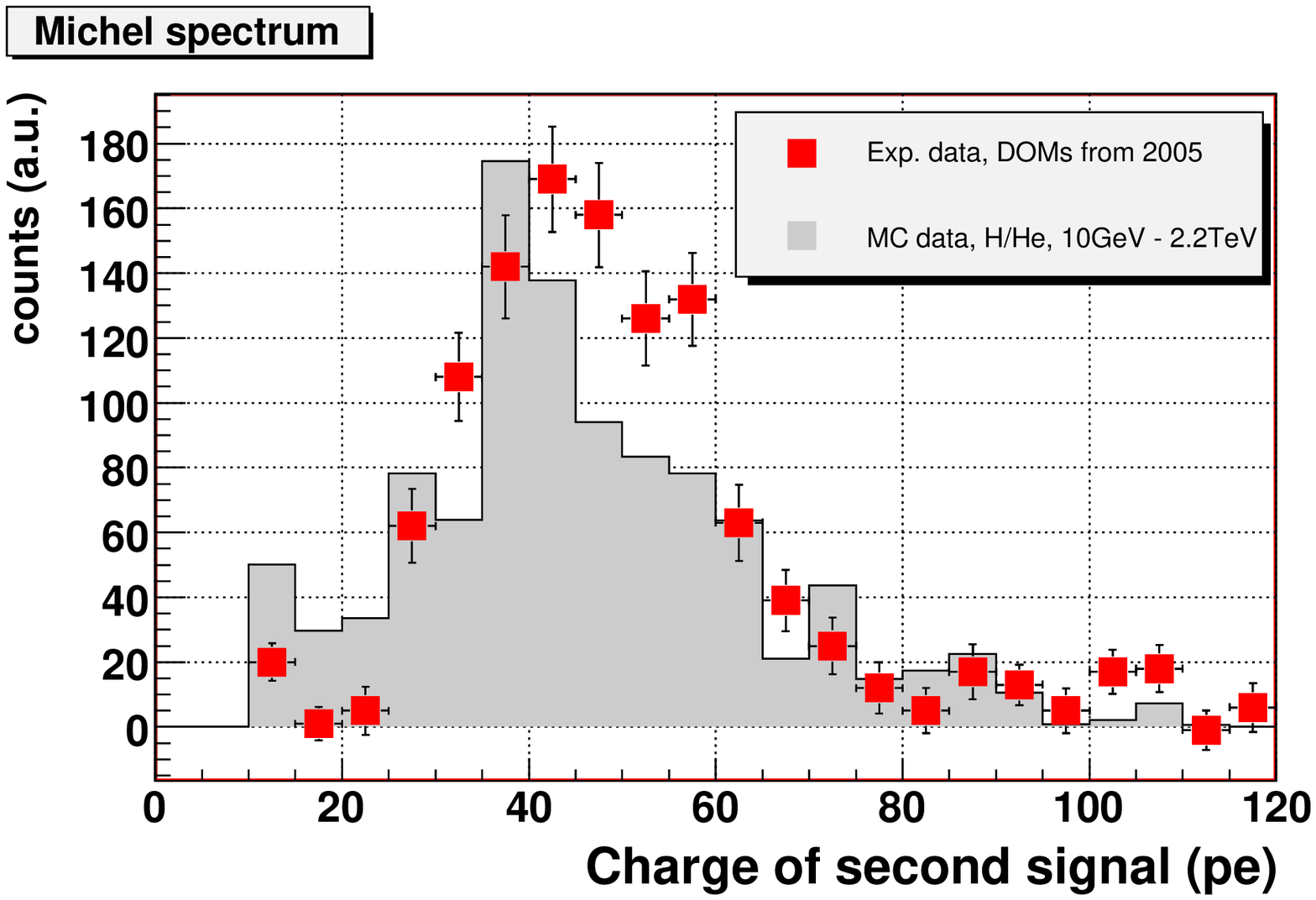}
  \caption{Measured Michel spectrum (symbols)
    in comparison with a simulated one.}
  \label{fig:icrc1059_fig07}
\end{figure}

%%%%%%%%%%%%%%%%%%%%%%%%%%%%%%%%%%%%%%%%%%%%%%%%%%%%%%%%%%%%%%%%%%%%%%%%
\section{Conclusion}
\label{sec:conclusion}

The VEM calibration of the IceTop air shower array with through--going
muons is a well established and well understood procedure.  The VEM is
measured and calibrated on a weekly to monthly basis and provides, in
conjunction with the single DOM rate and temperature, a basic set of
observables for monitoring the detector hardware. GEANT4 based
simulations agree well with the measured charge spectra and the muon
telescope data, showing that the input parameters describe the actual
tank properties rather well.

The stopping muon analysis has shown the feasibility of using the muon
decay signal as a supplementary calibration source. Already at this
stage, the GEANT4 based simulation shows a promising agreement with the
measured spectra. However, further improvements in both the analysis
and the simulation are needed to establish it as a standard
calibration method.

%%%%%%%%%%%%%%%%%%%%%%%%%%%%%%%%%%%%%%%%%%%%%%%%%%%%%%%%%%%%%%%%%%%%%%%%
\section{Acknowledgments}
\label{sec:acknowledgments}
This work is supported by the U.S. National Science Foundation, Grants
No. OPP-0236449 and OPP-0602679.

\setcounter{figure}{0}
\setcounter{table}{0}
%\documentclass{article}
%\usepackage{icrctc07, epsfig, wrapfig, amssymb, amsmath, times}
%\linenumbers
%\include{asp2004}
%\usepackage{../styles/rcsinfo/rcsinfo}
%\usepackage{../styles/dateiliste/dateiliste}
% $Id: low_energy_tank_response_icrc_2007.tex,v 1.33 2007/08/15 19:18:24 clem Exp $
\title{\bf Response of IceTop tanks to low-energy particles}
%$Revision: 1.33 $}
\shorttitle{Low energy response of IceTop}
\authors{J. M. Clem$^{1}$, P. Niessen$^{1}$, and S. Stoyanov$^{1}$ for
the IceCube Collaboration$^2$}
\shortauthors{J. M. Clem et al.}
\afiliations{$^1$University of Delaware, Dept. of Physics and
Astronomy, DE-19716 Newark, US of America\\
$^2$see special section of these proceedings}
\abstract{
Solar activity can cause variations in the cosmic-ray particle
flux measured at the Earth's surface. This manifests mostly in the
low-energy electromagnetic component of cosmic ray induced cascades.
The IceTop experiment detects these particles by their emission of Cherenkov
light in a contained ice volume through photo-multipliers.
We give the prediction of the response to the low-energy part
of cascades and compare to experiment.
}
%\begin{document}
\setlength\abovecaptionskip{1truemm}
\maketitle
\section{Introduction}
The IceTop Air Shower Array, located close to the geographical South
Pole (altitude 2835 m, 700g/cm$^{2}$), consists of tanks with reflective
liners using clear ice as a
Cherenkov medium. Light generated in the ice is observed by digital
optical modules (DOMs) which consist of a photo multiplier tube (PMT)
and digitising electronics assembled in a glass pressure sphere.
Thus, energy deposition of particles can be measured through the observed light yield. Each tank
has two DOMs running at different gain settings to increase the
dynamic range of the observations. Two tanks, placed at 10 metres from
each other, are combined into a station.
Currently, 26 stations, separated by typically 125 metres, forming
a diamond shaped triangular grid are deployed. In normal operation,
the high gain DOMs are run in coincidence to reject events not
associated with air showers. For this work, we use data from tanks
run in ``single mode'', in which the coincidence condition is disabled.

\section{Simulations}

Two separate simulations are utilised in this analysis, one based on
CORSIKA\cite{Heck1998} and another on
FLUKA/AIR\cite{Fasso2005,Clem2004}.

{\bf In the AIR model}, primary protons, alphas, carbon, silicon and
iron are generated within the rigidity range of 0.5GV-20TV uniform in
$\cos^2(\theta)$, $\theta$ being the zenith angle. The atmosphere
density profile (23.3\% oxygen, 75.4\% nitrogen and 1.3\% argon) was based on
the US Standard Atmosphere 1976 model. The primary cosmic ray spectrum
used in this calculation was determined through an analysis of
simultaneous proton and helium measurements made on high altitude
balloon flights (see refs. in \cite{Clem2004,Gaisser2001}).
The outer air-space boundary is radially separated
by 65 kilometres from the inner ground-air boundary and a single 1 cm$^2$
element on the air-space boundary is illuminated with
primaries. Particle intensity at various depths is determined by
superimposing all elements on the spherical boundary defining the
depth. Due to rotational invariance this process is equivalent to
illuminating the entire sky and recording the flux in a single
element at ground level. Although this approach provides a quick
result, it ignores the effects of multiple particle tracks entering the
IceTop tanks simultaneously.   

{\bf In the CORSIKA simulation}, the hadronic interaction model for
energies above 80 GeV is SIBYLL v2.1\cite{Fletcher1994}, for lower
energies FLUKA is applied.
%GHEISHA\cite{Fesefeldt1985}.
The electromagnetic interactions are treated with EGS4\cite{Nelson1985}.
Hydrogen as well as helium primaries are simulated with angles between
0 and 70 degrees. The angular spectrum is constant in $\cos^2(\theta)$,
like for the AIR simulation.
The cascades are generated with primary energies between 10 GeV and 468
GeV with a power-law $\sim (E/E_0)^{-1}$ and are re-weighted later
to the fluxes averaged from various experiments\cite{Gaisser2001}.
Two atmospheres for the austral winter and summer (1st of July/31st
of December) parametrised by the MSIS-90-E model\cite{MSIS_90_E} are
used. We find that the counting rate in the austral winter is approx. 6\%
higher compared to the summer.
In a second step, the cascade particles are inserted into the
detector simulation to generate the light yield in the photo
multiplier. The simulation is based on GEANT4\cite{GEANT4} and takes
into account the interactions of particles and the tracking of the Cherenkov
photons. This requires input of the optical parameters of the inside
of the tank. The reflectivity of the tank liner was measured as a
function of wavelength in the laboratory.
% yielding the results
% shown in
% figure \ref{fig:refl_plot}.
The first eight tanks of the experiment
are lined with Tyvek\texttrademark, while the later tanks
have an integrated coating using zirconium as reflective agent.
% corresponding to ``2004 Tyvekbag'', and ``Tank 203'', ``Tank 202''
%and ``Tank 201'' respectively.
The simulations are done using the optical properties of the
Tyvek\texttrademark liner.
% and ``Tank 203''.
%\begin{figure}[h!]
%\begin{center}
%\epsfig{file=plots/refl_plot.eps, width=0.95\columnwidth}
%\caption{\label{fig:refl_plot}Reflectivities of tank liners.}
%\end{center}
%\end{figure}

The tank is then modelled as a cylindrical polyethylene vessel of 0.93
metre radius and 1.00 metre height, filled with ice to a level of 0.90
metres. The tank is embedded in 0.3 metres of snow, simulated as water
of density 0.4 g/cm$^3$.
%The chemical composition and density of these materials are
%used in the simulation of particle interactions
%with the tank.
Regarding the optics of the ice, a refractive index of
1.33 is assumed and the absorption length is set to 200 metres,
based on measurements in the deep glacial ice 
and on comparisons of the simulations to the experimental data.
The ice is covered with 47 g/cm$^2$ of Perlite\texttrademark which is
modelled as opaque to light but reflective at the ice interface.
The light propagation in the DOM itself is simulated using the
geometry and optical properties of the pressure sphere, the PMT glass
and the optical gel coupling the two. The quantum efficiency of the
photo cathode is applied to yield individual photo electrons. However,
neither the amplification stages nor the signal processing electronics are
simulated. The final result of the simulation is the number 
of photo electrons (npe).

\section{Secondary particle spectra}
The resulting secondary particle spectra from simulation and
experiments\cite{Grieder2001} are shown in
Fig.~\ref{fig:E_log_E_by_particle_no_corsika}.
\begin{figure}[!t]
\begin{center}
\includegraphics[width=.95\columnwidth]{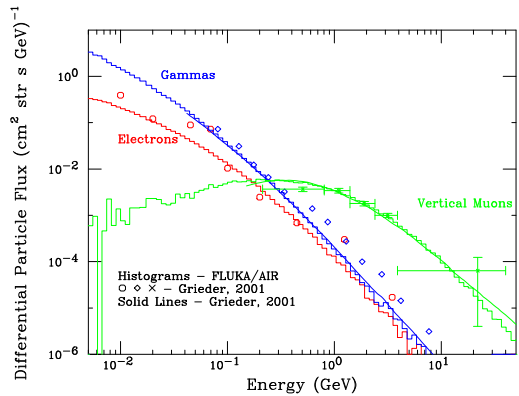}
%width=0.95\columnwidth
\caption{\label{fig:E_log_E_by_particle_no_corsika}Fluxes of secondary
electrons, muons and gammas from simulation for solar minimum, compared to experiments compiled in\cite{Grieder2001}.}
\end{center}
\vspace*{-5truemm}
\end{figure}
All measurements of the electrons, muons and gammas took place at solar minimum
and a low geomagnetic cutoff, comparable to South Pole conditions. The muon and electron measurements were
made by a balloon instrument while the gamma rays were measured from a mountain top.    
The agreement with the simulations is reasonable, however the differences will be investigated. 

\section{Response to electrons, muons and gammas}
The particles entering the tank are detected by the DOM either by
their own Cherenkov light (if they are charged) or by the light
emitted in stochastic processes (pair production, delta electrons,
etc.).
\begin{figure}[!h]
\begin{center}\includegraphics[width=.95\columnwidth]{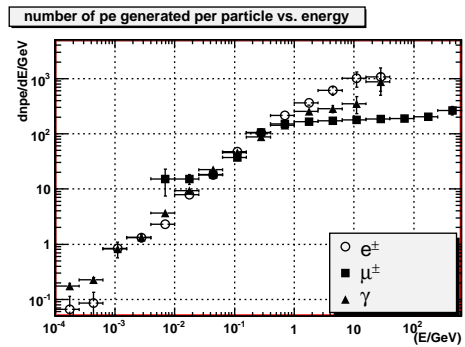}
\caption{\label{fig:dnpe_dE}Number of photons per particle
vs. particle energy.}
\end{center}
\vspace*{-5truemm}
\end{figure}
 The number of photo electrons seen per particle
as a function of the particle energy in the tank is shown in
Fig.~\ref{fig:dnpe_dE}.
It is averaged over all angles and impact parameters.
The light yield of the muons turns flat at around 1 GeV, where the
muons become minimal ionising plus a logarithmically rising
stochastic contribution.
For all three particle types, the
light generation threshold is around 1 MeV.

\section{Contributions to the photon electron rate}
The simulation allows one to study the contribution of the different
cascade secondary particles to the photo electron response from the
DOM. This is shown in Fig.~\ref{fig:npe_by_particles}.
\begin{figure}[!t]
\begin{center}
\includegraphics[width=.95\columnwidth]{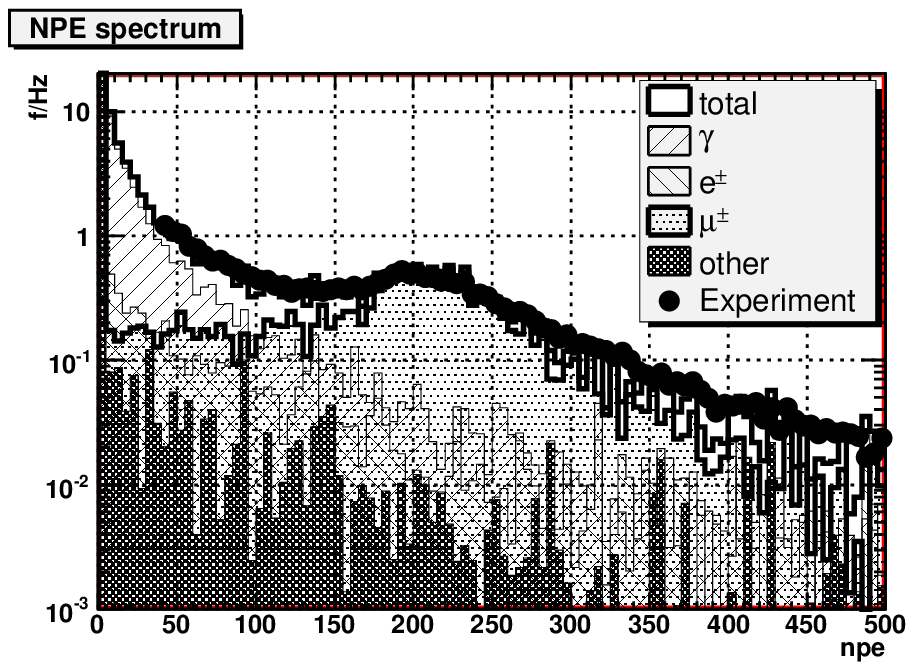}
\caption{\label{fig:npe_by_particles}Contribution of secondary 
particles to the overall npe signal. The experimental data for DOM 63
of station 29's Tyvek\texttrademark lined tank is shown as well.}
\end{center}
\vspace*{-5truemm}
\end{figure}
For different particles (gamma, electron, muon and other), the number
of photons seen by the DOM is summed up and histogrammed.

% The electromagnetic particles contribute mostly to the counting
% rates at lower light yields. In the experiment, they are suppressed by
% chosing a sufficiently high trigger threshold and requiring coincident
% signals from two tanks in a station. The muons produce a peak in the
% npe distribution at around 200 pe, the plateau to the left of the peak
% is due to so called corner clippers, in which the
% track does not cross the entire depth of the tank.

The dominant contributions come from gammas, electrons and muons, however the 
neutron component is significant at low primary energy.
Other particles contribute at the 1\% level.

The simulation is compared to measurements.
There are some variations in the position of the muon peak from
tank to tank and a tank fitting the simulation is shown. Since the
purpose of these data is to determine the position of the muon peak, a
threshold of about 40 npe is applied. There is good agreement between
experiment and simulation.

%The integral photon count rates for different
%combinations are shown in table \ref{table:count_rates}.

%\begin{table}[h]
%\begin{center}
%\begin{tabular}{lrr}
% & 01-JUL & 31-DEC \\
%\hline
%\hline
%Tyvekbag, $\lambda_{\mathrm{abs}}=200$ m & 848 Hz & 796 Hz \\
%Tank 203, $\lambda_{\mathrm{abs}}=200$ m & 796 Hz & 726 Hz \\
%Tyvekbag, $\lambda_{\mathrm{abs}}=80$ m & TBD Hz & TBD Hz \\
%Tank 203, $\lambda_{\mathrm{abs}}=80$ m & TBD Hz & TBD Hz \\
%\end{tabular}
%\caption{\label{table:count_rates}Integral photon count rates for
%different atmospheres and tank liners. {\bf Values are still too low
%because higher energy primaries (>500 GeV) are missing.}}
%\end{center}
%\end{table}

\section{Primary Cosmic Ray Single Mode IceTop Yield Function}

%In order to understand the IceTop tank in single mode as a primary particle
%detector, a relationship between the count rate and primary flux must be
%established.  As primary particles enter the atmosphere, they undergo
%multiple interactions resulting in showers of secondary particles, which may
%reach ground level and produce a signal in an IceTop tank.

The yield function $S (P, z)$ describes the primary cosmic ray detection
efficiency of a full sky illumination of particles averaged over all arriving angles (uniform in 
$\cos^2(\theta)$). It is related to the count-rate $N (P_C, z, t)$ by

\begin{eqnarray*}
N(P_C,z,t) % &=& \int\limits_{P_C}^{\infty} \sum_i (S_i(P,z)\, j_i(P,t)) dP
&=& \int\limits_{P_{C}}^{\infty} (S(P,z)\, j(P,t))  dP.
% &=& \int\limits_{P_{C}}^{\infty} (S(P,z)\, j(P,t))  dP.
\end{eqnarray*}

%\begin{eqnarray*}
%% N(P_C,z,t) &=& \int\limits_{P_C}^{\infty} \sum_i (S_i(P,z)\, j_i(P,t)) dP\\ 
%&=& \int\limits_{P_{C}}^{\infty} (S(P,z)\, j(P,t))  dP.
%\end{eqnarray*}

\begin{figure}[!h]
\begin{center}
\includegraphics[width=.95\columnwidth]{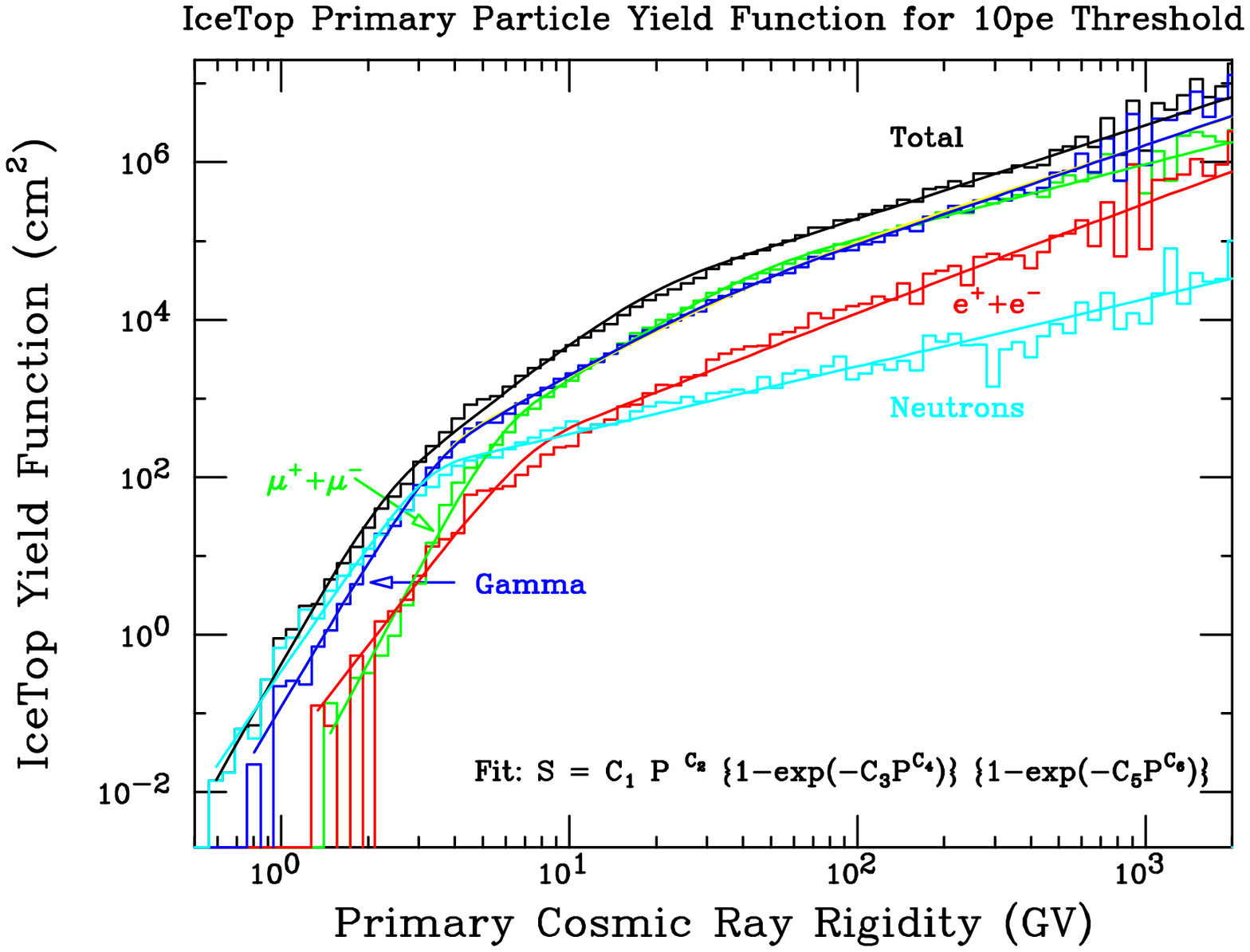}
\caption{\label{fig:yield_func} Primary cosmic ray yield function
  $S(P,z=700\mathrm{g}/\mathrm{cm}^2)$
for IceTop tank in singles mode. 
The individual contributions made by secondary components to the yield
function are separated into different curves.}
\end{center}
\vspace*{-5truemm}
\end{figure}
where $P$ is the particle's rigidity (momentum/charge), $z$ is the
atmospheric depth and $t$ represents time. 
$P_C$, the geomagnetic cutoff, is effectively 0 at the South Pole.
Using $S_i(P,z)$, the single mode IceTop yield function,
and $j_i(P,t)$, the primary rigidity spectrum for primaries
of particle type $i$, one can decompose the 
product of yield function and rigidity spectrum,
$S(P,z)\, j(P, z)$ into $\sum S_i(P, z)\, j_i(P, z)$.
Utilising the FLUKA/AIR model and a FLUKA Cherenkov optical model
assuming a zirconium lined IceTop tank, the IceTop yield function was
calculated for a 10pe threshold (Fig. \ref{fig:yield_func}). The
data are fit using a variation of the Dorman Function\cite{Clem2000}

\begin{eqnarray*}
S(P) = C_1 P^{C_2} & \times & (1-\exp\{-C_3 P^{C_4}\})
\\  & \times & (1-\exp\{-C_5 P^{C_6}\}),
\end{eqnarray*}

typically used to model Neutron Monitor latitude survey data.
The fit parameters extracted from the simulations in Fig.
\ref{fig:yield_func} for the total count rate as function of PE threshold are shown in Tab.~\ref{tab:yield_fit}.
\begin{table}[!h]
\begin{center}
\begin{tabular}{|c|cccc|r|}
        \hline
$C$	& 5pe	& 10pe &	25pe	&50pe \\
%$C$ &  Total    & Muons    & Gammas   & $e^++e^-$  \\
        \hline
%1   &  4024.24 & 6800.82 & 1433.477 & 1964.894 \\
%2   &  1.18865 & 0.946676 & 1.25253 & 1.06198\\
%3   & 7.90E-03 & 4.73E-03 & 4.95E-02  & 3.85E-02 \\
%4   & 1.791008 & 1.517915 & 0.99001 & 0.418627\\
%5   & 6.93E-02 & 6.06E-04 & 8.83E-03 & 1.86E-03 \\
%6   & 3.557329 & 4.368393& 3.934367  & 3.27841\\

%1&	164.05	&150.89	&108.95&	75.979\\
1&     32.81  &30.18 &21.79&     15.20\\
2&	4.8075	&4.8032	&4.7731&	4.7408\\
3    &0.0341	&0.0150&	 .00534&	.00232\\
4	&1.1849     &1.4696    &1.8457	&2.270\\
5	&30.588    &28.323    &30.874&	 33.54\\
6	&-3.6117  &-3.6184   &-3.6070&	-3.584\\

%1   &  140.2 & 6800.8 & 1433.8 & 1964.9 \\
%2   &  4.7589 & 0.9467 & 1.2525 & 1.0620\\
%3   & 1.54E-3 & 4.73E-3 & 4.95E-2  & 3.85E-2 \\
%4   & 1.403 & 1.5179 & 0.9900 & 0.4186 \\
%5   & 33.554 & 6.06E-4 & 8.83E-3 & 1.86E-3 \\
%6   & -3.5900 & 4.3684 & 3.9344  & 3.2784 \\
%1   &  4024.2 & 6800.8 & 1433.5 & 1964.9 \\
%2   &  1.1887 & 0.9467 & 1.2525 & 1.0620\\
%3   & $7.90\cdot10^{-3} $ & $4.73\cdot10^{-3}$ & $4.95\cdot10^{-2}$
%& $3.85\cdot10^{-2}$ \\
%4   & 1.7910 & 1.5179 & 0.9900 & 0.4186 \\
%5   & $6.93\cdot10^{-2}$ & $6.06\cdot10^{-4}$ & $8.83\cdot10^{-3}$ &
%      $1.86\cdot10^{-3}$ \\
%6   & 3.5573 & 4.3684 & 3.9344  & 3.2784 \\
        \hline
\end{tabular}
\caption{\label{tab:yield_fit}Fit values for the yield function}
\end{center}
\end{table}
\vspace*{-10truemm}

\section{Integral count rates}
The above information can now be used to predict counting rates above
a given threshold (Fig.~\ref{fig:npe_integral}). The
agreement between experiment and simulation is reasonably good for the
\begin{figure}[!h]
\begin{center}
\includegraphics[width=.95\columnwidth]{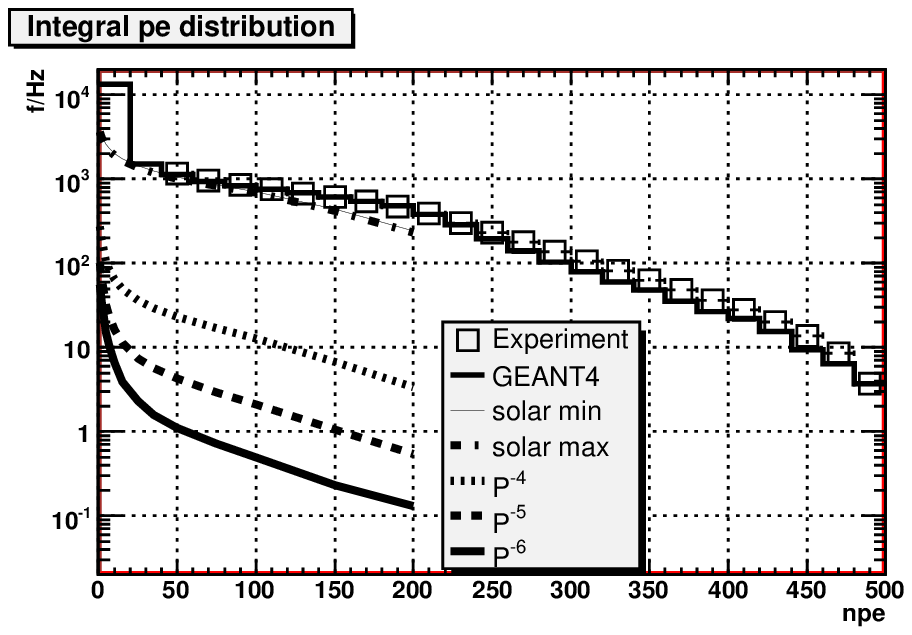}
\caption{\label{fig:npe_integral}Integrated photon counting rates for
various primary spectra, including $P^{-4, -5, -6}$ as expected for
solar activity. Note that solar minimum and maximum give approximately
the same rates.}
\end{center}
\end{figure}
solar minimum and maximum periods. The addition of $P^{-4, -5, -6}$ spectra,
which are typical for solar flares, to the galactic cosmic ray background is expected 
to yield a count rate enhancement by a few percent depending on the IceTop tank 
photo-electron threshold setting and solar particle intensity.

\section{Conclusion}
The IceTop tanks are sensitive to low energy particles produced in
cascades by cosmic radiation. The response of the IceTop detectors is
understood reasonably well in terms of the simulation, as shown by
comparison to experimental measurements. This allows predictions of
rate changes induced by changes in the primary particle spectrum. Furthermore 
these prediction suggest variations greater than that induced by atmospheric
variations, leading to good detectability of solar events.

%It is important to note
This analysis ignores the effects of 
multiple particle tracks entering the IceTop tanks simultaneously as each
particle track reaching the ground is treated as an uncorrelated event
regardless of arrival time. For low energy primaries this is a valid
approach, however at high energies this could be a source of systematic
errors. This effect will be investigated in order to quantify it.

\section{Acknowledgements}
This work is supported in part by the U.S. National Science
Foundation, Grants No. OPP-0236449 and OPP-0602679.

%\end{document}
 % epsfig !!!
%
%atmospheric neutrino papers
%
%icrc0373.pdf   (Kelley, Ahrens)
%IC9_Atmospheric_Neutrinos...
%icrc0190v5.pdf
%
\setcounter{figure}{0}
\setcounter{table}{0}
%document class
%\documentclass{article}

%\usepackage{icrctc07}

%\begin{document}

%title, author etc
\title{Testing alternative oscillation scenarios with atmospheric neutrinos using AMANDA-II data from 2000 to 2003}
\authors{J. Ahrens$^1$ and J.L. Kelley$^2$ for the IceCube Collaboration$^A$} %$^3$ 
\afiliations{$^1$ Institute of Physics, Mainz University, Staudinger Weg 7, D-55099 Mainz, Germany\\
$^2$ Department of Physics, University of Wisconsin, Madison, WI 53706, U.S.A.\\
$^A$ See special section of these proceedings}
\email{jens.ahrens@lycos.de, jkelley@icecube.wisc.edu}

%____________________________________________________________________________________________________________________________________________________
\abstract{The AMANDA-II neutrino telescope 
detects upward-going atmospheric muon neutrinos penetrating the Earth from the Northern Hemisphere 
via the Cherenkov light of neutrino-induced muons, allowing the reconstruction of the original neutrino direction.
Due to the high energy threshold of about $50\,\mathrm{GeV}$, the declination
spectrum is minimally affected by standard neutrino oscillations;
however, alternative oscillation models predicting subdominant effects can be tested and constrained.
Of particular interest are models that allow one to test Lorentz invariance and the equivalence principle.
Using the AMANDA-II data from the years 2000 to 2003, a sample of $3401$
candidate neutrino-induced events was selected.
No indication for alternative oscillation effects was found.
For maximal mixing angles, an upper limit is set on both the Lorentz violation parameter
$\delta c/c$ and the equivalence principle violation parameter
$2\vert\phi\vert\delta\gamma$ of $5.3\times10^{-27}$ at the 90\% confidence level.
}
%\begin{document}
\maketitle
%____________________________________________________________________________________________________________________________________________________
\section{Introduction and detector description}
Cosmic ray particles entering the Earth's atmosphere generate a steady flux of secondary
particles, including muons and neutrinos. High energy muons pass through the atmosphere and can
penetrate several kilometers of ice and rock, while atmospheric
neutrinos of energies only above roughly
$40\,\mathrm{TeV}$ start to be absorbed in the Earth.
Lower energy muon neutrinos penetrating the diameter of the Earth can
oscillate into tau neutrinos.
However, the oscillation maxima at
$30\,\mathrm{GeV}$~\cite{bib:Kamio} and below are beneath the AMANDA-II 
threshold. Departures from conventional mass-induced oscillations
could emerge at higher neutrino energies due to 
relativity-violating effects (see below).
Such mechanisms would distort the expected angular distribution and energy
spectrum of atmospheric neutrinos and could  
be detectable by AMANDA-II.

The AMANDA-II neutrino telescope is embedded $1500-2000\,\mathrm{m}$
deep in the transparent and inert ice of the Antarctic ice sheet, close to the geographic South Pole. 
AMANDA-II consists of $677$ optical modules (OMs) on $19$ vertical strings, which are arranged in three approximately concentric circles of 
$60\,\mathrm{m}, 120\,\mathrm{m}$ and $200\,\mathrm{m}$ diameter.
Muons produced in $\nu_{\mu}$-nucleon interactions can be directionally reconstructed by observing the
Cherenkov radiation that propagates through the ice to the array of photosensors. To ensure that the
observed muon is due to a neutrino interaction, the Earth is used as a filter against atmospheric muons, and only tracks
from the Northern Hemisphere (declination $\delta > 0^{\circ}$) are selected.
%____________________________________________________________________________________________________________________________________________________
\section{Phenomenology of standard and alternative neutrino oscillations}
\label{nuoszi}
It is commonly accepted that standard (mass-induced) $\nu_{\mu}\rightarrow\nu_{\tau}$ oscillations\footnote{In 
the regime of atmospheric neutrino oscillations, it suffices to consider a two-flavor system of eigenstates ($\nu_\mu,\nu_\tau$).}
are responsible for the measured deficit of atmospheric muon neutrinos
(see \textit{e.g.} \cite{bib:Kamio}).
Atmospheric neutrino data can also be used to test non-standard oscillation mechanisms that lead to observable
differences at higher neutrino energies. Various new physics scenarios can result in neutrino flavor mixing.
Two of these scenarios, which can be described in a mathematically analogous way, have been 
tested in this analysis. The underlying theories assume small deviations
from the principles of the theory of relativity and lead to measurable neutrino oscillations:
\begin{itemize}
\item In theories predicting violation of Lorentz invariance (VLI), a set of additional neutrino 
eigenstates with different maximal attainable velocities (MAV) $c_n/c$ is
introduced, violating special relativity \cite{bib:coleglas}.
\item In theories predicting violation of the weak equivalence principle (VEP),
gravitational neutrino eigenstates are introduced which couple with
distinct strengths $\gamma_n$ to a gravitational potential $\phi$,
conflicting with the universal coupling assumed in general relativity \cite{bib:gasp,bib:halp}.
\end{itemize}
The main difference between these oscillation scenarios and standard
oscillations is the linear energy dependence of
the oscillation frequency, shifting observable oscillation effects into the energy range of AMANDA-II.
For the sake of simplicity, we will focus on the VLI scenario. As both
theories are mathematically equivalent,
the results can be transferred to the VEP case by simply exchanging the relativity-violating oscillation 
parameters $\delta c/c \to 2\vert\phi\vert\delta\gamma$ and mixing angles $\Theta_c \to \Theta_\gamma$.

Combining standard and VLI oscillations,
one obtains three systems of neutrino eigenstates (flavor, mass, and MAV eigenstates),
resulting in a total of 5 oscillation parameters: the mass-squared
difference $\Delta m^2$, two mixing angles $\Theta_m$ and $\Theta_c$, the
VLI parameter $\delta c/c$, and a complex phase $\eta$.
Fixing $\Delta m^2 = 2.3\times10^{-3}\,\mathrm{eV}^2$ and
$\Theta_m=45^\circ$, the survival probability may then be written as:
\begin{equation}\label{survprob2fvo}
P(\nu_{\mu} \to \nu_{\mu}) = 1 - \sin^2 2\Theta\ \sin^2\left(\Omega\ L\right)
\end{equation}
\begin{equation}\label{globpara}
2\Theta  =  \arctan \left( s/t \right) \qquad \quad \Omega   =  \sqrt{s^2 + t^2}
\end{equation}
\begin{eqnarray}\label{vlimasspara2fvonurvli}
s & = & 2.92\times10^{-3}\,\vert\,1 / E_{\nu} \nonumber\ +\\
  &   & 8.70\times10^{20}\ \delta c/c\ \sin 2{\Theta_c}\ E_{\nu}\ \mathrm{e}^{i\eta} \vert\ , \nonumber\\
t & = & 2.54\times10^{18}\ \delta c/c\ \cos 2{\Theta_c}\ E_{\nu} \ .
\end{eqnarray}
Here the the muon neutrino path length $L$ is expressed in $\mathrm{km}$ and the neutrino energy $E_\nu$ in $\mathrm{GeV}$. 
For the given set of parameters, one can observe a significant effect
within the analyzed energy range ($100\,\mathrm{GeV}-10\,\mathrm{TeV}$) and
declination range \mbox{($\delta\geq20^\circ$)}, for certain values of $\Theta_c$
and $\delta c/c$. 
%____________________________________________________________________________________________________________________________________________________
\section{Data selection}
The data analyzed in this analysis are selected from $7.9\times10^9$ events
recorded from 2000 to 2003.  
Detector signals are recorded when 24 or more OMs report signals within a sliding window of $2.5$ $\mu$s.
Signals from unstable OMs, electronic and OM noise or cross-talk, as well as hits due to
uncorrelated muons coincident within the trigger time, are rejected. Also, data periods with reduced
data quality are discarded, corresponding to $87.8$ days. The $17.3\%$ deadtime of the data
acquisition system results in a total livetime of $807.2$ days used for the analysis.\\

The events are processed with a fast pattern recognition algorithm (A) to select tracks that are
likely to be upgoing ($\delta_{A}>-20^{\circ}$). The calculated track direction serves as a first
guess for 16-fold iterative maximum likelihood reconstruction algorithms (B), restricted to
upgoing tracks with $\delta > 0^{\circ}$. The alternative hypothesis of a downgoing track is
tested with a two-fold iterative fit requiring $\delta<-10^{\circ}$. In order to reduce the
probability of wrongly reconstructed tracks due to spurious hits, both fits are repeated after
rejecting hits with timing residuals larger than two standard deviations. Background rejection and
angular resolution are further improved by a 10-fold iterative fit (C) incorporating the
probabilities that modules registered hits for the given track. 
From an examination of the likelihood contours in declination and right
ascension~\cite{bib:Till}, an estimate of the median space angle resolution $\sigma_{\Psi}$ is
obtained for individual tracks. The following selection criteria are
applied, with $L_\mathrm{diff} \equiv \Delta \ln L$ being the
difference of up- and downgoing likelihood minima: 
(1) declinations $\delta_{A}> -20^{\circ}$, $\delta_{B}> 0^{\circ}$ and $\delta_{C}> 20^{\circ}$;
(2) space angle differences $\Psi(A,B)<30^{\circ}$, $\Psi(B,C)<7.5^{\circ}$;
(3) space angle resolutions $\sigma_{\Psi(B)}< 6^{\circ}$ and $\sigma_{\Psi(C)}<3.0^{\circ}$;
(4) likelihood difference $L_\mathrm{diff}(B,C) > 32.5$.

The oscillation probability depends on the neutrino flight length
(\textit{i.e.} declination) and the neutrino energy.  As an energy
estimator, we use a correlated observable, the number of OMs triggered in
an event ($N_\mathrm{ch}$). Using Monte Carlo simulations, declination-
and $N_\mathrm{ch}$-dependent selection criteria have been developed by
dividing the distribution of the angular resolution into equal declination and $N_\mathrm{ch}$ bins. For
each of these bins, a fixed, optimized percentage ($8\%$) of the events with poor angular resolution is rejected. 
The same was done for the likelihood difference
distributions.  These criteria improved the efficiency of the
data selection by $30\%$ compared to simple selections
in the angular resolution and the likelihood difference.  The resulting
number of selected neutrino candidate events is $3401$. From a study of 
the distribution of space angle difference, the background of wrongly reconstructed
atmospheric muons is estimated to be $4\%$.

A full simulation chain, including neutrino absorption in the Earth, neutral current
regeneration, muon propagation, and detector response is used to simulate the response of AMANDA-II
to atmospheric neutrinos~\cite{bib:Hill,bib:Chirkin}. The expected atmospheric muon
neutrino flux before oscillations is taken from
Lipari~\cite{bib:Lipari}.
\begin{figure}[t!]
\begin{center}
\includegraphics[width=7cm]{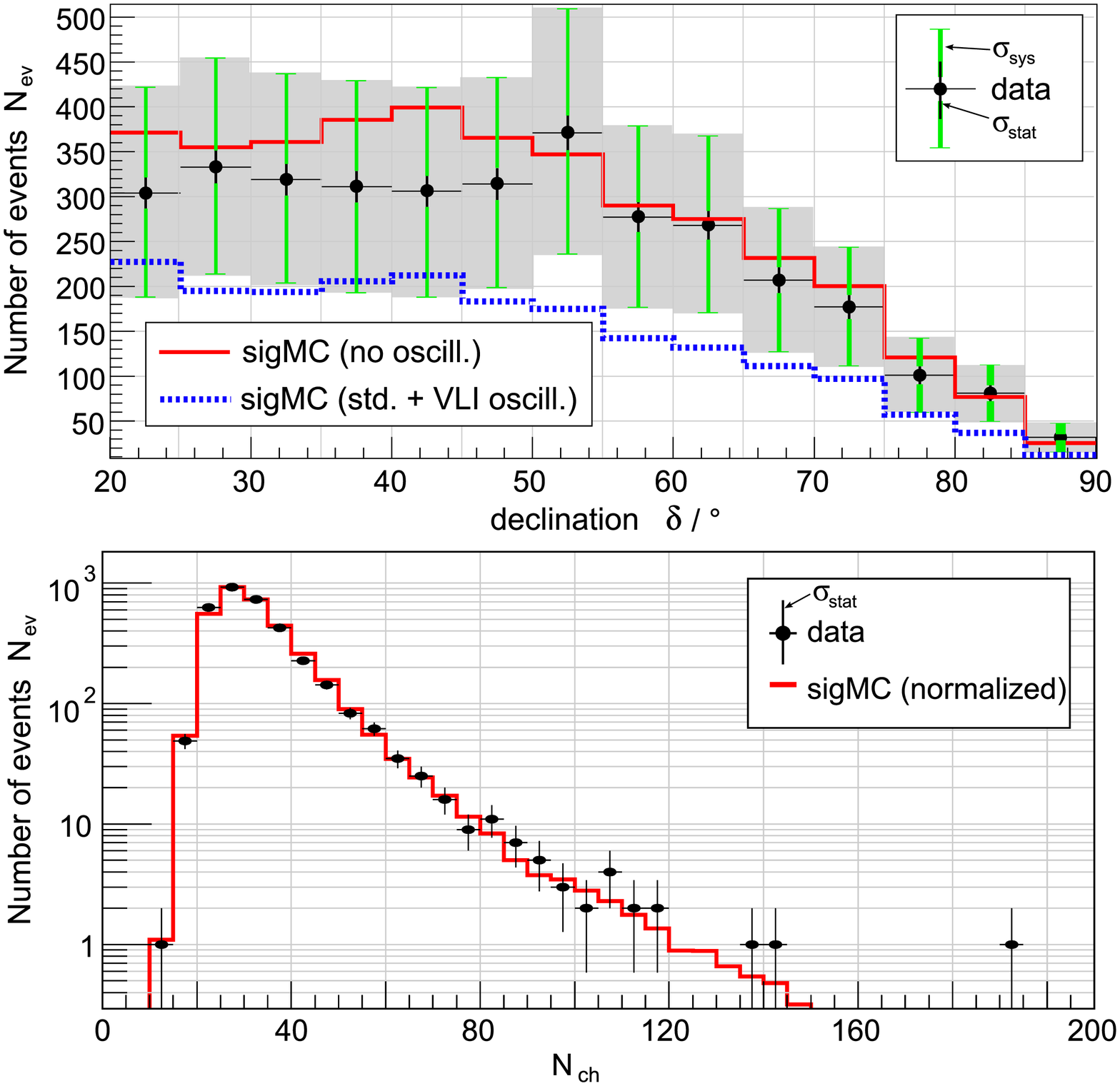}
\caption[declination and nch distribution of the final selection]{Top: Measured atmospheric neutrino 
declination distribution with statistical and systematic errors. Also shown are 
the predicted distributions without oscillation and with 
\mbox{$\delta c/c\!=\!10^{-24}$}, $\Theta_c=\pi/4$ and $\cos\eta=0$. 
Bottom: $N_\mathrm{ch}$ distributions of data 
(statistical errors only) and the predicted distribution without
oscillations, normalized to the data.}
\end{center}
\end{figure}

%____________________________________________________________________________________________________________________________________________________
\section{Analysis method and systematic errors}
The analysis method uses a $\chi^2$-test to compare the declination and $N_\mathrm{ch}$ distributions
of data with Monte Carlo simulations including VLI oscillation effects.
The systematic uncertainties affecting the Monte Carlo prediction are integrated
into the $\chi^2$ expression:
\begin{eqnarray}\label{chifinaldef}
\chi^2\left(\delta c/c,\Theta_c,\cos\eta\right) = & & \nonumber
\end{eqnarray}
\begin{eqnarray}
\sum^{N_\mathrm{Bins}}_{i=1}\frac{\left(N^\mathrm{D}_i\ - N^\mathrm{BG}_i\ - \ F\cdot N^\mathrm{MC}_i\left(\delta c/c,\Theta_c,\cos\eta\right)\right)^2}
{N^\mathrm{D}_i + N^\mathrm{BG}_i + \left(\sigma^\mathrm{MC}_i\right)^2} & & \nonumber
\end{eqnarray}
\begin{equation}
+ \left(\frac{\alpha}{\sigma_\alpha}\right)^2 + \left(\frac{\kappa}{\sigma_\kappa}\right)^2 + \left(\frac{\epsilon}{\sigma_\epsilon}\right)^2 \quad ,
\end{equation}
where $N^x_i$ denotes the number events in bin $i$ and $x$ denotes data (D), background (BG) and Monte Carlo (MC). The function $F$ represents the product of functions $f_\alpha\cdot f^i_\kappa\cdot f^i_\epsilon$ which are defined as:% (the letter i indicates the bin number):
\begin{eqnarray}\label{funcs}
f_\alpha & = & 1+\alpha,\qquad f^i_\kappa\ \ =\ \ c_i\cdot\kappa +1,\nonumber\\ 
f^i_\epsilon & = & 1+2\epsilon\ (0.5-\sin\delta_i) \ .
\end{eqnarray}
$\alpha$ parametrizes the systematic uncertainty in the overall normalization due to
uncertainties in the detector response and theoretical uncertainties of the atmospheric neutrino flux ($\sigma_\alpha = 30\%$). The uncertainty
due to the relative production rate between kaons and pions, which affects the shape of the 
declination distribution, is parametrized by $\epsilon$ and is estimated
as $\sigma_\epsilon=6\%$ in total \cite{bib:SuperKth}.
The uncertainty in the sensitivity of the optical modules is parametrized
by $\kappa$ ($\kappa\!=\!0$ for $100\%$ sensitivity) and was measured to be $\sigma_\kappa=11.5\%$. The function
$f^i_\kappa$ was derived from the changes in the declination distribution
generated by Monte Carlo distributions with different OM sensitivities.
In order to determine the optimal number of declination and $N_\mathrm{ch}$
bins and their optimal range, toy Monte Carlo samples of 10000 
events have been generated reflecting the simulated flux and systematic uncertainties as assumed above.
The mathematical properties of expression (4) were checked, and belts for 90\%, 95\% and 99\% confidence level were derived 
from a high statistics toy Monte Carlo sample.

The results of the toy Monte Carlo studies favor an analysis using the
following 4 bins:
($N_\mathrm{ch}\leq 49$, $\delta\leq55^\circ$), ($N_\mathrm{ch}\leq 49$,
$\delta > 55^\circ$), ($N_\mathrm{ch} > 49$, $\delta\leq55^\circ$), and
($N_\mathrm{ch} >   49$, $\delta > 55^\circ$).

The exclusion regions for alternative oscillation effects are obtained by scanning through 
the oscillation parameter space. For each point $\left[\delta c/c, \sin(2\Theta_c), \cos(\eta)\right]$ the
$\chi^2$ expression is minimized in the error variables $\alpha$, $\epsilon$ and $\kappa$.
%____________________________________________________________________________________________________________________________________________________
\section{Results and Outlook}
The analysis of the final atmospheric neutrino sample finds
no evidence for alternative oscillations, and a preliminary upper limit on the VLI
parameter $\delta c/c$ is set of $5.3\times10^{-27}$ 
at the $90\%$ confidence level, for nearly maximal mixing angles $\Theta_c \approx \pm \pi/4$. 
The dependence on the unconstrained phase $\eta$ is found to be small (see figure 2); the most
conservative limit is obtained for $\cos\eta=0$. The limit can also be
interpreted in the context of VEP theories,
leading to an upper limit of $2\vert\phi\vert\delta\gamma\leq 5.3\times10^{-27}$. 
This result
%competitive 
improves the limits obtained using data from
Super-Kamiokande~\cite{bib:gonz} and MACRO~\cite{bib:batti}. 
However, AMANDA is not sensitive to small mixing angles due to the systematic errors and its higher energy 
threshold.
\begin{figure}[t!]
\label{exclureg}
\begin{center}
\includegraphics[width=7cm]{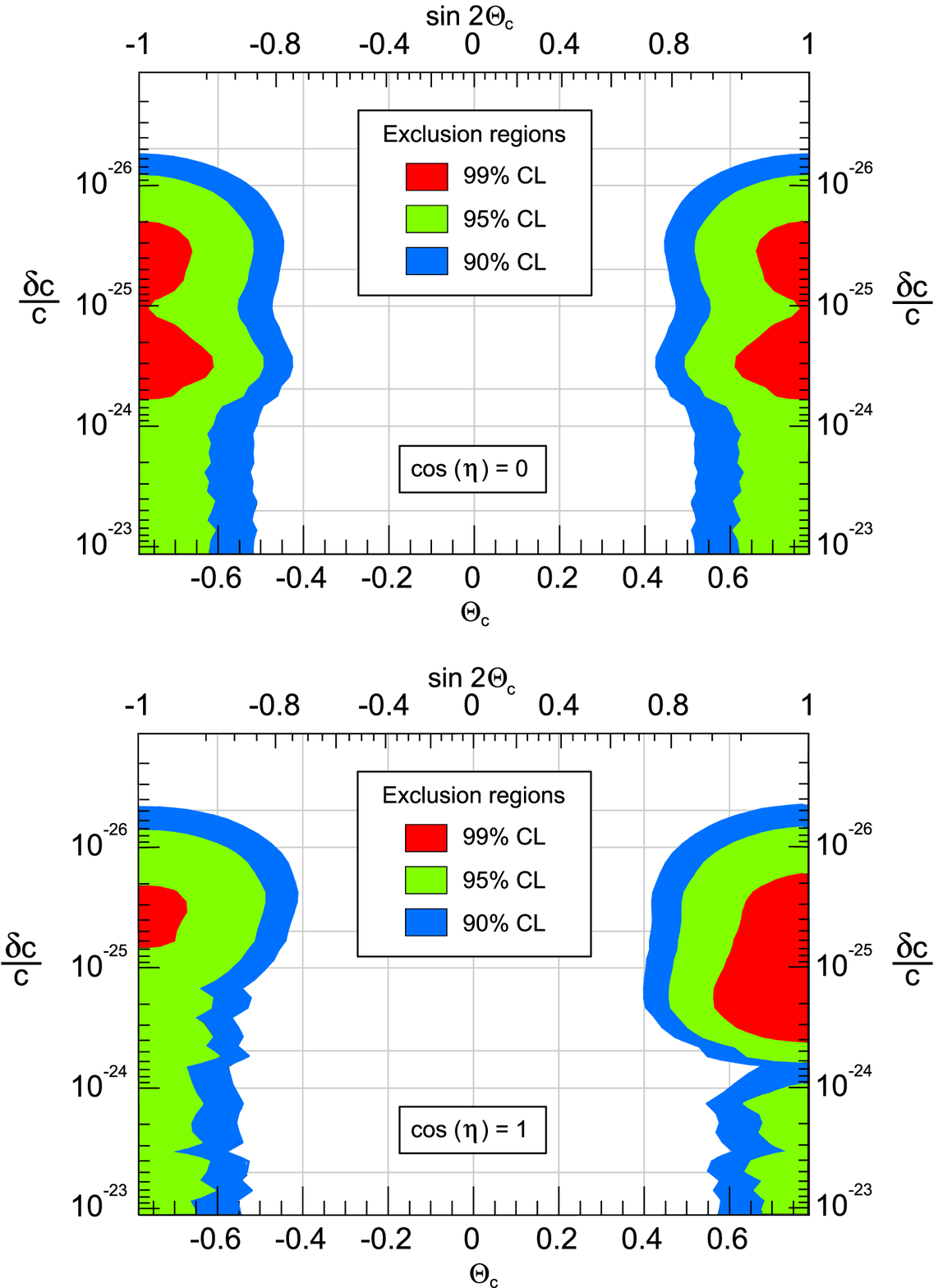}
\caption[Exclusion regions]{Shown are preliminary exclusion regions for VLI (VEP) oscillation effects, 
top for \mbox{$\cos\eta=0$}, bottom for \mbox{$\cos\eta=1$}.}
\end{center}
\end{figure}
A likelihood analysis of the 2000-2005 AMANDA-II data sample is in
progress, with
improved systematic error estimation and increased sensitivity~\cite{bib:kelley}.
This analysis will also extend the
technique to search for evidence of quantum decoherence resulting from
interaction of neutrinos with the background space-time foam~\cite{bib:morgan}.
The next-generation IceCube detector, when completed in 2010, will be
able to extend the sensitivity to VLI effects by about  
one order of magnitude~\cite{bib:Halzen}.
%____________________________________________________________________________________________________________________________________________________
\section{Acknowledgments}
J. A. thanks the German Research Foundation (DFG) and the German Federal Ministry of Education and Research (BMBF)
for financial support.
%____________________________________________________________________________________________________________________________________________________

%\end{document}
%\end

\setcounter{figure}{0}
\setcounter{table}{0}
%%
% International Cosmic Ray Conference 2007 Merida Yucatan Mexico
% In This file you will find detailed instructions to correctly
% typeset your document.
%
%
%

%Class Requeried
%\documentclass[dvips]{article}
%The ICRC Style
%\usepackage{icrctc07}

%The paper title
\title{Atmospheric muon neutrino analysis with IceCube}
%Short title to print in the headers to the final publication (Not showed in this print).
\shorttitle{Atmospheric muon neutrino analysis with IceCube}
%All paper authors
\authors{J. Pretz$^{1}$ for the IceCube Collaboration$^{2}$}
%Short title to print in the headers to the final puplication (Not showed in this print).
\shortauthors{J. Pretz for the IceCube Collaboration}
%All the affiliations.
\afiliations{$^1$Dept.~of Physics, University of Maryland, College Park, MD 20742, USA\\ $^2$ see special section of these proceedings}
\email{pretz@icecube.umd.edu}

%The abstract.
\abstract{
The heart of the IceCube neutrino observatory 
is a cubic kilometer Cherenkov detector 
being constructed in the deep ice under the geographic South Pole. IceCube is
sensitive to high-energy muon neutrinos and muon anti-neutrinos by detecting 
the secondary muon produced when the neutrino interacts in or near the 
instrumented volume. The principal source of muon neutrinos are 
neutrinos from the decay of hadrons in cosmic-ray air showers. 
IceCube operated during 2006 with 9 out of 80 anticipated strings in the 
ice. 
I will demonstrate that IceCube can find and 
reconstruct atmospheric neutrinos with high efficiency.
}

%%%%%%%%%%%%%%%%%%%% B E G I N   D O C U M E N T%%%%%%%%%%%%%%%%%%%%%%%
%\begin{document}
\maketitle
%Begin the section.

\section{Introduction}

The IceCube neutrino detector \cite{albrechtOverview}
is partially deployed at the geographic 
South Pole.  In 2006, the deep-ice detector consisted of 540 light-sensitive
Digital Optical Modules (DOMs), arranged 17 meters apart 
on 9 strings of 60 DOMs each.  The detector in this configuration is 
termed IC-9.  The strings are arranged on a hexagonal
grid and spaced 125 meters apart.  
DOMs are deployed in the deep ice between 1.5 and 2.5 kilometers below the 
surface.  
Figure \ref{fig:geometryOverhead} shows the location of strings making up the
IC-9 array
along with the relative position of the AMANDA detector.

\begin{figure}
\begin{center}
\includegraphics [width=0.48\textwidth]{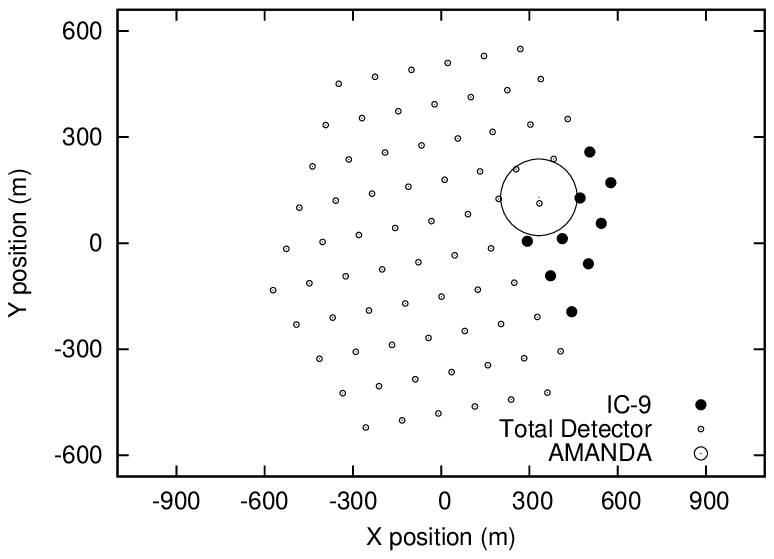}
\end{center}
\caption{
Shown are the locations
of strings for the 2006 IC-9 detector, and the location of the strings
in the completed detector.
The location of the AMANDA detector is also indicated.
}\label{fig:geometryOverhead}
\end{figure}

IceCube is sensitive to muon neutrinos (and anti-neutrinos) 
by observing the Cherenkov light
from the secondary muon produced when the neutrino interacts near the
detector volume.  
Atmospheric neutrinos, formed in the decay of mesons resulting from
a cosmic ray striking the atmosphere, dominate.
Since atmospheric neutrinos are relatively well-understood 
\cite{bartolAtmosphericNeutrino}, they 
serve as a verification and calibration tool for the new detector.
Muons from neutrino interactions are separated from muons produced in 
cosmic rays by selecting muons moving upward through the detector.
These muons must be the result of a neutrino interaction since neutrinos
are the only particle that can traverse the Earth without interacting.

\section{Data Acquisition and Filtering}

In 2006, we acquired 137.4 days of livetime with IC-9 suitable for analysis.
The waveform capture in a DOM was triggered whenever the DOM detected a signal
above a threshold of about 0.3 photoelectrons.
The DOMs were operated in Local Coincidence (LC) 
with their neighbors, 
meaning that a triggered DOM's waveform was only transmitted to the surface
if an adjacent DOM on the string also triggered within $\pm 1000$ ns.  
The surface data acquisition system 
set off a trigger
if 8 or more DOMs were read out in 5 $\mu s$.  When an event is formed, 
all DOM hits were read out within $\pm 8 \mu s$ around the trigger window.  

Because of limited bandwidth between the South Pole and the data center
in the North, the data is filtered in real time, and only candidates
for up-going events are sent North.

Hit cleaning algorithms were applied to the triggered events to remove light
from additional suprious muons, and to remove noise hits.
The photon arrival times are determined by a fit to the DOM waveform, with 
a variable number of photon arrivals.
The hit cleaning isolated the 4 $\mu s$
window in which the most hits occur,
and remaining DOM hits are kept only if another DOM hit occured within a 
radius of 100 meters and within a time of 500 ns.
At the pole, simple
first-guess algorithms were used to reject events that were down-going.
In addition, events with fewer than 11 DOMs hit were rejected 
to limit the data volume.
This filter reduced the data rate by approximately 95\%.  
The remaining events were 
transmitted to the data center via satellite for further study.

\section{Reconstruction and Event Selection}

In the North, we reconstructed the direction of events
using a maximum-likelihood technique similar to the AMANDA muon
reconstruction
\cite{muonReconstructionNIMPaper}.  Only the earliest arrival times were
used for reconstruction and no amplitude information was included in this
analysis.
The likelihood function
is based on a parametrization 
of the photon arrival time
distribution without any prior assumption of the relative likelihood of a
cosmic ray muon or neutrino event.
The likelihood function is formed with an analytic approximation to the 
photon arrival time probability density function, accounting for the
short ($\sim 20$ meter) scattering length of light in IceCube.  Events that 
reconstruct as down-going are discarded.  Despite the fact that remaining
events appear up-going, they are in fact dominated by mis-reconstructed
down-going events.  These mis-reconstructed events are removed with quality 
cuts and the remaining events constitute the neutrino candidate dataset.

The quality cuts are based on direct hits in the detector.  Direct hits
are those which arrive between $-15 ns$ and $+75 ns$ from the time expected
from unscattered Cherenkov photons radiated from the reconstructed muon.  
We cut
both on the number of recorded direct hits $N_{dir}$ 
and the largest distance of such hits along the track, $L_{dir}$.  
An event with
a large $N_{dir}$ and a large $L_{dir}$ is a better quality
event because the long lever arm of many unscattered photon arrivals
increases confidence in the event reconstruction.

We can fold these two cuts together into one dimensionless number, 
the cut strength
$S_{cut}$ which corresponds to cuts of $N_{dir} \geq S_{cut}$ and 
$L_{dir} > 25 \cdot S_{cut}$ meters. 

Table \ref{tab:eventPassingRates} shows the rates of events passing to the 
different levels of the analysis, for both experimental data and simulated
events.  Simulated events fall into three categories.  'Single shower'
events are events from single air-shower events in the atmosphere above 
the detector.  'Double shower' events come from two uncorrelated
air showers.  Finally 'atmospheric neutrino' events come
from $\pi$ and $K$ decay in the air showers in the
Northern hemisphere.  The CORSIKA air-shower \cite{corsikaReference}
simulation
was used to model down-going air shower events.  An extension to high
energies \cite{teresasCommuncationWithBarr} for the 
atmospheric neutrino model of \cite{bartolAtmosphericNeutrino}
with the cross-section parametrization of \cite{cteq5Reference}
was used to determine the expected up-going muon rate.  In estimating
the systematic error, we have included a 30\% uncertainty in the 
atmospheric neutrino flux modeling \cite{atmosphericNeutrinoUncertainties},
and a 20\% uncertainty due to uncertainties introduced in the modeling
of the depth-dependent ice properties and the DOM detection efficiency.

\begin{table*}
\begin{center}
\begin{tabular}{ccccc}
Criterion&Experimental&Single&Double&Atmospheric\\
Satisfied&Data&Shower&Shower&Neutrinos\\
\hline
Trigger Level&124.5&124.5&1.5&$6.6\rm{x}10^{-4}$\\
\hline
Filter Level&6.56&4.96&0.45&$3.7\rm{x}10^{-4}$\\
\hline
Up-going ($S_{cut}=0$)&0.80&0.49&0.21&$3.3\rm{x}10^{-4}$\\
\hline
Up-going ($S_{cut}=10$)&$(1.97\pm 0.12)\cdot 10^{-5}$&-&-&$(1.77\cdot \pm 0.63)\cdot 10^{-5}$\\
\hline
Up-going &&&&\\
($S_{cut}=10$ and $\theta > 120$)&$(1.19 \pm 0.10)\cdot 10^{-5}$&-&-&$(1.42\cdot 0.51)\cdot 10^{-5}$\\
\hline
\end{tabular}
\caption{
Event Passing Rates (Hz).  Shown are the
event passing rates through different processing levels for the
simulated event categories and for experimental data.  The trigger level 
comprises the
events triggering the detector after hit cleaning and re-triggering.  
The filter level comprises
events which passed the online filtering conditions.  
Rates are also shown for events which reconstruct as up-going with and without
the final quality cuts applied (see the text for cut definition).
Note that the rates from air-shower
events have been multiplied by $0.90$ so that the simulation and 
data agree at trigger level.  This is consistent with
an approximately 20\% uncertainty in the absolute cosmic-ray flux.
For the final sample, statistical errors are given for the data and
systematic errors are given for the atmospheric neutrino simulation.
}\label{tab:eventPassingRates}
\end{center}
\end{table*}

\section{Results}

Figure \ref{fig:dataVsCutStrength} shows the number of up-going events 
remaining as we tighten cuts.  
The contribution of the data is shown together with the expectation 
for atmospheric neutrinos and the total simulation prediction.
Below a cut strength of about 
$S_{cut} = 10$, the data is dominated by mis-reconstructed down-going
cosmic-ray shower muons.  For higher cut strengths, we have removed most
of these mis-reconstructed events and are dominated by atmospheric neutrinos.
The accurate simulation of the mis-reconstructed muon population 
requires excellent modeling of the depth-dependent ice properties and 
DOM sensitivity.  In this initial study, we observe a 60\%-80\% discrepancy
between data and simulation
for mis-reconstructed muons.  Nevertheless, over four orders of magitude,
the background simulation tracks the data, and we see a clear transition
to a population dominated by atmospheric neutrinos.

\begin{figure}
\begin{center}
\includegraphics [width=0.48\textwidth]{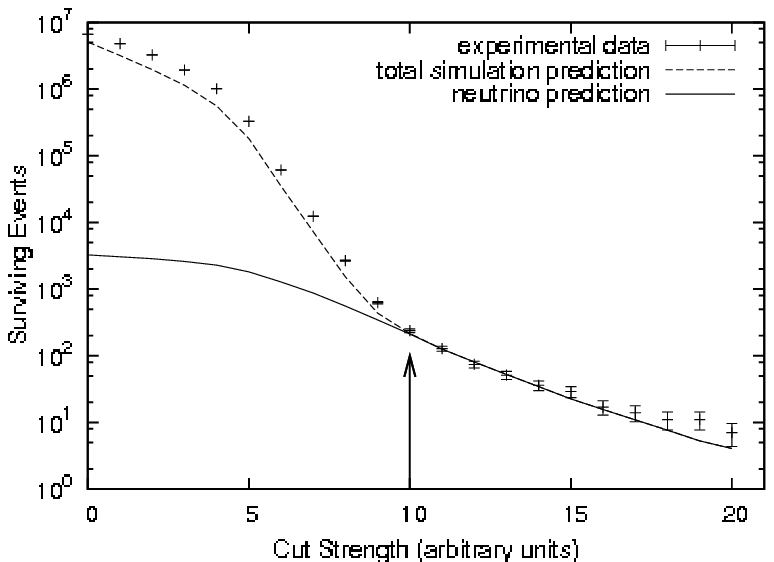}
\end{center}
\caption{
Data vs Cut Strength.  Shown is the remaining number of events 
as the cut strength $S_{cut}$ (defined in the text) 
is varied.  Curves are shown for the data
and the total simulation prediction.  Also shown is the prediction due to
atmospheric neutrinos alone.  The selection from the text corresponds
to a cuts strength of $S_{cut} = 10$, and is denoted by an arrow.  At
this point, the
data are dominated by atmospheric neutrinos.
}\label{fig:dataVsCutStrength}
\end{figure}

Figure \ref{fig:energyDistribution} shows the expected 
energy distribution of simulated
atmospheric neutrino events surviving to $S_{cut} = 10$.  The lower threshold
of about 100 GeV is set by the range of the secondary muons, 
and the dropoff at high energies is due to the decreasing flux of 
atmospheric 
neutrinos.  

\begin{figure}
\begin{center}
\includegraphics [width=0.48\textwidth]{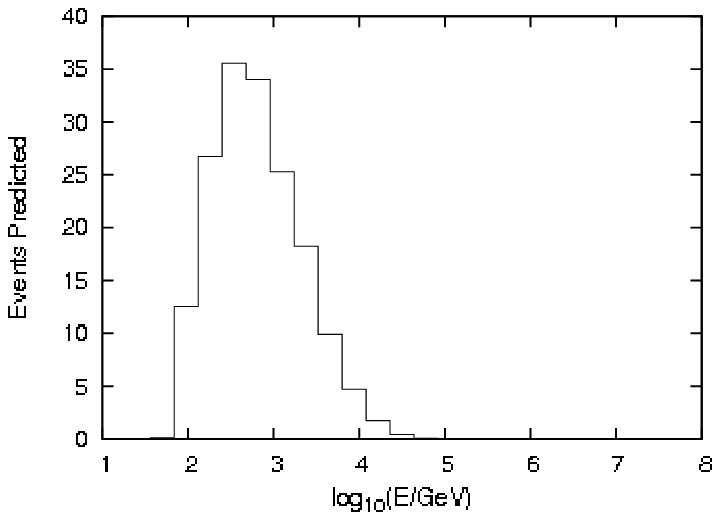}
\end{center}
\caption{
The distribution of neutrino energy for events surviving the
analysis cuts, as determined by the atmospheric neutrino simulation.
}\label{fig:energyDistribution}
\end{figure}

Figure \ref{fig:zenithDistribution} shows the zenith angle distribution
for events which survive at $S_{cut}=10$.  Above 120 degrees,
for vertical events, we have good agreement between experimental data 
and atmospheric neutrino simulation.  The excess at the horizon is
believed to be residual air-shower muon events.  This belief is reinforced
by the fact that excess data at the horizon is typically of lower quality
(as measured by $N_{dir}$, $L_{dir}$ and the number of hit DOMs)
than expected from atmospheric neutrino simulation.  The data above the 
horizon agrees well in these variables 
with a pure atmospheric neutrino expectation.

\begin{figure}
\begin{center}
\includegraphics [width=0.48\textwidth]{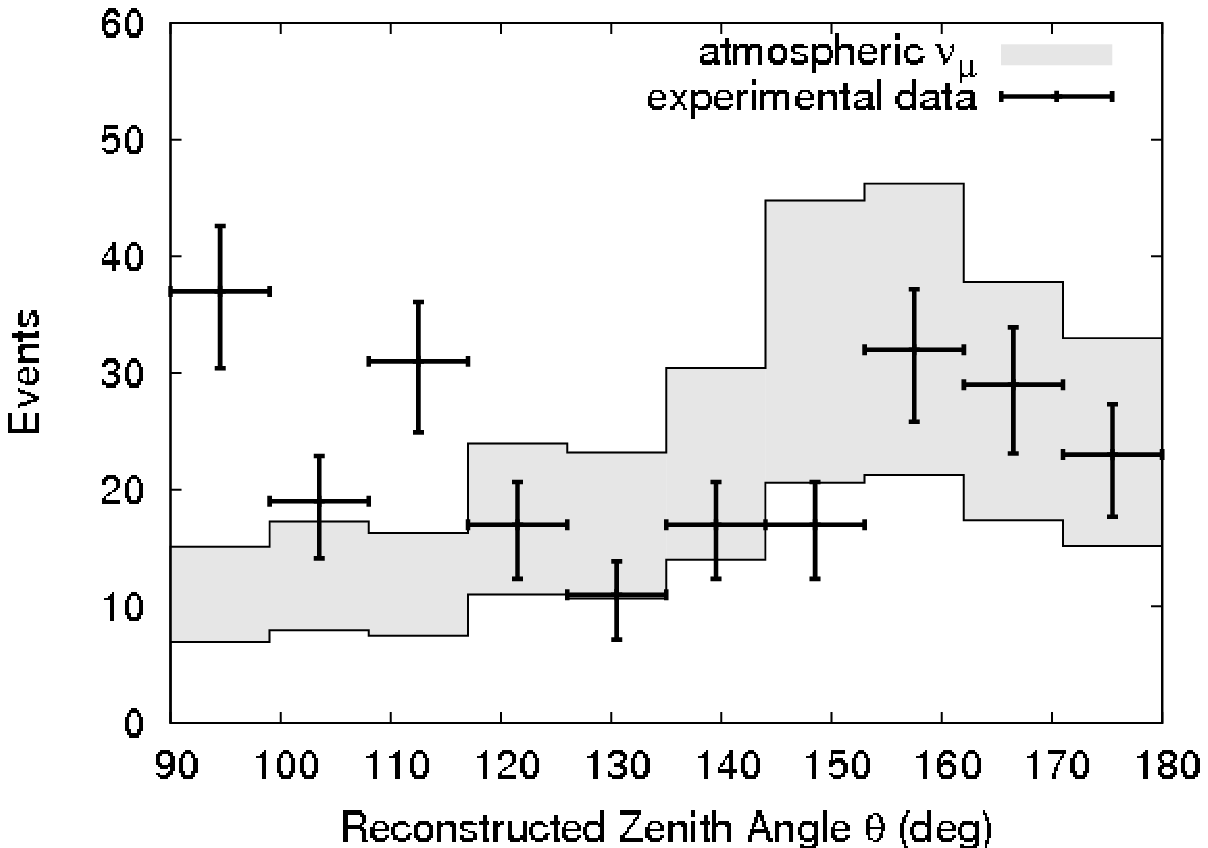}
\end{center}
\caption{
Distribution of the reconstructed zenith angle $\theta$
of the final event sample.  
A zenith angle of 90 degrees indicates a horizontal event, and a zenith of 180 
degrees
is a directly up-going event.
The band shown for the atmospheric neutrino simulation 
includes the systematic errors; the error bars on the data are
statistical only.
}\label{fig:zenithDistribution}
\end{figure}

In the recorded 137.4 days of livetime we measure 234 events surviving 
to $S_{cut} = 10$, compared to an expectation of 
$211 \pm 76(syst.) \pm 14(stat.)$ events from a pure atmospheric neutrino
signal.  Above a zenith of 120 degrees, where the background contamination
is small, we measure 142 events with an expectation of 
$169 \pm 60(syst.) \pm 13(stat)$ events.

\section{Conclusions}

IceCube is partially deployed and acquiring physics-quality data.  During
the 2006 season, we accumulated 137.4
days of livetime and observe an atmospheric neutrino signal consistent
with expectation.
We have identified 234 neutrino candidate events.  For zenith angles above
120 degrees, the background from misreconstructed muons is small 
and the sample is dominated by atmospheric neutrinos.
The selection of events was done within six months
of the beginning of data acquisition, demonstrating the viability of the full 
data acquisition chain, from PMT waveform capture at the DOM with 
nanosecond timing, to event selection at the South Pole
and transmission of that selected data via satellite to the North.

%\bibliography{ICRCPretz/icrc_ic9_atmonu}
%
%\bibliographystyle{unsrt}

%\end{document}

\setcounter{figure}{0}
\setcounter{table}{0}
%%

%Class Required
%\documentclass{article}
%The ICRC Style
%(This package is the last package in the usepackage list)
%If you need import other package you need write it first.
%\usepackage{icrctc07}

%The paper title
\title{Muon energy reconstruction and atmospheric neutrino spectrum unfolding with the IceCube detector}
%Short title to print in the headers to the final publication (Not showed in this print).
\shorttitle{Atmospheric neutrino spectrum with IceCube}

%All paper authors
\authors{Juan-de-Dios Zornoza$^{1,2}$, Dmitry Chirkin$^{3}$ on behalf of the IceCube collaboration$^{4}$}
%Short title to print in the headers to the final publication (Not shown in this print).
\shortauthors{J.D. Zornoza and D. Chirkin et al.}
%All the affiliations.
\afiliations{$^1$Department of Physics, University of Wisconsin,
Madison, Wisconsin, 53703, USA\\ $^2$IFIC (CISC-University of
Valencia), Ed. de Investigaci\'{o}n de Paterna, AC 22085, 46071, Valencia, Spain\\$^3$ Lawrence Berkeley National
Laboratory, Berkeley, California 94720, USA \\$^4$ See special section of these proceedings}
\email{zornoza@icecube.wisc.edu}

%The abstract.

\abstract{ Data collected during the year 2006 by the first 9 strings
of IceCube can be used to measure the energy spectrum of the
atmospheric muon neutrino flux. Atmospheric neutrinos, an important
scientific output by itself (for instance, to understand the
high-energy hadronic interaction models), are also fundamental in
order to check the performance of the detector and to estimate the
background for the extraterrestrial high-energy neutrinos searches. A
full reconstruction of the neutrino-induced muon tracks provides both
directional and energy information. The reconstructed
energy-correlated parameter, the photon density emitted by the muon
along its track, can be used to calculate the energy spectrum, which
is reconstructed by using unfolding techniques. We will discuss the
unfolding procedure to be applied to data from the 9-string
configuration of IceCube.}

%%%%%%%%%%%%%%%%%%%% B E G I N   D O C U M E N T%%%%%%%%%%%%%%%%%%%%%%%
%\begin{document}
\maketitle
%Begin the section.
\section{Motivation}

The IceCube collaboration is building a cubic kilometer neutrino
telescope in the Antarctic ice. Since neutrinos are neutral, stable
and weakly interacting, they are a unique probe to study the Universe
at high energies and IceCube will be the most powerful tool available
for observing them. The detector will be completed by 2011 and the
construction goal is 80 strings with 4800 photomultipliers, which will
detect the Cherenkov light emitted by the relativistic muons produced
in the CC interactions of high-energy neutrinos. IceCube can also
observe the cascades produced by CC $\nu_{e}$ and $\nu_{\tau}$
interactions and NC interactions of any flavor.

During the Austral summer 2006-07, a total of 22 strings were deployed
and the detector is working smoothly. In this paper we will study the data
corresponding to the previous season, when 9 strings were installed.

The scientific output of neutrino astronomy is very wide, including
the search of dark matter and the observation of astrophysical
neutrinos from a large variety of sources (gamma-ray bursts, active
galactic nuclei, microquasars, etc.) Therefore, it is very important
to study the background due to neutrinos from decay of pions and kaons
produced by the interaction of cosmic rays in the
atmosphere. Experiments like AMANDA~\cite{amanda} have measured the
neutrino atmospheric spectrum up to $\sim$100~TeV and IceCube will be
able to explore the region where the prompt neutrino component (due to
charmed meson decays) will dominate. The atmospheric muon background
can be severely reduced by selecting only up-going events and imposing
restrictive constrains in the quality of the reconstructed track. On
the other hand, atmospheric neutrinos cannot be rejected in this
way, so it is important to understand well the rates and spectrum of
this background.

A detailed study of the rates of the 9-string configuration of IceCube
can be found in~\cite{pretz}. In this
paper, we will focus on the reconstruction of the energy spectrum. This
spectrum cannot be reconstructed by just piling-up the energy of
individual events because of two factors. First, the
energy resolution is limited because we only see part of the muon
energy (which in turn is only part of the neutrino energy) and because
the muon energy loss is stochastic. Second, the spectrum falls
very quickly with energy (as E$^{-3.7}$), so the events for which the
energy is overestimated would bury the events at higher energy,
distorting the resulting spectrum. In order to overcome this problem,
a different approach is needed: the unfolding techniques.

The structure of this paper is as follows. In the next section, we
will describe the calculation of the variable used for the
unfolding. This variable has to be correlated with the neutrino energy
with the lowest possible spread. Among the different variables that
have been studied (number of hit optical sensors, total charge, photon
density along the track, etc.) the best results are obtained when
reconstructing the energy from the photon density along the track at
the point of closest approach to the center of gravity of hits in the
event. In the following section we make a brief description of the
unfolding procedure and test the robustness of the method. Finally, we
show the resulting unfolded spectrum.

\section{Energy reconstruction technique}

As a muon travels through ice, it emits about $3 \cdot 10^4$ Cherenkov
photons in the spectral range visible to the detector per meter along
its track, just from electromagnetic interaction of the bare muon. In
addition, the knock-on electrons, bremsstrahlung, electron pairs, and
photonuclear interactions caused by the muon traveling through ice,
generate short cascades along the muon track \cite{ICRC0190_mmc}. Particles
created in such cascades also emit Cherenkov radiation, increasing the
``effective length'' of the muon (which determines the total number of
Cherenkov photons using the above factor) by the amount proportional
to the energy of the cascade, on average by about
4~m$\cdot~E/$GeV \cite{wiebusch}. The number of additional Cherenkov photons emitted
by the passing muon due to cascades created along its path is
therefore proportional to the total energy deposited in form of such
cascades. In a well-known approximation of the muon energy losses,
$dE/dx=a+b\cdot E$, the second term is largely due to just such energy
deposits. Above the critical energy ($\sim$ 1 TeV), the second term
begins to dominate the energy losses, and the total number of
Cherenkov photons left by a muon per unit length of the muon track
becomes proportional to its energy:

\begin{equation}
N_c=3\cdot 10^4 \mbox{m}^{-1}(1.22+1.36 \cdot 10^{-3} E/\mbox{[GeV]})
\end{equation}

This ``photon density'' along the muon track enters naturally into the
muon track reconstruction through a term in a log likelihood function,
which describes how well the number of photons observed at a distance
$d$ from the track is described by the flux function (defined as the
lateral distribution of Cherenkov photons around the muon track given
as a function of the distance from the track). The flux function is
easily computed in the vicinity of the track, before the scattering of
light alters the original direction of photons in the Cherenkov cone
around the track. At large distances one may use the diffusive
approximation since the photons observed there have sustained many
direction-altering scattering events. In the intermediate distance
region these approximations are stitched together with a function,
chosen to describe all 3 regions. The shape of the function was
inspired by the eikonal small-angle scattering approximation of light
that may be used in low-scattering media, e.g., water. The chosen flux
function was verified against data and was found to perform extremely
well.

The photon density along the muon track thus becomes the
6th parameter in addition to two angles and 3 parameters describing a
point in space and time along the muon track, against which the
likelihood function is minimized. One may then calculate the energy by
either inverting equation (1) or performing a Monte Carlo study of
the correlation of the calculated photon density and energy (see
Figure \ref{fig51}). The second approach additionally results in a
smearing matrix, which can then be used for spectra
unfolding (next section). In all cases the energy of the muon is taken
at the point of closest approach to the center of gravity of hits left
by the muon in the detector (which yields better energy estimates than
alternatives).

\begin{figure}
\begin{center}
\noindent
%\fbox{\hbox{\vbox{\hsize=50mm \hfill \vspace{50mm}}}}
%uncomment next line to include real image
\includegraphics [width=0.45\textwidth]{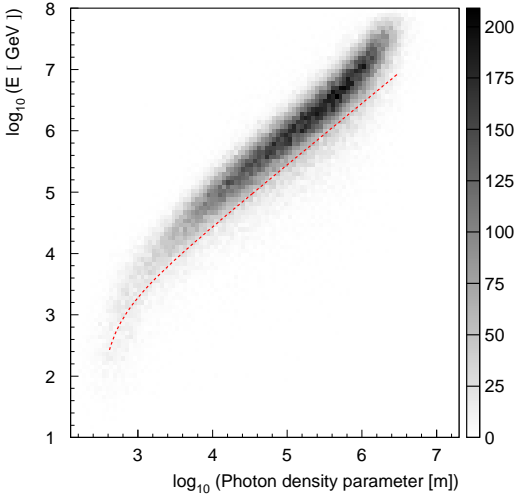}
\end{center}
\caption{Correlation of true muon energy and reconstructed photon
density parameter (photon density $N_c$ times optical sensor effective
PMT area).  The red line corresponds to the application of eq.\
(1). It overestimates the value of the photon density parameter
somewhat when compared to the detailed simulation.}\label{fig51}
\end{figure}

Figure~\ref{fig52} shows the resolution of the energy reconstructed
 with the method described here and with methods based on the
 calculation of the number of hit optical channels ($N_{ch}$) and
 total charge ($Q_{tot}$). For the isotropic fluxes in the energy
 range of $10^{4.4} - 10^{7.4}$ GeV a reconstruction precision of 0.3
 in log$_{10}$(E [GeV]) is achieved. This is close to the
 theoretically achievable (determined by the uncertainty related to stochastic
 nature of energy losses). For the atmospheric neutrino fluxes this
 energy range increases to $10^{3.6} - 10^{7.6}$ GeV (this may also be due to somewhat reduced statistics at lower energies). At low energies
 the resolution worsens due to a reduced dependence of muon energy
 losses on muon energy below the critical energy. This may potentially
 be improved by using the observed muon track length as an additional
 energy-correlated parameter.  At high energies one expects the nearby
 optical sensors to be saturated, leading to increased systematic
 uncertainties and, in turn, to reduced energy reconstruction
 precision. This will likely improve with more detailed corrections of
 the saturated behavior taken in the account.

\begin{figure}
\begin{center}
\noindent
%\fbox{\hbox{\vbox{\hsize=50mm \hfill \vspace{50mm}}}}
%uncomment next line to include real image
\includegraphics [width=0.4\textwidth]{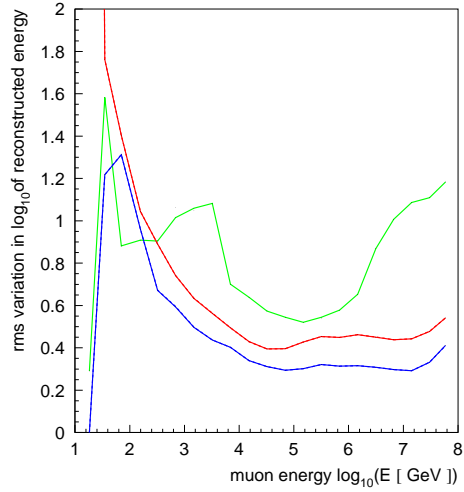}
\end{center}
\caption{Energy reconstruction precision: blue (lowest curve) for the
photon-density-based approach of this paper, red for the $Q_{tot}$ and
green (highest curve) for the $N_{ch}$-based
calculations.}\label{fig52}
\end{figure}

\section{Unfolding procedure}

There are different unfolding methods used in high energy
physics. Previous studies of the atmospheric neutrino unfolding have
been done both for AMANDA data~\cite{geenen,muenich} and ANTARES
simulation~\cite{zornoza}. For this analysis we have chosen the
Singular Value Decomposition algorithm~\cite{hoecker}, since it is
robust, efficient and easy to implement. The problem of unfolding can
be expressed, in matrix notation, by the expression $\hat{A} y=b$,
where $\hat{A}$ is the so-called smearing matrix (which has to be
generated by Monte Carlo), $y$ is the spectrum we want to measure (in
this case, the neutrino energy spectrum), and $b$ is the experimental
observable (reconstructed muon energy). Inverting the smearing matrix
does not give a useful solution because of the effect of statistical
fluctuations, which completely spoils the result. The SVD algorithm is
based on the decomposition of $\hat{A}$ as $\hat{A}=USV^{T}$, where
$U$ and $V$ are orthogonal matrices and $S$ is a non-negative diagonal
matrix whose diagonal elements are called ``singular values''. It can
be shown that this decomposition allows one to easily identify the
elements of the system that contribute to the statistical fluctuations
but provide useful information. Thus, these elements can be filtered
out in order to obtain a smoother solution.

Another interesting point of this method is that in practice we do not
try to solve directly the spectrum, but the deviations from a
reasonable assumption. This also helps to reduce the effect of
statistical fluctuations.

\section{Results and discussion}

The event selection used in this work is guided by one applied in the
atmospheric neutrino rates analysis~\cite{pretz}. The variables to
perform such a selection are the same but the values have been
somewhat relaxed in order to increase the statistics: $N_{dir}$
(number of unscattered photons) $\ge8$, $L_{dir}$ (length of the
track) $>$ 200 m and $\theta$ (zenith angle) $>$ 92 deg. These cuts
should still reject most of the background contamination, which is
still under study.  In order to check that the simulation is under
control, we have compared simulated and measured distribution of
several variables, finding good agreement.
%Figure~\ref{} shows one of these comparisons. 
%The agreement is reasonable good in all the cases.

%\begin{figure}
%\begin{center}
%\noindent
%\fbox{\hbox{\vbox{\hsize=50mm \hfill \vspace{50mm}}}}
%uncomment next line to include real image
%\includegraphics [width=0.2\textwidth]{figura.ps}
%\end{center}
%\caption{Comparison between the distributions of XXX for data and
%Monte Carlo.}\label{ICRC0190_fig1}
%\end{figure}

We have checked the robustness of the unfolded results as function of
the spectral index used when creating the smearing matrix and the
initial assumption made for the solution of the system. As a spectral
index of $\gamma=-2$ is far from $\gamma=-3.7$, the dependence on the
uncertainty of the smearing matrix is small. We used rather different
shapes for the initial assumption on the solution and could show that
the algorithm coverges towards the expected
solution. Figure~\ref{ICRC0190_fig2} compares the true (generated by Monte
Carlo) and unfolded distributions, showing a good agreement between
both (preliminary result).

\begin{figure}
\begin{center}
\noindent
%\fbox{\hbox{\vbox{\hsize=50mm \hfill \vspace{50mm}}}}
\includegraphics [width=0.5\textwidth]{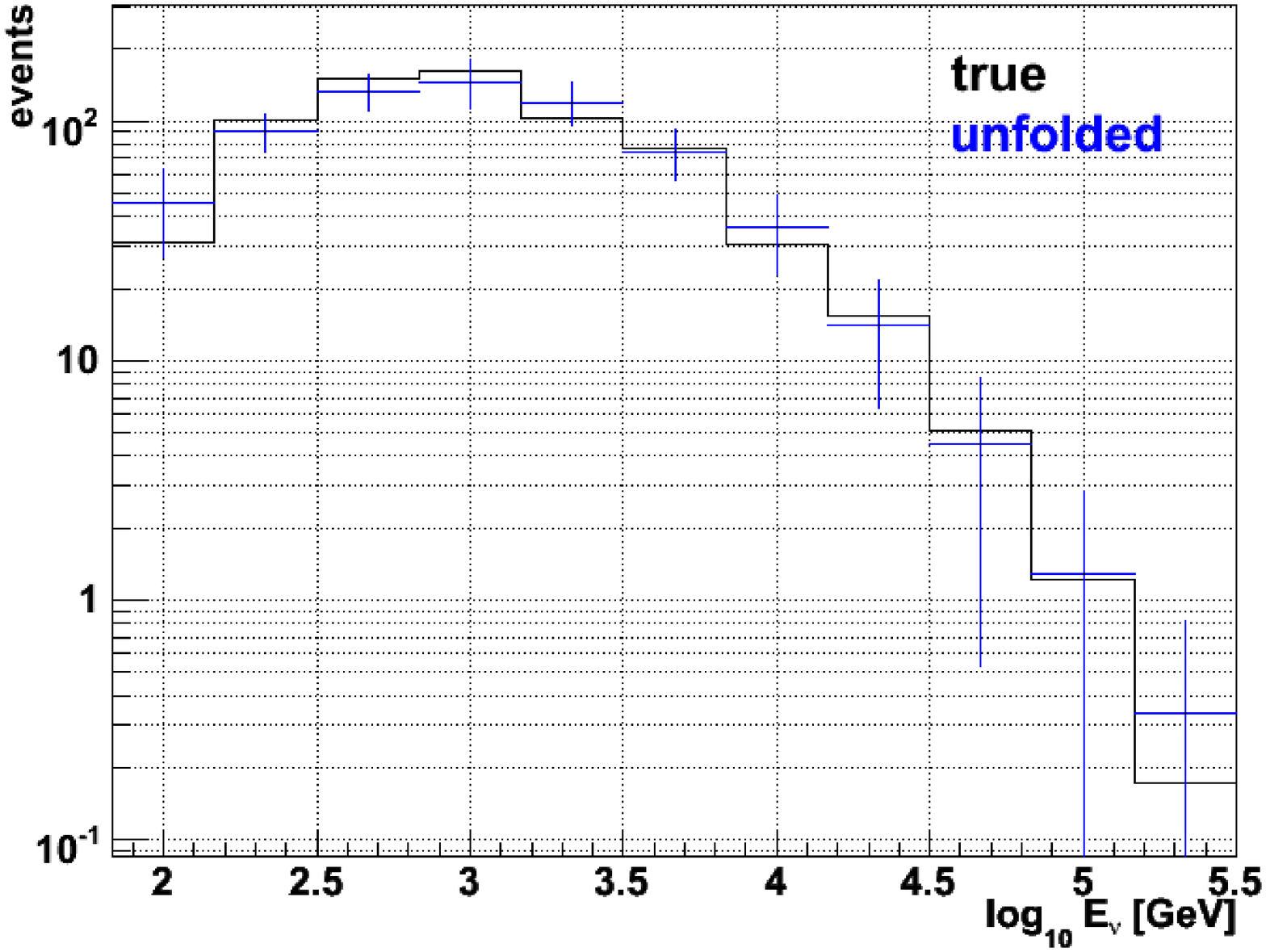}
\end{center}
\caption{True (black line) and unfolded (blue crosses) Monte Carlo
spectra (preliminary). It can be seen that the agreement between both
distributions is good. Errors include global unfolding uncertainties.}
\label{ICRC0190_fig2}
\end{figure}

\section{Conclusions}

The atmospheric neutrino spectrum is an important result both for its
intrinsic physics interest and because atmospheric neutrinos are the
main source of background in most of the analysis in neutrino
telescopes. In order to reconstruct this spectrum we have to use
unfolding techniques. In this paper we have described how to
reconstruct the muon energy (at the point of closest approach to the
center of gravity hits in the event), which is the variable found to
best correlate with the neutrino energy. Finally, the unfolded
spectrum is obtained, showing also that the algorithm works properly
when compared with Monte Carlo.

%\end{document}

%
%diffuse papers
%
%icrc1103.pdf
%leptons_icrc07_draft6.pdf
%icrc07_UHE-v0.5.pdf
%diffuse-likelihood2.pdf
%
\setcounter{figure}{0}
\setcounter{table}{0}
%%
% International Cosmic Ray Conference 2007 Merida Yucatan Mexico
% In this file you will find detailed instructions to correctly
% typeset your document.
%
% By: Victor De la Luz
% vdelaluz@inaoep.mx
% Mexico City

\newcommand{\diffunit}{$\mathrm{GeV\;cm^{-2}\;s^{-1}\;sr^{-1}}$}
\newcommand{\pointunit}{$\mathrm{TeV\;cm^{-2}\;s^{-1}}$}
\newcommand{\dNdE}{E^{2}_{\nu} \times dN_{\nu}/dE_{\nu}}
\newcommand{\Nch}{$N_{\mathrm{ch}}\;$}
\newcommand{\Ndir}{$N_{\rm dir}$}
\newcommand{\ea}{{\it et al} }
\newcommand{\ic}{IceCube}
\newcommand{\esqdnde}{$\mathrm{E^{2}_{\nu} \times dN_{\nu}/dE_{\nu}}$}
\newcommand{\puneicrc}{2005 Proc. 29th Int. Cosmic Ray Conf., Pune}
\newcommand{\ar}{Ahrens J {\it et al} }
\newcommand{\am}{Ackermann M {\it et al} }
\newcommand{\ab}{Achterberg A {\it et al} }

%Class Required
%\documentclass{article}
%The ICRC Style
%(This package is the last package in the usepackage list)
%If you need import other package you need write it first.
%\usepackage{color}
%\usepackage{icrctc07}

%The paper title
\title{Searches for a diffuse flux of extra-terrestrial muon neutrinos with AMANDA-II and IceCube}
%Short title to print in the headers to the final publication (Not showed in this print).
\shorttitle{Diffuse neutrino flux searches in AMANDA-II and IceCube}

%All paper authors
\authors{Kotoyo Hoshina$^1$, Jessica Hodges$^1$, Gary C. Hill$^1$ for the IceCube 
Collaboration$^2$}
%Short title to print in the headers to the final publication (Not shown in this print).
\shortauthors{K. Hoshina et al.}
%All the affiliations.
\afiliations{$^1$Dept.~of Physics, University of Wisconsin, Madison, WI 53706, USA\\
$^2$See special section of these proceedings.}
\email{kotoyo.hoshina@icecube.wisc.edu}

%The abstract.
\abstract{The AMANDA-II data collected during the period 2000--2003 have been analysed in a search
 for a diffuse flux of high-energy extra-terrestrial muon neutrinos from
 the sum of all sources in the Universe. With no excess of events seen, an
 upper limit of \esqdnde $ <7.4\times10^{-8}$ \diffunit was obtained. The
 astrophysical implications of this upper bound are discussed, in
 addition to results from the search for signals with other energy
 spectra. The sensitivity of the diffuse analysis of IceCube 9-string
%data and prospects for the current 22-string detector are
 is presented.}

%%%%%%%%%%%%%%%%%%%% B E G I N   D O C U M E N T%%%%%%%%%%%%%%%%%%%%%%%
%\begin{document}

\maketitle
%Begin the section.
\section{\label{intro}Introduction}
High energy photons have been used to paint a picture of the non-thermal
Universe, but a more complete image of the hot and dense regions of space
can potentially be obtained by studying astrophysical neutrinos. Neutrinos
can provide valuable information because they are undeflected by magnetic
fields and hence their paths point back to the particle's source. Unlike
photons, neutrinos are only rarely absorbed when traveling through
matter. However, their low interaction cross section also makes their
detection more challenging. The observation of astrophysical neutrinos
would confirm predictions that hadrons are accelerated in objects such as
active galactic nuclei or gamma-ray bursts~\cite{wb_bound,mpr}.

Instead of searching for neutrinos from either a specific time or location
in the sky, diffuse analyses search for extra-terrestrial neutrinos from
unresolved sources. If the neutrino flux from an individual source is too
small to be detected by point source search techniques, it is nevertheless
possible that many sources, isotropically distributed throughout the
Universe, could combine to make a detectable signal. This search method
assumes that the signal has a harder energy spectrum than atmospheric
neutrinos. When examining an energy-related parameter, an excess of events
over the expected atmospheric neutrino background would be indicative of an
extra-terrestrial neutrino flux.
\vspace{-1mm}
\section{Search Method}
Cosmic ray interactions in the atmosphere create pions, kaons and charmed
hadrons which can later decay into muons and neutrinos. The main background
for this analysis consists of atmospheric muons traveling downward through
the ice. Diffuse analyses use the Earth as a filter to search for upgoing
astrophysical neutrino-induced events. Once the background muons have been
rejected, the data set mainly consists of neutrino-induced upward
events. To separate atmospheric neutrinos from extra-terrestrial neutrinos,
we use an energy-related observable as a final filter. This procedure is
based on the assumption that the signal neutrinos follow a \mbox{$\Phi
\propto $ E$^{-2}$} energy spectrum resulting from shock acceleration
processes. The atmospheric neutrino flux has a much softer energy spectrum
(typically \mbox{$\Phi \propto $ E$^{-3.7}$} for light meson induced,
\mbox{$\Phi \propto $ E$^{-2.7}$} for charmed hadron induced).
\vspace{-1mm}
\section{AMANDA-II diffuse muon searches}
Searches for a diffuse flux have been performed with through-going muon
events from 1997 AMANDA-B10 data~\cite{1997diffuse} and 2000--2003
AMANDA-II data (807 days livetime)~\cite{hodges-diffuse}. 
A search based on a regularized
unfolding of the energy spectrum is also reported in these
proceedings~\cite{Munich}.  The energy estimator used by the 2000--2003
muon analysis was the number of optical modules (channels) that reported at
least one Cherenkov photon during an event (\Nch$\!$). Due to their harder
energy spectrum, extra-terrestrial neutrinos are expected to produce a
flatter \Nch distribution than atmospheric neutrinos (see Figure
\ref{ICRC1103_nch_sigrescaled}).

The search for an extra-terrestrial neutrino component used the number of
events above an \Nch cut, after subtracting a calculated contribution from atmospheric
neutrinos.  The cut was optimized to produce the best limit setting sensitivity~\cite{mrp}.
%Before biasing the analysis by examining the high energy data, an \Nch
%requirement (\Nch$>$100) was optimized using simulated events in order to
%best separate atmospheric background from astrophysical signal. The
%requirement was chosen to produce the best limit setting sensitivity~\cite{mrp}. 
In order not to bias the analysis, data above the resulting cut (\Nch $> 100$)
were kept hidden from the analyzer while the lower \Nch
events were compared to atmospheric neutrino expectations from Bartol~\cite{bartol2004} 
and Honda~\cite{honda2004}.  
The various atmospheric neutrino calculations (Bartol and Honda models, with and
without systematic uncertainties) were
normalized to the low \Nch data, and the resulting spread in the number of events predicted
with \Nch$> 100$ was figured as an uncertainty in the limit calculation. 
%Since multiple
%background simulations were considered (Bartol and Honda models, with and
%without systematic uncertainties), there were multiple predictions for the
%number of events in the \Nch$> 100$ region. The spread of these predictions
%gave an uncertainty on the expected number of high \Nch
%events. 

The observed \Nch distribution is compared to the atmospheric neutrino
background calculations in Figure~\ref{ICRC1103_nch_sigrescaled}.
%The results are shown in Figure \ref{ICRC1103_nch_sigrescaled} and compared
%to the data. 
For the \Nch$>100$ region, 6 events were seen, while 7.0 were
expected. Using the range of atmospheric uncertainty (shaded band in Figure
\ref{ICRC1103_nch_sigrescaled}) in the limit calculation~\cite{ch} leads to an upper
limit on a $\Phi \propto E^{-2}$ flux of muon neutrinos at Earth of
\esqdnde $= 7.4\times10^{-8}$ \diffunit. This upper limit is valid in the energy
range 16--2500 TeV. In comparison, an unfolding of the atmospheric
neutrino spectrum with this same data set leads to an upper limit of
\esqdnde $= 2.6\times10^{-8}$ \diffunit\ for the energy range 300--1000~TeV~\cite{Munich}. With this analysis, limits were also placed on
specific extra-terrestrial models and on the flux of prompt, charmed hadron
neutrinos from Earth's atmosphere~\cite{hodges-diffuse}.

Figure~\ref{sensitivity} shows the upper limit 
on the $\nu_{\mu}$ flux from sources with an E$^{-2}$ energy spectrum. 
The limit from the AMANDA-II 4-year analysis is a factor of four above the 
Waxman-Bahcall upper bound~\cite{wb_bound}. 

\begin{figure}[t!]
%\label{ICRC1103_nch_sigrescaled}
\begin{center}
\includegraphics[width=7cm]{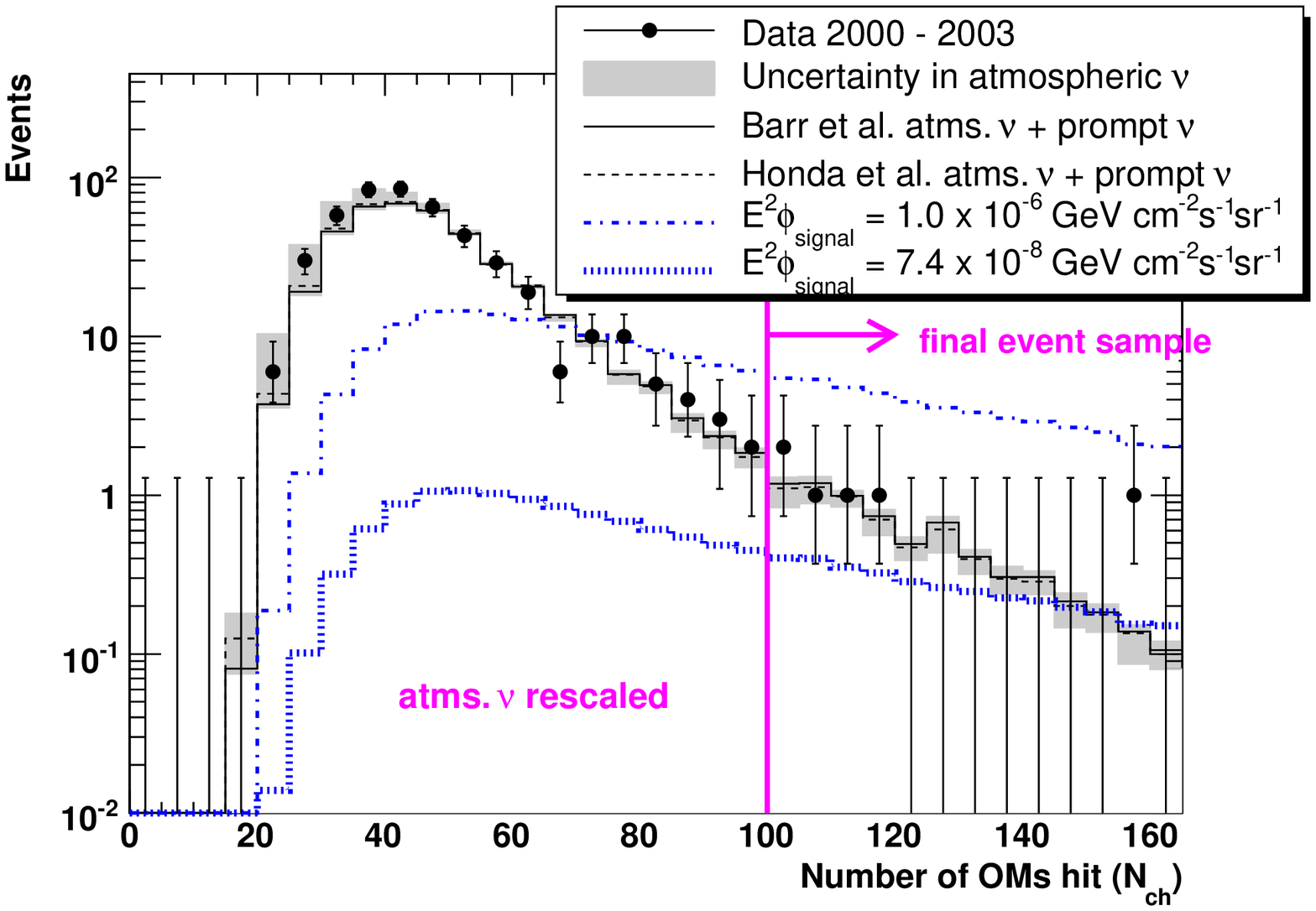}
\caption{
\Nch, the number of OMs triggered, for the AMANDA-II 2000--2003 diffuse muon
neutrino analysis. The data is compared to atmospheric neutrino
expectations~\cite{bartol2004,honda2004}. The signal prediction for a $\Phi
\propto E^{-2}$ flux is rescaled to reflect the upper limit derived from
this analysis. }\label{ICRC1103_nch_sigrescaled}
\end{center}
\end{figure}
\vspace{-1mm}
\section{IceCube 9 String}

The IceCube neutrino observatory is under construction and will be
completed within the next four years. 
In 2006, the first nine IceCube strings were operated as a physics detector for 137 days. 
The IceCube 9-string detector (IC9) has an instrumented volume four times
larger than AMANDA-II. Each string contains 60 digital optical modules
(DOMs) in ice, spaced in 17~m intervals
between depths of 1450 to 2450~m. The distance between strings is 125~m, 
approximately three times greater than in AMANDA-II. 

\begin{figure}
\begin{center}
\noindent
\includegraphics [width=0.45\textwidth]{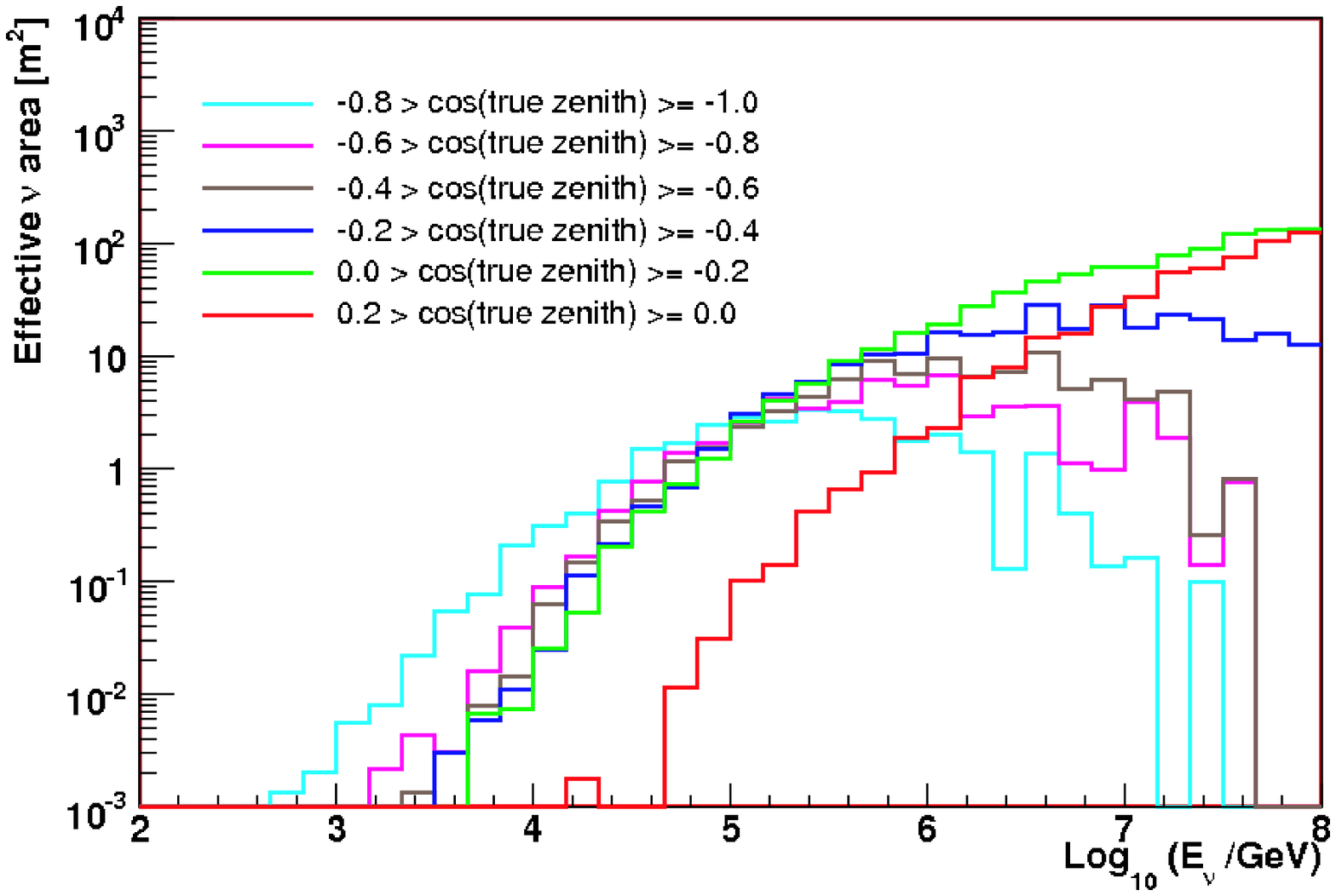}
\end{center}
\caption{Effective area after final background-rejection. 
%The discrepancy between cos(true zenith)$>$0 and others at lower energy
%comes from the zenith angle constraint ($>$80 degree) and 
%the energy-related constraint based on the average hit distance. 
The curve for $\rm cos(zenith)>0$ shows an increased energy threshold
because of the cut on average hit distance.
%The curve for $\rm cos(zenith)>0$ differs from the others at low energy
%because of the cut on average hit distance, which increases the energy threshold.
}\label{ic9eff}
\end{figure}
\vspace{-1mm}
\subsection{Muon Background Rejection}

Like the 2000--2003 AMANDA-II analysis, the IC9 analysis uses the
number of hit DOMs (\Nch$\!$) as an energy-related observable to distinguish atmospheric
neutrinos from extra-terrestrial neutrinos.
This method requires atmospheric muon backgrounds to be removed first.  
For IC9, the atmospheric muon rejection has been 
re-optimized to preserve more near-horizontal signal events (now covering 80--180 degrees in zenith)
and accommodate the new detector geometry.

For the background study, atmospheric muons were simulated using CORSIKA. 
In addition, coincident muon events were generated, in which 
muons from two independent atmospheric showers are detected
during the same trigger window.  For atmospheric neutrinos, $1.6 \times 10^7$ $\nu_{\mu}$
events were generated and re-weighted with the Bartol flux~\cite{bartol2004}.
%and the 1$\times$10$^{-6}$E$^{-2}$ GeV cm$^{-2}$ s$^{-1}$ sr$^{-1}$, 
%which correspond to the AMANDA-B10 limit. 

Atmospheric muons can enter the sample when they are mis-reconstructed as upgoing or when they
arrive from near the horizon.
One of the most effective parameters for rejecting mis-reconstructed events
is the number of direct hits (\Ndir).  These are hits close to the reconstructed track so they are assumed
to result mostly from unscattered Cherenkov photons.  The AMANDA-II analysis selected
well-reconstructed tracks based on an \Ndir\ cut and the distribution of hits along the length of the track.
With its larger string spacing, the IC9 analysis uses a relaxed \Ndir\ cut complemented by
new requirements on the calculated precision of the zenith angle reconstruction and the
number of strings hit.
Besides rejecting mis-reconstructed muons, these cuts lead to the energy threshold behavior
visible in Figure~\ref{ic9eff}.  Therefore lower energy atmospheric muons as well as atmospheric
neutrinos are further suppressed.

%After rejection of the obviously downgoing events which have less than 80 degree of reconstructed 
%zenith angle, the high level
%filter was optimized for IC9.

%AMANDA-II analyses have identified several parameters that are valuable tools
%for rejecting the atmospheric muon background. 
%One of the most effective parameters for rejection of the atmospheric muon background is the number
%of direct hits (Ndir). The direct hits are assumed to be generated by unscattered Cherenkov photons 
%yielded by a reconstructed track.
%The Ndir parameter has moderate correlation with the track energy and shows strong sensitivity
% against the mis-reconstructed track. By requiring larger Ndir, low energy events and 
%mis-reconstructed events are filtered out.

Preserving signal events near the horizon is important because the effective area for 
high energy $\nu_{\mu}$ is greatest there (Figure~\ref{ic9eff}).
This enhancement is strengthened in IC9 by the large height to width ratio.
However, atmospheric muon tracks at these zenith angles are generally well-reconstructed 
and often survive the other cuts.
Therefore another energy-related parameter was introduced, namely
the average perpendicular distance between all hit DOMs and
the reconstructed track.
The higher light yield for energetic tracks means light can reach far away DOMs, so a cut on the
average hit distance distinguishes strongly against the lower energy atmospheric muon events.
This cut is applied only for events above the horizon.

\begin{figure}
\begin{center}
\noindent
\includegraphics [width=0.45\textwidth]{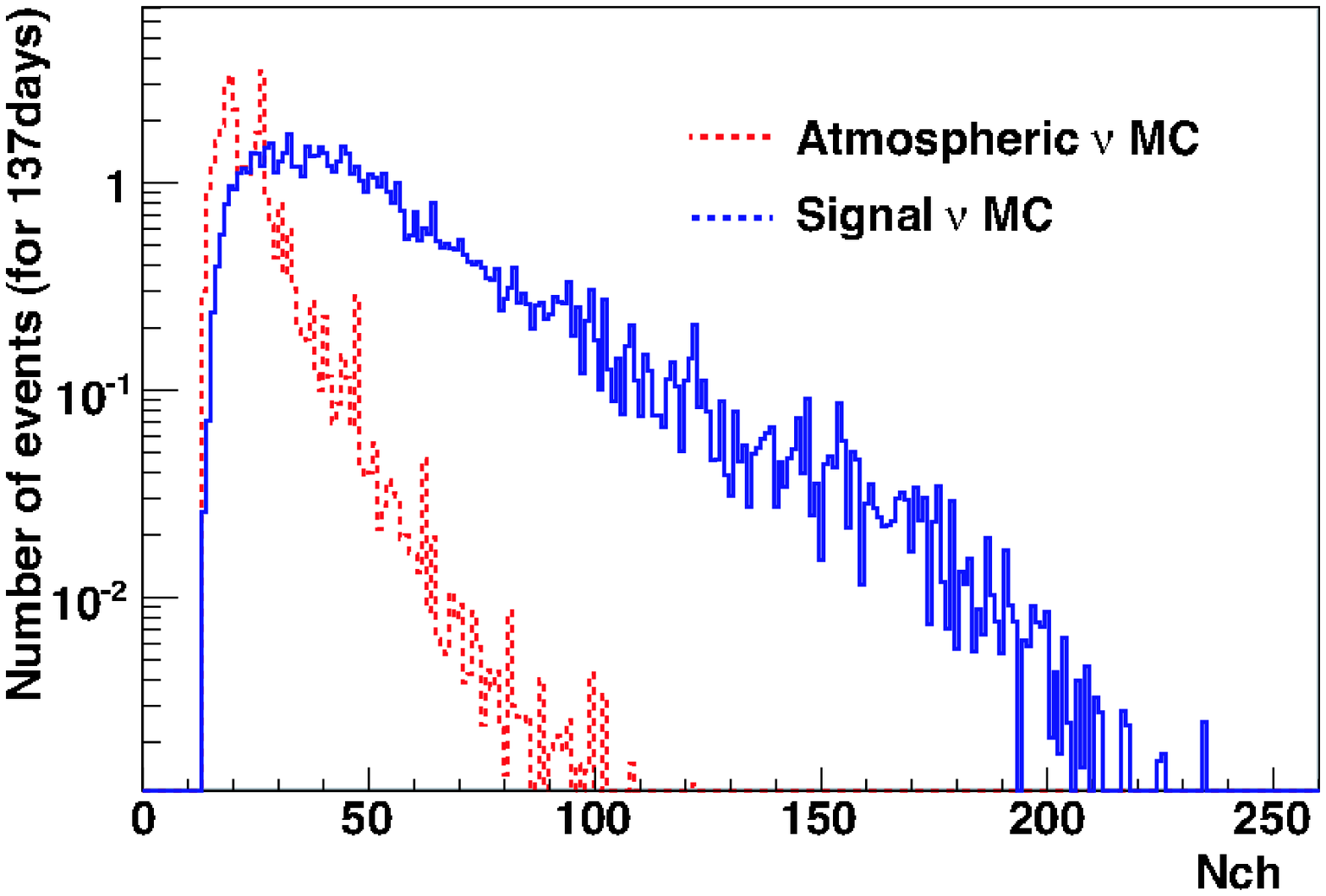}
\end{center}
\caption{\Nch distribution in IC9 after background atmospheric muon rejection.
The IC9 cuts raise the energy threshold relative to AMANDA-II, leading to a lower
atmospheric neutrino rate compared to Figure~1.  The signal curve corresponds
to a test flux of 1$\times$10$^{-6}$E$^{-2}$ GeV cm$^{-2}$ s$^{-1}$ sr$^{-1}$.
}\label{ic9nch}
\end{figure}

\subsection{Sensitivity}

\begin{figure*}[th]
\begin{center}
\noindent
%uncomment next line to include real image
\includegraphics [width=0.85\textwidth]{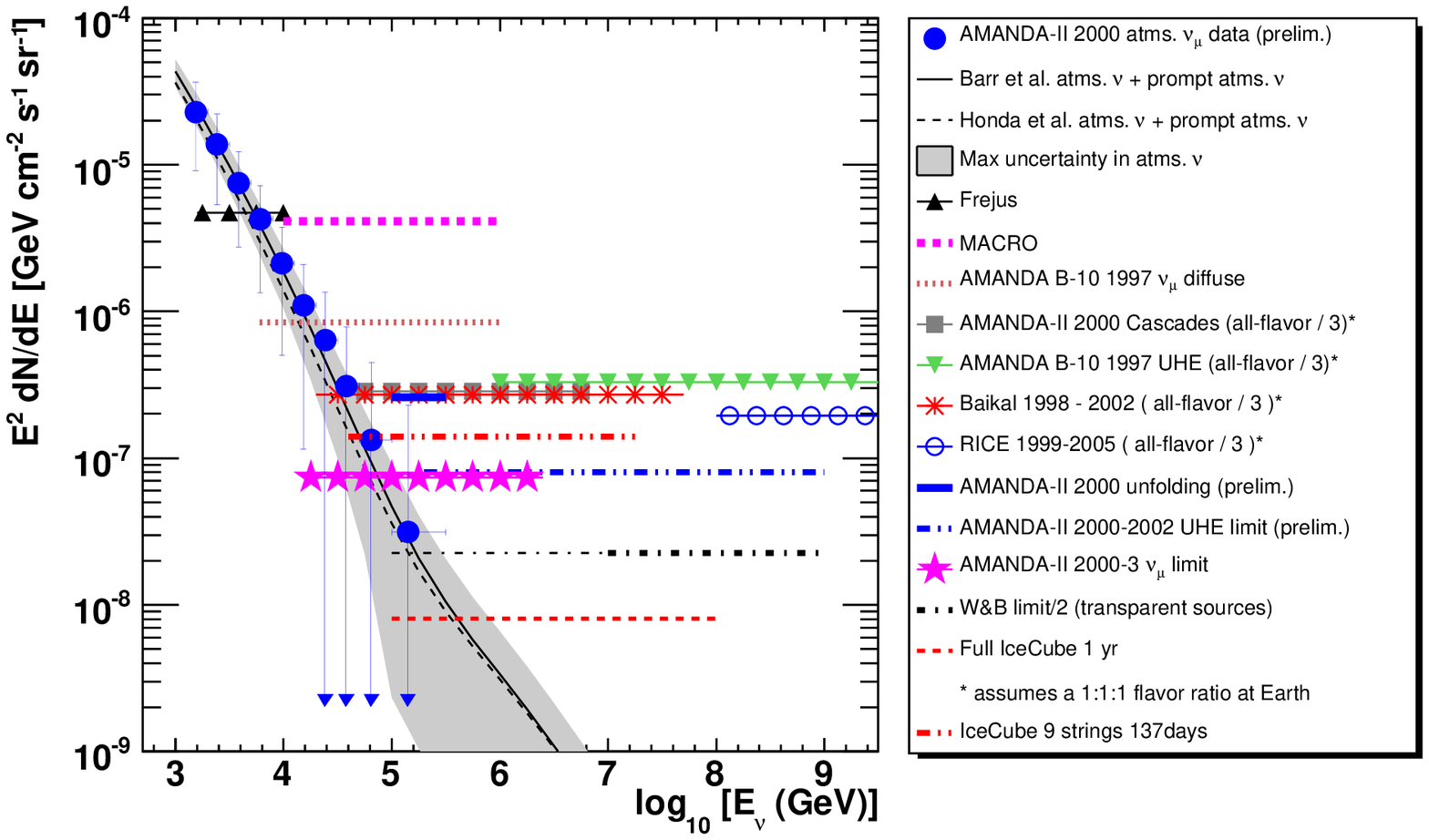}
\end{center}
\caption{Upper limit on the $\nu_{\mu}$ flux from sources with an $E^{-2}$ energy spectrum for the 2000--2003 AMANDA-II data, and expected sensitivity of IC9 for 137 days.}\label{sensitivity}
\end{figure*}

After the atmospheric muon rejection cuts, simulated events
are dominated by atmospheric and
extra-terrestrial neutrinos.  Figure~\ref{ic9nch} shows the \Nch distribution for these events.
%After the background rejection, the atmospheric neutrino induced events and the signal neutrino 
%induced events dominate the remaining events. Figure.\ref{ic9nch} shows \Nch for survived events. 
The best \Nch cut was determined to be 60 for IC9 (137 days) by optimizing the 
Model Rejection Factor~\cite{mrp}.
Assuming no extra-terrestrial signal, the expected upper limit was calculated using the 
Feldman-Cousins method~\cite{ch}, giving a sensitivity of 
1.4$\times$10$^{-7}$GeV cm$^{-2}$ s$^{-1}$ sr$^{-1}$. 
%The Feldman-Cousins method\cite{ch} was applied to calculate average upper limit event number. 
%The sensitivity is calculated by using Model Rejection Factor(MRF)\cite{mrp} which gives the 
%constraint factor to suppress the test flux. The MRF is the ratio of the number of signal events vs the 
%upper limit. By changing \Nch cut, the MRF curve is obtained as a function of cut \Nch. Least MRF 
%value restricts the test flux strongly, thus we get the best sensitivity.
%For IC9 137days, we obtained least MRF 0.14 at \Nch cut=60 for 
%E$^{2}$dN/dE = 1$\times$10$^{-6}$GeV cm$^{-2}$ s$^{-1}$ sr$^{-1}$ test flux. 
%The sensitivity is thus 1.4$\times$10$^{-7}$GeV cm$^{-2}$ s$^{-1}$ sr$^{-1}$. 
Figure~\ref{sensitivity} shows the IC9 sensitivity in relation to
sources with an E$^{-2}$ energy spectrum and the AMANDA-II search.
%Figure~\ref{sensitivity} shows the upper limit 
%on the $\nu_{\mu}$ flux from sources with an E$^{-2}$ energy spectrum. 
%The limit from the AMANDA-II 4-year analysis is a factor of four above the Waxman-Bahcall 
%upper bound. 
The IC9 sensitivity is only a factor 2 above AMANDA-II 4-year, despite its much lower integrated
livetime.
Further improvements may be expected, both from longer term operation of the full IceCube
detector and refinements of the analysis such as new energy reconstruction methods.

\section{Conclusion}
The AMANDA-II data collected during the period 2000--2003 have been analysed in a search
for a diffuse flux of high-energy extra-terrestrial muon neutrinos.
With no excess of events seen, an
upper limit of \esqdnde $<7.4\times10^{-8}$ \diffunit was obtained. 
The sensitivity of 9 IceCube strings for 137 days livetime was studied with simulated data,
making use of new cuts to improve acceptance near the horizon.
The expected sensitivity is 1.4$\times$10$^{-7}$GeV cm$^{-2}$ s$^{-1}$ sr$^{-1}$.
This analysis is ongoing and will be unblinded in the near future.

\section{Acknowledgements}
This work is supported by the Office of Polar Programs of the National Science Foundation.

%This is the reference to .bib file (Without .bib!)
%\bibliography{examplelibrary}
%This in the bibtex style, is ok.
%\bibliographystyle{plain}

%\end{document}
 %Jessica
\setcounter{figure}{0}
\setcounter{table}{0}
%%
% International Cosmic Ray Conference 2007 Merida Yucatan Mexico
% In This file you will find detailed instructions to correctly
% typeset your document.
%
%
%
%Class Requeried
%\documentclass{article}
%The ICRC Style
%\usepackage{icrctc07}
%\usepackage{umlaut}

%The paper title
\title{ Measurement of the atmospheric lepton energy spectra with AMANDA-II }
%Short title to print in the headers to the final publication (Not showed in this print).
\shorttitle{Neutrino energy spectrum}
%All paper authors
\authors{K. M\"unich$^{1}$, J. L\"unemann$^{1}$ for the IceCube Collaboration}
%Short title to print in the headers to the final puplication (Not showed in this print).
\shortauthors{K. M\"unich and et al}
%All the affiliations.
\afiliations{$^1$ Inst. of Physics, University of Dortmund, Dortmund, Germany }
\email{kirsten.muenich@udo.edu}

%The abstract.
\abstract{Extragalactic objects such as active galactic nuclei (AGN) and gamma-ray bursts (GRB) are potential sources for the ultra-high energy cosmic ray f\/lux. Assuming hadronic processes in these sources, a diffuse neutrino f\/lux might be produced together with the charged cosmic ray component. To measure this diffuse extraterrestrial neutrino f\/lux is one of the main goals of the Antarctic Muon and Neutrino Detector Array (AMANDA-II). The neutrino spectrum, based on a four year data set (2000-2003), is presented. The spectrum agrees with the atmospheric neutrino f\/lux predictions. Upper limits to isotropic extraterrestrial contributions are derived.}

%%%%%%%%%%%%%%%%%%%% B E G I N   D O C U M E N T%%%%%%%%%%%%%%%%%%%%%%%
%\begin{document}
\maketitle
%Begin the section.

\section{Introduction}

The search for extraterrestrial neutrino sources is the driving force behind the construction of large neutrino telescopes. Though all three neutrino species should arrive at Earth in equal number, muons from muon neutrinos have a distinct signature in the detector (a long path emitting Cherenkov light) that makes them a desirable focus for this analysis. The drawback of this signature is the existence of a large background of atmospheric muons entering the detector from the upper hemisphere. Atmospheric muons are suppressed by selecting only upgoing events as potential signal candidates. Muons from neutrinos produced in the atmosphere dominate even in this sample.

The search for extraterrestrial muon neutrinos within the data sample can be performed by multiple approaches, for instance by selecting local coincidences with proposed steady neutrino sources (AGN) or local and temporal coincidences with GRBs. Since the energy spectrum of extraterrestrial neutrinos is expected to be significantly harder than the atmospheric neutrino spectrum, another approach relies directly on the reduction of the atmospheric neutrino background by energy selection \cite{icrc07_jessica}. The analysis described here is based on the reconstruction of the energy spectrum of atmospheric muon neutrinos. Data taken with the AMANDA-II detector between 2000 and 2003 provide 2972 upgoing muons with a lifetime of 807 days. The criteria used for the selection of events are described in \cite{icecube5yr}. In addition a zenith angle veto at 10 degrees below the horizon is applied.

\section{Unfolding of the energy spectrum}

In this analysis, the problem of determining the energy spectrum from the observed detector response  is solved by applying a regularized unfolding method. The underlying Fredholm integral equation of first kind is reduced to a matrix equation system. The kernel is determined with Monte Carlo methods. Statistically insignificant contributions to the kernel are suppressed by regularization \cite{blobel,kirstenphd}. The observables used must be correlated to the neutrino energy. In total, eight observables are found to satisfy these conditions. Because the unfolding algorithm used for this calculation, RUN \cite{blobel}, allows only three input variables, six observables are combined into one energy-sensitive variable by a neural network application \cite{kirstenphd,icrc05_kirsten}. In \mbox{Figure \ref{nn}}, the Gaussian response of this variable to mono-energetic muons from the simulation is shown. The unfolded neutrino energy spectrum is compared to the f\/lux expectations from \cite{honda,ICRC0624_bartol} in \mbox{Figure \ref{honda_bartol}}. The error bars in the plot comprise both statistical and systematic uncertainties. The theoretical uncertainty of the atmospheric neutrino flux contributes with 25\% to the total systematic error of 30\%. For a detailed error discussion see \cite{icrc05_kirsten}. Good agreement is observed when the unfolded four-year neutrino spectrum is compared to the unfolded data from 2000 analysed in \cite{kirstenphd,icrc05_kirsten} (\mbox{Figure \ref{compare}}). 
\begin{figure}
\begin{center}
\includegraphics [width=0.50\textwidth]{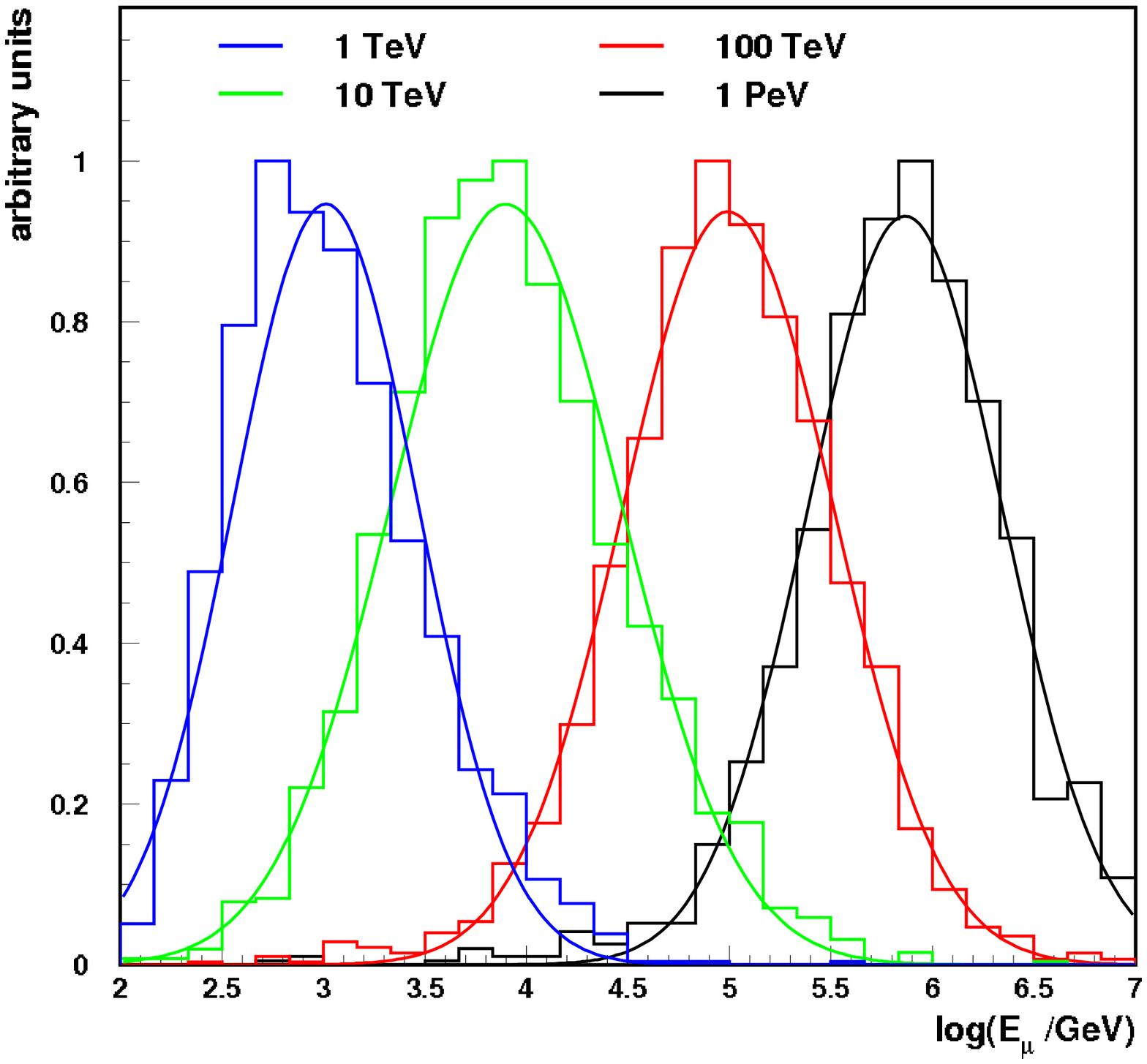}
\end{center}
\vspace{-0.5cm}
\caption{Neural network output for simulated mono-energetic muons fitted with Gaussian distributions.}\label{nn}
\end{figure}

%\begin{figure}
%\begin{center}
%\includegraphics [width=0.50\textwidth]{snn_Profile_plot_SNN_vs_E_nu.eps}
%\end{center}
%\caption{Comparison of the unfolded energy spectrum with flux expectations according to\cite{honda, bartol}. The shaded bands give the range between horizontal and vertical flux.}
%\end{figure}

\begin{figure}
\begin{center}
\includegraphics [width=0.50\textwidth]{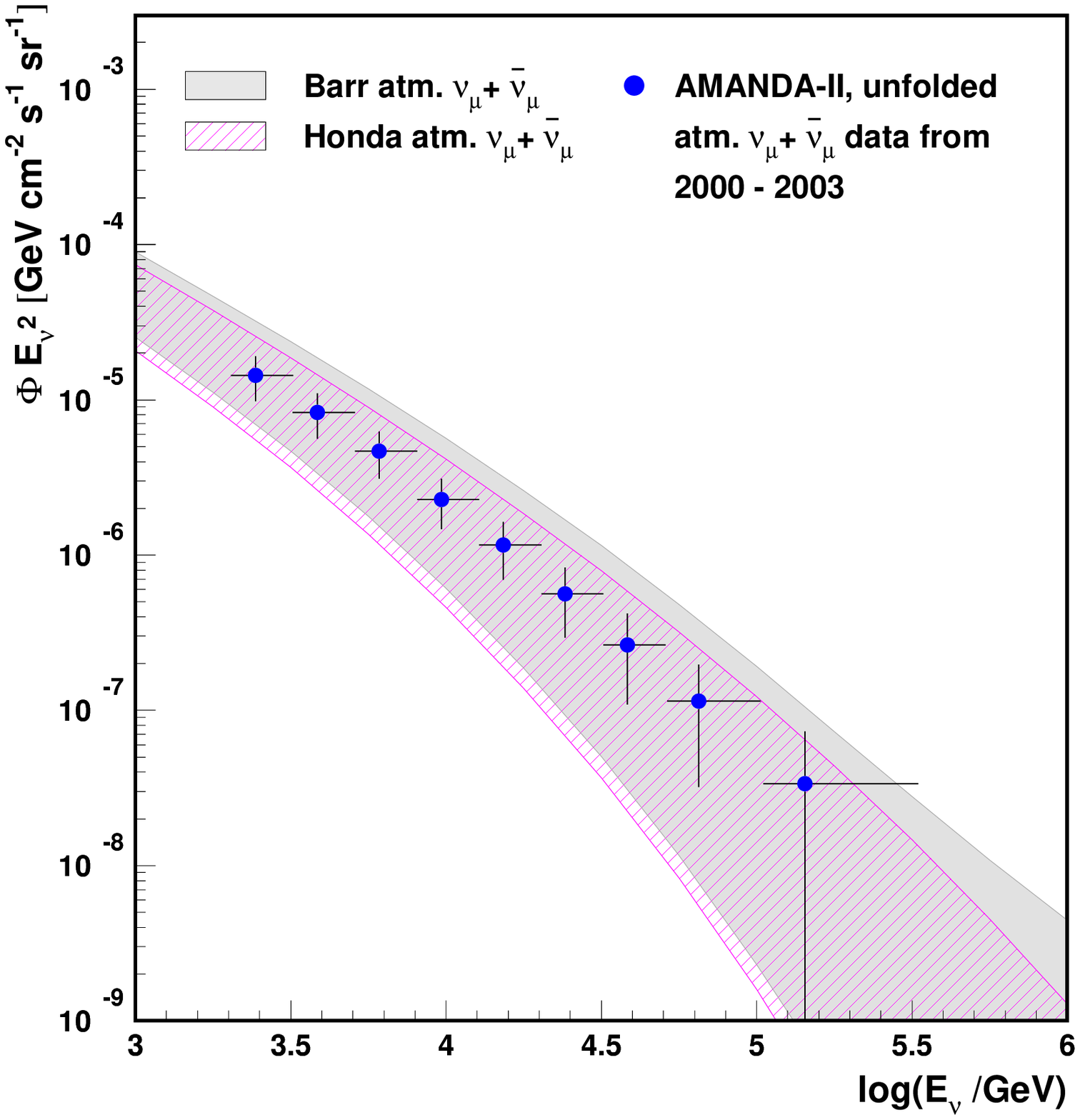}
\end{center}
\caption{Comparison of the unfolded energy spectrum with f\/lux expectations according to Ref.\cite{honda, bartol}. The shaded bands show the range between the horizontal (upper border) and vertical f\/lux (lower border).}\label{honda_bartol}
\end{figure}

\begin{figure}
\begin{center}
\includegraphics [width=0.50\textwidth]{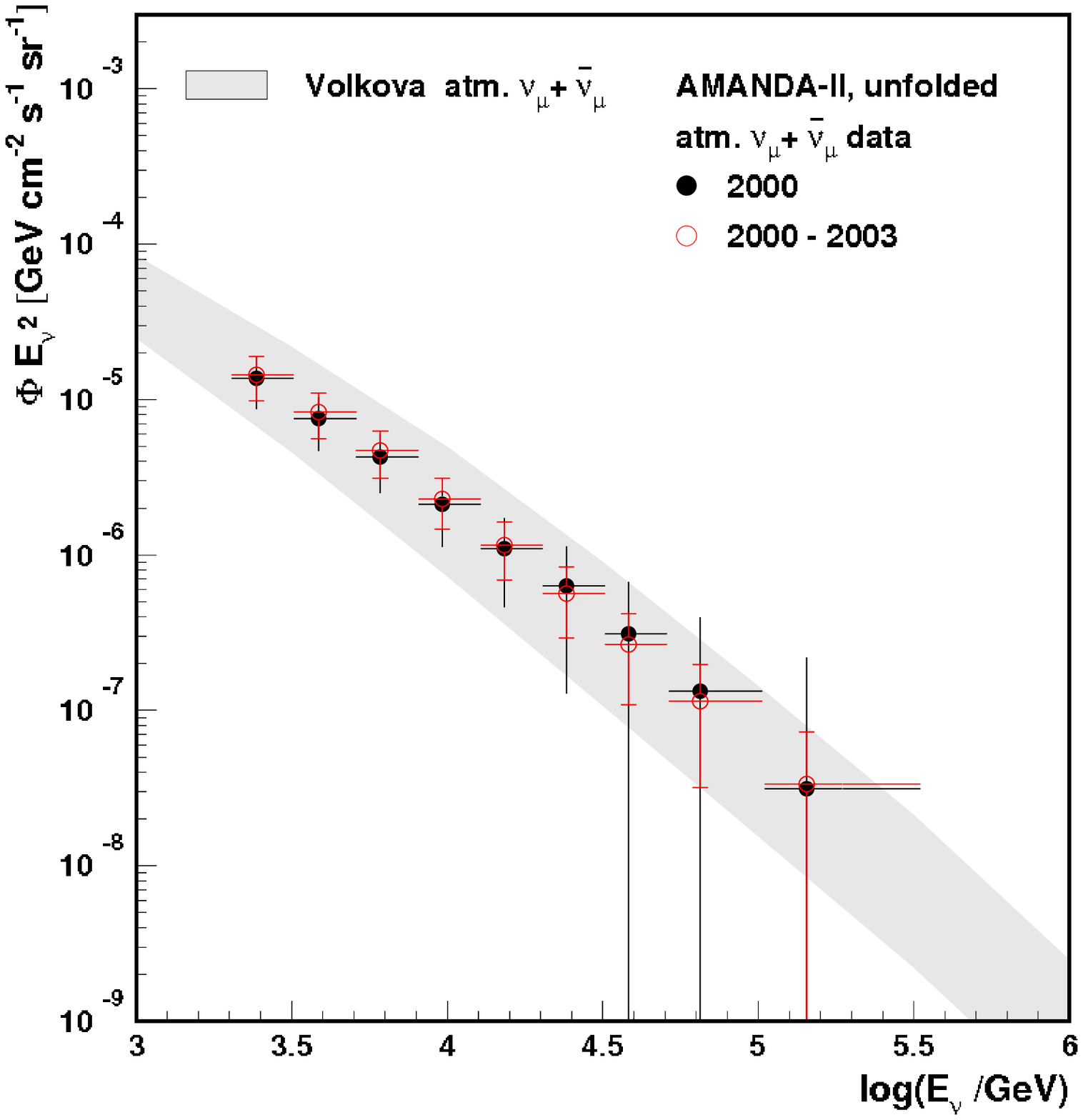}
\end{center}
\caption{Comparison of the unfolded energy spectrum for 2000 and 2000-2003.}\label{compare}
\end{figure}

\section{Upper limits to additional contributions to the neutrino f\/lux}

Two properties of the unfolded spectrum in \mbox{Figure \ref{honda_bartol}} should be noted. First, the variable binning with a width of about half of the resolution was optimized by Monte Carlo to obtain the best sensitivity to an $E^{-2}$ contribution of extraterrestrial neutrinos. The bins are statistically correlated to each other. This is taken into account in the error calculation. However, it is not obvious which kind of probability density function (pdf) the f\/lux errors obey and how upper limits to additional contributions to the atmospheric neutrino f\/lux have to be derived. Therefore, a confidence belt construction \cite{feldman} has been applied to the unfolding problem.
The second remark concerns the 2000-2003 data quality. During this period, small changes in the detector properties, such as the photomultiplier high voltage, resulted in different detector response in the observables used in this analysis. Since only the logarithm of these variables enters the unfolding procedure, these systematic effects concern only the low energy portion of the spectrum ($E<2$~TeV).\\
%In the Feldman-Cousins approach, the added statistical weight in singular bins of the unfolded energy spectrum has been used as input. In this approach the unfolding procedure is just serving as black box algorithm. The statistical weight in one bin is used as variable whose probability density function has to be determined. 
Assuming a diffuse signal energy spectrum with an energy dependence of $E^{-2}$, the unfolded response for 17 different signal contributions between \mbox{$10^{-8}$~{GeV~cm$^{-2}$~s$^{-1}$~sr$^{-1}$}} and \mbox{$4 \cdot 10^{-7}$~{GeV~cm$^{-2}$~s$^{-1}$~sr$^{-1}$}} has been calculated. For each signal contribution, the complete Monte Carlo and analysis chain has been applied. Finally, 1,000 Monte Carlo experiments each containing the equivalent of four years of \mbox{AMANDA-II} data have been used for each of the 17 signal contributions. The energy distributions of all 17,000 Monte Carlo experiments have been reconstructed. After applying an energy cut, the statistical weights, which corresponds to the weighted number of events, for a fixed signal distribution are summed, histogrammed and normalized to get the individual pdf. Using the pdfs for each signal contribution the Feldman-Cousins approach is applied. The resulting confidence belts are shown in \mbox{Figure \ref{cbelt}}. 
\begin{figure}
\begin{center}
\includegraphics [width=0.50\textwidth]{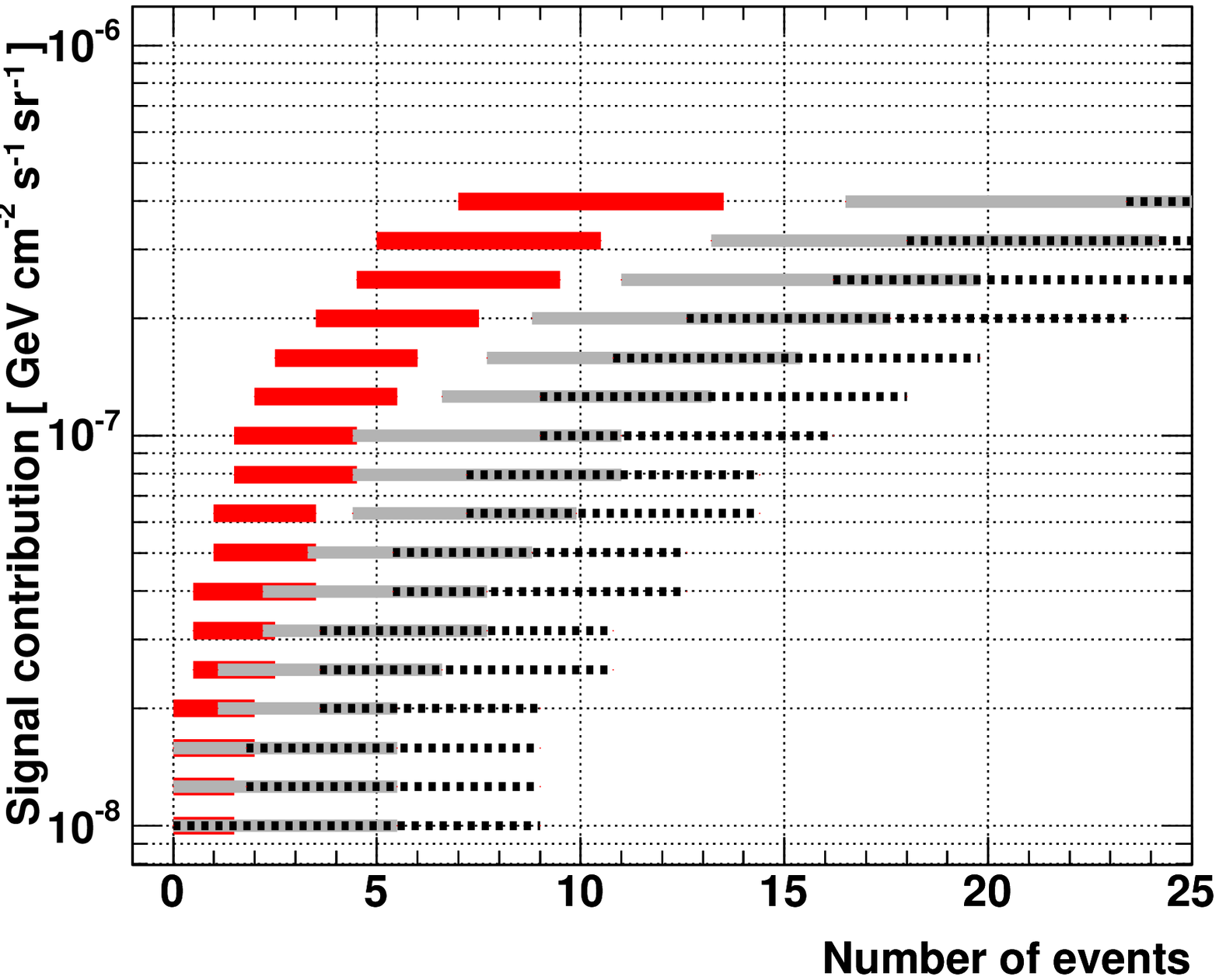}
\end{center}
\caption{The unified approach of Feldman and Cousins has been applied to the unfolding problem by calculating individual probability density functions. 90\% Feldman-Cousins confidence belts of three unfolding energy bins: $50$ to \mbox{$100$~TeV} (black dotted), $100$ to \mbox{$300$~TeV} (gray) and \mbox{$300$~TeV} to \mbox{$1$~PeV} (red) are displayed.}\label{cbelt}
\end{figure}
The upper limit is obtained from the confidence belt by reading off the f\/lux value that corresponds to the statistical weight of the unfolded data \mbox{(Figure \ref{weight})}. The statistical weight between $300$~TeV and $1$~PeV is $0.005$.
\begin{figure}
\begin{center}
\includegraphics [width=0.50\textwidth]{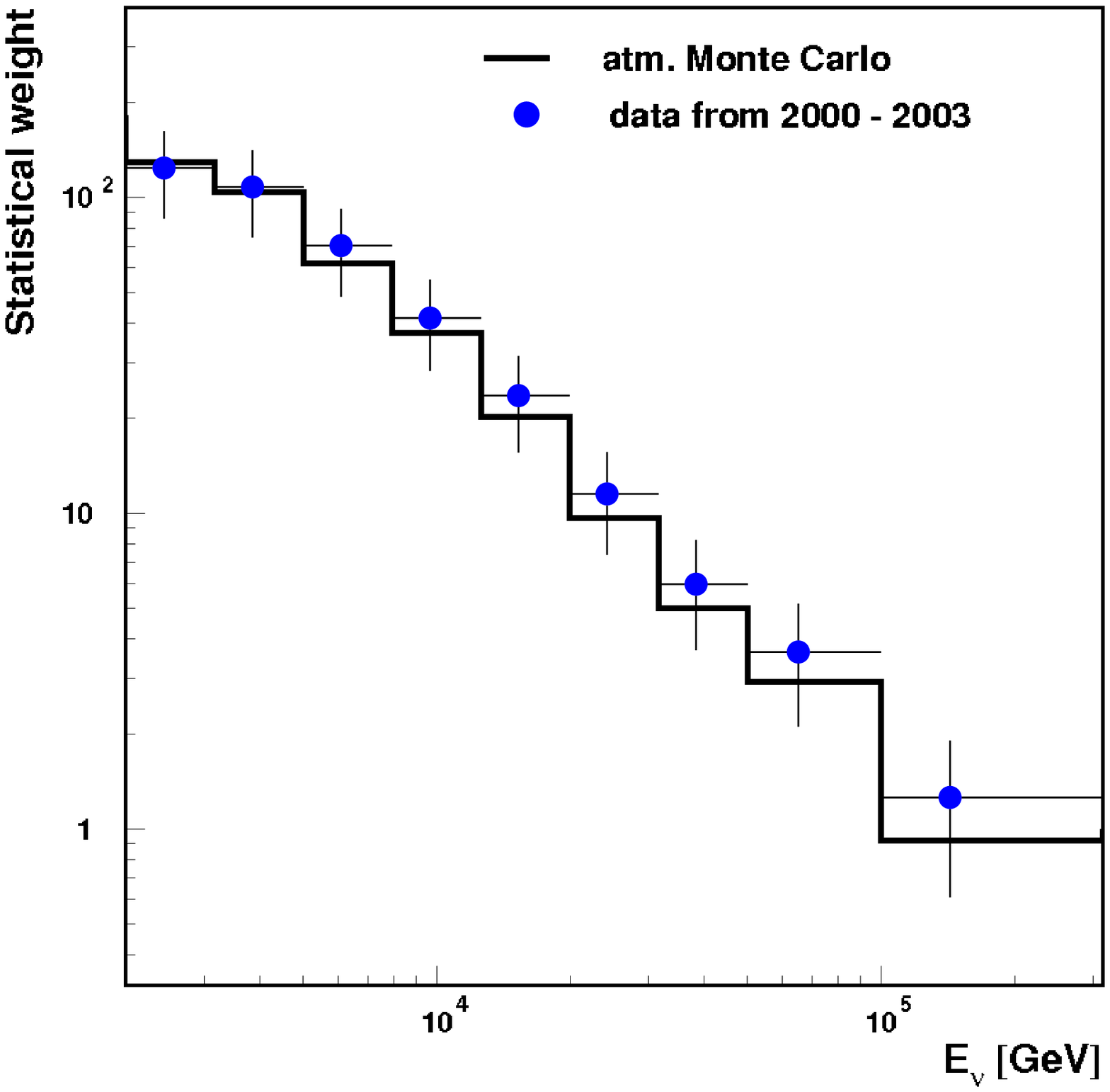}
\end{center}
\caption{Statistical weight of the unfolded data.}\label{weight}
\end{figure}
The error bars can be used to calculate an upper limit. Assuming normal distribution for the pdfs, the 90\% upper limit on the sum of atmospheric plus extraterrestrial flux is given by 1.28 times the standard deviation. By subtracting the atmospheric portion (gained by fitting the Volkova prediction \cite{Volkova} to the unfolded spectrum) from the total upper limit, an upper limit on the extraterrestrial contribution can be calculated, see \cite{kirstenphd}. In Figure \ref{limit} the unfolded neutrino spectrum (blue circles) for data from 2000-2003 as well as the resulting upper limits are shown. The upper limits obtained by the Feldman-Cousins procedure (blue lines) are compared to those upper limits (pink lines) obtained by using the normal distributed pdf and the atmospheric fit. 
%The upper limits obtained with both methods are in good agreement. Therefore the errors in the unfolding procedure are correctly treated. 
Since the upper limits obtained from the two different methods are in agreement, this is a good indication  that the statistic errors in the procedure have been treated properly.
The upper limits derived by calculating the individual pdfs in combination with the Feldman-Cousins approach deliver slightly more restrictive bounds. The resulting limits are compared with different f\/lux models (see Figure \ref{limit}). MPR-max represents the maximum neutrino f\/lux from blazars in photo-hadronic interactions. An upper bound on the f\/lux from AGN was estimated in~\cite{mannheim_bound}, which is indicated in the figure as shaded region (MPR-bound). The upper border of that region represents the limit for sources that are optically thick to $n\gamma$ interactions, $\tau_{n\gamma}\gg 1$. The bound for optically thin sources ($\tau_{n\gamma}< 1$) is given by the lower bound of the shaded region. 

\begin{figure}
\begin{center}
\includegraphics [width=0.50\textwidth]{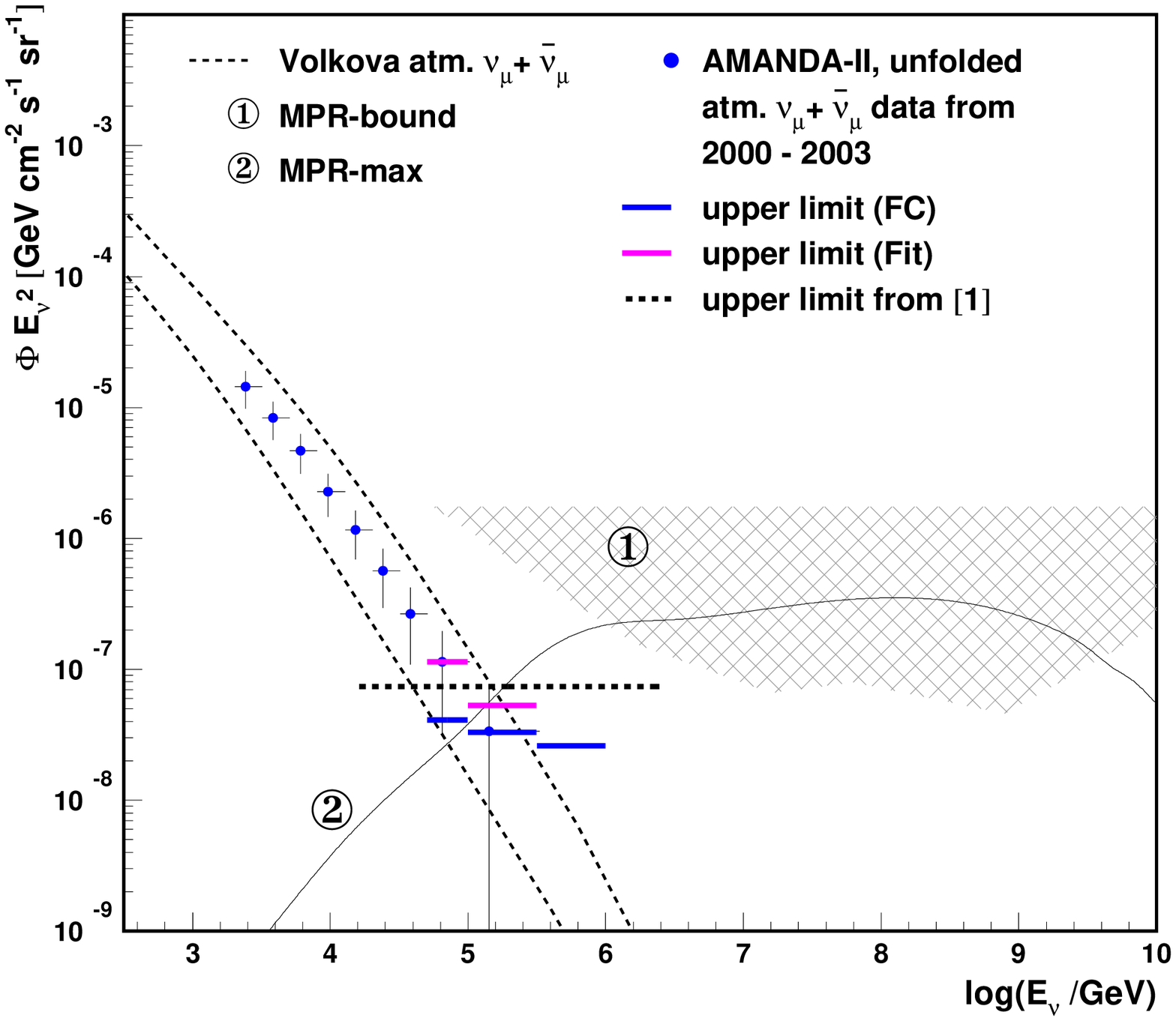}
\end{center}
\vspace{-0.5cm}
\caption{Reconstructed neutrino spectrum and resulting upper limits (blue and pink lines) for data from \mbox{2000-2003. The} results are compared with different f\/lux models \cite{mannheim_bound} and the result from \cite{icrc07_jessica}. For the FC upper limit we added a bin from 300 TeV to 1 PeV which is not shown in Figures \ref{honda_bartol}, \ref{compare} and \ref{weight} as only $0.005$ events were observed in this range and the corresponding flux value is out of the displayed flux range. }\label{limit}
\end{figure}
%The blue upper limits are derived by calculating individual pdf and apply to the Feldman-Cousins approach.

\section{Conclusion}

The energy spectrum of atmospheric muon neutrinos has been reconstructed with a regularized unfolding method in the energy range between \mbox{$1$~TeV} and \mbox{$300$~TeV}. In this energy range, no f\/lattening of the spectrum is observed, as would be expected if a significant extraterrestrial neutrino contribution was presented. Upper limits to additional contributions of \mbox{$\phi \cdot E^2 = 4.1 \cdot 10^{-8}$~{GeV~cm$^{-2}$~s$^{-1}$~sr$^{-1}$}} to the energy bin between $50$~TeV and $100$~TeV, \mbox{$\phi \cdot E^2 = 3.3 \cdot 10^{-8}$~{GeV~cm$^{-2}$~s$^{-1}$~sr$^{-1}$}} between $100$~TeV and $300$~TeV and \mbox{$\phi \cdot E^2 = 2.6 \cdot 10^{-8}$~{GeV~cm$^{-2}$~s$^{-1}$~sr$^{-1}$}} between $300$~TeV and $1$~PeV  are obtained. This is presently the most restrictive upper limit in this energy range and at the given energies well below the theoretical upper bound by Mannheim et al. \cite{mannheim_bound}. This upper limit restricts the parameter range of the source models for AGN classes with flat luminosity distributions (FRII) \cite{becker07}. A comparison of these upper limits to the upper limits obtained with independent methods in \mbox{AMANDA-II} \cite{icrc07_jessica} shows good agreement. All results shown here are preliminary.

\section{Acknowledgements}
This work is partially supported by the German agencies BMBF under contract
number \mbox{05~CI5PE1/0} and DFG under number \mbox{LU1495/1-1}.

%This is the reference to .bib file (Whitout .bib!)
%\bibliography{ICRC0624/icrc0624}
%This in the bibtex style, is ok.
%\bibliographystyle{plain}
%\bibliographystyle{unsrt}

%\end{document}
 %Kirsten
\setcounter{figure}{0}
\setcounter{table}{0}
%%
% International Cosmic Ray Conference 2007 Merida Yucatan Mexico
% In This file you will find detailed instructions to correctly
% typeset your document.
%
%
%

%Class Requeried
%\documentclass{article}
%The ICRC Style
%\usepackage{icrctc07}

%The paper title
\title{Multi-year Search for UHE Diffuse Neutrino Flux with AMANDA-II}
%Short title to print in the headers to the final publication (Not showed in this print).
\shorttitle{UHE Neutrino Search with AMANDA-II}
%All paper authors
\authors{L. Gerhardt$^{1}$ for the IceCube Collaboration$^{2}$}
%Short title to print in the headers to the final puplication (Not showed in this print).
\shortauthors{Gerhardt}
%All the affiliations.
\afiliations{$^1$University of California, Irvine, Irvine, CA,
USA\\$^2$See special section of these proceedings}
\email{gerhardt@hep.ps.uci.edu}

%The abstract.
\abstract{AMANDA-II is a high volume neutrino telescope designed to search for 
astrophysical neutrinos. Data from 2000 - 2002 has been searched for a diffuse flux of 
ultra-high energy (UHE) neutrinos with energies in excess of 10$^{5}$ GeV. Due to 
absorption of UHE neutrinos in the earth, the UHE signal is concentrated at the horizon 
and has to be separated from the background of large muon-bundles induced by cosmic ray 
air showers. No statistically significant excess above the expected background is seen in 
the data, and a preliminary upper limit is set on the diffuse all-flavor neutrino flux of 
E$^{2}$ $\Phi$$_{\mathrm{90\% CL}}$ $<$ 2.4 $\times$ 10$^{-7}$ GeV cm$^{-2}$ s$^{-1}$ 
sr$^{-1}$ valid over the energy range of 2 $\times$ 10$^{5}$ GeV to 10$^{9}$ GeV. A 
number of models which predict neutrino fluxes from active galactic nuclei are 
preliminarily excluded at the 90\% confidence level.}

%\email{aastex-help@aas.org}

%%%%%%%%%%%%%%%%%%%% B E G I N   D O C U M E N T%%%%%%%%%%%%%%%%%%%%%%%
%\begin{document}
\maketitle
%Begin the section.

\section{Introduction}
AMANDA-II is a large volume neutrino telescope with the capability to search for neutrinos 
from astrophysical sources \cite{Casc-AMA}. In a previous publication \cite{UHE-AMA} 
it was shown that AMANDA-II is able to search for UHE neutrinos (neutrinos with energy 
greater than 10$^{5}$ GeV). UHE neutrinos are of interest because they are associated 
with the potential acceleration of hadrons by AGNs \cite{P96,St05}, are produced by the 
interactions of exotic phenomena such as topological defects \cite{Sigl98} or Z-bursts 
\cite{Yosh98}, and are guaranteed by-products of the interaction of high energy cosmic 
rays with the cosmic microwave background \cite{Engel01,kal02b}. 

Above 10$^{7}$ GeV the Earth is essentially opaque to neutrinos \cite{Klein}. This, 
combined with the limited overburden above AMANDA-II (approximately 1.5 km, for a 
description of the AMANDA-II detector see \cite{Casc-AMA}), means that UHE neutrinos 
will be concentrated at the horizon. The background for this analysis consists of bundles 
of down-going, high-energy muons from atmospheric cosmic ray showers. The muons from these 
bundles can spread over cross-sectional areas as large as 200 m$^{2}$. 

\section{Experimental and Simulated Data}
This analysis used AMANDA-II data collected between February 2000 and November 2002, 
with an integrated lifetime of 571 days after offline retriggering and correcting for 
dead time and periods where the detector was unstable. Of this data 20\% from each year 
was used to develop selection criteria, while the rest, with a lifetime of 456.8 days, 
was set aside for the final analysis. Cosmic ray air shower background events were 
generated using CORSIKA \cite{ICRC1062_CORSIKA}. The UHE neutrinos were generated with energies 
between 10$^{3}$ GeV and 10$^{12}$ GeV using ANIS \cite{ANIS}. For more details on AMANDA 
simulation procedures see \cite{Casc-AMA,UHE-AMA}.
\section{Method}
This analysis exploits the differences in light deposition from the background of bundles 
of many low energy muons and single UHE muons or cascades from UHE neutrinos. A muon 
bundle with the same total energy as a UHE neutrino spreads its light over a larger 
volume, leading to a lower light density in the array. Both types of events have a large 
number of hits, but for the same number of hit optical modules (OMs), the muon bundle has 
a lower total number of hits (each OM may have multiple, separate hits in one event). 
Background muon bundles also have a higher fraction of OMs with a single hit, while 
the UHE neutrino generates more multiple hits. In addition to selecting on variables 
which correlate with energy, selecting on the reconstructed direction of the lepton track 
separates the primarily horizontal UHE neutrinos from down-going muon bundles (Fig. 
\ref{figzen}). 
\begin{figure}[ht]
\begin{center}
\includegraphics [width=0.48\textwidth]{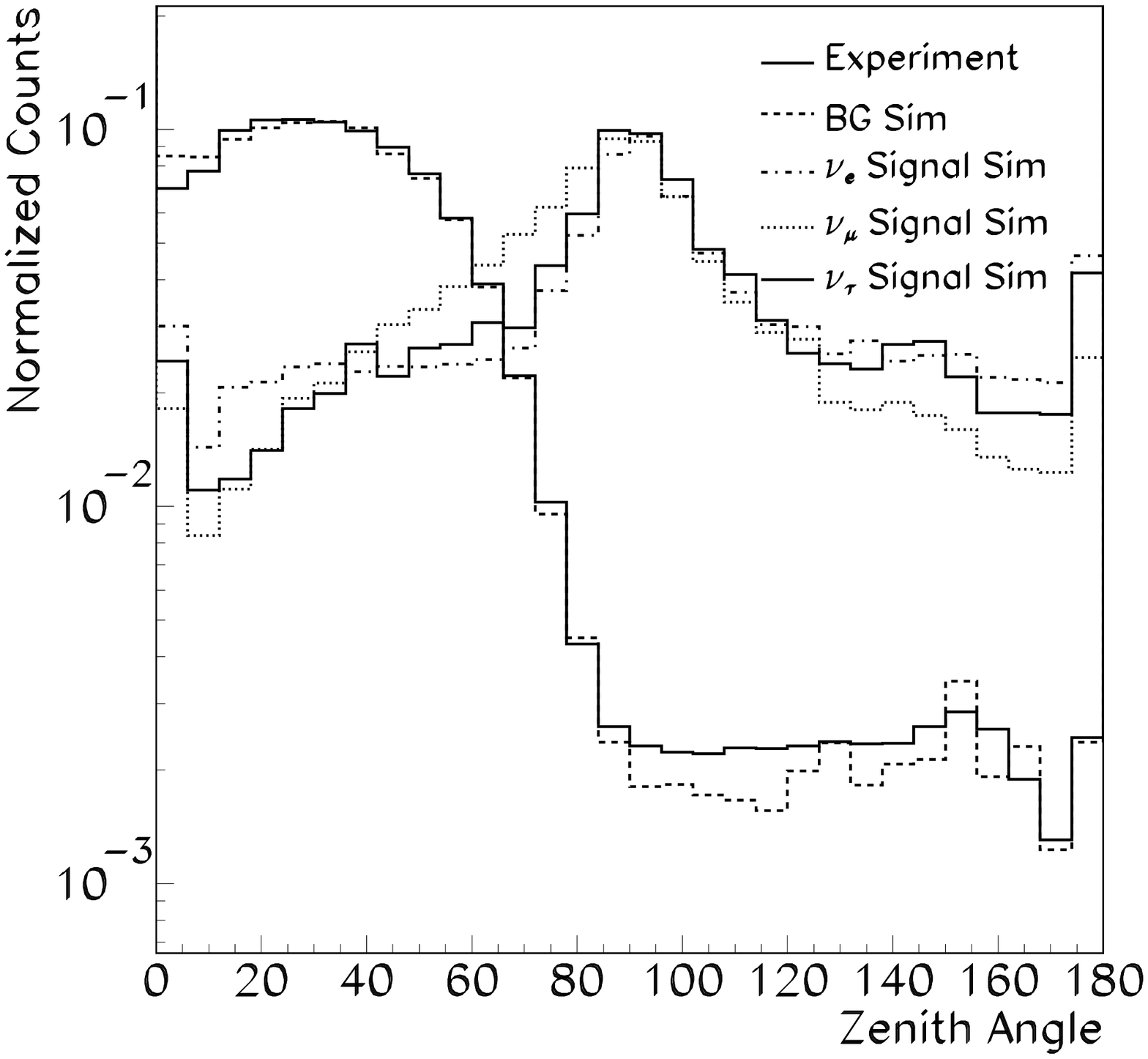}
\end{center}
\caption{Reconstructed zenith angle for the experiment, background muon bundle and 
E$^{-2}$ electron, muon, and tau neutrino signal simulations. The majority of signal 
events are expected at the horizon, while the background is primarily downgoing. 
}\label{figzen} \end{figure}
Reconstruction algorithms optimized for cascade light deposition \cite{Casc-AMA} 
are also used to select UHE neutrinos with an energy deposit from stochastic process (i.e. 
bremsstrahlung or e$^{+}$/e$^{-}$ pair creation) many orders of magnitude brighter than 
the depositions from background muon bundles. 
\section{Systematic and Statistical Uncertainties}
The sensitivity of AMANDA-II is determined from simulation. The dominant sources of 
uncertainty in this calculation are listed below.
\begin{description}
\item \textbf{Normalization of Cosmic Ray Flux:} The average energy of simulated cosmic 
ray primaries at the penultimate selection level is 4.4 $\times$ 10$^{7}$ GeV. Estimates 
of the error in the normalization of the cosmic ray flux range from 20\% \cite{hor03} 
to a factor of two \cite{pdg04}. This analysis uses the more conservative uncertainty of 
a factor of two.
\item \textbf{Cosmic Ray Composition:} There is considerable uncertainty in the cosmic ray 
composition above the knee \cite{pdg04}. The difference between background passing rates 
at the penultimate selection level for iron- and proton-dominated spectra is 30\%; this is 
taken as the uncertainty due to cosmic ray composition.
\item \textbf{Detector Sensitivity}
The optical properties of the refrozen ice around each OM, the absolute sensitivity of 
individual OMs, and obscuration of OMs by nearby power cables can effect the detector 
sensitivity. Variations of these parameters can cause a 15\% variation in the 
background and E$^{-2}$ signal passing rate.
\item \textbf{Neutrino Cross Section:} The uncertainty in the standard model neutrino 
cross section is as large as a factor of two at high energies depending on the model 
assumed for the proton structure \cite{gan96}. This causes a maximum variation in number 
of expected signal events for an E$^{-2}$ spectrum of 8\%.
\item \textbf{Statistical:} Due to the very demanding computational requirements, 
background simulation statistics are somewhat limited. A statistical error of 1$\sigma$ 
for a Poissonian distribution with $\mu =$ 0 is assumed for each year at the final 
selection level. The signal simulation has an average statistical error of 5\% for each 
neutrino flavor.
\end{description}
Summing the systematic errors of the signal simulation in quadrature gives a systematic 
uncertainty of 17\%. Combining this with the statistical uncertainty of 5\% per neutrino 
flavor gives a total uncertainty of 18\%. Following a similar method for the background 
simulation, the systematic uncertainty is 105\%, and the maximum background expectation 
is fewer than 2.1 events for three years. These uncertainties are included in the final 
limit using a method outlined in \cite{teg05}.
\begin{figure}[ht]
\begin{center}
\includegraphics [width=0.48\textwidth]{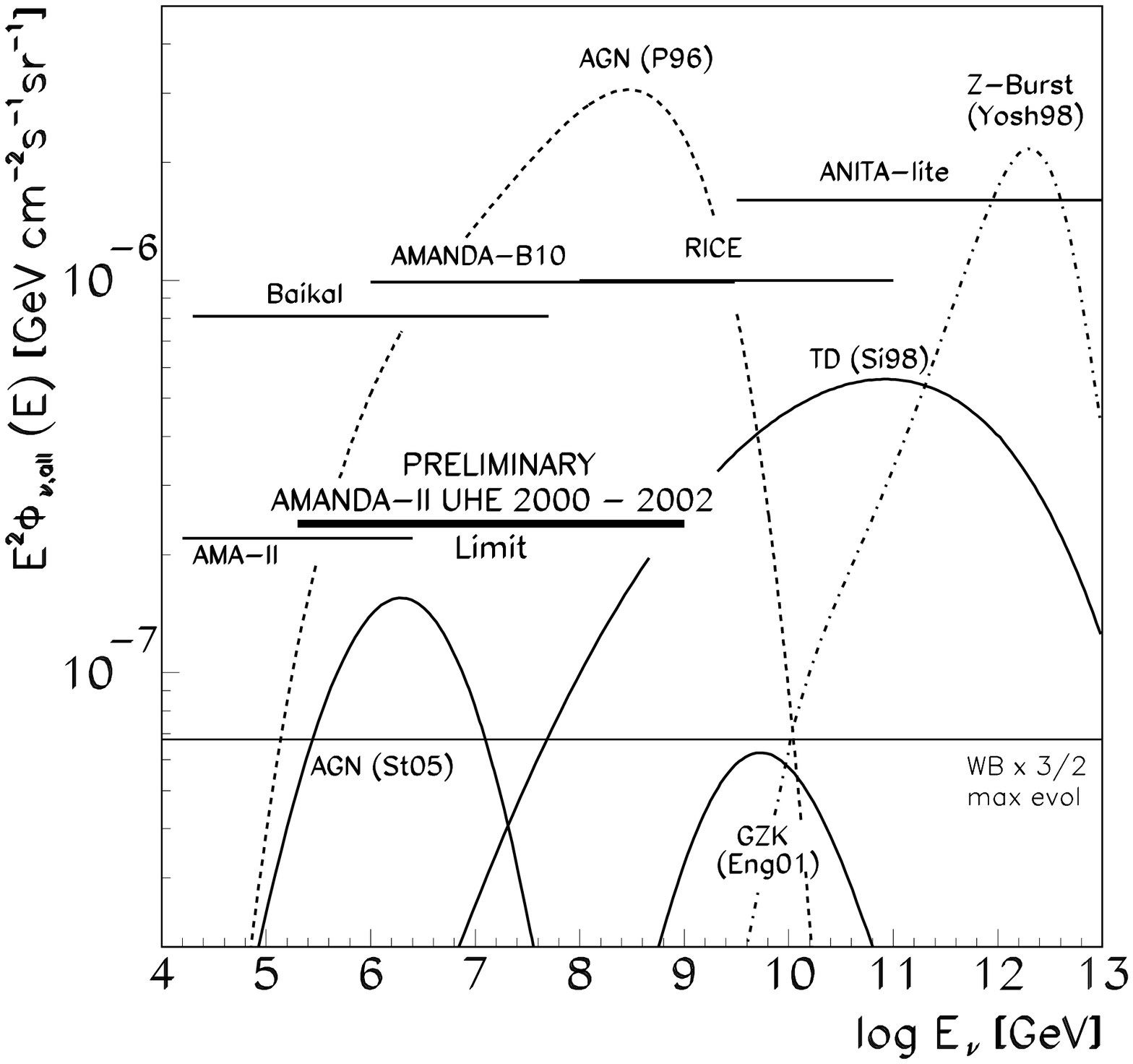}
\caption{Preliminary all-flavor neutrino flux limit and sensitivity for 2000 - 2002 over 
the range which contains 90\% of the expected signal with an E$^{-2}$ spectrum. Also 
shown are several representative models:  St05 from \cite{St05}, P96 from \cite{P96}, 
Eng01 from \cite{Engel01}, Si98 from \cite{Sigl98}, Yosh98 from \cite{Yosh98} and the 
Waxman-Bahcall upper bound \cite{bah98}. Existing experimental limits shown are from
RICE \cite{kra06}, ANITA-lite \cite{bar06}, Baikal \cite{ayn06}, AMANDA-B10 
\cite{UHE-AMA} and AMANDA-II lower energy diffuse search \cite{ach07}.}\label{fig-rep}
\end{center}
\end{figure}

\section{Results}
The effective area after applying all selection criteria is shown in Fig. \ref{fig-area}.
After applying all selection criteria two events were found in the 456.8 days of data 
between 2000 - 2002. The background expectation for the same time period is fewer than 2.1 
events, after including simulation uncertainties. This yields a 90\% confidence level 
average event upper limit \cite{fel98} of 4.74 and a preliminary upper limit on the 
all-flavor neutrino flux of \begin{equation}
\mathrm{E^{2}\Phi_{90\% CL} \le 2.4 \times 10^{-7} GeV\ cm^{-2}\ s^{-1}\ sr^{-1}}
\end{equation}
\noindent
with 90\% of the E$^{-2}$ signal found between the energies of 2 $\times$ 10$^{5}$ GeV 
and 10$^{9}$ GeV. This is the most stringent limit at these energy ranges to date (Fig. 
\ref{fig-rep}). A number of neutrino flux predictions are eliminated at the 90\% 
confidence level (see Table \ref{tbl-mrf}).
\begin{figure}[ht]
\begin{center}
\includegraphics [width=0.48\textwidth]{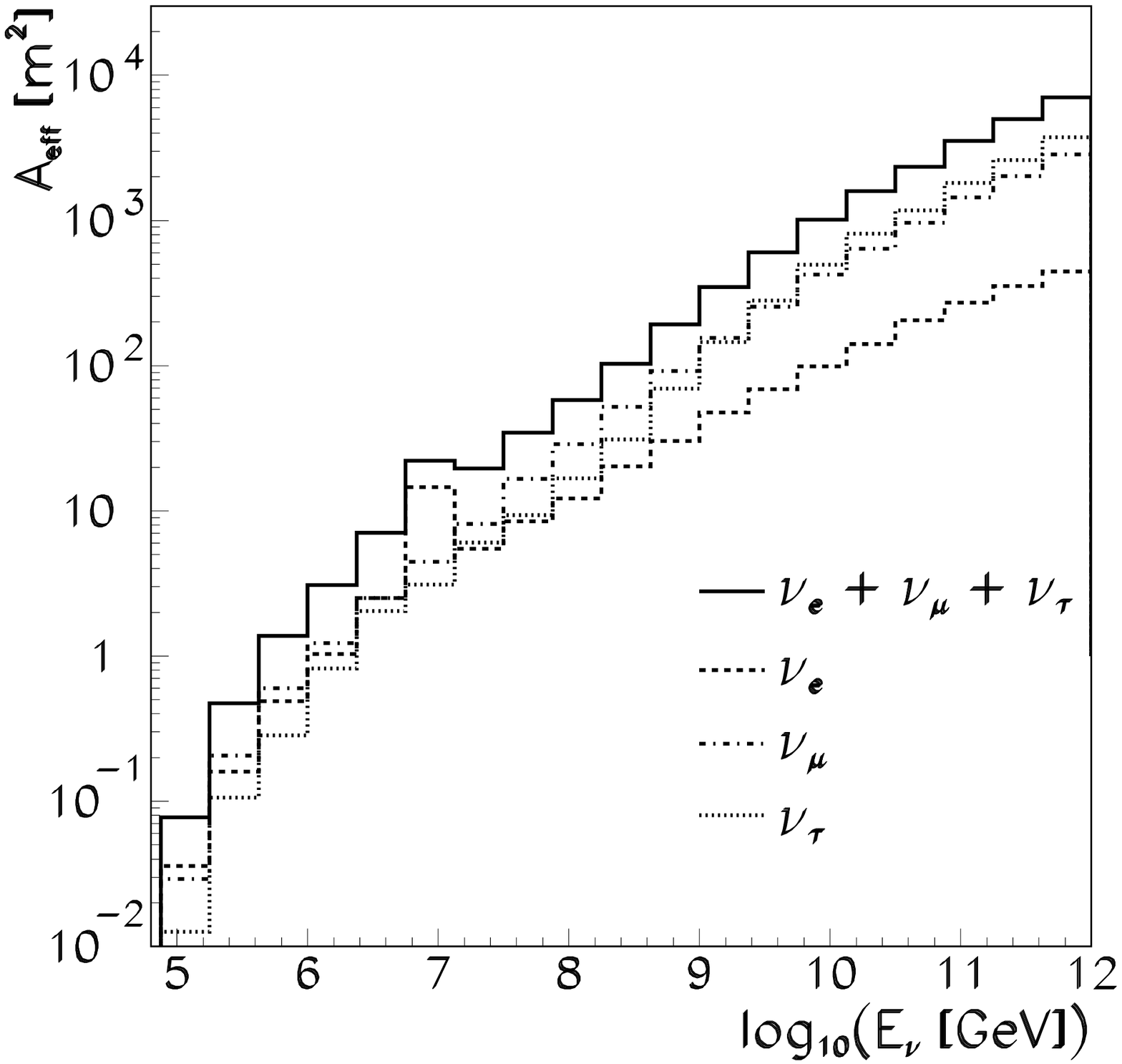}
\caption{Angle-averaged neutrino effective area for 2000 - 2002 after application of all 
selection criteria. The peak at $\sim$10$^{7}$ GeV in the $\nu_{e}$ effective area is 
due to the Glashow resonance.}\label{fig-area}
\end{center}
\end{figure}

\section{Future Prospects}
AMANDA-II hardware upgrades which were completed in 2003 should lead to an 
improvement of the sensitivity at ultra-high energies \cite{sil05}.
AMANDA-II is now surrounded by the next-generation IceCube detector which is currently 
under construction. The sensitivity to UHE neutrinos will further increase as the IceCube 
detector approaches its final size of 1 km$^{3}$ \cite{ahr04}.

\section{Acknowledgements}
L. Gerhardt recognizes the support of of the University of California, Irvine MPC 
Computational Cluster and Achievement Rewards for College Scientists.
\begin{table} [ht]
\begin{center}
\caption{Flux models, the number of neutrinos of all flavors expected at the Earth at the 
final selection level and the preliminary MRFs for 456.8 days of livetime. A MRF of less 
than one indicates that the model is excluded with 90\% confidence.\label{tbl-mrf}}
\begin{tabular}{l|r|r}
Model & $\nu_{all}$ & MRF \\
\hline
AGN \cite{P96} &20.6 &0.23\\
AGN \cite{St92} &17.4 &0.27\\
AGN \cite{hal97} &8.8 &0.54\\
AGN \cite{mpr00} &5.9 &0.80\\
AGN RL B \cite{man95} &4.5 &1.05\\
Z-Burst \cite{kal02a} &2.0 &2.37\\
AGN \cite{St05} &1.8 &2.63\\
GZK $\nu$ norm AGASA \cite{ahl05} &1.8 &2.63\\
GZK $\nu$ mono-energetic \cite{kal02b} &1.2 &3.95\\
GZK $\nu$ a$=$2 \cite{kal02b} &1.1 &4.31\\
GZK $\nu$ norm HiRes \cite{ahl05} &1.0 &4.74\\
TD \cite{Sigl98} &0.9 &5.27\\
AGN RL A \cite{man95} &0.3 &15.8\\
Z-Burst \cite{Yosh98} &0.1 &57.4\\
GZK $\nu$ \cite{Engel01} &0.06 &79.0\\
\end{tabular}
\end{center}
\end{table}
%This is the reference to .bib file (Whitout .bib!)
%\bibliography{ICRC1062/icrc1062}
%This in the bibtex style, is ok.
%\bibliographystyle{unsrt}

%\end{document} % Lisa
\setcounter{figure}{0}
\setcounter{table}{0}
%_________________________________________________document class
%\documentclass{article}
%\newcommand{\diffunit}{$\mathrm{GeV\;cm^{-2}\;s^{-1}\;sr^{-1}}$}
%\newcommand{\pointunit}{$\mathrm{TeV\;cm^{-2}\;s^{-1}}$}
%\newcommand{\dNdE}{E^{2}_{\nu} \times dN_{\nu}/dE_{\nu}}
%\newcommand{\Nch}{$N_{\mathrm{ch}}\;$}
%\newcommand{\ea}{{\it et al} }
%\newcommand{\ic}{IceCube}
%\newcommand{\esqdnde}{$\mathrm{E^{2}_{\nu} \times dN_{\nu}/dE_{\nu}}$}
%\newcommand{\puneicrc}{2005 Proc. 29th Int. Cosmic Ray Conf., Pune}
%\newcommand{\ar}{Ahrens J {\it et al} }
%\newcommand{\am}{Ackermann M {\it et al} }
%\newcommand{\ab}{Achterberg A {\it et al} }
%\usepackage{icrctc07}
%\linenumbers
%_______________________________________________________________________________packages
%\usepackage{epsfig,graphicx, amssymb,amsmath,times,color}
%\setcounter{page}{1}
%
%_______________________________________________________________________________defines
%\def\thefootnote{\fnsymbol{footnote}}
%\begin{document}
%
%_______________________________________________________________________________document begin
%_______________________________________________________________________________title, author etc
\title{Likelihood deconvolution of diffuse prompt and extra-terrestrial neutrino
 fluxes in the AMANDA-II
detector }
\shorttitle{Likelihood deconvolution of diffuse prompt and extra-terrestrial neutrino
 fluxes}
\authors{Gary C. Hill for the IceCube Collaboration}
\shortauthors{Gary C. Hill for the IceCube Collaboration}
\afiliations{
 Department of Physics, University of Wisconsin, Madison, WI 53706, U.S.A.}
%\presenter{Presenter: J. Kelley (jkelley@icecube.wisc.edu)}
\email{ghill@icecube.wisc.edu}

%____________________________________________________________________________________________________________________________________________________
\abstract{
The AMANDA-II detector at the South Pole station, Antarctica, has been used in several
searches for a flux of extra-terrestrial neutrinos from the sum of all sources in the
universe. These searches are complicated by uncertainties in the expected fluxes
of background neutrinos, both those from cosmic-ray  pion and kaon meson production 
(conventional atmospheric neutrinos) and those from charm-containing mesons (prompt
atmospheric neutrinos). In this work, we explore the use of a full likelihood 
analysis on flux sensitive distributions in order to account for the uncertainties and
place simultaneous constraints on the fluxes of interest. 
The method is illustrated using simulated data sets, with application to the real AMANDA-II data
to come. 
}
%\begin{document}
\maketitle
\section{Introduction}
The search for an extra-terrestrial diffuse flux
is one of the most challenging tasks of a neutrino
detector. In contrast to a point source search, where backgrounds are
measured from off-source data, a diffuse search requires a good understanding 
and prediction of the expected backgrounds.
In the case of a diffuse neutrino search, the backgrounds are atmospheric
neutrinos. There are two components to this flux, one thought to be 
well understood, and another less certain. The conventional atmospheric
neutrinos\cite{ICRC1104_bartol2004,ICRC1104_honda2004} are due to decay of pions and kaons produced by cosmic
radiation interacting with the earth's atmosphere. Prompt atmospheric
neutrinos\cite{martin_gbw,naumov_rqpm_a,naumov_rqpm_b,prompt_lepton_cookbook,zhv_charm},
 from the production and decay of mesons containing charm quarks,
 have never been identified and predictions of this flux span orders of magnitude. 
The prompt component should follow the spectral index of the primary cosmic rays, whilst
the conventional component has a spectrum about one power steeper. The expected
flux of extra-terrestrial neutrinos from, for example, the sum of all active 
galaxies in the universe, is expected to have a harder spectrum $(\sim E^{-2})$ than either of the
atmospheric neutrino components. The low expected event rates and similarity of the spectra
of prompt and extra-terrestrial neutrinos will make their independent identification 
difficult\cite{nusim}. The AMANDA-II detector data from the years 2000-03 have been
searched for prompt and extra-terrestrial components\cite{ICRC1104_hodges-diffuse,HHH}. Spectral
differences in the neutrino fluxes would manifest themselves in
different expected energy distributions of detected events in the AMANDA-II neutrino
detector. The  number of optical modules ($N_{\rm ch}$)
 registering at least one photon was
used as an energy estimator. A diffuse extra-terrestrial signal would appear as an excess
of events at higher values of the $N_{\rm ch}$ parameter. In order not to bias the analysis,
a blind analysis, and a simulation based unbiased optimum limit setting technique were
used to choose the best cut appropriate for each signal spectrum. The atmospheric neutrino 
background simulation was normalised to observed data below the cut in order to constrain some
of the uncertainties.
The prompt neutrinos were treated in two ways, firstly, they were included as a background for
the extra-terrestrial searches, and secondly, they were treated as an unknown signal, to be constrained
by the observed data.  The final limit on an $E^{-2}$ flux was set at a level of 
\esqdnde $= 7.4\times10^{-8}$ \diffunit, valid over an energy range  16-2500 TeV. This is the best limit
to date on extra-terrestrial neutrino fluxes. 
Despite this success, the cut and count method does suffer from some drawbacks. Primarily, the shape of
the $N_{\rm ch}$ distribution is not used in the analysis, only the integrated number of events above the
cut value. A likelihood analysis can be used to take advantage of the full shape of the $N_{\rm ch}$ distribution.
In addition, such an analysis can simultaneously constrain all the parameters, both those
of direct interest (the numbers of prompt and extraterrestrial neutrinos) and those of
indirect interest - known as ``nuisance parameters'' (normalisation and shape of the conventional atmospheric neutrinos).
Another key point is that if an entire distribution is used in a likelihood analysis, then there is no need to
optimise a selection cut on that parameter, removing discussion of what is the optimal cut criterion.
These likelihood methods with nuisance parameters are  standard for neutrino oscillation
analyses\cite{KelleyAhrens}, and for ``unbinned'' astrophysical point source searches\cite{Till,JB,CF}.

\section{Methodology}

The likelihood function in this analysis is the product over a binned version of the
 $N_{\rm ch}$ distribution of the bin-by-bin Poisson probabilities of events observed given
events expected.
\begin{equation}
      P(\left\{n_i\right\} \mid \left\{\mu_i\right\}) =
     \Pi_{i} \frac{(\mu_i)^{n_i}}{n_i!} \exp(-\mu_i) + \frac{\Delta\epsilon^2}{\sigma_\epsilon}
\end{equation}
For each bin, the expectation $\mu_i$ is the sum of conventional and prompt atmospheric
neutrinos, and extra-terrestrial neutrinos
\begin{equation}
    \mu_i = \epsilon(A_c \mu_{ci}(\Delta\gamma) + A_p \mu_{pi} + A_e \mu_{ei})
\end{equation}
where subscripts c, p and e stand for conventional, prompt and 
extra-terrestrial neutrino fluxes respectively. 
 As an example, the term $A_e \mu_{ei}$
 is the number of events expected in bin $i$ after convolving an extra-terrestrial
flux, normalised to a total of $A_e$ events, with the effective area of the 
detector (which includes absorption effects in the earth).
 The parameter
$\Delta\gamma$ of the conventional atmospheric flux allows for changes in the
spectral shape relative to the prediction. Full calculations\cite{ICRC1104_bartol2004,ICRC1104_honda2004}
 of the angular and spectral dependence of the flux have been made, here we allow for 
deviations away from the exact form $\Phi_0(E,\theta)$ by using $\Phi(E,\theta) = \Phi_0(E,\theta) 
E^{\Delta\gamma}$.
  Since the spectrum only
approximately follows a power-law (and this varies with angle) we choose to fit
for deviations away from the actual spectrum, rather than fit for a simple power
law $\gamma$.
 Fitting for $\Delta\gamma$ would allow statements to made such as ``the data 
favour a similar/harder/steeper spectral form than that calculated theoretically,'' rather than
simply fitting for a single value of $\gamma$.
 The parameter $\epsilon$ is
an
efficiency term reflecting
 uncertainties in the effective area 
of the detector. While this is strictly energy- and thus bin-dependent, 
with strong bin-to-bin correlations, here we simplify to a constant
form for this initial illustration of the method. 
 Epsilon is constrained to a Gaussian
form with width $\sigma_\epsilon$ by the penalty term in the likelihood
function, with $\Delta\epsilon$ being the difference between the tested
value of the efficiency, $\epsilon$, and the notional best fit value for
the efficiency, $\epsilon_0 = 1$. 
To test a given hypothesis, e.g. that $A_p = 20.0$ and $A_e = 10.0$, the
likelihood is maximised, fixing $A_p$ and $A_e$ to the desired values and
allowing $\epsilon$, $A_c$ and $\Delta\gamma$ to float. This likelihood,
denoted $\mathcal{L}$, is then compared to the likelihood $\hat{\mathcal{{L}}}$
 where all parameters are free
to float in the fit.  The tested hypothesis is then rejected at a 
confidence level set by the probability of observing a greater likelihood
ratio, given the truth of the null hypothesis $A_p$ and $A_e$, than the 
specific one that was observed.
The distribution of the likelihood ratio  statistic under the null hypothesis
 is known approximately from 
Wilks' theorem. Asymptotically, the likelihood ratio defined by $-2\log\mathcal{L}/\hat{\mathcal{{L}}}$
follows a chi-square distribution with degrees of freedom equal to the number of
fixed parameters in the $\mathcal{L}$ fit. The confidence
level at which the hypothesis is then rejected is found from checking the ratio
$-2\log\mathcal{L}/\hat{\mathcal{{L}}}$ against the appropriate chi-square value (e.g. a
90\% c.l. corresponds to a chi-square of 4.6 for two degrees of freedom).
In order to compute the exact confidence level for each null hypothesis, the likelihood
ratio may be compared to its expected distribution,  generated from many random event distributions
drawn from the null hypothesis\cite{feldcous}. In this paper, we use the chi-square approximation
for simplicity, leaving the full interval constuction for final analysis.

Having written down the form of the likelihood function, the details of the
components must be determined. Here, we take the shape of the conventional
atmospheric neutrino detector response, $\mu_c(\Delta\gamma)$ as the 
convolution of  the Bartol flux\cite{ICRC1104_bartol2004}, with the detector effective area, multiplied
by the factor $E^{\Delta\gamma}$. There are two
primary sources of uncertainty in the prediction of the atmospheric neutrino
flux - the cosmic ray primary spectrum and the interaction model. Together, these manifest
themselves as overall uncertainties in the normalisation (fitted by $A_c$), and as an
increasing uncertainty in the flux as a function of energy (see figure 12 of \cite{ICRC1104_hodges-diffuse}).
This energy dependent uncertainty can be approximately parameterised as a change of slope in the neutrino spectrum.
 The prompt 
flux is the ``Charm D'' model\cite{zhv_charm}, an older prediction, but with a spectral shape
similar to more recent preditions. The extra-terrestrial flux follows an $E^{-2}$ power law.
The value of the effective area uncertainty, $\sigma_\epsilon$, is taken as 10\%, effectively
bounding (95\% region) it to extrema of plus/minus  20\%. 

\section{Example fitting of a test data set}
To demonstrate the power of the likelihood method, we derive a random test data set by 
sampling 450 events from the Bartol $N_{\rm ch}$ distribution. These event are then 
treated as though they are the real data set. Figure \ref{fitted} shows the result of
the fitting procedure, where the data set is best fit by 446.5 atmospheric neutrinos and
3.6 extra-terrestrial neutrinos. The normalisation and $\Delta\gamma$ of the atmospheric
neutrinos, and the effective area parameter $\epsilon$, were allowed to float during this fit. 
The potential to constrain the atmospheric neutrino parameters is shown in figure \ref{acc-atmos}, 
where an acceptance region was found while allowing the effective area uncertainty to float. 
The size of this experimentally determined allowed region
is similar to the theoretical uncertainties of flux. This simple $N_{\rm ch}$ 
fitting procedure is not powerful enough to constrain the theory with only AMANDA-II.  However, 
with increased exposure (more AMANDA-II data and the 
larger IceCube detector) the experimental observations will begin to constrain the theory, allowing for
proper measurements of the flux. 
In figure \ref{acc-s-c}  the allowed regions for prompt and extra-terrestrial fluxes are shown. Since there
is only background in the test data set,  the allowed region includes the background only corner of the
plane. The upper bounds of the allowed regions define combinations of allowed amounts of the two 
components. The 90\% confidence level
count on the extra-terrestrial axis (25 events) corresponds to a flux
level of \esqdnde $= 1.2\times10^{-7}$ \diffunit. Since this result is just for one specific test data set, 
a meaningful comparison to the standard analysis\cite{ICRC1104_hodges-diffuse} cannot be made, without determining a
sensitivity over many repeated random experiments. 
It is expected that the likelihood method
will lead to an improvement in the sensitivity.
  The actual predicted  level of the CharmD prompt flux
corresponds to 8 events in this acceptance region.

\section{Future work}

To properly estimate the sensitivity and discovery potential, many test sets, drawn from mixtures of
backgrounds and signals must be processed and the acceptance regions combined. This will be done
using the median likelihood ratios at each point in the plane.
% For example, suppose we test an 
%ensemble of background-only events. Then for any point in the acceptance plane there will be a 
%distribution of likelihood ratio values.  The medians across the plane then define the sensitivity
%contours of the search, describing the confidence level at which  each point would be excluded  in 
%more than 50\% of cases. To characterise discovery, one would generate event sets containing  
%various combinations of signals, then for each combination see what the median likelihood ratio for the
%background only acceptance point is. If the median were 11.8, then that combination of signal
%rates would result in a 3 sigma discovery (of something) about 50\% of the time.
 Required signal combinations for definite
 discovery of
either or both of the signal fluxes could also be determined.

The nature of the parameterisations of the fluxes can be  further developed and 
improved. In principle, the atmospheric 
neutrinos could be parameterised in ways more directly connected to the physics of the cosmic ray fluxes
 and interaction
models, for instance to fit for the charm production cross-sections, and to allow for the charm
spectral index to float. Instead of using an $E^{-2}$ extra-terrestrial
spectrum, the spectral index of this additional component flux could be a fit parameter. The uncertainties on 
the detector response could be treated in a proper  bin-to-bin correlated  manner.

%The sensitivity of this procedure, using the median likelihood ratio method over many repeated
%tests, is shown in figure \ref{atmosfit}. We sampled and fit many repeated experiments, drawn from the 
%Bartol flux.  As expected, the procedure fits to the best fit values of an event rate of 459 events, and
%a spectral slope change of zero. Various confidence level bounds on the parameters are shown, corresponding
%to $1\sigma$, 90\%, $2\sigma$, and $3\sigma$. The sensitivity to the parameters of the atmospheric
%neutrino flux is similar to that dervied from theoretical accounting of the uncertainties. 

\section{Conclusions}
A likelihood ratio fitting method, incorporating nuisance parameters, has been developed for application to 
a neutrino search with the AMANDA-II detector. This method allows for the simultaneous constraint of background
and signal flux parameters. The use of an entire distribution in the analysis removes the need for 
optimisation of a selection cut, and allows all the available information to be incorporated into the 
confidence interval construction.

\begin{figure}[t!]
\begin{center}
\includegraphics[width=7cm]{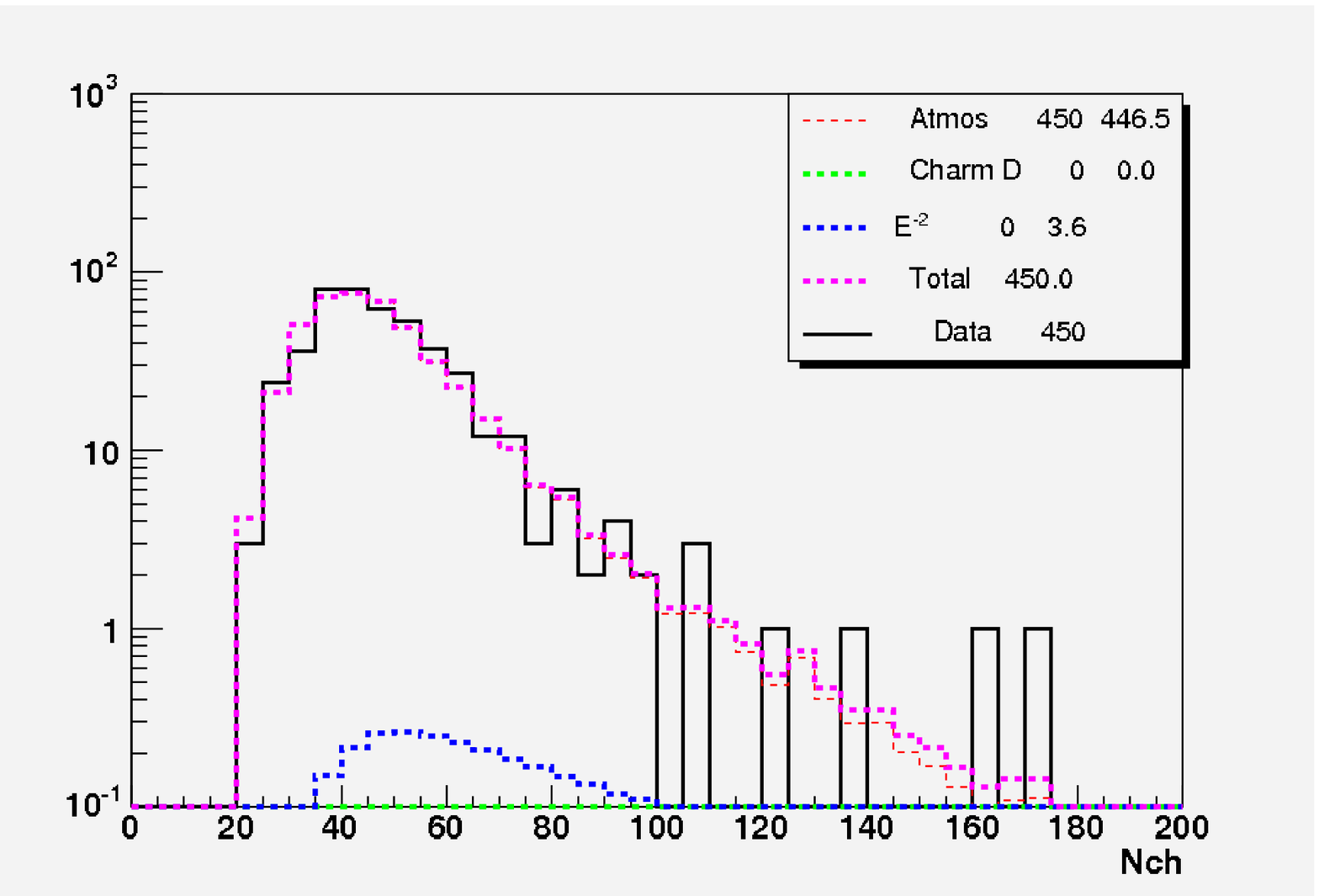}
\caption{ \label{fitted}
Fitting of a test data set with the likelihood procedure. The data set,
drawn from the Bartol atmospheric neutrino distribution, is best fit by
a near pure atmospheric neutrino contribution, plus 3.6 extra-terrestrial
events. The allowed regions for the additional components are shown in
figure \ref{acc-s-c}.}
\end{center}
\end{figure}

\begin{figure}[t!]
\begin{center}
\includegraphics[width=7cm]{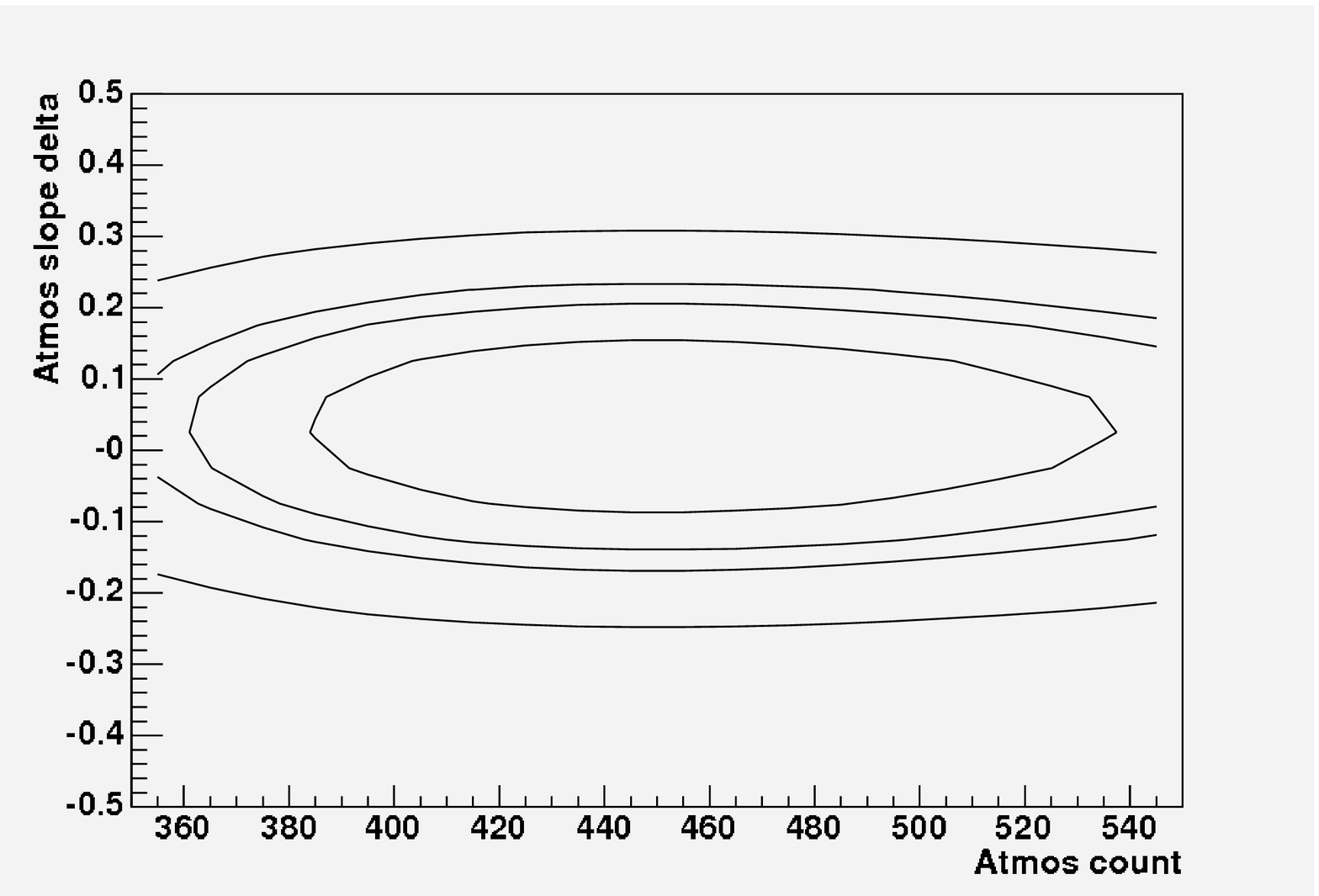}
\caption{ \label{acc-atmos}
Test data set allowed regions of the atmospheric neutrino total event count, and 
spectral slope difference $\Delta\gamma$. The confidence level contours
correspond to one-sigma, 90\%, two and three sigma, moving outward from the 
best fit point.
}
\end{center}
\end{figure}

\begin{figure}[t!]
\begin{center}
\includegraphics[width=7cm]{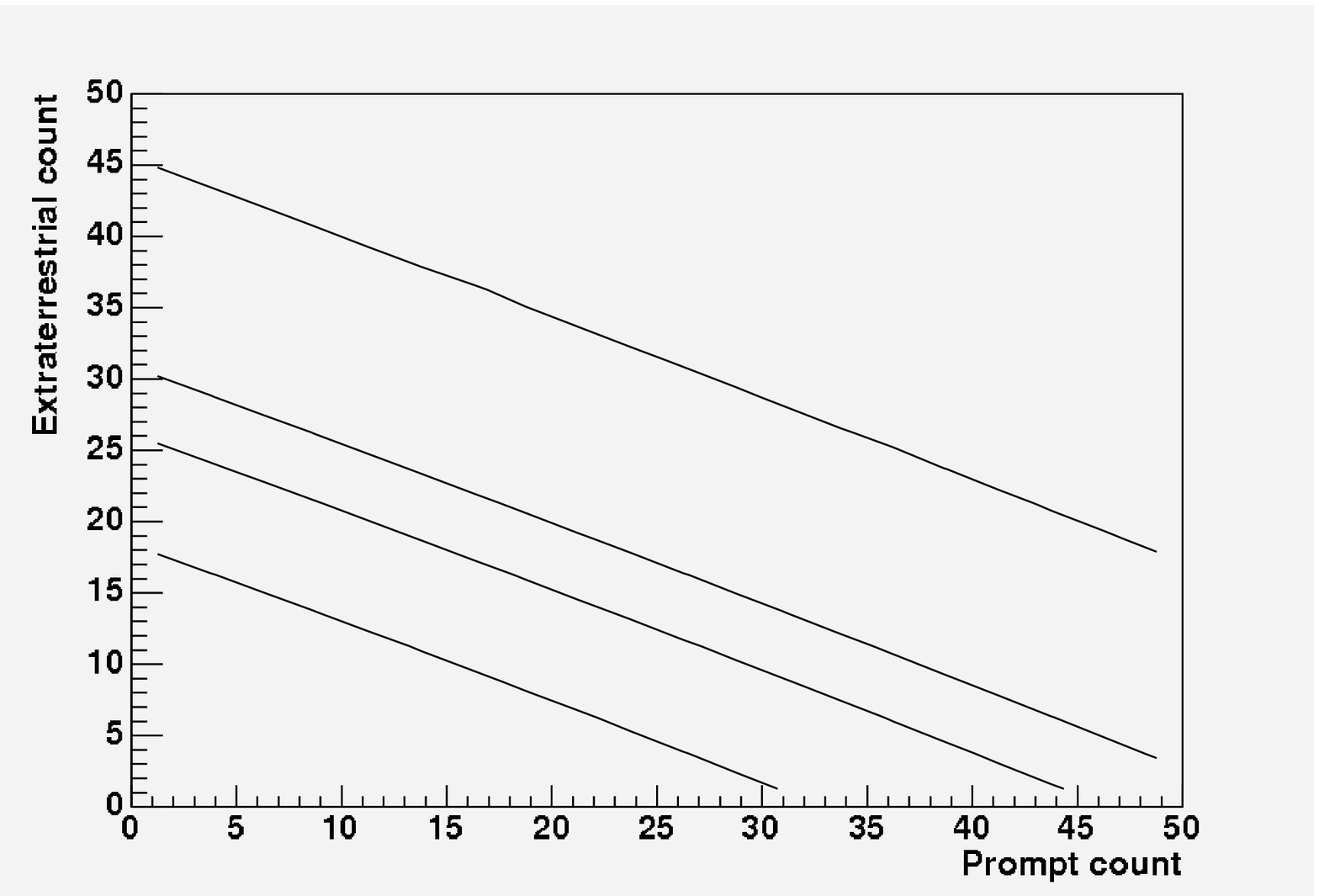}
\caption{ \label{acc-s-c}
Allowed regions of the prompt and extra-terrestrial neutrino 
contributions for the test data set, allowing the atmospheric neutrino and detector
effective area parameters to float. The 90\% confidence level
count on the extra-terrestrial axis (25 events) corresponds to a flux 
level of \esqdnde $= 1.2\times10^{-7}$ \diffunit. 
 }
\end{center}
\end{figure}

%____________________________________________________________________________________________________________________________________________________
%____________________________________________________________________________________________________________________________________________________

%\end{document}
 % Gary
%
%
%cascade papers
%
%cascades_icrc..  (Oxana)
%icrc_ehe0.fin  (Aya)
%LPM_0843  (LPM effect)
%icrc0761.pdf  (light sources)
%
\setcounter{figure}{0}
\setcounter{table}{0}
%%
% International Cosmic Ray Conference 2007 Merida Yucatan Mexico
% In This file you will find detailed instructions to correctly
% typeset your document.
%
%
%

%Class Requeried
%\documentclass{article}
%The ICRC Style
%\usepackage{icrctc07}

%The paper title
\title{Search for Neutrino-Induced Cascades with AMANDA data taken in 2000-2004}
%Short title to print in the headers to the final publication (Not showed in this print).
\shorttitle{Search for Neutrino-Induced Cascades with AMANDA data taken in 2000-2004}
%All paper authors
\authors{O.\,Tarasova$^a$, M.\,Kowalski$^b$, M.\,Walter$^a$ 
for the IceCube collaboration$^c$}
%Short title to print in the headers to the final puplication (Not showed in this print).
\shortauthors{O.Tarasova and M.Kowalski and M.Walter  and et al}
%All the affiliations.
\afiliations{$^a$DESY, D-15735, Zeuthen, Germany \\ $^b$Institut f\"ur Physik, Humboldt Universit\"at zu Berlin, D-12489 Berlin, Germany \\ $^c$see special section of these proceedings}
\email{tarasova@ifh.de,marek.kowalski@physik.hu-berlin.de,walter@ifh.de}

%The abstract.
\abstract{The Antarctic Muon And Neutrino Detector Array (AMANDA) is a Cherenkov
detector deployed in the Antarctic ice cap at the South Pole \cite{ICRC1124_ref1}.
The charged-current interaction of high-energy electron or tau neutrinos,
as well as neutral-current interactions of neutrinos of any flavor, can
produce isolated electromagnetic or hadronic cascades.
There are several advantages associated with the cascade channel in the
search for a "diffuse" flux of astrophysical neutrinos.
The energy resolution of AMANDA allows us to distinguish
between a hard astrophysical spectrum and a soft atmospheric spectrum.
In addition, the flux of atmospheric electron neutrinos is lower by an
order of magnitude relative to atmospheric muon neutrinos, while the
background from downward-going atmospheric muons can be suppressed due
to their track-like topology.
The low background in this channel allows us to attain $4\pi$
acceptance above energies of $\sim50$\,TeV.
We present the analysis of AMANDA data collected during 2000-2004.
Compared to our previous analysis, this data
set is a factor of five larger, resulting in a correspondingly improved
sensitivity for the flux of astrophysical neutrinos.}

%%%%%%%%%%%%%%%%%%%% B E G I N   D O C U M E N T%%%%%%%%%%%%%%%%%%%%%%%
%\begin{document}
\maketitle
%Begin the section.

\section{Introduction}
There are several theoretical predictions that cosmic neutrinos are produced by accelerated protons within high-energy astrophysical objects such as Active Galactic Nuclei (AGN) and Gamma Ray Bursts (GRB). Neutrinos can propagate in straight lines through the universe as they are not effected by magnetic fields of the galaxy and essentially do not interact with particles on the way to the earth. They are expected to be produced in the source with a ratio $\nu_{e}:\nu_{\mu}:\nu_{\tau}\sim 1 : 2 : 0$ but due to flavor-mixing during propagation a $1 : 1 : 1$ ratio is expected at the detector. However, due to the very small cross-section neutrinos are also difficult to detect. In order to perform a search for galactic and extragalactic neutrinos, the AMANDA telescope was installed in the antarctic ice cap at the geographical South Pole and has been operating since 2000. It consists of 677 optical modules (OM) which are attached to 19 strings and buried at depths from 1500\,m to 2000\,m under the ice surface. Each optical module contains a photomultiplier suited to register Cherenkov light emitted by a charged particle which is produced in the neutrino interaction. The signature of a charged-current interaction of $\nu_{e}$ and $\nu_{\tau}$ is an electromagnetic and a mainly lower energetic hadronic cascade. Via neutral-current interaction, neutrinos of any flavor can produce isolated hadronic showers. This analysis is focused on a search for neutrinos from unresolved sources (diffuse flux) which have a cascade-like signature in the AMANDA detector. The muon-like events are the main background for this analysis. In the cascade channel the direction of the incoming neutrino is poorly reconstructed, however, the energy resolution of the detector for cascade reconstruction is $\cal{O}$$(\rm log(E_{\nu}))=0.18$. By removing track-like events, one can eliminate most of the background from atmospheric muons. In addition, the flux of atmospheric electron neutrinos is much lower than the flux of muon neutrinos.  

\section{Experimental data and MC simulation}
The experimental data used in this analysis were collected between 2000 and 2004.
After excluding bad and unstable runs from the analysis we end up with a lifetime of 1000.1 days, where in total $8.8\times10^{9}$ triggered events were recorded. The main contribution are muons from meson decays in the atmosphere.

The atmospheric muon background was simulated with CORSIKA \cite{ICRC1124_ref2}. 
To reach large statistics for the high energy part of the background spectrum with acceptable computing time, about 5000 days of downgoing atmospheric muons were generated with energies above 5\,TeV. For comparison a smaller sample of standard CORSIKA events was produced.

The cascades were simulated with ANIS \cite{anis} generating all three neutrino flavors ($\nu_{e}, \nu_{\tau}$ and $\nu_{\mu}$) at energies between 100\,GeV and 100\,PeV assuming an $E^{-1}$ energy spectrum. The resulting muons were further propagated  using MMC (see \cite{ICRC1124_mmc} for details). The signal spectrum was reweighted afterwards to a hypothetical $E^{-2}$ flux of $\nu_{e}$. Atmospheric $\nu$ were simulated by reweighting the same neutrino events to a steeper $\sim E^{-3.7}$ spectrum \cite{atm}.  

\section{Analysis Optimization}

The analysis consists of several filter levels including reconstruction of the cascade vertex and energy as well as a few quality cuts to select high quality events. The reconstruction algorithms based on the likelihood minimization method are described in \cite{mmc,reco}.  The vertex resolution of cascade-like events is about 4\,m. Quality cuts were performed using the likelihood values $L_{\rm vertex}$ and $L_{{\rm energy}}$, given by reconstruction algorithms.  

\begin{figure}[ht]
\begin{center}
\includegraphics[width=0.5\textwidth]{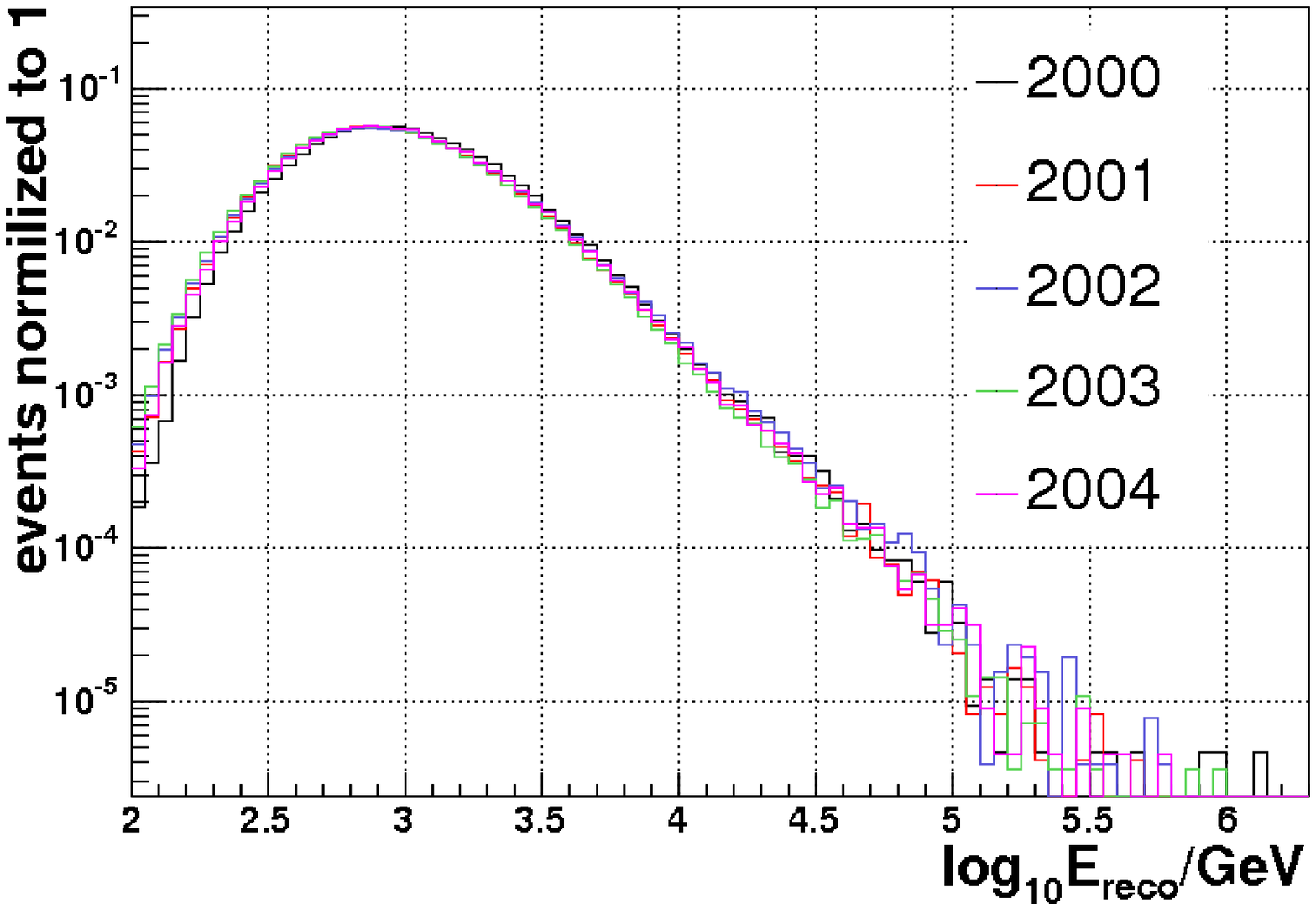} 
\caption{The reconstructed energy distribution of cascade candidate events for the five years (2000-2004) used in this analysis. }
\end{center}
\end{figure}

In order to reduce events with a mis-reconstructed vertex, the cut on the vertex likelihood function has been applied, $L_{\rm vertex}<7.1$. The cut on the energy likelihood $L_{{\rm energy}} $ was performed as
function of the reconstructed energy.  Another energy-dependent cut was applied on the radial distance of the reconstructed vertex position $\rho_{xy}$. For  ${E_{\rm reco} <1.25}$\,TeV this cut was set such that only events with the reconstructed vertex position within a 100\,m radius from the detector center (fiducial volume of AMANDA) were used in the analysis. Taking into account that the higher energetic events are often reconstructed at distances outside of the detector and the fact that the anticipated background is rather small for these energies, we allow an energy dependent increase of the volume above ${E_{\rm reco} >1.25}$\,TeV.      

By this filter, the set of experimental data is reduced by a factor of $10^{5}$.
Fig.\,1 shows the reconstructed energy of cascade candidates for the different years. Small variations arise from slightly different hardware configuration for different periods.

At the final filter level, two additional cuts were  performed and optimized for the analysis. In addition to a cut on $E_{\rm reco}$ we
introduced a discriminating parameter $Q_{\rm s}$ that involves the following set of three variables: 

\begin{itemize}
\item vertex likelihood value $L_{\rm vertex}$,
\item ${\rm cos(\theta_{\mu})}$ taken from muon track likelihood reconstruction,
\item radial distance, $\rho_{xy}^{60}$, between the vertex position of two likelihood vertex fits; the second fit is thereby not using hits  
within a 60\,m sphere around the vertex position determined by the first fit.   
\end{itemize}

\begin{figure*}[th]
\begin{center}
\includegraphics[width=0.8\textwidth]{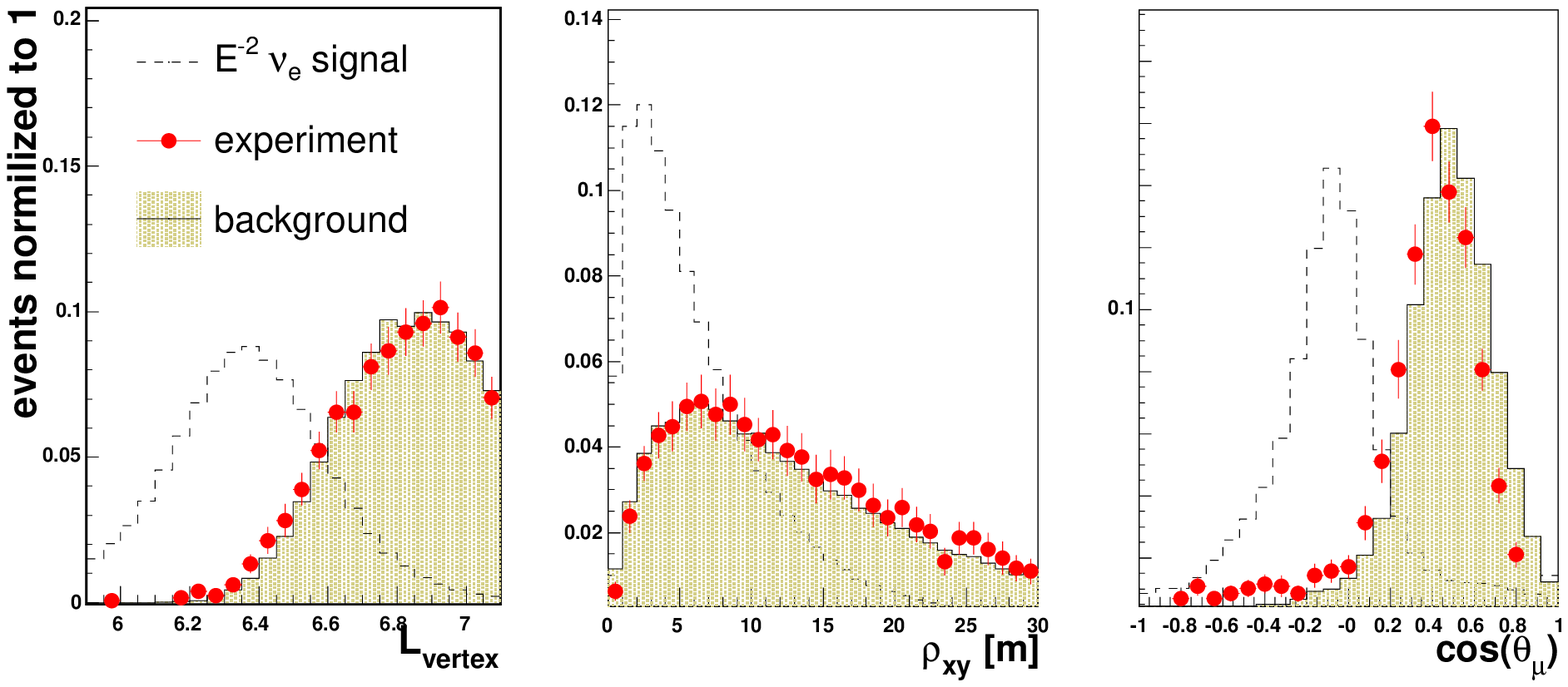} 
\caption{Distributions of the three variables used to construct the discriminating parameter $Q_{\rm s}$ for the experimental data, the background and the signal MC. Left: vertex likelihood distribution for signal and background. Middle: $\rho_{xy}^{60}$ distributions (see text for details). Right: ${\rm cos(\theta_{\mu})}$ distribution taken from the iterative muon likelihood reconstruction. }
\end{center}
\end{figure*}

The method to construct the discriminating parameter $Q_{\rm s}$ is described in more detail in \cite{Casc2004}. 
The three distributions are shown in Fig. 2 for signal and background Monte Carlo and for experimental data. All distributions for data and background MC are in a good agreement apart some discrepancy in the $\rm cos(\theta_{\mu})$ distribution. The reason for this could be an incorrect simulation of the ice properties and it needs to be taken into account in the systematic error. 
To maintain blindness we used only $20\%$ of the experimental data to perform the final cut optimization. However, the optimization was done assuming the statistics of the full data sample i.e. the data were re-scaled by a factor of five.
In Fig.\,3 one sees the energy spectra for signal and background Monte Carlo and for experimental events which passed through the cascade filter.  Here the background distribution was normalized to the experiment. 

\begin{figure}
\begin{center}
\includegraphics[width=0.5\textwidth,angle=0,clip]{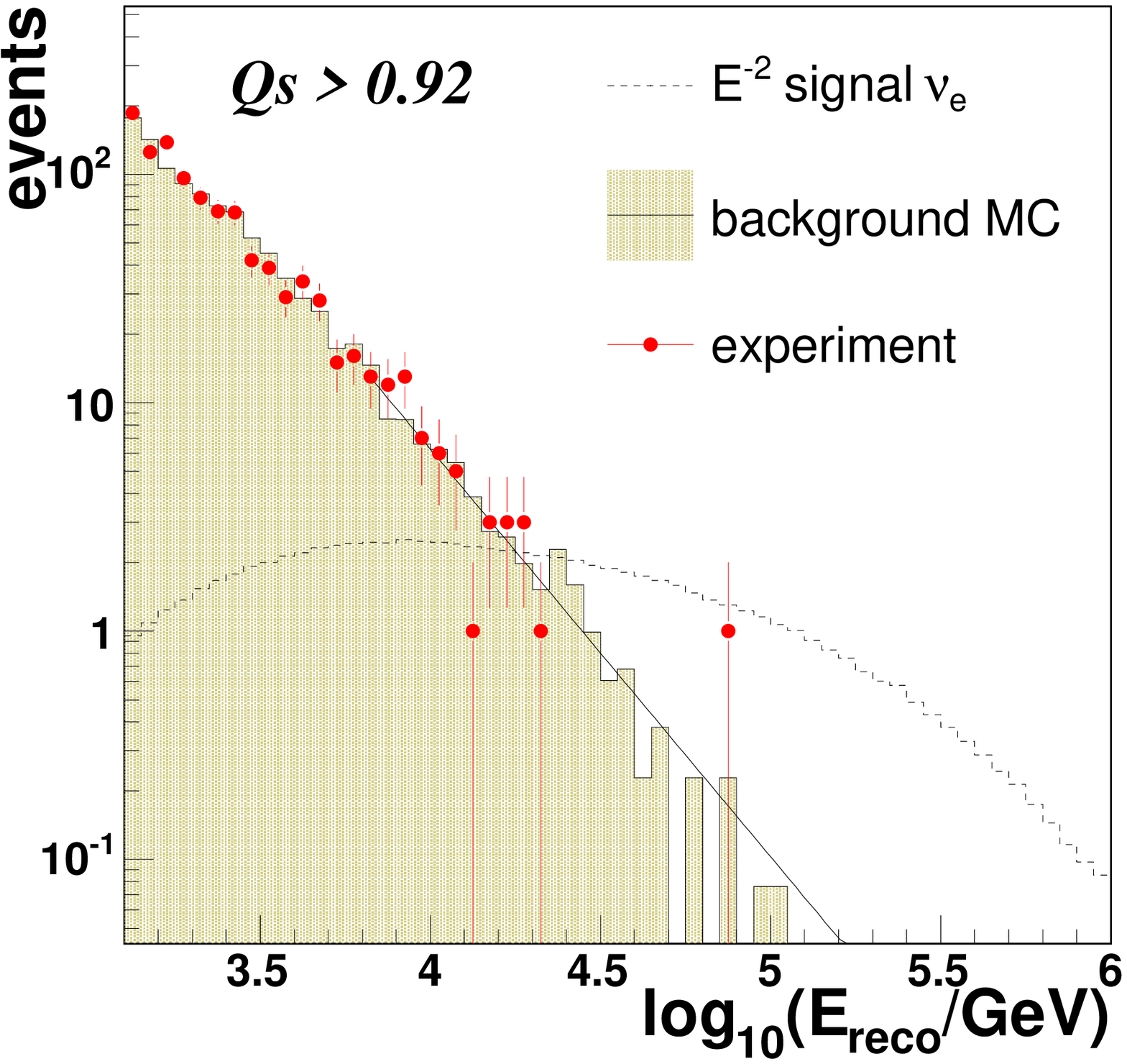}
\caption{The reconstructed cascade energy distribution $E_{\rm reco}$. Shown are experimental data as well as background and signal Monte Carlo simulation after application of all quality cuts and a cut on the discriminating parameter $Q>0.92$. The smooth line is a result of the power-law fit to the background simulation.}
\end{center}
\end{figure}

The final cuts on the reconstructed energy were applied following the optimization method described in \cite{hill}. This cut was performed in order to separate the potential signal from the background. Both cuts $Q_{\rm s}$ and $E_{\rm reco}$ were chosen to result in the highest sensitivity to an astrophysical neutrino flux. The sensitivity is defined here as the average upper limit \cite{FC} which was obtained in an ensemble of identical experiments in absence of the signal. In Fig.\,4, the average upper limit $\bar{\phi}$ is shown as a function of $E_{\rm cut}$ for $Q_{\rm s}>0.92$. This procedure was repeated for a large range of $Q_{s}$ values in order to obtain the optimal discriminating parameter and energy cut. To make a smooth background interpolation possible, the background distribution was fitted with a power-law function (see Fig.\,3). For the discriminating parameter the optimal cut is at $Q_{s} > 0.92$. The energy cut obtained from the optimization is ${\rm log(E)>4.65}$. The corresponding sensitivity on the flux of $\nu_{e}$ is $2.7\times10^{-7} {\rm (E/GeV)^{-2} / (GeV s \ sr \  cm^{2})}$.

There is 1 event from the $20\%$ experimental data subset passing this cut. We expect 1.3 background events from atmospheric muons.  The expectation for the 
atmospheric $\nu_{e}$ and  $\nu_{\mu}$ which passed all cuts is 0.02 events for the $20\%$ sample. 
No systematic uncertainties have been estimated yet, however, the uncertainties in the detector response and in the predictions of the atmospheric muon and neutrino fluxes are expected to be substantial.

\begin{figure}[ht]
\begin{center}
\includegraphics[width=0.48\textwidth,angle=0,clip]{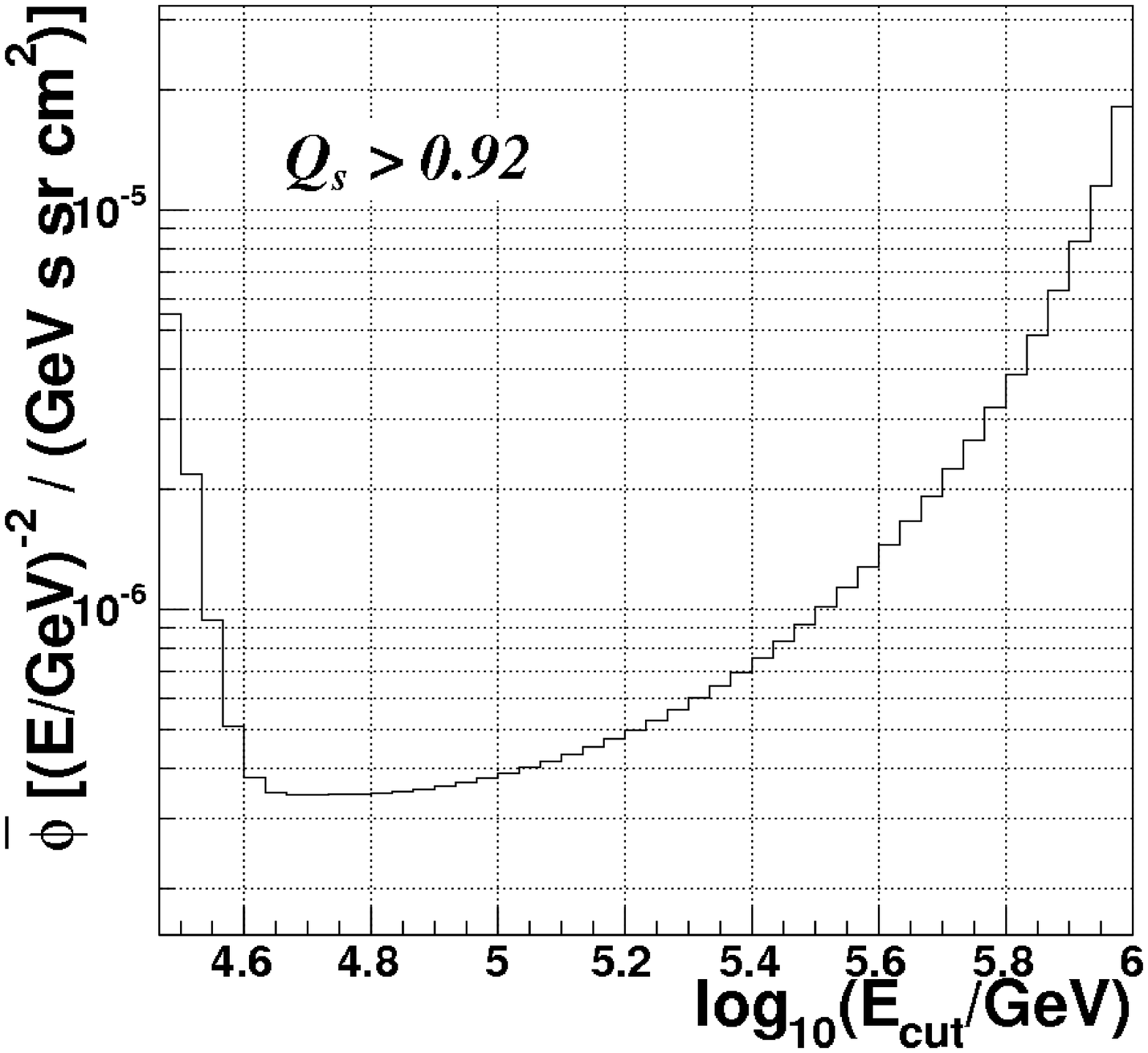} 
\caption{The average upper limit as a function of energy cut. The best  sensitivity is reached for ${\rm log(E_{cut})=4.65}$. }
\end{center} 
\end{figure}

\section{Results}

Analyzing a $20\%$ sub-sample of the 5 years AMANDA data, a search for cascade-like events was performed. The observed events from experimental data are statistically consistent with the background expectation. The expected number of signal events from a diffuse flux assuming a $E^{-2}$ spectra and a strength of $10^{-7} {\rm (E/GeV)^{-2} / (GeV s\  sr\ cm^{2} )}$ is 2.1 $\nu_{e}$ events, leading to a preliminary sensitivity on the $\nu_{e}$ flux of  $2.7\times10^{-7} {\rm (E/GeV)^{-2} / (GeV\ s \ sr \  cm^{2})}$.  
Fig.\,5 shows the effective areas after all selection cuts combined for neutrinos and anti-neutrinos. The effective area for tau neutrino is larger at high energy due to tau regeneration. Anti-electron neutrinos show a large increase in the effective  
area near 6.4 PeV due to the Glashow resonance. 

\begin{figure}[ht]
\begin{center}
\includegraphics*[width=0.48\textwidth,angle=0,clip]{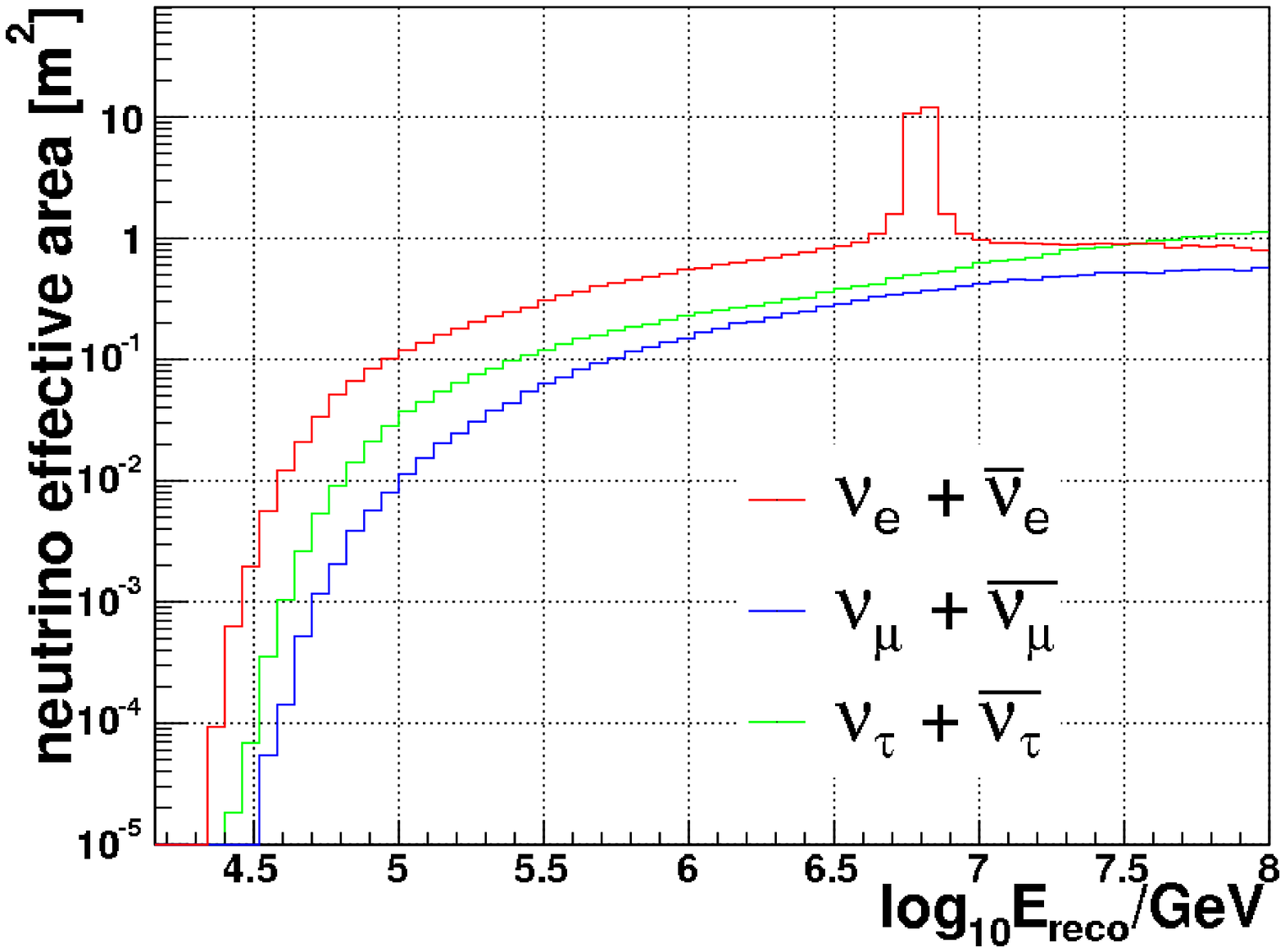}
\caption{The effective neutrino areas for $\nu_{e},\nu_{\mu}$ and $\nu_{\tau}$ are shown as a function of neutrino energy after all selection criteria have been applied.  }
\end{center}
\end{figure}

%This is the reference to .bib file (Whitout .bib!)
%This in the bibtex style, is ok.
%\bibliographystyle{unsrt}
%\bibliography{ICRC1124/libros}

%\end{document}

\setcounter{figure}{0}
\setcounter{table}{0}
\title{EHE Neutrino Search with the IceCube 9 String Array}
\shorttitle{IceCube EHE Neutrino Search}
\authors{Aya Ishihara$^{1}$ for the IceCube Collaboration$^{2}$}
\afiliations{$^{1}$ Department of Physics, Chiba University,\
Chiba 263-8522, Japan\\
$^{2}$ See special section of these proceedings}
\email{aya@hepburn.s.chiba-u.ac.jp}

% The abstract
\abstract{
 The performance of the partially ($\sim$10\%) constructed IceCube neutrino
 detector on the search for extremely high energy (EHE) neutrino in data
 taken in 2006 is presented.
 Background event numbers are estimated based on 
 an empirical model which reasonably describes a part of the same
 experimental sample.
Following this background estimate an upper limit of the neutrino fluxes
at 90\% C.L. would be placed at $E^2 \phi_{\nu_e+\nu_\mu+\nu_\tau}\simeq 1.6\times
10^{-6}$ GeV/cm$^2$ sec sr for neutrinos with an energy of $10^8$ GeV in
the absence of signals in the 2006 sample.
The corresponding neutrino effective area is also presented.
  }
%

%=====================================================
%%%%%%%%%%%%%%%%%%%% B E G I N   D O C U M E N T%%%%%%%%%%%%%%%%%%%%%%%
%\begin{document}
\vspace{-0.4cm}
\maketitle
\vspace{-0.4cm}
%=====================================================
%=====================================================
\vspace{-0.4cm}
\section{\label{sec:intro} Introduction}
\vspace{-0.4cm}
%=====================================================
%=====================================================
%
Extremely high energy (EHE) neutrinos are expected to fill a key role in
connecting the observed EHE cosmic-rays to their birthplaces, which may shed light
on the long standing puzzles of the origin of EHE cosmic-rays. 
Because of their low intensity, the detection of EHE neutrino requires a
huge effective detection volume.
The IceCube neutrino observatory~\cite{IceCubePhysics}, located at the
geographic South Pole, will consist of a km$^3$ fiducial volume of clean
glacier ice as a Cherenkov radiator and an array of photon detectors.  
The initial IceCube 9 string array (\mbox{IC-9}) was deployed by February 2006.
Each string was positioned with a spacing of approximately 125~{\rm m} and with 60 optical
sensors attached to it at intervals of $\sim$17~{\rm m}. 
The \mbox{IC-9} detector was operational from June through November of 2006.
% anddata taken in this period is a subject of the first analysis in search
% for signatures of EHE cosmic neutrinos by the IceCube observatory. 
%
The high energy events sample used in this analysis is a part of the full dataset
taken with \mbox{IC-9} satisfying the condition that a minimum of 80
out of 540 \mbox{IC-9} optical sensors (DOMs) record Cherenkov pulses within
5~{\rm $\mu$sec}. 
The effective livetime corresponding to this dataset is 124 days after rejecting events
taken during times of unstable operation.%while operation were unstable.

We report here for the first time on the expected sensitivity of this
\mbox{IC-9} detector configuration for neutrinos with energies $10^7$ GeV
and above. 
%
%
%=====================================================
\vspace{-0.4cm}
\section{\label{sec:events} EHE events in IceCube}
\vspace{-0.4cm}
%=====================================================
\begin{figure*}[hbt]
\begin{center} 
\includegraphics*[width=0.6\textwidth,angle=0,clip]
{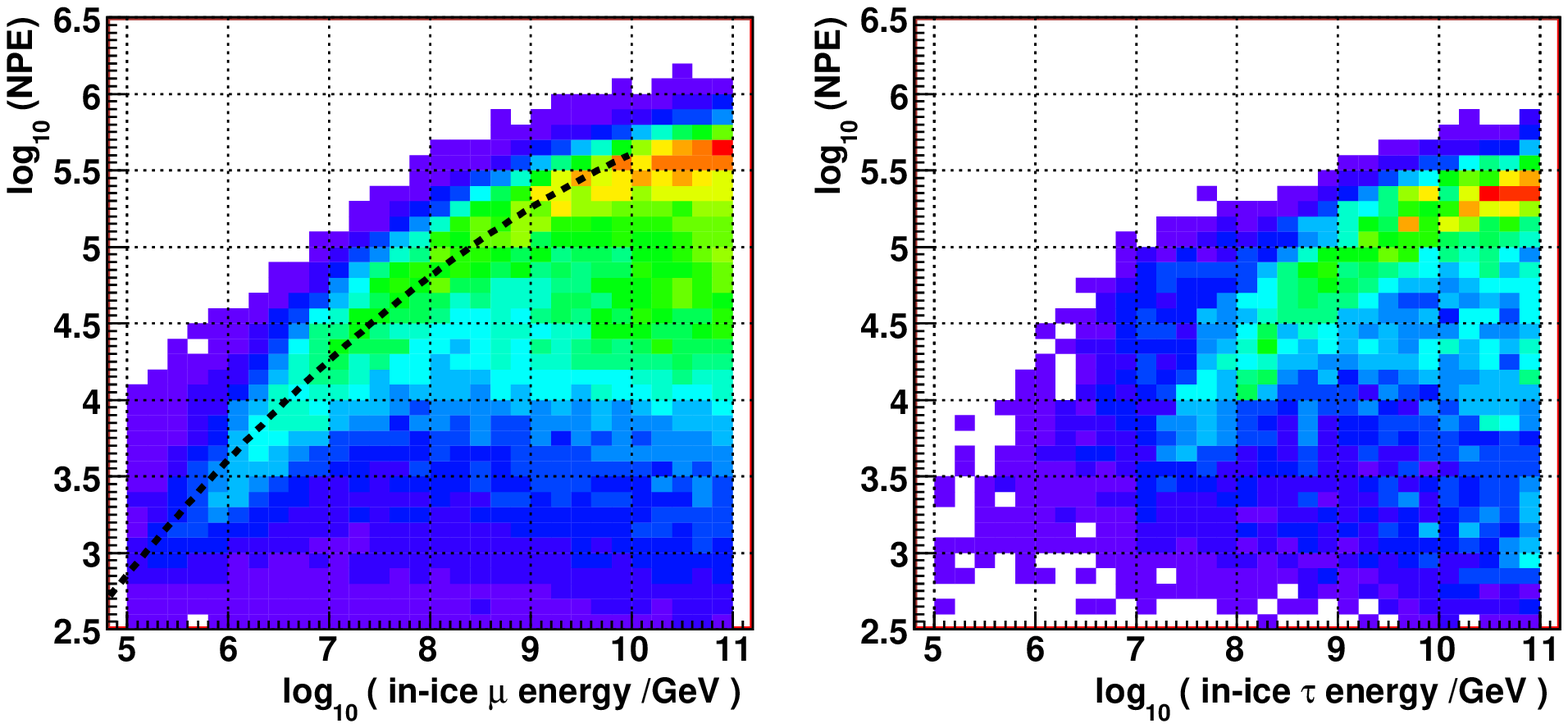}
\vspace{-0.1cm}
\caption{
\label{fig:EnergyNPE}
 Event from the Monte Carlo simulation of the \mbox{IC-9}
 detector in a plane of NPE and charged lepton energy measured at 880 meters
 from the IceCube center. 
 Events passing within 880~{\rm m} of the center of IceCube are considered in
 the plots and more distant events do not contribute to the data sample.
 The distribution in the left plot is for muons.
 The plot for taus on the right illustrates the suppression of energy loss compared to that of muons 
 and the contributions from tau-decays.
 The charged lepton energy distribution is assumed to follow $E^{-1}$
 in these plots for illustrative purposes.
 }
\end{center}
\vspace{-0.6cm}
\end{figure*}
%=====================================================
At extremely high energies, neutrinos are mainly detected by secondary
muons and taus generated during propagation of EHE neutrino in the
Earth~\cite{yoshida04}. The propagation of particles has been simulated in
detail by the JULIeT package~\cite{juliet}. 
Particles are seen in the detector as series of energetic cascades from
radiative energy loss processes rather than bare tracks. 
These radiative energy losses are proportional to the energies of muons and
taus and so is the Cherenkov light deposit in the IceCube
detector.
Figure~\ref{fig:EnergyNPE} shows
distributions of the total number of photoelectrons (NPE) detected by
the 540 DOMs as function of muon and tau energies from the full IceCube Monte Carlo
simulation. 
The trigger condition of 80 or more recorded DOM signals has been applied.
A clear correlation between NPE and the energy of particles measured at 880~{\rm m}
from the IceCube center is observed.  
The \mbox{IC-9} DOM response to a large NPE signal is limited mainly due to 
its readout configuration and PMT performance.
Taking fully into account these effects in the simulation,
the visible departure from linearity stems from the saturation of the
detector during signal capture. 
%in the higher NPE region is visible due
%to the saturation of the detector in the signal charge capture. 
%
Particles traversing far away from the detector leave low NPE signals regardless
of their energy.
From these observation, we use NPE as a robust estimator of the particle
energy - together with the zenith angle - as main selection criterion. 
%
%In the following, NPEs and reconstructed zenith angle are used as event selection
%criteria. 

%=====================================================
\vspace{-0.4cm}
\section{\label{sec:atm_muon_bg}
Background modeling}
\vspace{-0.4cm}
%=====================================================
%=====================================================
%
EHE neutrino induced muons and taus enter mostly from near or above the horizon with
down-going geometry because of the increase of neutrino cross section
with energy. Therefore, atmospheric muon bundles, penetrating the detector
from above, constitute a major background. 
However, the estimation of the atmospheric muon event rate in the relevant energy
range is highly uncertain, as it involves poorly characterized hadronic interactions and a knowledge
on the primary comic ray composition at energies
where there is no direct measurement available.
%
%In the present analysis, we employ a part of the experimental \mbox{IC-9} high
%energy event sample fitted by an empirical formula to build the
%atmospheric muon background model which is extrapolated to higher
%energies to estimate background intensity in the signal region.  
In the present analysis, we fit a part of the experimental \mbox{IC-9} high
energy event sample by an empirical formula to build the atmospheric
muon background model. 
The model is then extrapolated to higher energies to estimate background
intensity in the signal region.  

%
%=====================================================
\begin{figure}[hbt]
  \begin{center}
\includegraphics*[height=6cm,width=7cm]
{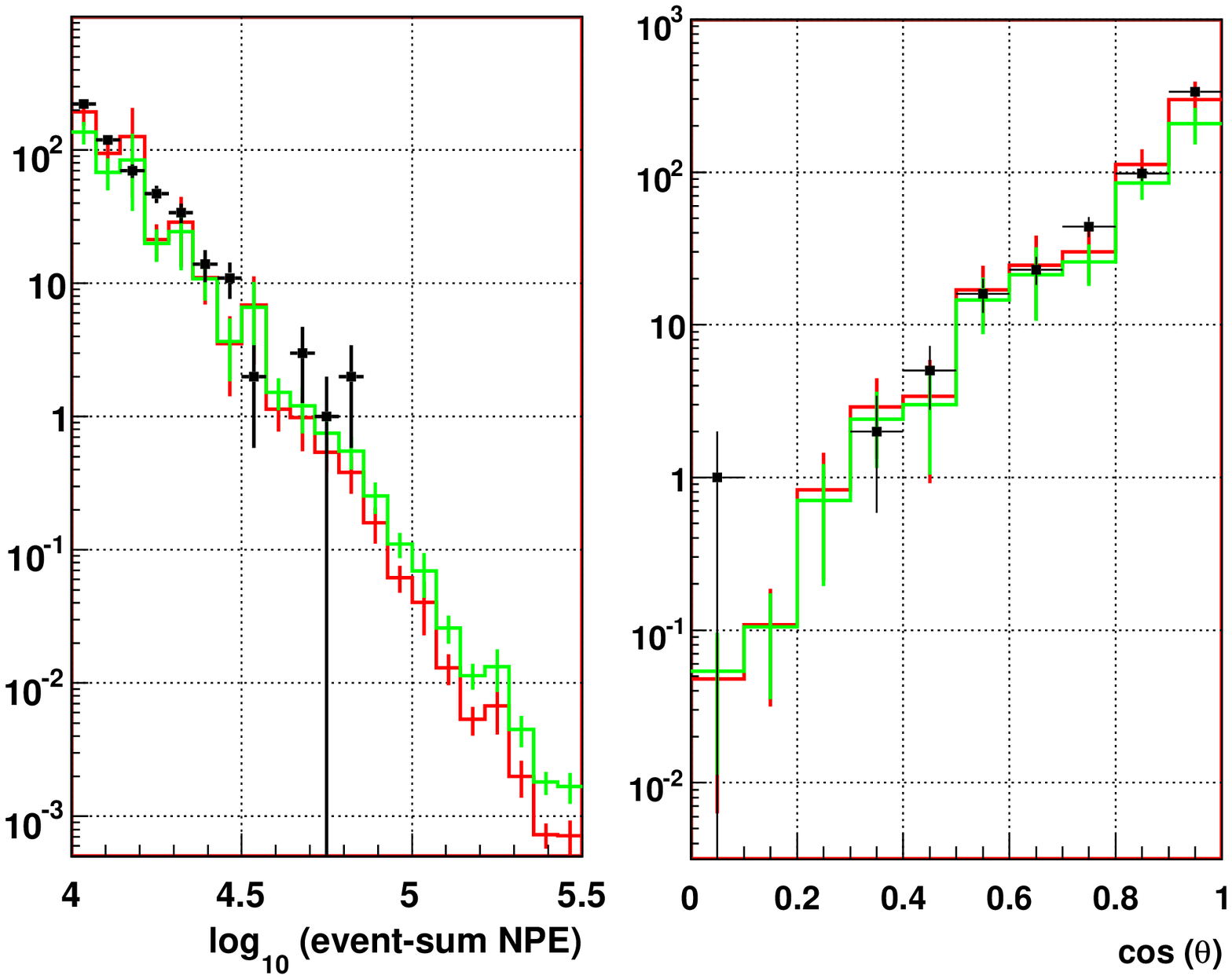}
\vspace{-0.1cm}
\caption[]{\label{fig:npe_distribution_IC9}
    Event distributions of data as a function of NPE (left) and zenith angle
   $\theta$ (right). Black dots with error bars denote the \mbox{IC-9} high energy event
   sample in $4.0 \leq$ $log_{10}$ NPE $\leq 5.0$.
   Red and green histograms are from the Monte Carlo simulation of the
   empirical atmospheric muon bundle model with two sets of parameters
   that gives similarly good agreement with experimental data.}
\end{center}
\vspace{-0.6cm}
\end{figure}
%=====================================================
%
%This study used an event sample with NPEs in the region between
This study used an event sample with $10^4$ $\le$ NPE $\le$ $10^5$
% in the region between
%$10^{4}$ and $10^{5}$ 
in which the bias in the high energy event dataset
from the filter requirement of 80 DOMs is minimal. Events are dominated
by atmospheric muons over possible cosmic neutrino events by more than 2 orders of
magnitude as shown in Ref.~\cite{aya06}. 
The empirical model is based on
the Elbert formula~\cite{gaisser90} 
that describes the number of muons with energies
greater than a energy threshold initiated in a cosmic ray air shower cascade.
The energy weighed integration of the formula relates the total energy
carried by a muon bundle to the primary cosmic ray energy.
The relation associates muon bundle event rate to given primary
cosmic-ray flux which was taken from the compilation in
Ref.~\cite{nagano01}. 
In other words, the background muon event rate is
governed by the intensity of the cosmic ray flux and depends on the
fraction of energy that goes to a muon bundle in an air shower.
The two parameters of the model,
the coefficient to determine multiplicity of muons in a bundle
and the lowest energy of muons in a bundle to leave detectable signal in the IceCube detectors,
are estimated by comparing model simulation and experimental sample in the plane of NPE and
reconstructed zenith angle for NPE below $10^5$.
The comparison of the model and experimental data is shown in 
Fig.~\ref{fig:npe_distribution_IC9}. %Denoted by the black dot is a
The black dots show a mid-NPE subsample of the data. Colored lines indicate
the model simulation with two sets of parameters that give similar goodness
in fits in terms of ${\chi^2}/{d.o.f}$ with respect to the experimental sample but with extreme
cases of the low muon multiplicity coefficient (green line) and the low threshold
energy coefficient (red line) in a bundle. 
%
%As seen in Fig.~\ref{fig:npe_distribution_IC9}, the models outline the
Obviously, the models represent the experimental NPE and declination dependence well.

%=====================================================
%=====================================================
%The IceCube Monte Carlo simulation package then generates events with
%energies and intensities following the obtained fluxes.
%
%
%For the atmospheric muon flux, which is considered our main background,
%we take the the empirical model obtained by the comparison with
%the IceCube-9 data described previously.
%=====================================================

\vspace{-0.4cm}
\section{\label{sec:results}Results}
\vspace{-0.4cm}
%=====================================================

%=====================================================
\begin{figure*}[hbt]
  \begin{center}
\includegraphics*[width=0.8\textwidth,angle=0,clip]
{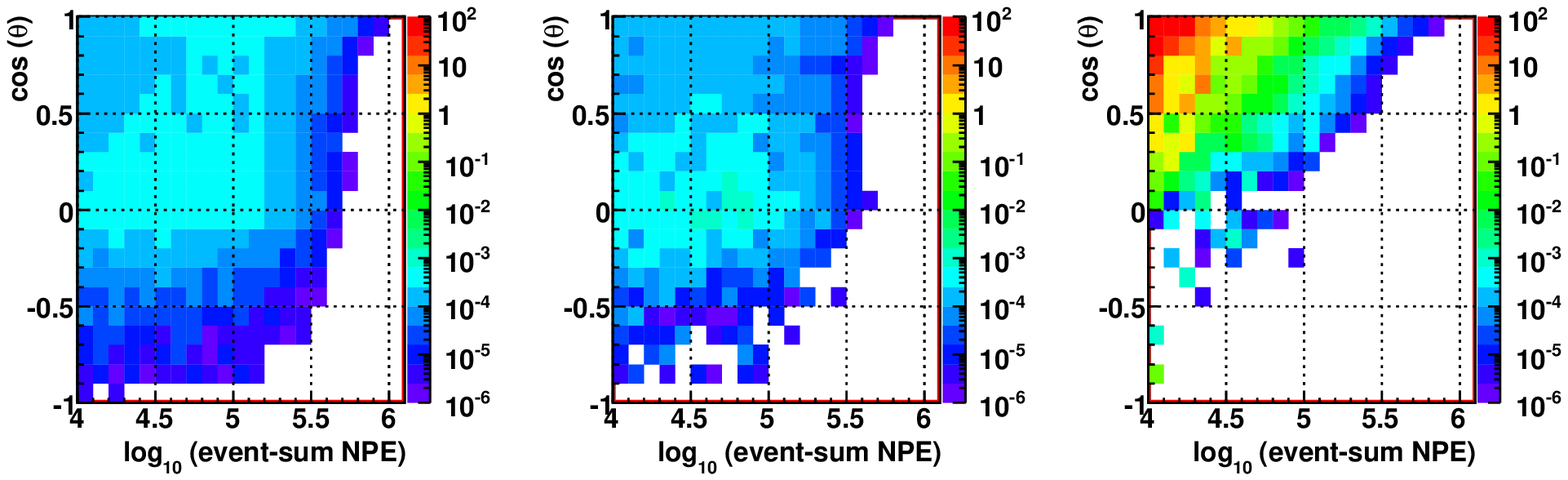}
   \vspace{-0.1cm}
   \caption[]{\label{fig:NPECosTheta} 
   Event distribution in the plane of NPE and cosine of zenith angle
   obtained by Monte Carlo simulations.
   Plotted on the left and middle are those for GZK neutrino-induced 
   muon and tau signals, respectively. The background atmospheric muon
   bundle model is shown on the right. Projections of the atmospheric muon bundle
   distribution is represented by green lines in Fig.~\ref{fig:npe_distribution_IC9}.
   }

  \end{center}
 \vspace{-0.4cm}
\end{figure*}
Event distributions for signal and muon-bundle induced background are
shown in Fig.~\ref{fig:NPECosTheta}.
For the signal we chose a GZK cosmogenic neutrino
model~\cite{berezinsky69} as calculated in Ref.~\cite{yoshida93}.
%
%plane of NPE and reconstructed zenith angle $\theta$ from the signal neutrino represented
%by 
%atmospheric muon bundle background model described in the previous
%section are shown 
%
%of muon (left), tau (middle)
%
The plots show that the atmospheric muon bundle model has a steeper
distribution in NPE compared with that of the signal GZK model.
%
%It is also observed that 
The number of muons and taus originating from the
propagation of the signal neutrino in the earth exceeds that of
atmospheric muon bundles at directions near the horizon as well as at the higher NPE.
These observations suggest that the background can be rejected by
excluding events with low NPE values and vertical reconstructed directions.
%retaining events from the signal neutrino.  
%
%
%
%As down-going GZK muons and taus have a higher tendency 
%to arrive from near the horizontal direction than
%the atmospheric background~\cite{yoshida04},
%one can expect a difference in the distribution of 
%zenith angles and NPE for EHE signals and background.
%
The signal domain is defined by the following conditions:
%======================================
%\begin{eqnarray}
%1.1(\cos\theta-0.1)-0.9(\log {\rm NPE} - 4.7) \leq 0 \nonumber\\
%\log {\rm NPE} \geq \log{\rm NPE}_{low} \quad {\rm if}\ \cos\theta \leq 0.1.
%\label{eq:domain}
%\end{eqnarray}
%======================================
%\begin{equation}
%{\rm log_{10} NPE }\geq \Bigl\{
%\begin{tabular}{cc}
% log$_{10}$ NPE$_{low}$  &  and \\
%$4.7+\frac{1.1}{0.9} (\cos \theta - 0.1)$ &\\
% if $\cos\theta\geq$ 0.1&
%\end{tabular}\label{eq:domain}
%\end{equation}
%======================================
\begin{equation}
{\rm log_{10} NPE} \geq {\rm log_{10} NPE_{low}},
 \label{eq:domain1}
\end{equation}
and if $\cos\theta\geq 0.1$,
\begin{equation}
{\rm log_{10} NPE} \geq 4.7+\frac{1.1}{0.9} (\cos \theta - 0.1).
\label{eq:domain2}
\end{equation}
%======================================
%---------------------------
\begin{figure*}[hbt]
  \begin{center}
    \begin{tabular}{c}
\vspace{-0.1cm}
     \includegraphics*[height=5.cm]{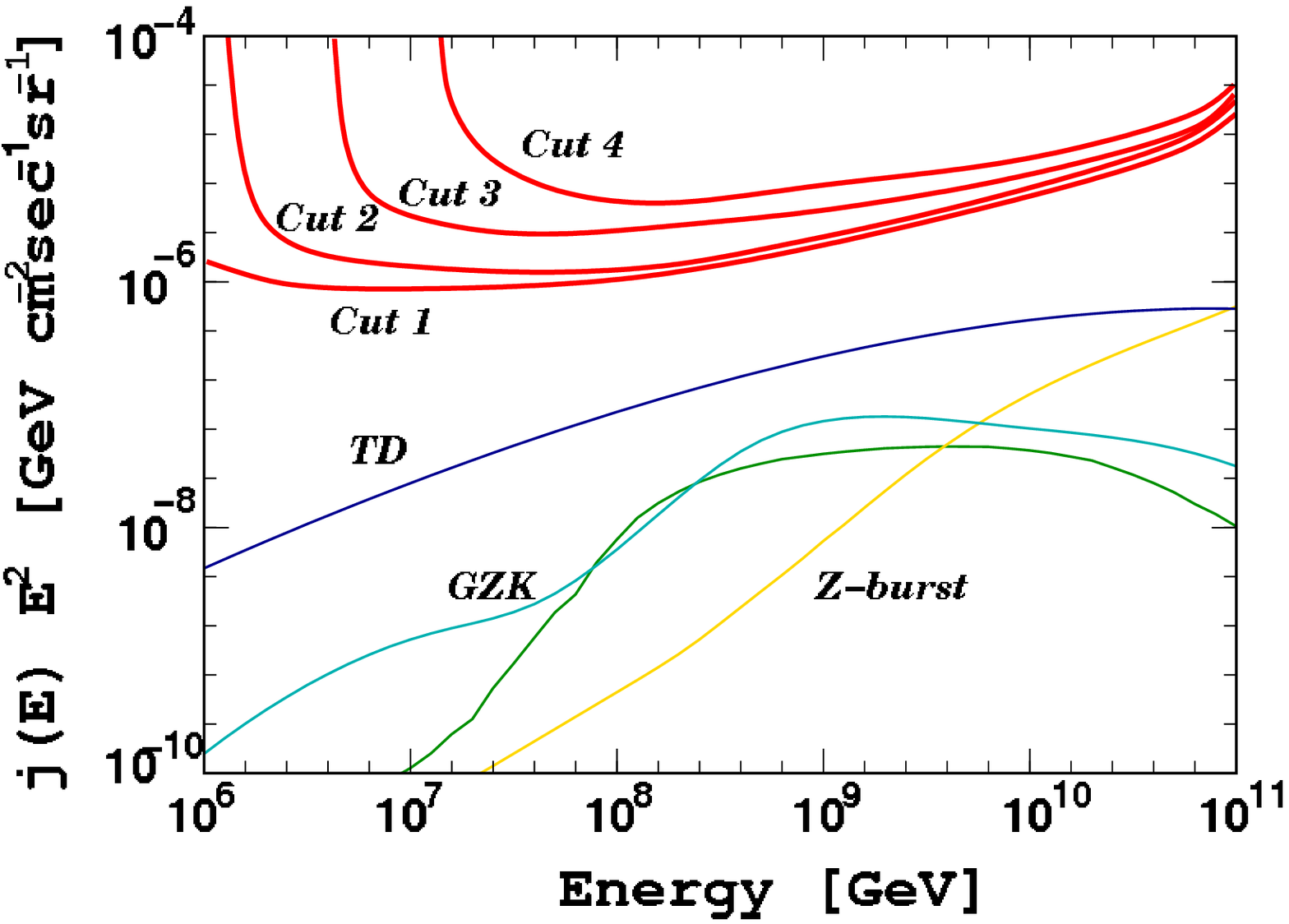}
     \includegraphics*[height=5.cm]{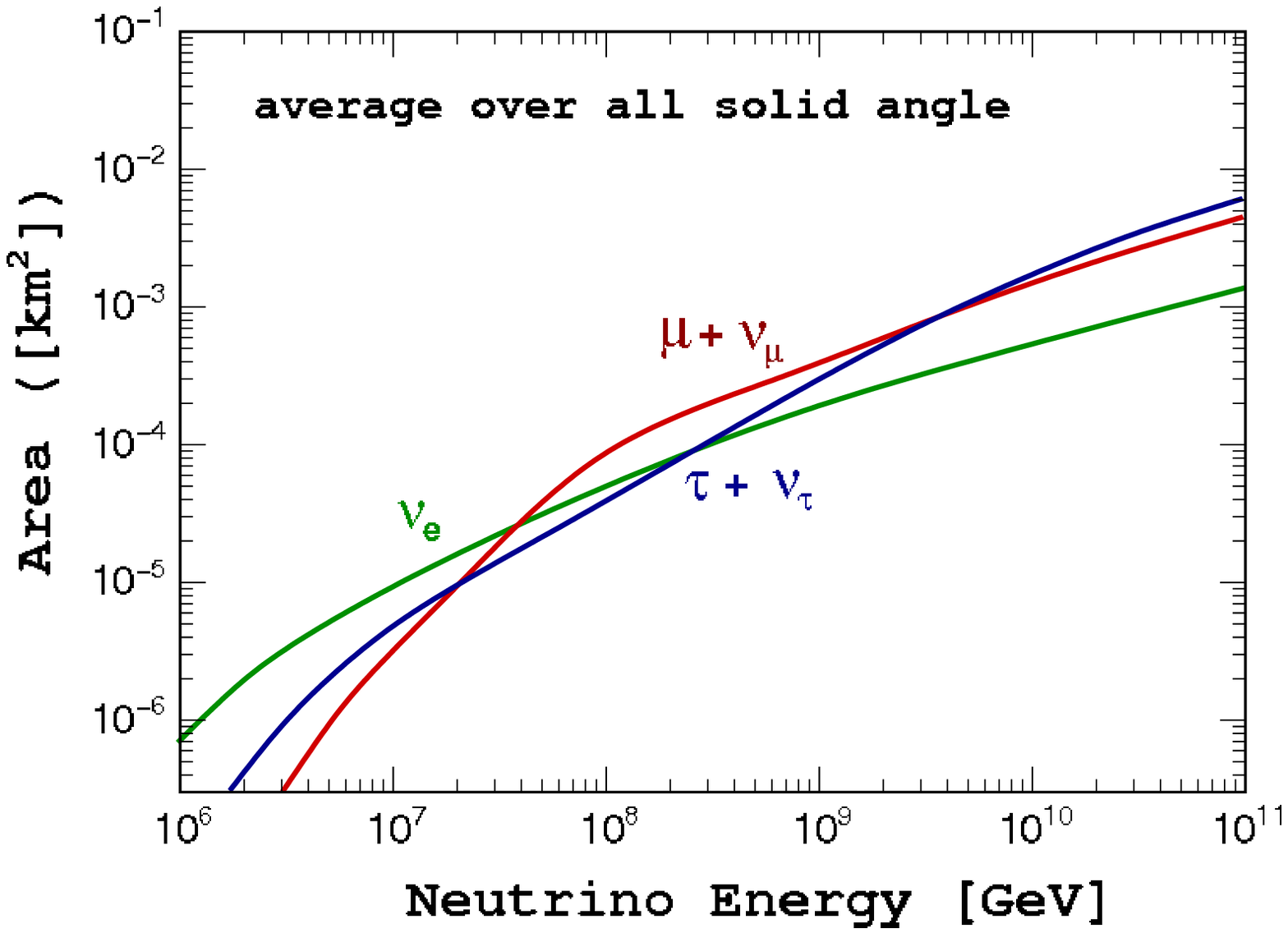}
    \end{tabular}
%NeutEffectiveAreaFlavor_June3.eps
%IceCube9StringSensitivityWithCaption_June3.eps
    \caption[]{
\label{fig:eventrate}
The 2006 \mbox{IC-9} sensitivity curves at 90\% C.L. on the EHE neutrino model
   fluxes is shown on the left.
 The fluxes of the three neutrino flavors $\nu_e$, $\nu_\mu$, $\nu_\tau$ are summed up.
   GZK refers to the GZK model from  Ref.~\cite{yoshida93} for the lower
   curve and Ref.~\cite{kalashev02} for the upper curve. 
   The TD and Z-burst predictions are
   from Ref.~\cite{SLBY} and Ref.~\cite{yoshida98}, respectively. 
   Plotted on the right is the corresponding neutrino effective area of
   three neutrino flavors for cut level 2.
}
  \end{center}
\vspace{-0.6cm}
\end{figure*}
%---------------------------

\begin{table}[ht]
\vspace{-0.6cm}
\small
\caption{
Preliminary numbers of expected IceCube EHE events of 
muons and taus produced from the GZK model~\cite{yoshida93} 
and from the background atmospheric muon model prediction. The
 predictions are normalized to a livetime of 124 days and presented for different $\log_{10}{\rm
 NPE}_{low}$ values in Eqs.~(\ref{eq:domain1}) and (\ref{eq:domain2}). 
GZK $\mu + \tau$ indicates the number of events with muons and taus induced
 by the GZK neutrino outside the IceCube detector volume
 defined by a sphere of 880 {\rm m} radius. GZK $\nu_{e + \mu +
 \tau}$ indicates contributions from charged particles created inside the sphere.
}
\label{table:eventrate}
\vspace{2mm}
\newcommand{\m}{\hphantom{$-$}}
\newcommand{\cc}[1]{\multicolumn{1}{c}{#1}}
\begin{tabular}{@{}lllll}
\hline
{cut level}       & $1$ & $2$ & $3$ & $4$ \\
\hline
 $\log_{10}{\rm NPE}_{low}$      & $4.4$   & $4.6$   & $4.8$   & $5.0$ \\
 GZK $\mu + \tau$           & $0.033$ & $0.027$ & $0.020$ & $0.011$  \\
 GZK $\nu_{e + \mu + \tau}$ & $0.028$ & $0.024$ & $0.020$ & $0.015$  \\
 atmospheric $\mu$          & $0.003$ & $\leq10^{-4}$ & $\leq10^{-4}$ & $\leq10^{-4}$ \\
\hline
\end{tabular}
\end{table}
%======================================
%  --------------------------------------------------
%	          nu-e	nu-mu	nu-tau	muon	tauon
% cut#4 	0.012	0.0080	0.0078	0.017	0.016
% cut#6 	0.010	0.0072	0.0068	0.014	0.013
% cut#8 	0.0080	0.0061	0.0054	0.011	0.0087
% cut#10	0.0061	0.0048	0.0037	0.0073	0.0044
%--------------------------------------------------
%                0.0278,   0.033  
%                0.024,    0.027
%                0.0195,   0.0197
%                0.0146,   0.0117
%%%%%%%%%%%%%%%%%%%%%%%%%%%%%%%%%%%%%
Summarized in Table~\ref{table:eventrate} are the expected numbers of signal and background events above
cut levels defined with different values of $\log_{10} {\rm NPE}_{low}$. 

The resulting sensitivity to the all flavor EHE neutrinos is calculated
independent of the neutrino flux models with the quasi-differential
method based on the flux per energy decade. 
A similar approach is found in Ref.~\cite{auger}. 
%======================================
The first year \mbox{IC-9} sensitivity curves at 90\% confidence level are
shown in the left plot of Fig.~\ref{fig:eventrate} for the four cut
levels in Table~\ref{table:eventrate} with an assumption of negligible
background.
It is also shown that this EHE neutrino search is sensitive to the
neutrinos with energies on the surface ranging between  $\sim$$10^7$ and  $\sim$$10^9$ GeV. 
%
%With the 
Choosing cut level number 2,
the 90\% C.L. upper limit of
EHE neutrino fluxes by the 2006 \mbox{IC-9} observation
would be placed at
$E^2 \phi_{\nu_e+\nu_\mu+\nu_\tau}\simeq 1.6\times 10^{-6}$~GeV/cm$^2$ sec sr
for neutrinos at an energy of $10^8$~GeV;
the corresponding
neutrino effective area with our preferred cut level 2 is also shown on the right plot of Fig.~\ref{fig:eventrate}.

%=====================================================
\vspace{-0.4cm}
\section{Discussion}
\vspace{-0.4cm}
The sensitivity estimate has been obtained with the assumption of negligible background based
on the empirical model prediction. 
%
%It must be noted that one should pay attention to the systematic
%uncertainties in the low background expectation from the model, however.
The systematic uncertainties in the background estimation must be
further considered, however.
%
%A sizable systematic error may arise from 
Possible contributions from fluctuations in the hadronic interaction processes in the air shower cascades and fluctuations
in the muon bundle spatial distribution at IceCube detector
depths (1450-2450~{\rm m}) are disregarded in the current study. 
The estimation of these effects must be performed before the cuts are
finalized. 
%
%Let us also remark the possible emergence of the prompt muon component.
We would like to also remark that estimations of the contribution from
the prompt muon in the present background model are uncertain.
While the \mbox{IC-9} high energy sample below $10^5$ NPE (corresponding roughly to
$E\leq 10^{7-8}$ GeV) shows no indication of a significant prompt muon
contribution, a potential excess of events beyond the atmospheric muon bundle model
could either be due to prompt muons, cosmic neutrinos or due to events of exotic physics origin. 
%
%
%
%Events in the signal
%domain would be interpreted either by unconventional prompt muons, 
%cosmic neutrinos, or exotic physics. The maximum energy of primary cosmic
%rays should limit the first possibility, which will be discussed elsewhere.
%
%=====================================================
\vspace{-0.6cm}
\section{Acknowledgments}
\vspace{-0.4cm}
%=====================================================
We acknowledge 
the Office of Polar Programs of the U.S. National Science Foundation, and
all the agencies to support the IceCube project.
This analysis work is particularly supported by
the Japan-US Bilateral Joint Projects program in the Japan Society for 
the Promotion of Science.

%\end{document}

\setcounter{figure}{0}
\setcounter{table}{0}
%%
% International Cosmic Ray Conference 2007 Merida Yucatan Mexico
% In This file you will find detailed instructions to correctly
% typeset your document.
%
%
%

%Class Requeried
%\documentclass{article}
%The ICRC Style
%\usepackage{icrctc07}

%The paper title
\title{Very high energy electromagnetic cascades in the LPM regime with IceCube}
\shorttitle{Very High Energy cascades with IceCube}
%All paper authors
\authors{J. Bolmont$^1$, B. Voigt$^1$ and R. Nahnhauer$^1$,
        for the IceCube Collaboration$^2$.}
%Short title to print in the headers to the final puplication (Not showed in this print).
\shortauthors{J. Bolmont et al. for the IceCube Collaboration}
%All the affiliations.
\afiliations{$^1$ DESY, D-15738 Zeuthen, Germany\\ $^2$ see special section of these proceedings}
\email{julien.bolmont@desy.de}

%The abstract.
\abstract{With a volume of $\sim$1~km$^3$, IceCube will be able to detect very high energy
neutrinos above $\sim$100~PeV. At these energies, bremsstrahlung and pair
production are suppressed by the Landau-Pomeranchuk-Migdal (LPM) effect.
Therefore, $\nu_e$ and $\nu_{\tau}$ interactions in the ice can produce several hundred
meter long cascades. We present an analysis of IceCube sensitivity to $\nu_e$ events. It includes cascade simulation in the LPM regime and makes use of preliminary algorithms for incident angle reconstruction. We give the obtained effective area for the 22 string configuration and discuss IceCube angular reconstruction precision.}

%%%%%%%%%%%%%%%%%%%% B E G I N   D O C U M E N T%%%%%%%%%%%%%%%%%%%%%%%
%\begin{document}
\maketitle
%Begin the section.

\section{Introduction}

Different models predict a significant flux of high energy neutrinos above $\sim$100~PeV. Topological defects, superheavy relics of the Big-Bang, the GZK mechanism or gamma ray bursts could produce such high energy neutrinos (see \cite{gandhi} for a review).

The IceCube neutrino detector is under construction at the South Pole \cite{karle}. Currently, it is made of 22~strings each holding 60~optical detectors, instrumenting a volume of $\sim$0.3~km$^3$. Strings are separated by 125~m and modules on a string are separated by 17~m. By its completion in 2011, there will be up to 80 strings and the corresponding volume will be $\sim$1~km$^3$. 

At low energies, $e^\pm$ produced by charged current interactions produce small cascades compared to the spacing between two optical modules. The produced light is emitted in the direction of the Cherenkov cone but it is scattered in the ice so that when observed from a distance, it can be considered to be emitted almost isotropically from the centre of the cascade. Therefore, the angular resolution for cascades is poor.

However, for a 100~PeV neutrino, the secondary particle energy is high enough for brems\-strah\-lung and pair-produc\-tion to be supressed by the Landau-Pomeranchuck-Migdal (LPM) \cite{lpm1,lpm2} effect. This leads to an elongation of the cascades, which in turn could result in better angular resolution for cascade events in IceCube. 

Here, we focus on electromagnetic cascade ana\-lysis for $\nu_e$ events at ener\-gies above $\sim$100~PeV where the LPM effect has to be taken into acount. The LPM effect also affects hadronic cascades but as the input energy is distributed over a large number of secondary particles, their length does not increase as dramatically. Hadronic cascades will not be discussed in this paper.

In the next section, we describe two simulation tools for high energy cascades in the LPM regime. The longitudinal profiles obtained are used to estimate the Cherenkov light output in the ice. Following that, effective area is computed for the 22 string detector. Finally, the precision of incident angle reconstruction is evaluated.

The results shown are preliminary.

\section{Simulation of high-energy cascades}\label{sec:sim}

To study high-energy cascades in ice, two simulation packages have been developed. 
One allows the rapid simulation of cascade profiles and uses a parameterisation of bremsstrahlung and pair-production cross sections in the LPM regime, and a parameterisation of energy deposition for the low energy products of the cascade. The other package is a full Monte Carlo simulation of cascades based on \texttt{CORSIKA}. It is devoted to fine studies of the development of cascades.

\subsection{Hybrid approach}

\begin{figure}[t!]
\centering
    \includegraphics[width=0.5\textwidth]{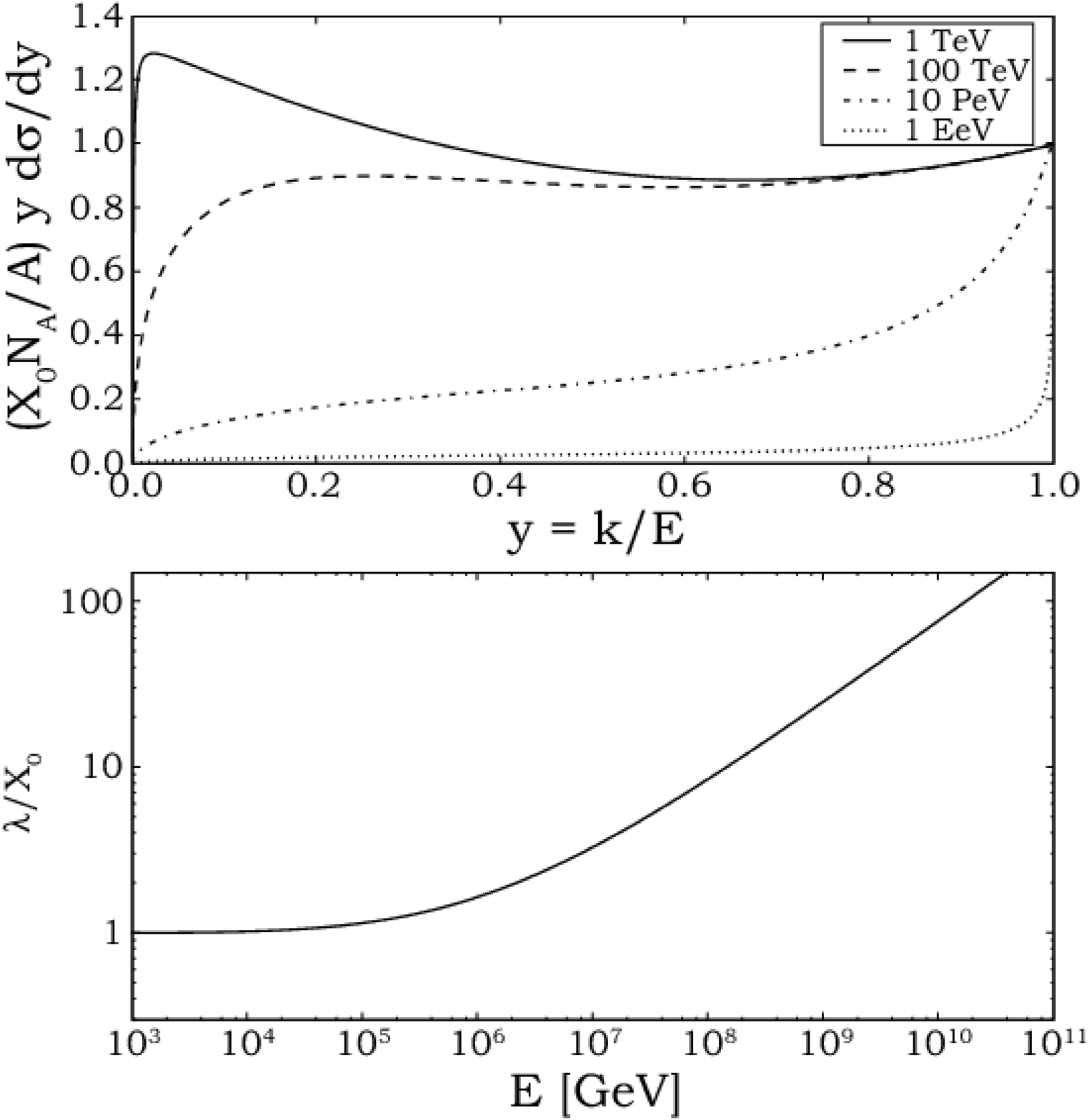}
\caption{Top panel: differential energy cross section for 
bremsstrahlung as a function of $y = k/E$ where $k$ is the energy of the photon and $E$ the energy of the electron. Bottom panel: mean free
path for bremsstrahlung as a function of energy.}
\label{fig:crosssec}
\end{figure}

Following Niess and Bertin \cite{niess}, a simulation of the cascade 
development has been implemented. This simulation takes into account bremsstrahlung and pair
production interactions and works only in one dimension. 
The suppression of both processes by the LPM effect is included. Parameterisations 
of bremsstrahlung and pair production cross sections are taken from \cite
{klein}. Fig.~\ref {fig:crosssec} shows the differential cross section and radiation
length parameterisation for bremsstrahlung. The increase of the mean 
free path above ~1 PeV is due to the LPM effect.

\begin{figure}[t!]
\centering
    \includegraphics[width=0.5\textwidth]{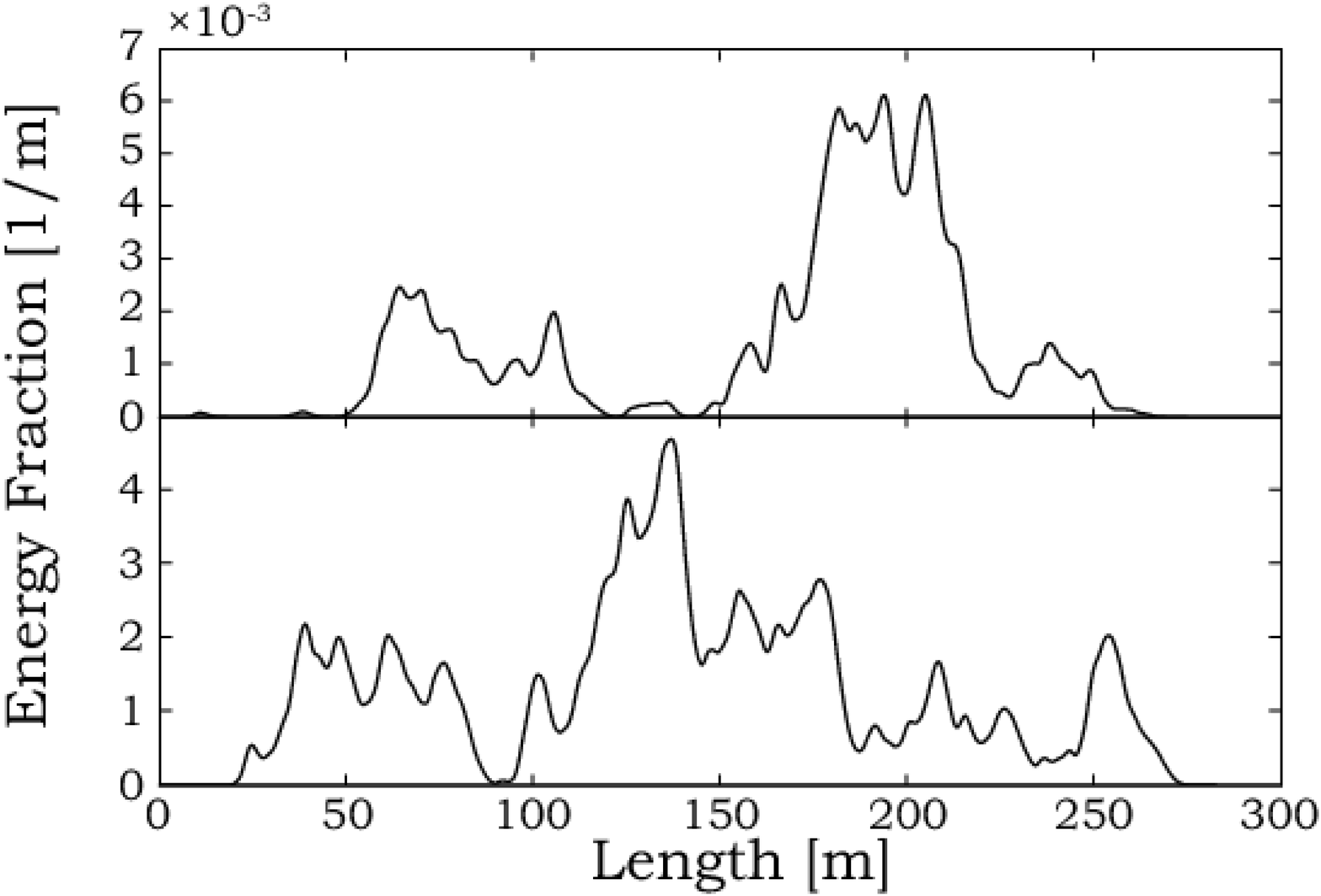}
\caption{Longitudinal energy profiles of two 10~EeV electromagnetic 
cascades.}
\label{fig:LPMprofile}
\end{figure}

In the simulation, high energy particles are propagated until their energy falls
below a cut-off energy on the order of 1 TeV and the energy loss profile of these 
particles is computed using a parameterisation. The individual
energy loss profiles of these low energy particles are summed to obtain the total energy
deposit profile of the full shower.

The fractional energy of the secondary particles is generated 
randomly from the differential cross section using a Metropolis-Hastings Algorithm \cite{hastings}. This allows the quick generation
of random samples. For instance, when the cut-off energy is on the order of 1~TeV, a single cascade with energies in 
the PeV range can be simulated in a few milliseconds. A 10~EeV
cascade is simulated within less than 3 minutes when the cut-off is
set to 50~TeV.

Fig.~\ref{fig:LPMprofile} shows longitudinal energy profiles of two 10~EeV cascades. Their length is about 200~m. Many different sub-cascades contribute to this profile. The figure also shows that the shape and length of the longitudinal profile can vary significantly from one cascade to another.

\begin{figure}[t!]
	 \centering
   \includegraphics[width=0.45\textwidth]{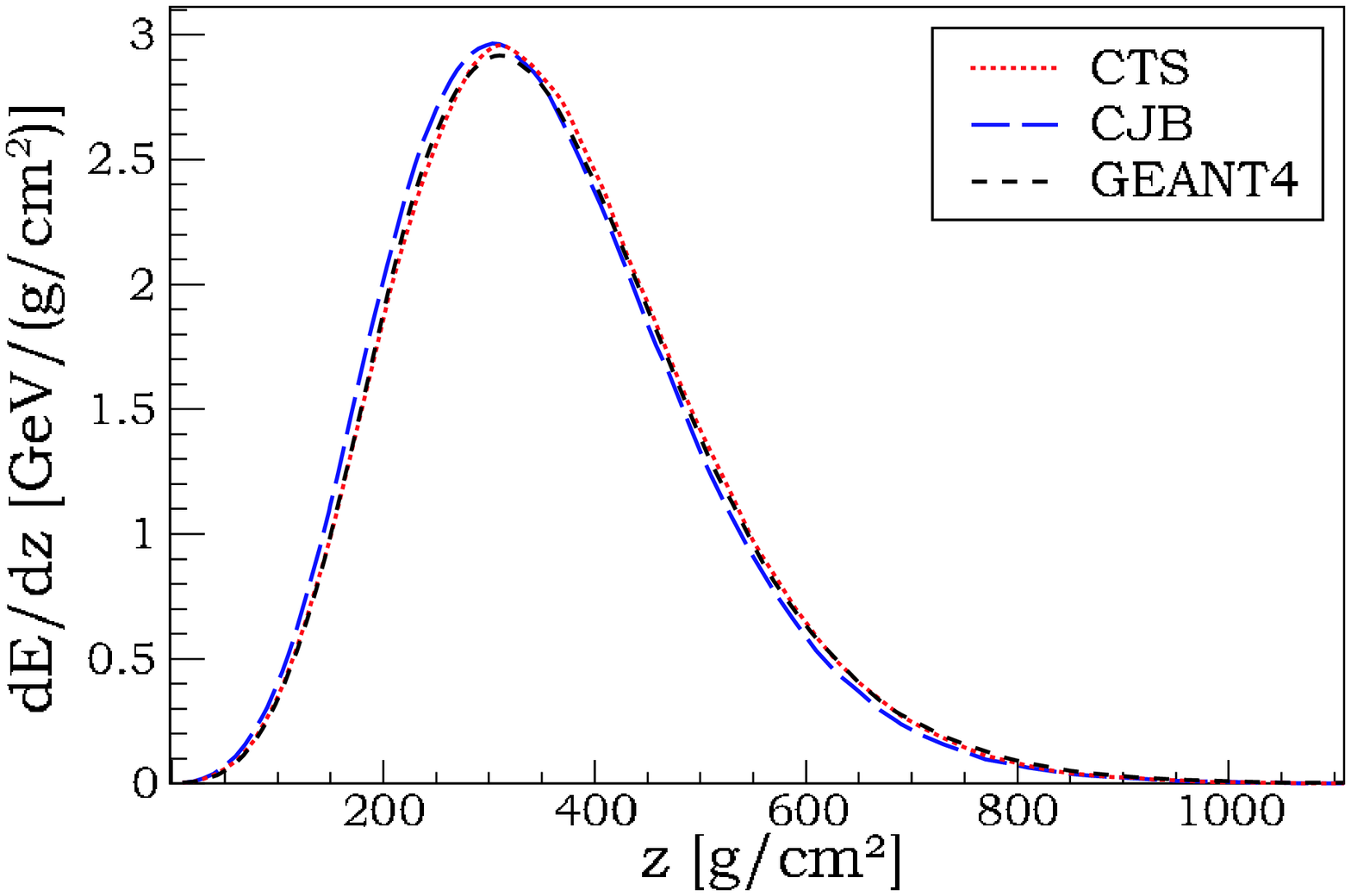} 
   \caption{Average longitudinal profile of one hundred 1~TeV cascades. Comparison between \texttt{CTS}, \texttt{CJB} and \texttt{GEANT4}.}
   \label{fig:corsencomp}
\end{figure}

\subsection{Monte Carlo approach}

In addition to the simple hybrid approach presented in the previous section, we have also developed a more realistic simulation tool, able to provide more precise information on cascade development. This tool is based on the well-known atmospheric cascade simulation tool \texttt{CORSIKA} \cite{ICRC0843_corsika}. 

As \texttt{CORSIKA} is devoted to cascade simulations in the atmosphere, several modifications had to be made in order to adapt the code to a uniform density medium. This work was initially done by T. Sloan for the ACoRNE collaboration with \texttt{CORSIKA 6204} \cite{acorne}. Hereafter, this version will be denoted \texttt{CTS}.

We have used these modifications and taken them a step further to get more functionality and more flexibility. The new modifications allow us to:
\begin{itemize}
\item switch the medium from air to ice during the configuration step,
\item use the different simulation packages (\texttt{VENUS}, \texttt{QGSJET} and others) available with \texttt{CORSIKA},
\item use all the other options available in \texttt{CORSIKA}, whenever they are relevant to a simulation in water/ice.
\end{itemize}

The changes were made starting from the most recent version of \texttt{CORSIKA} (\texttt{CORSIKA 6502}).

To check the validity of this software (denoted \texttt{CJB}), we simulated 1~TeV electrons using both versions \texttt{CJB} and \texttt{CTS} with the same input parameters and the same random generator seeds. The results were also compared with \texttt{GEANT4} \cite{geant}.

Fig.~\ref{fig:corsencomp} shows the energy deposition profiles for the two versions of \texttt{CORSIKA} and \texttt{GEANT4}. The profiles are very similar. The small difference between \texttt{CJB} and \texttt{CTS} comes mainly from minor revisions in the \texttt{EGS4} code \cite{egs} between \texttt{CORSIKA} releases \texttt{6204} and \texttt{6502}.

A 10~EeV cascade can be simulated in less than 2 minutes for default values of cut-off energies, provided the thinning option is enabled.

\section{Reconstruction and effective areas}\label{sec:rec}

Electron neutrinos with energies between 10~TeV and~10~EeV were generated, propagated through the Earth and forced to interact in the vicinity of the instrumented detector volume using a software package based on the \texttt{ANIS} \cite{ICRC0843_anis} neutrino generator. The development of the cascades in the ice has been done with the hybrid simulation tool described previously. The detector response includes the simulation of light propagation through the ice, optical module responses and a trigger simulation requiring 8 modules hit within a time window of 4~$\mu$s. 

\begin{figure}[t!]
   \centering
   \includegraphics[width=0.45\textwidth]{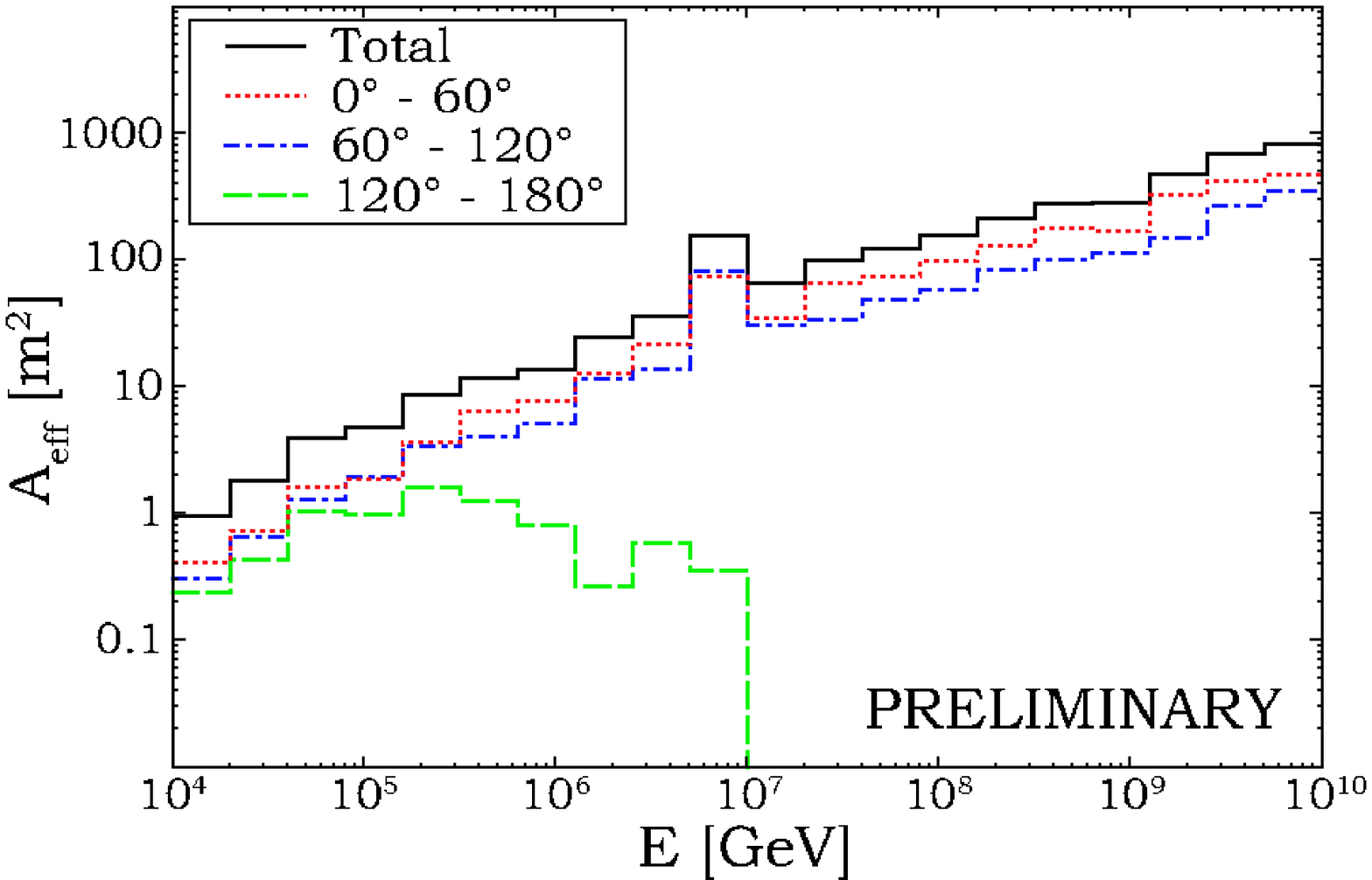}
   \caption{Effective area for different incident angles and for the 22 string configuration.}
   \label{fig:effarea}
\end{figure}

A basic analysis method typically used to reject muon background was applied to the pure $\nu_e$ sample in order to calculate the effective area, taking into account the reconstruction efficiency. The selection is done by computing the ratio between the longitudinal and lateral size of the light distribution, using the fact that cascades are more spherical than muon tracks.

The number of passing events was used to calculate the neutrino effective area for three different zenith angle ($\theta$) bands for the 22 string configuration (Fig.~\ref{fig:effarea}). The effective area generally increases with energy due to the rising cross section of neutrino interactions. However, for neutrinos with energies above $\sim$1~PeV the earth becomes opaque and the effective area for neutrinos coming from below the horizon (120$^\mathrm{o}$ $< \theta <$ 180$^\mathrm{o}$) falls off. The peak between 5~PeV and 10~PeV is caused by resonant $\overline{\nu_e}$ + $e^-$ scattering at energies around 6.3 PeV (the Glashow resonance).

\section{Precision of incident angle reconstruction}

Cascade-like events passing the trigger conditions are reconstructed using a simple \textit{line-fit} algorithm \cite{amandarec} usually used for muon track reconstruction. It uses the hit times to produce a track defined by a vertex point and a direction.

The difference $\phi$ between generated and reconstructed directions is computed. Fig.~\ref{fig:recangle} shows the cumulative fraction of events reconstructed with an arbitrary precision $\phi$. At 1~EeV, the proportion of cascades reconstructed with a precision better than 20$^\mathrm{o}$ is $\sim$5\%. At 10~EeV, when the LPM effect is taken into account, this proportion is $\sim$20\%.

\section{Conclusions}

At very high energies, the LPM effect can increase the length of cascades to several hundred meters. This could lead to better angular resolution for high energy cascades. We have developed two new tools in order to study these events. 

A very simple \textit{line-fit} method seems to indicate a significant improvement of angular reconstruction precision at high energies. However, this improvement, due to cascade lengthening, still leads to insufficient resolution for possible high energy neutrino source identification.

Achieving this goal will require dedicated algorithms to fully exploit the cascade lengthening and to obtain improved angular resolution. Such methods are under development.

\section{Acknowledgements}

The authors would like to thank T. Sloan from University of Lancaster, Lancaster, UK, D. Heck and T. Pierog from Institut f\"ur Kernphysik, Karlsruhe, Germany, for their help during the development of our simulation tool.

This work is supported by the Office of Polar Programs of the National Science Fundation.

\begin{figure}[t!]
   \centering
   \includegraphics[width=0.45\textwidth]{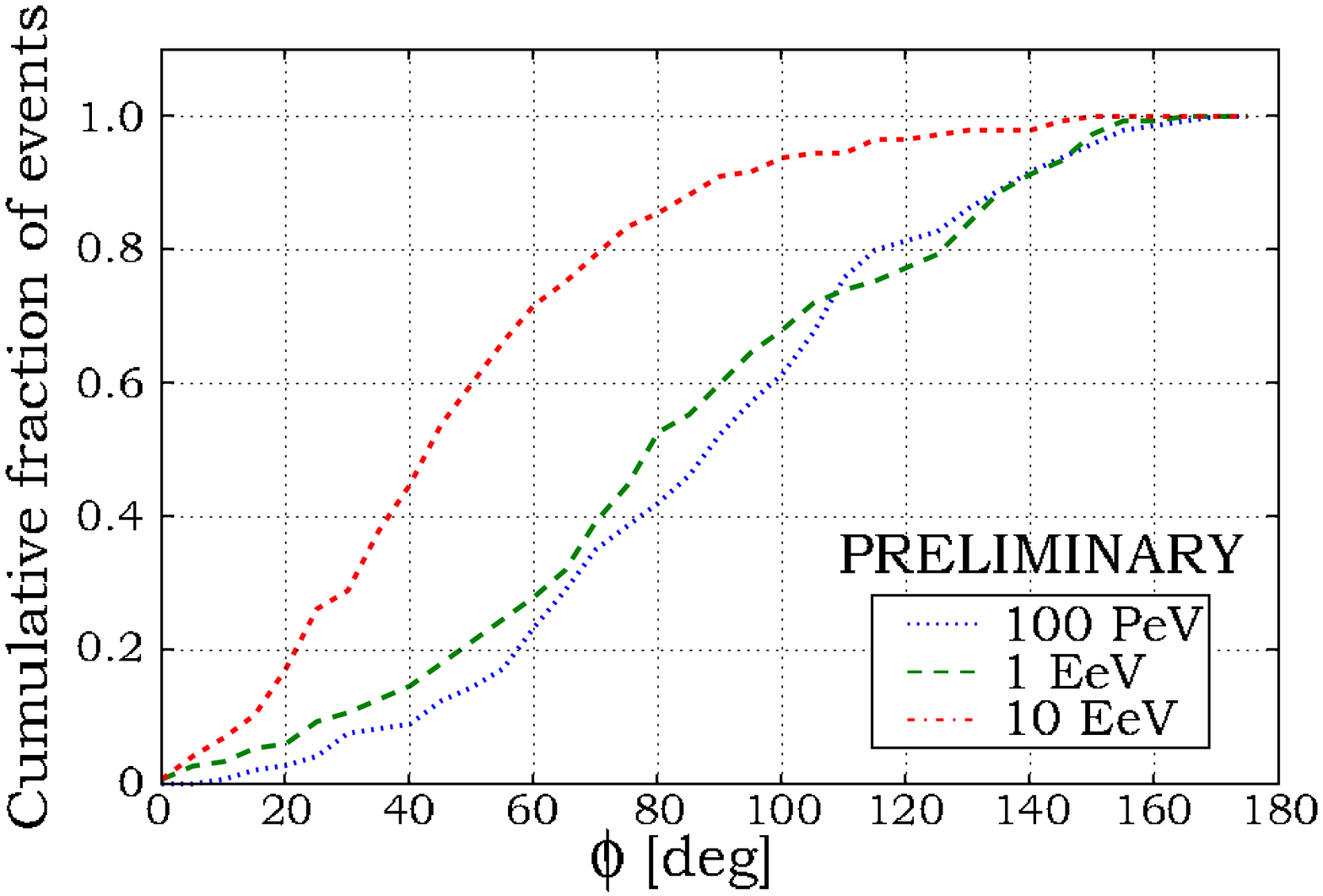}
   \caption{Cumulative fraction of events reconstructed with an arbitrary precision $\phi$, for the 2007 22 string configuration.}
   \label{fig:recangle}
\end{figure}

%\end{document}

\setcounter{figure}{0}
\setcounter{table}{0}
%%
% International Cosmic Ray Conference 2007 Merida Yucatan Mexico
% In This file you will find detailed instructions to correctly
% typeset your document.
%
%
%

%Class Requeried
%\documentclass{article}
%The ICRC Style
%\usepackage{icrctc07}

%The paper title
\title{IceCube Performance with Artificial Light Sources:
the Road to Cascade Analyses
}
%Short title to print in the headers to the final publication (Not showed in this print).
\shorttitle{IceCube Artificial Light Sources}
%All paper authors
\authors{J. Kiryluk$^{1}$, M.V. D'Agostino$^{2}$, S.R. Klein$^{1}$, C. Song$^{3}$, and D.R. Williams$^{4}$ 
for the IceCube Collaboration$^{5}$} 
%Short title to print in the headers to the final puplication (Not showed in this print).
\shortauthors{J. Kiryluk {\it{et al.}} }
%All the affiliations.
\afiliations{
$^{1}$Lawrence Berkeley National Laboratory, Berkeley, CA 94720, USA\\
$^{2}$ Dept. of Physics, University of California, Berkeley,  CA 94720, USA\\
$^{3}$ Dept. of Physics, University of Wisconsin, Madison, WI 53706, USA\\
$^{4}$ Dept. of Physics,  Pennsylvania State University, University Park, PA 16802, USA\\
$^{5}$ see special section of these proceedings} 
%\afiliations{Afiliations}
\email{JKiryluk@lbl.gov}
%\email{e-mail}

%The abstract.
\abstract{
The IceCube one km$^3$ neutrino observatory will collect large samples of
neutrino interactions, allowing for observations with small statistical
errors.   To make maximum use of this statistical power,  it is also being
designed to minimize systematic errors, via a variety of different calibration
techniques.  LED and laser light sources are a key part of many of these
calibration techniques.  To a significant extent, they mimic cascade
($\nu_{e}$) interactions, allowing fairly direct tests of cascade reconstruction
techniques.  This contribution will survey the light sources and discuss
selected calibration studies.  
}

%\email{aastex-help@aas.org}

%%%%%%%%%%%%%%%%%%%% B E G I N   D O C U M E N T%%%%%%%%%%%%%%%%%%%%%%%
%\begin{document}
\maketitle

\section{Introduction}

The main goal of IceCube~\cite{icecube} is to detect cosmic neutrinos of all flavors
in a wide energy range, from $\sim$100 GeV to $\sim$100 EeV and
search for their sources. 
When complete,  the IceCube detector will be composed of up to  $4800$ Digital Optical Modules (DOMs)
on $80$ strings spaced by $125$ m. 
The array covers an area of one km$^{2}$  from 1.45 to 2.45 km below the surface~\cite{icecube-nim}.

High energy neutrinos are detected by observing the Cherenkov radiation from secondary
particles produced in neutrino interactions inside or near the detector.
Muon neutrinos in charged current (CC) interactions
are identified by the final state muon track ~\cite{amanda-muon}.
Electron and tau neutrinos in 
CC interactions, as well as all flavor neutrinos initiating neutral current (NC)
interactions are identified by observing electromagnetic or hadronic showers (cascades).
For example,  up to $\sim$10 PeV, electromagnetic showers initiated by the final state electron 
can be  approximated as expanding light spheres originating from a point source.
A $10$ TeV cascade triggers IceCube optical modules out to a radius of about $130$ m~\cite{francis}.
Cascade reconstruction  is expected to have limited pointing capability
but good energy resolution, $0.11$ in $\log_{10}(E)$~\cite{icecube-proposal}.   
The good energy resolution and
low background from atmospheric neutrinos makes cascades attractive for diffuse
extraterrestrial neutrino searches~\cite{kowalski-fluxes}.

Artificial light sources are of particular importance in IceCube.
Each DOM includes $12$ LEDs (flashers) as a calibration source.
As shown in Fig.\ref{ICRC0761_fig1}, one string also holds a nitrogen laser
with absolute calibration that serves as a ``standard candle''. 
The flashers and standard candle (SC) are used for a wide variety of purposes:
timing, charge amplitude and geometry calibrations, to measure the optical properties
of the ice (a key problem for IceCube), and to mimic cascades.  The flasher light output is
comparable to cascades with energies up to about $500$ TeV, while the standard candle output
is comparable to cascades with energies up to about $30$ PeV.  For the ice studies,
the availability of flashers at different depths is critical,  allowing comparisons of ice
properties at different depths.  
These studies build on the lessons learned from AMANDA, which pioneered the use of
artificial light sources~\cite{AMANDA2}. 

In this report we present the results of a few selected studies performed with the flashers
and standard candle: geometry and timing calibrations, and the position
resolution of cascade algorithms.
This work uses data collected during 2006 with
9 strings that had been deployed in IceCube at that time.

\begin{figure}
\begin{center}
\includegraphics[width=0.4\textwidth]{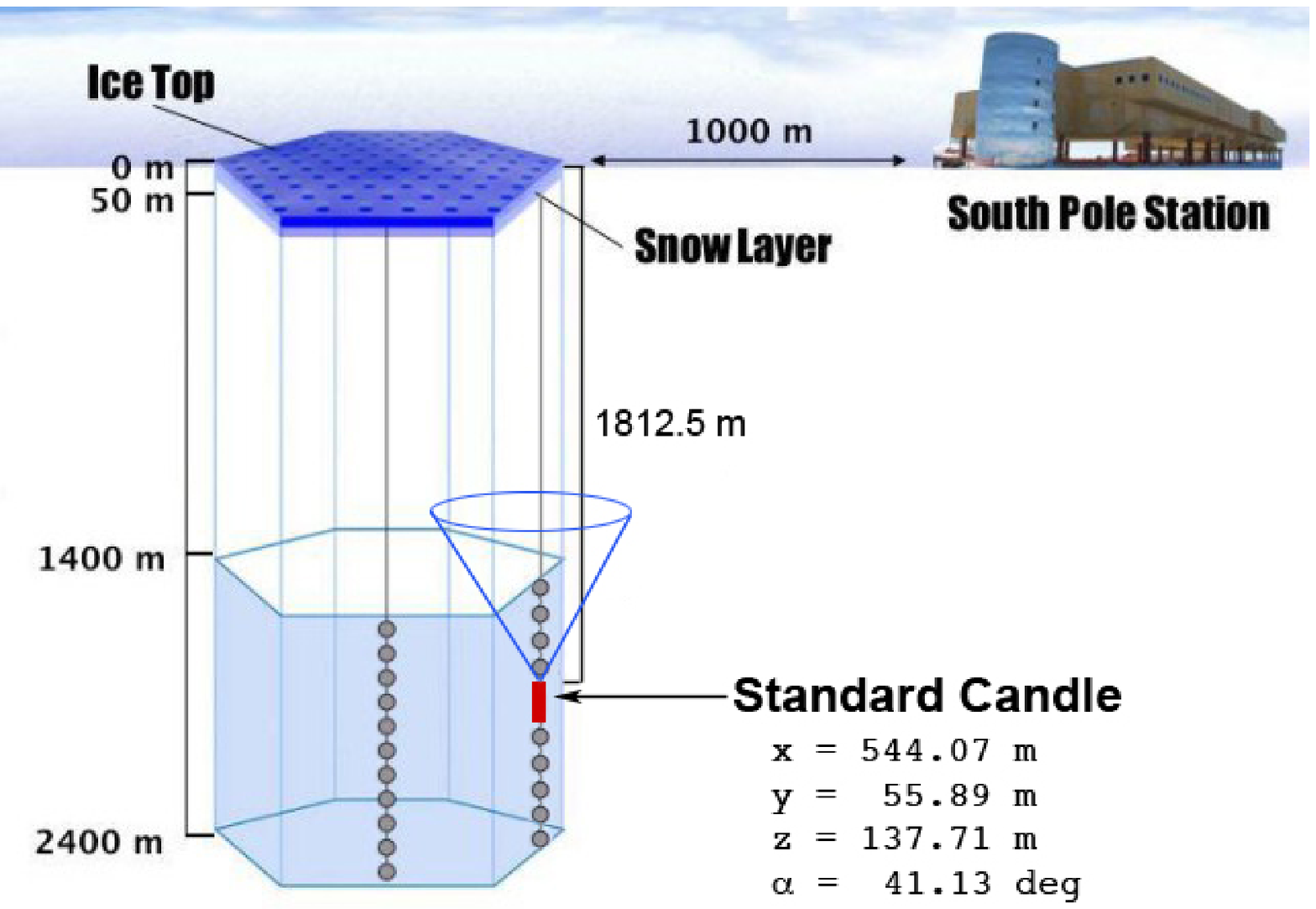}
\end{center}
\caption{
Schematic of the IceCube detector showing the location of the Standard Candle. For clarity, only two out of $80$ strings are shown.}\label{ICRC0761_fig1}
\end{figure}

\section{LED Flashers} 

\noindent
Each DOM contains a flasher board which
holds  twelve $405$ nm LEDs. 
Six of them point horizontally outward and six point upwards at $\sim$48 degrees. 
They are mounted on the top and the bottom of the flasher board respectively, cf. Fig.\ref{ICRC0761_fig2}. 
The LEDs are individually flashed with a programmable pulse width and amplitude. 
Typical flasher runs last $500$ s with the LEDs firing at $10$~Hz at full brightness, 
with nominal width of $10$~ns. 
\subsection{Geometry Calibration}
 \begin{figure}
\includegraphics[height=0.3\textwidth]{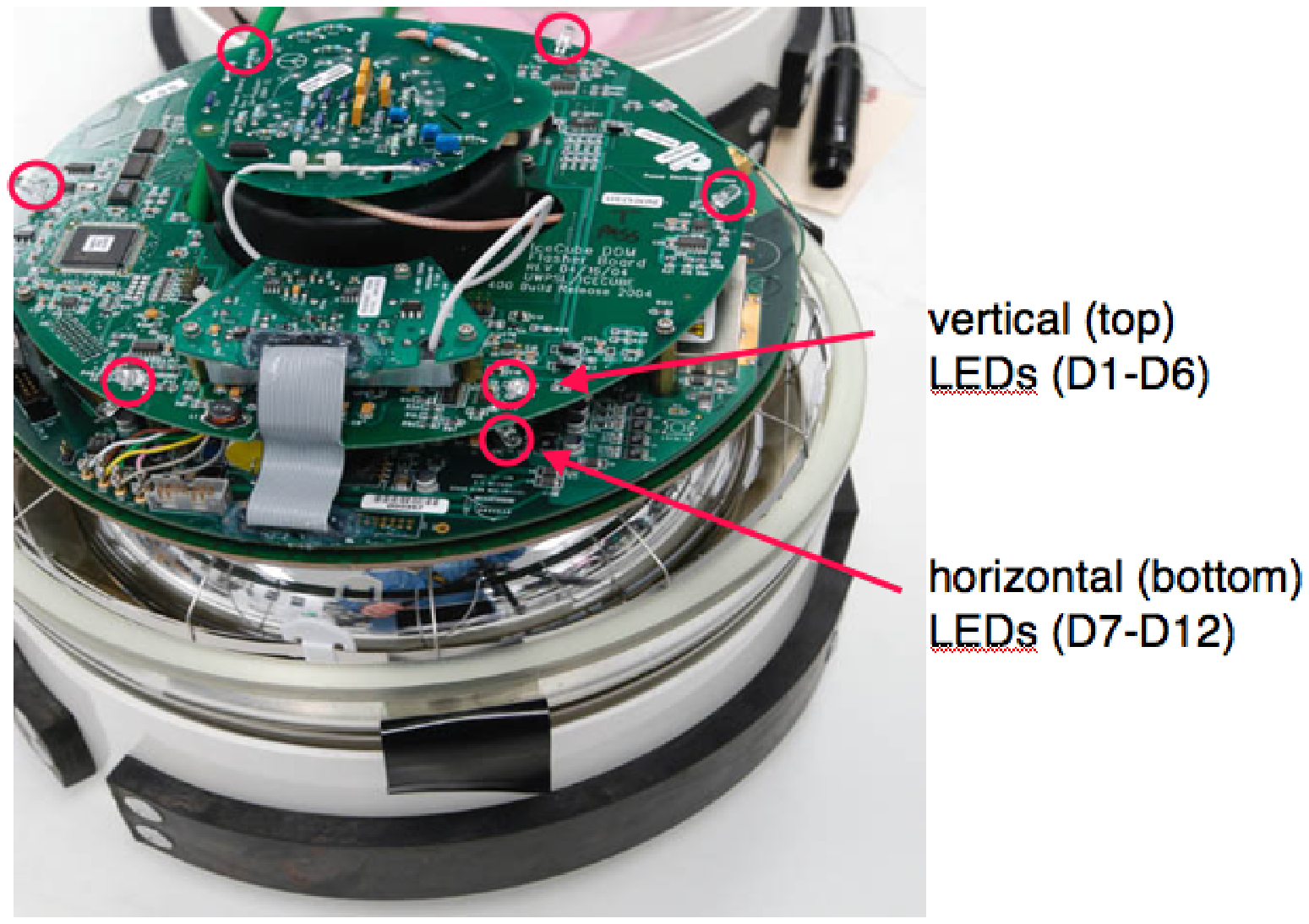}
\caption{
Digital Optical Module with six horizontal and six vertical LED flashers. 
}\label{ICRC0761_fig2}
\end{figure}
The LED flashers were used to calibrate the position of the DOMs.
Figure~\ref{ICRC0761_fig3}a) shows a schematic of one study that was used to measure the relative 
depth of DOMs on different strings.  The LEDs in a DOM on one string were pulsed and 
arrival times for nine nearby DOMs on a neighboring string were analyzed.  
\begin{figure}
\begin{center}
\includegraphics[width=0.4\textwidth]{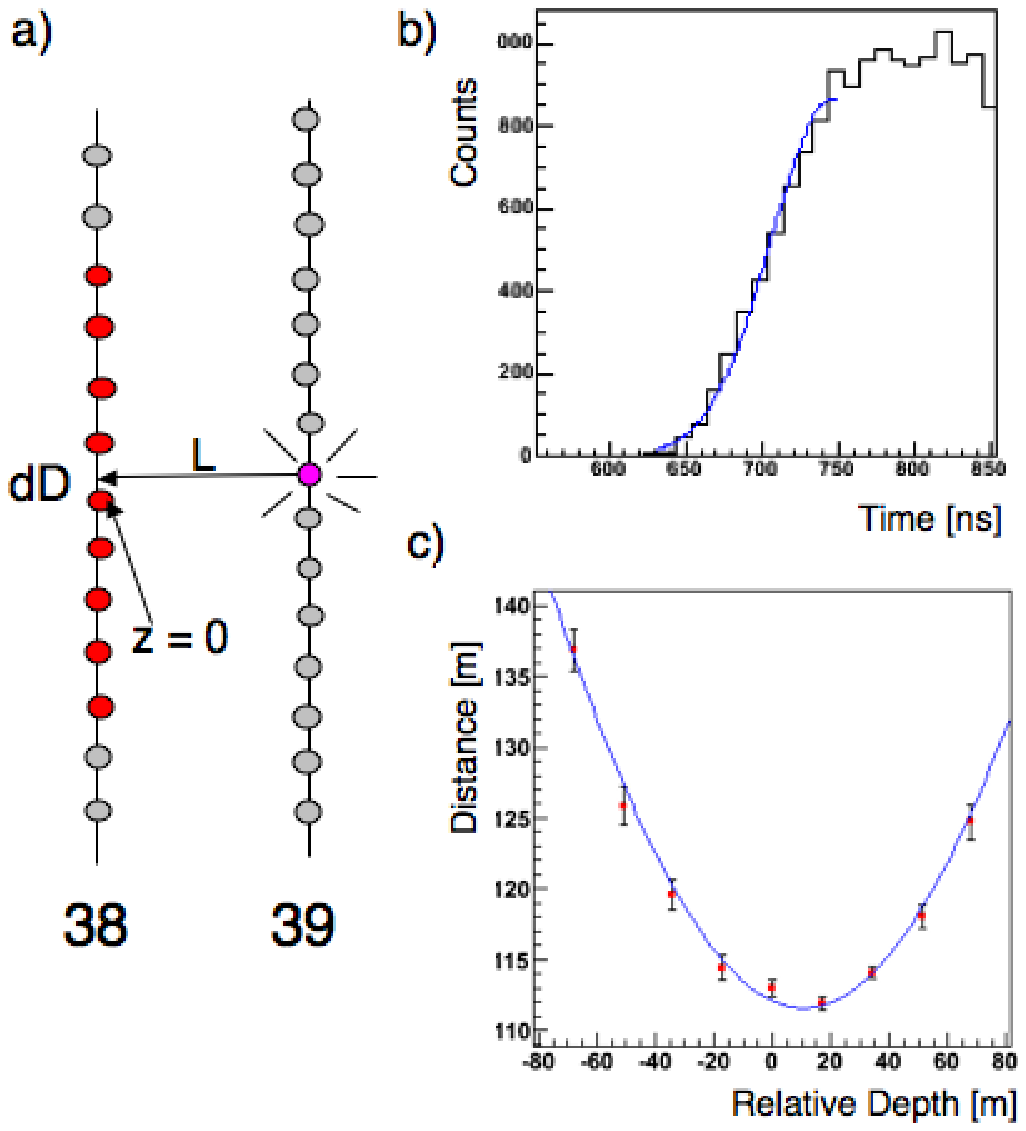}
\end{center}
\caption{
a) Schematics of the interstring detector geometry measurement. The 
flasher light from LEDs on DOM $39$-$15$ is seen on a neighboring string.
b) The earliest hit time distribution for light detected at DOM $38$-$10$.  
c)  The earliest hit time observed at DOMs on string $38$ shown as a 
function of the relative depth between the observing DOMs and the flashing DOM. 
}\label{ICRC0761_fig3}
\end{figure}
The time of the earliest hit $t_{0}$ was derived from the photon arrival time distribution, cf.  Fig.~\ref{ICRC0761_fig3}b), by 
fitting a Gaussian in the turn-on region: $t_{0}=\mu-3\sigma,$
where $\mu$ and $\sigma$ are the mean and sigma of the Gaussian. 
The uncertainty is determined by propagating the errors on the fit parameters.
The arrival times of the earliest hits are converted to distances (assuming that
there is no scattering, appropriate for the first photon seen).  Figure~\ref{ICRC0761_fig3}c)
shows these distances versus the relative depth from the deployment records. 
This distribution was fitted with a hyperbola to determine the relative 
depth and lateral separation between the two strings.
The position of the minimum gives the relative depth
and is used to correct the string position determined from deployment and survey data.
Systematic uncertainties in the determination of the lateral separation are under study.

\subsection{Timing Calibration}
Flasher data are used also to verify the system timing resolution.
The method is to flash an LED on a DOM and measure the arrival time of light 
reaching a nearby DOM, as shown in Fig.~\ref{ICRC0761_fig4}a). 
The earliest photons are likely not scattered, hence the difference in timing 
between the two DOMs  reflects the time in ice. 
The distance between DOMs on the same string ( $\sim$17m) is smaller than 
the light scattering length in ice ($\sim$25m) and the light intensity 
is high enough so that direct light is seen on neighboring DOMs.
The resolution is dominated by electronics and timing uncertainties.
A distribution of the first photon arrival time for a single receiving DOM
is shown in Fig.~\ref{ICRC0761_fig4}b). 
The resolution for most DOMs was found better than $2$~ns, as shown in Fig.~\ref{ICRC0761_fig4}c),  confirming 
the precision of the time synchronization procedure. 
The results are consistent with an alternative method which uses muon tracks~\cite{icecube-nim}. 
\begin{figure}
\begin{center}
\includegraphics[width=0.4\textwidth]{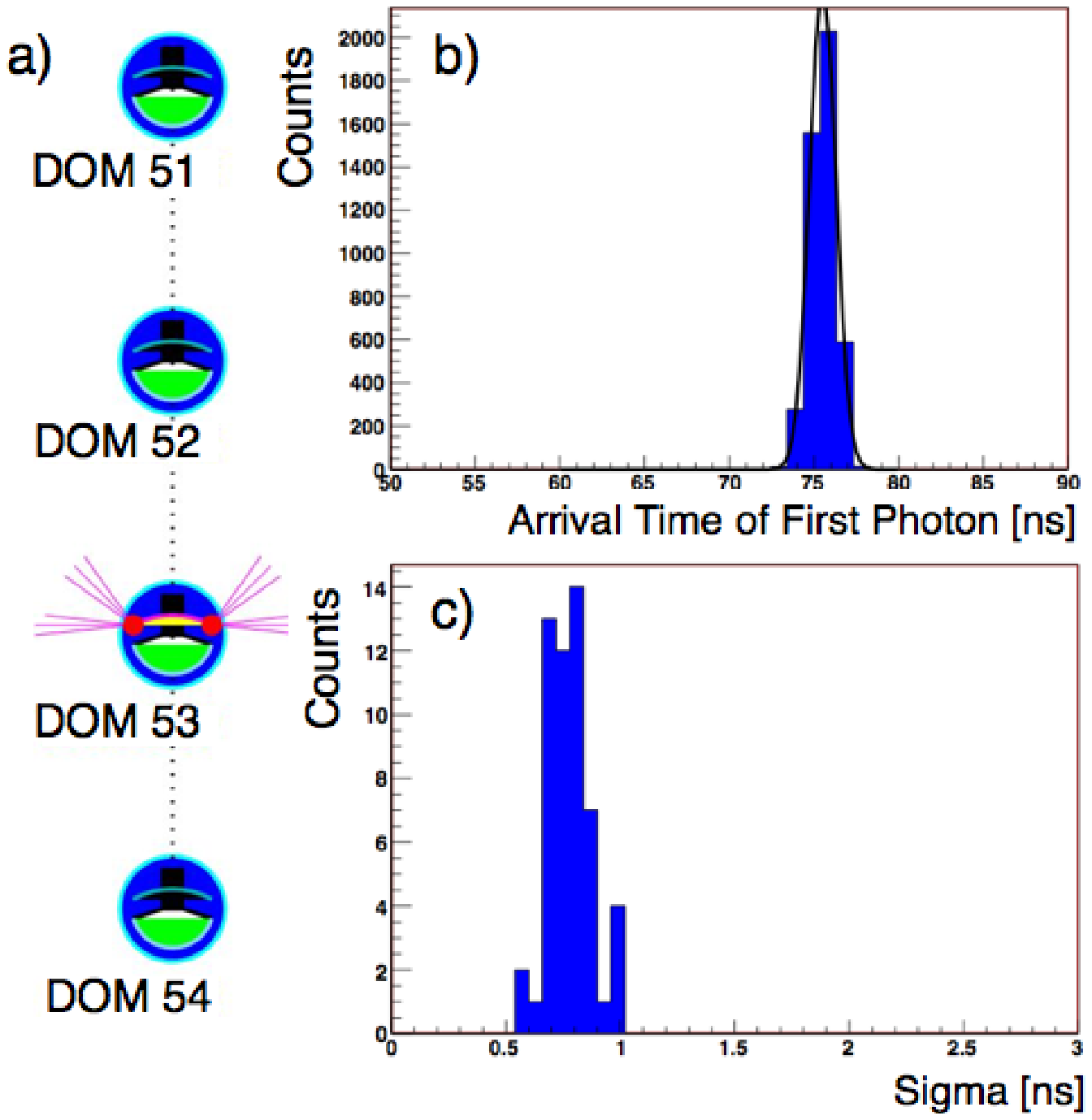}
\end{center}
\caption{
a) DOM $53$ is flashing.  b) Photon arrival time delay at DOM $52$ when DOM $53$ is flashing.  
c) RMS variation of time delay measured with flashers for $59$ DOM pairs on an IceCube string.
}\label{ICRC0761_fig4}
\end{figure}

\section{Standard Candle}

The Standard Candle (SC) is an in-situ calibrated $N_{2}$ pulsed laser,
which emits light with a wavelength of $337$ nm. 
It is used to study cascade reconstruction, and to provide a method for calibrating
the cascade energy scale which is independent of Monte Carlo simulations.
At 100 \% 
intensity, the SC generates $(4.0\pm0.4)\times10^{12}$  photons 
which are emitted at an angle of 41$^{\circ}$ with respect to the candle axis, as is shown in
Fig.~\ref{ICRC0761_fig1}. The $41^{\circ}$ angle was
chosen to  approximately match the Cherenkov radiation from a cascade.  Although the light
distribution initially matches that of a cascade, the wavelength of $337$ nm is shorter
than most of the Cherenkov radiation observed in IceCube.
This results in $\sim$10 \%  shorter absorption and scattering lengths,
and requires  adjustments to the amplitude calibrations. Pre-deployment calibration 
and internal power measurement contribute to 10\% uncertainty in light output. 
The light intensity is determined on a pulse-by-pulse basis.
 
The SC is equipped with  an adjustable attenuator that can reduce the light 
output down to 0.5\% of the full scale.  This is used to study detector (especially photo-multiplier tubes) 
non-linearities.  We plan to deploy one additional standard 
candle, which will point downwards or to the side, allowing
different cascade geometries to be studied in future. 

\subsection{Reconstruction Results}

Figure~\ref{fig5} shows an event with the SC at full laser intensity.
\begin{figure}
\begin{center}
\includegraphics[width=0.22\textwidth]{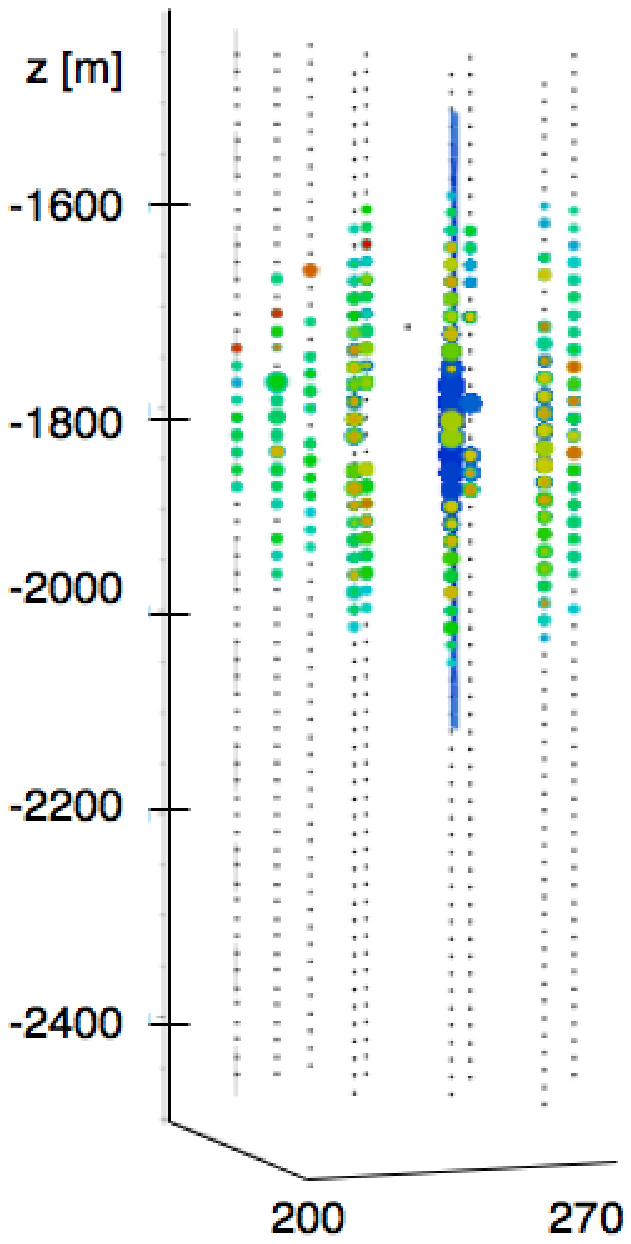}
\end{center}
\caption{
Standard Candle event display with 162 DOMs hit.  The size of the circles is proportional to the signal 
amplitude, while the color distinguishes between relative photon arrival times in the DOMs. 
}\label{fig5}
\end{figure}
Results from the SC laser events reconstruction as cascades 
are shown in Fig.~\ref{fig6}.   The dashed histogram shows the center-of-gravity (COG) $x$ position.
The COG is calculated for each event as the signal amplitude weighted mean of all hit DOM positions.  
The mean COG $x$ position, about $512$ m, is about $30$ meters from the actual SC $x$ position
of $544$ m (shown as a dashed-dotted line).
The reason for this discrepancy is that the SC is on a string at the edge of
the $9$ string array, and the COG is pulled toward the center of the array. 
The COG is used as a first approximation for a full maximum-likelihood reconstruction
algorithm~\cite{dima}.  This algorithm considers the photon arrival times at all of
the other DOMs.  It finds an $x$ position (continuous histogram) within 10 m of the actual SC position for about
99\% of the events. Similar results have been obtained for $y$ and $z$ vertex positions.
The fact that the algorithm can find the position so well for asymmetric events (with DOMs
on only one side of the SC) gives us confidence in the reconstruction algorithm
accuracy.

\section{Summary}

The IceCube flasher LEDs and standard candle laser are used for a variety of calibration and
verification studies, including geometry and timing calibrations, and studies of ice properties.  
It has been demonstrated that for most DOMs the timing resolutions is better than $2$~ns and 
the DOM positions are known to 1~m.
These studies will help IceCube reduce the systematic
errors for various physics analyses.  
\begin{figure}
\begin{center}
\includegraphics[width=0.4\textwidth]{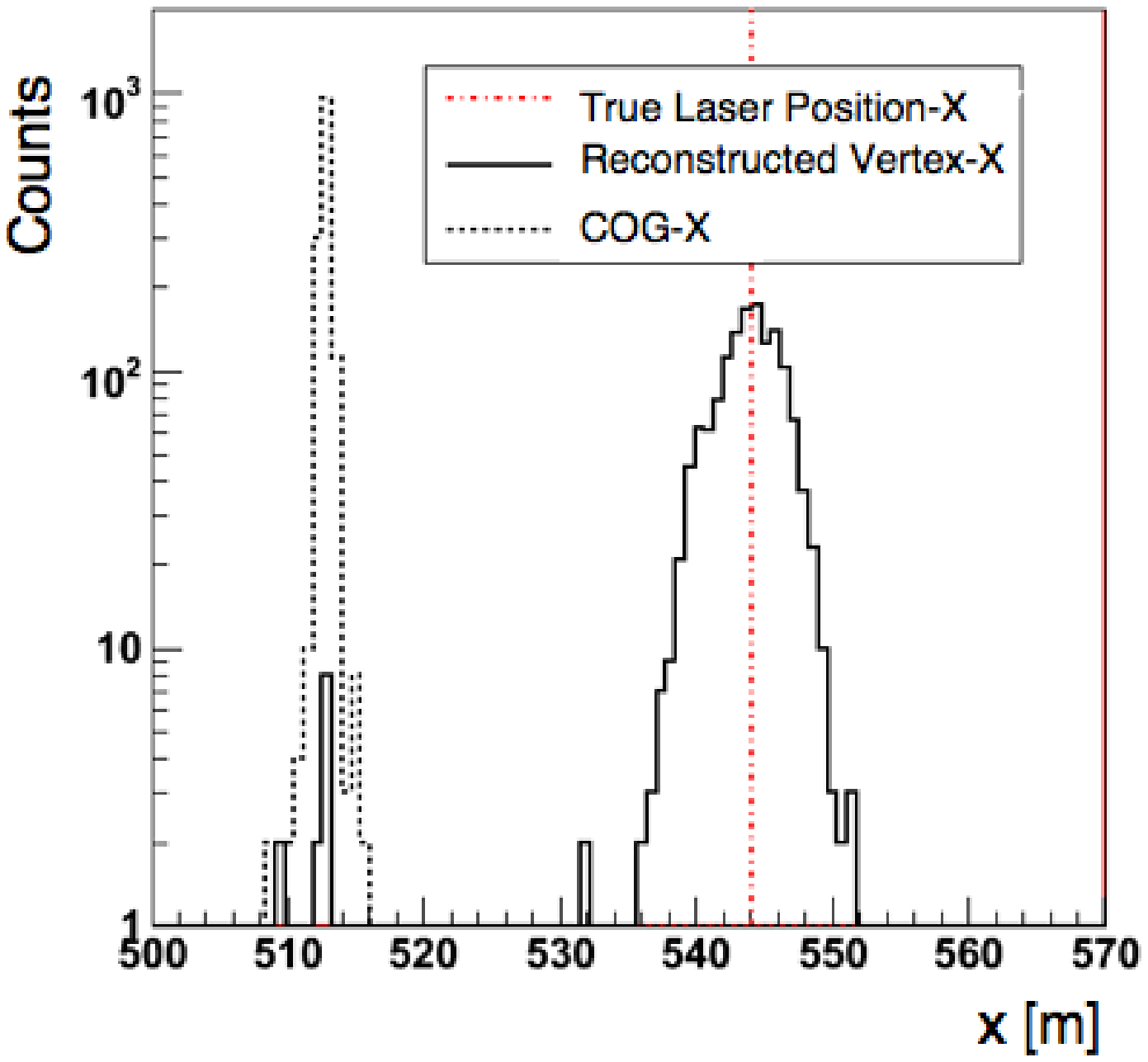}
\end{center}
\caption{
The reconstructed $x$-vertex position of 'cascades' from SC events.  
The dashed histogram shows the center-of-gravity (COG) position,
and the continuous histogram shows the reconstructed vertex position. The dashed-dotted line is the 'true' SC laser 
position in the detector ($x=544.1$ m).  }\label{fig6}
\end{figure}
Artificial light sources have also been used to study the position reconstruction
performance of cascade reconstruction algorithms, and to study the absolute energy scale of the
detector.  In future, they will be used to study also the energy and directional reconstruction
of more advanced algorithms.  

\vspace*{0.2cm} 

We acknowledge the support from the following agencies:
National Science Foundation-Office of Polar Program,
National Science Foundation-Physics Division, 
University of Wisconsin Alumi Research Foundation, 
and Division of Nuclear Physics-Department of Energy.

\bibliographystyle{plain}

%\end{document}

% pointsource papers
%
%icrc_V20.pdf (2005 search)
%icrc_2-0.pdf (PeV search)
%icrc0764_v5.pdf (IceCube point source analysis)
%icrc0851_06.pdf (multipole search)
%icrc07_KS_V7 (cluster search, Konstancja)
%icrc_porrata_2007.pdf (all sky transient search)
%NtoO_Mexico_ICRC...  (target of opportunity)
%
\setcounter{figure}{0}
\setcounter{table}{0}
%%
% International Cosmic Ray Conference 2007 Merida Yucatan Mexico
% In This file you will find detailed instructions to correctly
% typeset your document.
%
%
%

%Class Requeried
%\documentclass{article}
%The ICRC Style
%\usepackage{icrctc07}

%The paper title
\title{Neutrino Point Source Search Strategies for AMANDA-II and Results from 2005}
%Short title to print in the headers to the final publication (Not showed in this print).
\shorttitle{AMANDA-II 2005 Neutrino Point Source Search}
%All paper authors
\authors{J. Braun, A. Karle, and T. Montaruli for the IceCube Collaboration}
%Short title to print in the headers to the final puplication (Not showed in this print).
\shortauthors{J. Braun et al.}
%All the affiliations.
\afiliations{Physics Department, University of Wisconsin, Madison, WI 53706, USA}
\email{jbraun@icecube.wisc.edu  For a full authorlist, see the special section in these proceedings.}

%The abstract.
\abstract{Current point source searches mostly utilize only direction and time
of the reconstructed event; furthermore, they reduce available information
by grouping events into sky bins. In this analysis we use a search based on
maximum likelihood techniques, utilizing both event angular resolution and energy,
to enhance our ability to detect point sources. Especially, use of energy
information allows us to fit the spectral index of a hypothetical source
simultaneously with flux. This method improves both sensitivity and discovery
potential of the AMANDA-II array by greater than 30\%. The method
can naturally be applied to IceCube and allows superposition of data from detectors with
different sensitivity and angular resolution, such as the IceCube array
which changes and improves with each season of construction.

}

%%%%%%%%%%%%%%%%%%%% B E G I N   D O C U M E N T%%%%%%%%%%%%%%%%%%%%%%%
%\begin{document}
\maketitle
%Begin the section.

\section{Introduction}
\vspace{-2mm}
Pinpointing the origin of high energy cosmic rays is one of the
most important goals of neutrino astrophysics.  Observation of a
high energy neutrino source would provide clear indication of hadronic
processes associated with cosmic rays.  Neutrinos are neither deflected
by magnetic fields nor significantly attenuated on transit to Earth,
making them excellent astronomical messengers in the $>$TeV universe.

The Antarctic Muon And Neutrino Detector Array (AMANDA), a subdetector of
the IceCube Observatory, is composed of 19 strings with 677 total optical
modules located 1500 m -- 2000 m below the ice surface at the Geographic South Pole.
Muons produced by charged-current $\nu_{\mu}$ and $\bar{\nu}_{\mu}$ interactions
produce tracks of \v Cerenkov light and are reconstructed with 1.5$^o$--2.5$^o$
median angular resolution \cite{ICRC1090_ps}.  The large background of muons from cosmic
ray interactions in the atmosphere precludes  $\nu_{\mu}$ and $\bar{\nu}_{\mu}$
searches in half of the sky, but for $\delta >$ 0 cosmic ray muons are attenuated
by Earth leaving a relatively pure atmospheric neutrino background.

Detection of an extraterrestrial high energy neutrino source has so far
eluded the neutrino telescope community.  To probe lower fluxes, either larger
neutrino telescopes must be built, more sophisticated point source analysis
techniques \cite{ICRC1090_t2} \cite{ICRC1090_aart} must be developed to better utilize data
from existing experiments, or both \cite{ICRC1090_chad}.

\section{Method}
\vspace{-1mm}
An unbinned maximum likelihood search method is used in contrast to previous
AMANDA point source analyses \cite {ICRC1090_ps}.  The past binned
search method makes use of a single statistic, namely ``How many events
are within bin radius `b'" and a background estimation to make a
statement about the existence of a source at any particular
position in the sky. It is reasonable to think the use of additional
information must enhance ability to search for point sources.
Additional information includes:
\begin{itemize}
\item Events outside the search bin
\item The distribution of events within the search bin
\item Event energy estimation.
\end{itemize}
The energy distribution of a hypothetical E$^{-2}$ source is drastically different
from that of the atmospheric neutrino background. If high energy events are observed,
such events are not very compatible with atmospheric neutrino background and enhance
discovery potential. Conversely, if high energy events are not observed, the method is able
to reject the signal hypothesis with higher confidence. In AMANDA, the number of optical modules,
or channels, hit by at least one photon during an event correlates with event energy.  By using the
difference in the distribution of number of hit channels, shown in figure \ref{Fig:6}
for various energy spectra, events are more accurately classified as signal
or background.

\begin{figure}
\mbox{\includegraphics[width=2.8in]{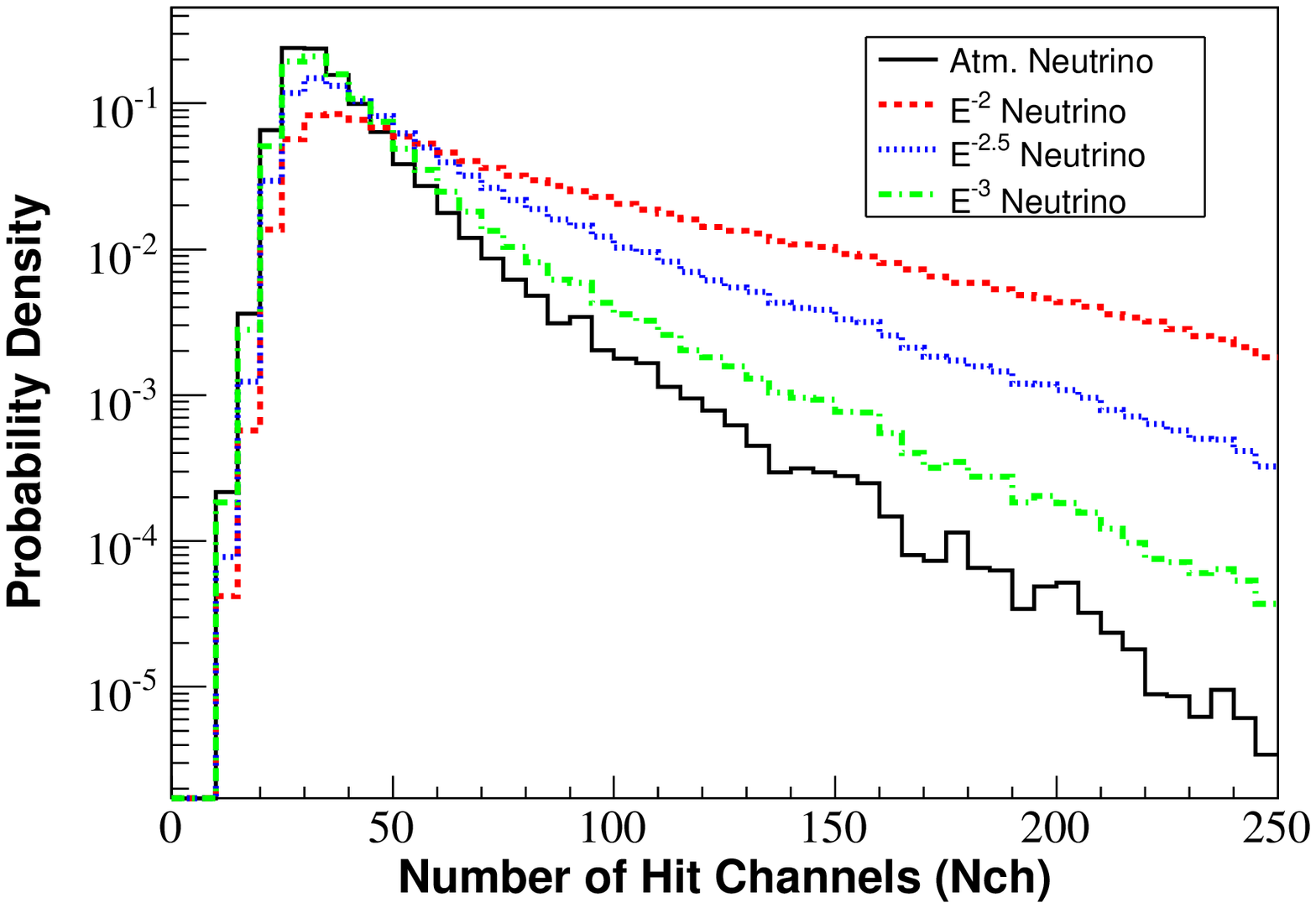}}
\caption{\label{Fig:6} Number of hit channels (Nch)
PDF for simulated atmospheric neutrinos and various signal spectra}
\end{figure}

At a hypothetical source position $x_o$, the data is modeled as an
unknown mixture of background and events produced by the source.  Each
event near the source declination is assigned a likelihood of belonging
to the source.  This source PDF is the product of probability functions describing the
detector point spread, which is zenith dependent, and number of channels hit (Nch):

\begin{displaymath}
\mathcal{S}_{i}(x_i, x_o, \theta, N_{ch}, \gamma) = P(x_i|x_o, \theta)P(N_{ch}|\gamma),
\end{displaymath}

where $\gamma$ is the source spectral index. The detector point spread is modeled
as a two dimensional Gaussian:

\begin{displaymath}
P(x_i|x_o, \theta) = \frac{e^{-\frac{|x_i-x_o|^2}{2\sigma^2(\theta)}}}{2\pi\sigma^2(\theta)}.
\end{displaymath}

The Gaussian width $\sigma$ is fitted to simulation.  The background PDF depends on $P(N_{ch}|Atmos. \nu)$,
the probability of obtaining the observed Nch value from atmospheric neutrinos, and event density within the band.
The full likelihood function is a combination of signal and background probabilities $\mathcal{S}$
and $\mathcal{B}$ over all events in the declination band ranging $\pm$5$^o$ of the source position $x_o$ and
containing N total events:

\begin{displaymath}
\mathcal{L} = \prod^{N}_{i}\bigg(\frac{n_s}{N}\cdot\mathcal{S}_{i}(x_i, x_o, \theta, N_{ch}, \gamma) +
(1 - \frac{n_s}{N})\cdot\mathcal{B}_{i}(N_{ch})\bigg).
\end{displaymath}

The signal and background PDF are normalized such that
the free parameter $n_s$ describes the number of signal events present.
The quantity $-log(\mathcal{L})$ is minimized with respect to $n_s$ and $\gamma$, obtaining best
estimates of signal strength $\hat{n}_s$ and spectral index $\hat{\gamma}$.  The logarithm of the likelihood
ratio

\begin{displaymath}
\lambda = log\frac{\mathcal{L}(\hat{n}_s, \hat{\gamma})}{\mathcal{L}(n_s=0, Atmos. \nu)}
\end{displaymath}

is used to determine significance and flux limits for each observation.

Significance is calculated by comparing the observed value of $\lambda$ to
the distribution obtained from randomized data.  Adding a simulated signal
flux shifts the distribution of $\lambda$ to higher values, corresponding
to higher significance.  Discovery potential is measured by calculating
the signal flux necessary to increase $\lambda$ such that a given significance
is exceeded in a given percentage
of trials.  Feldman-Cousins confidence intervals \cite{ICRC1090_FC} are constructed
knowing the response of $\lambda$ to increasing signal flux and are used
to calculate sensitivity and flux upper limits.  A 30\% improvement in
sensitivity and discovery potential using the unbinned maximum likelihood method
is shown in both sensitivity and discovery potential in figure \ref{Fig:2}.

Since signal spectral index is estimated simultaneously with flux, the
obtained value of $\hat{\gamma}$ serves as an estimate of spectral
index.  The value -2log$\mathcal{L}/\hat{\mathcal{L}}$ approximately follows
a chi-square distribution with two degrees of freedom when signal strength $n_s$ and
spectral index $\gamma$ are simultaneously varied.  Using this approximation, confidence contours in signal
strength and spectral index are shown in figure \ref{Fig:1}.  The signal strength $n_s$
is typically overestimated by approximately 10\% due to mismatch between the true point spread
function and the Gaussian approximation used in this analysis.  This effect is measured using
detector signal Monte Carlo and is calibrated away.
\begin{figure}[t]
\mbox{\includegraphics[width=2.8in]{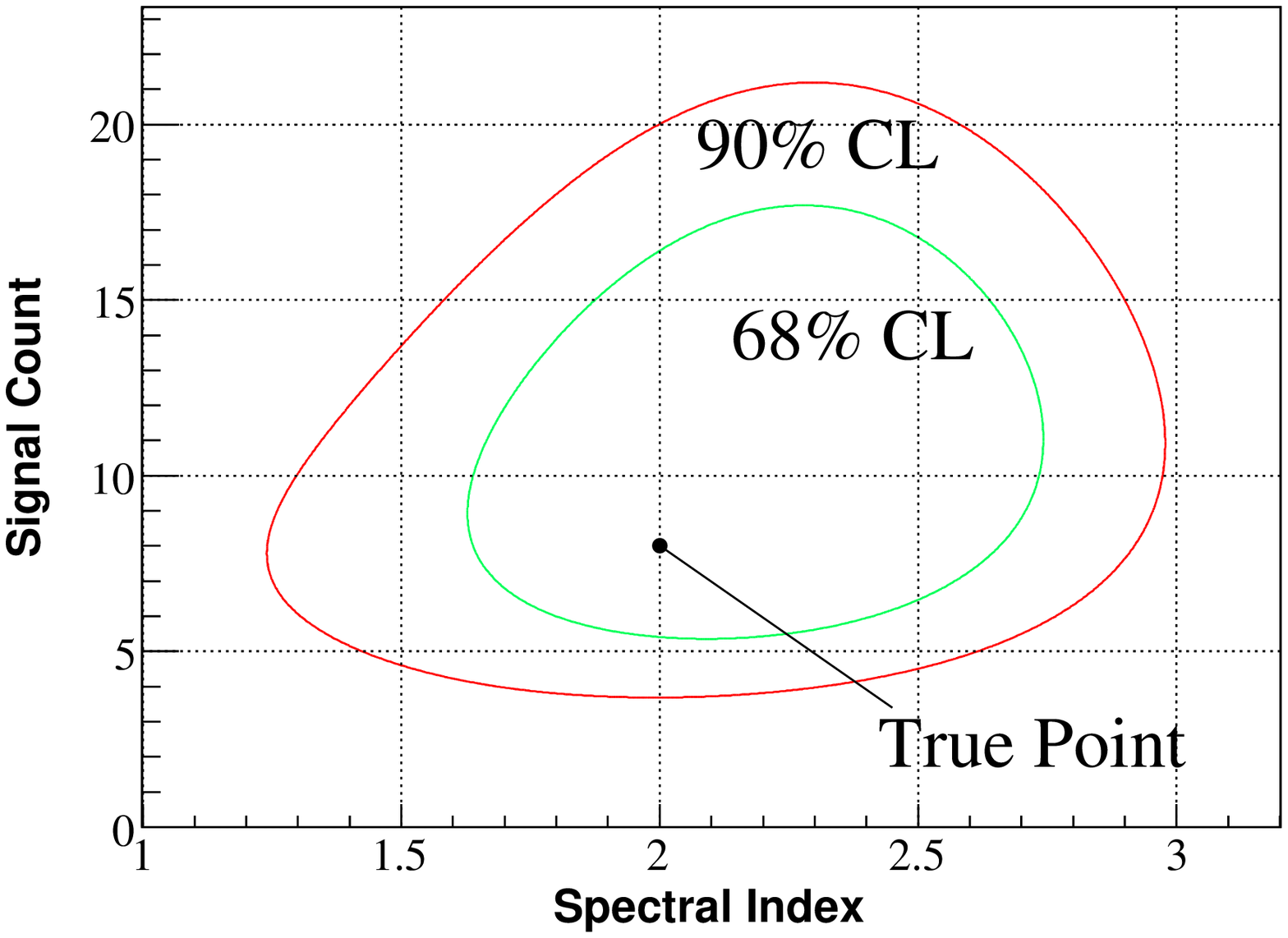}}
\mbox{\includegraphics[width=2.8in]{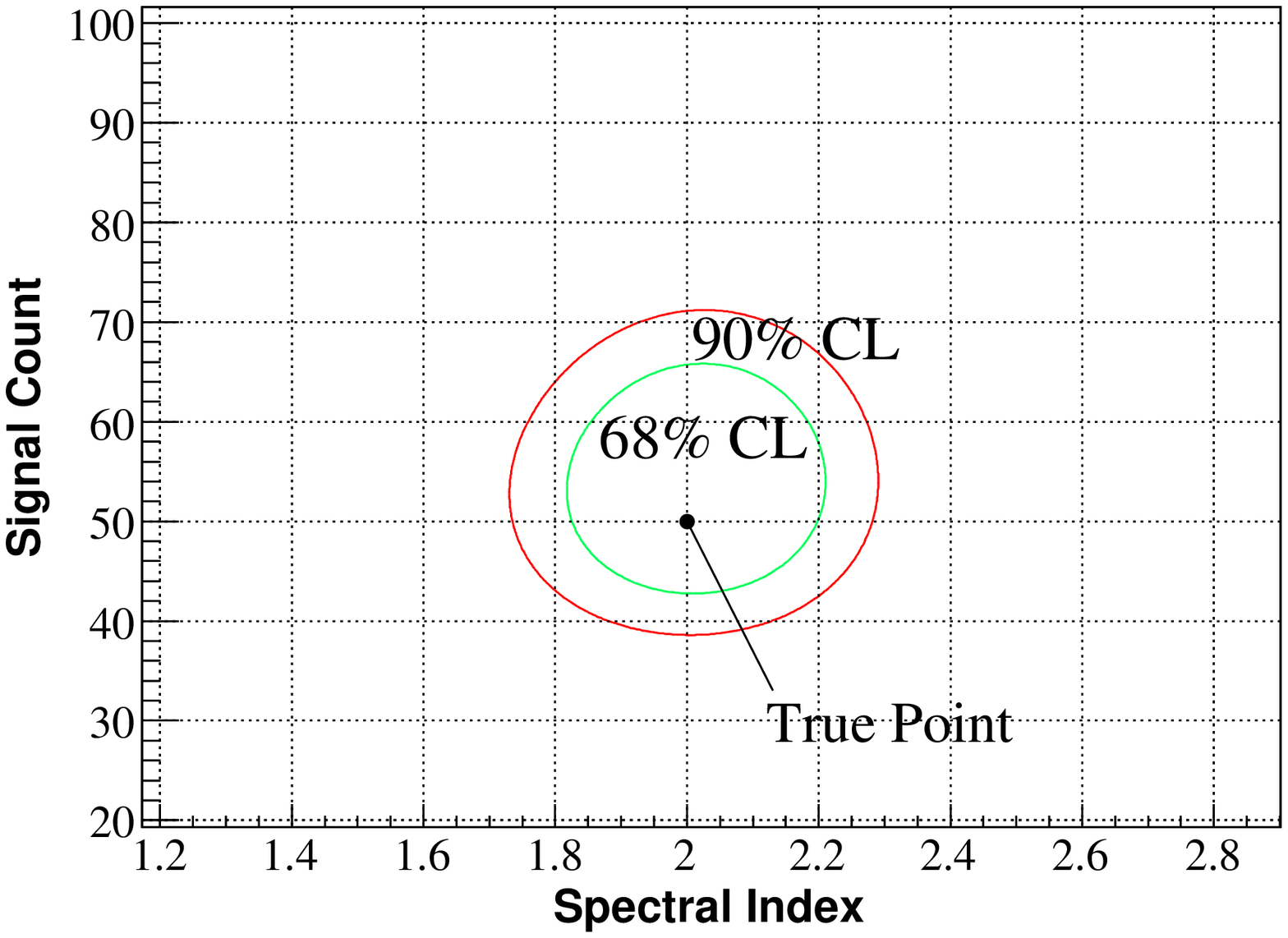}}
\caption{\label{Fig:1} Confidence estimates in source strength and
spectral index for a case of 8 E$^{-2}$ signal events (top) and 50 E$^{-2}$ signal events (bottom)}
\end{figure}
As an example, suppose Markarian 421 ($\delta$ = 38.2$^o$) produces 8 events in the detector with an
E$^{-2}$ energy spectrum. Application of this method to the coordinates of Markarian 421 would yield a
53\% chance of discovery at 5$\sigma$ confidence level.  Preliminarily, 1$\sigma$ spectral index
confidence bounds for this source would be better than $\pm$0.5 around $\hat{\gamma}$ for an energy
spectrum near E$^{-2}$.

Another benefit is the ability to combine data from detectors
with different angular resolution.  A binned search regards each
event equally and bin radius must be optimized given the combination of datasets; however,
this method can recognize which dataset the event is from and use
the appropriate point spread distribution to more accurately describe the event.
This benefit is particulary important during the construction phase of IceCube,
as detector resolution will improve each year.

\section{Data Sample}
\vspace{-2mm}
Data are taken during the austral winter from mid-February 2005 through October 2005.
Accounting for the time the detector is down and a brief time the detector is dead
following each event yields 199.3 days of detector livetime and 1.8$\cdot$10$^9$ events.  Most events are recorded
from a multiplicity trigger requiring at least 24 optical modules register photon hits within 2.5 $\mu$s.
False hits produced by crosstalk, isolated hits caused by PMT dark noise,
and hits from 154 modules with either an abnormal dark noise rate or position outside the main detector
volume are removed.  Remaining hits from 523 optical modules are
reconstructed as muon tracks with increasing accuracy and cpu requirements \cite{ICRC1090_reco},
and zenith filters are applied to remove the majority of cosmic ray muon background.
Filtering is divided into levels to maximize CPU efficiency while retaining the vast majority
of neutrino events \cite{ICRC1090_ps}.  5.2 million events remain in the final filtered sample,
mostly misreconstructed muons.
Neutrino events are chosen from this sample to minimize average flux upper limit \cite{ICRC1090_mrf} based on
reconstruction and topological criteria including a track angular resolution estimate \cite{ICRC1090_Till},
the ratio of upgoing reconstruction likelihood to downgoing likelihood, the distribution
of hits along the track, and track length.  Events are divided into $5^{o}$ declination bands, and
optimization is performed simultaneously on all parameters for E$^{-2}$ and E$^{-2.5}$ source spectra.
A compromise cut is applied between the E$^{-2}$ -- E$^{-2.5}$ optimization.
Optimized point source sensitivity (figure \ref{Fig:2}) shows a $\sim$30\% improvement against the
binned method uniform over the sky.  Discovery potential is similarly
improved.  After the cut, 887 events remain above $\delta=10^o$, with any 10$^{o}$
declination band containing 50-150 events.  A large number of misreconstructed
muons add to atmospheric neutrinos in the final sample below $\delta=10^o$.

\begin{figure}
\mbox{\includegraphics[width=2.8in]{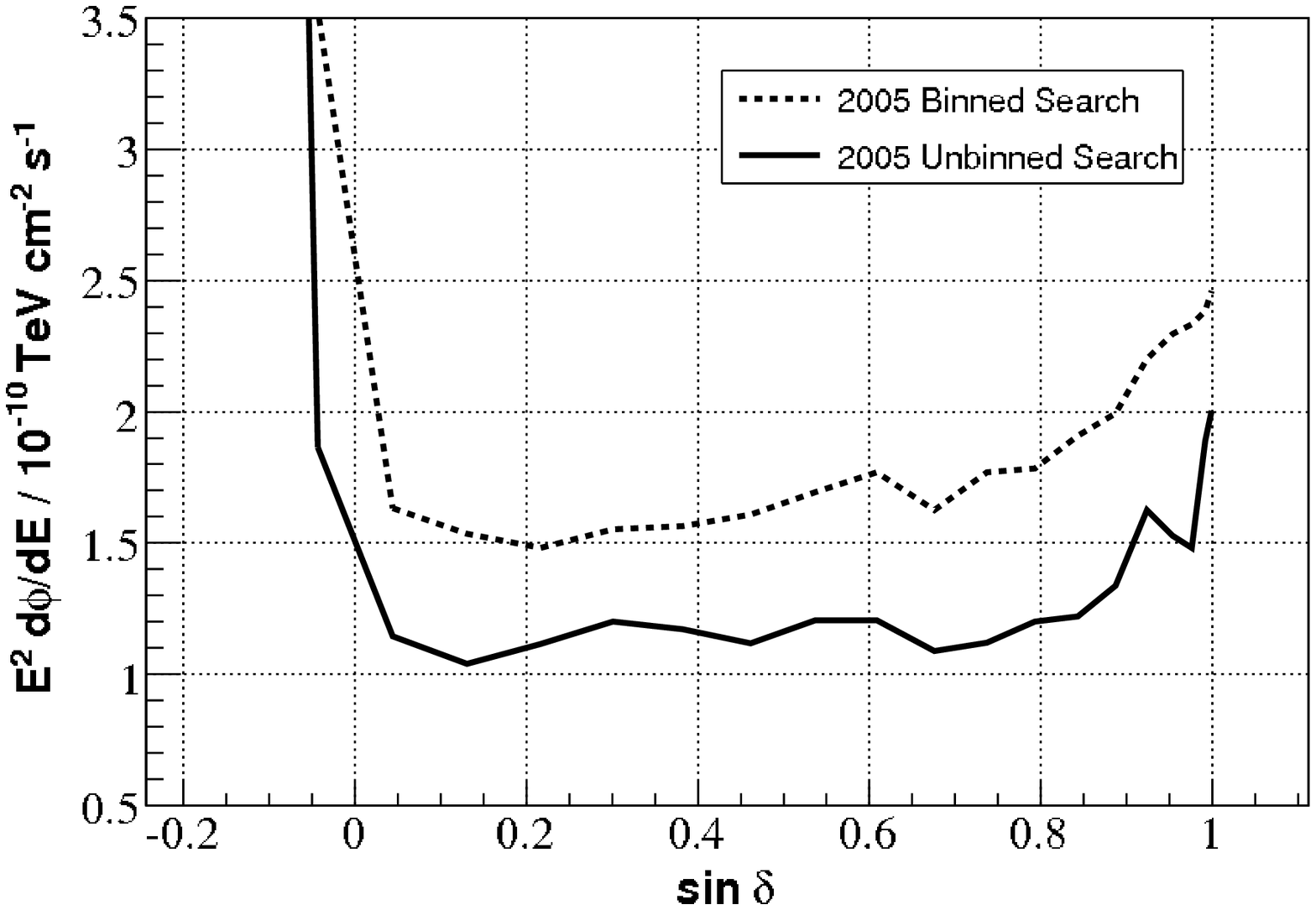}}
\mbox{\includegraphics[width=2.8in]{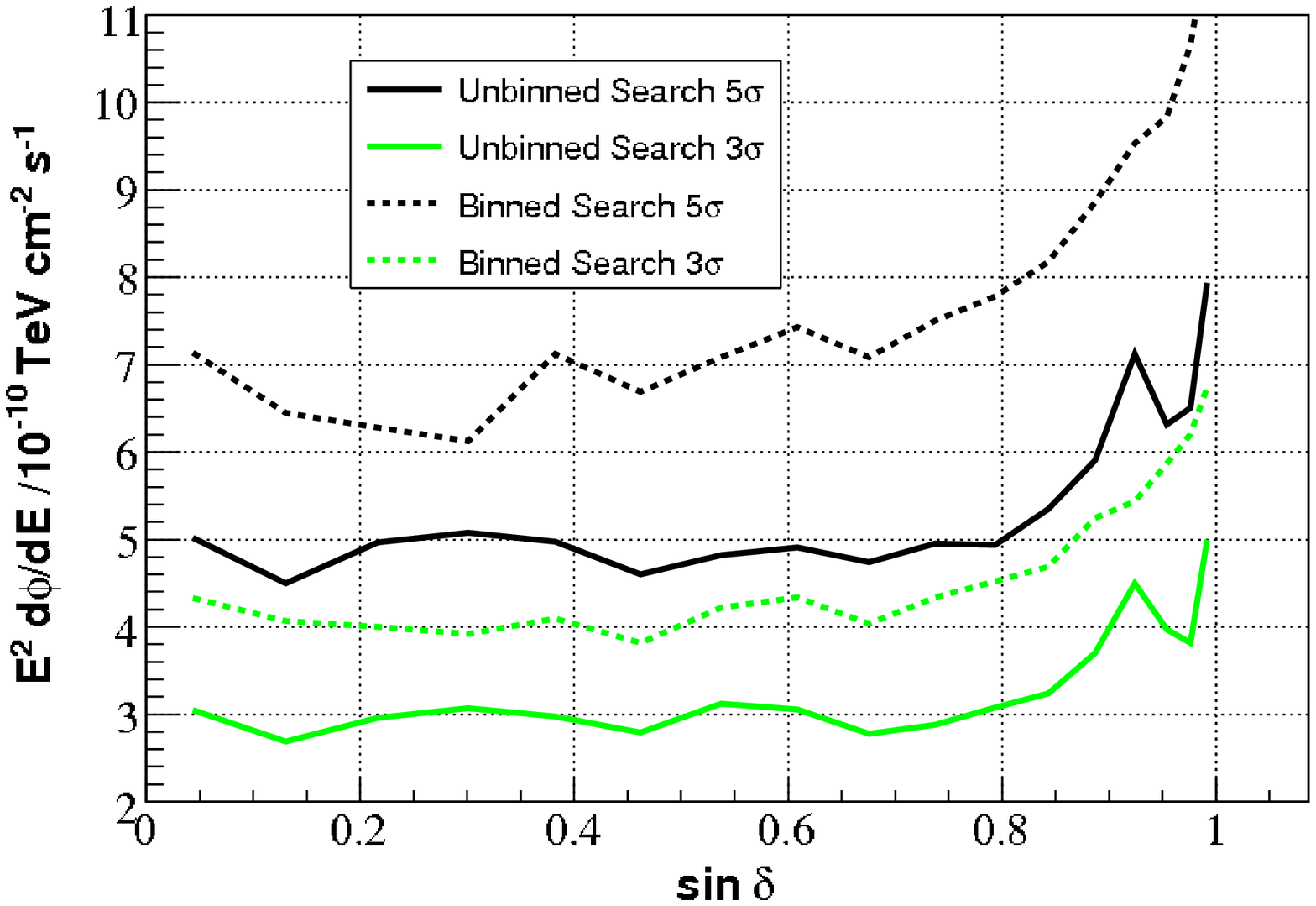}}
\caption{\label{Fig:2} Preliminary point source sensitivity to E$^{-2}$
energy spectra (top), and discovery flux for E$^{-2}$ energy spectra (bottom).
90\% of sources with this flux are detected at the stated significance,
excluding trial factors.}
\end{figure}

\section{Results}
\vspace{-2mm}
The method is applied to a catalog of candidate neutrino sources including
microquasars, supernova remnants, TeV blasars, and other objects of interest.
Results for a selected subset of objects are summarized in table \ref{Tab:1}.
\begin{figure}
\mbox{\includegraphics[width=2.9in]{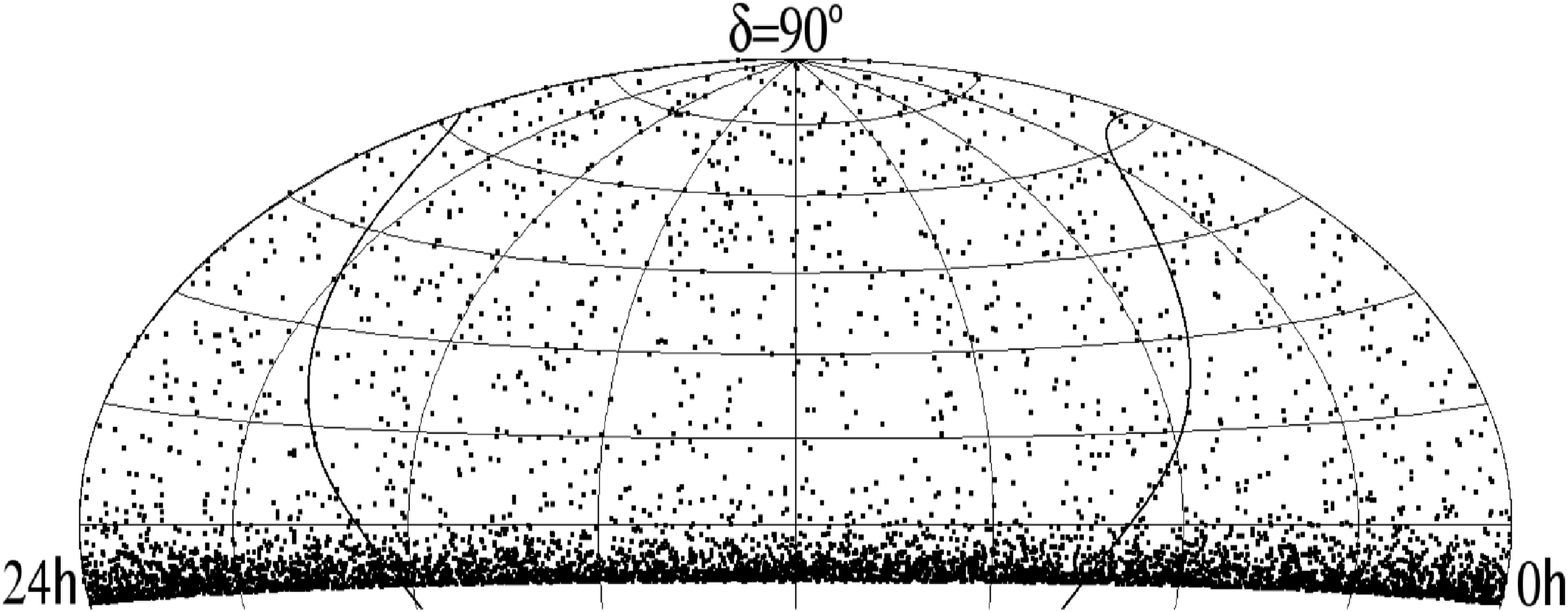}}
\mbox{\includegraphics[width=2.9in]{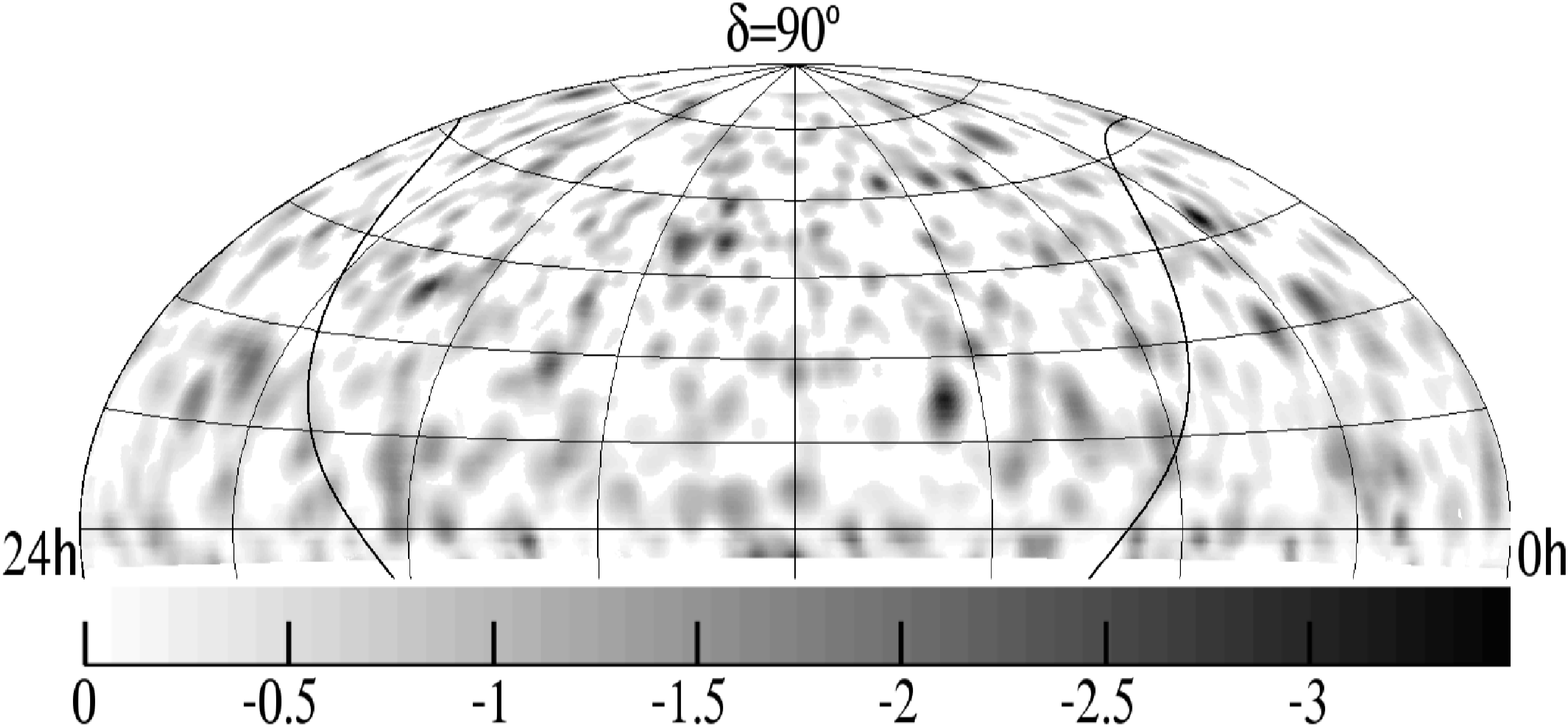}}
\caption{\label{Fig:4} Preliminary sky map of neutrino candidate events (top),
and of log$_{10}$(p-value) (bottom)}
\end{figure}
A scan of the entire sky at points spaced by 0.25$^{o}$ is also performed using this method.  The resulting
p-value map is shown in figure \ref{Fig:4}.
The highest obtained p-value corresponds to 3.6$\sigma$.  The probability of this
deviation due to background alone is evaluated by comparing against 100 simulated experiments
with randomized right ascension, and is found to be 69\%.

\begin{table}
\begin{tabular}{lcccc}
\hline
Candidate
&$\delta$($^o$)
&${\mu}_\mathrm{90}$
&$p$
\\\hline
Markarian 421 &38.2 &5.87 &$\sim$1\\
Markarian 501 &39.8 &18.1 &0.184\\
Cygnus X-1    &35.2 &12.9 &0.414\\
Cygnus X-3    &41.0 &11.0 &0.458\\
LS I +61 303  &61.2 &3.81 &$\sim$1\\
Crab Nebula   &22.0 &9.24 &$\sim$1\\
MGRO J2109+37 &36.8 &20.1 &0.152\\
\hline
\end{tabular}
\caption{\label{Tab:1}
        Preliminary flux upper limits for selected neutrino source candidates over 199.3 days livetime:
        source declination $\delta$ in degrees,
         flux 90\% confidence level upper limits for E$^{-2}$ spectra \\(E$^{2} \cdot \phi <
        \mu_{90} \cdot 10^{-11}\,\mathrm{TeV}\,\mathrm{cm}^{-2}\,\mathrm{s}^{-1}$), probability of
        observed or higher likelihood given random chance}
\end{table}

{\bf Acknowledgements:} We wish to thank Aart Heijboer for constructive discussion
regarding the unbinned maximum likelihood method.  This work is supported by the Office
of Polar Programs of the National Science Foundation.

%This is the reference to .bib file (Whitout .bib!)
%\bibliography{ICRC1090/icrc1090}
%This in the bibtex style, is ok.
%\bibliographystyle{plain}

%\end{document}

\setcounter{figure}{0}
\setcounter{table}{0}
%%
% International Cosmic Ray Conference 2007 Merida Yucatan Mexico
% In This file you will find detailed instructions to correctly
% typeset your document.
%
%
%

%Class Requeried
%\documentclass{article}
%The ICRC Style
%\usepackage{icrctc07}

%The paper title
\title{Point source analysis for cosmic neutrinos beyond PeV energies with
AMANDA and IceCube}
%Short title to print in the headers to the final publication (Not showed in
%this print).
\shorttitle{Neutrino Point Sources beyond PeV Energies}
%All paper authors
\authors{R. Franke$^{1}$, R. Lauer$^{1}$, M. Ackermann$^{1,2}$, E.
Bernardini$^{1}$ for the IceCube Collaboration$^{3}$}
%Short title to print in the headers to the final puplication (Not showed in
%this print).
\shortauthors{R. Franke, R. Lauer, M. Ackermann, E. Bernardini}
%All the affiliations.
\afiliations{$^1$ DESY, D-15735 Zeuthen, Germany\\ $^2$ Now at Stanford Linear
Accelerator Center, Stanford, California 94305-4060, USA\\ $^{3}$ For a complete authorlist
see special section in these proceedings.} \email{robert.franke@desy.de}

%The abstract.
\abstract{The Antarctic neutrino telescope AMANDA-II, part of the IceCube
observatory, can be used for searches for cosmic point sources of neutrinos
with a wide range of energy. The highest of these energy bands spans from
about $10^5$ to $10^{10}\,$GeV. Several source models predict a significant
neutrino flux in this part of the spectrum, for example from active galactic
nuclei. Since the interaction length of neutrinos with energies above $5\cdot
10^4$ GeV is smaller than the diameter of the Earth, the observable area lies
mainly in the southern sky, in contrast to point source searches at lower
energies. Nonetheless, the low atmospheric muon background at these energies
makes such an analysis feasible, and it would comprise some interesting source
candidates. We present the methods and sensitivity of this analysis as applied
to data collected with the AMANDA-II detector during the year $2004$. We comment
also on the status of an equivalent analysis being developed for data from
IceCube in its nine string configuration of $2006$.}

%\email{aastex-help@aas.org}

%%%%%%%%%%%%%%%%%%%% B E G I N   D O C U M E N T%%%%%%%%%%%%%%%%%%%%%%%
%\begin{document}
\maketitle
%Begin the section.

\section{Introduction}

Active galactic nuclei (AGN), and blazars in particular, are promising sources
of high energy neutrinos detectable with the Antarctic Neutrino Telescope
AMANDA-II, part of the IceCube observatory. Being candidates for the production of an
observed flux of charged particles with energies up to a few hundred EeV,
there is reason to expect a measurable neutrino flux beyond PeV energies from
this class of objects. Additionally, theoretical models for several of these
extra-galactic sources predict their neutrino spectra to be peaked in the PeV
to EeV energy range, as for example presented in \cite{Neronov:2002xv},
\cite{Protheroe:2002gv}.\\ An analysis with the aim to find neutrino point
sources in this very high energy range is different from other point source
analyses, as for example~\cite{Achterberg:2006vc}. The usual approach to
reduce the background of atmospheric muons is by selecting up-going neutrinos
only, i.e. neutrinos which have traversed the Earth before interacting in the
ice or bedrock near the detector. This effectively limits the accessible
neutrino spectrum due to the increase of neutrino cross section with energy.
For multi-PeV neutrinos, the interaction length is much smaller than the
diameter of the Earth and thus prevents most of the up-going neutrinos in this
energy range from reaching the detector. On the other hand, down-going
neutrinos from the southern sky high above the horizon have only the ice above
the detector as target material and hence a significantly reduced interaction
probability. Thus, a dedicated ultra high energy neutrino analysis must
utilize a zenith angle band around the horizon, where the sensitivity of a
standard search is limited by atmospheric muons. At higher energies, these muons form
a much smaller background due to their soft spectrum. Bringing part of the sky
in the southern hemisphere into the field of view also gives the possibility
to observe candidate objects not included in other neutrino searches, thus
enlarging the angular window where AMANDA-II is sensitive to point source
signals.
%This way of enlarging the angular window where AMANDA-II is sensitive to
%point source signals is an additional motivation for such an analysis.\\

\section{Source Candidates}

The main class of objects which are expected to emit a comparatively large
flux of neutrinos at ultra high energies are blazars, particularly the
GeV-blazars detected by EGRET and the TeV-blazars discovered by various air
\v{C}erenkov telescopes. The analysis is also sensitive to the galactic center
as a possible source, lying in a region less than $30^{\circ}$ above the
horizon. The third EGRET catalog contains 39 confirmed AGN gamma ray sources
with declinations between $+20^{\circ}$ and
$-30^{\circ}$ \cite{Hartman:1999fc}. The strongest sources have gamma ray
fluxes of the order of $10^{-6}$ photons cm$^{-2}$ s$^{-1}$, integrated for
energies above $100$ MeV.\\ In the final analysis, we will select a subset of
these objects to avoid reducing the statistical significance by trial factors.
As a first approach to find a suitable
classification and identify the blazars with the highest potential as neutrino
point sources, we extrapolate each gamma ray flux distribution to higher
energies. The flux distribution is approximated with a power law
$F(E)=F_0\cdot E^{-\Gamma}$ where $E$ is the photon energy and $F_0$ the flux
normalization, making use of the spectral index $\Gamma$ as measured by EGRET.
For our current purposes of comparing the candidates, we assume a direct
correlation between photons and neutrinos. We calculate the integrated photon
flux $F_I = \int_{E_{th}}^{\infty}F(E)\,dE$ with $E_{th}=100\,$TeV as the
lower energy threshold for this analysis. The resulting maximum values for
individual sources lie in the order of $10^{-10}$ photons cm$^{-2}$ s$^{-1}$.\\
We work on improving this first classification by using the
parametrization of spectral energy distributions for blazars as presented in
\cite{Fossati:1998zn}, with the plan to perform a more detailed study of flux
predictions by individually fitting the observed EGRET spectra to the hadronic
model used in \cite{Aharonian}. In addition to the GeV blazars, the source
list will include the galactic center and a sample chosen from 20 objects, located in the
chosen zenith band, from which TeV gamma rays have been observed.

\section{Reconstruction Methods}

The point source analysis for neutrinos beyond PeV energies we present here is
developed for data from the AMANDA-II detector taken during the year $2004$.  The
detector consists of $677$ optical modules (OMs) on $19$ strings, most of
which are deployed at depths between $1.5$ and $2$ km in the deep ice located
at the Gepgraphic South Pole. For this analysis we use $540$ OMs that show
a stable performance. The analysis strategy is based on identifying tracks
from neutrino-induced muons passing through the detector and emitting
\v{C}erenkov radiation.\\
%For this, $540$ optical modules (OMs) showing stable performance are used,
%while the whole detector array consists of $677$ OMs on $19$ strings, most of
%which are deployed at depths between $1.5$ and $2$ km in the deep ice located
%at the Geographic South Pole. 
To account for photon scattering in the ice, it is necessary to use likelihood
algorithms to reconstruct particle tracks. An iterative maximum likelihood fit
of the photon arrival times in the OMs finds the most probable muon
track~\cite{Ahrens:2003fg}. As a parametrization of the light propagation in
ice we use an empirical model of the ice properties.
%Using a parametrization of light propagation based on an empirical model of
%the ice properties, an iterative fit of photon arrival times in the OMs
%calculates the likelihoods for several track hypotheses and identifies the
%most probable one~\cite{Ahrens:2003fg}.
The standard version of this likelihood approach includes only the timing
information of the first photon hit in each photomultiplier. Monte Carlo
simulations show, however, that the angular resolution of AMANDA-II with this
reconstruction method degrades for higher energies. A high energy muon emits
more photons per track length than one at lower energies. As photons are
scattered independently in the ice, the order of arrival of multiple photons
in one OM is not identical to their sequence of emission from the track. As a
remedy for this we use an improved version of the likelihood fit. The
likelihood is given by the probability that any of the detected photons in an
OM arrives at the time of the first hit recorded in that OM and all other
photons arrive at a later time~\cite{Ahrens:2003fg}. This requires a numerical
integration over the probability density function which is computationally
expensive.
%Accounting
%for the random order of multiple detected photons in a given OM, a summation
%over all possibilities for an individual photon to arrive first is performed,
%including an integration over the probabilities for the other photons to
%arrive at later times~\cite{Ahrens:2003fg}.\\
%The detection of one hit as the first of several photons in a single OM gives
%no direct information about their relative times of emission from the track,
%since their timing is dominated by different propagation lengths due to
%individual scattering in the ice. The advanced likelihood fit allows for the
%fact that each of the detected photons has a probability of being the one
%emitted first. Summing over all possibilities and integrating over the
%probabilities for the other photons to arrive at later times, this method
%finds the track fit with the highest likelihood. Only by chance one of these
%photons reaches the OM first being the least scattered one. The advanced
%likelihood fit allows for the fact that in case of $n$ detected photon hits
%there are $n$ ways to choose the photon emitted first and multiplies each of
%these probabilities with the probability to detect the other $n-1$ photons
%later than the first detected arrival time.\\
%A disadvantage of this method is the additional computation time necessary for
%the integration which can only be done numerically.
For this reason, it is not
possible to run the improved fit iteratively for each event, but instead the
track result of the standard likelihood fit is taken as the initial hypothesis
for the improved likelihood maximization.\\
In Monte Carlo simulations of a
signal neutrino flux between $10^5$ and $10^{10}$ GeV this method shows an
improvement in median angular resolution.
%, defined as the 50\% quantile of the angular difference between the
%simulated and the reconstructed track.
For an $E^{-2}$ spectrum the angular resolution obtained with the improved
fit is $3.87^{\circ}$, compared to $6.9^{\circ}$ for the standard approach.
The resolution as a function of primary neutrino energy for the standard fit
and improved fit method is shown in Fig.~\ref{resvsenergy}. The whole analysis
was performed using the IceCube software framework to simplify the use and
exchange of tools and method implementations~\cite{Franke}.

\begin{figure}[tb]
	\begin{center}
		\noindent
% 		\fbox{\hbox{\vbox{\hsize=50mm \hfill \vspace{50mm}}}}
		\includegraphics [width=0.35\textwidth]{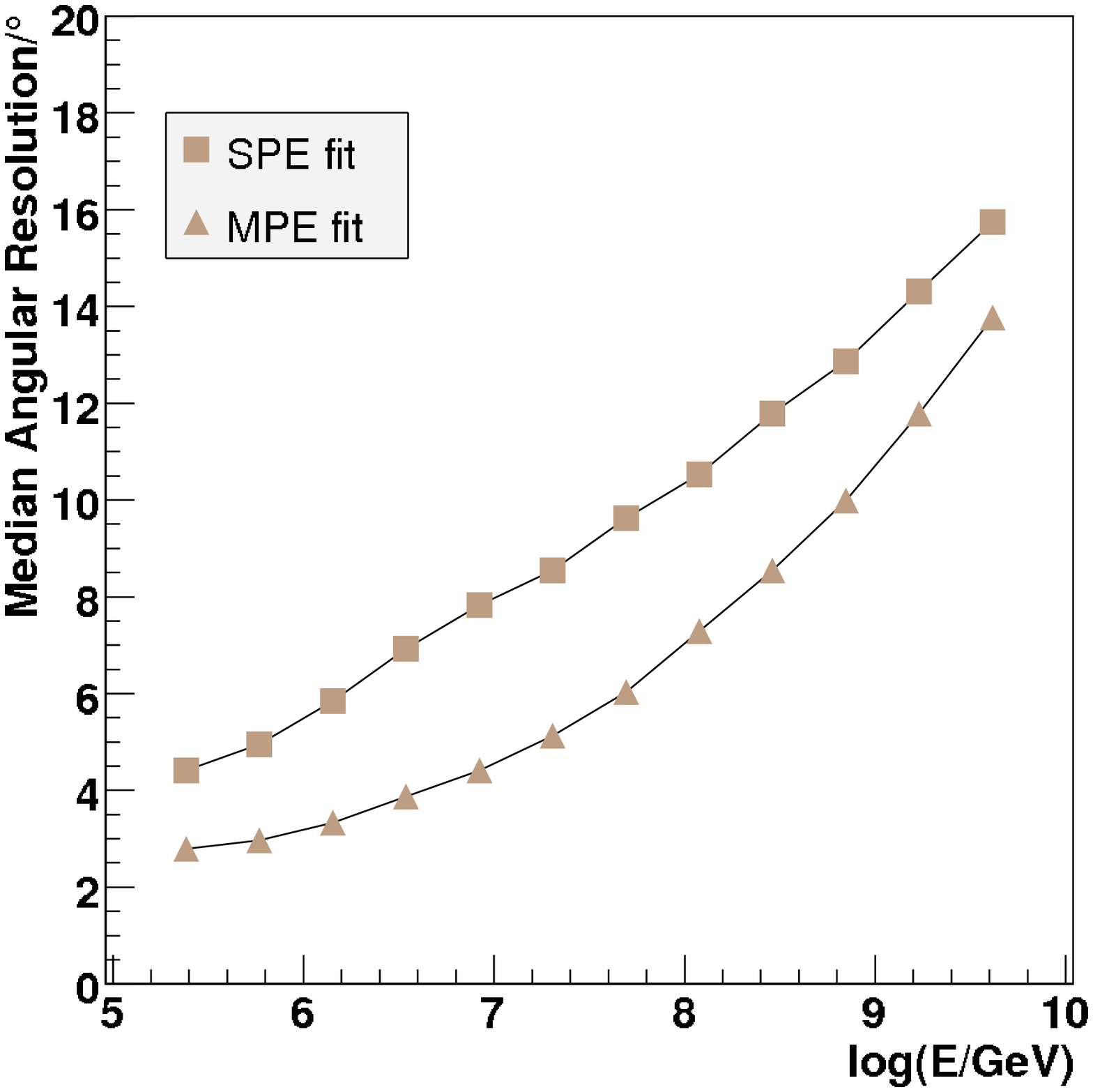}
	\end{center}
    \vspace{-0.3cm}
	\caption{Median angular resolution in degrees as a function of primary
    neutrino energy from Monte Carlo simulation, reconstructed with the simple
    (SPE) and improved fit (MPE) accounting for multiple scattered photons.
    These resolutions are based on the discussed event sample with more than
    $140$ hits and a one-photoelectron fraction smaller than $0.72$.}
	\label{resvsenergy}
\end{figure}

% The problem with this likelihood is that one has to know the integral of the Pandel function or do the integration 
% numerically. This is the only possibility for the convoluted Pandel function. As this is very slow (the 
% minimisation of one seed track takes about 1 minute) a lookup table was implemented in ipdf to speed this up. This 
% brought the track fitting time down to about 10-15 s. 

\section{Event Selection}

\begin{figure}[tb]
	\begin{center}
		\noindent
% 		\fbox{\hbox{\vbox{\hsize=50mm \hfill \vspace{50mm}}}}
		\includegraphics [width=0.4\textwidth]{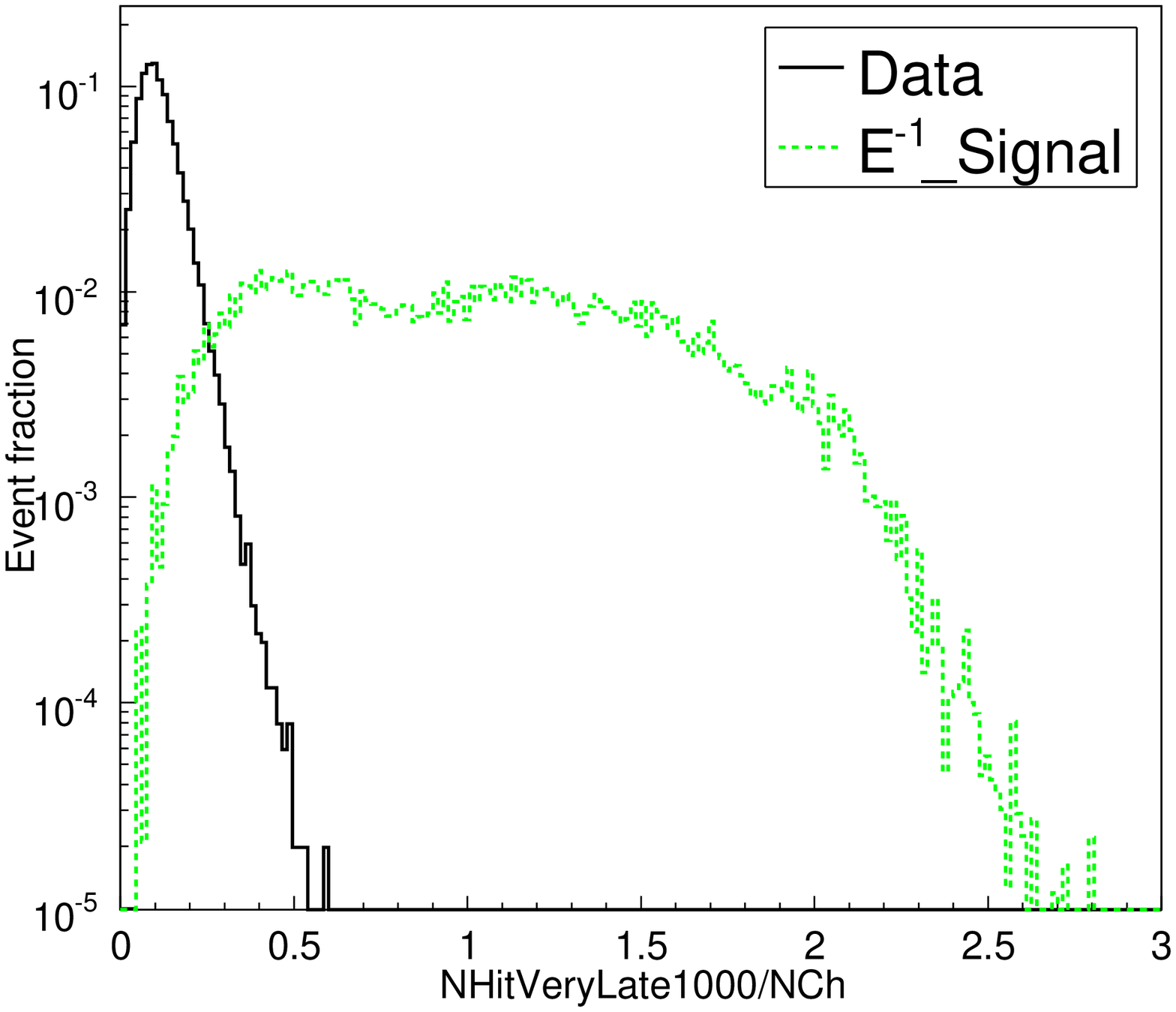}
	\end{center}
    \vspace{-0.35cm}
	\caption{Ratio of late hits (hits ocurring more than 1000 ns after the
    first hit in that OM) to the number of hit OMs for an E$^{-1}$ signal spectrum and experimental data.}
	\label{latehits}
\end{figure}
From the data collected with AMANDA-II in ca. $195$ days of lifetime during $2004$
we select events with a large light output that is likely to be caused by high energy events.
We require at least $140$ hits in the detector and a fraction of one-photoelectron hits smaller than $0.72$.
This results in a data sample of approximately $1.5\cdot 10^{7}$ events.
Standard cleaning procedures are applied to the sample to eliminate isolated
hits and reduce electronically induced cross-talk.\\
The main background dominating the data sample after this first selection is
intense muon bundles from energetic cosmic ray air showers, which can fake the
signature of a single muon of higher energy. However, the light from intense
muon bundles is expected
to be distributed more evenly through the detector as it is emitted from
multiple tracks instead of a single one as in the case of a signal event.  A
multi-PeV neutrino-induced muon emits significantly more photons through
stochastic energy losses and Monte Carlo simulations show that this leads to a
higher fraction of very late hits. We define very late hits as hits occurring
more than 1000 ns after the first hit in the same OM. These can be caused by scattered photons or afterpulses in the
photomultipliers.
%, thus in general also including some photons with very long scattering
%lengths.
Normalizing the number of OMs with very late hits to the number of hit OMs
yields a useful basic discrimination variable between expected signal and
background, see Fig.~\ref{latehits}.\\ Due to the long computation time of the
improved likelihood method, this selection is also motivated by reducing the
number of events before reconstruction. Hence, choosing a cut value for the
afterpulse fraction is based on the aim to keep approximately 20~\% of the
(background dominated) data. Monte Carlo simulations of signal and background
show that this implies a signal passing rate of 94~\% for an $E^{-1}$ spectrum
and 98~\% for an $E^{-2}$ muon-neutrino spectrum. Therefore we select events with a
fraction of OMs with very late hits larger than $0.15$.\\
To estimate a sensitivity for this
analysis, a number of background-signal discrimination variables have been
examined. In a first iteration three variables sensitive to the light
distribution in the detector and with respect to the track fit were chosen.
These variables are the number of photons registered outside a $50$ m cylinder
around the track fit, the ratio of hit channels to the total number of hits
and the ratio of late hits to the total number of hits. The cuts on these
variables were optimized for sensitivity in different zenith
bands, using the data as a
background estimate. The achieved preliminary sensitivity versus zenith angle
for this analysis can be seen in Fig.~\ref{ICRCPEV_sensitivity}.
\begin{figure*}
%\begin{center}
%\noindent \fbox{\hbox{\vbox{\hsize=130mm \hfill \vspace{50mm}}}}
%\includegraphics [width=0.2\textwidth]{logoblack.eps}
%\begin{figure}[tb]
%	\begin{center}
		\noindent
 		%\fbox{\hbox{\vbox{\hsize=50mm \hfill \vspace{30mm}}}}
		\includegraphics [width=0.5\textwidth]{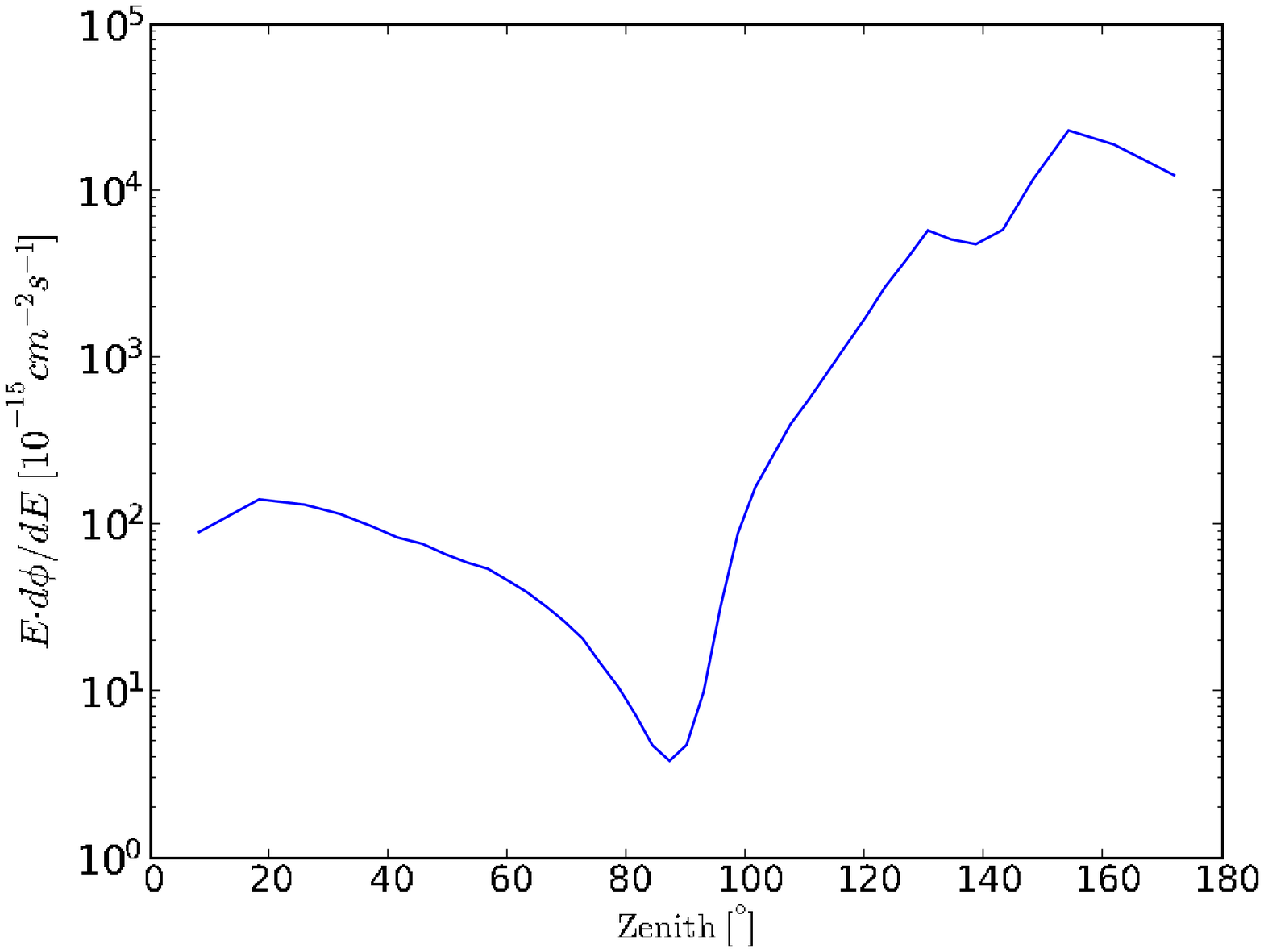}
%	\end{center}
%    \vspace{-0.35cm}
%	\label{ICRCPEV_sensitivity}
%\end{figure}
%\begin{figure}[tb]
%	\begin{center}
%		\noindent
 		%\fbox{\hbox{\vbox{\hsize=50mm \hfill \vspace{30mm}}}}
		\includegraphics [width=0.5\textwidth]{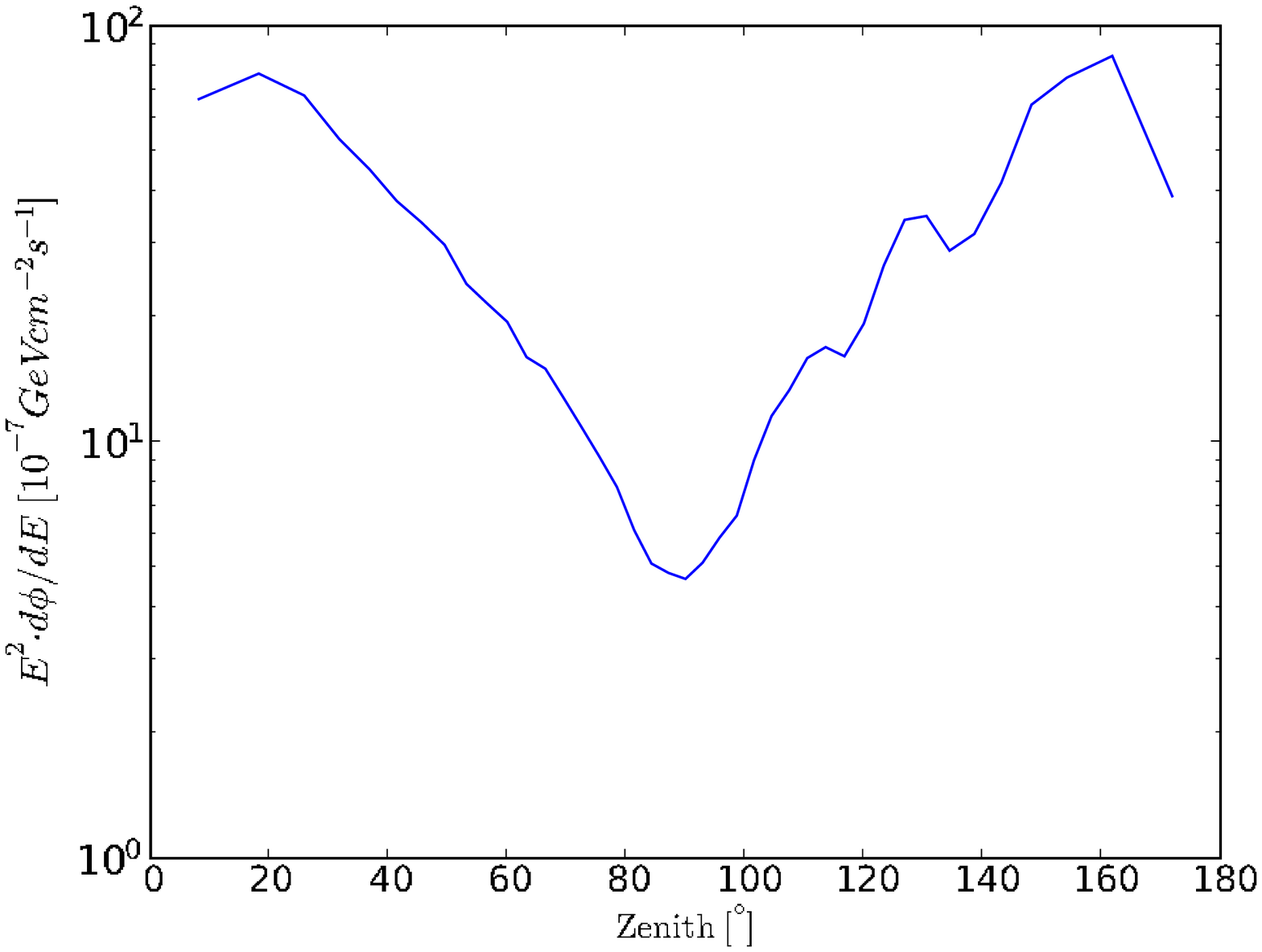}
%	\end{center}
%    \vspace{-0.35cm}
%	\caption{}
%	\label{ICRCPEV_sensitivity_E2}
%\end{figure}

%\end{center}
\caption{Preliminary sensitivities for this analysis for an $E^{-1}$ (left) and $E^{-2}$ (right) flux of muon neutrinos in the energy range from $10^5\;\mbox{GeV}$ to $10^{10}\;\mbox{GeV}$ vs. zenith angle. The upper limit is shown as a limit to the normalization constant $\Phi_0$ of the differential flux $d\Phi/dE=\Phi_0 E^{-\gamma}\;,\gamma=1,2$.
%The horizontal lines indicate the width of the overlapping zenith bands were the cuts were independently optimized for sensitivity.
}
\label{ICRCPEV_sensitivity}
\end{figure*}
For the $E^{-1}$ signal spectrum $90\%$ of the events over the whole zenith range have an energy
between $3.6\cdot 10^7\,$GeV and $8.9\cdot10^9\,$GeV after the cuts. At the horizon this energy range is $5.9\cdot 10^7\,$GeV
to  $9.0\cdot 10^9\,$GeV. For the $E^{-2}$ spectrum the energy range which contains $90\%$ of the events extends from $1.4\cdot 10^5\,$GeV to $1.2\cdot
10^8\,$GeV over the whole sky and from $2.0\cdot 10^5\,$GeV to
$4.0\cdot10^8\,$GeV at the horizon.

\section{Conclusions and Outlook}

Presented here is a dedicated analysis for the search for point-like sources
of cosmic neutrinos beyond PeV energies. Our strategy enlarges the window for
potential discoveries with AMANDA-II to parts of the southern sky and improves
the methods for detecting neutrino events at the highest energies.\\ The
concept of this analysis is currently being developed further with the aim to
be applied to the data taken with IceCube in the nine string configuration of
$2006$. A preliminary study of reconstruction methods after a basic selection
of high multiplicity hits shows an angular resolution of approximately
$2^{\circ}$. Due to the asymmetric detector configuration the sensitivity of
the analysis is not expected to improve much compared to the results presented
here for AMANDA-II. A significant improvement of the sensitivity for
point-like neutrino sources with extremely high energies can be expected with
the 22-string configuration of IceCube in $2007$.
%A significant step towards a better detection capability
%for neutrino point-like sources will be an analysis of data collected with the
%IceCube configuration of 2007, which has 22 deployed strings.

{\small{\bf Acknowledgements} We thank the Office of Polar Programs of the
National Science Foundation, as well as DESY and the Helmholtz 
Association.}

%This is the reference to .bib file (Whitout .bib!)
%\bibliography{ICRCPEV/icrc_wo-table}
%This in the bibtex style, is ok.
%\bibliographystyle{plain}

%\end{document}

\setcounter{figure}{0}
\setcounter{table}{0}
%%
% International Cosmic Ray Conference 2007 Merida Yucatan Mexico
%%

%Class Requeried
%\documentclass{article}
%The ICRC Style
%\usepackage{icrctc07}

\newcommand{\comment}[1]{}

%The paper title
\title{Nine-String IceCube Point Source Analysis}
%Short title to print in the headers to the final publication (Not showed in this print).
\shorttitle{Nine-String IceCube Point Source Analysis}
%All paper authors
\authors{C. Finley$^{1}$, J. Dumm$^{1}$, T. Montaruli$^{1,2}$, for the IceCube Collaboration$^{3}$.}
%Short title to print in the headers to the final puplication (Not showed in this print).
\shortauthors{C. Finley et al.}
%All the affiliations.
\afiliations{
$^1$Dept. of Physics, Universtiy of Wisconsin, Madison, WI, 53706, USA\\ 
$^2$on leave from Universit\`a di Bari, Dipartimento di Fisica, I-70126, Bari, Italy\\
$^3$see special section of these proceedings
}
\email{chad.finley@icecube.wisc.edu}

%The abstract.
\abstract{
The construction of the IceCube Neutrino Observatory began during the austral summer of 2004-05, and is expected to continue through 2011.  During 2006, nine of the projected 80 strings were already deployed and taking data, making IceCube an operational neutrino observatory while still at about 10\%  
of its final size.  We present the first results of a point-source search based on the analysis of this year of data, and characterize the angular resolution and effective area of the nine string configuration.  With 137.4 days of detector livetime, 233 neutrino candidate events were selected in the analysis; the sky-averaged point-source sensitivity for an $E^{-2}$ spectrum is $\frac{d\Phi}{dE} = 12\times 10^{-11}\ \mathrm{TeV^{-1}\ cm^{-2}\ s^{-1}}\ (E/\mathrm{TeV})^{-2}$.  No significant point-source is found.  We also discuss how the performance is expected to improve as the detector moves toward completion.
}

%\email{aastex-help@aas.org}

%%%%%%%%%%%%%%%%%%%% B E G I N   D O C U M E N T%%%%%%%%%%%%%%%%%%%%%%%
%\begin{document}
\maketitle
%Begin the section.

\section{Introduction}
The IceCube Neutrino Observatory is a cubic kilometer-scale detector under construction at the geographic South Pole.  Its primary mission is the search for high energy extraterrestrial neutrinos, which may reveal the origin of cosmic rays and offer insight into the most energetic phenomena in the universe.  The detector consists of an array of digital optical modules: 60 modules are connected on one string, and a planned total of up to 80 strings are to be deployed in the Antarctic ice at depths between 1.5 and 2.5 kilometers beneath the ice surface.  Construction is limited to the austral summer time, and therefore is spread over a number of years.  As the detector grows, commissioning of the new strings and data-taking occur during the rest of the year.  

\begin{figure}
\begin{center}
\includegraphics [width=0.48\textwidth]{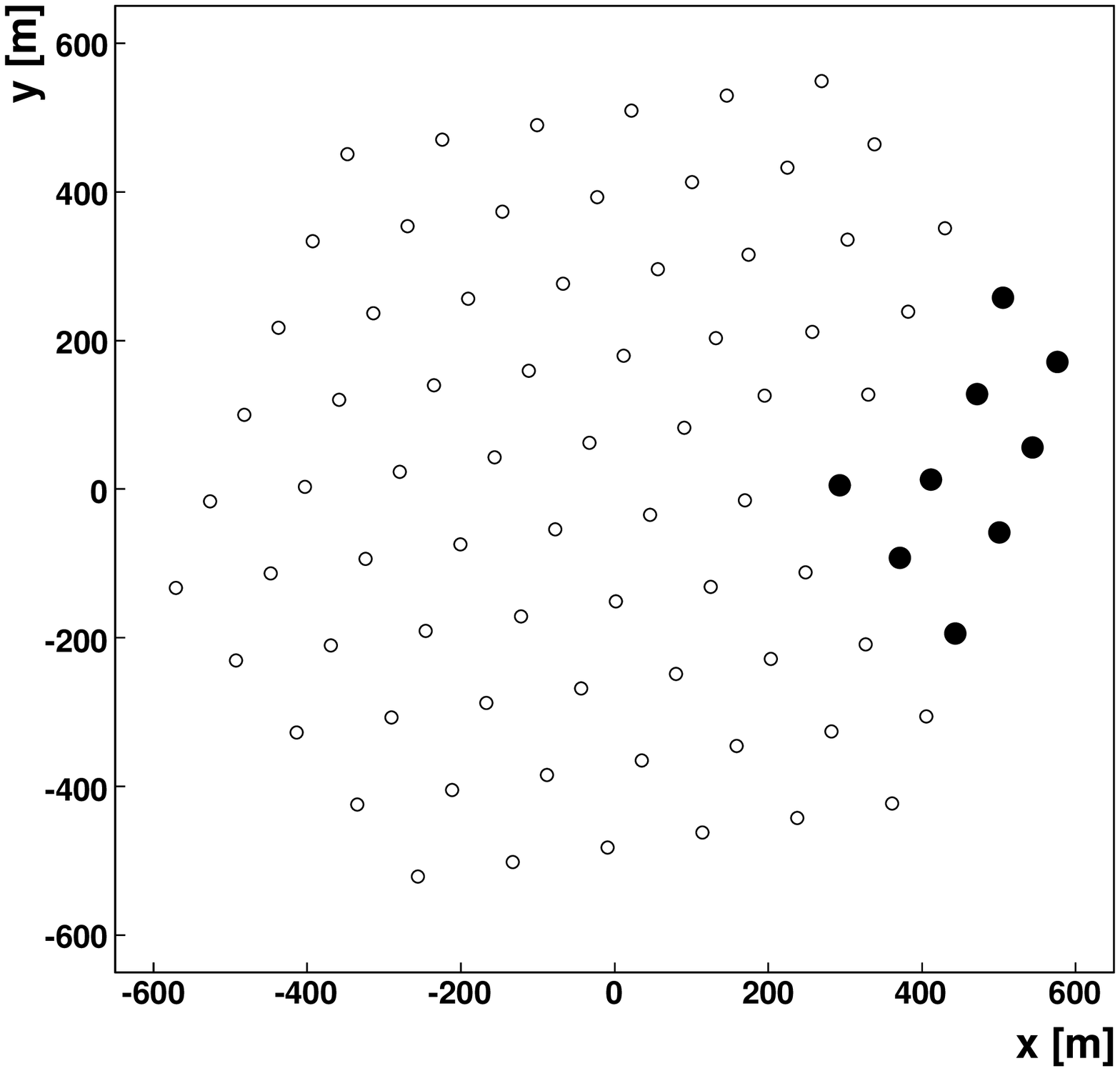}
\end{center}
\caption{
Configuration of IceCube strings; filled markers indicate the location of the nine strings already deployed and taking data in 2006.
}
\label{IC9_config}
\end{figure}

Nine strings were in operation during 2006, shown in Figure~\ref{IC9_config}.  
At about 10\% of its completed size, the partial detector configuration is not optimal: muon tracks which traverse the long axis of the detector can be reconstructed much more accurately than those which pass through from other directions.  Nevertheless, high-quality data was obtained between June and November, providing the first opportunity to perform a search for extraterrestrial neutrinos with the IceCube detector.  
Point source searches like the one presented here are the simplest and most direct way to distinguish an extraterrestrial neutrino signal from the experimental backgrounds.  Discovery of point sources would also directly indicate the sites of cosmic ray acceleration.

\section{Method}

An unbinned maximum likelihood method is used to search for point sources.  For a specified, hypothetical source location $x_s$ and total number of events $n_{\mathrm{tot}}$, the source hypothesis is that the data set is a mixture of $n_s$ signal events (distributed around the source according to their individual angular uncertainty) and $n_{\mathrm{tot}}-n_s$ background events (distributed over the sky according to the detector background distributions).  This can be expressed as the partial probability $P_i$ of each event:
$$
P_i(x,x_s,n_s) = \frac{n_s}{n_{\mathrm{tot}}} S_i(x,x_s) + \left(1-\frac{n_s}{n_{\mathrm{tot}}}\right) B_i(x)
$$
where $S_i(x,x_s)$ is the source pdf of the event (determined by its angular uncertainty) and $B_i(x)$ is the background pdf.  The background pdf is determined by using the declination distribution of the real data set.

The likelihood $\mathcal{L}$ is defined as the product of all individual event pdf's evaluated at the event and source coordinates:

$$\mathcal{L}(x_s,n_s) = \prod P_i(x_i,x_s,n_s)$$

The best estimate for the number of signal events $\hat{n}_s$ is found by maximizing the log likelihood ratio $\lambda$ with respect to the null hypothesis $n_s = 0$:

$$\log \lambda = \log \frac{\mathcal{L}(x_s,\hat{n}_s)}{\mathcal{L}(x_s,n_s = 0)}.$$

$\log\ \lambda$ is the test statistic which determines the significance of an observed deviation from the null hypothesis.

\begin{figure}[ht]
\begin{center}
\includegraphics [width=0.48\textwidth]{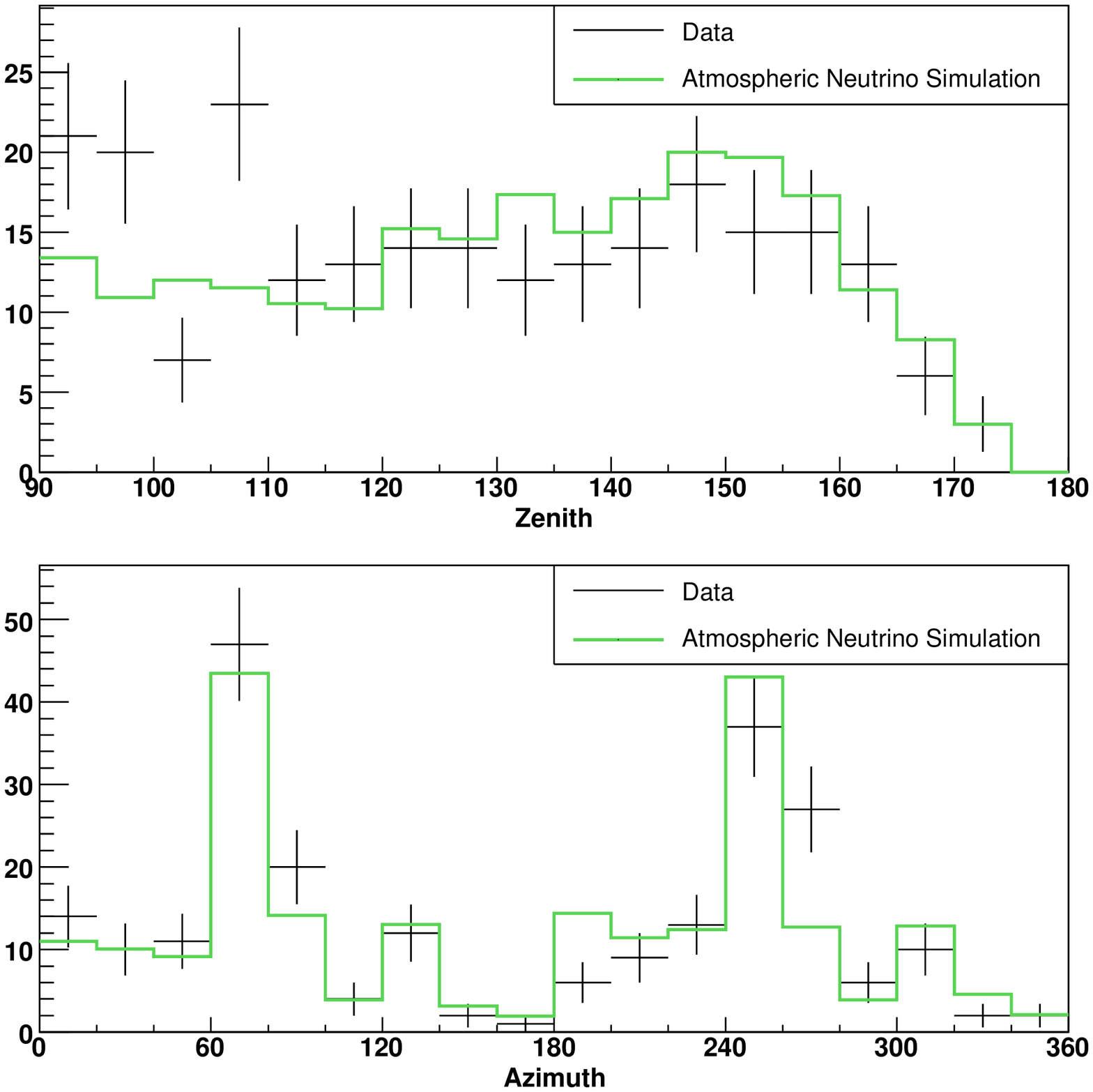}
\end{center}
\caption{
Distribution of data events in zenith and azimuth after analysis cuts have been applied, compared with simulated atmospheric neutrino events.  (Note: declination $\delta = \theta_{\mathrm{zen} }- 90^{\circ}$.)
}
\label{Data_Dist}
\end{figure}

\section{Event Selection}

Data in this analysis first passed two levels of filtering to reject down-going muon events; these filter levels are described in \cite{Achterberg:2007bi}.  The remaining events were reconstructed using a likelihood algorithm that also provides an angular uncertainty estimate by evaluation of the likelihood function around the direction of the best fit.  After filtering, the main background is still mis-reconstructed down-going muons and muon bundles from cosmic ray showers.  To reduce this mis-reconstructed background, a tight cut on each track's angular uncertainty was used, and only tracks which reconstructed as up-going (zenith angle greater than $90^{\circ}$) are kept in the analysis.  A second cut on the minimum number of modules which were hit by direct Cherenkov photons (as estimated for the reconstructed track, using a time window of $-15$ to $+75$~ns around the expected arrival time) provides additional background rejection, primarily of down-going muons from two different cosmic ray showers which trigger the detector in coincidence and reconstruct as a single upward-going event.  What remains after tight cuts on both of these parameters is  the ``irreducible'' background of well-reconstructed upward-going atmospheric neutrino events, the product of cosmic ray showers in the northern hemisphere.

To determine the final cut values, the point source analysis was performed on simulated data sets, consisting of simulated source events added to real data scrambled in right ascension.  The cuts were optimized for discovery potential: the combination of cuts which could detect the smallest source flux at $5\,\sigma$ significance in 50\% of the trials.  For most possible source declinations and a range of spectra ($\frac{d\Phi}{dE} \propto E^{-\gamma}$ for the range $\gamma = 2$ to $\gamma=3$) the optimal cuts were the same.

\begin{figure}[t]
\begin{center}
\includegraphics [width=0.5\textwidth]{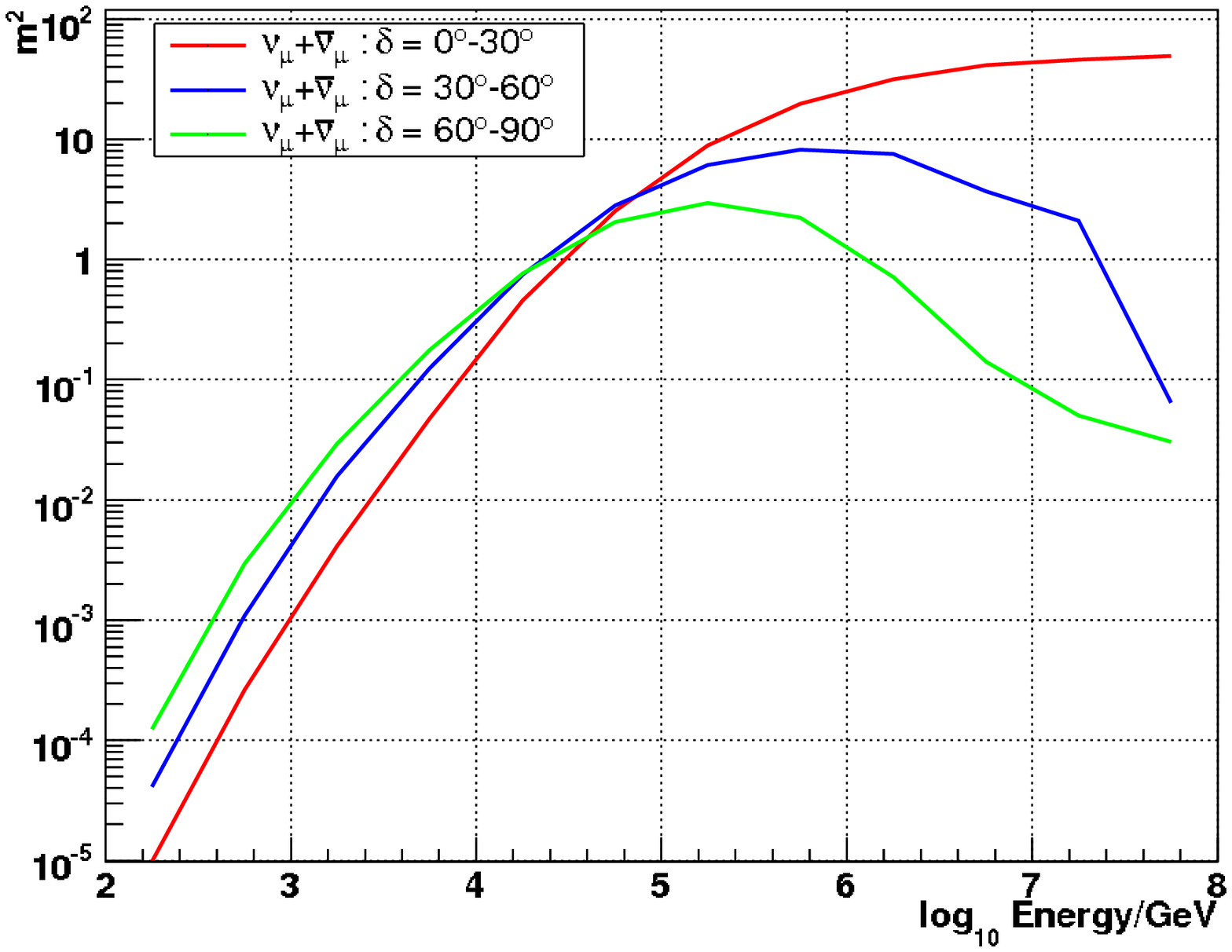}
\end{center}
\caption{
IC-9 effective area to a flux of $\nu_{\mu} + \bar{\nu}_{\mu}$, averaged over different declination ranges.
}
\label{Eff_Area}
\end{figure}

\section{Data Sample}

Data taking occurred between June and November 2006.  The detector livetime was 137.4 days.
The zenith distribution of data events is shown after final cuts and compared with simulated atmospheric neutrino events (using the spectrum predicted by the Bartol group\cite{Barr:2004br}) in Figure~\ref{Data_Dist}.  The final sample is restricted to events with declination less than $85^{\circ}$, because the right-ascension scrambling technique for estimating background does not work near the pole, where statistics are low and the events cannot be scrambled.  After cuts, there are 233 events in the data sample, and 227 predicted atmospheric neutrino events.  The excess of data events at low zenith angles is most likely mis-reconstructed down-going muons, which are increasingly hard to reject near the horizon.  Because the cut optimization was performed using scrambled real data, this residue of mis-reconstructed events indicates that harder cuts, which could eliminate these events entirely, would ultimately decrease the discovery potential.

The azimuth distribution of data and simulation is also shown in Figure~\ref{Data_Dist}.  The two directions corresponding to the long axis of the nine-string detector are clearly visible.  For other directions, it is more difficult to reconstruct tracks with high accuracy and to reject background.

\begin{figure}
\begin{center}
\includegraphics [width=0.5\textwidth]{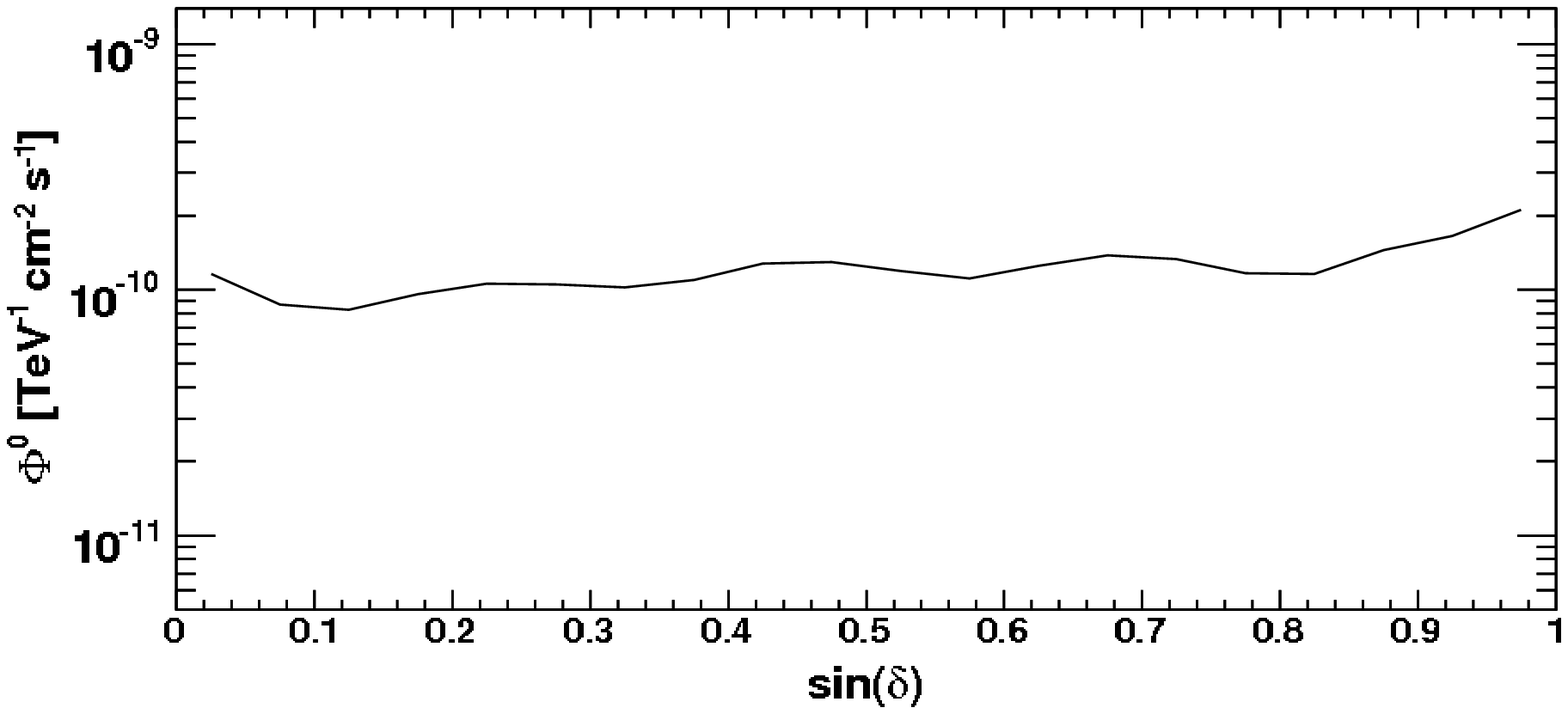}
\end{center}
\caption{
IC-9 sensitivity as a function of declination to a point source with differential flux $\frac{d\Phi}{dE} = \Phi^0 (E/\mathrm{TeV})^{-2}$.  Specifically, $\Phi^0$ is the minimum source flux normalization (assuming $E^{-2}$ spectrum) such that 90\% of simulated trials result in a log likelihood ratio $\log \lambda$ greater than the median log likelihood ratio in background-only trials ($\log \lambda = 0$).
}
\label{Sensitivity}
\end{figure}

\begin{figure*}
\begin{center}
%\noindent \fbox{\hbox{\vbox{\hsize=130mm \hfill \vspace{50mm}}}}
\includegraphics [width=1.\textwidth]{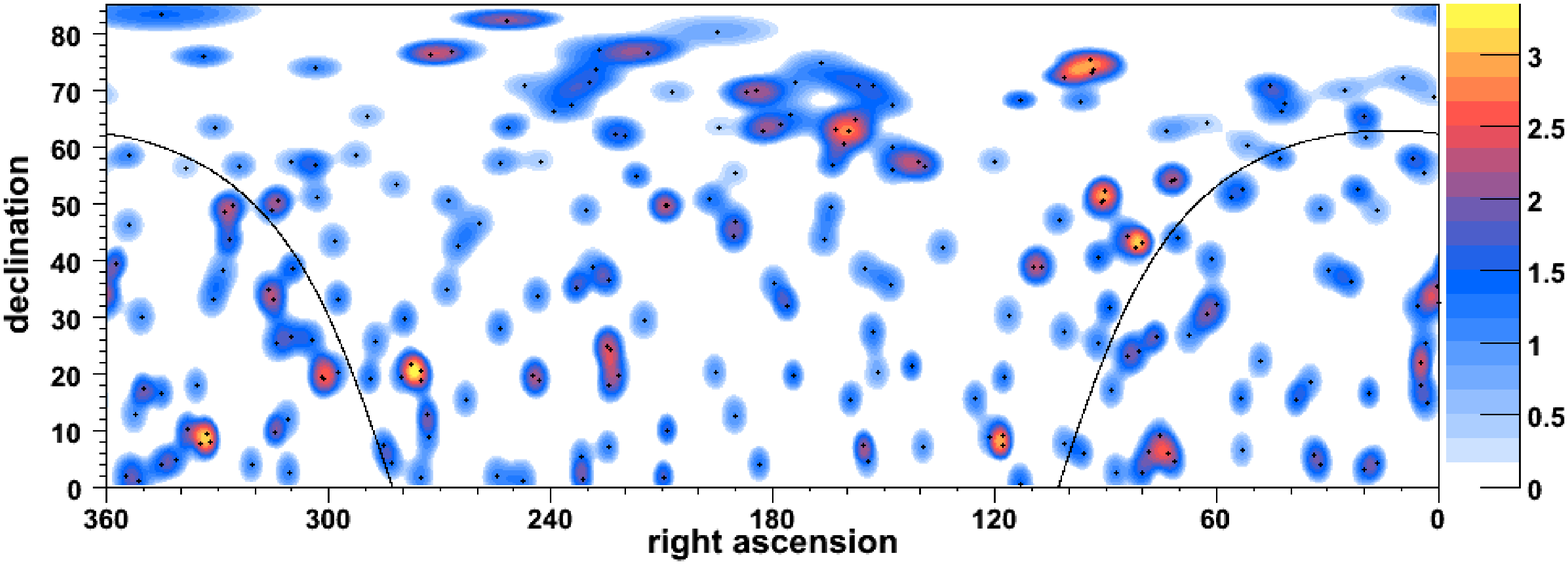}
\end{center}
\caption{Sky map of the significance [$\sigma$] of deviations from background, estimated from the maximum likelihood point-source search. (Black points are reconstructed event directions.)}
\label{Skymap}
\end{figure*}

The effective area to an equal-ratio flux of $\nu_{\mu}+\bar{\nu}_{\mu}$ is shown in Figure~\ref{Eff_Area}.
In Figure~\ref{Sensitivity}, the sensitivity (median flux upper limit) is shown as a function of declination.  
The sky-averaged point-source sensitivity to an $E^{-2}$ source spectrum is $\frac{d\Phi}{dE} = 12\times 10^{-11}\ \mathrm{TeV^{-1}\ cm^{-2}\ s^{-1}}\ (E/\mathrm{TeV})^{-2}$.
The median angular reconstruction error is $2.0^{\circ}$.

\section{Results}

The analysis consists of an all-sky point source search, and individual point source searches using a pre-defined source list.  The result of the all-sky search is shown in Figure~\ref{Skymap}.  The maximum upward deviation from background is at r.a.~$= 276.6^{\circ}$, dec~$= 20.4^{\circ}$, with $3.35\,\sigma$ significance.  This is consistent with random fluctuations: in simulations of background-only data sets (data scrambled in right ascension), 60\% have a maximum deviation (anywhere) of $3.35\,\sigma$ or greater.

Twenty-six galactic and extragalactic objects were included in the pre-defined source list.  Of these, the most significant excess over background was $1.77\,\sigma$, found for the Crab Nebula.  This is also consistent with random fluctuations: the probability for at least one out of 26 source directions to have an excess of $1.77\,\sigma$ or greater is 65\%.  The 90\% confidence level flux upper limit for the Crab Nebula is $\frac{d\Phi}{dE} = 22\times 10^{-11}\ \mathrm{TeV^{-1}\ cm^{-2}\ s^{-1}}\ (E/\mathrm{TeV})^{-2}$.

\section{Discussion}

Nine IceCube strings (out of a projected total of 80) were operating and taking data in 2006.  Analysis of this first year of data indicates that the point-source sensitivity of the nine string detector is comparable to an equivalent livetime of the AMANDA-II detector.  This is a promising result, given that the configuration of the nine-string detector is far from optimal.  For example, as seen in Fig.~\ref{Data_Dist}, more than half of the well-reconstructed events arrive from less than 10\% of the full range of azimuth.  Therefore as  construction continues, enlarging the array will not only increase the detector volume, but also greatly improve the angular resolution in all directions.  This should become apparent with the 22-string configuration which began operating this year.  Continued software development should also deliver more advanced track reconstruction algorithms and background rejection techniques.  The current analysis can serve as a benchmark for evaluating the performance of these new tools.  Extrapolating the present rate of growth, the IceCube Neutrino Observatory will begin to deliver results of unsurpassed sensitivity well before detector construction is completed.

\section{Acknowledgments}

This work is supported by the Office of Polar Programs of the National Science Foundation.

%This is the reference to .bib file (Whitout .bib!)
%\bibliography{ICRC0764/icrc0764}

%This in the bibtex style, is ok.
%\bibliographystyle{plain}

%\end{document}

\setcounter{figure}{0}
\setcounter{table}{0}
\title{Search for Signatures of Extra-Terrestrial Neutrinos with a Multipole Analysis of the AMANDA-II Sky-map}
\shorttitle{Search for Signatures of Extra-Terrestrial Neutrinos ...}
\authors{J.-P. H\"ul\protect\ss{}$^{1}$, Ch. Wiebusch$^{1}$ for the IceCube Collaboration$^a$}
\shortauthors{J.-P. H\"ul\protect\ss{} et al.}
\afiliations{$^1$ RWTH Aachen University, Aachen, Germany; $^a$ see special section of these proceedings}
\email{huelss@physik.rwth-aachen.de}
\abstract{In this analysis 3329 neutrino events detected by AMANDA-II during the years 2000-2003 are analysed for anisotropies or unexpected structures in their arrival direction. The structures could arise due to the presence of a signal from many weak and therefore unresolved cosmic neutrino sources, a few brighter sources or extended sources (e.g. a diffuse flux from the galactic plane). The sky-distribution of arrival directions (sky-map) is expanded in a series of spherical harmonics and the power in each multipole moment is calculated. Compared to previous AMANDA-II analyses, it provides a new complementary approach, in particular in the search for very weak individual astro-physical sources. No excess from extra-terrestrial sources is found. Statistical errors as well as systematic errors related to the uncertainty of the angular distribution of the atmospheric neutrinos are quantified using the Feldman-Cousins unified approach. Limits for contributions from extra-terrestrial sources to the sky-map are derived as function of the average source strength and the spectral index of the energy spectrum for different sky-distributions: weak sources isotropically distributed in the northern sky, sources located in the galactic and super-galactic plane. The tested average flux per source varies between $\phi_{low} = 5\cdot10^{-13}\;$cm$^{-2}$s$^{-1}$ and $\phi_{high}=5\cdot10^{-11}\;$cm$^{-2}$s$^{-1}$ at the earth, assuming an $E^{-2}$ power spectrum in the sensitive energy range between $1.6\;$TeV and $1.6\;$PeV. The number of sources in the sky can be limited at 90\% C.L. to be less than 3524 for the assumed $\phi_{low}$ and less than 28 for $\phi_{high}$.}
%\begin{document}
\maketitle

\section{Introduction}
There are several proposed candidate objects which could be neutrino sources in the universe, e.g. Active Galactic Nuclei, Supernova Remnants or Micro Quasars.

A direct measurement of these neutrinos is not possible. However, they produce high energetic muons in charged current interactions. Which points into the initial neutrino direction. The charged muons produce Cherenkov-Light passing through the deep ice at the South Pole. The emitted light is measured with the AMANDA-II detector \cite{5YEAR} using photomultipliers and the direction and energy of the muon is reconstructed.

The AMANDA-II detector was completed in 2000 and is taking data since then. This analysis uses 4 years of AMANDA-II data (2000 to 2003, 807 days lifetime). The main background are muons produced in the atmosphere. To reject these events only up-going events are included in this sample. This reduces the field of view to the northern sky. The final event sample consists of $N=3329$ muon neutrino events. The measured data is reconstructed and filtered as described in \cite{5YEAR}. The background of miss-reconstructed down-going muons in this sample is below $5\%$.
\section{Angular Power Spectrum}
This analysis compares the angular power spectrum of the measured data to the background expectation of neutrinos produced in the atmosphere. The data is expanded by means of spherical harmonics $Y_l^m(\theta,\phi)$. The multipole index $l$ characterises the angular scale ($\delta\approx\pi/l$) and $m$ the orientation of the angular structures. Small $l$ correspond to large angular scales (e.g. overall sky-distribution). Small structures appear at large $l$ (e.g. angle between sources). Orientation averaged observables are the multipole moments $C_l$ (power components):
\begin{eqnarray*}
C_l=\frac{1}{2l+1}\sum_{m=-l}^l|a_l^m|^2,\\
a_l^m=\int_\Omega \sum_{i=1}^N\delta(\theta_i,\phi_i) \bar{Y}_l^m(\theta,\phi) d\Omega.
\end{eqnarray*}
$\Omega$ stands for the integration over the unit sphere. The software GLESP \cite{GLESP} is used to calculate the integral.

The accuracy of the calculated $C_l$ values from GLESP is limited by the event statistics. The obtained values for $C_l$ which are expected to be zero are found to be non-zero but to scale about as $C_l\sim C_{0}/N$ and $C_0$ is normalised to $\pi$. 
The AMANDA-II point source resolution is about $3^{\circ}$ corresponding to $l\approx60$. An estimate for the maximum $l$ is provided by the mean angle between the data points: $29\;$mrad corresponding to $l\approx116$. A limitation for the maximum $l$ is derived from the degrees of freedom. This is $l=57$ for 3329 events.

Correlations between the multipole moments due to the limited aperture are taken into account in the statistical analysis.
\section{Data and Background Simulation}
The angular power spectrum for the background (atmospheric neutrinos) and different signals is estimated by simulations. Each simulated data set has 3329 events (same as the experimental sample) and contains atmospheric background as well as signal events. The neutrinos are distributed according to the angular acceptance of AMANDA-II. This acceptance is energy dependant. The directions of all simulated neutrinos are varied randomly according to the angular resolution function of AMANDA-II.

The simulation of the atmospheric neutrinos  is done according to their angular zenith distribution. Theoretical uncertainties are considered by varying the assumed distribution randomly within it's uncertainties for each simulated data set. For the azimuth angle a flat distribution is assumed due to the rotation of the detector.

Source neutrinos are simulated with a Poisson-distributed number of events per source at the earth and an power law energy spectrum. The mean number of events varies between $\mu=0.1$ (corresponding to $\phi\approx5~\cdot~10^{-13}\;$cm$^{-2}$s$^{-1}$) and $\mu=10$ (corresponding to $\phi\approx5~\cdot~10^{-11}\;$cm$^{-2}$s$^{-1}$). $\phi$ is the integrated flux per source at the earth in the sensitive range between $1.6\;$TeV and $1.6\;$PeV assuming an $E^{-2}$ energy spectrum. Source locations are simulated isotropically distributed in the northern hemisphere or located in the (super) galactic plane.

Figure \ref{fig_cl} shows the angular power spectrum for atmospheric neutrinos compared with an example spectrum for extra-terrestrial neutrinos. The steep falling of the spectrum for $l<6$ appears due to the restriction to the northern sky while the flat tail corresponds to  the statistical limitation of GLESP (see above). Error bars are derived from the RMS spread found for 1000 simulated and analysed data sets. The tested multipole moments $C_l$ for the analysis are chosen by simulation according to their sensitivity for a certain signal \cite{HUELSS}: $C_{2/3/5}$ for isotropic distributed sources with a flux below $\phi = 5\cdot10^{-12}\;$cm$^{-2}$s$^{-1}$, $C_{1-40}$ for a higher flux and $C_{1-15}$ for the (super) galactic plane. For weak sources ($\mu\le1$) using only $C_{2/3/5}$ restricts the sensitivity to the overall distribution of the neutrinos.
\section{Experimental Result}
The analysis steps have been optimised using simulation without referring to the data (blind analysis). 
The angular power spectrum of the experimental data is calculated in the same way as for the simulated data. Figure \ref{fig_cl} shows the result. The experimental moments are generally within the errors of the background expectation and no general deviation is observed. For further analysis
$$d_l\equiv(C_l^{exp.}-C_l^{sim.})/\sigma_l$$ 
is defined as the difference between measurement and simulation normalised to the combined uncertainty from statistics and the model dependence. The average $<d_l>$ over $l$ for the experimental data and the purely atmospheric expectation is $<d_l>=0.2\pm0.14$ with a $RMS=1.0\pm0.3$. The value $D^2=\sum_{l=1}^{40} d_l^2 = 57.2$ is calculated. The probability to obtain a larger $D^2$ is 7\% (from the simulations). This is the probability for consistency of experimental data and the purely atmospheric expectation. It does not show a good agreement. However, this probability does not reject the purly atmospheric assumption.
\begin{figure*}
\begin{center}
\noindent 
\includegraphics [width=0.6\textwidth]{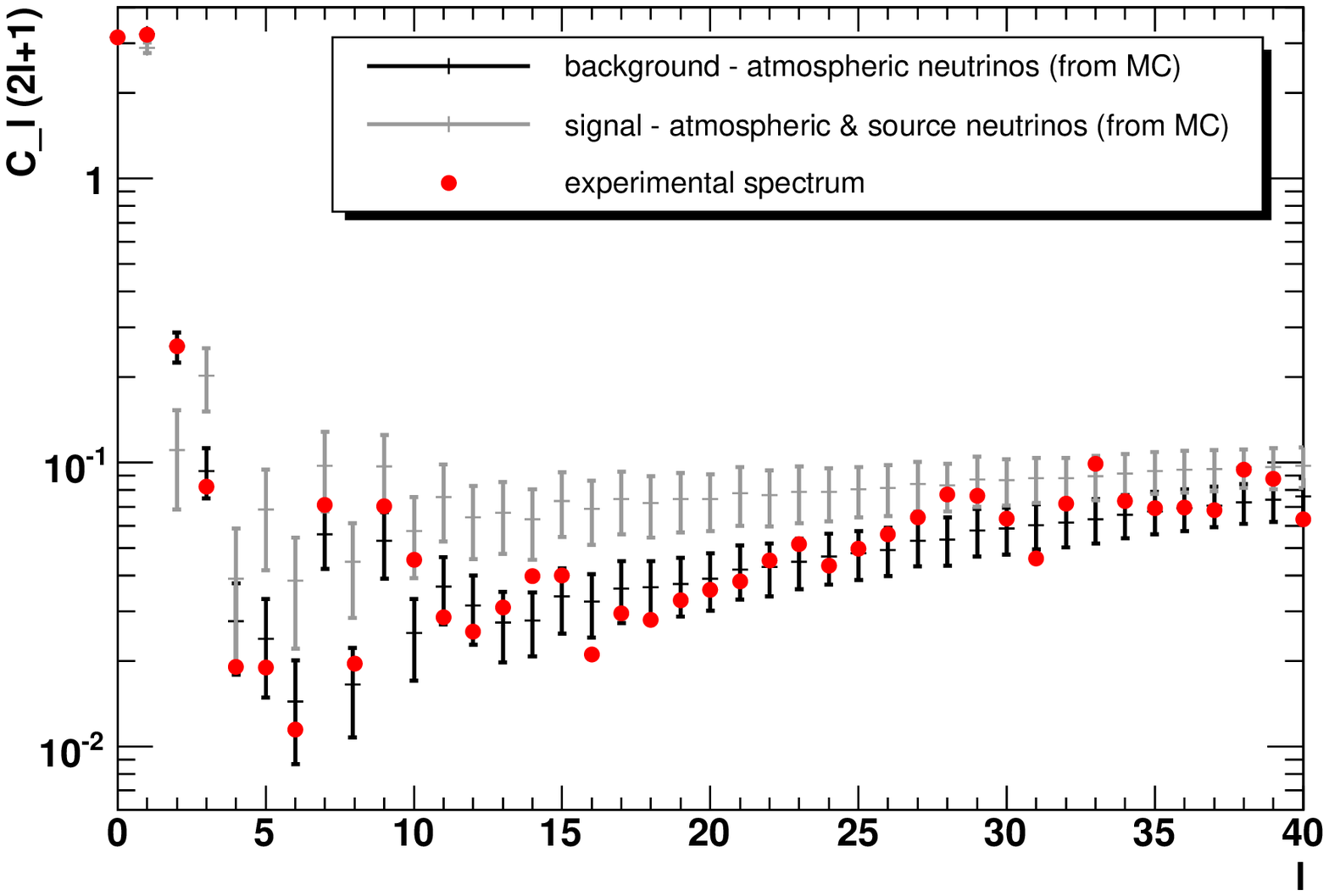}
\end{center}
\vspace{-4mm}\caption{Expected angular power spectrum for atmospheric neutrinos (black) and one signal example (gray) (both from Monte Carlo Data) compared to the experimental spectrum (red dots). The example data is simulated for 110 sources in the northern sky with a flux per source $\phi = 5.23\cdot10^{-11}\;$cm$^{-2}$s$^{-1}$. The error bars include systematic errors as described in the text.}\label{fig_cl}
\end{figure*}

\section{Limits on Cosmic Contributions}
The contribution of signal events in the experimental sample is tested by means of the observable $D^2=\sum d_l^2$. Upper limits on these are derived by constructing confident belts according to \cite{FELMANN_COUSINS} as a function of the number of signal neutrinos in the data sample.

The derived upper limits for an energy range from $1.6\;$TeV to $1.6\;$PeV and an $E^{-2}$ energy spectrum are shown in table \ref{tab_lim} and in figure \ref{fig_lim}. The limit on the total number of signal neutrinos in the data sample is almost independent of the source strength. We limit the contribution from isotropically distributed sources to be less than about 300 events total and less than about 200 events for the (super) galactic plane. As expected the results for the galactic and super galactic plane are nearly identical. The step in the limits for the isotropically distributed sources corresponds to the change in the used $l$ (see above).
\begin{table}[ht]
\begin{center}
\begin{small}
\begin{tabular}{|c|c||c|c|c|c|c|} \hline
&$\mu$&0.1&1&2&3&10\\\hline
\multicolumn{7}{|l|}{isotropically distributed sources}\\\hline
\raisebox{-.5ex}{$E^{-2}$}&$\phi$&0.52&5.2&10&16&52\\\hline
&$N_\nu$&290&295&490&380&230\\\hline
&$N_S$   &3524&358&298&154&28\\\hline
\raisebox{-.5ex}{$E^{-2.5}$}&$\phi$     &&&29&43&\\\hline
&$N_\nu$   &&&775&625&\\\hline
&$N_S$   &&&625&298&\\\hline\hline
\multicolumn{7}{|l|}{sources in the galactic plane}\\\hline
\raisebox{-.5ex}{$E^{-2}$}&$\phi$&0.76&7.6&15&23&76\\\hline
&$N_\nu$&162&175&182&190&250\\\hline
&$N_S$   &1968&213&111&77&30\\\hline
\raisebox{-.5ex}{$E^{-2.4}$}&$\phi$     &&&36&53&\\\hline
&$N_\nu$   &&&168&172&\\\hline
&$N_S$     &&&115&78&\\\hline\hline
\multicolumn{7}{|l|}{sources in the super galactic plane}\\\hline
\raisebox{-.5ex}{$E^{-2}$}&$\phi$&0.74&7.4&15&22&74\\\hline
&$N_\nu$&172&183&197&195&240\\\hline
&$N_S$&2090&222&120&79&29\\\hline
\raisebox{-.5ex}{$E^{-2.4}$}&$\phi$     &&&33&49&\\\hline
&$N_\nu$   &&&175&178&\\\hline
&$N_S$   &&&119&81&\\\hline
\end{tabular}
\end{small}
\caption{Derived 90\% CL limits on the number of measured extra terrestrial neutrinos $N_\nu$ and the number of sources in the northern hemisphere $N_S$ depending on the source flux $\phi$ (in $10^{-12}$cm$^{-2}$s$^{-1}$) and the energy spectrum.}\label{tab_lim}
\end{center}
\end{table}
\begin{figure*}
\begin{center}
\noindent 
\includegraphics [width=0.7\textwidth]{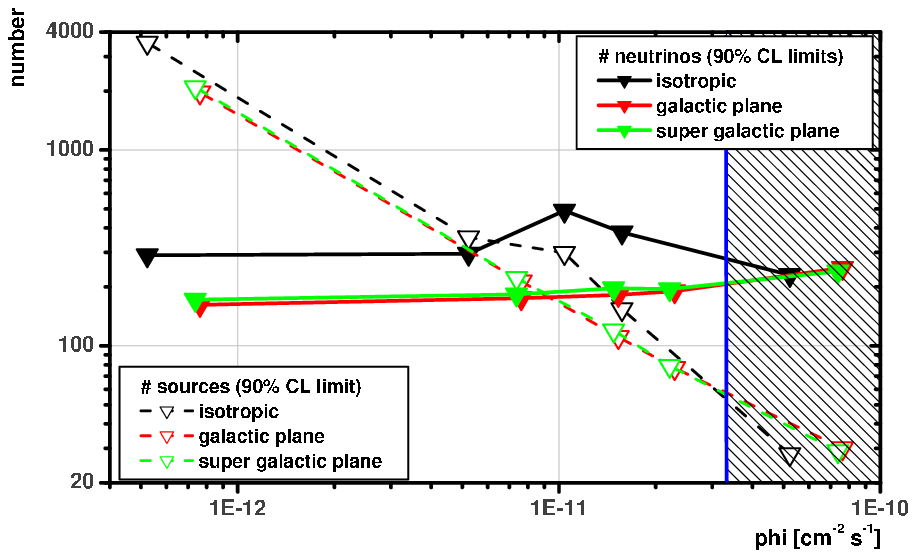}
\end{center}
\vspace{-4mm}\caption{90\% CL upper limits on the number of neutrinos from extra terrestrial sources in the data sample (full lines) and on the number of sources in the northern sky (dashed lines) for different distributions assuming an $E^{-2}$ energy spectrum. Complementary to this analysis the direct search for point sources excludes any source above a flux of $\phi = 4.38\cdot10^{-11}\;$cm$^{-2}$s$^{-1}$ \cite{5YEAR}. This restriction is indicated as the shaded region in the graph.}\label{fig_lim}
\end{figure*}

The limits on the number of neutrinos can be converted to limits on the number of sources (fig. \ref{fig_lim}). The number of sources is decreasing with increasing strength.
The tested flux per source in this analysis is chosen to be below the flux limit for resolved sources $\phi = 4.38\cdot10^{-11}\;$cm$^{-2}$s$^{-1}$ derived by \cite{5YEAR}. However, limits on the number of sources presented here depend on the assumed sky distribution of the sources and the equal source strength.

With this analysis further limits are derived \cite{HUELSS}. Table \ref{tab_lim} shows limits for other power law energy spectra.% Additionally limits for diffuse neutrino fluxes are obtained. 
The limit on a diffuse neutrino flux ($E^{-2}$) is about $5\cdot10^{-7}\;$GeV$^{-1}$cm$^{-2}$s$^{-1}$sr$^{-1}$. This is a factor 2 to 3 worse compared to the actual limit set by AMANDA-II.
A diffuse flux ($E^{-2.7}$) from the galactic plane is limited to be below $E^{2.7}\cdot d\phi/dE<3.4\cdot10^{-3}\;$GeV$^{1.7}$cm$^{-2}$s$^{-1}$sr$^{-1}$ with 90\% C.L.. 
$C_{2,3}$ are sensitive to atmospheric neutrino oscillation. Figure \ref{fig_cl} shows no difference between the expectation and the experimental result. We derive a limit $\Delta m^2_{atms.} < 5\cdot10^{-3}\;$eV$^2$ (90\% C.L.) for maximum mixing.
\section{Conclusion}
For the first time the technique of a multipole analysis, well known from CMBR, is applied to the AMANDA-II data. It is found suitable to search for a signal of extra-terrestrial neutrinos. The analysis is not well optimised yet. For the future with increased statistics and improved analysis we expect a substantially increasing sensitivity.

%\bibliography{icrc0851}
%\bibliographystyle{plain}
%\end{document}

\setcounter{figure}{0}
\setcounter{table}{0}
%%
% International Cosmic Ray Conference 2007 Merida Yucatan Mexico
% In This file you will find detailed instructions to correctly
% typeset your document.
%
%
%

%Class Requeried
%\documentclass{article}
%The ICRC Style
%\usepackage{icrctc07}

%The paper title
\title{Cluster Search for neutrino flares from pre-defined directions}
%Short title to print in the headers to the final publication (Not showed in this print).
\shorttitle{Cluster Search for neutrino flares from pre-defined directions}
%All paper authors
\authors{K.Satalecka$^1$, E.Bernardini$^1$, M.Ackermann$^1$$^2$, M.Tluczykont$^1$ for the IceCube Collaboration$^3$}
%Short title to print in the headers to the final puplication (Not showed in this print).
\shortauthors{Satalecka,K. et al}
%All the affiliations.
\afiliations{$^1$ DESY Zeuthen, Plataneneallee 6, D-15738, Germany\\
$^2$ Stanford Linear Accelerator Center, Stanford, California 94305-4060, USA\\
$^3$ see special section of these proceedings
}
\email{elisa.bernardini@desy.de}

%The abstract.
\abstract{We present here a method to search for clusters of high energy neutrinos from pre-defined
directions, a study of the background rate over short time scales, and report
novel results obtained from AMANDA data from years 2004 to 2006. 
The time structures we search for must indicate an occasional
deviation from the background hypothesis while not contradicting
observations from time-integrated searches. In the context of the
multi-messenger approach, where the information from high energy neutrinos and
different electromagnetic wavelengths (e.g., high energy gamma-rays) is combined,
we look for correlations between the high energy neutrinos and high 
states of $\gamma$-ray emission of selected sources. This test is performed before 
the cluster search in order to prevent a posteriori findings of coincidences of neutrino events 
with $\gamma$-ray flares once a significant cluster is found.} 
%We also discuss the perspectives for IceCube.}

%\email{aastex-help@aas.org}

%%%%%%%%%%%%%%%%%%%% B E G I N   D O C U M E N T%%%%%%%%%%%%%%%%%%%%%%%
%\begin{document}
\maketitle
%Begin the section.

\section{Introduction}
 Different observations of some candidate neutrino sources indicate that
 their electromagnetic emission is very variable and often shows a flare-like behavior. 
According to several models one can expect that the neutrino emission from those 
sources have a similar character. Time integrated analyses \cite{5yrps} \cite{IC9} \cite{Jim05}
are not always sensitive to this behavior: if signal events are emitted in flares,
for an equivalent signal efficiency the integrated background is higher over longer 
exposures. We therefore developed a dedicated time variability analysis with the goal 
of improving the discovery chance.\\
Using a time-clustering algorithm, we look for time 
structures (clusters) in the time distribution of the neutrino events from 
certain directions. This approach has the advantage of being independent 
of any a priori assumption on the time structure of the potential signal, 
but is affected by a high trial factor.
An issue for this type of analysis is the reliability of the background estimation 
over short time scales. So far the background was estimated from the event 
density as a function of the declination (similar to the ON/OFF-source approach of 
$\gamma$-ray astronomy) \cite{5yrps}. This method however fails when 
applied to short time scales due to the limited event statistics. 
To address this problem we developed a parametrization 
of the background which reduces its statistical uncertainty. In the next section
we describe in more detail the principle of this analysis, 
discuss its performance in comparison to previous analyses and give results obtained 
on data collected with AMANDA-II in 2004 to 2006.\\
The analysis presented here is realized in two steps.
In order to prevent a posteriori observations of coincidences with $\gamma$-rays
we first test the event sample for a coincident $\gamma$-ray emission for those sources 
and periods when the $\gamma$-ray data is available. The outcome of this test 
is declared positive 
if an excess significance equal or higher than 5$\sigma$ is found. 
%\begin{enumerate}
%\item \textbf{search for neutrino events in coincidence with $\gamma$-ray 
%flares:} observed events and expected background are compared only for those time periods in 
%which the energy-integrated $\gamma$-ray flux from a selected source exceeds an earlier 
%established threshold.\\
%This allows us to avoid a situation in which correlations in between neutrino and $\gamma$-ray
%flares are found a posteriori, preventing a correct estimation of the significance.\\
If in the first step none of the observations shows
a significance of 5$\sigma$ or higher (or if there are not enough $\gamma$-ray data for 
a coincidence study) we apply the time-clustering algorithm to the whole analysis period 
for a set of selected sources.
Three types of sources were chosen for this analysis: blazars, XRBs and radio laud AGNs 
\cite{ICRC05}. The selection criteria required: a variable 
character of the source in one or more wavelengths and indications of non-termal emission. 
%availability of the source's $\gamma$-ray light-curve for the years 2004 and 2006.\\

\section{Time-clustered search for neutrino bursts}  
For each preselected direction all combinations (clusters) of the arrival time of events 
within a certain angular bin are constructed. For each cluster its multiplicity ($m$) 
is compared to the expected background
($\mu_{bg}^{loc}$) and the significance of the cluster ($S_{bg}$) is calculated.
% integrated Poisson probability to observe a multiplicity $m$ or higher due to a fluctuation of the background, is calculated ($P_{bg}$). 
The cluster with
the highest significance ($S^{best}_{bg}$) is chosen as the "best". The overall probability 
($P$, trial factor corrected) to observe a cluster of significance $S^{best}_{bg}$ or higher is
calculated based on 10,000 Monte Carlo (MC) experiments.
The main difference between this analysis and what was presented in \cite{ICRC05} is 
that in this work no assumption is made on the duration of signal flares.
Moreover, a correct background
estimation over short time scales is necessary, in order to properly calculate the cluster's 
significance and its compatibility with the background hypothesis. The method used 
previously in the time integrated analysis \cite{5yrps} is simple and fast. 
However due to the low statistics it is affected by large uncertainties in a 
case of short time scales (e.g. $\Delta t<10$ days).\\
A different approach for a background estimation has been developed for this work.
We first tabularise the detector up-time development. This takes into account 
the inefficiency periods and data gaps after the data quality selection.
Once corrected for the detector exposure we calculate the expected neutrino rate from the
whole northern sky\footnote[1]{We did not observe any dependency of the results for 
different choices in the binning of the event rates or angular 
regions of the sky (e.g. estimating the expected rate for different declination regions).} 
by fitting the event rate versus time. We obtained 4.13$\pm$0.13/3.7$\pm$0.13/4.30$\pm$0.13 
events per day ($\mu_{bg}^{year}$) for 2004/2005/2006 respectively. For each sky angular bin the number 
of expected events in the whole data period (i.e year 2004, 2005 or 2006) is then 
calculated as:\\
\begin{equation}
\mu_{bg}^{loc}=\mu_{bg}^{year}\times\frac{N_{band}\times A_{bin}}{N_{all}\times A_{band}}
\end{equation}
where: $N_{band}$ the number of 
events in the declination band in the sky defined by the bin size, $N_{all}$ the number of 
all events in the sample for the analysed year, $A_{bin}$ the area of the angular search bin,
and $A_{band}$ the area of the declination band defined by the size of the angular search 
bin. The ratio $N_{band}/N_{all}$ allows to account for the different background density 
at different declinations.\\
The result of equation (1) is what we expect when we neglect the variation of the 
efficiency with the azimuth angle caused by the asymmetrical shape of the detector, 
(shown in \ref{ICRC0582_fig1}) and assume a continuous up-time.
\begin{figure}[ht]
\begin{center}
\includegraphics*[width=0.48\textwidth]{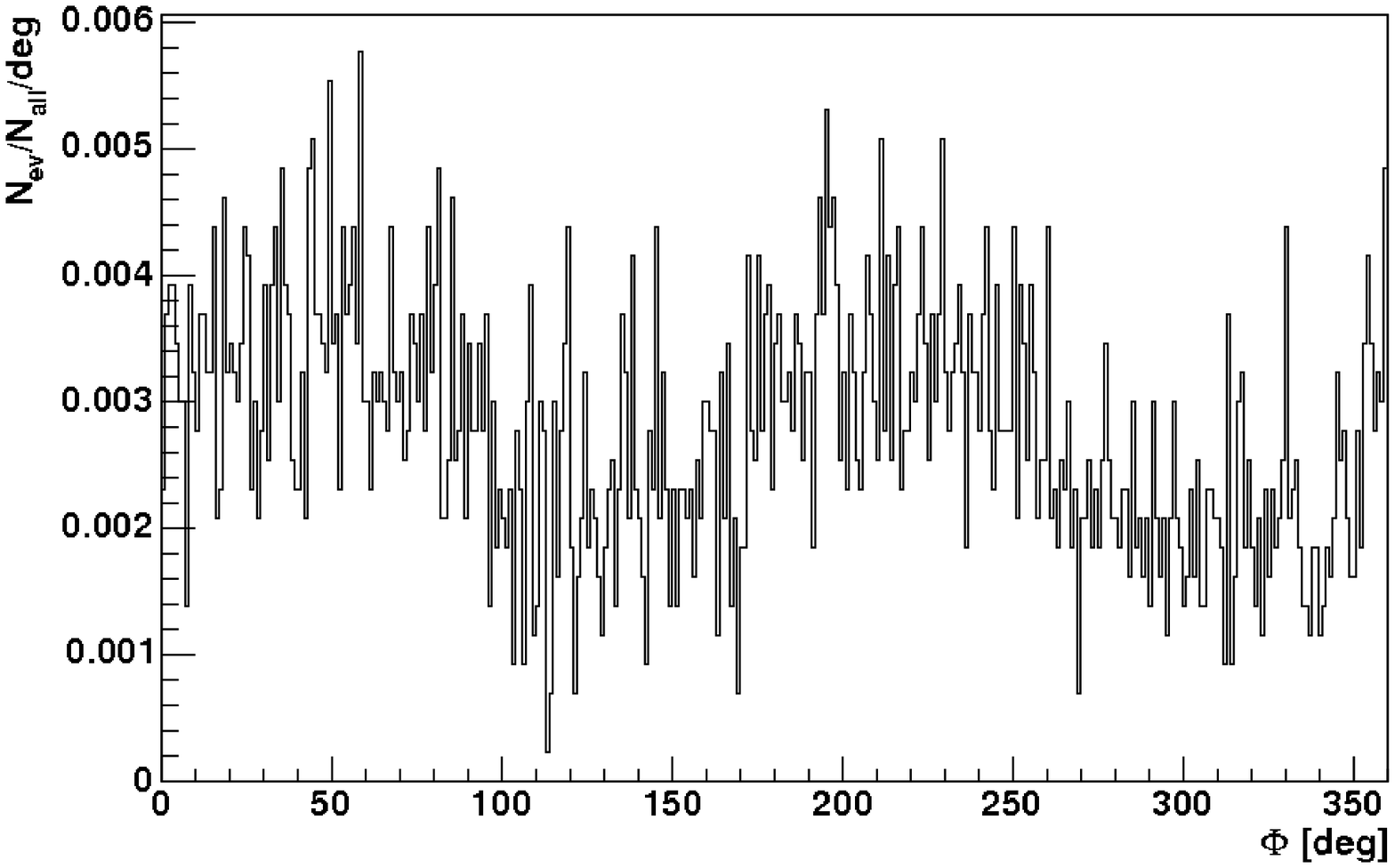}
\vspace{-0.5cm}
\caption{\label {ICRC0582_fig1}
The normalized azimuth distribution of the data sample reported in \cite{5yrps}.}
\end{center}
\end{figure}
The variability averages out for long
periods of data. However, it plays a role for very short periods of 
integration. We therefore correct the value estimated in equation (1)
for the effective azimuth exposure, calculated for each individual time cluster. 
The correction applied is given by the integral of the azimuth efficiency over the time 
period of a cluster, taking into account the effective time coverage (up-time) of the
azimuth bin. The overall error in the background estimation is a combination of 
a statistical uncertainty, the error of the fit and the uncertainty introduced 
by the azimuth corrections.\\         
A comparison of the outcome of this method to previous results shows 
that for short time scales the new method yields much smaller uncertainties while for
longer time periods they are in very good agreement. For example for $\Delta t=3$ days, 
we could achieve in this analysis an error of 20\% compared to 30\% in previous works.\\ 
%old method:$\mu^{loc}_{bg}=0.015\pm 0.005$, new method: $\mu^{loc}_{bg}=0.011\pm 0.002$).
%(e.g. $\Delta t=140$ days, old method: $\mu^{loc}_{bg}=0.7\pm 0.08$ 
%new method:$\mu^{loc}_{bg}=0.68\pm 0.07$ ).\\
%\fbox{\parbox{\columnwidth}{PLACE HOLDER FOR ALGORITHM PERFORMANCE STUDY\\
%\rule{0pt}{5cm}\\
%}}
Fig.~\ref{ICRC0582_fig2} shows a study of the neutrino flare detection chance depending on the 
strength and duration of the signal. We produced about 10,000 MC experiments simulating 
a variable neutrino point-source of different signal strengths and durations, 
on a background $\mu_{bg}$, characteristic for a chosen region of the sky. 
Positions in the sky of the on-source events 
were generated randomly, corresponding to the Point Spread Function, while the 
number of signal and background events were generated using corresponding Poisson 
averages. We performed this study for the cases of fixed and variable angular search 
bin size (chosen among the set of angular distances of the events relative to the sky positions 
of the pre-selected sources). We found that the best detection chance is obtained with
a variable bin size.\\ 
The results of the cluster search for neutrino flare for 2004 to 2006 are reported in Table~\ref{ICRC0852_tab2}.
\begin{figure}[ht]
\begin{center}
\includegraphics*[width=0.48\textwidth]{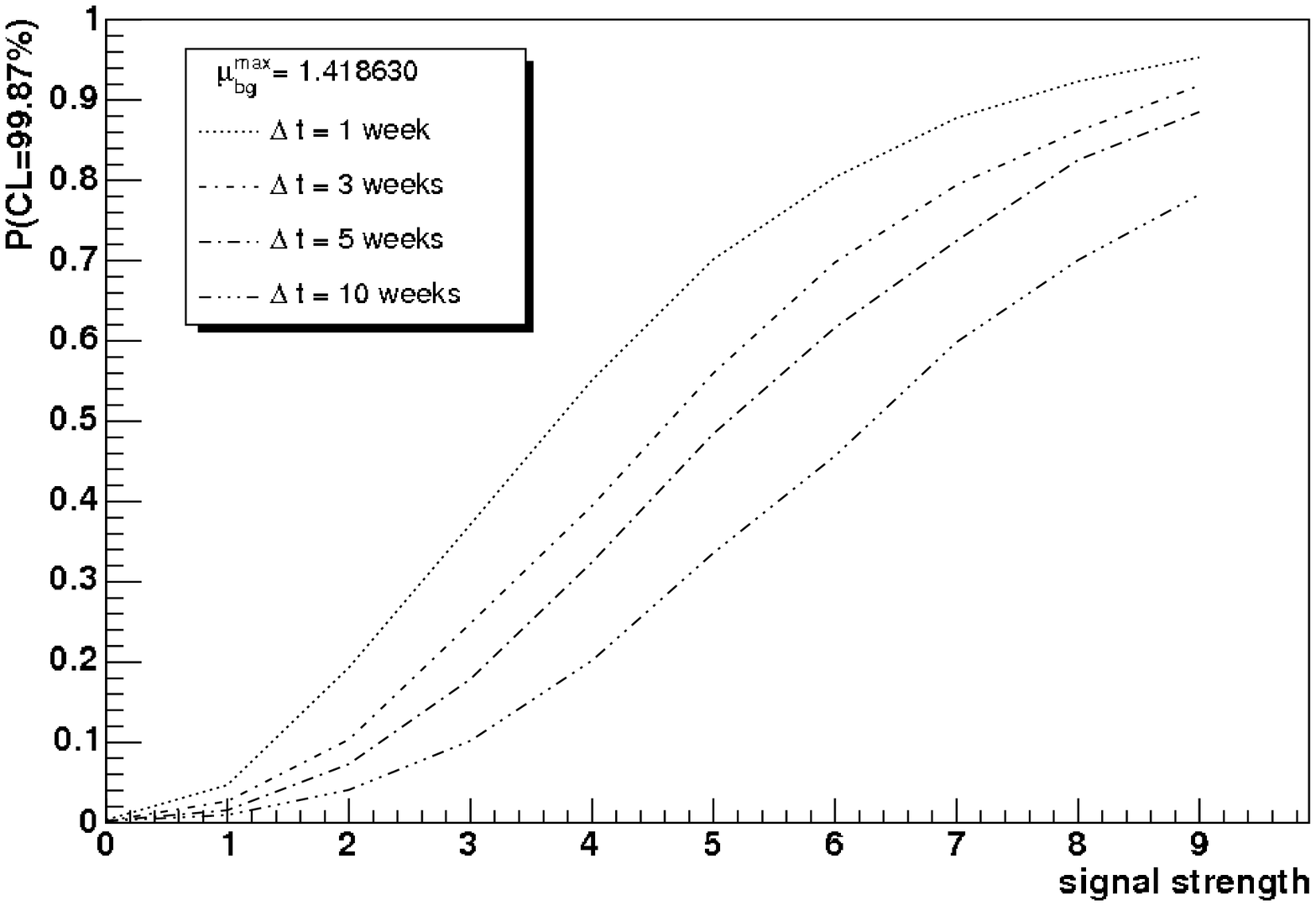}
\vspace{-0.5cm}
\caption{\label {ICRC0582_fig2}
Probability of detecting a neutrino flare with a significance of 99.87\% or 
higher in one year of data (2004 in this example). The X-axis shows the signal 
strength (mean number of signal events). The 
curves indicate different time duration of the signal (1, 3, 5 
or 10 weeks).
}
\end{center}
\end{figure}

\section{Search for neutrino events in coincidence with $\gamma$-ray flares}
Observations of strong variability in the high energy (TeV) $\gamma$-ray emission exist for 
various TeV neutrino candidate sources. However, often there is no long coverage of 
their flux and also a very limited knowledge exists on the frequency of $\gamma$-ray flares, 
as well as on their eventual time correlation with neutrino flares.
Nevertheless, a search for coincidences between high energy neutrinos and $\gamma$-rays
can possibly increase the discovery chance. Here we present the results of a test 
for correlation of neutrinos with high state of $\gamma$-ray emission for a sub-sample of 
objects for which $\gamma$-ray data for the 
years 2004 to 2006 are published. For each selected source we established a flux threshold 
for the selection of periods of interest. 
The number of neutrino events observed - $n_{obs}$ - in the whole period was compared 
to the expected background - 
 $\mu_{bg}$ - and the significance of this observation was calculated.\\       
The threshold to the gamma-ray flux was chosen based on an analysis of
combined light-curves \cite{MT06TeV}. For each source we considered the integral
flux above variable thresholds (S) and optimized the latter for the best
S/sqrt(B), where B is proportional to the time coverage of the periods
above threshold. We exclude periods of measurement gaps longer than one week as well as 
periods with upper limits on the flux only. In a case of Cygnus X-1
only one day of significant measurement was available so we took the sensitivity of the
experiment as the flux threshold.\\
The results of the search for neutrino events in coincidence with $\gamma$-ray flares are
reported in Table~\ref{ICRC0852_tab1}.
No significant excess was found.\\
\vspace{-0.5cm}
\setlength{\tabcolsep}{0.1cm}
\begin{table}[h]   %[H] add [H] placement to break table across pages
\caption{\label{ICRC0852_tab1} Results of the search for neutrino events in coincidence with $\gamma$-ray flares. Column ''Selected periods'' give the year and integrated up-time of the detector in days.}
\vspace{-0.2cm}
\begin{center}
\begin{tabular}{c|c|c} 
~Source~  & Sel. periods & $n_{obs}$ / $\mu_{bg}$ \\
                             \hline
Mkn421         & 2004 (7.6) & 0 / 0.057$\pm$0.007 \\ 
              & 2005 (1.0)  &0 /0.0067$\pm$0.0008 \\    
             & 2006 (10.8) &  0 / 0.078$\pm$0.009 \\ 
Mkn501         & 2005 (21.1) & 1 / 0.13$\pm$0.02 \\
1ES1959+650    & 2005 (0.95) & 0 / 0.0040$\pm$0.0007   \\
BL Lac    & 2005 (2.0) & 0 / 0.008$\pm$0.001  \\
H1426+428    & 2006 (3.0) & 0 / 0.018$\pm$0.002   \\
                            \hline          
Cyg X-1    & 2006 (1.0) & 0 / 0.0070$\pm$0.0008 \\
                             \hline          
M87    & 2005 (4.7) & 0 / 0.033$\pm$0.004   \\
                            
\end{tabular}
\end{center}
\end{table} 

\section{Results}
The input data sample for this analysis for 2004 and 2005 was taken from \cite{5yrps} and
\cite{Jim05} respectively. 
For 2006 we used the results of the AMANDA on-line event
reconstruction and filtering chain, which was implemented following the scheme
reported in \cite{5yrps}. After excluding periods of IceCube calibration with an artificial 
light source and selecting high quality data we used 247.5/199.9/239.5 effective days of 
data taking for the year 2004/2005/2006.\\ 
Table~\ref{ICRC0852_tab2} reports the results of the cluster search for neutrino flares for combined data 
sets of 2004, 2005 and 2006. The highest excess observed (for Cygnus X-3 ) corresponds 
to 3.56$\sigma$. The overall probability to observe a cluster of this 
significance or higher at any time in the whole periods analyzed equals 5.9\% 
(not including the trial factors due to looking on several sources) and is well compatible 
with the background hypothesis.
\vspace{-0.3cm}
\begin{table}[ht]   %[H] add [H] placement to break table across pages
\caption{\label{ICRC0852_tab2} Results of the search for neutrino clusters: duration $\Delta t$ [days], 
angular bin size $\Delta\psi$ [deg], significance of the
best cluster found $S_{bg}^{best}$ [$\sigma$] and the overall probability to observe a cluster of this 
significance or higher at any time in the whole periods analyzed $P$ [\%] .}
\vspace{-0.3cm}
\begin{center}
\begin{tabular}{c|c|c|c|c} 

~Source~  & $\Delta t$  & $\Delta\psi$ & $S_{bg}^{best}$  & $P$  \\
                             \hline
Mkn 421         & 3.9 & 5.2 & 1.6 & 95.0 \\                   
Mkn 501         &  26.5  & 4.8 & 3.2 & 14.5  \\
Mkn 180         &   0.35 & 2.2 & 2.92 & 30.0  \\
1ES 1959+650    &  11.2   & 2.8 & 2.82 & 29.0   \\
1ES 2234+514    &  42.2  & 3.4 & 2.7 & 35.0  \\
1ES 1218+30.4    & 5.0   & 6.0 & 1.4  & 95.0 \\
BL Lac     &  51.6  & 4.6 & 2.45  & 46.0  \\
H1426+428    &  4.4  & 5.2 & 1.5 & 92.0  \\
3C 66A    & 7.7   & 5.0 &  2.45  & 44.0   \\
3C 454.3    & 8.1   & 4.8 & 2.7  & 33.0  \\
                             \hline          
GRO J0422+32    & 19.5   & 5.8  & 1.75  & 90.0 \\
GRS 1915+150   &  94.4  & 2.0  & 3.2  & 8.4 \\
LSI+61 303      &  0.2   & 4.5 & 2.9 & 31.0 \\
Cyg X-1    &  27.5  & 6.37  & 3.2  & 15.0  \\
\textbf{Cyg X-3} & \textbf{8.8} & \textbf{4.3} &\textbf{3.56}  &\textbf{5.9} \\
XTE J1118+480   & 31.1 & 4.5  & 2.25  & 64.0 \\
                             \hline          
3C 273    &  194.5  & 6.1 & 2.88 & 9.1  \\
M87    &  11.1  & 6.6 & 2.0 & 69.0  \\                           
\end{tabular}
\end{center}
\end{table}
\section{Summary}
We have presented the first search for neutrino flares from pre-selected sources in AMANDA-II 
with no a priori assumption on the time structure of the signal. In order to prevent 
a posteriori findings of coincidences with $\gamma$-ray flares a 
pre-test was performed, to look for correlations between the high energy neutrinos and high 
states of $\gamma$-ray emission of selected sources. In both cases no significant excess 
was found above the expected background.  
To accomplish the time-clustered search we 
have developed a new background estimation method which allows to reduce the statistical 
uncertainties as compared to the classical ON/OFF-source approach. The method here presented also 
properly takes into account the effects due to the detector asymmetries arising from a 
non-homogeneous detector. This approach becomes relevant when analysing data for IceCube, 
a detector under construction with a non-homogeneous distribution of the strings before 
completion.\\
 
{\small{\bf Acknowledgments} We thank the Office of Polar Programs of
the National Science Foundation, DESY and the Helmholtz
Association, as well as MAGIC, HESS and VERITAS Collaborations which
provided us with the necessary VHE $\gamma$-ray data.}\\
 
%This is the reference to .bib file (Whitout .bib!)
%\bibliography{ICRC0582/icrc0582}
%This in the bibtex style, is ok.
%\bibliographystyle{plain}

%\end{document}

\setcounter{figure}{0}
\setcounter{table}{0}
%%
% International Cosmic Ray Conference 2007 Merida Yucatan Mexico
%
%

%Class Requeried
%\documentclass[dvips]{article}
%The ICRC Style
%\usepackage{icrctc07}

%The paper title
\title{All-Sky Search for Transient Sources of Neutrinos Using Five Years of AMANDA-II Data}
%Short title to print in the headers to the final publication (Not showed in this print).
\shorttitle{All-Sky Search for Autocorrelated Neutrino Transients}
%All paper authors
%\authors{Rod\'{\i}n Porrata$^1$ for the IceCube Collaboration$^2$}
\authors{Rod\'{\i}n Porrata$^1$ for the IceCube Collaboration}
%Short title to print in the headers to the final publication (Not showed in this print).
\shortauthors{Rod\'{\i}n Porrata and et al}
%All the affiliations.
\afiliations{$^1$University of California at Berkeley, 
	Department of Physics,
%       366 Le Conte Hall \#7300, 
        Berkeley, CA 94270, USA}
%	$^2$see special section of these proceedings}
\email{porrata@berkeley.edu}

%The abstract.
\abstract{Up to now, analyses of AMANDA data have been limited to searches for diffuse astrophysical sources, 
time-integrated searches for point sources, and searches for flares and bursts from preselected sources 
(AGN and GRB) over very limited timescales. However, 
multi-wavelength studies have shown that AGN and GRB emissions generally
occur in exponential flares or bursts with strengths that can be much greater than that of the corresponding quiescent 
emission, and that the timescales for these violent outbursts can vary from milliseconds to months.
%While these same multi-wavelength studies are much needed to determine the full complexity of the character of 
%astrophysical sources, and can sometimes be used to pinpoint when to search for neutrino flares, their 
%present lack of coverage and low duty cycle make them inadequate to take fully advantage of IceCube's high duty 
%cycle two-pi coverage. Furthermore, it is possible that neutrino flares occur without contemporaneous emission at 
%other wavelengths. 
Therefore, we are performing an all-sky search 
for transient sources of neutrinos with AMANDA data taken from the years 2000 to 2004~\cite{five}, surveying the largest range 
of timescales for which an improved signal to noise ratio can be obtained. 
In this report we describe a new analysis technique that utilizes an 
unbinned two-point correlation function which separates pairs of signal events from the atmospheric
neutrino background by taking into account the probabilities
for observing the given spatial separation, time separation,
and total number of hit channels (NCH)
of the events given both signal and background hypotheses. At the shortest timescales probed, this analysis
achieves a differential fluence sensitivity,
$\bar{\mathcal{F}}^{0} = \left( \frac{\mathrm{E}}{\mathrm{1TeV}} \right)^{\gamma} \frac{d\bar{\mathcal{F}}}{\mathrm{dE}}$,
to flaring FR-I galaxies that is almost a factor of three better than the 5-year stacked point source sensitivity, assuming a spectral
	index, 
$\gamma = 2$, and  a $\mathcal{F}_{\nu_{\mu}+\bar{\nu}_{\mu}} / \mathcal{F}_{\nu_{\tau}+\bar{\nu}_{\tau}}$
flavor ratio of one.
%Although the cosmological source distribution
%is assumed to be isotropic, sources are more likely to be detected in directions
%and at timescales at which the detector is most sensitive, since the fluence from these
If they produce events in the detector at all, fluences from such sources must be critically close to the detection threshold to avoid
having been observed in other surveys, thus a pair search could provide
the earliest detection of astrophysical neutrinos.
}

%
%\begin{document}

\maketitle

\section{Introduction}

%We are performing an all-sky search for flaring and bursting sources of neutrinos using
%data collected from AMANDA-II during the years 2000 to 2004~\cite{five}. 
%The search focuses on the possible observation of pairs of neutrinos that 
%are both spatially and temporally correlated, and further, whose 
%combined NCH\footnote{The total number of hit detector channels in the event reflects
%	the energy of the source neutrinos. Here, the total NCH for the pair is used, corrected for detector geometry.} value is more
%likely to have resulted from neutrinos produced in an astrophysical source
%than by the atmospheric neutrino background. 
Studying the space-time-energy properties of pairs of neutrinos has
several advantages over other methods of searching for astrophysical sources. 
%The first is that we 
1.~We can search
the whole sky for astrophysical sources in disregard of the scarcity of multi-wavelength information that could
potentially aid such a search, allowing the possible detection of source classes that are dark at
other wavelengths. 
%Thirdly, 
%\footnote{The total number of hit detector channels in the event reflects
%	the energy of the source neutrinos. Here, the sum of the NCH values for the pair is used, corrected for detector geometry.}
2.~Since the search utilizes the energetic information that can be inferred from NCH data, it 
uses the same advantage that a diffuse search does to observe a faint astrophysical signal, however, unlike a diffuse
analysis,
%comes closest to being a diffuse search, however,
correlated event searches
%with separations in time greater than a few nanoseconds 
are unaffected by a charm component.
%, thus able to distinguish a true astrophysical signal from the harder atmospheric component
%that is so far unknown.
%Fourthly, 
3.~If the number is sufficient, a pair search has the greatest sensitivity to detect very weak classes of astrophysical point sources,
	and is more powerful than searching for other multiplicities,
	e.g., pairs probe 
84 \% more space than triplets, while 
%in the case of observation, 
if present, a neutrino triplet will be counted as 3 pairs, 
%so the significance is numerically very close.
with comparable significance.\footnote{ Classes of point sources with $N_{e}$
	events per source and $N_{s}$ objects will be expressed
in this analysis as 
$
N_{p} = N_{s} N_{e} (N_{e} - 1 )/ 2$
pairs, e.g., if a source like Mrk 421,
for which there is 7 events observed in the integrated point source analysis, 
emits those neutrinos on a timescale less than one day, then
it will appear in this analysis as 21 pairs -- a situation that
has a chance better than 50~\% of producing a 5-$\sigma$ discovery.}
	
\subsection{The Time Variability of AGN and GRB}
In summary, the activity of AGN and GRB~\cite{galdos}\cite{kniffen}\cite{hurley}\cite{thompson}\cite{nolan}\cite{valtaoja}\cite{albert}, 
	the primary astrophysical candidates
associated with theorized hadronic processes inducing  neutrino
emission, suggests that we 
could
observe astrophysical neutrinos arriving in bursts with almost any 
imaginable time difference. We assume that on a logarithmic scale, flare
timescales are uniform. However, considering the differences
in the astrophysical processes over different timescales, the pair search is 
separated into searches over several different timescales based upon the
classes of objects that might be observed. Tab.~\ref{times} lists a possible way
to construct the search categories.
Optimization of the search strategy 
and calculation of the significance of the search results is 
conducted separately for each search timescale.  
\begin{table}[ht]
\centering
\begin{tabular}{|l|c|c|}
\hline 
Flare Category & T$_{l}$ / T$_{u}$  & L \\
 \hline
 GRB / TeV SN / TeV AGN &  1 s / 2 hr & A  \\
 GRB afterglow / TeV AGN  & 2 hr / 3 dy &  B  \\
Large scale AGN flare  & 3 / 30 dy &  C \\
\hline
\end{tabular}
\caption{Timescales over which we will search for various astrophysical categories: 
(1) category of objects, (2) minimum/maximum time between events 
(T$_{l}$/T$_{u}$), 
(3) label for this discussion. 
\label{times}}
 \end{table}

\subsection{Constraints and Background}
%The observation of individual point sources performed in the 5-year integrated analysis
The diffuse analysis~\cite{hodges} %\cite{diffuse}
imposes the most stringent limit on the total number of neutrinos 
from all astrophysical sources integrated over the
entire sky that 
could possibly be observed. 
For the 2000-2004 point source dataset, this translates to $\sim$~80 neutrinos
assuming an E$^{-2}$ source spectrum. We have used this number as the maximum
number of neutrinos that we simulate for any of the source classes we consider. 

%\subsection{Background Samples}

 While studying the
response of the detector to background and potential astrophysical sources, the time-integrated
point source analyses are able to average the detector efficiencies over time and right ascension (RA). 
However, this analysis probes the detector down to timescales of a second or less, so it can
be strongly affected by the asymmetries of the detector. 
%The standard method of randomizing the RA
%of the data
%to study the background does not properly take these asymmetries into account.
A new method of randomizing the data was developed that properly takes into account 
the asymmetries in the combined zenith (ZEN) and azimuth (AZ) distribution of events, 
the preferential occurrence of NCH values from certain AZ and
ZEN directions, and the granularity of the detector on-periods. 
%were accounted for, so background
%nd signal events are simulated only at times when the detector was recording data and was functioning
%in a stable mode. 
The ZEN and AZ of background events are sampled from the data itself and a smearing
function is applied. The smearing function, determined by MC, is the point spread function of the
detector given an atmospheric
neutrino spectrum. Times are sampled from a list of all possible detector on-periods for the entire 5 year analysis.
Having obtained a map of the ZEN, AZ, and time of the events according
to the efficiencies of the detector, the RA 
is calculated using the transformation
\[
\mathrm{RA} = ( \mathrm{MJD} \cdot 24.06571 \cdot 15 - \mathrm{AZ} ) \% 360
\]
which is valid for a pair analysis performed on data obtained at South Pole. MJD is the Modified Julian Date.
This method, comparable to \textit{direct integration}~\cite{atkins},
keeps all detector efficiencies intact, while
producing a randomized sky-map of the data that is complete.

\section{Search Technique}
\label{method}
%Having settled upon the minimum and maximum time differences for the separation
%in time of a pair of events for the class objects that we consider (see Sec.~\ref{timescales}), 
%and upon $\zeta$, our function providing evidence in favor or against a pair being an observation
%of signal, 
%to be discussed shortly, all possible pairs of neutrino events are compared. For each pair that falls within
All pairs of neutrino events are compared. For each pair that falls within
the minimum and maximum time differences given by the search class (see Tab.~\ref{times}), $\zeta$ is
calculated and if its value surpasses a predetermined threshold, $\zeta_{c}$, 
the count of observed events
%, $n_{obs}$, 
is incremented. Once the tally is complete, the significance of
the observation is determined using the Poisson p-value.
%\[
%\alpha = \sum_{n=n_{obs}}^{\infty} \frac{ \mu_{bg}^{n} }{ n!}  e^{-\mu_{bg}} 
%	\label{pvalue}
%\]
%where $\mu_{bg}$ is the average number of pairs expected from pure background. 
To derive $\zeta$, 
consider the likelihood ratio for the $i$th pair of events:
\[
\mathcal{LR}_{i}  = \mathrm{\frac{ P( NCH_{i} | S) \  P(\log_{10}[\Delta t_{i}]| S)\  P(\psi_{i} | S) }
	{ P( NCH_{i} | B) \  P(\log_{10}[\Delta t_{i}]| B) \  P(\psi_{i} | B) }}
	\label{llh}
\]
where:

%\begin{description}
%\item[$\mathbf{P( NCH_{i} | B)}$]
$\mathbf{P( NCH_{i} | B)}$ - The probability distribution for
	        NCH, given a pair of background events. This is obtained by 
		calculating the distribution of all combinations of NCH values from the data itself. Before
		this distribution is calculated, the values are standardized across 8 different declination
		bands by subtracting the median
		of the distribution and dividing by the inter-quartile difference, both declination dependent quantities. 
		The standardization process removes to first order the geometric component of the
		variation of NCH values as a function of ZEN, leaving the spectral energy dependence intact.
%\item[$\mathbf{P(\log_{10}[\Delta t_{i}]| B) }$]

$\mathbf{ P(\log_{10}}\mathrm{\Delta}\mathbf{ t_{i}  | B)} $  - The distribution of the logarithm of time differences of background pairs of events.
This is obtained by fitting time differences
of the data
from 0.001 to 30 days with a power law. The result of the fit gives
$
\mathrm{P(\log_{10}[\Delta T]|B) \propto \ \Delta T^{0.98} }
$.
This is in agreement with expectations that it should increase proportionally with the time difference.
The fit is used to obtain the probability of observing pairs of events that occur 0.01 days apart
or less. For longer timescales the randomized time distribution is used directly.
%\item[$\mathbf{ P(\psi_{i} | B) }$]

$\mathbf{ P(\psi_{i} | B) }$ - 
The probability of observing a given spatial separation of background events is $\sim \rho_{bg} \sin{\psi/2}$, where
$\rho_{bg}$ is the local spatial density of background events.
Since the background does not vary too quickly, and
since an average over all directions is obtained when moving away from the point in question, this approximation is good
enough.
%\item[$\mathbf{P( NCH_{i} | S)}$] The probability distribution for

$\mathbf{P( NCH_{i} | S)}$ - The probability distribution for
	NCH given a pair of signal events. This is obtained from 
source MC, weighted according to an E$^{-2}$ spectrum. This probability distribution is 
standardized using the same
quantities used to standardize P(NCH$\mathbf{_{i}} |$B).
%\item[ $ \mathbf{ P(\log_{10}  \Delta t_{i}  | S)} $] 

$ \mathbf{ P(\log_{10}}\mathrm{\Delta}\mathbf{ t_{i}  | S)} $ -
Based upon reviews of AGN and GRB activity, the central assumption of this
work is that the distribution of flare/burst timescales is 
the scale invariant Jeffrey's prior~\cite{gregory}, i.e.,
a \textit{constant} for logarithmically sized bins. 
%The prior probability of observing a given 
%logarithm of the separation in time of a pair of signal
%events, $\log_{10}[\Delta t_{i}]$, is 
%the central assumption of this work. Based upon reviews of AGN and GRB activity,
%	the distribution of flare/burst timescales is scale invariant, i.e.,
%\textit{constant} for logarithmically sized bins\cite{gregory}. 

%\item[$ \mathbf{P(\psi_{i} | S)} $] 
$ \mathbf{P(\psi_{i} | S)} $ -
The probability of observing a given separation in 
space of a pair of signal
events from the same source. This is obtained from source MC data, 
weighted according to an E$^{-2}$ spectrum. The PSF takes into account both the intrinsic
$\nu_{\mu} \rightarrow \mu $ mismatch angle and the mismatch angle between the reconstructed muon
and its true direction. Note that
the PSF is evaluated at an angle, $\Psi$,  which is half the separation
angle, $\psi$,  between the events.
%\end{description}
\begin{figure}[th]
\includegraphics*[width=0.5\textwidth,angle=0,viewport=95 185 550 565,clip]{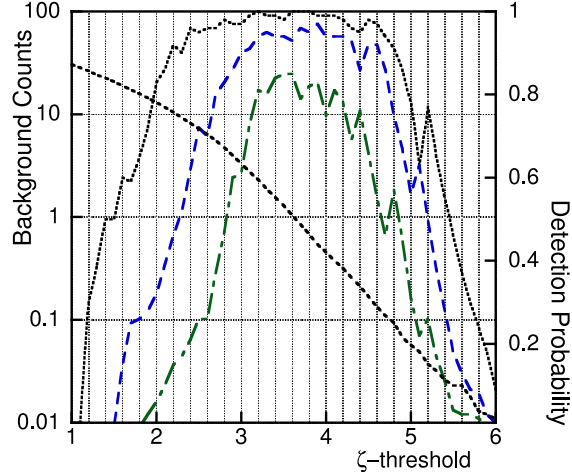}
\caption{\textbf{Preliminary} probabilities for (from top to bottom) 3, 4, and 5-$\sigma$ detections are plotted (units on r.h.s.) for a 2 hour search as a function of cut 
on $\zeta$. 
The monotonically decreasing curve is the expected counts from background only. 
The time distribution of the signal is logarithmically uniform, 
5.5 source pairs per $\log_{10}(\mathrm{\Delta T / Days})$ extending from 10 seconds to 2 hours. 
\label{pl:detprob}}
\end{figure}

According to the Neyman-Pearson Lemma, the quantity,
$
\zeta_{i} = \log_{10}( \mathcal{LR}_{i} )
$,
represents one of the best possible ways of utilizing all the information we have discussed 
in order to provide evidence to decide whether the $i$th pair of events is an observation
of signal or background.
The choice of the value of $\zeta_{c}$ is
optimized in the following manner: 
The analysis is run on 10,000 simulated experiments
with pure background and 1000 simulated experiments that contain a small
amount of simulated signal.  For each experiment, the background and signal pairs are counted as a function of the cut on $\zeta$ and the
significance of the observation is calculated. Then the median signficance 
is calculated from the set of experiments and this is used to determine the best evidence threshold.
For a search for GRB-timescale flares (A in Tab.~\ref{times}), where the signal
consists of $\sim$16 sources distributed isotropically
on the sky and distributed log-uniformly from 10 seconds to two hours, each source contributing two events to the dataset,
the expected background is plotted in Fig.~\ref{pl:detprob}, as 
are the 3, 4, and 5-$\sigma$ detection probabilities. 
Here it is seen that the 5-$\sigma$ detection probability is 
maximal at $\zeta_{c} = 3.6$ where its value is better than 80\%. 

\section{Results and Conclusions}
The fluence sensitivity
is given by 
\[
\bar{\mathcal{F}}^{0}_{L} = \frac{\bar{\mu}_{90}}{n_{s}} \mathcal{F}^{0}_{s}
\]
where $\mathcal{F}^{0}_{s}$ is the normalization constant on the differential fluence 
(the differential flux integrated over the duration of flaring events) of the signal model studied, $n_{s}$ is the number of neutrino
events that would be observed given such a fluence, and $\bar{\mu}_{90}$ is the Feldman-Cousins average upper
limit given the background and no source events~\cite{fc}. For classes of objects with more than one member, 
$n_{s}$ is the sum of contributions from each. 
Preliminary differential fluence sensitivities, as well as detection probabilities, are presented in Tab.~\ref{tabsens} for
two classes of objects that meet the requirements of the diffuse analysis. 
\begin{table}[h!]
\centering
\begin{tabular}{l|c|c|c|c|c}
\hline 
Cat. & N$_{\mathrm{s}}$ & N$_{\nu ps}$ & $\varepsilon_{\mathcal{T}}$ & $\bar{\mathcal{F}}^{0}_{L}$ & P$_{\mathrm{5-}\sigma}$ \\
\hline 
%A &10 & 2 & 0.78 & 5.34 & 14 \\ % good, May 29, 2007
A & 20 & 2 & 0.78 & 2.70 & 99.6 \\
%\multicolumn{5}{|c|}{$T_{min} = 2$ hr --  $T_{max} = 3$ days}  \\
%B & 40 & 2 & 0.49 & 5.0 & 42 \\ % good May 29 2007.
%B & 10 & 8 & 0.41 & 30.0 & 1.0 \\ %May 15, 2007 ok
B & 27 & 3 & 0.44 & 8.2 & 86 \\ 
%B & 16 & 5 & 0.43 & 18.1  & 1.0 \\ %may 15 ok
% 20 & 4 & 0.46 & 13.6  & 0.996\\  %may 15 ok
%B& 27 & 3 & 0.49 & 9.4  & 0.80  \\ %May 15, 2007
\hline
\end{tabular}
\caption{\textbf{Preliminary} differential fluence sensitivities and 
detection probabilities for representative source classes: 
(1) Category of objects, (2) no. of sources, (3) no. of neutrinos per source,
(4) signal efficiency, 
(5) $\nu_{\mu} + \nu_{\tau}$ fluence sensitivity in units of $10^{-4}$~TeV$^{-1}$~cm$^{-2}$,
(6) 5-$\sigma$ detection probability (\%), excluding our $N_{T} = 4$ trials factor.
\label{tabsens}}
 \end{table}
In Fig.~\ref{pl:fluence} we plot the differential fluence sensitivities of
GRB-timescale source classes for which the
$n$th source has a relative strength given by~\cite{stacking}
$
\mu_{n} = \mu_{0} n^{-\alpha} e^{-n/n_{c}}
$
.  Our sensitivity to FR-I-like 
objects~\cite{FRI} (not including the brightest source, which is  3C-274,
		$\alpha = \mathrm{0.65}$)
%~\cite{FRI}\footnote{Not including the brightest source, which is  3C-274,
%$\alpha = \mathrm{0.65}$.} 
that produce Cat.~A flares is 1.3$\times 10^{-3}\ \mathrm{TeV}^{-1} \mathrm{cm}^{-2}$ compared to 
the integrated stacking analysis~\cite{stacking} result of 3.4$\times 10^{-3} \ \mathrm{TeV}^{-1} \mathrm{cm}^{-2}$.
%\section{Discussion}
%The 5-Year integrated analysis~\cite{five} places only mild constraints on what could possibly be observed 
%by this analysis. For that analysis, only upper limits were derived from observations of selected source
%positions where as much a 10 events were observed. They also performed a simple pair correlation analysis,
%without selecting events for evidence of signal. For the latter analysis a source class with 20 sources
%contributing 10 events on average would produce a result that is less than 3-$\sigma$. In comparison,
%Table~\ref{results} shows that if a similar source class emits those neutrinos on timescales less than
%3 days, also with 20 sources, but  contributes only 4 neutrinos on average, then it would have
%99.6 \% chance of producing a 5-$\sigma$ discovery. 
\begin{figure}[th]
\includegraphics*[width=0.48\textwidth,angle=0,viewport=55 165 575 650,clip]{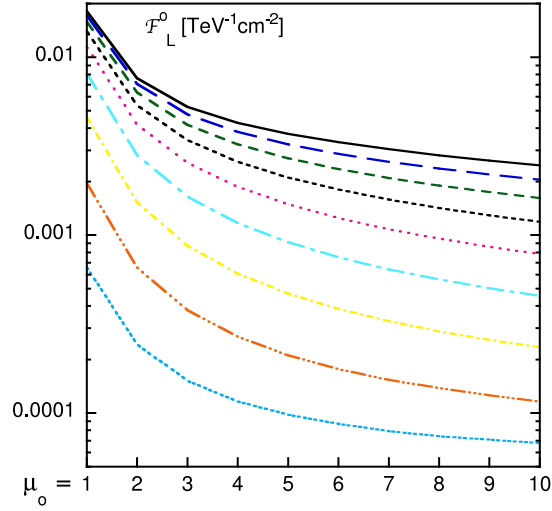}
\caption{\textbf{Preliminary} fluence sensitivities for
	Cat.~A objects characterized by $\alpha$ and $\mu_{o}$. 
	The curves, from top to bottom, are for $\alpha=2$ to $\alpha=0$ in steps of $0.25$, assuming $n_{c}=50$.
\label{pl:fluence}}
\end{figure}
The results of this survey are presented in Tab.~\ref{results}.
\begin{table}[h]
\centering
\begin{tabular}{c|c|c|c|c}
\hline 
Cat. & $\zeta_{c}$ & $\mu_{bg}$  & $n_{obs}$ & p-value\\
 \hline
 A &  3.885 & 0.573 & 1 & 0.44 \\
B  &  1.94   &  31.5  & 37 & 0.19 \\
C  &  0.63 &  431 & 457        & 0.11 \\
\hline
\end{tabular}
\caption{\textbf{Preliminary} results of survey.  
(1) Category of objects, (2) optimized evidence threshold, (3) no. background pairs expected,
(4) no. pairs observed, (5) significance, excluding $N_{T} = 4$.
\label{results}}
 \end{table}
%The most significant observation was for the Cat.~X flare timescales for which X events were observed
%with X expected from background, giving an X-sigma observation taking into account $N_{T}$. Thus we
%infer that there ????????????? classes of flaring neutrino objects observable in the AMANDA-II dataset. However, our
Although none of the observations
were significant, the
results of the sensitivity study show the potential of this technique to search for weak astrophysical sources that flare. This study will serve as the
starting point for all-sky transient searches performed with the full IceCube detector.

%\end{document}

\setcounter{figure}{0}
\setcounter{table}{0}
%%
% International Cosmic Ray Conference 2007 Merida Yucatan Mexico
% In this file you will find detailed instructions to correctly
% typeset your document.
%
% By: Victor De la Luz
% vdelaluz@inaoep.mx
% Mexico City

%Class Required
%\documentclass{article}
%The ICRC Style
%(This package is the last package in the usepackage list)
%If you need import other package you need write it first.
%\usepackage{icrctc07}

%The paper title
\title{Neutrino Triggered Target of Opportunity (NToO) test run with AMANDA-II and MAGIC}
%Short title to print in the headers to the final publication (Not showed in this print).
\shorttitle{Neutrino Target of Opportunity}

%All paper authors
\authors{M. Ackermann$^1,5$, E. Bernardini$^1$, N. Galante$^2$, F. Goebel$^2$, M. Hayashida$^2$, K. Satalecka$^1$, M. Tluczykont$^1$, R.~M. Wagner$^2$, for the IceCube$^3$ and MAGIC collaborations$^4$}
%Short title to print in the headers to the final publication (Not shown in this print).
\shortauthors{Ackermann and et al.}
%All the affiliations.
\afiliations{$^1$DESY, Platanenallee 6, 15738 Zeuthen\\ 
             $^2$MPPMU, F\"ohringer Ring 6, 80805 M\"unchen\\
             $^3$See special section of these proceedings\\
             $^4${\tt http://magic.mppmu.mpg.de/collaboration/members}\\
             $^5$Now at SLAC, Stanford University, USA}
\email{elisa.bernardini@desy.de} 

%The abstract.
\abstract{Kilometer scale neutrino telescopes are now being
constructed (IceCube) and designed (KM3NeT). While no neutrino flux of
cosmic origin has been discovered so far, the first weak signals are
expected to be discerned in the next few years. Multi-messenger 
(observations combining different kinds of emission) 
{investigations can enhance the discovery chance for neutrinos in case
of correlations}. One possible application is the search for time
correlations of high energy neutrinos and established signals. We show
the first adaptation of a Target of Opportunity strategy to collect
simultaneous data of high energy neutrinos and gamma-rays. 
Neutrino events with coordinates close to preselected candidate sources 
are used to alert gamma-ray observations. The detection
of a positive coincidence can enhance the neutrino discovery
chance. More generally, this scheme of operation can increase the
availability of simultaneous observations. If cosmic neutrino signals
can be established, the combined observations will allow time
correlation studies and therefore constraints on the source
modeling. A first technical implementation of this scheme involving
AMANDA-II and MAGIC has been realized for few pre-selected sources in a
short test run (Sept. to Dec. 2006), showing the feasability of the concept.
Results from this test run are shown.
%The principles of the NToO test run and its first outcomes will
%be shown and the potential for IceCube will be discussed.  
}

%%%%%%%%%%%%%%%%%%%% B E G I N   D O C U M E N T%%%%%%%%%%%%%%%%%%%%%%%
%\begin{document}

\newcommand{\Nobs}{$n_\mathrm{obs}\,$}
\newcommand{\Nbck}{$n_\mathrm{bck}\,$}
\newcommand{\Ncoinc}{$n_{\gamma}\,$}
\newcommand{\pgam}{${p_{\gamma}}\,$}
\newcommand{\tevflux}{$\mathrm{F_{VHE}}$}
\newcommand{\asmflux}{$\mathrm{F_{ASM}}$}
\newcommand{\rhs}{${R_{HS}}$}
\newcommand{\thigh}{$\mathrm{T_{high}}$}
\newcommand{\tlow}{$\mathrm{T_{low}}$}

\newcommand{\signeu}{P$_\nu$}
\newcommand{\sigtoo}{S$_{ToO}$}

\maketitle
%Begin the section.

\section{Introduction}
\label{NToO_intro}
%\vspace{-0.2cm}
The major aim of neutrino astrophysics is to contribute to the understanding 
of the origin of high energy cosmic rays. 
A point-like neutrino signal of cosmic origin would be an unambiguous signature 
of hadronic processes, unlike $\gamma$-rays which can also be created in
leptonic processes.
%Charged particles and high energy gamma rays have been 
%detected up to energies of few hundred EeV and a few tens TeV respectively. 
%Neutrinos are therefore expected to be emitted in several astrophysical 
%scenarios, both extragalactic~\cite{learned} and galactic~\cite{gal-review}.
%
Neutrino telescopes are ideal instruments to monitor the sky and 
look for the origin of cosmic rays
because they can be continuously operated. 
%Typically, half of the sky is accessible by a neutrino telescope when selecting
%up-going tracks as neutrino candidates (muon channel analysis).
The detection of cosmic neutrinos is however very challenging because
of their small interaction cross-section and because of a large 
atmospheric background.
%The expected small signals have to compete with an irreducible
%background of neutrinos and muons produced in cosmic ray interactions
%in the atmosphere.
%
%To date no significant extraterrestrial high energy neutrino signal has been 
%observed.
%
{Parallel measurements using neutrino and electro-magnetic observations
(multi-messenger)
can increase the chance} to 
discover the first signals by reducing the trial factor penalty 
arising from observation of multiple sky bins and over different time periods. 
%Input from established electromagnetic observation 
%windows is %change FG
%used for this purpose.
%
In a longer term perspective, the multi-messenger approach also aims 
at providing a scheme for the phenomenological interpretation of the 
first possible detections.
The Antarctic Muon and Neutrino Detector Array (AMANDA) was built
with the aim to search for extraterrestrial high energy neutrinos 
\cite{NToO_amanda}. The Major Atmospheric Gamma Imaging Cherenkov telescope 
(MAGIC) is a current generation $\gamma$-ray telescope that
operates in the northern hemisphere at a trigger energy threshold of 60\,GeV 
\cite{magic}.

\vspace{-0.2cm}
\subsection{Neutrino Target of Opportunity test run}
\label{sec:NToO}

%\subsection{Basic idea}
% basic idea
% AMANDA-II / MAGIC
The neutrino target of opportunity (NToO) test run described here was 
defined as a cooperation between the AMANDA (neutrinos) and MAGIC
($\gamma$-rays) collaborations \cite{bernardini:2005a}.
Each time a neutrino event was detected from the direction 
of a predefined list of objects,
a trigger was sent to the $\gamma$-ray telescope.
MAGIC then tried to %change FG
observe the object within a predefined time window
after the neutrino trigger. %change FG
{The primary goal of the NToO approach is to achieve
simultaneous neutrino/$\gamma$-ray observations.
This can be realized by triggering
follow-up observations of interesting neutrino events,
such as multiplets within a short time window or 
very high energy events, therewith assuring 
$\gamma$-ray coverage for these neutrino events.
Multiplets are very seldom in AMANDA-II observations (low statistics). 
We therefore implemented a test run based on single high energy neutrino events
from pre-defined directions.} These events are most likely due 
to atmospheric neutrino background.
%\subsection{NToO test run}
The test run took place between 27th of September and 27th of November 
2006 and its purpose was to test the
technical feasability of the NToO strategy. 
%The AMANDA-II sub-array was chosen 
%AMANDA-II \cite{finley:2007a}.
The AMANDA-II DAQ data at the South Pole passed through 
an online reconstruction
filter that selected up-going muon tracks and provided 
a monitoring of the data quality. 
Whenever a neutrino event was reconstructed within a few degrees of
one of the selected sources and passed the
data quality criteria, a message was sent 
via e-mail to the MAGIC shift crew. The message contained the time of the
event, the source name and the reconstructed angular distance from the source.
If possible (day/night duty cycle), the object was then observed 
with the MAGIC telescope within 24 hours {for a duration of 1 hour}.
A coincidence is counted when a $\gamma$-ray high state (flare) is measured
in these observations.
A $\gamma$-ray flare can be defined as an
observation above a predefined threshold flux ${F_{thr}}$.
The individual thresholds were chosen either based on the MAGIC
sensitivity or in case of Mrk\,421 to a conservatively low value for which
the probability to observe a high state as defined above
would be {of the order of few percent}.

%\vspace{-0.2cm}
\subsection{An example analysis: Blazars}
\label{analysis}
%A large number of excess events above the expected background
%is usually necessary for a significant detection.
%For instance, an excess of several events clustered in time
%which is incompatible with the expected atmospheric background
A stand-alone neutrino analysis can only yield a significant
result if an excess above the expected atmospheric background is observed. 
%e.g. significantly time-clustered events
%(see e.g. \cite{satalecka:2007a}).
In the multi-messenger framework, 
the observation of a number of neutrino events in coincidence
with gamma-ray high states can be an indication
for a neutrino/$\gamma$-ray correlation.
If this correlation is incompatible with
the chance probability for coincidence
with atmospheric neutrinos
such an observation would be evidence at the same time
for a cosmic origin of the neutrino events and
a hadronic nature of the gamma-ray signal.
In this scheme for the interpretation of data a statistical 
test was defined before the measurements. 
Under the hypothesis that all the neutrinos 
detected from the direction of the source are atmospheric,
the chance probability of detecting at least \Nobs neutrinos 
and observing at least \Ncoinc coincident gamma-ray flares
is given by:
{
\begin{equation}\label{NToO_prob}
%\small
P=\sum_{i=n_\mathrm{obs}}^{+ \infty}\frac{(n_\mathrm{bck})^i}{i!}
\mathbf{e^{-n_\mathrm{bck}}}
\sum_{j=n_{\gamma}}^{i}\frac{i!}{j!(i-j)!}p_{\gamma}^j (1-p_{\gamma})^{i-j},
\end{equation}
}
where the first term describes the Poisson probability of observing 
at least \Nobs neutrinos with \Nbck expected background events,
and the second term describes the probability of observing 
at least \Ncoinc coincident gamma-ray flares
{out of the $j\,\ge$\,\Nobs triggers}.
\pgam is the probability to observe a gamma-ray {high state
above a certain threshold ${F_{thr}}$ within a given time window}. 
$P$ defines the post-trial
significance of a set of coincidences observed from one source. 
Trial factors to account for the number of sources 
considered can be easily included using Binomial statistics.
For illustration of Equation~\ref{NToO_prob}, let us assume that we observe
\Nobs = 10 neutrinos with a background expectation of \Nbck = 10.
In itself this measurement would not be significant. 
However, if coincidences with $\gamma$-ray high-states are observed
the significance increases as shown in Figure~\ref{significance} for
different $\gamma$-ray probabilities.
\begin{figure}[ht]
\centering
\includegraphics*[width=0.42\textwidth,angle=0,clip]{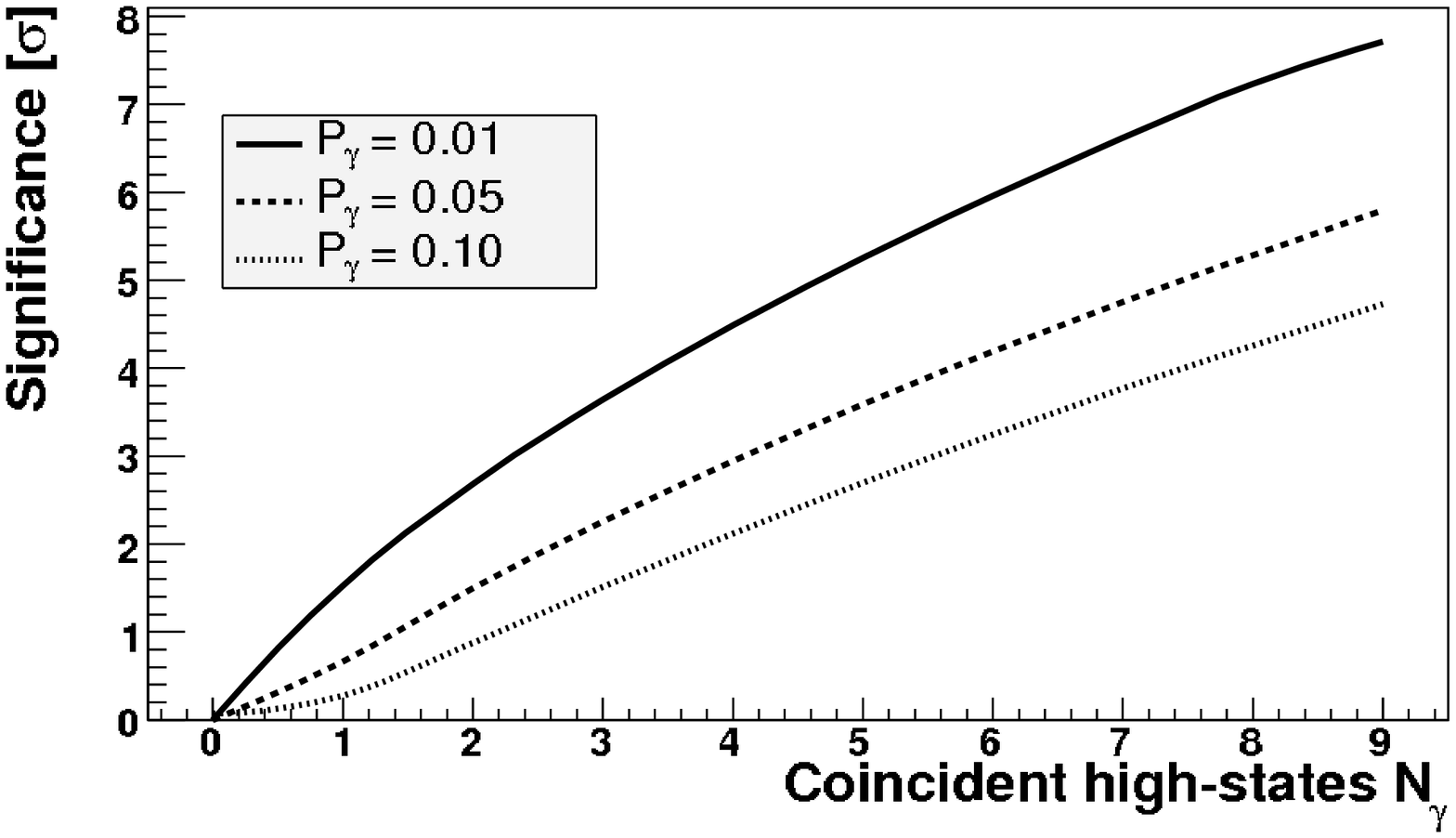}
\vspace{-0.3cm}
\caption{\label{significance}Significance of simultaneous neutrino/gamma-ray observations
         vs. the number of observed coincidences, given for
         different values of \pgam (Equation~\ref{NToO_prob}). Here, \Nobs = \Nbck = 10 was assumed.}
\end{figure}
So far, limited knowledge is available on \pgam. 
Efforts are on-going to address the issue of estimating an 
upper limit on \pgam for a few interesting sources, 
from a compilation of gamma-ray observations~\cite{tluc} 
and from random or long term monitoring observations 
(e.g. performed by the VERITAS and the MAGIC telescopes). 
We notice that a compilation of existing data is likely biased 
from the availability of measurements triggered by high states 
of emission observed at different wavelengths, which would 
tend to give an overestimation of \pgam and therefore an 
underestimation of the significance of the coincidences.
The probability \pgam is, on average, equal to the average 
high-state rate of an object.
One method for the estimation of the high-state rate is
{based on} the flux frequency distribution of the object,
shown in Figure~\ref{fluxstates} for Mrk\,421. {This distribution
can be interpreted as a stochastic flux-state distribution and can
be well fit by an exponential}. The high-state rate \rhs(${F_{thr}}$)
{above a threshold ${F_{thr}}$
is then given by}
\begin{equation}\label{hsr}
  R_{HS}(F_{thr}) = \frac{
                        \int_{F_{thr}}^\infty e^{b x} dx
                         }
                         {
                        \int_{F_{0}}^\infty e^{b x} dx
                         }
                  = \frac{e^{b F_{thr}}}{e^{b F_{0}}}
\end{equation}
where ${F_{0}}$ is the baseline flux of the object and b is the slope of
the flux distribution.
The relative high-state rate
of Mrk\,421 as derived from this formula is shown 
in Figure ~\ref{hsrate} 
as a function of the chosen threshold $F_{thr}$.
{Due to the bias to high states of the available Mrk\,421 observations,
the high state rate is systematically overestimated here.}
%The collection of light curves, the combination of different lightcurves,
%the definition of a common lightcurve format and the derivation of high 
%state rates 
These results will be described in detail in \cite{tluc}.
The estimation for \pgam can be used in Equation~\ref{NToO_prob} 
in the case of Blazars, for which $\gamma$-ray data exist
and long-term lightcurves have been compiled.
The expected background rate is the rate of atmospheric neutrinos 
in the sky bins around the selected sources. 
Depending on the source declination and on the choice of the bin size, 
this rate ranges from about 1 to 4 events per year and per source based 
on the AMANDA-II event information and according to the current scheme 
of event reconstruction and selection \cite{achterberg:2006a}.
\begin{figure}[ht]
 \centering
 \includegraphics*[width=0.42\textwidth,angle=0,clip]{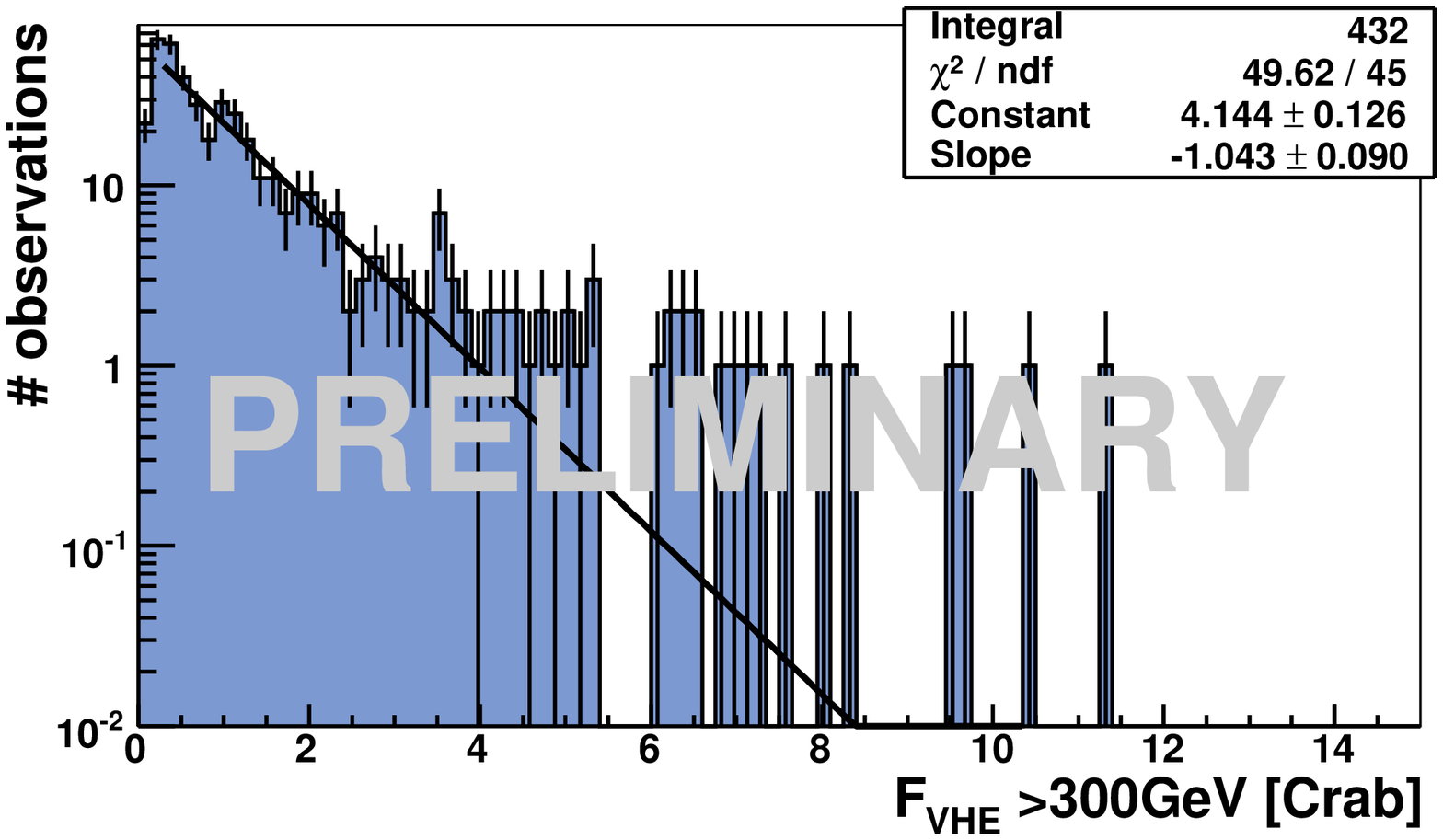}
 \vspace{-0.3cm}
 \caption{\label{fluxstates}Distribution of flux states above 300\,GeV 
          of 15 years of VHE observations of Mrk\,421 \cite{tluc}.}
\end{figure}
\begin{figure}[ht]
 \centering
 \includegraphics*[width=0.42\textwidth,angle=0,clip]{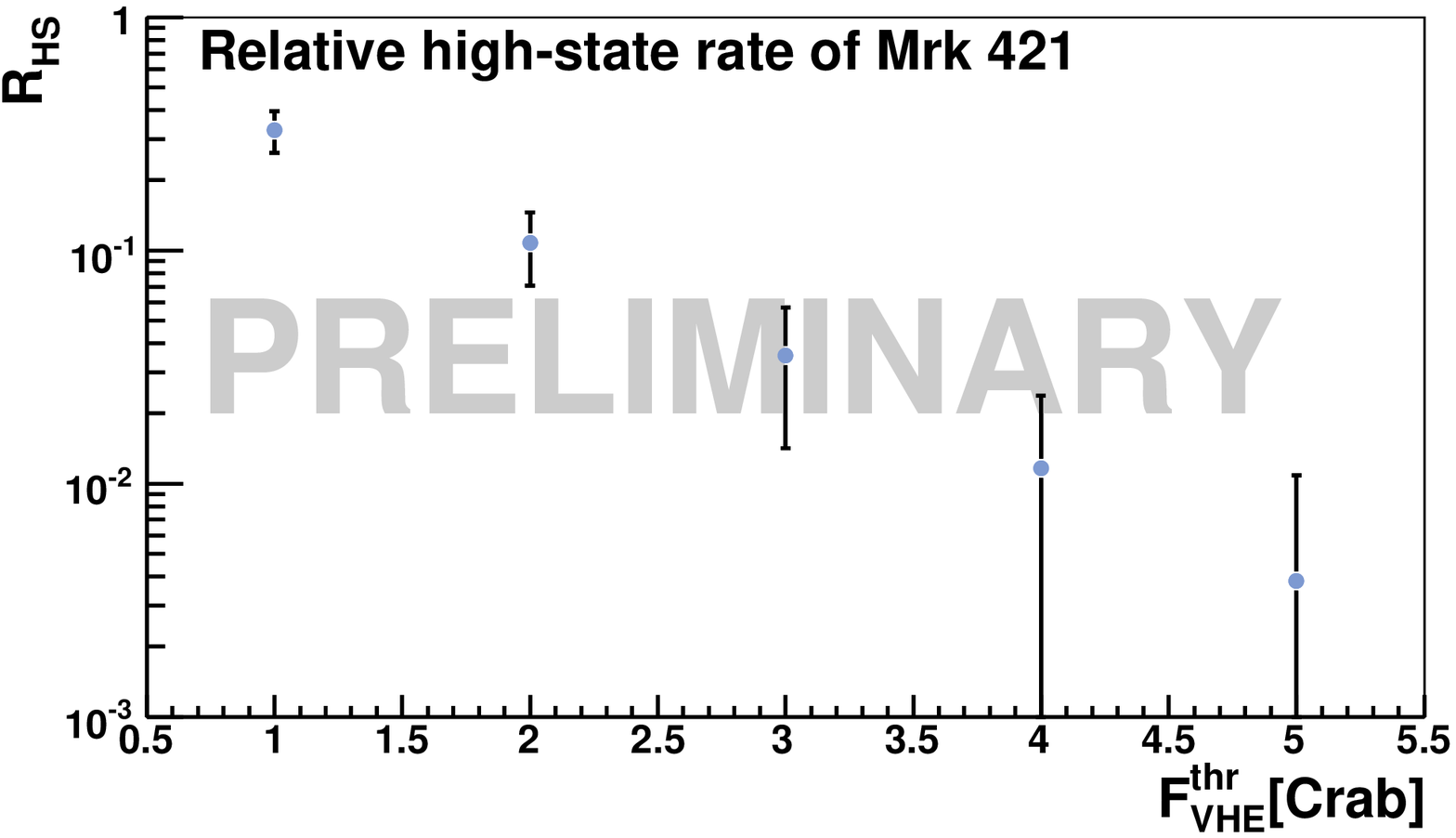}
 \vspace{-0.3cm}
 \caption{\label{hsrate}High-state rate calculated by applying equation~\ref{hsr} to
          the fit of the distribution of fluxstates in Figure~\ref{fluxstates}.}
\end{figure}
\vspace{-0.2cm}
\subsection{List of selected sources}
{The first criterion for the selection of sources for the NToO
test run is their variability.
Only sources known or expected to be variable were chosen 
for the test run.
Other criteria are}
the minimal impact on the scientific
plans of 
MAGIC and the possibility to efficiently
organize the independent observation plans.
Target sources are therefore preferably selected among those which
are already included in the scheduled observation program (MAGIC).
Further criteria are their potential for high-energy neutrino emission,
good visibility for MAGIC during the time period of the test run (September--December) and
previous detections at high-energy $\gamma$-rays {or} high probability for
$\gamma$-ray emission.
%\begin{trivlist}
%\setlength{\itemsep}{0.1ex}
%\item potential sources of high energy neutrinos, correlated to enhanced states of high energy gamma-ray emission
%\item established sources of high energy gamma rays
%\item objects with an independent observation plan by the partner telescopes
%\end{trivlist}
Sources meeting these requirements are 
Blazars and X-ray binaries.
For these sources the level of correlation between 
high energy neutrinos and gamma-rays 
can be different under different scenarios 
(see for example the cases discussed in~\cite{mq2}). 
%Table~\ref{sourcelist} lists for each object considered here
%the numbers for expected background events,
%observed neutrino events, follow-up observations by the MAGIC telescope,
%observed coincidences with high states (along with the predefined 
%flux threshold) and corresponding high state probabilities.
%Mrk421  1.51 +/- 0.13
%1ES1949 1.0 +/- 0.10
%1es2344 0.92 +/- 0.11
%lsi61+303 0.86 +/- 0.10
%grs1919 1.26 +/- 0.11
%
%source     nobs   nbg   intPoisProb
%Mrk421      3     1.51   0.193669
%1es2344     1     0.99   0.628423
%1es1959     0     0.92   1.     
%lsi 60+303  0     0.86   1.
%grs1915     1     1.26   0.716346
%
\begin{table}
\begin{center}
\setlength{\tabcolsep}{1.4mm}
\renewcommand{\arraystretch}{0.8}
\begin{tabular}{|llllll|}\hline
%		& \multicolumn{2}{c}{X-ray binaries} & \multicolumn{3}{c|}{blazars} \\
		& \rotatebox{90}{\bf\tiny LSI+61\,303}& \rotatebox{90}{\bf\tiny GRS\,1915+105}& \rotatebox{90}{\bf\tiny 1ES\,2344+514}	& \rotatebox{90}{\bf\tiny 1ES\,1959+650} & \rotatebox{90}{\bf\tiny Mrk\,421} \\\hline
\Nbck		& 0.86&1.26 &0.99&0.92&1.51 \\
\Nobs		& 0 & 1 & 1 & 0 & 3 \\
Follow ups 	& 0 & 0 & 1 & 0 & 1 \\
\Ncoinc		& -- & -- & 0 & -- & 0 \\
$F_{thr}$ $[$C.U.$]$& 0.2 & 0.2 & 0.5 & 1.0 & 4.0 \\
\pgam		& -- & -- & -- & $<0.15$ & $<0.05$ \\\hline
\signeu		& 1.0 & 0.7 & 0.6 & 1.0 & 0.2 \\\hline
\end{tabular}
\caption{\label{sourcelist}List of selected sources for the NToO test run. 
Given are preliminary numbers for expected (\Nbck) and observed (\Nobs) neutrino 
triggers, the number of observed coincidences (\Ncoinc), the $\gamma$-ray
high-state probability and the probability \signeu\ for observing \Nobs neutrinos or more.
The error on \Nbck is typically 0.1.
%         The number of expected and observed triggers are given along 
%         with the number of observed coincidences with a $\gamma$-ray high state by MAGIC.
%         The probability \pgam {to observe a $\gamma$-ray high state (i.e. a flux $F_{thr}$ above a pre-defined threshold) within a given time window (1 day after the trigger)} as estimated from Eq. \ref{hsrate} is also given.
}
\end{center}
\end{table}

%\vspace{-0.2cm}
\subsection{Results and Interpretation}
%\fbox{\parbox{\columnwidth}{
During the two months of data taking for the NToO program a total of 5
neutrino event triggers were initiated by AMANDA-II and sent to the MAGIC observatory.
% for the objects of the list in Table~\ref{sourcelist}.
In two cases follow-up observations were performed with the MAGIC
telescope lasting for 1 hour each. For the remaining 3 triggers, the source was not observable
with MAGIC within 24~h following the trigger due to unfavourable
astronomical, moon or weather conditions. 
In Table~\ref{sourcelist} the individual neutrino event counts \Nobs
are given along with
the number of expected neutrino background events \Nbck, the number of
coincident observations with MAGIC, the number \Ncoinc of observed coincident
$\gamma$-ray flares (as defined above) and the $\gamma$-ray flare probability
\pgam derived from Equation~\ref{hsr}.
%The sources considered as targets for a test run based on AMANDA-II are
%listed in Table~\ref{sourcelist}.
%Also given are the number of expected and observed neutrino events and the number 
%of observed coincidences between neutrino events and $\gamma$-ray flares
%{as defined above}.
%
%
The MAGIC follow up observation data has been analyzed with the
standard MAGIC analysis chain \cite{magicanalysis}. 
The sensitivity of MAGIC is sufficient
to detect a $\gamma$-ray flux level of 30\% Crab Units (C.U.) with 5 sigma
significance within 1 hour. It is therefore enough to determine
whether the 2 triggered sources Mrk421 and 1ES2344 were in flaring
state (according to the definition of
flaring state in Table~\ref{sourcelist})
 during the NToO observations.
The analysis yielded an upper limit for 1ES2344 (16\% C.U.) and
a low flux state for Mrk421 (30$\pm$10\% C.U.). No coincident
$\gamma$-ray flaring state has thus been observed.

\section{Discussion and Perspectives}
The NToO strategy was implemented
in a test run involving the AMANDA-II and
the MAGIC telescope for a time period of two-months.
No coincident events have been observed during this test run.
However, the technical feasibility of a NToO strategy
%The communication exchange infrastructure 
was successfully tested. 
The neutrino trigger information
sent via e-mail has initiated follow-up observations, whenever the sources
were visible and the weather and astronomical (moon/sun) 
conditions allowed the operation of the
MAGIC telescope. 
At the end of the test run, a different communication
infrastructure was also implemented, based on a test client/server
connection, which allows the queuing of follow-up observations using a
similar pipeline as that already used by MAGIC to follow-up GRB alerts.
Perspectively, different event selections will be developed for IceCube.
A search for multiplets with pre-defined significances 
will provide a means for the selection of
flare-like neutrino events.
Furthermore, work is in progress for the analysis
of high-energy neutrino events with the IceCube 22-string 
detector (2007) and with extensions in subsequent years.
These analyzes will possibly be implemented in an IceCube
NToO program in 2008.

%{\small{\bf Acknowledgements} We thank the Office of Polar Programs of the
%National Science Foundation, DESY, the Helmholtz Association.
%We would like to thank the IAC for excellent working conditions. The
%support of the German BMBF and MPG, the Italian INFN and the Spanish
%CICYT is gratefully acknowledged.}

%\vspace{-0.2cm}

%\bibliography{ICRCNToO/litbank_mexico}
%\bibliographystyle{plain}

%\end{document}

%
%´GRB
%
\setcounter{figure}{0}
\setcounter{table}{0}
%%
% International Cosmic Ray Conference 2007 Merida Yucatan Mexico
% In this file you will find detailed instructions to correctly
% typeset your document.
%
% By: Victor De la Luz
% vdelaluz@inaoep.mx
% Mexico City

%Class Required
%\documentclass{article}
%The ICRC Style
%(This package is the last package in the usepackage list)
%If you need import other package you need write it first.
%\usepackage{icrctc07}

\newcommand{\eVdist}{\kern-0.06667em}
\newcommand{\Pev}{{\mathrm{Pe}\eVdist\mathrm{V\/}}}
\newcommand{\pev}{{\,\mathrm{Pe}\eVdist\mathrm{V\/}}}
\newcommand{\Tev}{{\mathrm{Te}\eVdist\mathrm{V\/}}}
\newcommand{\tev}{{\,\mathrm{Te}\eVdist\mathrm{V\/}}}
\newcommand{\Gev}{{\mathrm{Ge}\eVdist\mathrm{V\/}}}
\newcommand{\gev}{{\,\mathrm{Ge}\eVdist\mathrm{V\/}}}
\newcommand{\met}{{\,\mathrm{m}}}
\newcommand{\cm}{{\,\mathrm{cm}}}
\newcommand{\km}{{\,\mathrm{km}}}
\newcommand{\Km}{{\mathrm{km}}}
\newcommand{\scnd}{{\,\mathrm{s}}}
\newcommand{\ns}{{\,\mathrm{ns}}}
\newcommand{\ms}{{\,\mathrm{ms}}}
\newcommand{\yr}{{\,\mathrm{yr}}}

%The paper title
\title{Detecting GRBs with IceCube and optical follow-up observations}
%Short title to print in the headers to the final publication (Not showed in this print).
\shorttitle{Detecting GRBs with IceCube and optical follow-up observations}

%All paper authors
\authors{A. Kappes$^{1,2}$, M. Kowalski$^3$, E. Strahler$^1$, I. Taboada$^4$ for the IceCube Collaboration$^5$}
%Short title to print in the headers to the final publication (Not shown in this print).
\shortauthors{A. Kappes$^{1,2}$, M. Kowalski$^3$, E. Strahler$^1$, I. Taboada$^4$ for the IceCube Collaboration$^5$}
%All the affiliations.
\afiliations{
  $^1$Physics Dept. University of Wisconsin, Madison WI 53703. USA\\ 
  $^2$on leave of absence from Universit\"at Erlangen-N\"urnberg, D-91058 Erlangen. Germany\\
  $^3$Institute f\"ur Physik. Humboldt Universit\"at zu Berlin, D-12489 Berlin. Germany\\
  $^4$Physics Dept. University of California, Berkeley CA 94720. USA\\
  $^5$see special section of these proceedings
 }
\email{itaboada@berkeley.edu}

%The abstract.
\abstract{We present a summary of AMANDA results obtained in searches for
neutrinos from Gamma-Ray Bursts (GRBs). Using simulations, we show how
the IceCube detector, which is currently being constructed at the
South Pole, will improve the sensitivity of the search. In order to
improve the prospects for detections of gamma-ray dark bursts (e.g. choked
bursts), as well as core collapse Supernovae (SNe), we discuss a novel
follow-up scheme of high energy neutrino events from IceCube. Triggered by
neutrino events from IceCube, a network of small optical telescopes is meant
to monitor the sky for SNe rising lightcurves and GRB afterglows. The
observing program is outlined and its status discussed.}

%%%%%%%%%%%%%%%%%%%% B E G I N   D O C U M E N T%%%%%%%%%%%%%%%%%%%%%%%
%\begin{document}
\maketitle

\section{Introduction}

GRBs have been proposed as one of the most
plausible sources of ultra-high energy cosmic rays \cite{prl:75:386}
and high energy neutrinos \cite{prl:78:2292}. In addition to being
a major advance in astronomy, detection of high energy
neutrinos from a burst would provide corroborating evidence for the
acceleration of ultra-high energy cosmic rays within GRBs. It has
been noticed that so-called long GRBs are often accompanied by SNe type
Ib/Ic \cite{grb030329}. The prevalent interpretation is that the 
progenitors of these SNe and GRBs are very massive stars that undergo core
collapse that leads to the formation of a black hole. The material accreted by
the black hole can form highly relativistic jets which then produce the
observed burst of $\gamma$-rays and accelerate particles to high energy. The
connection between SNe and GRBs has inspired the speculation that a fraction
of core collapse SNe which do not lead to GRBs may still be the source of TeV
neutrinos \cite{ando-beacom}.

For the purposes of establishing the sensitivity of IceCube to GRBs we will
use the Waxman-Bahcall GRB \cite{prl:78:2292} model as a benchmark. We will
assume a flavor flux ratio of 1:1:1 at Earth motivated by neutrino
oscillations. We use a flux normalization of 1.35$\times
10^{-8}$~GeV~cm$^{-2}$~s$^{-1}$~sr$^{-1}$ at 1~PeV at the Earth for all 
neutrino flavors combined, a break energy of 100~TeV and a synchrotron break 
energy of 10~PeV. Finally, we assume that the Waxman-Bahcall GRB model
corresponds to 670 GRBs per year over the full sky resulting in an 
average neutrino fluence per burst of $F_\mathrm{burst} = 1.3 \times
10^{-5}$~erg/cm$^2$ for all flavors combined. However, it should be noted 
that fluctuations in the characteristics of GRBs, notably redshift and
$\gamma$-ray fluence, lead to significant fluctuations in the expected number
of neutrinos from burst to burst \cite{guetta}.

IceCube is a high energy ($E>1\tev$) neutrino telescope currently
under construction at the South Pole \cite{icecube-icrc2007}. When
completed, the deep ice component of IceCube will consist of 4800
digital optical modules (DOMs) arranged in up to 80 strings frozen into the
ice, at depths ranging from $1450\met$ to $2450\met$. Each DOM
contains a photomultiplier tube and supporting hardware inside a glass
pressure sphere. The total instrumented volume of IceCube will be
$\sim 1\km^3$. The DOMs indirectly detect neutrinos by measuring the
Cherenkov light from secondary charged particles produced in
neutrino-nucleon interactions. AMANDA-II \cite{nat:410:441}, now
integrated in IceCube as a sub-detector, was commissioned in the year
2000 and consists of a total of 677 optical modules. These are
arranged on 19 strings with the sensors at depths ranging from
$1500\met$ to $2000\met$ in a cylinder of $100\met$ radius. Its
instrumented volume is about 70 times smaller than that of the deep
ice component of IceCube.

The two main channels for detecting neutrinos with IceCube
and AMANDA are the $\nu$-induced muon and the $\nu$-induced cascade
channels. For the muon channel the detectors are mainly sensitive to up-going
muons as the Earth can be used to shield against the much larger flux of
down-going atmospheric muons. Searches for neutrinos from GRBs in the muon
channel benefit from good angular resolution ($\sim 1^\circ$ for $E_\nu >
1$~TeV ) and from the long range of high energy muons. For cascade channels
the detectors are sensitive to all neutrino flavors through various
interaction channels. Here, analyses benefit from good energy resolution
($\sim 0.1$ in $\log_{10}E$) and from 4$\pi$~sr sensitivity to high energy
neutrinos. The number of expected detected events can be calculated by
convoluting the neutrino flux $\Phi$ with the corresponding effective neutrino
area $A_{\nu}^\mathrm{eff}$: 
\begin{equation}
N_\mathrm{evts} = T \int \mathrm{d}\Omega \mathrm{d}E
A_{\nu}^\mathrm{eff}(E,\theta) \frac{\mathrm{d}\Phi}{\mathrm{d}E}(E,\theta) \ ,
\end{equation}
where $T$ is the observation time.

Searches for neutrinos in coincidence with GRBs have been conducted
with the AMANDA detector in the muon and the cascade channels. The
muon search was performed on over 400 bursts reported by BATSE and
IPN3 between 1997 and 2003 \cite{apj:07b}. Additionally, a dedicated
search for neutrinos in coincidence with GRB030329 was performed
\cite{stamatikos}. Using the cascade channel, one analysis \cite{apj:07a}
focused on 73 bursts reported by BATSE in 2000 and another analysis
searched for a statistical excess of cascade-like events during a
rolling period of $1\scnd$ and $100\scnd$ for the years 2001-2003
\cite{apj:07a}. So far no evidence for neutrinos from GRBs has been
found. With the muon search the 90\% c.l. limit set by AMANDA is 1.3 times
higher than Waxman-Bahcall's prediction as defined above.

The main source of GRB data to be studied by IceCube will be GLAST, but also
other satellites, e.g. Swift, will contribute. Since finding an
unusually bright nearby GRB is key for the detection of neutrinos, the ideal
satellite has a very large field of view (FoV). GLAST's FoV is $\sim$9~sr and
Swift's $\sim$1.4~sr. In our calculations we assume a yearly detection rate of
200 GRBs by GLAST which are distributed uniformly over the sky. IceCube can be
operated while being built. By the end of 2008 the accumulated exposure will
be 0.75~km$^2\cdot$yr; 1.3~km$^2\cdot$yr by 2009 and 1.9~km$^2\cdot$yr by
2010. IceCube is currently taking data with 22 strings.

\section{IceCube sensitivity to muon neutrinos}

In order to calculate the sensitivity of the IceCube detector to muon
neutrinos from GRBs, bursts are simulated uniformly distributed over
the northern sky. Each burst is assumed to produce a muon neutrino
fluence of $1/3\,F_\mathrm{burst}$. The muon neutrino effective area as
a function of neutrino energy and zenith angle is obtained from a full
Monte Carlo simulation including a detailed ice and detector
simulation. Displayed in Fig.\ref{fig:effArea} is the effective area
at trigger level as a function of energy averaged over zenith angles
above $90^\circ$ (up-going neutrinos). Past searches with AMANDA \cite{apj:07b}
have shown that 25--75\% efficiency can be obtained with respect to trigger
level once selection criteria are applied to data in order to remove the
down-going muon background. IceCube has the potential of even higher
efficiency. Thus we consider trigger level effective area representative
(upper limit) of what the  detector will be able to achieve.

The narrow constraints on the position and the timing of neutrinos
from a GRB combined with the good angular and time resolution of IceCube lead
to a very low expected background. For this first sensitivity estimate we
therefore assume a background free observation.

In the following we estimate the number of GRBs required to reach the
GRB flux predicted by Waxman-Bahcall and exclude it at 99.73\%
C.L. ($3\sigma$). With the observation of no events and a mean expected
background of zero events the Feldman-Cousins method \cite{ICRC1132_FC} yields an
event upper limit of 6.0. In order to reach this number about 70 bursts in the
northern hemisphere must be observed which can be expected after about 1 year
of operation of the full detector.

\begin{figure}
\begin{center}
\includegraphics[width=0.48\textwidth,angle=0,clip]{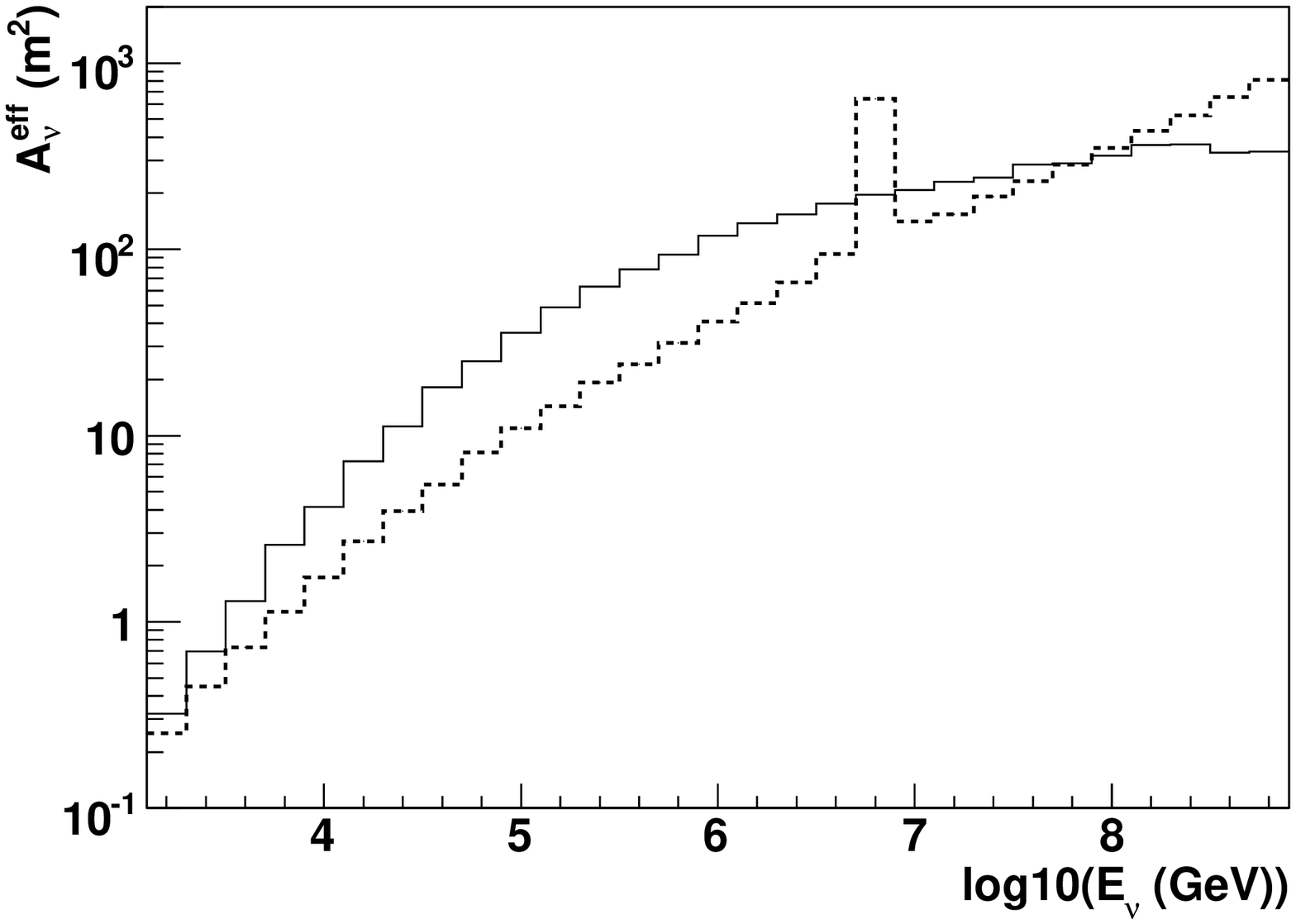}
\end{center}
\caption{IceCube's effective neutrino area as a function of the neutrino
  energy at trigger level for $\nu_\mu$ (solid) and $\nu_e$ (dashed). The
  $\nu_\mu$ effective area is the average for neutrinos from the northern sky
  (2$\pi$~sr) whereas that for $\nu_e$ is averaged over the whole sky
  (4$\pi$~sr). The peak at 6.3~PeV is the Glashow resonance.}\label{fig:effArea}
\end{figure}

\section{IceCube sensitivity to electron neutrinos}

For $\nu_e$-induced cascades the effective neutrino area
$A^\mathrm{eff}_{\nu_e}$ is calculated for a uniform distribution of $\nu_e$
and $\bar{\nu}_e$ over 4$\pi$~sr. We have taken into account charged current
and neutral current interactions as well as the Glashow resonance. The
effective area, averaged over an equal mixture of $\nu_e$ and $\bar{\nu}_e$,
can be seen in Fig.~\ref{fig:effArea}. As with the muon channel, the effective
area presented here is at trigger level.

The narrow constraints on the timing of neutrinos from a GRB combined with the
good cascade energy and time resolution of IceCube lead to a very low expected
background. We calculate the sensitivity supposing a background free
search.

In the following we estimate the number of GRBs required to reach the
GRB flux predicted by Waxman-Bahcall and exclude it at 99.73\% c.l.
(3$\sigma$). With the observation of no events and a mean expected background
of zero events the Feldman-Cousins method yields an upper limit of
6.0. In order to reach this number about 560 bursts must be monitored. This can
be expected after almost 3~yr of operation of IceCube in coincidence with
GLAST. Including the contributions from $\nu_\mu$ and $\nu_\tau$-induced
cascades would roughly double the number of expected events \cite{apj:07a}
\footnote{Given the improved capabilities of IceCube it may be possible to
  treat high energy $\nu_\tau$-induced events as a separate channel.} Thus the
number of bursts required is $\sim$~280 or 1.5~yr of satellite coincident
observations. 

\section{An optical follow-up for neutrino events}

\begin{figure}
\begin{center}
\includegraphics*[width=0.48\textwidth,angle=0,clip]{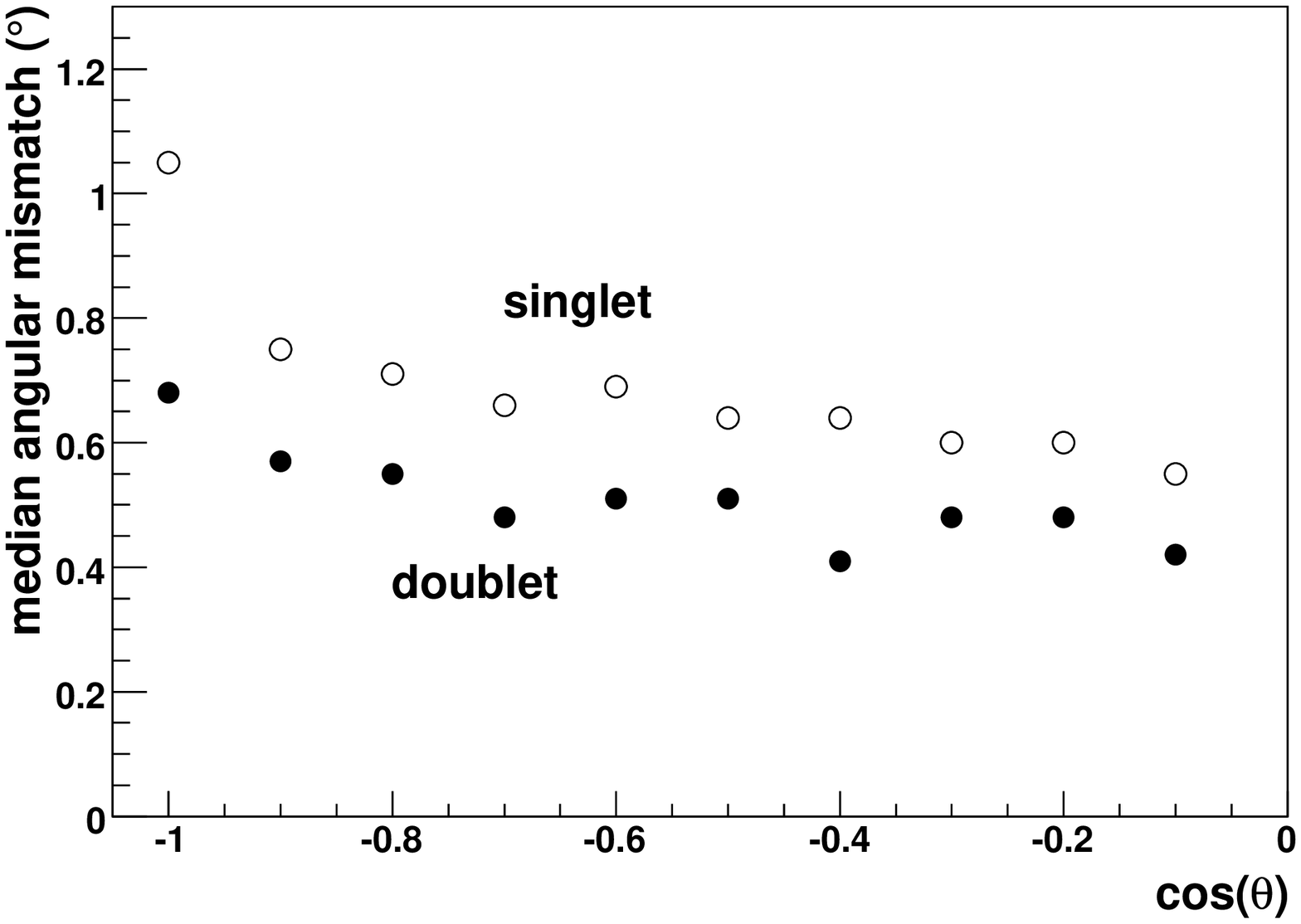}
\end{center}
\caption{Median angular mismatch for single neutrino events and 
neutrino doublets
  (with $\Delta \Psi <3^\circ$). Quality cuts have been applied.}\label{ICRC1132_fig2}
\end{figure}

So far we have discussed the search for a neutrino signal in coincidence with
a GRB identified through satellites. For the future, we plan to
complement this by performing an automated optical follow-up observations of
selected neutrino events from IceCube \cite{optical_followup}. First, the
direction of neutrinos detected with IceCube is reconstructed online and if
energy or multiplicity pass a certain threshold, a notice is sent to a network
of optical telescopes. Then, within minutes after reception of the notice,
automated optical telescopes monitor the corresponding part of the
sky. Transient objects are thereby identified, e.g. through detection of GRB
afterglows (on a timescale of minutes to hours) or rising SNe light-curves (on
a timescale of days). These multi-messenger observations significantly improve
the discovery potential of IceCube by providing a chance to detect and
identify the source of the high-energy neutrinos. In what follows, we discuss
the IceCube neutrino-burst trigger and the corresponding telescope
requirements for the optical follow-up observations.

The rate of atmospheric $\nu_\mu$ events observed in the full IceCube 
detector is too high to perform individual follow-up observations: the 
trigger rate of neutrinos with zenith angle $> 80^\circ$
 will be $\sim$~$7\times~10^5$ events per year. After imposing quality cuts 
similar to those used in Ref.~\cite{icecube_simulation}, the 
rate is $\sim$~$7\times~10^4$ per year but remains high. 
However, the background rate can be significantly reduced by
triggering on multiplets, i.e. two or more neutrinos  detected
within a short time window, $\Delta t$, and within a spacial window,
$\Delta \Omega$. Here $\Delta t$ is determined by the typical burst
duration. An adequate time scale which covers the duration of most
GRBs and SNe models is 100~s. The optimal size $\Delta \Omega$
is determined by the pointing resolution of IceCube, which is of the
order of one degree. Using simulation, the rate of doublets
for a maximal angular separation of $3^\circ$ as well
as $\Delta t = 100$~s is 30 per year. This number is low enough that
individual follow-up observations can be performed.

Once the alert will be issued, the corresponding part of the sky has to
be searched for transient sources. By averaging the reconstructed
directions of the neutrinos in a multiplet, one can improve the
localization of the potential source. The median angular mismatch
between the average and the true direction as a function of
the declination band is shown in Fig.\ \ref{ICRC1132_fig2}. The median angular mismatch
is  $0.5^\circ-0.6^\circ$. 

An optical telescope with a FoV of $2^\circ \times 2^\circ$
would cover more than 80~\% of the IceCube's point spread function for
doublets. Already now optical telescopes with such a FoV exist. For example,
the ROTSE-III network consists of 4 fully automated telescopes, each with a
0.45~m diameter mirror and a $1.85^\circ \times 1.85^\circ$  FoV.

\section{Conclusions}

We have presented a preliminary summary of the capabilities of
IceCube. With several search strategies in place, IceCube will be ready to
discover a range of different phenomena related to GRBs, such as prompt
emission of PeV neutrinos, precursor and afterglow neutrinos from GRBs and
neutrinos from core-collapse SNe.

Within a few years IceCube will be able to detect the neutrino flux predicted
by Waxman-Bahcall with high significance or set limits well below any
current prediction. Follow-up observations with optical telescopes as
suggested in this paper will further enhance and complement the satellite
triggered searches by enabling IceCube to observe the potentially large
fraction of bursts where no $\gamma$-ray signal is detected by satellites
(dark GRBs). An optical follow-up program for IceCube will possibly be
implemented in 2008.

Both in the case of detection or in the case of the derivation of an upper
limit, the results from IceCube will boost our understanding of GRBs, one of
the most puzzling phenomena in our universe, and contribute to the resolution
of the mystery of the origin of cosmic rays at the highest energies.

\section{Acknowledgments}
A. Kappes acknowledges the support by the EU Marie-Curie OIF
Program. M. Kowalski acknowledges the support of the DFG.

%This is the reference to .bib file (Without .bib!)
%\bibliography{ICRC1132/icrc1132}
%\bibliographystyle{plain}\begin{thebibliography}{10}

%\end{document}

%
% Wimps
%
%icrc0379.pdf (search for neutralino dark matter with AMANDA)
%icrc0690_submitted.pdf (Prospects of dark matter search)
%
\setcounter{figure}{0}
\setcounter{table}{0}
%\documentclass[dvips]{article}
%\usepackage{icrctc07,lineno,graphicx,amssymb,amsmath,color}

% new commands: units
%\newcommand{\unit}[1]{\, \mathrm{#1}}	% units: roman + small space before
%\newcommand{\eV}{\unit{eV}}		% electronvolt
%\newcommand{\MeV}{\unit{MeV}}		% mega-electronvolt
\newcommand{\GeV}{\unit{GeV}}		% giga-electronvolt
\newcommand{\MGeV}{\unit{GeV}/c^2}	% giga-electronvolt/c^2
\newcommand{\TeV}{\unit{TeV}}		% tera-electronvolt
\newcommand{\m}{\unit{m}}		% meter
\newcommand{\mus}{\unit{\mu s}}		% microsecond
\newcommand{\degg}{\mathrm{^\circ}}     % deg

\hyphenation{sub-sample lo-ca-tion an-aly-sis cos-mic energies atmo-sphere atmo-spheric con-taminate eli-minate electro-weak neu-tra-li-no energy matter products array showers sample under re-duction angles azi-muthal ge-ne-ra-ted filter griest akerib limits anni-hilate anni-hilation samples possible}

%The paper title
\title{Search for neutralino dark matter with the AMANDA neutrino telescope}
%Short title to print in the headers to the final publication (Not showed in this print).
\shorttitle{Search for neutralino dark matter with the AMANDA neutrino telescope}
%All paper authors
\authors{D. Hubert$^{1}$ and A. Davour$^{2}$ for the IceCube Collaboration$^3$}
%Short title to print in the headers to the final puplication (Not showed in this print).
\shortauthors{D. Hubert and A. Davour for the IceCube Collaboration}
%All the affiliations.
\afiliations{$^1$Dienst ELEM, Vrije Universiteit Brussel, B-1050 Brussels, Belgium \\
	     $^2$Division of High Energy Physics, Uppsala University, S-75121 Uppsala, Sweden \\
	     $^3$See special section of these proceedings}
\email{dhubert@vub.ac.be}

%The abstract.
\abstract{If non-baryonic dark matter exists in the form of neutralinos, a neutrino flux is expected from the decay of neutralino pair annihilation products inside heavy celestial bodies.  Data taken with the AMANDA neutrino telescope located at the South Pole can be used in a search for this indirect dark matter signal.  We present the results from searches for neutralinos accumulated in the Sun using AMANDA data from $2001$, and improved new limits on the flux of muons from $50$\---$250 \MGeV$ neutralino annihilations in the Earth obtained with data from $2001$\---$2003$.}

%\begin{document}

%% title/abstract stuff
\maketitle

\vspace{-1.0pc}
\section{Introduction}
\vspace{-1.0pc}
Cosmological observations have suggested the presence of non-baryonic dark matter on all distance scales.  The WMAP results~\cite{wmap} confirmed our current understanding of the Universe, summarized in the concordance model.  In this model the Universe contains about $23\%$ non-baryonic cold dark matter, but nothing is predicted about the nature of this dark matter.  
A massive, weakly interacting and stable particle appears in Minimally Supersymmetric extensions to the Standard Model that assume R-parity conservation.  Indeed, the supersymmetric partners of the neutral electroweak and Higgs bosons mix into a dark matter candidate, the neutralino, whose mass is expected in the GeV-TeV range~\cite{neutralinoDM}.  
On their trajectory through the Universe these particles will scatter weakly on normal matter and lose energy.  Eventually, dark matter particles will be trapped in the gravitational field of heavy celestial objects, like the Earth and the Sun~\cite{trap}.
The particles accumulated in the center of these bodies will annihilate pairwise.  The neutrinos produced in the decays of the Standard Model annihilation products can then be detected with a high energy neutrino detector as an excess over the atmospheric neutrino flux.
In this paper we present the results of searches with the Antarctic Muon And Neutrino Detector Array (AMANDA) for neutralino dark matter accumulated in the Earth ($2001$\---$2003$ data set) and the Sun ($2001$ data set).  We also discuss current improvements and preliminary results from ongoing analyses on higher statistics data samples accumulated during recent years.

The AMANDA detector~\cite{ICRC0379_amanda} at the South Pole uses the polar ice cap as a Cherenkov medium for the detection of relativistic charged leptons produced in high energy neutrino interactions with nuclei.  The $500 \m$ high and $200 \m$ wide detector was completed in $2000$ and totals $677$ light sensitive devices distributed on $19$ strings.  The detector is triggered when at least $24$ detector modules are hit within a sliding $2.5 \mus$ window.  Since $2001$ an additional, lower multiplicity, trigger (referred to as \emph{string trigger}) is operational that exploits not only temporal information but also the space topology of the hit pattern.  This lowers the energy threshold of the detector and is especially beneficial for the sensitivity to neutralinos with $m_\chi < 200\MGeV$.

Reconstruction of muons, with their long range, offers the angular resolution required to reject the background produced by cosmic ray interactions with the atmosphere and search for a neutralino-induced signal, which, due to the geographic location of AMANDA, yields vertical upward-going (Earth) or near horizontal (Sun) tracks in the instrumented volume.  Indeed, it is possible to eliminate the dominant background, downward-going atmospheric muons.  However, upward-going and horizontal atmospheric neutrinos will always contaminate the final, selected data sample.

\vspace{-1.0pc}
\section{Signal and background simulation}
\vspace{-1.0pc}
We have used the \textsc{DARKSUSY} program~\cite{darksusy} to generate dark matter induced events for seven neutralino masses $m_\chi$ between $50 \MGeV$ and $5000 \MGeV$, and two annihilation channels for each mass: the $W^+W^-$ channel produces a hard neutrino energy spectrum ($\tau^+\tau^-$ for $m_\chi < m_W$), while $b\bar{b}$ yields a soft spectrum.
The cosmic ray showers in the atmosphere, in which downward-going muons are created, are generated with \mbox{\textsc{CORSIKA}~\cite{ICRC0379_corsika}} with a primary spectral index of $\gamma$=2.7 and energies between $600 \GeV$ and $10^{11} \GeV$.  The atmospheric neutrinos are produced with \textsc{ANIS}~\cite{ICRC0379_anis} with energies between $10 \GeV$ and $325 \TeV$ and zenith angles above $80 \degg$.

\begin{figure*}[t]
\vspace{-1.pc}
\begin{center}
\includegraphics*[width=0.46\textwidth,height=0.285\textheight,angle=0,clip=true,viewport=0 -20 500 465]{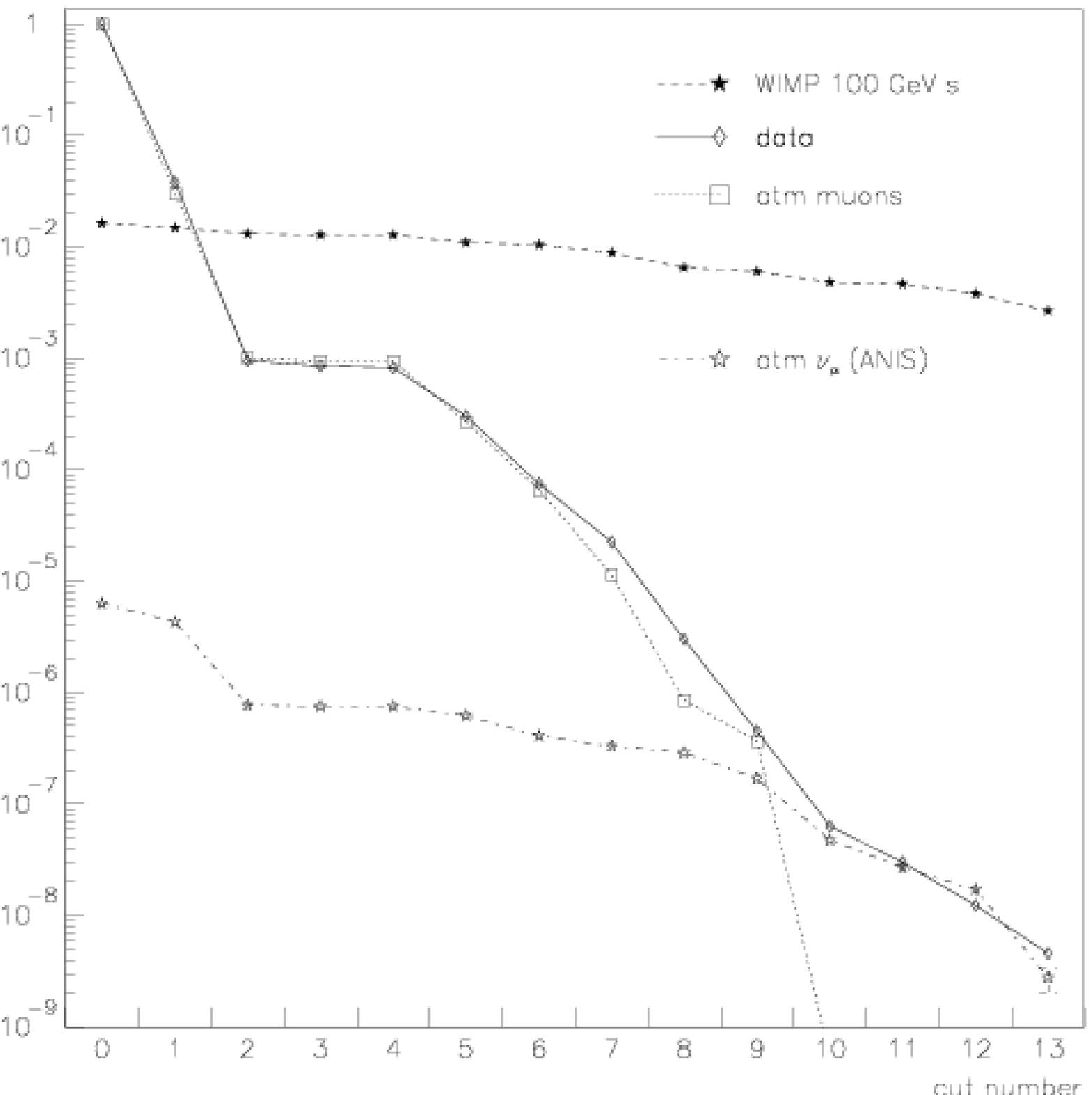}
\includegraphics*[width=0.49\textwidth,height=0.305\textheight,angle=0,clip=true]{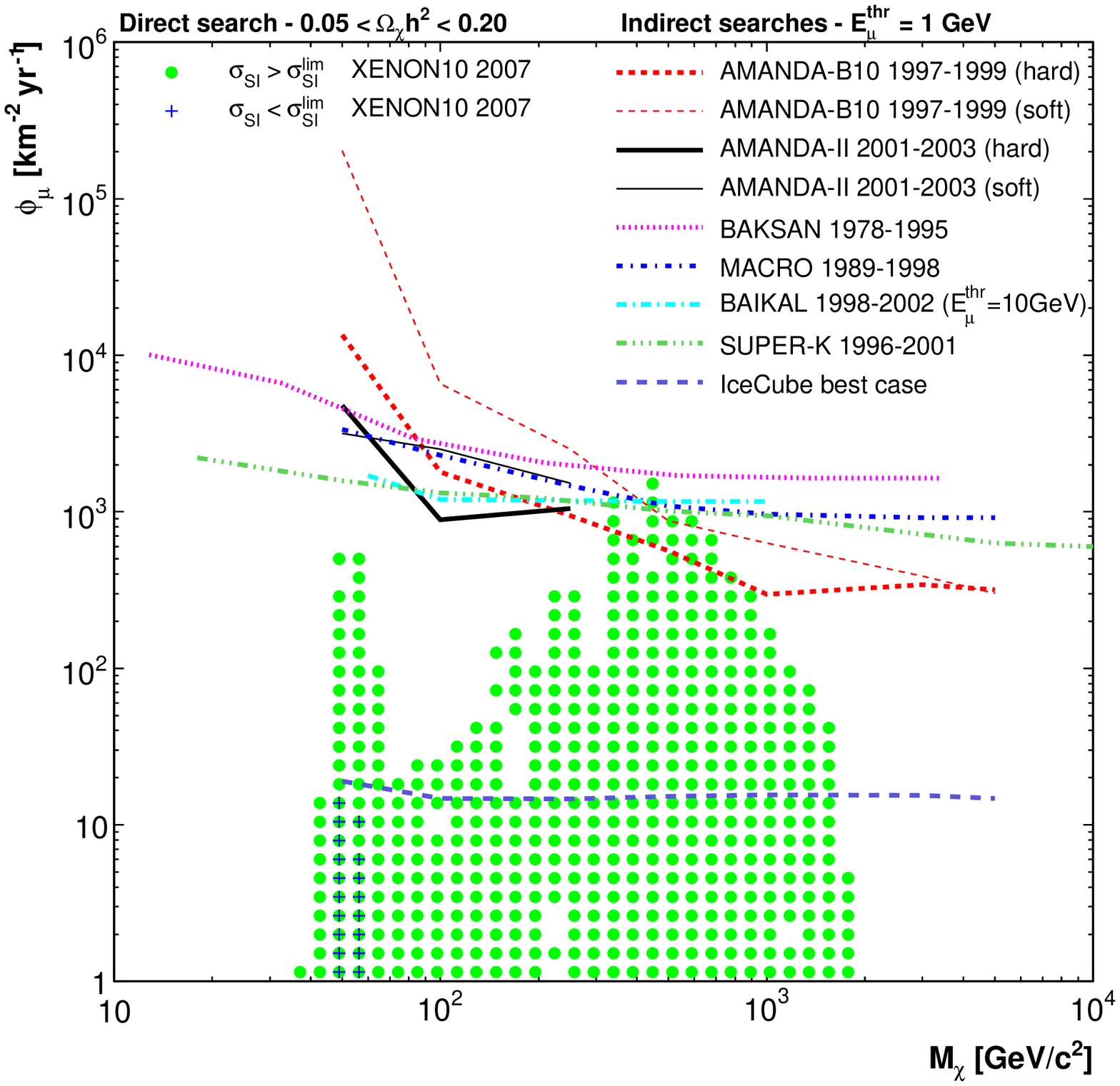}
\vspace{-.7pc}
\caption{\label{earth_fig} (a) Detection efficiencies relative to trigger level for the different filter levels in an Earth neutralino analysis ($m_{\chi} = 100 \MGeV$, soft spectrum) for $2001$\---$2003$ data.  (b) As a function of neutralino mass, the $90\%$ CL upper limit on the muon flux from hard (bottom) and soft (top) neutralino annihilations in the center of the Earth compared to the limits of other indirect experiments~\cite{indirect} and the sensitivity estimated for a best-case IceCube scenario~\cite{icecube_wimp}.  Markers show predictions for cosmologically relevant MSSM models, the dots represent parameter space excluded by XENON10~\cite{xenon10}.}
\end{center}
\vspace{-2.pc}
\end{figure*}

\vspace{-1.0pc}
\section{\label{earth_section}Search for low mass neutralinos in the center of the Earth}
\vspace{-1.0pc}
A neutralino-induced signal from the center of the Earth is searched for in AMANDA data collected between $2001$ and $2003$, with a total effective livetime of $688.0$ days.  This search focuses on improving the sensitivity for low mass neutralinos, with $m_\chi \leq 250 \MGeV$, and includes events triggered with the string trigger.
The complete data set of $5.3 \times 10^9$ events is divided in a $20\%$ subsample, used for optimisation of the selection procedure, and a remaining $80\%$ sample, on which the selection is applied and final results calculated.  Detector data are used as background for the optimisation, and compared to simulated background events to verify the understanding of the background and the simulation.  The simulated atmospheric muon sample contains $3.6 \times 10^7$ triggered events (equivalent to an effective livetime of $4.5$ days).  The sample of atmospheric neutrinos totals $2.4 \times 10^5$ events, which corresponds to $2.5 \times 10^4$ triggers when scaled to the livetime of the data sample used for calculation of the final results.

The characteristics of the signal differs depending on the neutralino model under study.  Hence, the selection criteria are tuned separately for each neutralino model.  Between $2001$ and $2002$ the detector was upgraded and the trigger settings changed slightly.  The event selection is therefore developed separately for $2001$.

\begin{figure*}[t] 
\vspace{-1.pc}
\begin{center}
\includegraphics*[width=0.49\textwidth,height=0.305\textheight,angle=0,clip,viewport=0 5 575 535]{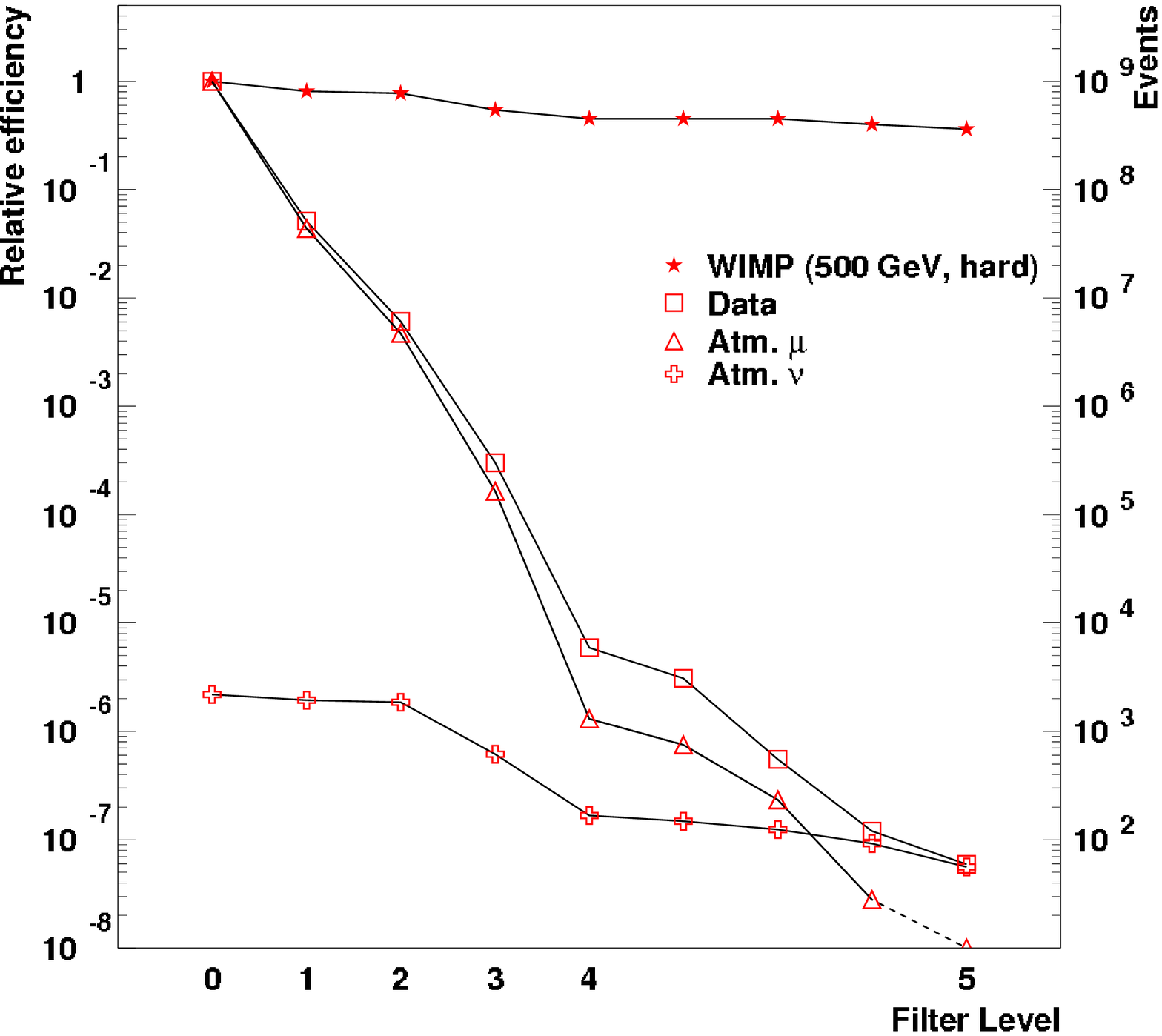}
\includegraphics*[width=0.49\textwidth,height=0.305\textheight,angle=0,clip]{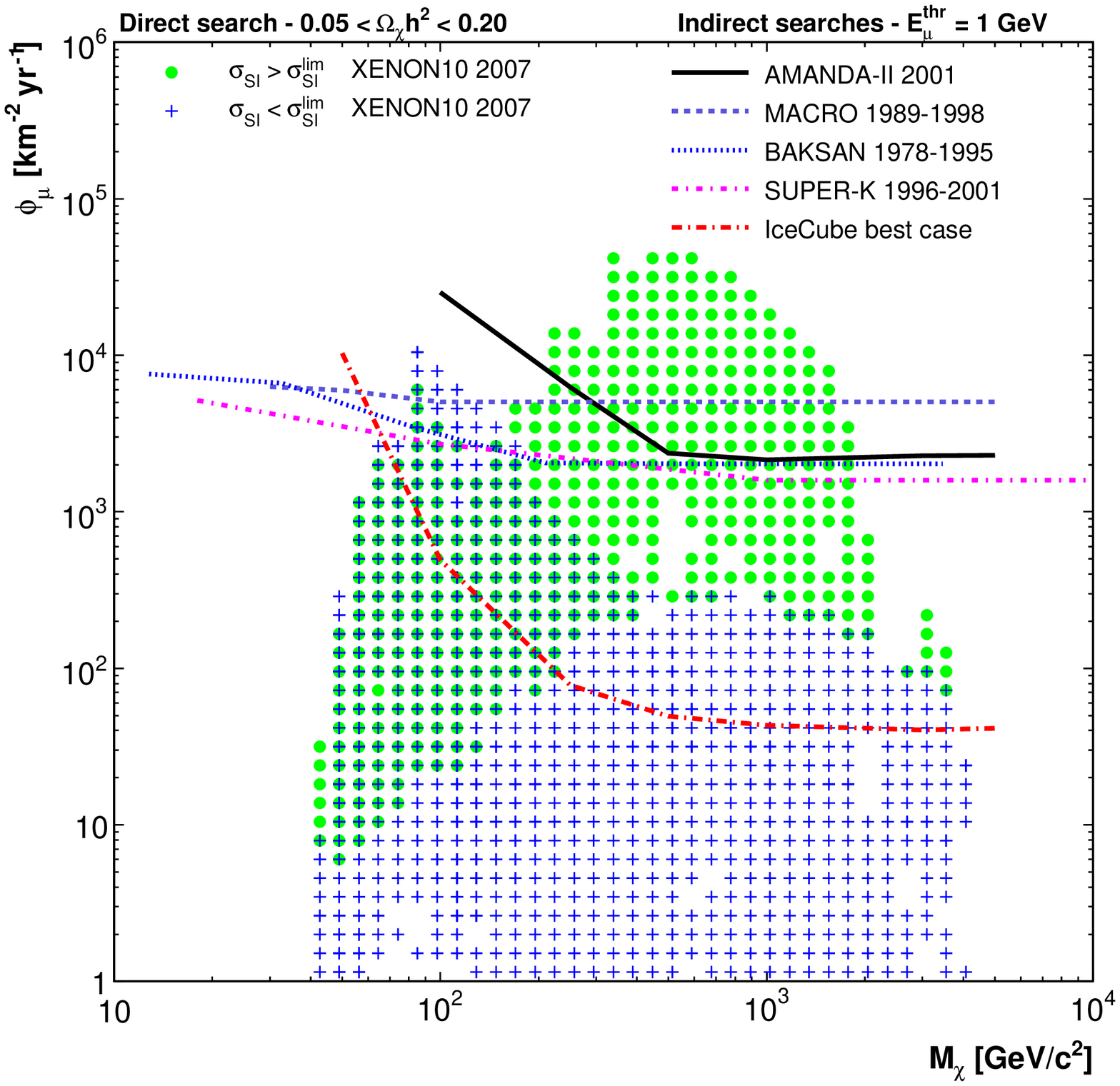}
\vspace{-.7pc}
\caption{\label{sun_fig} (a) Detection efficiencies relative to trigger level for the different filter levels in a Sun neutralino analysis ($m_{\chi} = 500 \MGeV$, hard spectrum) for $2001$ data.  (b) As a function of neutralino mass, the $90\%$ CL upper limit on the muon flux from hard neutralino annihilations in the center of the Sun compared to limits of other indirect experiments~\cite{indirect} and the sensitivity estimated for a best-case IceCube scenario~\cite{icecube_wimp}.  Markers show predictions for cosmologically relevant MSSM models, the dots represent parameter space excluded by XENON10~\cite{xenon10}.}
\end{center}
\vspace{-2.pc}
\end{figure*}

First, we reduce the total background by selecting events with upward-going signature.  Then the data are tested against reconstruction criteria to remove events unlikely to be correctly reconstructed.  After this, the search is limited to the events with reconstructed angle differing less than $40\degg$ from straight upward-going.  At this level (cut number $4$ in Fig.~\ref{earth_fig}a), the sample is still dominated by misreconstructed atmospheric muon events, more than $10^3$ times more abundant than the atmospheric neutrino background.  The background is then further reduced by a series of sequential cuts on reconstruction quality parameters and energy parameters.

After about ten cuts (depending on mass and annihilation channel), the data sample is dominated by atmospheric neutrinos.  All of the data from the three years are combined at this analysis level, and the final selection is applied on the three years together.  With no significant excess of vertical tracks observed, the final selection on reconstructed zenith angle is optimised for the average lowest possible $90\%$ confidence level upper limit on the muon flux.  From the number of observed events and the amount of estimated background in the final angular search bin, we infer the $90\%$ confidence level upper limit on the number of signal events for each of the considered neutralino models.  Combined with the effective volume at the final cut level and the livetime of the collected data, this yields an upper limit on the neutrino-to-muon conversion rate, which can then be related to the muon flux~\cite{newearth}, see Fig.~\ref{earth_fig}b.

\vspace{-1.0pc}
\section{Search for neutralino annihilations in the Sun}
\vspace{-1.0pc}
The data collected in $2001$ is also used for the search for solar neutralinos and corresponds to $143.7$ days of effective livetime.  The total event sample contains $8.7 \times 10^8$ events, but does \emph{not} include events triggered only by the string trigger.  In contrast to the neutralino search in the Earth, the background level can be reliably obtained from randomization of the azimuthal angle.  The advantage of this procedure is that it allows the use of the full data set for cut optimisation.  The azimuthal angles are restored once the optimisation is finalised and results are calculated.  The simulated atmospheric background sample at trigger level totals $1.6 \times 10^8$ muons (equivalent to $32.5$ days of effective livetime) and $1.9 \times 10^4$ neutrinos.

The solar neutralino analysis suffers the same backgrounds as the terrestrial neutralinos, but the signal is expected from a direction near the horizon, due to the trajectory of the Sun as seen from the South Pole.  This analysis was only possible after completion of the full detector, whose $200 \m$ diameter size provides enough lever arm for robust reconstruction of horizontal tracks.

A similar analysis strategy as in previous section is adopted.  First, events are selected with well-reconstructed horizontal or upward-going tracks.  The remaining events are then passed through a neural network that was trained separately for the neutralino models under study and used data as background (filter level $4$).  Although a data reduction of \mbox{$\sim 10^{-5}$} compared to trigger level is achieved, the data sample is still dominated by misreconstructed downward-going muons, see Fig.~\ref{sun_fig}a.  Finally, these are removed by cuts on observables related to reconstruction quality.

There is no sign of a significant excess of tracks from the direction of the Sun in the final data sample.  The background in the final search bin around the Sun is estimated from off-source data in the same declination band, which eliminates the effects of uncertainties in background simulation.  Combining this with the number of observed events, the effective volume and the detector livetime, we obtain $90\%$ confidence level upper limits on the muon flux coming from annihilations in the Sun for each considered neutralino mass~\cite{newsun}, as shown in Fig.~\ref{sun_fig}b.

\vspace{-1.0pc}
\section{\label{discussion}Discussion and outlook}
\vspace{-1.0pc}
Figures~\ref{earth_fig}b and~\ref{sun_fig}b present the AMANDA upper limits on the muon flux from neutralino annihilations in the Earth and the Sun (only hard channel) respectively, together with the results from other indirect searches~\cite{indirect}.  Limits have been rescaled to a common muon threshold of $1 \GeV$ using the known energy spectrum of the neutralinos.  Also shown are the cosmologically relevant MSSM models allowed (crosses) and disfavoured (dots) by the direct search from XENON10~\cite{xenon10}.

Compared to the previously published AMANDA results from searches for neutralinos in the Earth~\cite{newearth} the analysis of $2001$\---$2003$ data benefits from the larger detector volume and the addition of the string trigger with its lower energy threshold.  This makes it possible to improve the sensitivity especially for low energy Earth neutralino models;  for masses above $250\MGeV$ the effect is expected to be less pronounced.  The new limits on the neutralino-induced muon flux are up to a factor~$60$ stronger than our earlier result.

A similar improvement (with respect to~\cite{newsun}) is expected for the solar neutralino analysis of $2001$\---$2003$ data, thanks to the increased detector exposure, improved reconstruction techniques and the string trigger.  A preliminary analysis shows a factor $10$\---$100$ improvement of the effective volume at early analysis level for low energy models, mainly due to the inclusion of the string triggered events.

The neutralino searches will be continued on a larger set of AMANDA data from $2000$\---$2006$.  Since $2007$ AMANDA is embedded as a high-granularity subdetector of the IceCube neutrino telescope, currently under construction.  This offers additional opportunities for the dark matter searches, as described in~\cite{icecube_wimp}.

%This is a test sentence.  It serves to estimate the amount of pagespace is left for this proceedings.  So far no page limit is reached yet.  Testing test test test test test test test test test test test test test test.

%\end{document}

\setcounter{figure}{0}
\setcounter{table}{0}
%%
%Contribution 0690. Prospects of dark matter detection in IceCube.
%Gustav Wikstrom for the IceCube collaboration
%
%\documentclass[dvips]{article}

%\usepackage{amssymb}
%\usepackage{amsmath}
%\usepackage{icrctc07}

\hyphenation{sub-sample location cosmic energies atmo-sphere atmo-spheric con-taminate eli-minate electro-weak neu-tra-li-no energy matter products array showers sample under re-duction angles azi-muthal ge-ne-ra-ted filter griest akerib limits anni-hilate anni-hilation samples possible}

%The paper title
\title{Prospects of dark matter detection in IceCube}
%Short title to print in the headers to the final publication (Not showed in this print).
\shorttitle{Prospects of dark matter detection in IceCube}

%All paper authors
\authors{G. Wikstr\"om$^{1}$ for the IceCube collaboration$^{2}$}
%Short title to print in the headers to the final publication (Not shown in this print).
\shortauthors{G. Wikstr\"om}
%All the affiliations.
\afiliations{$^{1}$Department of physics, Stockholm University, AlbaNova, S-10691 Stockholm, Sweden.\\ $^{2}$ See special section of these proceedings.}
\email{wikstrom@physto.se}

%The abstract.
\abstract{The IceCube neutrino telescope, under construction at the South Pole, currently consists of 22 IceCube strings and 19 AMANDA strings. Combining the two arrays leads to a large instrumented volume with AMANDA as a dense core, an ideal situation for indirect detection of WIMP dark matter annihilations in the Sun. From simulations we calculate the current detector's sensitivity for solar WIMP neutrinos and find that it improves considerably compared to AMANDA-II. The improvement is due to a combination of reduced trigger thresholds and larger detector volume which permits the use of veto against muonic background.}

%%%%%%%%%%%%%%%%%%%% B E G I N   D O C U M E N T%%%%%%%%%%%%%%%%%%%%%%%
%\begin{document}
\maketitle

\section{Introduction}
We investigated the possibilities of detecting a neutrino signal from neutralino WIMP dark matter annihilations in the Sun. The studied neutralino masses were $m_{\chi}=$ 50, 100, 250, 500 and 1000 \rm{GeV} and the annihilation channels were $W^{+}W^{-}$ (\textit{hard channel}) and $b\overline{b}$ (\textit{soft channel}). For $m_{\chi}=50$ \rm{GeV} the $\tau^{+}\tau^{-}$ channel is defined as the hard channel. Neutrinos produced in the Sun from the decay and interactions of the neutralino annihilation products can reach the detector and produce muons in CC reactions $\nu_{\mu}(\overline{\nu}_{\mu}) + N \rightarrow \mu^{-}(\mu^{+}) + X$. These signal muons traversing the ice sheet produce Cherenkov light, detectable by the Optical Modules (OM) of the IceCube detector. The WIMP neutrino zenith angle will follow the Sun's position over the year, $\theta_{\odot}\in[67^{\circ},113^{\circ}]$, and the mean muon energy will be around $\langle E_{\mu}\rangle \sim m_{\chi}/3$ for hard channels and $\langle E_{\mu}\rangle \sim m_{\chi}/6$ for soft channels.

Muons produced in cosmic ray interactions in the atmosphere have a zenith angle range of $\theta_{\mu}\in[0^{\circ},90^{\circ}]$ since muons cannot traverse the whole Earth. These \textit{atmospheric} $\mu$ constitute the main background. Another background is that of muon neutrinos produced in the atmosphere, \textit{atmospheric} $\nu_{\mu}$, which have a near-isotropic angular distribution.

The 2007 IceCube detector \cite{ICRC0690_icecube} consists of 41 strings of which 19 constitute AMANDA \cite{ama}. The two arrays have separate trigger and data aquisition systems (DAQs) which record events autonomously. However, a trigger in AMANDA will force a readout of the IceCube strings, even if IceCube did not have a trigger.

To reject atmospheric $\mu$ background we searched for contained events, i.e. neutrino events with the CC vertex inside a fiducial volume, as defined in figure \ref{ICRC0690_fig1}. We demanded the events to either have no OMs hit in the veto region or that the first OM hit in the veto region came later than the OM hits in the fiducial region. This aimed at ensuring that the muon was created inside the detector, and did not come from the atmosphere. To reduce the number of atmospheric $\mu$ events leaking in between veto strings, we also demand that the average downwards motion of hits should be less than 50 \rm{m}. Events that did not fulfill these conditions were still accepted provided that they had track reconstructions with $\theta_{\it rec}\geq 70^{\circ}$ and more than 10 hit OMs. These conditions together constitute the low-level filtering that will run at South Pole.

\begin{figure}
\begin{center}
\noindent
\includegraphics [width=0.5\textwidth]{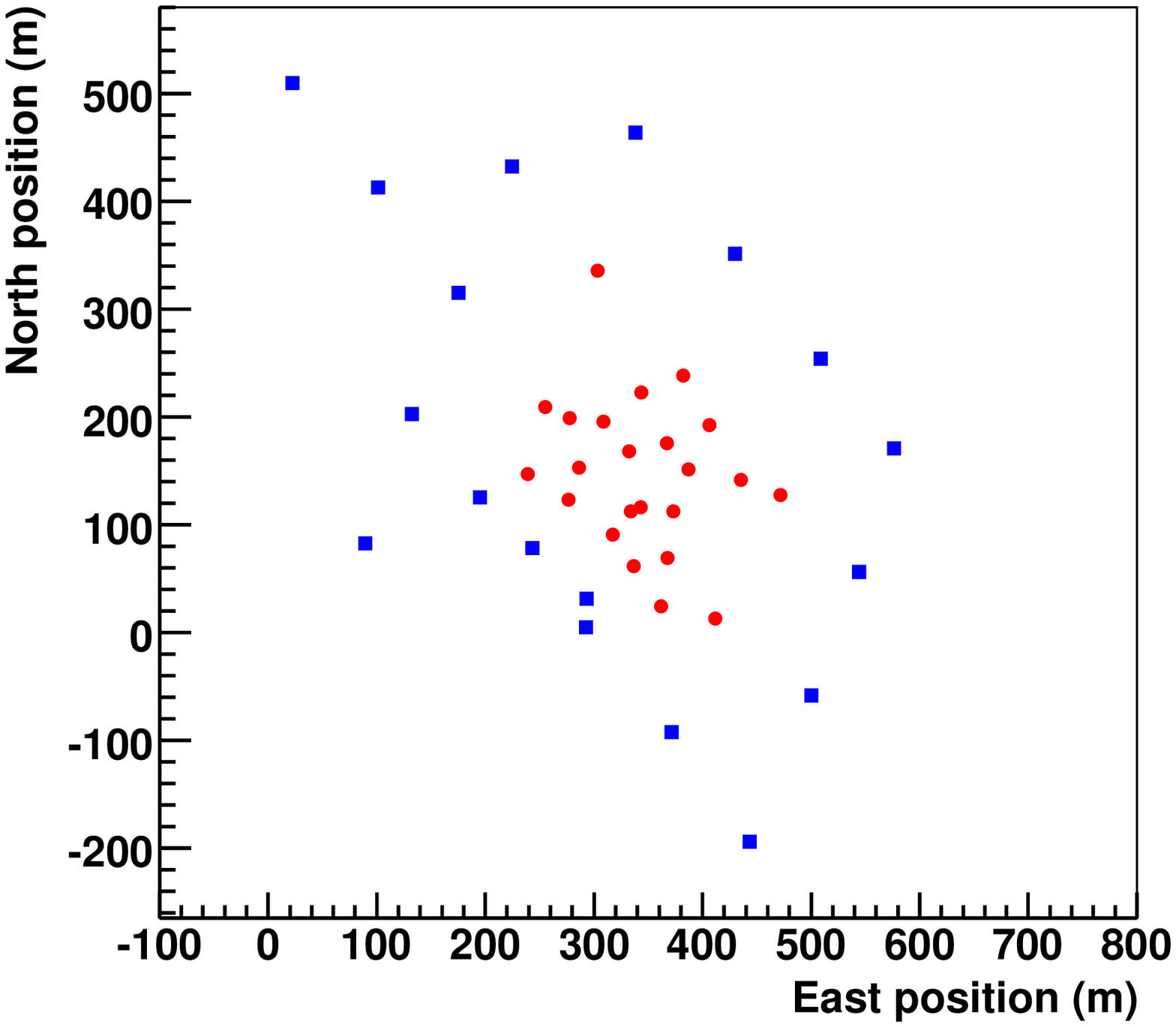}
\end{center}
\caption{Top view of the 2007 IceCube detector consisting of 41 strings. The inner strings (dots) define the fiducial region, surrounded by veto strings (squares). Uppermost OMs of the fiducial strings belong to the veto region.}\label{ICRC0690_fig1}
\end{figure}

\section{Simulations}
A sample of atmospheric $\mu$ background events corresponding to $\sim1$ hour of detector livetime ($2.3\cdot 10^{6}$ events triggering) was simulated using \texttt{CORSIKA} \cite{cors} with the H\"orandel CR composition model \cite{ICRC0690_hoerandel}. For the atmospheric $\nu_{\mu}$ background, a sample corresponding to $\sim0.5$ years of detector livetime ($4.2\cdot 10^{4}$ events triggering) was generated according to the Bartol spectrum \cite{ICRC0690_bartol}.

The solar WIMP signals were simulated with \texttt{WimpSim} \cite{eds}, which uses \texttt{DarkSUSY} \cite{susy} and \texttt{PYTHIA} \cite{pyt} to calculate annihilation rates and neutrino production. The neutrinos were propagated through the Sun and to the Earth with standard full flavour oscillations \cite{maltoni}. A charged lepton and a hadronic shower were then generated in the ice. For this analysis only simulated muon events with the Sun under the horizon, $\theta_{\odot}\in[90^{\circ},113^{\circ}]$, were used.

Muon propagation through the ice was simulated with \texttt{MMC} \cite{ICRC0690_mmc}. Cherenkov light propagation through the ice to the OMs, taking into account the ice properties \cite{iceprop}, was done with \texttt{Photonics} \cite{pho}. The detector response was simulated with the IceCube simulation package \texttt{icesim}.

\section{Filtering}
Events were first selected based on a log-likelihood (LLH) reconstruction, by demanding $\theta_{\it LLH}\in[90^{\circ},120^{\circ}]$. Half of the atmospheric $\mu$ and the WIMP events passing this cut were then used to train and half to test a neural network (NN) using two hidden layers and eight event observables based on hit topology as well as the LLH reconstructed track parameters. A cut was made on the NN output value, the hit multiplicity and the reconstruction quality. This cut removed all simulated atmospheric $\mu$ background, and only WIMP events and atmospheric $\nu_{\mu}$ background remained.

Among the remaining events we selected the neutrino candidates originating from the Sun's direction within a cone with half opening angle varying between $3^{\circ}$ and $10^{\circ}$ depending on $m_{\chi}$ and annihilation channel. At this final analysis stage $V_{\it eff}$ for the observation of WIMP signals were calculated as 
\begin{equation}
V_{\it eff}=\frac{N_{\it det}\cdot V_{\it gen}}{N_{\it gen}},
\end{equation}
where $N_{\it gen}$ is the number of generated CC interactions, $V_{\it gen}$ is the generation volume, and $N_{\it det}$ is the number of WIMP events in the search cone. Results are given in figure \ref{ICRC0690_fig2} (squares).

\begin{figure}
\begin{center}
\noindent
\includegraphics [width=0.5\textwidth]{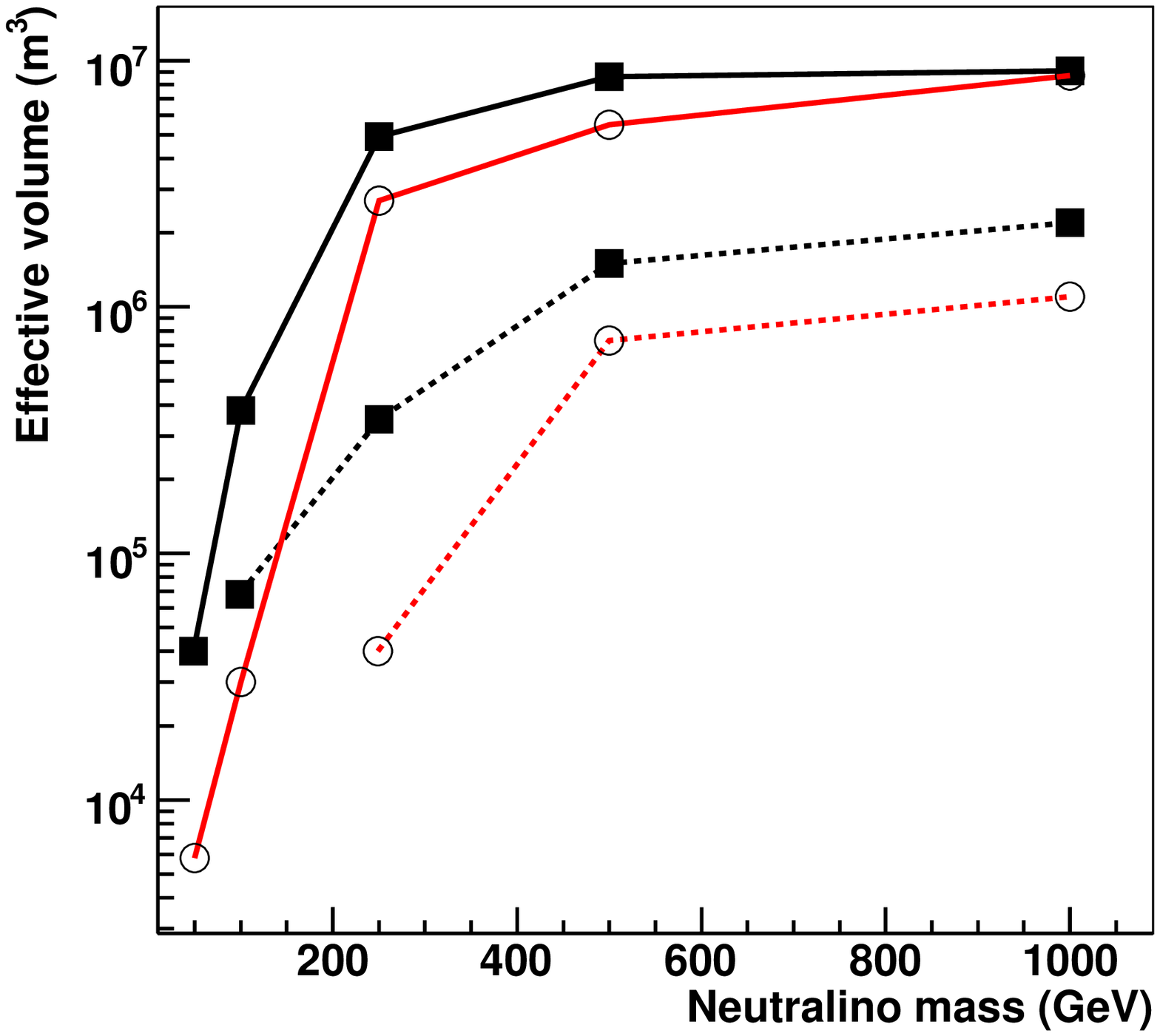}
\end{center}
\caption{WIMP effective volume as a function of neutralino mass for hard (solid) and soft (dashed) annihilation channel, for analysis done on IceCube-22 + AMANDA (squares) and IceCube-22 only (circles).}\label{ICRC0690_fig2}
\end{figure}

\section{Sensitivity}
From the expected number of surviving atmospheric $\nu_{\mu}$ events $\mu_{b}$ we calculated the mean expected Feldman-Cousins $\mu_{s}^{90\%}(n_{\it obs},\mu_{b})$ \cite{f-c} signal upper limit from all possible outcomes $n_{\it obs}$ as the Poisson weighted sum
\begin{equation}
\bar{\mu}_{s}^{90\%}=\sum_{n_{\it obs}=0}^{\infty}\mu_{s}^{90\%}(n_{\it obs},\mu_{b})\frac{(\mu_{b})^{n_{\it obs}}}{n_{\it obs}!}e^{-\mu_{b}}.
\end{equation} 
From the WIMP $V_{\it eff}$ and $\bar{\mu}_{s}^{90\%}$ we then calculated the mean expected upper limit on the neutrino to muon conversion rate
\begin{equation}
 \bar{\Gamma}_{\nu\rightarrow\mu}^{90\%} = \frac{\bar{\mu}_{s}^{90\%}}{V_{\it eff}\cdot t},
\end{equation} 
where $t$ is the analysis livetime, which here is 0.5 years. These values (one for each WIMP signal) were then used to calculate the mean expected upper limit on the muon flux from neutralino annihilations in the Sun, $\bar{\Phi}_{\mu}^{90\%}$ \cite{sun-muon, flux}, which is a measure of the detector's WIMP sensitivity. Comparing these values with the mean expected values from an earlier analysis with AMANDA-II \cite{yulia,ICRC0690_newsun}, we see improvements of an order of magnitude for lower WIMP masses, see figure \ref{ICRC0690_fig3}.

Repeating the analysis with only the IceCube-22 array, we found that AMANDA stands for the major contribution to the sensitivity at lower $m_{\chi}$ and that the IceCube strings dominates for $m_{\chi}\geq$ 500 \rm{GeV} (see figure \ref{ICRC0690_fig2}). The increase in sensitivity for the AMANDA array is due to lowered trigger multiplicity thresholds thanks to the new TWR-DAQ \cite{twr}, whereas for higher $m_{\chi}$ the increased sensitivity comes from the increased detector volume in IceCube compared to AMANDA. The forced readout of IceCube when AMANDA has a trigger makes it much easier to distinguish a neutrino event from an atmospheric $\mu$ event.

\begin{figure}
\begin{center}
\noindent
\includegraphics [width=0.5\textwidth]{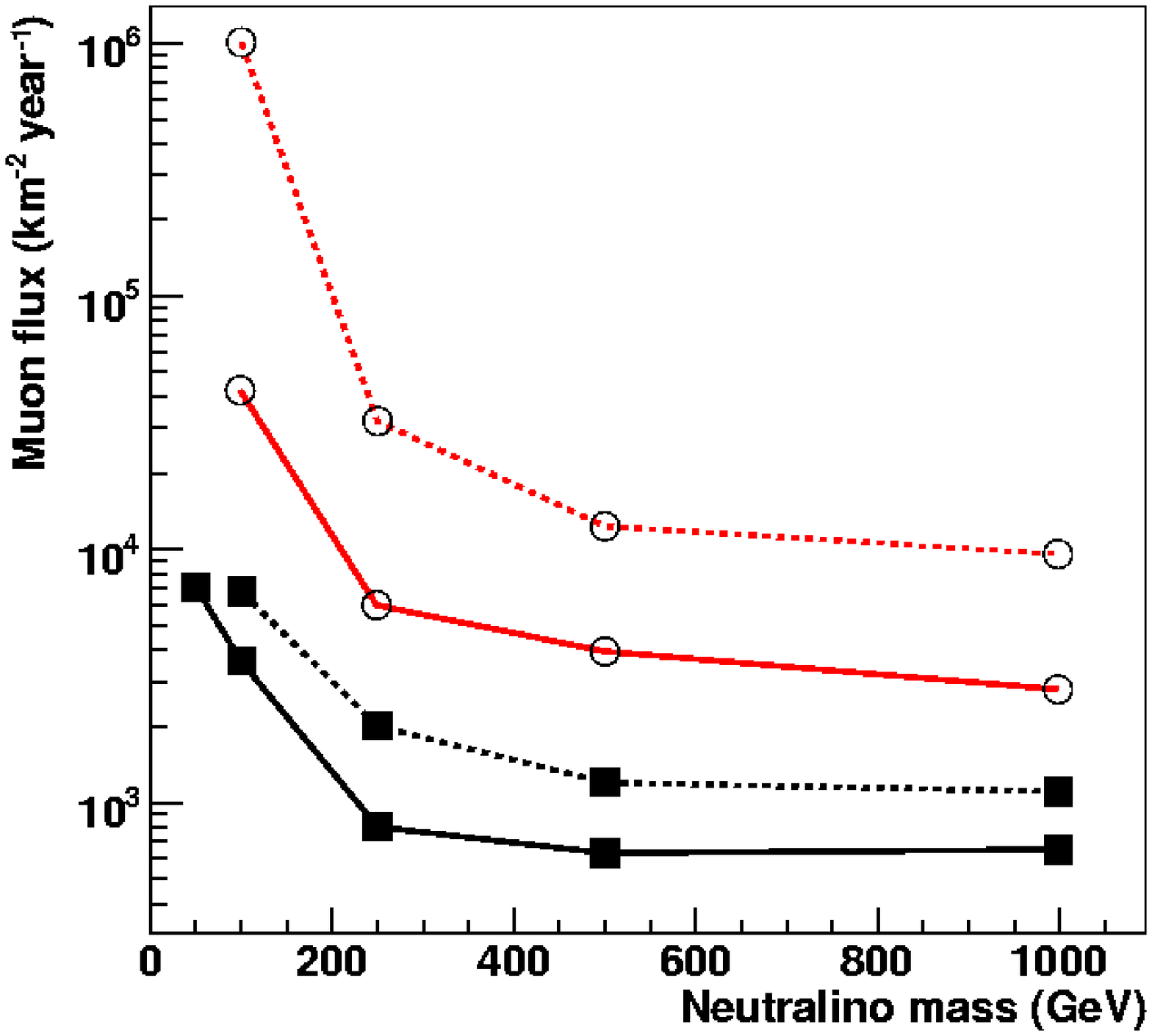}
\end{center}
\caption{Expected sensitivity to muon flux from neutralino annihilations in the Sun as a function of neutralino mass for hard (solid) and soft (dashed) annihilation channel for this analysis using IceCube-22 and AMANDA (squares), compared to the 2001 AMANDA analysis \cite{yulia,ICRC0690_newsun} (circles). Systematic uncertainties are not included.}\label{ICRC0690_fig3}
\end{figure}

\section{Outlook}
This sensitivity estimation demonstrates the feasibility of a WIMP analysis on experimental data from 2007 using the combined IceCube-22 and AMANDA detector.

%\end{document}

\newpage
\newpage
%
% Exotics
%
%icrc0870.pdf (Henrike)
%pohl__hardke
%icrc0788.pdf (Christy, IceCube)
%
\setcounter{figure}{0}
\setcounter{table}{0}
%%
% International Cosmic Ray Conference 2007 Merida Yucatan Mexico
% In This file you will find detailed instructions to correctly
% typeset your document.
%
%
%

%Class Requeried
%\documentclass{article}
%The ICRC Style
%\usepackage{icrctc07}

%The paper title
\title{Search for Relativistic Magnetic Monopoles with the AMANDA-II Detector}
%Short title to print in the headers to the final publication (Not showed in this print).
\shorttitle{Relativistic magnetic monopoles in AMANDA-II}
%All paper authors
\authors{H. Wissing$^{1}$ for the IceCube Collaboration$^\star$}
%Short title to print in the headers to the final puplication (Not  the ed in this print).
\shortauthors{H. Wissing}
%All the affiliations.
\afiliations{$^1$III Physikalisches Institut, RWTH Aachen University, D-52056 Aachen, Germany\\
$^\star$see special section of these proceedings}
\email{hwissing@physik.rwth-aachen.de}

%The abstract.
\abstract{Cherenkov emissions of magnetic charges moving through matter will exceed 
those of electric charges by several orders of magnitude. The AMANDA
neutrino telescope is therefore capable of efficiently detecting relativistic magnetic monopoles 
that pass through its sensitive volume. We
present the to date most stringent  limit on the flux of relativistic magnetic monopoles based on the 
analysis of one year of data taken with AMANDA-II.}

%%%%%%%%%%%%%%%%%%%% B E G I N   D O C U M E N T%%%%%%%%%%%%%%%%%%%%%%%
%\begin{document}
\maketitle
%Begin the section.

\section{Introduction}

The existence of magnetic monopoles is mandatory 
in a large class of Grand Unified Theories. 
Predicted monopole masses range from 10$^8$ to 10$^{17}$\,GeV, depending 
on the symmetry group and unification scale of the underlying theory \cite{tHooft:1974qcw}. 
The monopole magnetic charge will be an integer multiple of  
the {\it Dirac Charge}  $g_D=e/(2\alpha)$, where $e$ is the electric elementary charge and 
$\alpha=1/137$ is the fine structure constant. 
Since magnetic monopoles are topologically stable, 
they  should still be present in
today's universe and can be searched for in cosmic radiation.   
Once created, monopoles can efficiently be accelerated by large 
scale magnetic fields. 
Monopoles with masses below $\sim10^{14}$\,GeV are expected 
be relativistic \cite{Wick:2000yc} and neutrino telescopes
could detect their direct  Cherenkov emissions.
The number of Cherenkov photons $N_{\gamma}$  emitted  per path length $dx$ 
and photon wavelength $d\lambda$ radiated from a monopole with magnetic charge $g$ passing 
through matter with index of refraction $n$ is \cite{Tompk65}
\begin{equation} 
\frac{dN_{\gamma}}{dxd\lambda} = \frac{2\pi\alpha}{\lambda^2}\left(\frac{gn}{e}\right)^2 
\left( 1-\frac{1}{\beta^2n^2} \right), 
\end{equation} 
where $\beta$ is the speed of the monopole as a fraction of the speed of light in vacuum. 
The Cherenkov light intensity in ice radiated from a relativistic magnetic monopole carrying 
one Dirac charge is enhanced by factor $(g_D \cdot n/e)^2=8300$ compared to the intensity 
radiated from a particle with electric charge $e$ and the same speed.

\section{Search Strategy}
The AMANDA-II neutrino telescope consists of 677 light sensitive optical modules (OMs)
embedded in the ice under the geographic South Pole at 
depths between 1500 and 2000 meters. 
Each OM contains a photomultiplier tube (PMT) and supporting electronics enclosed in  
a transparent pressure sphere. The OMs are deployed on 19 vertical strings 
arranged in three concentric circles (see Figure \ref{topview}).
The inner ten strings    
are read out electrically via coaxial or twisted-pair cables, while the outermost 
strings use optical fiber transmission. 
For each triggered event, leading and trailing edges of up to eight
PMT pulses and one peak amplitude can be recorded per OM.
\begin{figure}[h]
\centering
\includegraphics[width=0.2\textwidth]{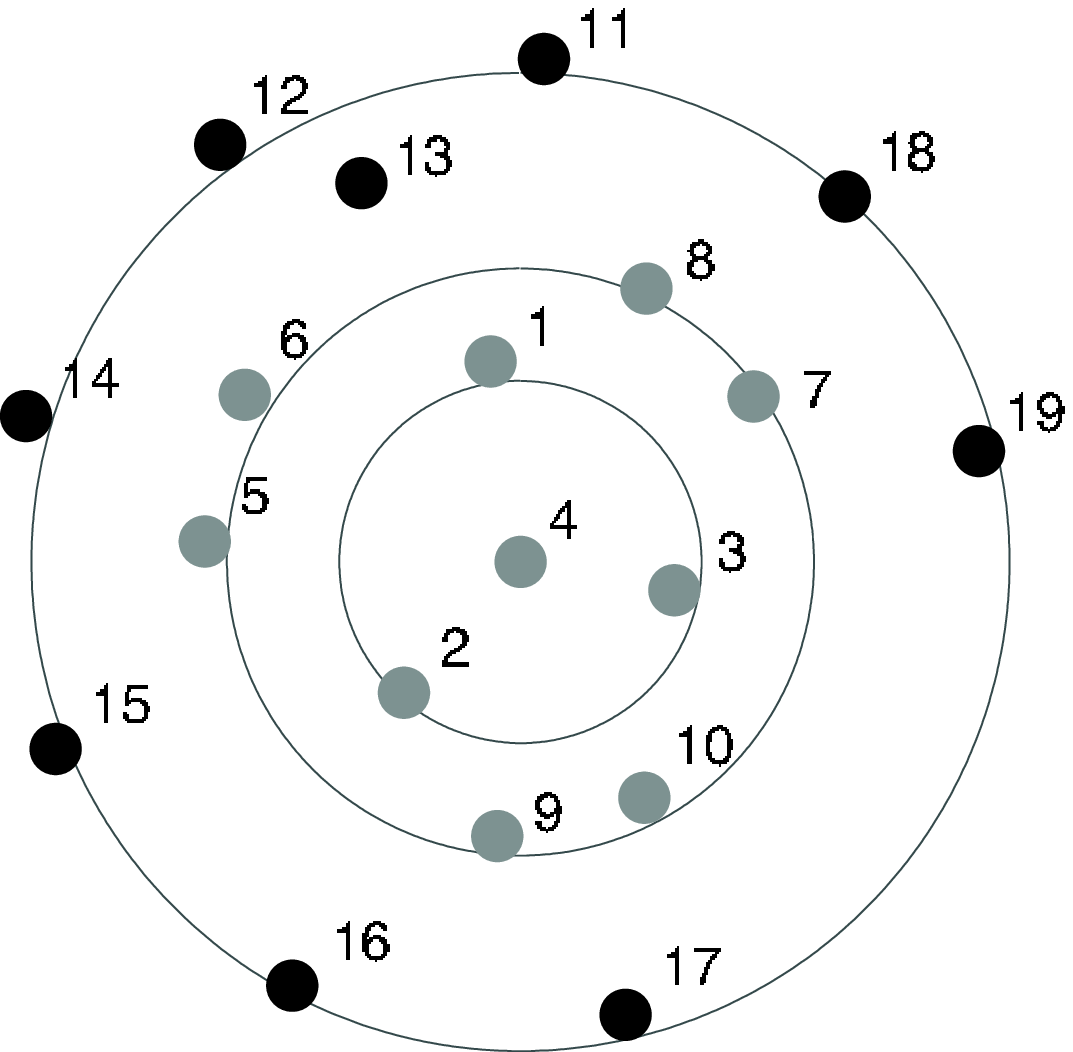}
\caption{Arrangement of the 19 strings of the AMANDA-II detector in the horizontal plane.}
\label{topview}
\end{figure}
AMANDA-II has been taking data since the beginning of the year 2000. 
This analysis concerns data taken
between February and November 2000.

We have simulated the detector response to relativistic magnetic monopoles carrying 
one Dirac Charge with four different speeds: $\beta = v/c = 0.76, 0.8, 0.9,$ and 
$1.0$. Down-going atmospheric muons, which form the principal background to 
this search, were simulated with the air-shower simulation package 
{\tt CORSIKA} \cite{ICRC0870_corsika}. 
Following a  ``blind'' analysis procedure, the data selection chain is optimized 
on simulated data, and only a subset of 20\% of 
the experimental data is used to verify the simulation of the detector response. 
This 20\% is later discarded, and the developed selection criteria 
applied to the remaining 80\% of the data. The blinded data comprise 
about 154 days of livetime, after correction for dead-time and rejection of 
periods of low data quality.

A  relativistic magnetic monopole passing through the detector's sensitive volume will
stand out as an extremely bright event relative to the atmospheric muon background. 
Observables that provide a measure of the light yield in the detector
are the number of OMs hit during an event, the total 
number of pulses (or \emph{hits}) recorded, the fraction of OMs which 
registered only a single hit (as opposed to those which recorded multiple hits), and 
the sum of the recorded PMT pulse amplitudes. These observables are used to reject the bulk of 
low energy atmospheric muons, either as one-dimensional cut parameters or
as input to a discriminant analysis \cite{Fisher:1936et}.

\section{Up-going Monopoles}

\begin{figure}[th]
\centering
\includegraphics[width=.5\textwidth]{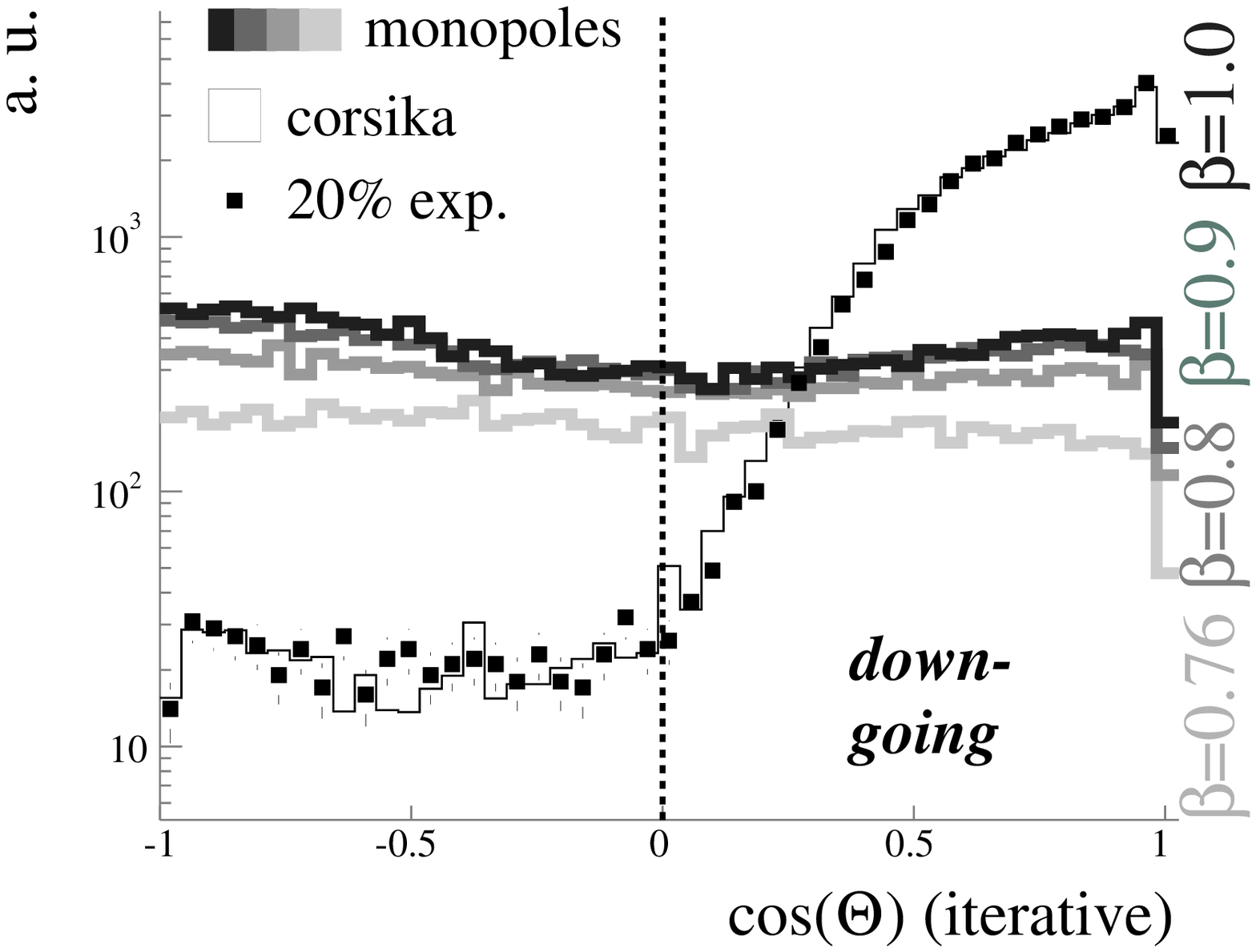}
\caption{Cosine of the reconstructed zenith angle for simulated atmospheric muon 
background (light black histogram), 20\% experimental data (black markers), 
and simulated monopole signal (heavy grey histograms). Each of the four 
grey histograms corresponds to one of the simulated monopole speeds
from $\beta=0.76$ (lightest grey) to $\beta=1.0$ (darkest grey).
A zenith angle of $\Theta = 0^\circ$ corresponds to  vertically 
down-going.}
\label{fig:llhZenith}
\end{figure}

% key obsac ervable is reco zenith
\begin{figure}[th]
\centering
\includegraphics[width=.5\textwidth]{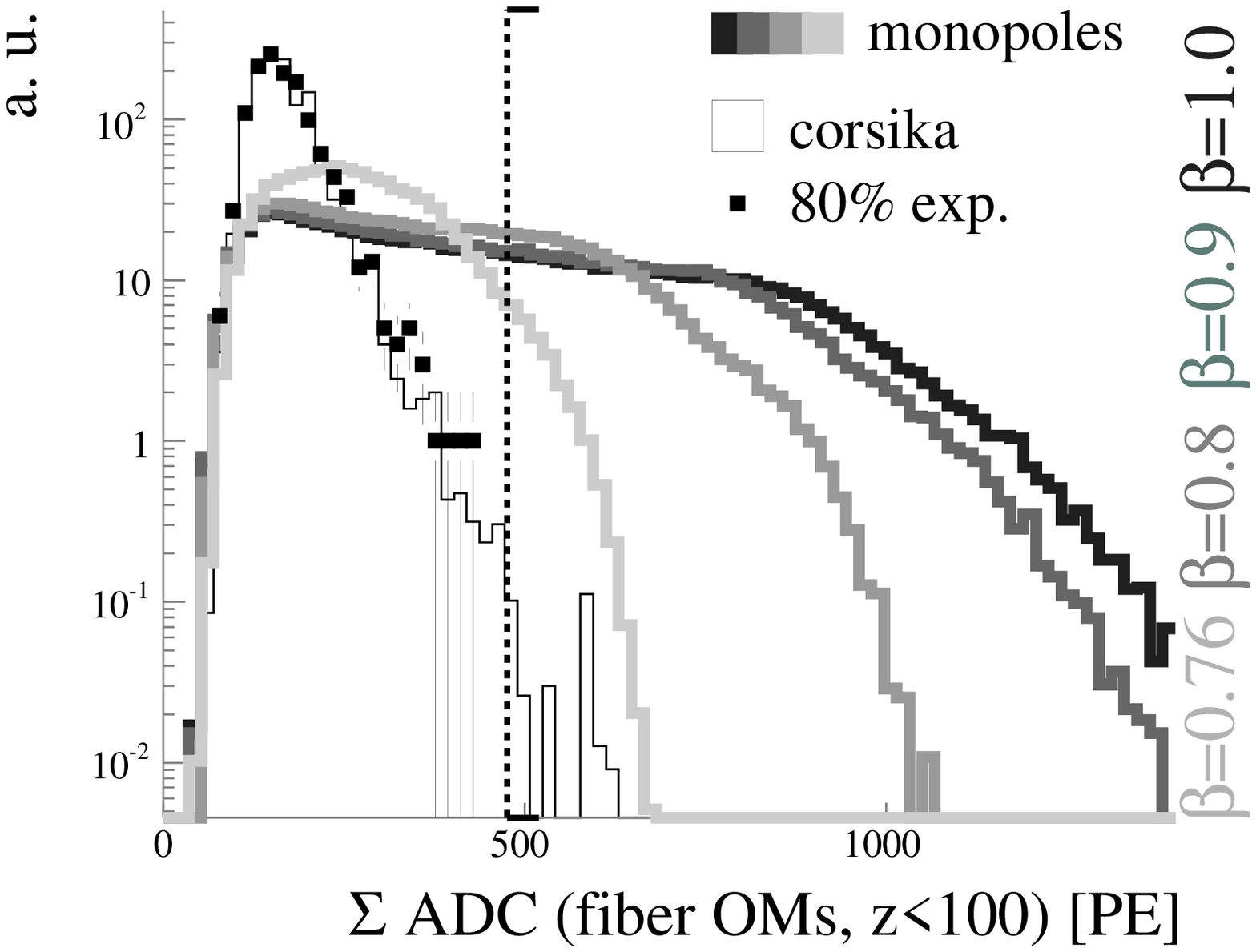}
\caption{Sum of the PMT pulse amplitudes measured in the outer strings
below a depth of 1630\,m. The final cut (dashed line) requires the 
amplitude sum to be bigger than 476 photo electrons.}
\label{fig:upxxx}
\end{figure}

Magnetic monopoles with masses in excess of $10^{11}$\,GeV can cross the entire earth
and enter the detector from below \cite{Derkaoui:1998uv}. The search for up-going
relativistic magnetic monopoles is in principle background free, 
since up-going charged leptons induced by atmospheric neutrinos from
the northern hemisphere will produce significantly less Cherenkov light 
than would magnetic monopoles. However, track reconstruction algorithms sometimes
fail to identify large atmospheric muon bundles as down-going, and those 
misreconstructed events will remain as residual background.
Figure \ref{fig:llhZenith} shows the zenith angle distribution of reconstructed 
particle tracks obtained from an iterative
likelihood reconstruction \cite{ICRC0870_Ahrens:2003fg} for simulated signal and background as well as 
for 20\% of the experimental data.
For the up-going monopole search we reject all events for 
which the reconstructed zenith angle is smaller than 90 degrees.
The background of misreconstructed atmospheric muon bundles
is rejected by a final cut on the sum of PMT 
pulse amplitudes ({\bf$\Sigma ADC$}). 
At this level of the analysis, an excellent simulation of the OM response
to large amounts of light is required. This involves an accurate modeling 
of the sensitivity of individual OMs as well as the probability with which  
OMs \emph{``overflow''}, \emph{i.e.}, record more than eight hits during one event
(in which case a fraction of the hits is discarded by the data acquisition system).
These requirements dictate that we use the amplitude sum of only a subset of OMs as final 
cut parameter, including only those OMs for which the detector simulation 
provides an exact description. This is the case for the OMs which are read out 
via fiber optics and which are located at depths below 1630\,m.
The fiber OMs have a substantially better time and double pulse resolution than 
the electronically read-out OMs. Thereby the simulation of their response to 
multiple photons is more reliable.
Using only this subset of OMs does not affect sensitivity,
since the fiber OMs are attached to the outer nine strings of the 
array and define the surface area of the detector. Rather, 
the obtained flux limit improves as a result of the reduced systematic error.

Figure \ref{fig:upxxx}  shows the  distribution of the final cut parameter. 
The exact value of the final cut is determined 
by optimizing the \emph{sensitivity} of the analysis, \emph{i.e.},
the cut is placed where we expect to obtain the most stringent flux limit.  
The background simulation predicts 0.23 events from
misreconstructed atmospheric muons to remain in the 
80\% data set. After unblinding the data, no events are observed.

The 90\% C. L. flux limits in units of $10^{-16}$cm$^{-2}$s$^{-1}$sr$^{-1}$ 
obtained for monopoles with various 
speeds $\beta$ are listed in Table \ref{tab:uplll}.

\begin{table}[b]
\centering
\begin{tabular}{lcccc}
\hline
$\beta$ & 0.76 & 0.8 & 0.9 & 1.0 \\
\hline
$\Phi_{90\%C.L.}$  &  8.6 & 0.66  & 0.42 & 0.37 \\
\hline
\end{tabular}
\caption{90\% C. L. upper limits in units of $10^{-16}$cm$^{-2}$s$^{-1}$sr$^{-1}$ 
on the flux of relativistic magnetic monopoles with masses $>10^{11}$\,GeV.}
\label{tab:uplll}
\end{table}
The limits are valid for monopoles with masses greater than $10^{11}$\,GeV.
A systematic uncertainty of 20\% in both background rate and signal efficiency  is 
incorporated into the calculation of the confidence belts 
according to \cite{Conrad:2002kn}. 

\section{Down-going Monopoles}

\begin{figure}[t]
\centering
\includegraphics[width=.43\textwidth]{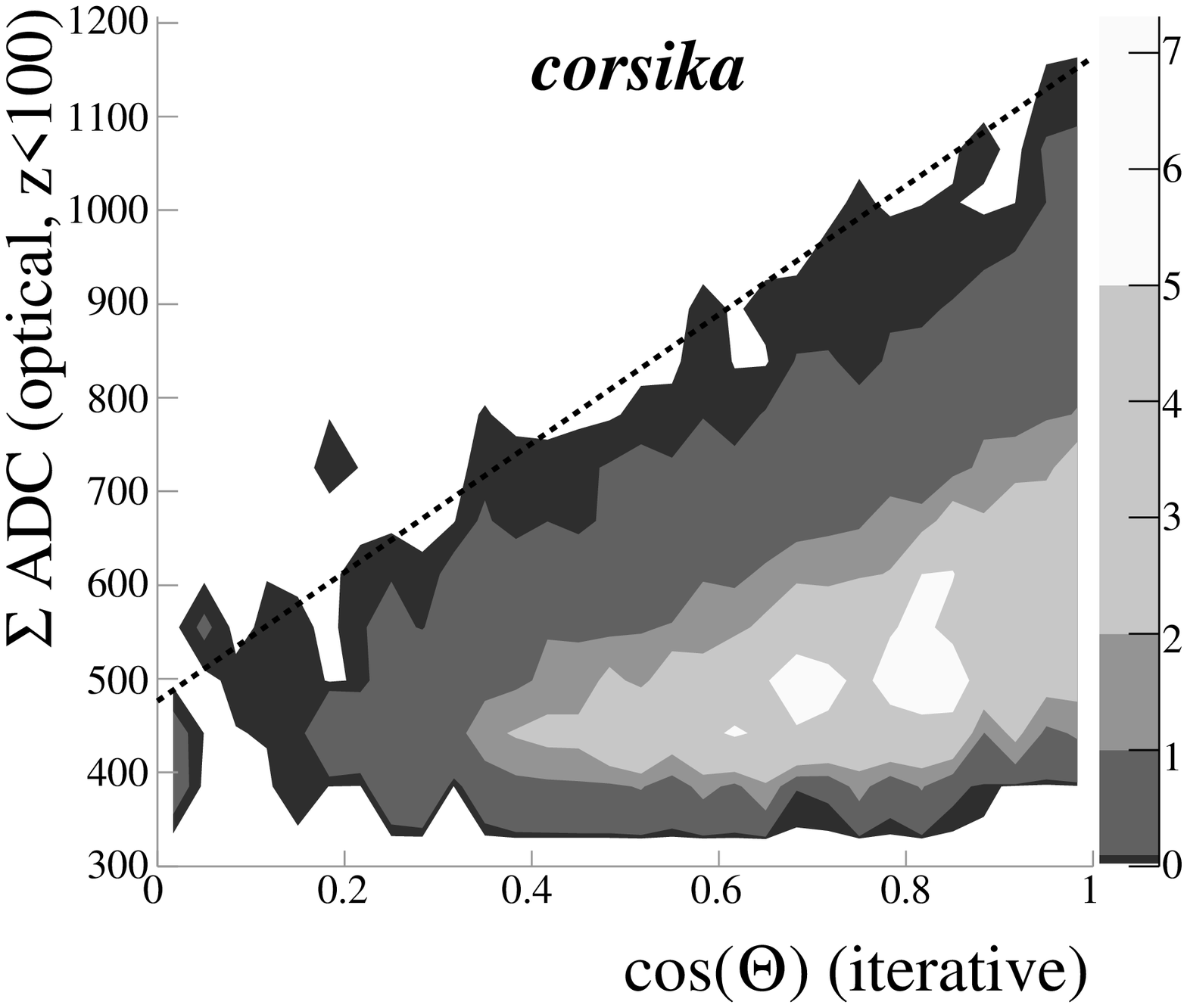}
\caption{Expected background from atmospheric muons in the 
$\cos(\Theta) - \Sigma ADC$ plane before placing the final cut (dashed line).}
\label{fig:adcthetaplane} 
\end{figure}

\begin{figure}[t]
\centering
\includegraphics[width=.5\textwidth]{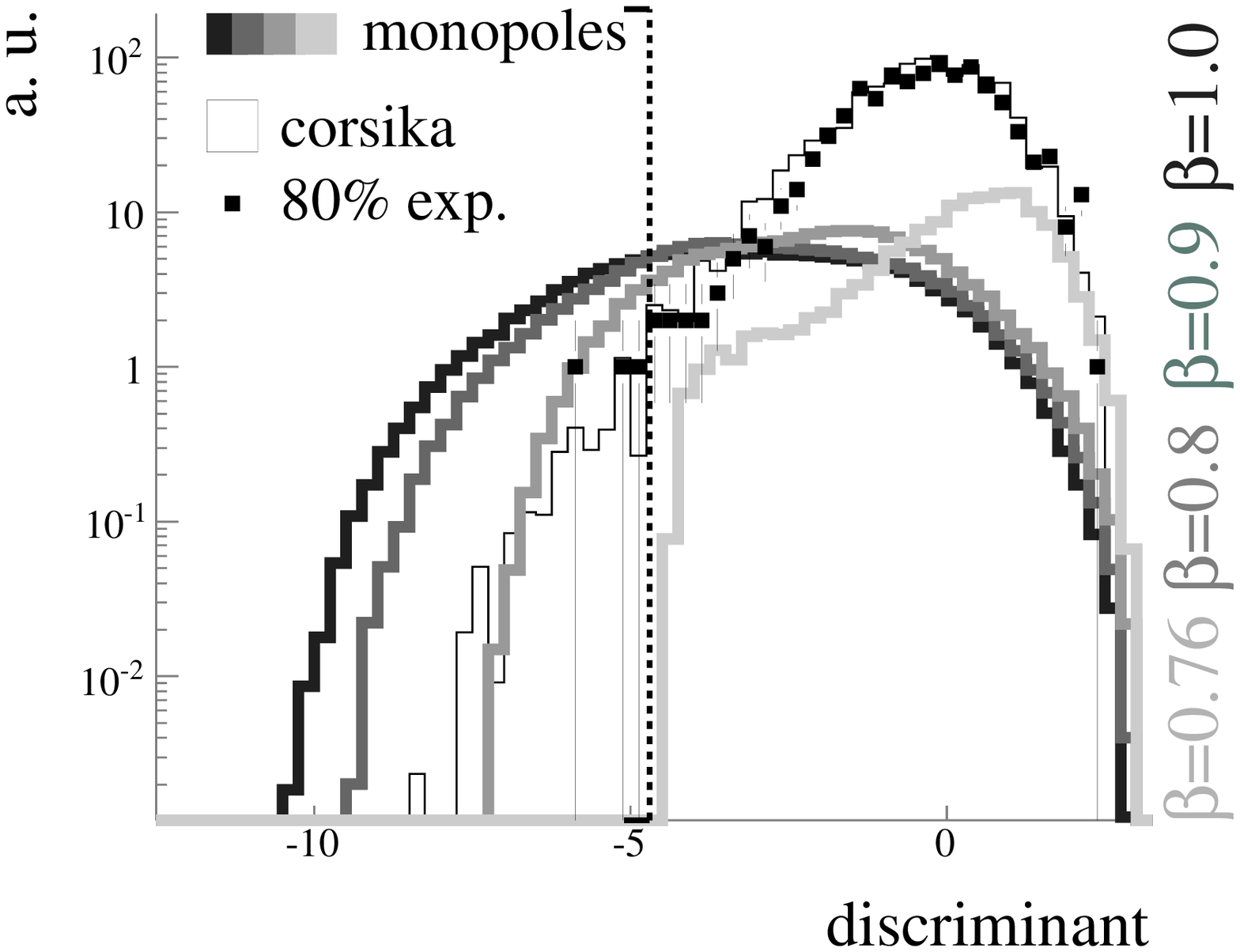}
\caption{Final cut parameter (obtained from a discriminant analysis 
using $\cos(\Theta)$ and  $\Sigma ADC$ as input observables) 
for the unblinded 80\% experimental data (black markers), expected atmospheric muon background 
(black histogram), and simulated monopole signal (grey heavy histograms).
The dashed black line marks the final cut.
}
\label{fig:downFinal}
\end{figure}

The search for down-going magnetic monopoles  
is subject to a much higher background rate. 
In order to preserve sensitivity 
over $4\pi$\,sr, we use linear combinations of the 
reconstructed zenith angle and 
observables that are sensitive to the light deposition in the detector
as cut parameters. The coefficients of each observable are found 
by a discriminant analysis. This optimization naturally results 
in cuts that require a greater light deposition for vertical tracks 
(smaller zenith angles), while the requirement is relaxed towards the
horizon.

The final cut parameter for the down-going monopole search is a 
linear combination of the cosine of the reconstructed zenith angle ($\cos\Theta$) 
and  the sum of pulse amplitudes recorded by the OMs on the outer strings at depths 
below 1630\,m ($\Sigma ADC$). 
Figure \ref{fig:adcthetaplane} shows the expected distribution
of background events in the  $\cos(\Theta) - \Sigma ADC$ plane.
The final cut parameter is shown in Figure \ref{fig:downFinal}. 
Like in the up-going monopole search, the 
final cut is optimized such that we expect to obtain the most stringent limit.
The background simulation predicts 2.6 events to remain in the experimental data set,
and three events are observed after unblinding the data. The limits 
on relativistic monopoles with various speeds obtained
from this observation are listed below (Table \ref{tab:downXXX}).
The limits are valid for monopoles with masses greater than $10^{8}$\,GeV.
Systematic uncertainties are accounted for according to \cite{Conrad:2002kn}.

\begin{table}[ht]
\centering
\begin{tabular}{lcccc}
\hline
$\beta$ & 0.8 & 0.9 & 1.0 \\
\hline
$\Phi_{90\%C.L.}$ & 16.33 & 4.1 & 2.8 \\
\hline
\end{tabular}
\caption{90\% C. L. upper limits in units of $10^{-16}$cm$^{-2}$s$^{-1}$sr$^{-1}$ 
on the flux of relativistic magnetic monopoles with masses greater than $10^{8}$\,GeV.}
\label{tab:downXXX}
\end{table}

\section{Conclusions}

The analysis of data taken with the AMANDA-II neutrino 
telescope during the year 2000 permits constraint of the  
flux of relativistic magnetic monopoles 
with speeds $\beta=v/c > 0.76$. For monopole speeds
greater than $\beta=0.8$ and monopole masses greater than $\sim10^{11}$\,GeV,
the flux limit is presently the most stringent experimental limit.    
The search for lighter monopoles is possible, but 
less sensitive. With the analysis of one year of AMANDA data, the flux of
magnetic monopoles with masses as low as $10^{8}$\,GeV and speeds 
close to $\beta=0.8$ can be constrained to a level below the Parker 
Bound \cite{Parke82}. Figure \ref{fig:limits} shows the flux limits set by AMANDA  
compared to those set by some other experiments.

\begin{figure}[ht]
\centering
\includegraphics[width=.47\textwidth]{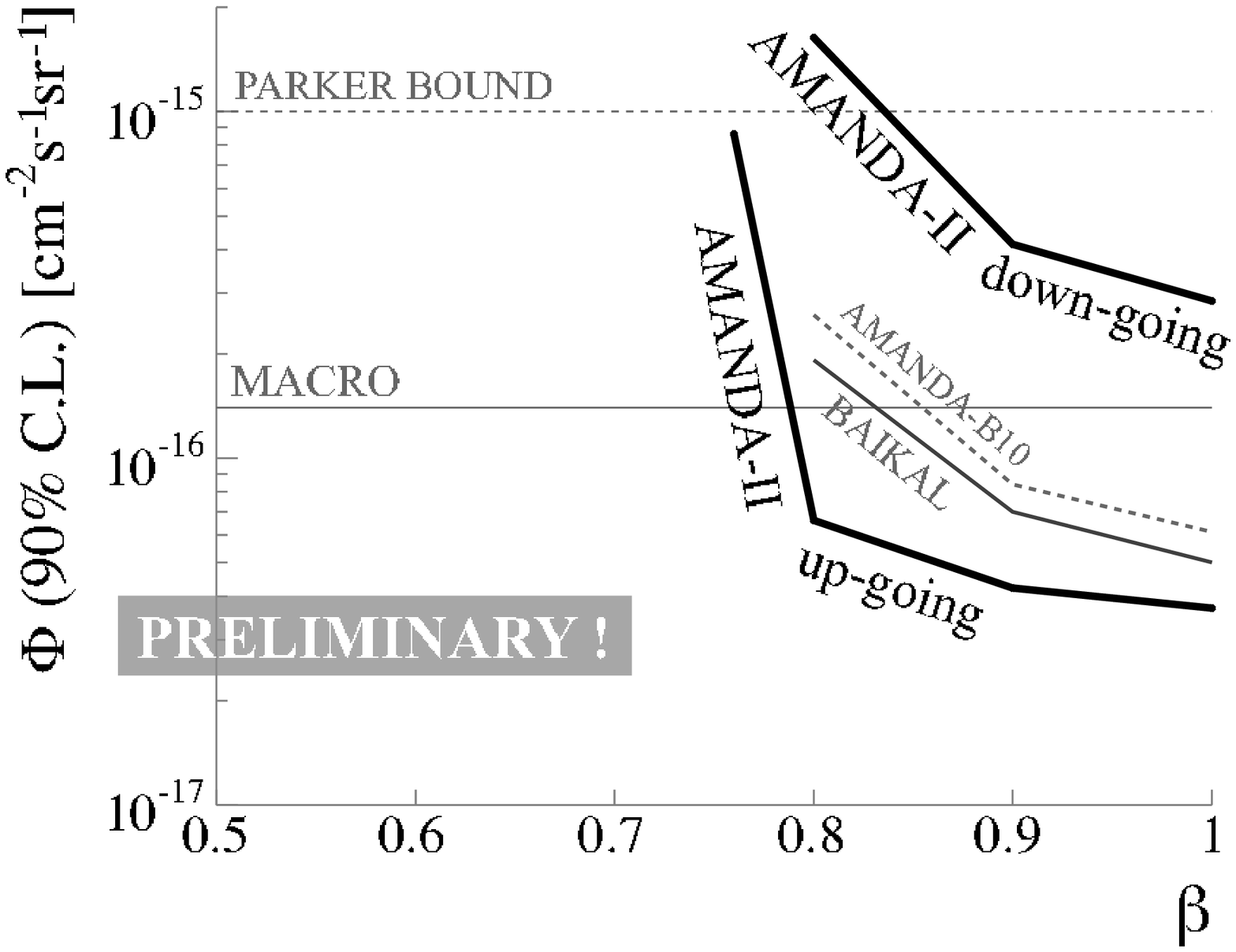}
\label{fig:limits}
\caption{Limits on the flux of relativistic magnetic monopoles set AMANDA-II (this work),
by MACRO \cite{Ambrosio:2002qq}, and by BAIKAL \cite{Aynutxdinov:2005sg}.}
\end{figure}

%\section{Acknowledgements}
%This work is partially supported by the Brazilian agencies FAPESP and CNPq, and the Argentinian agencies
%CNEA and ANPCyT.
%\nocite{ref4}
%\nocite{ref5}
%\nocite{ref6}
%\nocite{ref7}
%This is the reference to .bib file (Whitout .bib!)
%This in the bibtex style, is ok.

\section{Acknowledgments}
{\small
This work was supported by the National Science Foundation \-- Office of Polar Programs, and the German Ministry of Education and Research (BMBF).
}

%\bibliography{refs}
%\bibliographystyle{plain}

%\end{document}

\setcounter{figure}{0}
\setcounter{table}{0}
%%
% International Cosmic Ray Conference 2007 Merida Yucatan Mexico
% In this file you will find detailed instructions to correctly
% typeset your document.
%
% By: Victor De la Luz
% vdelaluz@inaoep.mx
% Mexico City

%Class Required
%%% for classical LaTeX
%\documentclass[dvips]{article}
%%% for PDFLaTeX
%\documentclass[pdftex]{article}
%The ICRC Style
%(This package is the last package in the usepackage list)
%If you need import other package you need write it first.
%\usepackage{icrctc07}

%The paper title
\title{Subrelativistic Particle Searches with the AMANDA-II detector}
%Short title to print in the headers to the final publication (Not showed in this print).
\shorttitle{Subrelativistic Particles in AMANDA-II}

%All paper authors
\authors{A. Pohl$^{1,2}$, D. Hardtke$^{3}$ for the IceCube collaboration$^{4}$ }
%Short title to print in the headers to the final publication (Not shown in this print).
\shortauthors{A. Pohl et al.}
%All the affiliations.
\afiliations{$^1$Division of High Energy Physics, Uppsala University, S-75121 Uppsala, Sweden\\ 
$^2$School of Pure and Applied Natural Sciences, University of Kalmar, S-39182 Kalmar, Sweden \\
$^3$Department of Physics, University of California, Berkeley, CA  94720, USA \\
$^4$see special section of these proceedings. }  
\email{arvid.pohl@hik.se}    
% see special section of these proceedings.

\abstract{Supermassive particles like magnetic monopoles, Q-balls and 
nuclearites may emit light at subrelativistic speeds through different 
suggested mechanisms. One of them is nucleon decay catalysis by magnetic 
monopoles, where the decay products would emit Cherenkov radiation 
along a monopole trajectory. The emitted secondary light from subrelativistic 
particles could make them visible to the AMANDA-II neutrino telescope, 
depending on the resulting luminosity. We present first experimental 
results from a search with AMANDA-II for events of this kind.}

%%%%%%%%%%%%%%%%%%%% B E G I N   D O C U M E N T%%%%%%%%%%%%%%%%%%%%%%%
%\begin{document}
\maketitle
\section{\label{ICRC0970_sec:intro} Introduction}
The Grand Unified Theories (GUT) predict the existence of magnetic monopoles with expected mass 
of the order of $10^{16} - 10^{17}$ GeV\cite{tHooftPolyakov74}.
These supermassive monopoles might become accelerated above virial velocities due to 
magnetic fields, but not relativistic \cite{Wick03}. 

Rubakov and Callan have indepentendtly proposed a mechanism by which SU(5) GUT monopoles 
are able to catalyse nucleon decay with a detectable cross section \cite{Rubakov81, Callan82}. 
The main decay channels would be $e^+\pi^0$, $\mu^+K^0$ for protons and $e^+\pi^-$, $\mu^+K^-$ for neutrons, 
see \cite{ICRC0970_MACRO} and refs. therein. The catalysis cross section has been suggested to be 
$\sigma = \sigma_0 \beta^{-1}$ \cite{Rubakov81} or, at sufficiently low speeds, \mbox{$\sigma = \sigma_0 \beta^{-2}$} 
\cite{Rubakov83, Arafune83}, where $\sigma_0$ is a cross section 
typical of strong interactions. Nuclear attenuation factors have also been proposed, 
expressing nuclear spin effects on the decay catalysis \cite{Arafune83}. 
The expected mean distance between nucleon decays catalysed along a monopole trajectory in ice, 
reaches down to submillimeter scales (following the cross sections above). 
Above the meter scale, the signal falls below our detector threshold.  

In a neutrino telescope, the signature of these catalyzing monopoles would be 
a series of closely spaced light bursts produced along the monopole trajectory.   
Each burst would be Cherenkov radiation from an electromagnetic shower whose energy is close to the proton mass.

Other massive particles have also been hypothesized to exist in cosmic radiation. 
Two that might be detectable with neutrino telescopes are:
Nuclearites (nuggets of strange dark matter) \cite{derujula84} and Q-balls (supersymmetric coherent states of squarks, 
sleptons and Higgs fields, predicted by supersymmetric generalizations of the standard model) \cite{kusenko98b}.

Electrically neutral Q-balls would dissociate nucleons, emitting pions, which give them the same 
experimental signature in a neutrino telescope as catalyzing monopoles. 
Their cross section for nucleon dissociation is their geometric size. By limitations given in \cite{bakari01}, 
it ranges from $\sim\!10^{-26}\,\mathrm{cm^2}$ and many orders of magnitude upwards. 

Nuclearites and charged Q-balls might also be detectable, as, travelling through matter, 
they would generate a thermal shock wave which emits blackbody radiation 
at visible wavelengths \cite{derujula84,arafune00}.
Their luminosity as given by \cite{derujula84} is determined by their geometric size, which is atomic or larger, 
and would exceed that of magnetic monopoles and neutral Q-balls by several orders of magnitude.

So far we have only considered magnetic monpoles. 

\section{\label{sec:detector} The AMANDA-II Neutrino Telescope}

AMANDA-II is a neutrino telescope  located  at a depth between 1500\,m and 2000\,m 
under the ice at the geographic South Pole. A cylindrical volume of roughly 200\,m diameter 
of the Polar ice was instrumented with a total of 677 optical modules (OMs), 
consisting of a photomultiplier tube (PMT) and supporting electronics enclosed in  
a transparent pressure sphere. The OMs were deployed on 19 vertical strings.

A variety of triggers are used. 
First, the 24-fold multiplicity trigger requiring a minimum of 24 OMs hit within a fixed 
coincidence window of 2.5\,$\mu$s, and second, a so-called correlation trigger,
requiring $n$ OMs to be hit in any group of $m$ adjacent OMs on the same string ($m,n$ typically $\sim\!6,9$). 
For each triggered event, PMT pulse data is recorded 
over a time window of $\sim\!33\,\mu$s. 
The vast majority of triggers are due to down-going atmospheric muons, yielding an 
average event rate of roughly 90\,Hz. 

\section{\label{ICRC0970_sec:sim} Simulation}

The detection of slow particles builds on the fact that relativistic muons emit light during $\sim\!3\,\mu$s, 
whereas slow particles emit during a large fraction of the 33\,$\mu$s time window. 
A comparison is shown in Fig. \ref{signature}. The upper picture shows a background event with the 
triggering muon at time 19\,$\mu$s, and an accidental early non-triggering muon at 
9\,$\mu$s. The lower picture shows a simulated signal event. The signal separation from background 
is based on hits at times when no light from triggering muons is expected, 
the \textit{early and late hits} outside the interval 16-24\,$\mu$s.

\begin{figure}
\begin{center}
\noindent
\includegraphics[width=0.4\textwidth]{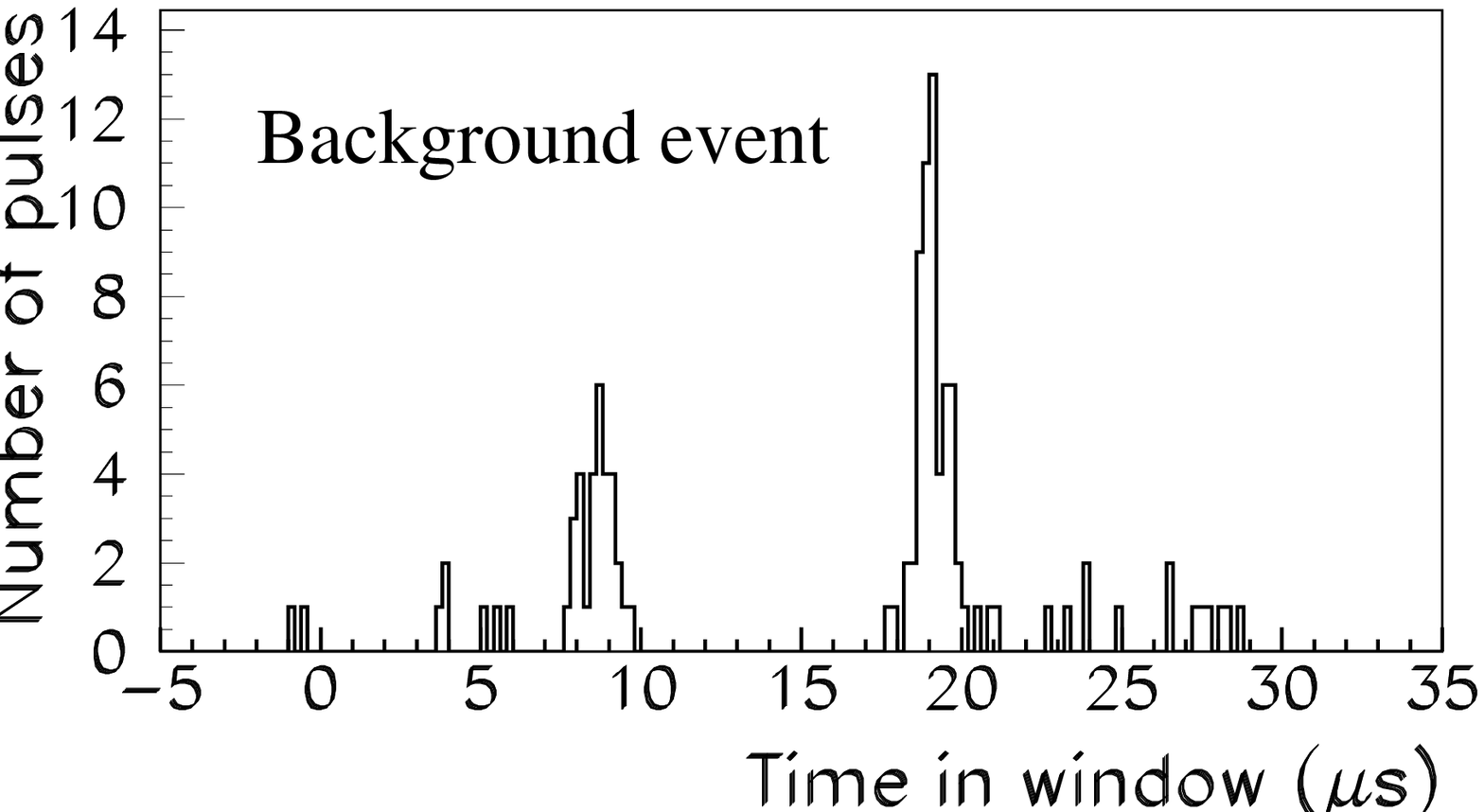}
\includegraphics[width=0.4\textwidth]{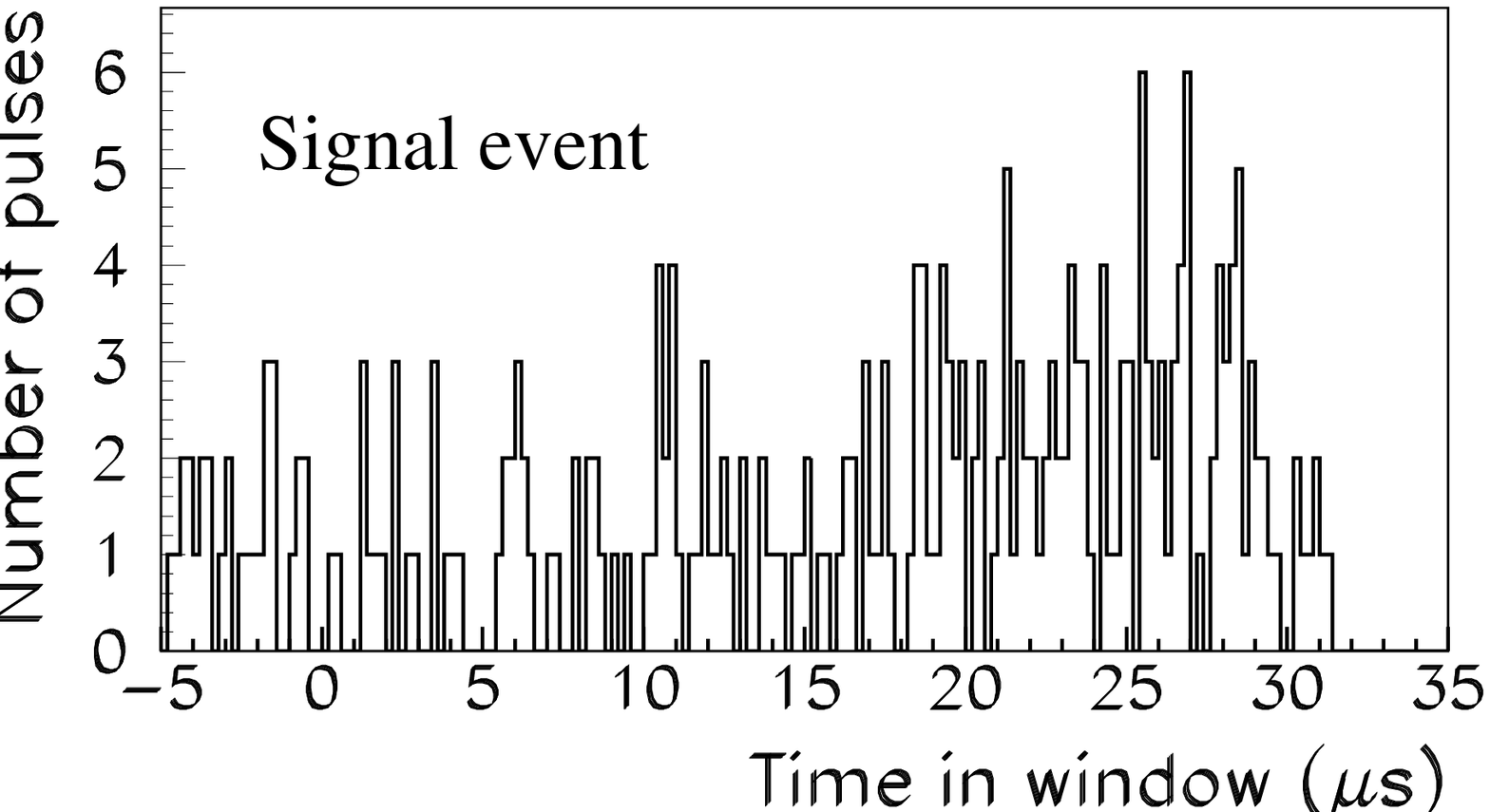}
\end{center}
\caption{Upper: a background event with a non-triggering muon (left) and a triggering muon (right).
Lower: a simulated signal event from a particle traveling at speed $\beta=v/c=0.01$.}
\label{signature}
\end{figure}

In the simulation of sub-relativistic particles, all light output was expressed as Cherenkov radiation 
from electromagnetic showers arising from nucleon decay. 
All slow particles were simulated with isotropic directions and with speed \mbox{$\beta=v/c=10^{-2}$}. 
In the simulations, the luminosity was expressed as the mean distance $\lambda$ between two 
electromagnetic showers. So far, the simulated $\lambda$ were in the range 2\,mm - 60\,cm. 

For monopoles, only the decay of hydrogen protons was considered, and only the catalysis 
decay channel $p\rightarrow e^+\pi^0$ (with a branching ratio of 0.9 or higher \cite{bais83}). 
It creates an electromagnetic shower with energy close to the proton mass, 
whereas other channels lose some of their shower energy to neutrinos.

If a slow particle would approach the detector, atmospheric muons would cause contributing hits and 
possibly fire a trigger. These muons were included in the simulation. 

The catalysis cross sections $\sigma$ that correspond to the chosen $\lambda$ 
are \mbox{$3\cdot 10^{-25}\,\mathrm{cm}^2 - 9\cdot 10^{-23}\,\mathrm{cm}^2$}. 
These are at the upper edge of what appears to be allowed by theoretical considerations. 
                                                                                
\section{\label{sec:analysis} Data analysis and results}

A period of 113 days in 2001 when a constant correlation trigger definition was used, is considered here. 
It required a multiplicity of 6 within any 9 adjacent OMs in four strings and a multiplicity 
of 7 within any 11 adjacent OMs in the remaining strings. The simulations show that the correlation trigger 
was substantially more sensitive to this type of signal than the multiplicity trigger. 

The background properties and a preliminary expected sensitivity was determined using 20\% of the data. 
A first filter reduced the data by 99\%, requiring a total of at least 14 early and late hits. 

Non-triggering muons contribute largely to early and late hits. The aim of the final filtering was to separate them from 
possible signal events. Hits from non-triggering muons arise within a short time span compared to hits from slow particles, 
as can be seen in Fig. \ref{signature}. We defined hit clusters as collections of early hits that were separated by less than 2\,$\mu$s. 
Each event was characterized by its cluster with most hits. 

After trigger cleaning, we performed second level filtering using two cluster based cuts and one based on the events' geometries, 
as signal events are fairly well localized. The remaning events after filtering have an exponential distribution in the number of early hits. 
It is shown in Fig. \ref{exp_tail}, along with an exponential fit. 

\begin{figure}
\begin{center}
\noindent
\includegraphics[width=0.4\textwidth]{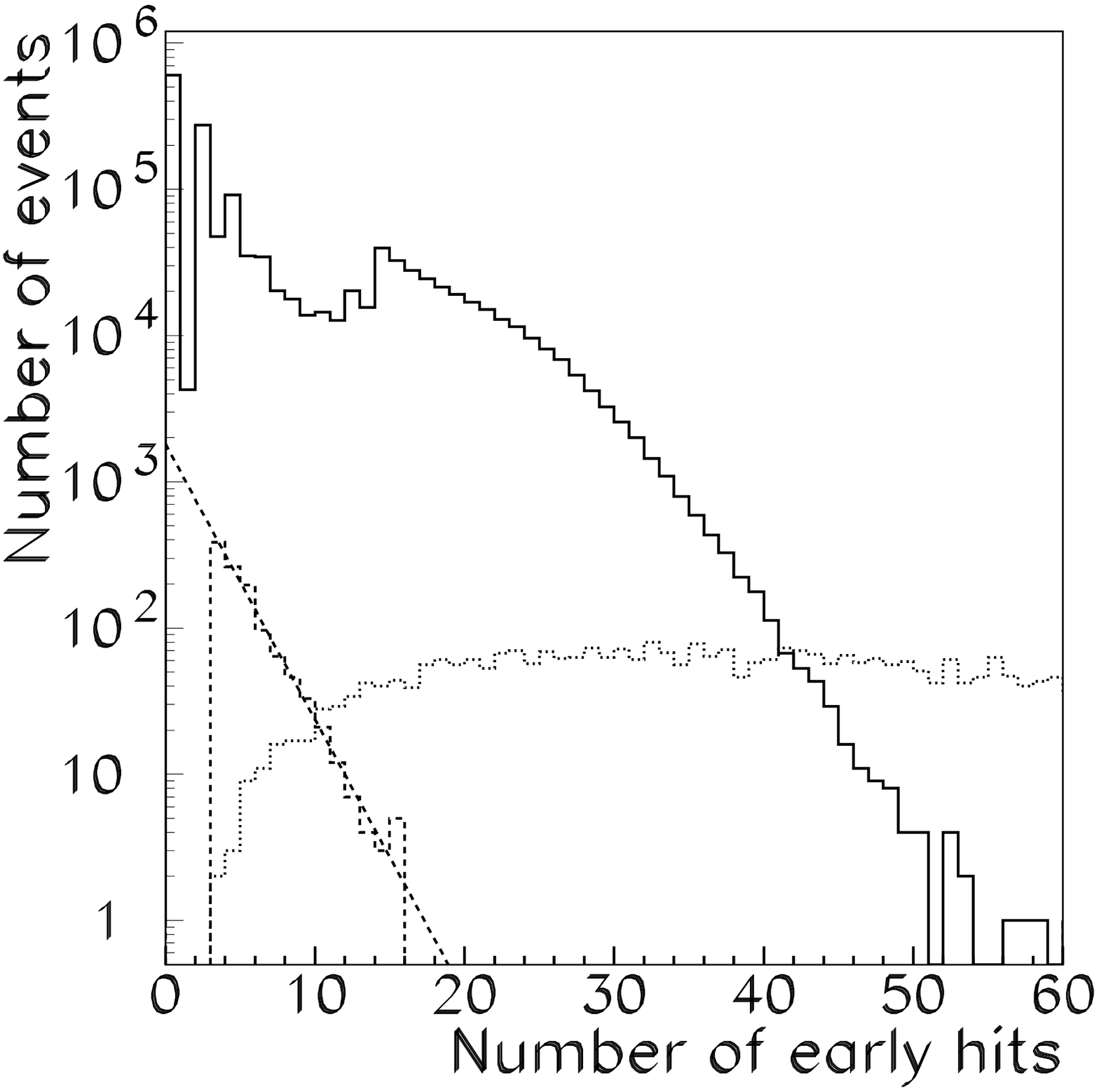}
\end{center}
\caption{Number of early hits. \emph{Solid}: Experimental data after first level filtering and cleaning. \emph{Dashed}: Experimental data after second level filtering with an exponential fit. \emph{Dotted}: Simulated signal after second level filtering ($\beta=10^{-2}$, $\lambda=2$\,cm). }
\label{exp_tail}
\end{figure}

About 80\% of signal events would be expected to have more than 20 early hits (cf. Fig. \ref{exp_tail}). 
Since none were found in the filtered data, the data must be almost signal free. 
Thus, the fit parameters are suitable for background estimation. 
They were used for calculating the expected number of background events at varying cuts in the number of early hits.

We optimized the final cut following the scheme described by \cite{hillrawlins} in order 
to achieve the optimum sensitivity, which is the 90\% C.L. flux upper limit that we would obtain if no true signal were present. 

The optimal final cut for the 80\% sample requires $>27$ early hits. The resulting sensitivities, without systematic uncertainties, are given in Fig. \ref{sens}. For comparison, limits at similar particle speed are included: the MACRO limit based on nucleon catalysis from \cite{ICRC0970_MACRO} 
%, the Gyrlyanda limit from \cite{belolaptikov98} 
and the IMB limit from \cite{becker94}. Limits at lower velocities have been presented by Baikal and Kamiokande \cite{belolaptikov97,kajita85}.

\begin{figure}
\begin{center}
\includegraphics[width=0.4\textwidth]{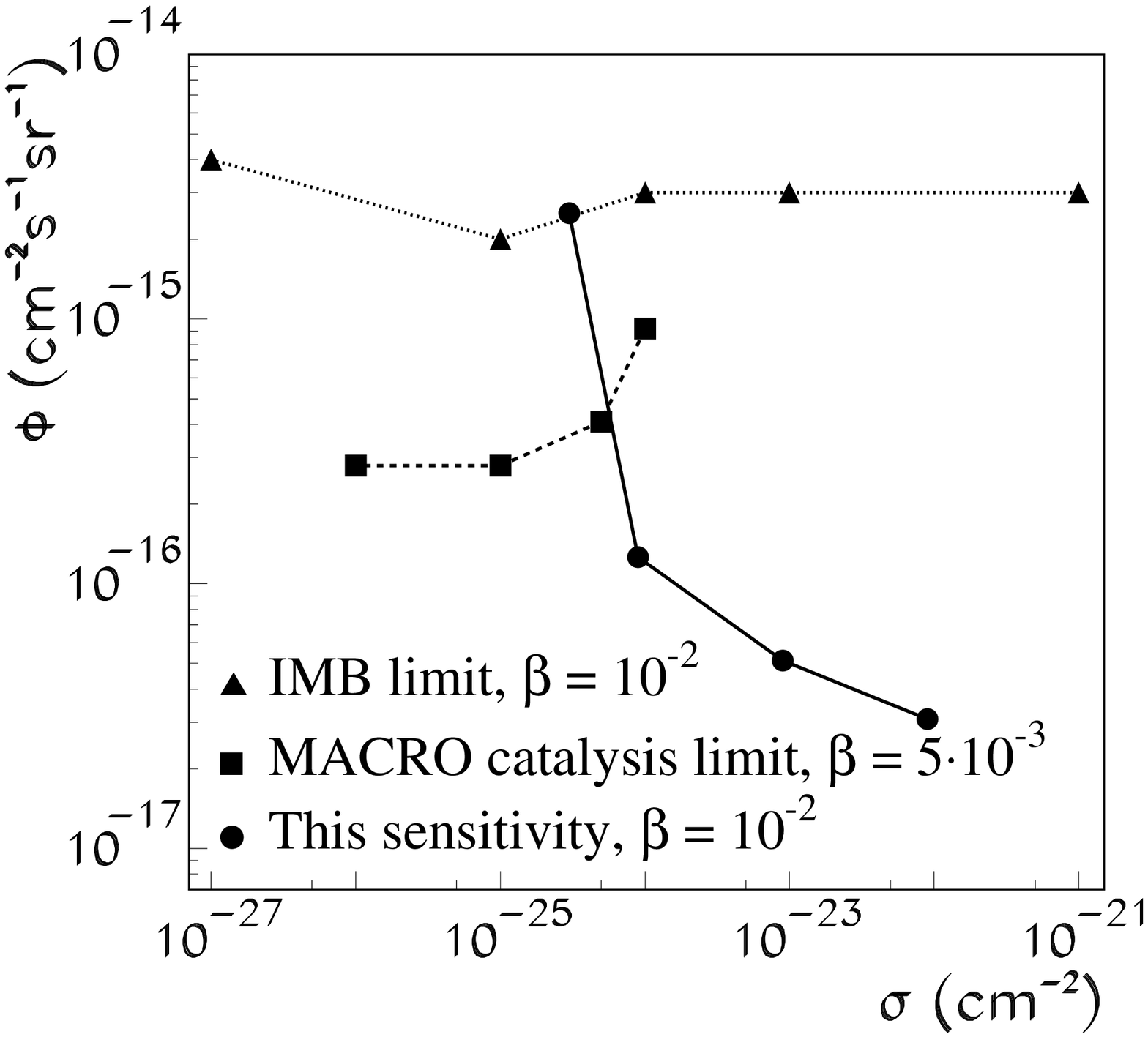}
\end{center}
\caption{Flux limits (90\% C.L.) and preliminary sensitivity (expected flux limit) at varying catalysis cross section. }
\label{sens}
\end{figure}

\section{Discussion and Outlook}
The AMANDA neutrino telescope is an excellent instrument to search for several postulated super heavy exotic particles. In this document, we present first studies of the sensitivity of AMANDA to sub-relativistic particles.
The given sensitivities are still  preliminary. Specifically, systematic uncertainties are not yet included. So far, we have used relatively small sub-sets of the available AMANDA data in order to outline our analysis strategies. The sensitivity of the analysis will improve substantially with more data.

This analysis used data from the original AMANDA data acquisition system (DAQ).
For each channel, the analog signal from the PMT is recorded using Time To Digital Converters
(TDCs) and Peak Sensing Analog to Digital Converters (ADCs). The   
original AMANDA DAQ system is
unable to precisely characterize multi photoelectron events.  In addition, the DAQ suffers from a $\sim\!1$ millisecond dead time after each triggered event while the ADCs/ TDCs are read out.
For events with slowly moving particles, this means that the DAQ system is unable to record the bulk of the signal.

Beginning in 2003, the AMANDA data acquisition system was upgraded to include full waveform readout and to reduce the detector deadtime.
Each channel is now connected to a Transient Waveform Recorder (TWR), a flash ADC that samples at 100 MHz with 12
bit resolution.   Although the readout window for the upgraded DAQ is
shorter than for the original DAQ (10.24\,$\mu$s vs. 33\,$\mu$s), the upgraded DAQ is able to record nearly continuously.  In addition to the improved characterization of each event using the waveforms, the new DAQ allows for
a reduction in the detector trigger threshold.    Prior to 2004, AMANDA
was generally run requiring a 24 channel coincidence in a 2.5\,$\mu$s period, The upgraded DAQ can operate with a threshold of 18 optical  
modules.   Additionally, events with between 13 and 17 hits
are processed separately using a software trigger algorithm that looks for events where nearby optical modules  are hit. 
The ability to almost continuously monitor the trajectory of a slowly moving particle, 
combined with the reduced trigger threshold, will greatly improve the sensitivity of AMANDA detector to such particle events.

AMANDA is now integrated with IceCube, and will continue
to take data for several years.   The analysis of the data from the
integrated detector should give the best limits on the fluxes of slowly moving massive particles.
                                                                           
This work has been supported by the Office of Polar Programs of the National Science Foundation.

%\end{document}

\setcounter{figure}{0}
\setcounter{table}{0}
%\documentclass{article}

%The ICRC Style
%\usepackage{icrctc07}

\title{Exotic Particles Searches with IceCube }
\shorttitle{Exotics with IceCube}

\authors{Brian Christy$^{1}$, Alex Olivas$^{1}$, and David Hardtke$^{2}$ For The IceCube Collaboration$^{3}$}
\shortauthors{Brian Christy and et al}
\afiliations{$^1$Dept. of Physics, University of Maryland, College Park, MD, 20742, USA\\ 
  $^2$Dept. of Physics, University of California, Berkeley, CA, 94720, USA\\
  $^3$see special section of these proceedings\\ }
\email{bchristy@icecube.umd.edu}

\abstract{The IceCube neutrino observatory, currently under construction at
the South Pole, offers a novel environment to search for particles
beyond the Standard Model. With IceCube nearly 20\% complete it is
currently the largest operating neutrino telescope. The large instrumented
volume and clear glacial ice allows for a big improvement of the sensitivity
to many types of exotic cosmological relics. Exotic particles that IceCube is 
sensitive to include magnetic monopoles, nuclearites, and Q-balls. Estimated 
sensitivities for these particles will be presented.}

%\begin{document}
\maketitle

\section{Introduction}

In 1931, Dirac \cite{ICRC0788_ref1} quantified the charge of a magnetic monopole
by demonstrating that $g=Ne/2\alpha$, where $\alpha$ is the fine structure
constant. Forty-three years later, t'Hooft and Polyakov independently
found solutions to certain groups of Grand Unified Theories (GUTs)
that matched the charge of the Dirac Monopole \cite{ICRC0788_ref3,ICRC0788_ref2}. This
allowed estimates for the masses of magnetic monopoles to be $\sim\Lambda/\alpha$,
where $\Lambda$ is the symmetry breaking scale. This results in a
mass range from $10^{8}$ GeV to $10^{17}$ GeV for various GUT models.
A lower limit is set by choosing $\Lambda$ to be the electroweak unification 
scale, leading to a mass of $10^{4}$ GeV.
IceCube will expand the search for magnetic monopoles in two regimes.
A magnetic monopole traveling through the detector above the Cherenkov
threshold ($\beta>0.76$) will emit radiation roughly
8300 times that of the bare muon \cite{ICRC0788_ref4}.

At very large masses, monopoles may move with virial velocities ($\beta\sim10^{-3}$).  A slow-moving, super massive magnetic monopole will not emit Cherenkov radiation, but may be observed in other ways.  Rubakov proposed that supermassive magnetic monopoles will catalyze nucleon decay.  The nucleon decay products (primarily pions) will produce relativistic electrons that produce Cherenkov radiation. If the catalysis cross-section is sufficiently high, the supermassive magnetic monopole will appear as a slow moving track in the detector.

A similar signature would accompany the passage of a electrically neutral supersymmetric Q-ball though the IceCube array.  A Q-ball is a soliton produced during the decay of the proposed Affleck-Dine condensate in the early universe.  Sufficiently massive Q-balls would be absolutely stable and could account for some or all of the required dark matter in the universe.  A neutral Q-ball passing near a nucleon will absorb the baryon number and emit $\sim$1GeV of energy in the form of pions \cite{ICRC0788_ref9}.  The cross-section for this process is governed by the size of the Q-ball and can therefore be quite large \cite{ICRC0788_ref10}.

It is also possible that novel forms of nuclear matter could be absolutely stable for very large baryon number \cite{ICRC0788_ref11}.  Strangelets are a hypothetical state of nuclear matter with nearly equal up, down, and strange quark content.  If such a state is the ground state of dense nuclear matter, cosmic-ray strangelets (aka nuclearites) could be produced in neutron star collisions.  These heavy strangelets would have atomic sizes but nuclear densities.  As they pass through the South Pole ice, they would produce a thermal shock emitting black-body radiation.  This black-body radiation would register in the IceCube photomultiplier tubes and cause the strangelet to appear as a slowly moving track.

\section{Detector}

IceCube is a kilometer-scale neutrino telescope currently being built
between 1450 to 2450 meters below the Antarctic ice surface.
It is designed for up to 80 strings of 60 Digital Optical Modules
(DOMs), spaced out in a hexagonal pattern. For the data presented,
we use the configuration of IceCube as of 2006, that is a total of 540 DOMs
in 9 strings. The instrumented volume is $\sim$0.625 $km^{3}$, compared to 
the AMANDA instrumented volume of $\sim$0.016 $km^{3}$
The DOM is the cornerstone of the detector \cite{ICRC0788_ref5}. It is configured
to detect photon signals via a Hamamatsu 10 inch Photomultiplier Tube(PMT).
Onboard electronics contain two waveform digitizers, a fast Analog to
Digital Converter (FADC) and an Analog Transient Waveform Digitizer
(ATWD). The FADC has a nominal sampling rate of 25 ns/sample and can read up to
256 samples of the incoming waveform produced by the PMT. The ATWD
digitizes the wavefrom across 3 channels representing different gain
values. It runs with the nominal sampling rate of 3.3 ns/sample and can read up
to 128 samples.
In 2006, the number of samples was limited to reduce bandwidth.
The highest gain ATWD channel was set to keep all 128 samples, while the two 
lower gains were only set to record the first 32 samples. Meanwhile, the FADC 
only kept the first 50 samples for a time window of 1.25$\mu$s.
Since monopole events are extremely bright, their waveforms largely saturate 
the highest gain and hence information from the ATWD beyond 100 ns is greatly 
reduced. Though the FADC saturates before any ATWD channel, the longer time 
scale provides greater distinction between the signal and background.  
Hence, this study uses data provided by the FADC.

\section{Signal and Background Simulations}

\subsection{Relativistic Magnetic Monopoles}

The simulation of relativistic magnetic monopoles is done in three stages.

Magnetic monopoles are generated uniformly on a disk located 600m from the
center of the detector pointing back towards it at various orientations.
For this study, 10,000 monopole events were generated at binned angles
theta and phi of 45 degrees, for a total of 260,000 events per dataset.
A dataset was generated for four different speeds, $\beta$= 0.99,
0.9, 0.8, and 0.76. 

Energy loss of the magnetic monopoles as they pass through the ice
is modeled using the Bethe-Bloch formula as adapted by Ahlen \cite{ICRC0788_ref6}.
Future plans are to extend this to include delta electrons, which
will add to the overall light deposition in the detector.

The light output and propagation is modeled by a version of 
\texttt{PHOTONICS} \cite{ICRC0788_ref8} specifically generated to work with cone 
angles associated with the different speeds simulated. The light amplitude 
is scaled up using the formula of Tompkins \cite{ICRC0788_ref4}.

\subsection{Background}

For this study, a 20\% sample of the data for 2006 is used as the
background. This sample consists of every fifth data event that passed
the online high energy filter, in place to reduce the data rate 
over the satellite. The filter is set to accept events with the number
of hit DOMs greater than 80.  This filter reduced the number of triggered 
data events from $\sim3.5*10^{8}$ to $\sim3*10^{5}$.

\section{Estimated Sensitvity to Relativistic Monopoles}

The brightness of the magnetic monopole is the primary distinguishing
feature. Therefore, we use parameters associated with the light yield
in the first level of cuts. The two chosen are the number of hit DOMs
(NDOM) and the total integrated FADC waveform (FCHARGE). The event
rates are normalized to the expected rate for the 137.4 days of live
time recorded by IceCube in 2006. For the monopole signal, a flux
of $5*10^{-17}cm^{-2}s^{-1}str^{-1}$ is used, representing the lowest
limit set by BAIKAL \cite{ICRC0788_ref7}. To get a conservative estimate on
the sensitivity of the detector, a tight cut is made to eliminate
all the 20\% data sample. Figures 1 and 2 plot the signal and data
for FADC vs NDOM at the largest and smallest values of $\beta$ studied.
The following linear cut is chosen: \[
(FADC>10^{6}+7500*(NDOM-125))\]
 OR \[
(FADC>3*10^{6})\]
Table 1 shows the effective area of the signal resulting from
this cut for each of the four monopole speeds. Assuming no events
are seen, the flux sensitivity is calculated for the 90\% C.L.

\begin{figure}
\begin{centering}\includegraphics[width=0.48\textwidth]{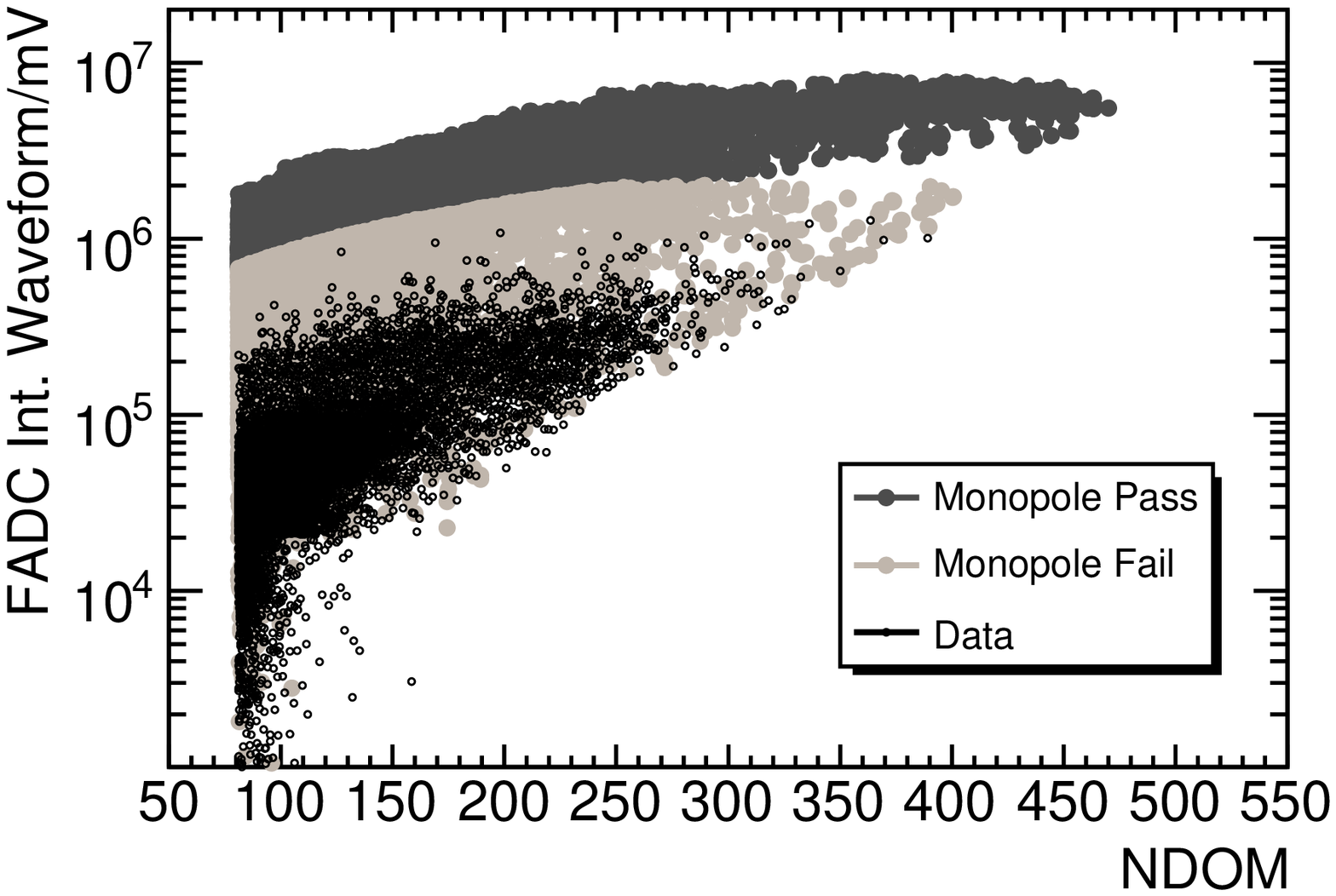} \par\end{centering}
\caption{The effect of applying the linear cut to the integrated charge versus the number of hit DOMs distribution.  Shown are the monopole signal simulation for $\beta=0.99$ and data.  The dark grey dots are signal events that pass the cut while the light grey signal dots and data (black) are rejected.}
\label{ICRC0788_fig1} 
\end{figure}

\begin{figure}
\begin{centering}\includegraphics[width=0.48\textwidth]{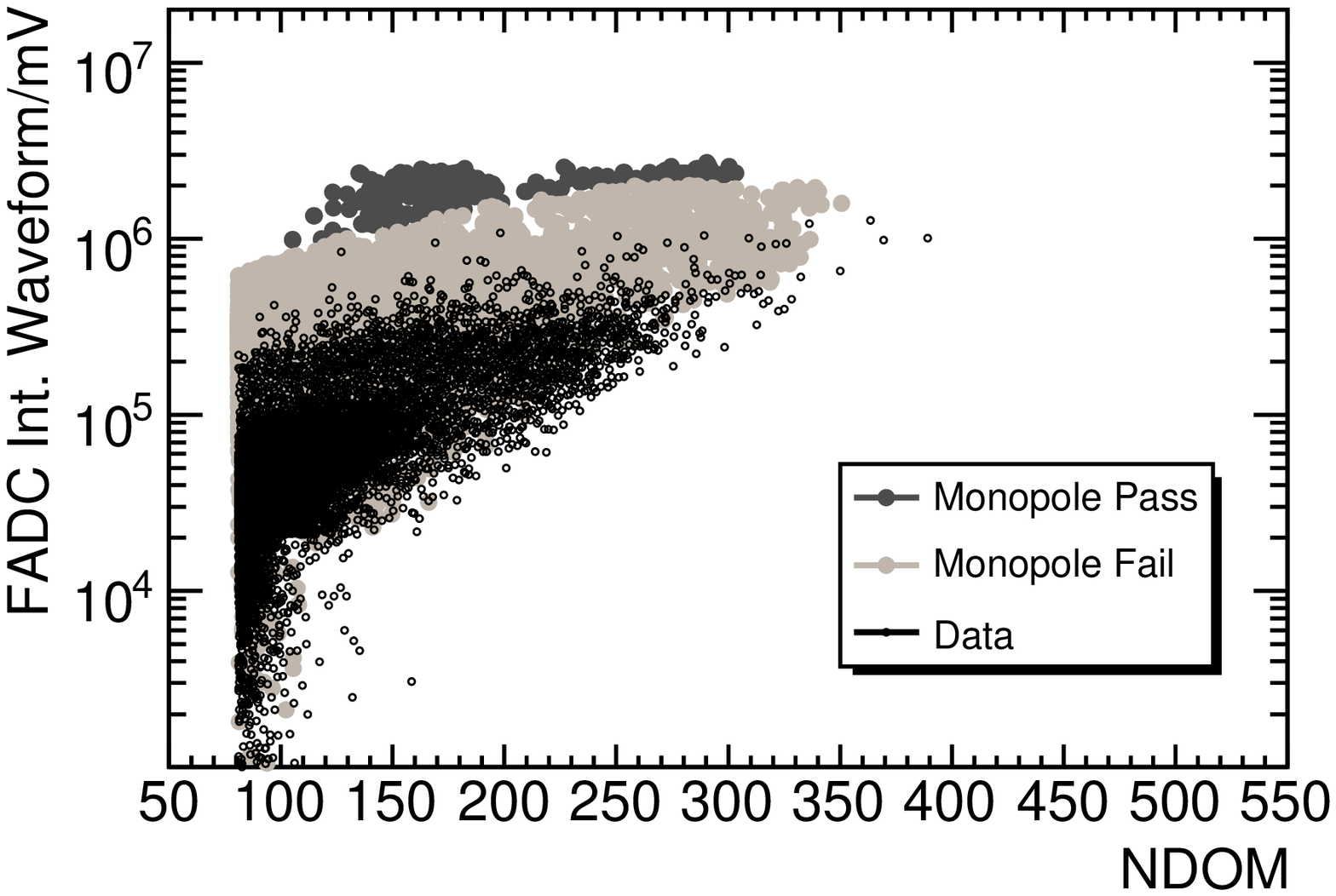} \par\end{centering}
\caption{The effect of applying the linear cut to the integrated charge versus the number of hit DOMs distribution.  Shown are the monopole signal simulation for $\beta=0.76$ and data.  The dark grey dots are signal events that pass the cut while the light grey signal dots and data (black) are rejected.}
\label{ICRC0788_fig2} 
\end{figure}
\begin{table}[t]
\begin{centering}\begin{tabular}{l|ccc}
\hline 
$\beta$ &
$A_{eff}(km^{2})$ &
Exp Signal &
$\Phi_{90}$ \tabularnewline
\hline 
0.99 &
0.3 &
19.05 &
$7*10^{-18}$ \tabularnewline
0.9 &
0.26 &
16.34 &
$7*10^{-18}$ \tabularnewline
0.8 &
0.08 &
4.92 &
$3*10^{-17}$ \tabularnewline
0.76 &
$10^{-3}$ &
0.09 &
$2*10^{-16}$ \tabularnewline
\hline
\end{tabular}\par\end{centering}
\caption{Passing rates for linear cut. Expected signal and sensitivity
for a full year of data.}
\label{Table1} 
\end{table}

\section{Estimated Sensitvity to Subrelativistic Particles}

Slowly moving particles that traverse IceCube will appear as a connected series of small electromagnetic showers.  The defining characteristic of the events is the length of time that photons remain in the detector.  For a typical downgoing muon event, the mean event length is $\sim$1-2 $\mu$s, whereas a slowly moving particle will last hundreds of microseconds or even milliseconds.  IceCube DOMs run as autonomous data collection devices and events are selected using a software trigger based on the individual DOM data.  This makes IceCube very sensitive to slowly moving particle events.  As long as the light output remains sufficient, the trigger will continue to add the DOM data to the triggered event.  Currently, the IceCube sensitivity to Q-balls, Rubakov monopoles, and supermassive strangelets is limited by the high trigger threshold (8 DOMs in 5$\mu$s).  Investigations are underway, however, of topological and tracking algorithms in the IceCube trigger system.  Such a trigger will improve the sensitivity to slowly moving particles that produce less light.

Figure \ref{ICRC0788_fig3} shows the expected flux sensitivity to slowly moving massive particles ($\beta\sim10^{-3}$) for the 2007 IceCube configuration (1320 DOMs in 22 strings) and the eventual full IceCube array.  With the full IceCube array, we expect sensitivities more than two orders of magnitude better than the current experimental limits.  

\begin{figure}
\begin{center}
\includegraphics [width=0.48\textwidth]{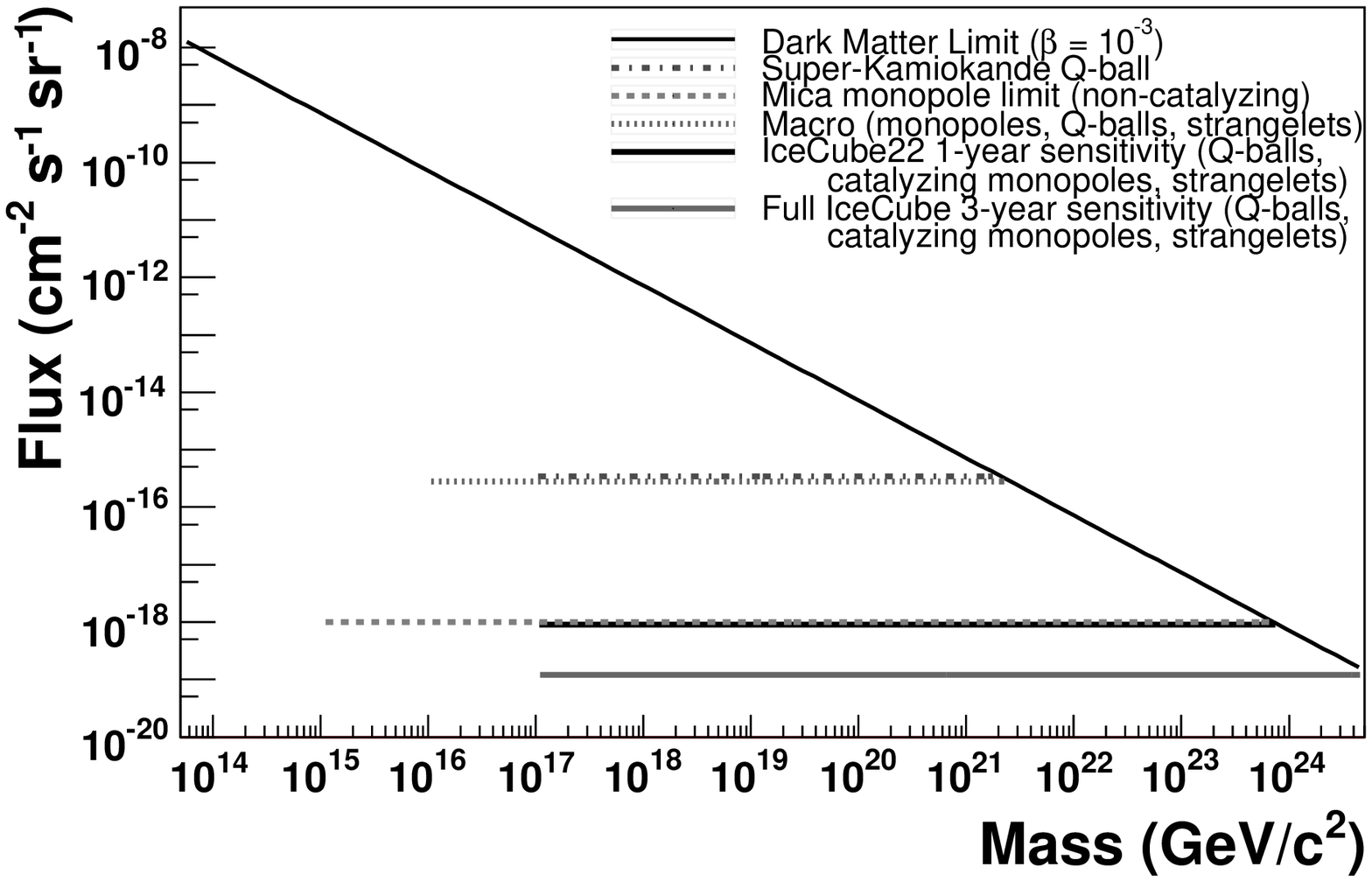}
\end{center}
\caption{Current Limits \cite{ICRC0788_ref12,ICRC0788_ref13,ICRC0788_ref14} and Projected Sensitivities for Slowly Moving Massive Particles that may be seen by IceCube}\label{ICRC0788_fig3}
\end{figure}

\section{Outlook and Conclusion}

Each year, IceCube's capability to search for exotic particles
will increase dramatically. With the 9 string detector alone, competitive
limits on the flux of relativistic magnetic monopoles are achievable.
However, these results are preliminary and will be refined. Background simulation will
start with cosmic ray air showers produced by \texttt{CORSIKA}. Since only
the high energy events are considered, weighting methods will be used.
The asymmetry of the detector will require further analysis of the
signatures produced at different angles. Finally, a log likelihood 
or neural network analysis may be employed to refine and optimize the cuts. 
With the additional analysis for slow moving exotics, IceCube will become a
valuable tool in the search for these particles.

%\bibliography{ICRC0788/icrc0788}
%\bibliographystyle{plain} 
%\end{document}

%
% Others 
%
%icrc1231.pdf (IceCube improve DAQ)
%ph_icrc07:su (Phtoniics)
%wf_draft.pdf (waveform improvements)
%aura.ICRC.V3
%
\setcounter{figure}{0}
\setcounter{table}{0}
%%
% International Cosmic Ray Conference 2007 Merida Yucatan Mexico
% In this file you will find detailed instructions to correctly
% typeset your document.
%
% By: Victor De la Luz
% vdelaluz@inaoep.mx
% Mexico City

%Class Required
%\documentclass{article}
%The ICRC Style
%(This package is the last package in the usepackage list)
%If you need import other package you need write it first.
%\usepackage{icrctc07}

%The paper title
\title{Effect of the improved data acquisition system of IceCube on its neutrino-detection capabilities}
%Short title to print in the headers to the final publication (Not showed in this print).
\shorttitle{new approach to neutrino detection with IceCube}

%All paper authors
\authors{Dmitry Chirkin$^{1}$ for the IceCube Collaboration$^{2}$ }
%Short title to print in the headers to the final publication (Not shown in this print).
\shortauthors{Chirkin and et al.}
%All the affiliations.
\afiliations{$^1$Lawrence Berkeley National Laboratory, Berkeley, CA, U.S.A.\\
$^2$see special section of these proceedings }
\email{dchirkin@lbl.gov}

%The abstract.
\abstract{The IceCube data acquisition system is capable of recording information about all photons registered by its photomultiplier tubes for up to 13 microseconds for each sensor with high precision. A time resolution of 3 ns and charge resolution of 30\% of all one photoelecton pulses within each sensor's event record is achieved. The improvement in quality of the data reconstruction due to the improved design of the experiment is estimated and its effect on the IceCube capabilities as a neutrino detector is discussed. }

%%%%%%%%%%%%%%%%%%%% B E G I N   D O C U M E N T%%%%%%%%%%%%%%%%%%%%%%%
%\begin{document}
\maketitle
%Begin the section.
\section{Introduction}

The ability of IceCube optical sensors to record information about all photon registered by their PMTs has not yet been fully utilized in the data analysis (see, e.g., \cite{ICRC1231_pretz}). While the much improved timing and energy resolution are being used to improve upon the energy resolution of the detected muon events \cite{juande}, this contribution attempts to demonstrate the improvement in muon neutrino analysis due to the ability to separately detect individual photoelectrons with their respective times and charges (shown in Figure \ref{ICRC1231_fig51}).

The goal of selecting muon neutrinos in the presence of a $10^6$ times higher background of atmospheric muons is to maximize the signal yield at a low background level, while achieving the best possible resolution with least mis-reconstruction of signal events.

In this preliminary study we present the analysis of one month of data collected by a 9-string IceCube detector configuration in year 2006. Data reconstruction algorithms using only the first photon per sensor were compared with those incorporating the full multi-photon information. The angular resolution achieved in both cases is very similar; however, the number of badly mis-reconstructed signal events is lower for multi-photon reconstruction. Using the additional information available from all recorded photons leads to the correspondingly improved separation of signal and background and allows one to  to achieve the required background reduction while retaining a higher signal yield.

\begin{figure}[h]
\begin{center}
\noindent
%\fbox{\hbox{\vbox{\hsize=50mm \hfill \vspace{50mm}}}}
%uncomment next line to include real image
\includegraphics [width=0.45\textwidth]{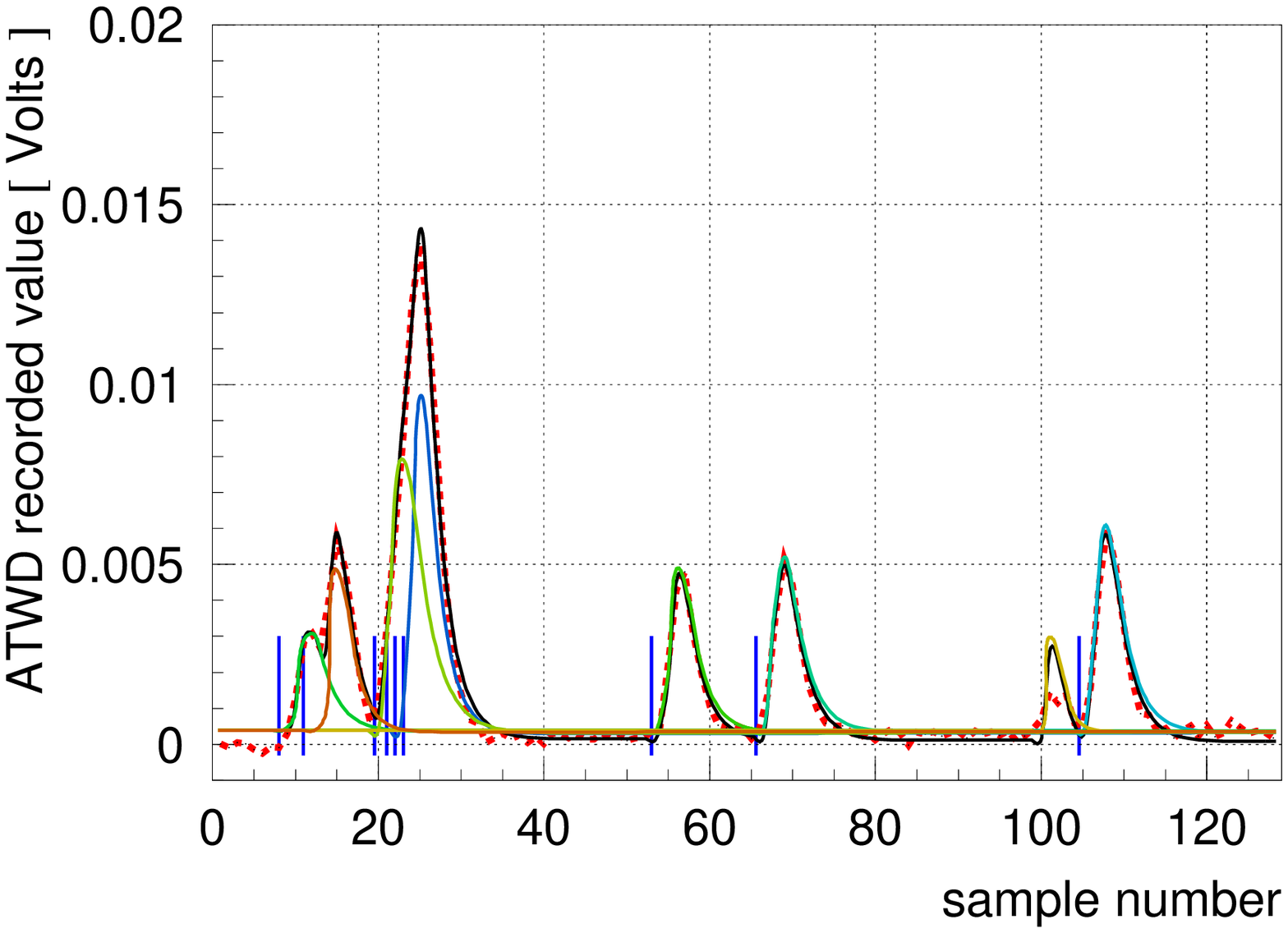}
\end{center}
\caption{A typical PMT signal trace recorded by the faster digitizer of an IceCube optical sensor. The trace contains 128 samples, 3.3 ns.\ per sample. Results of 2 different photon deconvolution methods shown agree well. Blue vertical lines denote the hit times of the first method. The black fit line with colored lines denote deconvolved pulses of the second method. The data is shown with a red dashed line.}\label{ICRC1231_fig51}
\end{figure}

A new method of combining cuts to optimize background reduction is presented. First, a robust definition of angular resolution of reconstructed muon direction in simulated data is introduced. The cuts are optimized to maximize the angular resolution of the remaining events, and then are tightened to remove the background of misreconstructed events.

\subsection{Angular resolution and cut optimization}

The precision of the track reconstruction methods is determined from the deviation of the reconstructed result from the true track direction from the simulation (typical distribution shown in Figure \ref{ICRC1231_fig52}). It was not possible to describe all such distributions at different reconstruction quality levels with a single shape depending only on the distribution width. Therefore the following very general definition was introduced instead: the {\it angular resolution} $\alpha$ of a given simulated data sample is chosen so that 2/3 of the data have reconstructed result deviate from the true track direction by less than the resolution, and 1/3 by more. This simple definition allows one to calculate the angular resolution $\alpha$ easily for all data quality levels, providing a good measure of the effectiveness of the quality cuts.

\begin{figure}
\begin{center}
\noindent
%\fbox{\hbox{\vbox{\hsize=50mm \hfill \vspace{50mm}}}}
%uncomment next line to include real image
\includegraphics [width=0.45\textwidth]{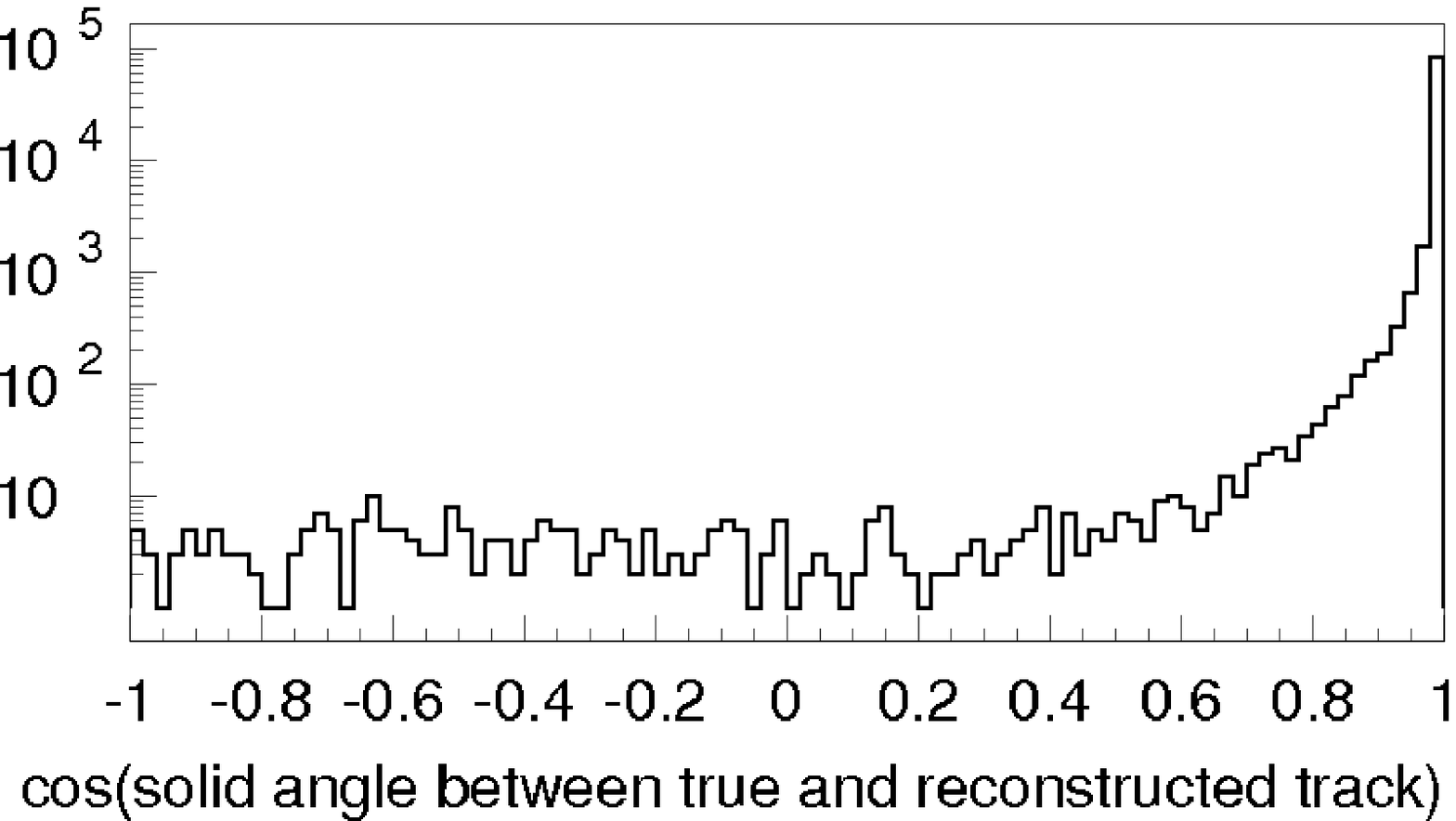}
\end{center}
\caption{ Distribution of the deviation of the reconstructed from the true direction for the studied simulated data sample shown after some cuts.}\label{ICRC1231_fig52}
\end{figure}

The cut parameters were chosen to have the following property: as the value of the cut on the parameter is lowered (i.e., the cut becomes stronger), the angular resolution $\alpha$, as defined above, improves.

Several reconstructions were performed in succession. These differed by the ice description used in the calculation of the photon scattering probabilities, by whether the muon energy was allowed to vary during the reconstruction, and whether all recorded photons or only first recorded photons were used.

For each reconstruction several quantities have the ``cut property'' defined above: minus reduced log likelihood of the reconstructed result, closest approach distance from the reconstructed track to the center of gravity of hits, relative uncertainty and variation of the energy measure, and uncertainty in the zenith and azimuth angles (defined as the range in the parameter in which the log likelihood stays above its maximum minus 0.5). Additionally the differences in the direction of different reconstruction results were formed. One more parameter appeared necessary: 1 over the total length of the track defined as the distance between two farthest from each other projections of hits on the reconstructed track. Parameters with similar distributions were grouped together, resulting in 7 groups. In each group the maximum value of the parameters in the group was chosen as the parameter of the group.

Cuts $c_i$, $i=1,...,7$ were applied to the parameter groups defined above in such a way as to maximize the angular resolution $\alpha$ for each given fraction of events $r$ left after the cuts. The fastest decent approach was chosen to optimize the cuts: starting with a full dataset, at each step reducing the fraction of the events left by the amount $\delta r$ the cuts were adjusted by the amount proportional to $\partial \alpha / \partial c_i$.

Since the relative and overall cut strength depends on the number of degrees of freedom available during the reconstruction, cuts were optimized individually for event groups with different number of sensors with signal (here called channels) $N_{ch}$ from the simulated dataset. This resulted in a set of cuts, one representation of which, describing achievable efficiencies (fractions of events left, $r_n$) for given $\alpha$ and $N_{ch}$, here called {\it efficiency matrix}, is shown in Figure \ref{fig53}. In order to determine the cut sets needed to achieve a certain angular resolution $\alpha$ the efficiency matrix is consulted to determine the fractions of events $r_n$ with given $n=N_{ch}$. The set of cuts strong enough to leave only a fraction $r_n$ of events that were used in the efficiency matrix evaluation are then the cuts that reduce the data to a set with the desired angular resolution $\alpha$.

\begin{figure}
\begin{center}
\noindent
%\fbox{\hbox{\vbox{\hsize=50mm \hfill \vspace{50mm}}}}
%uncomment next line to include real image
\includegraphics [width=0.45\textwidth]{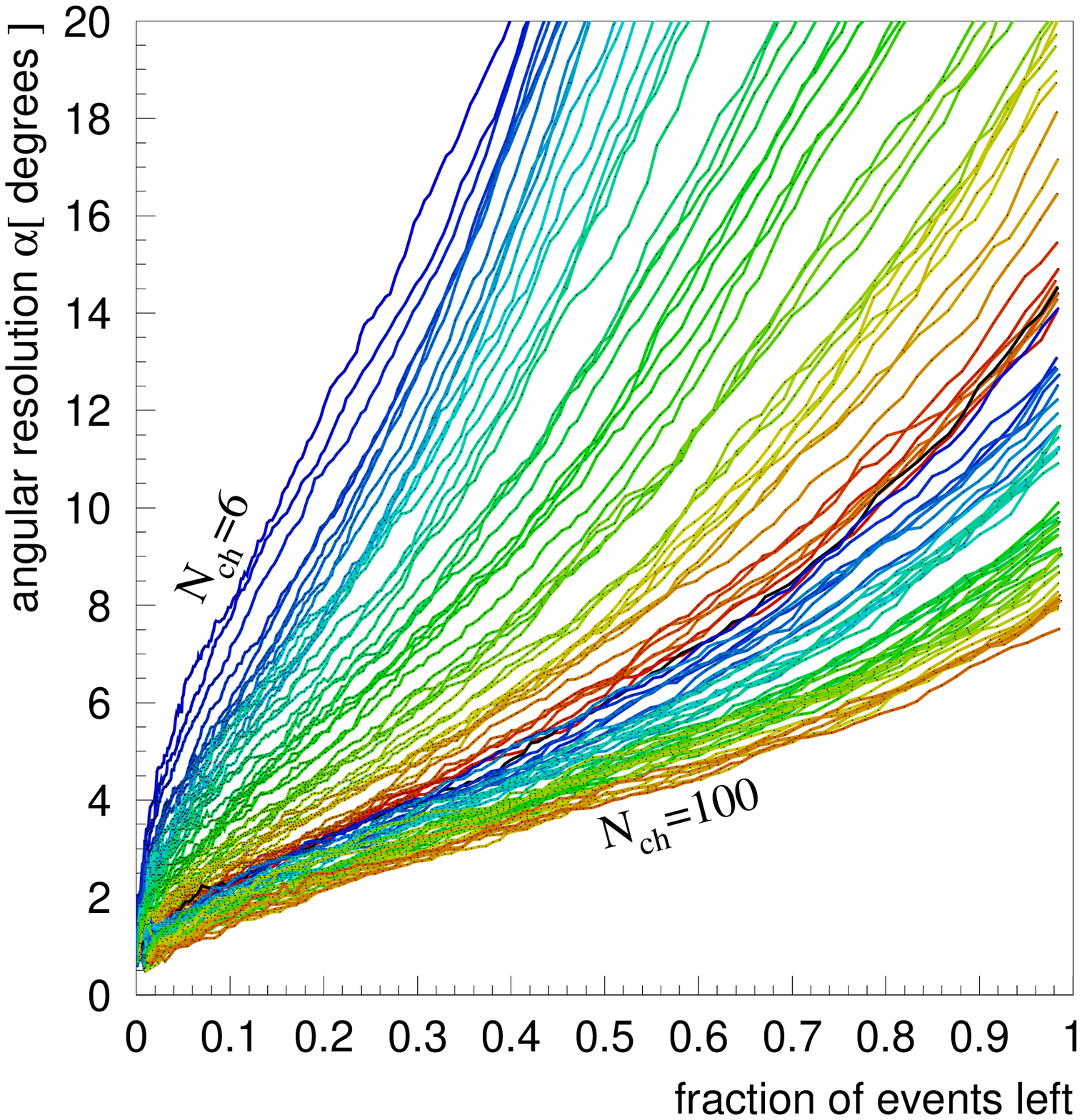}
\end{center}
\caption{Efficiency matrix shows for each $N_{ch}$ the best achievable angular resolution $\alpha$ at each given fraction of events left after applying cuts (this definition is equivalent to that given in the text).}\label{fig53}
\end{figure}

To study the improvement in data analysis due to the availability of information about multiple pulses from each sensor the parameters corresponding to the multi-photon reconstructions were removed from the cut groups defined above. The resulting efficiency matrix looks nearly identical to the one shown in Figure \ref{fig53} except that points on the lines correspond to somewhat more constrained cut values as compared to the multi-photon-enabled efficiency matrix. Therefore the first-photon-only cuts are just as effective as the complete cut set in improving the angular resolution $\alpha$ for a given data reduction fraction. This, however, is to be expected for a self-sufficient cut set, meaning that more cuts do not improve the angular resolution for a given fraction with the used angular resolution definition.

Nevertheless, as shown in the following section, at the final neutrino selection cut levels there is a substantial improvement in both the angular resolution $\alpha$ of the final sample and the fraction of events retained, indicating that the outliers of the angular distribution are reduced in the analysis employing the full cut set, showing the clear advantage of the method utilizing all recorded pulses.

\begin{figure}
\begin{center}
\noindent
%\fbox{\hbox{\vbox{\hsize=50mm \hfill \vspace{50mm}}}}
%uncomment next line to include real image
\includegraphics [width=0.45\textwidth]{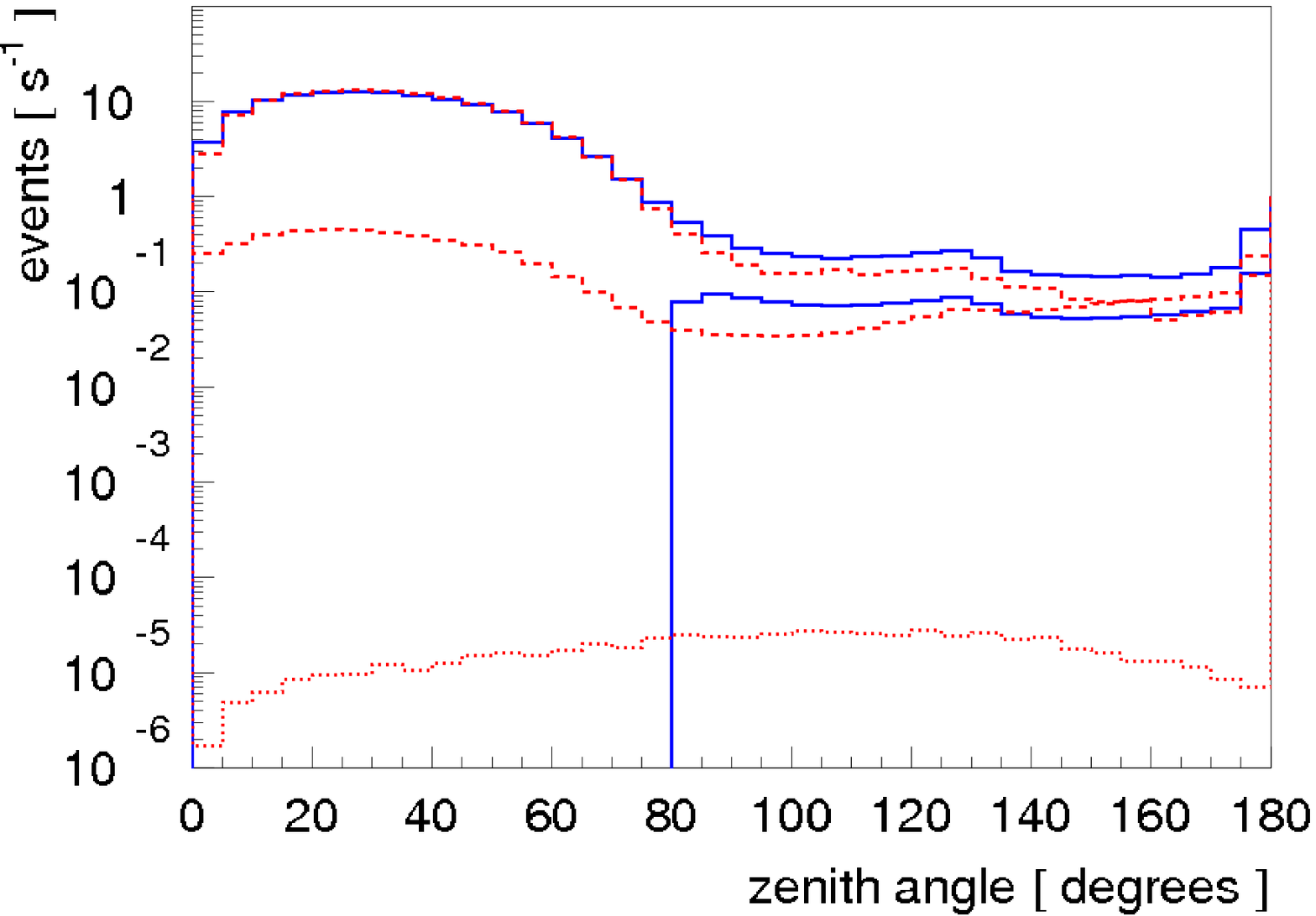}
\end{center}
\caption{Initial zenith angle distribution (no cuts): red dashed lines: upper: downgoing muon background, lower: coincident downgoing shower background; red dotted line: muons from atmospheric muon neutrinos; upper blue line: reconstructed data; lower blue line: reconstructed data with cut of zenith angle above 80 degrees applied to all reconstructions.}\label{fig54}
\end{figure}

\subsection{Atmospheric neutrino search}
Figure \ref{fig54} shows the zenith angle distribution of reconstructed tracks in real and simulated data. The data remaining after the cuts on the zenith angle for all reconstructions are applied contains mostly poorly reconstructed downgoing background events that fall into the tail of events reconstructed with wrong direction shown in Figure \ref{ICRC1231_fig52}. The data shown in Figure \ref{ICRC1231_fig52} is at the cut level corresponding to an angular resolution $\alpha$ of 4 degrees; misreconstructed events are suppressed by more than 4 orders of magnitude at this cut level. Without any cuts the level of misreconstructed events is higher, about 2 orders of magnitude below the peak, matching the level of misreconstructed events in Figure \ref{fig54}. By applying successively stronger cuts corresponding to lower values of angular resolution $\alpha$ the background of misreconstructed events can be reduced until most of the events reconstructed as upgoing are, indeed, upgoing.

\begin{figure}
\begin{center}
\noindent
%\fbox{\hbox{\vbox{\hsize=50mm \hfill \vspace{50mm}}}}
%uncomment next line to include real image
\includegraphics [width=0.45\textwidth]{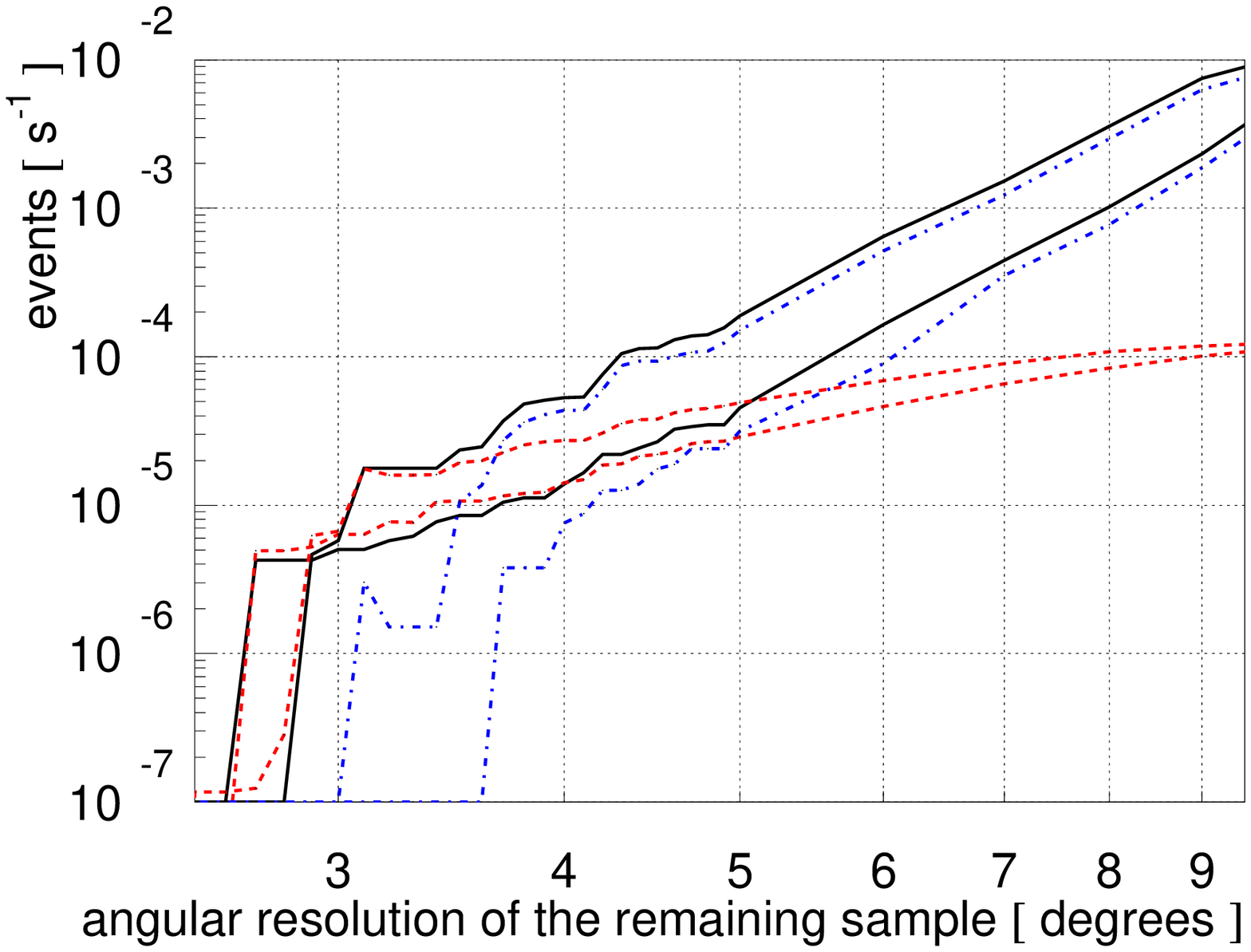}
\end{center}
\caption{Events remaining at different cut levels corresponding to requested values of angular resolution $\alpha$. Black solid: data; red dashed: muon neutrino simulation; blue dashed-dotted: background simulation. Upper curves are for full cut set; lower curves are for first-photon-only cut set. }\label{fig55}
\end{figure}

To determine the angular resolution $\alpha$ required to suppress the background of misreconstructed events below the signal of upgoing events successively stronger cuts are applied. At each cut level the data left after the cuts is compared to simulation of both background and signal, as shown in Figure \ref{fig55}. The cut level required to achieve the desired signal purity can thus be selected. The 50\% purity is achieved at the intersection points of simulated background and neutrino lines in Figure \ref{fig55}: at angular resolution $\alpha$=3.7 with 96 events left for the full cut set, and at angular resolution $\alpha$=4.9 degrees with 90 events left for the first-photon-only data set. At the same signal purity level the angular resolution $\alpha$ of neutrino events in the remaining sample is 30\% better for the full set.

It is more difficult to estimate the purity and number of events left as the cuts are tightened more, due to the limited amount of simulated data at the time this paper was written. However, following the lines of Figure \ref{fig55}, one could estimate the angular resolution $\alpha$ and number of events left at $\sim$ 90\% purity level of 3.4 degrees and 46 events (shown in Figure \ref{fig56}) for the full cut set, and 3.6 degrees and 22 events for the first-photon-only cut set. This indicates that at the highest signal purity levels the number of neutrino events is more than doubled when incorporating the full information about all pulses into the analysis.

\begin{figure}
\begin{center}
\noindent
%\fbox{\hbox{\vbox{\hsize=50mm \hfill \vspace{50mm}}}}
%uncomment next line to include real image
\includegraphics [width=0.45\textwidth]{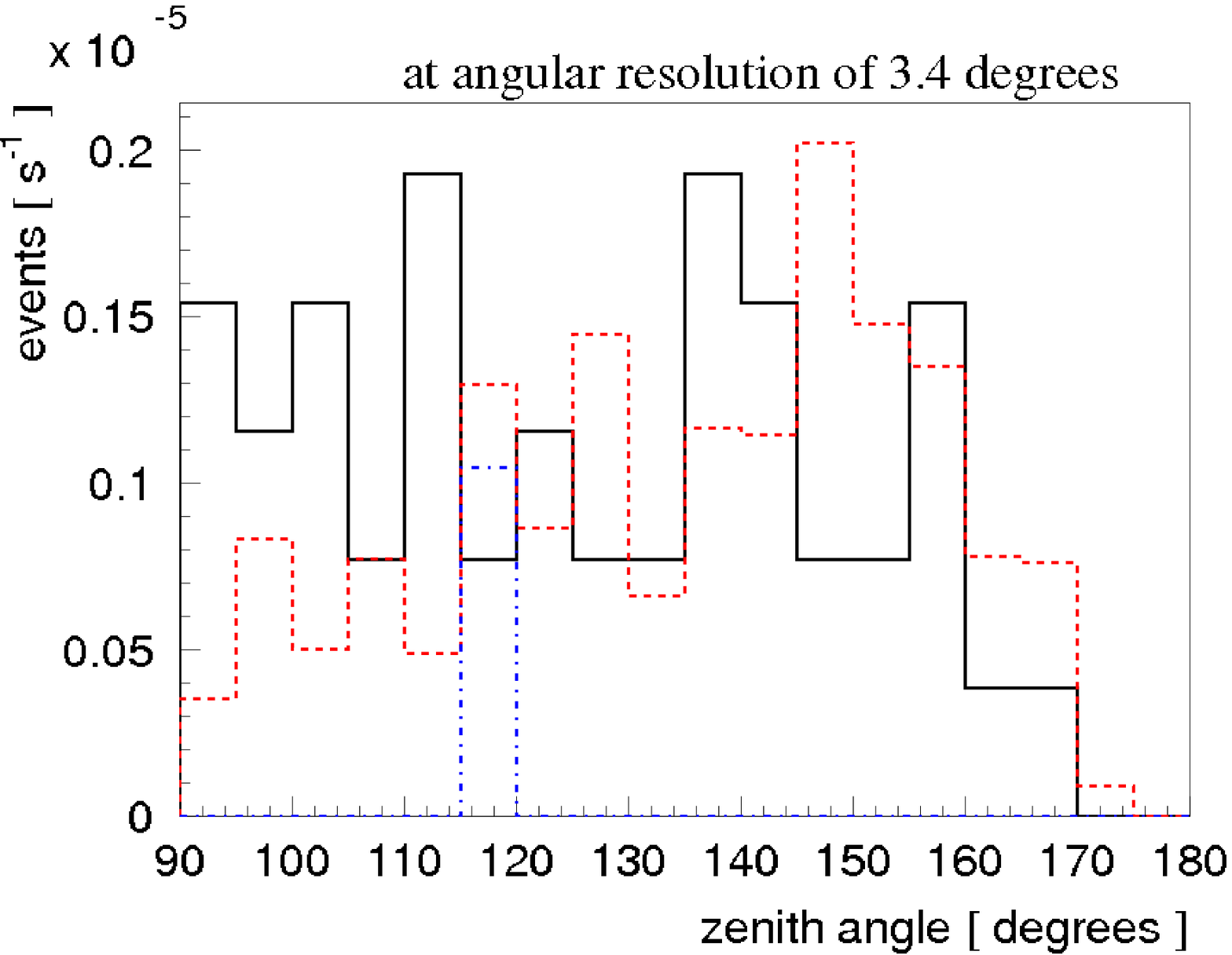}
\end{center}
\caption{Zenith angle distribution at final cut level. Black solid line: data; red dashed: muon neutrino simulation; blue dashed-dotted: one event remaining from the background simulation.}\label{fig56}
\end{figure}

\subsection{Conclusions}

A new approach to background rejection in IceCube is taken: instead of optimizing cuts to maximize signal over background, cuts are first optimized to maximize the angular resolution $\alpha$ of single muon tracks while retaining as many of the events as possible. Then the cuts corresponding to successively better values of angular resolution $\alpha$ are applied until the desired signal purity is achieved.

This approach allowed us to study the effect of including the complete information on all pulses recorded by the optical sensors of the detector. The number of signal events retained at the highest purity levels doubled (important for diffuse analysis), while the angular resolution $\alpha$ at somewhat relaxed cuts improved by 30 \%.

We thank the U.S. National Science Foundation and Department of Energy, Office of Nuclear Physics, and the agencies listed in Ref. \cite{ICRC1231_karle}.

%This is the reference to .bib file (Without .bib!)
%\bibliography{examplelibrary}
%This in the bibtex style, is ok.
%\bibliographystyle{plain}

%\end{document}

\setcounter{figure}{0}
\setcounter{table}{0}
%%
% International Cosmic Ray Conference 2007 Merida Yucatan Mexico
% In This file you will find detailed instructions to correctly
% typeset your document.

%Class Requeried
%\documentclass{article}

%\usepackage{float}
%\usepackage[latin1]{inputenc}       %% Använd latin1 (europeisk text)
%\usepackage[english]{babel}        
%\usepackage[dvips]{graphicx}
%\usepackage{psfrag}
%\usepackage{amssymb}
%\usepackage{amsbsy} %bold math
%\usepackage{subfigure}

\providecommand{\etal}{{et\hspace{.9ex}al\@.\hspace{.95ex}}}
\newcommand{\vect}[1]{\boldsymbol{#1}} %% you need \usepackage{amsbsy}
\newcommand{\de}{\mathrm{d}}        %% derivatatecken. se även \partial
\newcommand{\px}{P{\textsc{hotonics}}}
\newcommand{\cpp}{\mbox{\textsc{c}}{\tiny\raisebox{.5ex}{++}}}
\newcommand{\iA}{\mathrm{A}}
\newcommand{\iB}{\mathrm{B}}
\newcommand{\iAB}{\mathrm{AB}}
\newcommand{\eff}{\mathrm{e}}
\newcommand{\abb}{\mathrm{a}}
\newcommand{\sct}{\mathrm{s}}
\newcommand{\ngr}{n_\mathrm{\mathrm{g}}}
\newcommand{\nph}{n_\mathrm{\mathrm{p}}}
\newcommand{\src}{\mathrm{s}}
\newcommand{\rec}{\mathrm{rec}}
\newcommand{\ppdf}{\mathrm{pdf}}
\providecommand{\chv}{Cherenkov }
\providecommand{\degr}{\mbox{$^\circ$}}
\providecommand{\usp}{\hspace*{0.2em}}
\providecommand{\amanda}{\textsc{Amanda~}}
\providecommand{\sqr}[1]{(#1)^2}

%The ICRC Style
%\usepackage{icrctc07}
\title{Improved Cherenkov light propagation methods for the IceCube neutrino telescope}
\shorttitle{}
\authors{J. Lundberg$^1$, for the IceCube collaboration$^2$ }
\shortauthors{}
\afiliations{$^1$\mbox{Division of High Energy Physics, Uppsala University,  Uppsala, SE}}
\email{$^1$johan.lundberg@tsl.uu.se, $^2$see special section of these proceedings}

%The abstract.
\abstract{ In the field of neutrino astronomy, optically transparent
media like glacial ice or deep ocean water are commonly used as detector
medium. Elementary particle interactions are studied using in situ light
detectors recording time distributions and fluxes of faint photon fields
of \chv radiation, typically generated by ultra-relativistic muons. For
simulations of such photon fields, the IceCube collaboration uses a
versatile software package, \px, which was developed to determine photon
flux and time distributions throughout a large volume with spatially
varying optical properties. Photons are propagated and time
distributions are pre-calculated as binary photon tables for fast and
transparent access from event simulation and reconstruction. This is the
main tool by which IceCube event simulations take into account how depth
and wavelength dependent variations of the optical properties of the
South Pole glacier distort the footprints of elementary particle
interactions. }

%%%%%%%%%%%%%%%%%%%% B E G I N   D O C U M E N T%%%%%%%%%%%%%%%%%%%%%%%
%\begin{document}
\maketitle

\section{Introduction}
In optical high energy neutrino astronomy, light from charged particle
interactions is observed using a large number of sensors (photomultipliers) placed in
transparent natural media like glacial ice, lake water, or deep ocean water.
Successful simulation and reconstruction of such events relies on
accurate knowledge of light propagation within the detector medium.
The typical scattering lengths in these detector media are of the order
of tens to hundreds of metres. Since this scale is comparable to the
typical sensor spacing for neutrino telescopes, scattering effects
can not be ignored, and analytical calculations do not suffice.
The problem is further complicated by the
anisotropy of the light emitted in particle interactions and the
heterogeneity of natural detector media.

The software package \px\cite{pxcode} contains routines for detailed
photon
simulations in heterogenous media like the South Pole glacier.
Photon simulation results are pre-calculated  and used in
event simulation and reconstruction through interpolation of lookup
tables for fast and accurate access
to photon signal timing and amplitude probabilities.

\section{Photon flux simulation technique}

At any location throughout the medium, the local optical properties for a given
wavelength are described by the absorption length $\lambda_{\abb}$, the
geometrical scattering length $\lambda_{\sct}$, and the scattering phase
function which is the probability density function for angular
deviations at each scatter. For ice, the Henyey-Greenstein (HG)
phase function\cite{hengreen} is used to describe the strongly forward peaked scattering. It is
completely characterized by a single parameter, the mean of the
cosine of the scattering angle, $\tau = \langle\cos\theta\rangle$. 
For
most physical media, a strong correlation between $\lambda_{\sct}$ and
$\tau$ is observed. One therefore considers the effective
scattering length, $\lambda_{\eff} \equiv \lambda_{\sct}/(1-\tau)$. 

Photons are generated according to emission spectra specific
         for the given light source (particle physics events or
         calibration light sources) and propagated
throughout the medium in accordance
with the heterogeneous propagation medium description. Each photon's
spatial and temporal path is calculated and its contribution to the
overall light field is recorded in a cellular grid throughout a user
defined portion of the simulation volume. The locations of sensors are
not fixed, but can be dynamically specified when accessing the
simulation results.

The detector efficiency as function of angle and wavelength, as well as
the effects of absorption, is accounted for by applying appropriate weights during the
photon recording.

The local photon flux is calculated in each recording cell with one of
two
independent methods. In the volume-density method, photons are
propagated in small (typically equidistant) steps between scattering
points, so that the contribution to each recording cell is related to
the number of photon steps taken in that particular cell. In the
surface-crossing method, photons are instead interrupted only at
scattering points and recording cell boundaries. The flux contribution
is then related to the number of cell boundary crossings, taking into
account the projected cell surface area of each cell
boundary crossing. The two methods typically give compatible results at a
comparable simulation speed, depending slightly on the layout of the
simulation grid and the optical parameters.

To improve the speed of the Monte Carlo simulation, importance-weighted
scattering is supported; Photons can be propagated using scattering
parameters ($\lambda_{\eff'},\tau'$), different from those of the physical
scattering situation at hand. For example, straighter
paths can be oversampled by choosing scattering angles $\theta$ from a HG phase function $f_{\tau'}$ with $\tau'$ closer to $0$, while
applying a weight of $f_\tau(\theta)/f_{\tau'}(\theta)$. \label{hgphase}

The result of the photon simulation is multidimensional binary photon
tables, containing the expectation number of photo-electrons produced at
photomultiplier tubes and the corresponding differential or cumulative
time distributions.

\section{Modeling of glacial ice and applications to neutrino astronomy}
\label{watermodels}
\label{icemodels}

\begin{figure}[htbp!!!]
\psfrag{hsp}[r]{\footnotesize \raisebox{-13pt}{horizontal position $x$ [m]}}
\psfrag{ver}[r]{\footnotesize \raisebox{4pt}{vertical position $z$ [m]}}
\psfrag{lig}[r]{\footnotesize \raisebox{-14pt}{light flux [m$^{-2}$ns$^{-1}$]}}
\psfrag{pro}[r]{\footnotesize \hspace*{1em}\raisebox{-14pt}{probability [ns$^{-1}$]}}
\psfrag{residualtime}[c]{\footnotesize {\rule{0pt}{1.3em}\rule{3em}{0pt}Residual time [ns]}}
\psfrag{timedistribution}[c]{\footnotesize \raisebox{1.3em}{\rule{4em}{0pt}Time distribution [s$^{-1}$]}}
\psfrag{e}{emitter} \psfrag{mu}{muon} \center
\psfrag{icemodelicemodeicemodel}{\footnotesize\scalebox{0.95}{ Heterogeneous model}}
\psfrag{specialspecialsicemodel}{\footnotesize\scalebox{0.95}{ Single histogram model}}
\mbox{
\subfigure[Nd:YAG laser, 532 nm in ice
]{\hspace*{-0.05\columnwidth}\includegraphics[width=1.05\columnwidth]{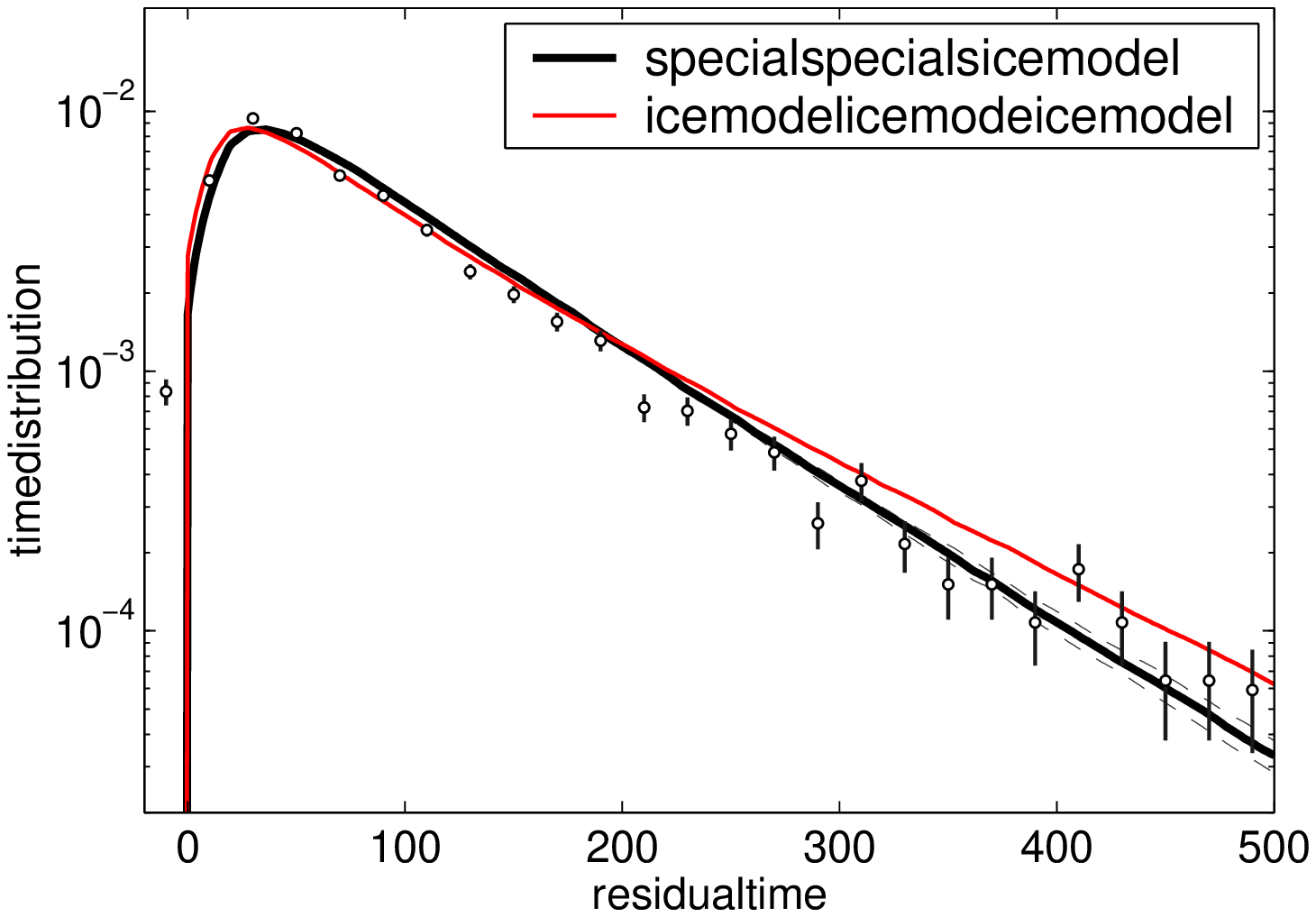}}
}
\mbox{
\subfigure[Blue LED beacons, 470 nm in ice
]{\hspace*{-0.05\columnwidth}\includegraphics[width=1.05\columnwidth]{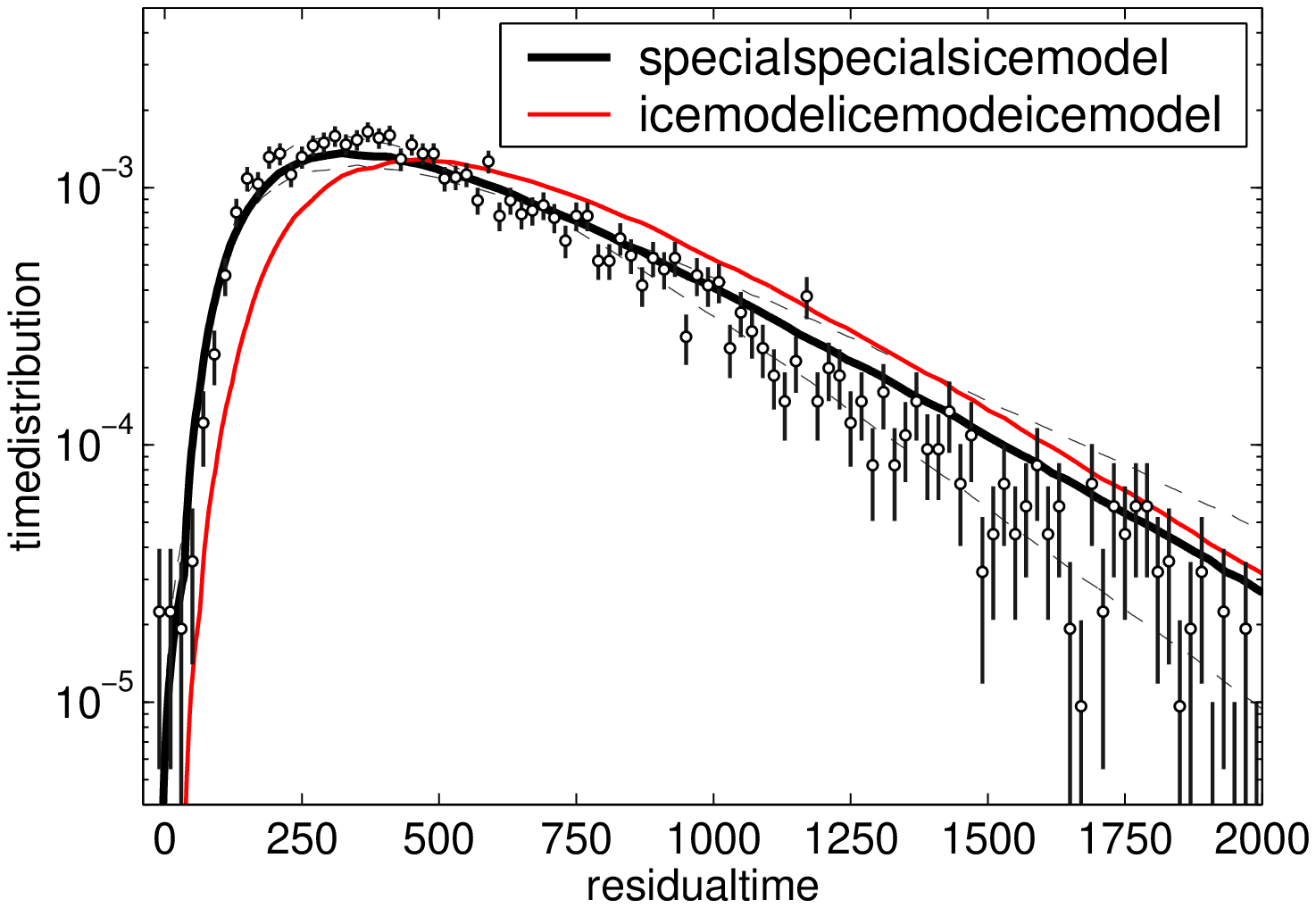}}
}
\\
%[-1.3em]
\caption{
 \label{iceplots}Residual time distributions of simulated light pulses in
 deep glacial ice. In (a), for a 532 nm Nd:YAG isotropic laser pulse, emitted at a depth
 of 1825 meter, as seen from a horizontal distance of 75 m. In (b), for an
 upward pointing 470 nm LED emitter located at a depth of 1580 m as 
seen from a horizontal distance of 140 m. 
The black dots show two time distributions of glacial ice
 surveys\cite{amandaice}, with vertical Poissonian error bars. The thick black lines show
 our results using the scattering and absorption parameters of these
 particular source--receiver combinations, and thin dashed lines
 represents the model uncertainty. 
The thin (red) lines show the
 simulation results with the heterogeneous ice model\cite{amandaice} which was
 constrained by other data.}
\end{figure}

\begin{figure*}[hp!!!]
\psfrag{hsp}[r]{\footnotesize \raisebox{-13pt}{horizontal position $x$ [m]}}
\psfrag{ver}[r]{\footnotesize \raisebox{4pt}{vertical position $z$ [m]}}
\psfrag{lig}[r]{\footnotesize \raisebox{-14pt}{light flux [m$^{-2}$ns$^{-1}$]}}
\psfrag{pro}[r]{\footnotesize \hspace*{1em}\raisebox{-14pt}{probability [ns$^{-1}$]}}
\center \mbox{ \subfigure[\mbox{$\Phi(t)$}]%
{\includegraphics[width=0.46\textwidth]{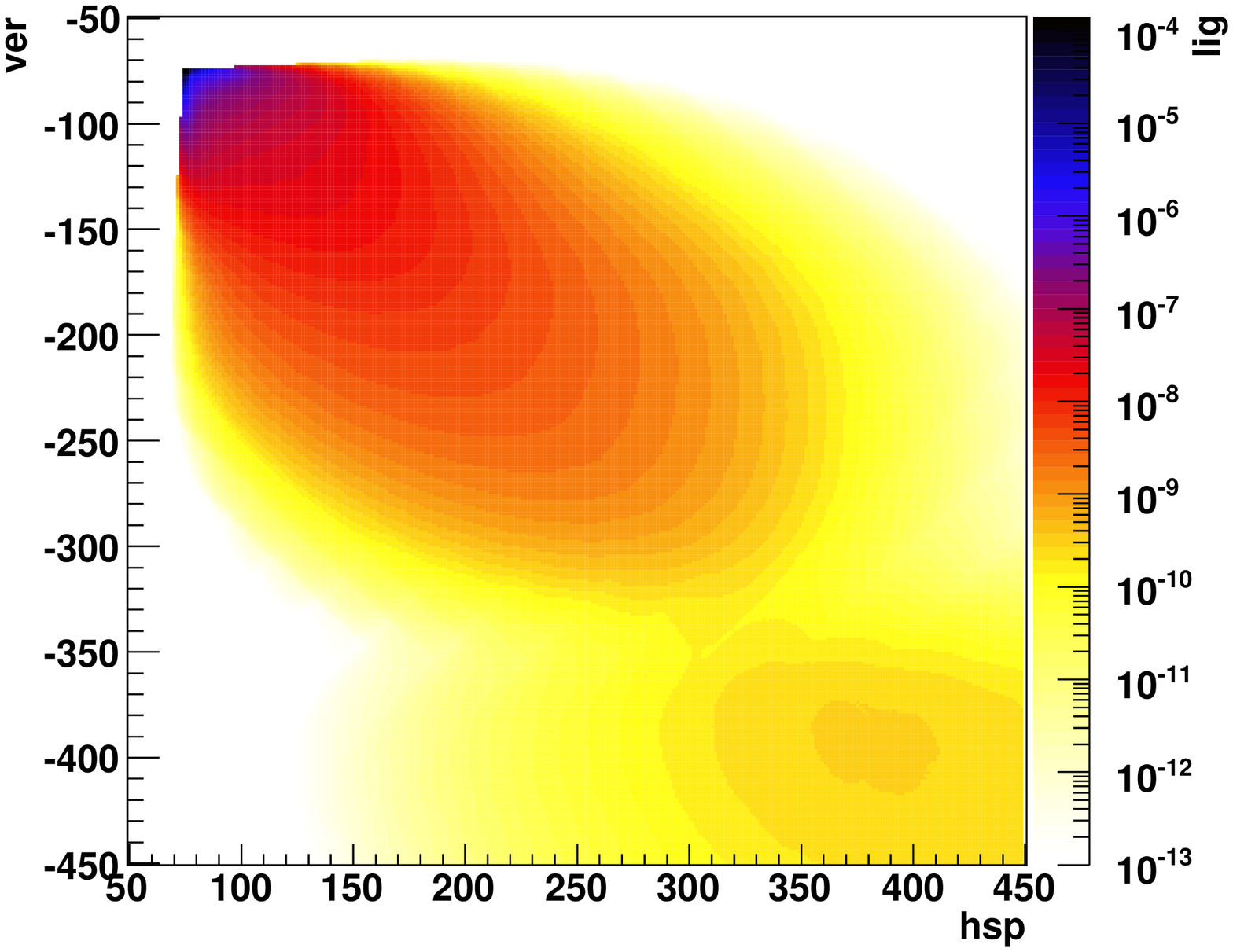}}
\hspace*{1em}\subfigure[\mbox{$f_{\ppdf}(t)$}]%
{\includegraphics[width=0.46\textwidth]{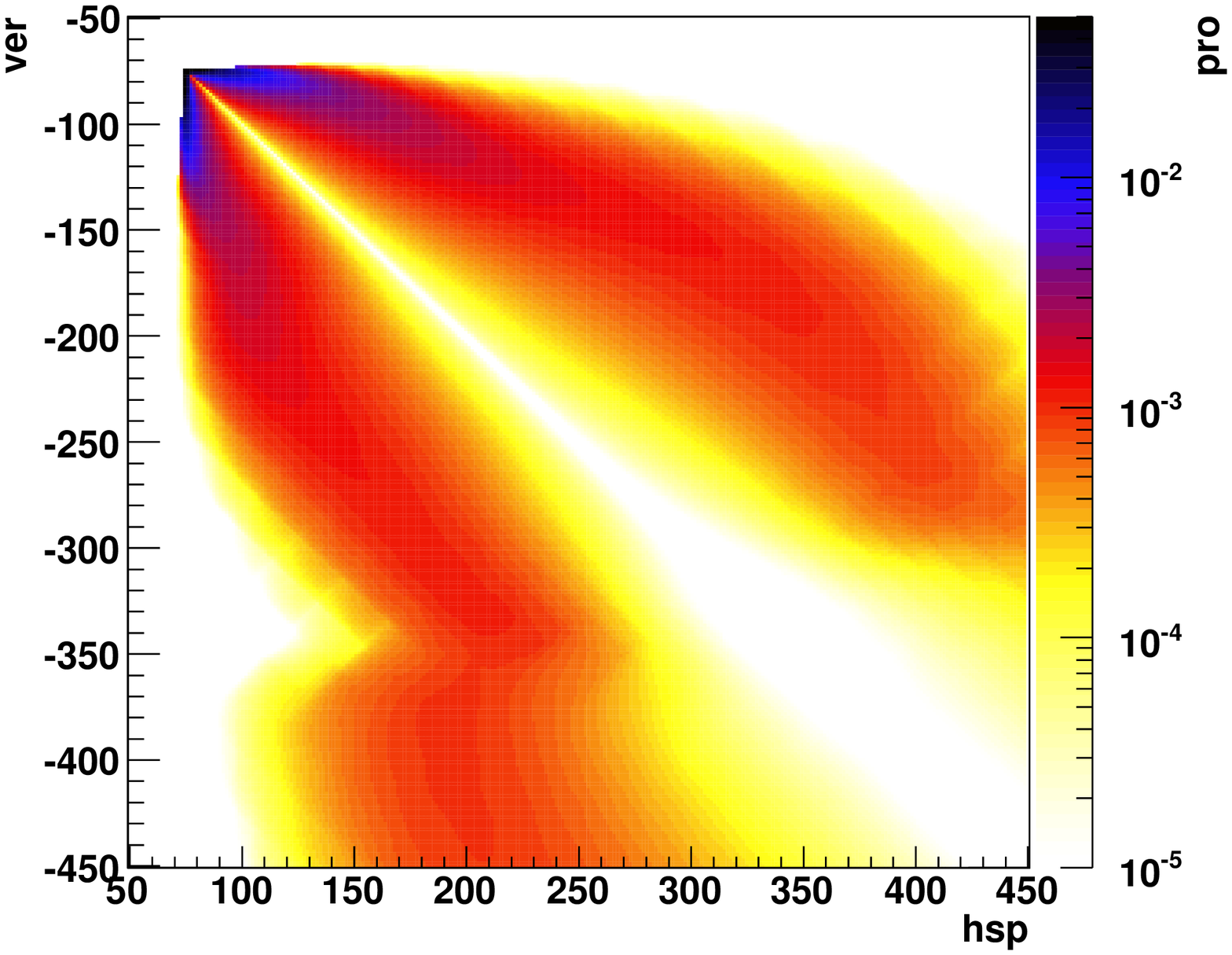}}}
 \caption{
\label{infmuonpic} A snapshot of the light distribution produced by a
simulated ultra-relativistic muon which entered from below, at an angle
$\Theta_{\src}=135$\degr~diagonally towards the glacier surface
 ($\Theta_{\src}=180$\degr~would be straight upwards) passing through the origin at a depth of 1730~m below the glacier surface. Stronger fluxes $\Phi(t)$ are
observed both above and below the particularly dusty region around
$z=-350$~m which has stronger scattering and absorption. Scattering
causes a bending of the Cherenkov light cone, most easily seen in 
the differential probability distribution $f_{\ppdf}(t)$, whose
time integral is by definition normalized to unity at each spatial
location.
Inhomogeneities in the optical properties of the medium cause the
additional structure seen in the figure, especially \mbox{around -350~m}.
}

\end{figure*}
\begin{figure*}[hp!!!]
\center
\psfrag{hsp}[r]{\footnotesize \raisebox{-13pt}{horizontal position $x$ [m]}}
\psfrag{ver}[r]{\footnotesize \raisebox{4pt}{vertical position $z$ [m]}}
\psfrag{lig}[r]{\footnotesize \raisebox{-14pt}{light flux [m$^{-2}$ns$^{-1}$]}}
\psfrag{pro}[r]{\footnotesize \hspace*{1em}\raisebox{-14pt}{probability [ns$^{-1}$]}}
\mbox{
\subfigure[\mbox{$\Phi(100$ ns$)$, $Z_{\src}=0$~m}\label{a350t100}]%
{\includegraphics[width=0.46\textwidth]{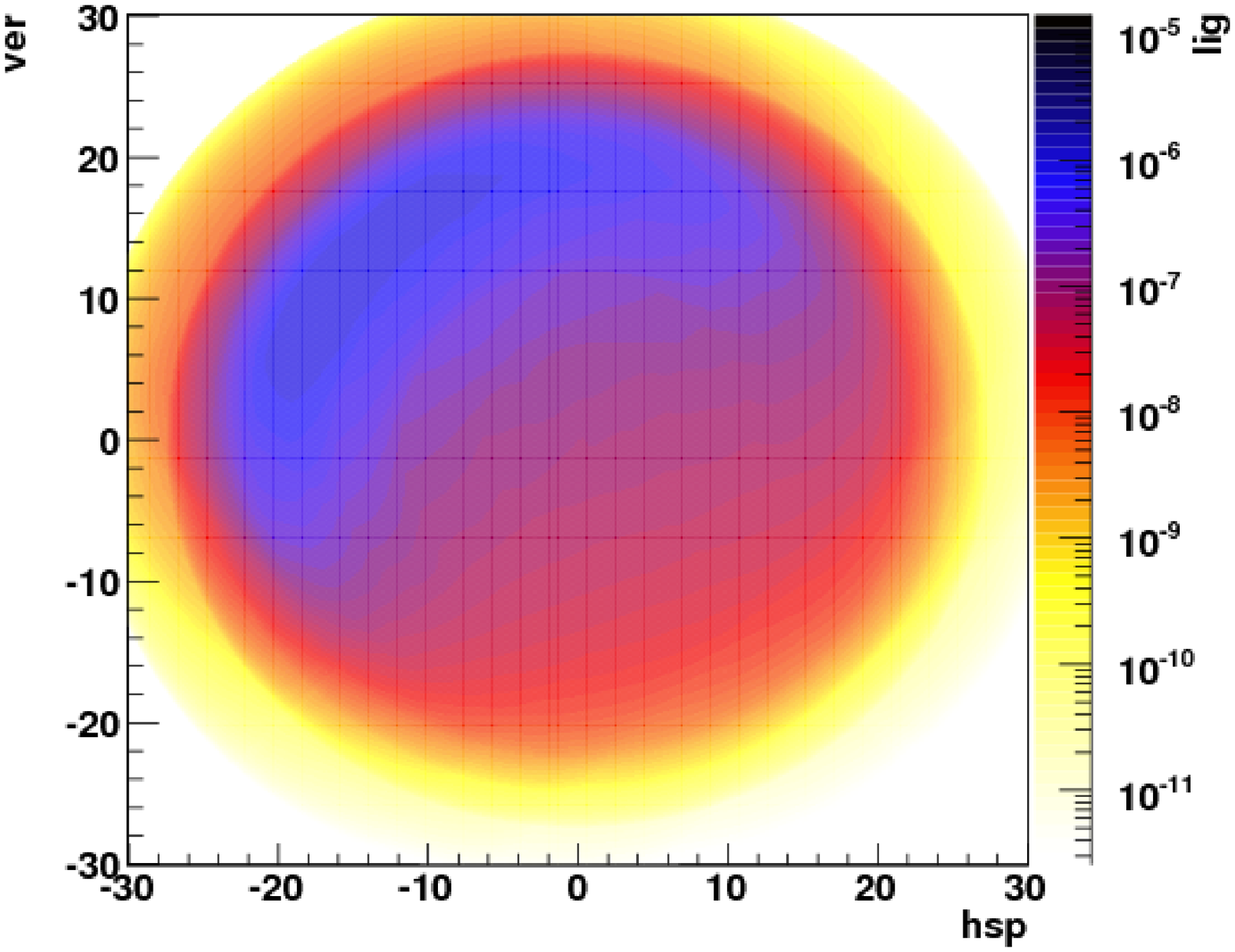}}
\hspace*{1em}\subfigure[\mbox{$f_{\ppdf}(t=100$ ns$)$, $Z_{\src}=0$~m}]%
{\includegraphics[width=0.46\textwidth]{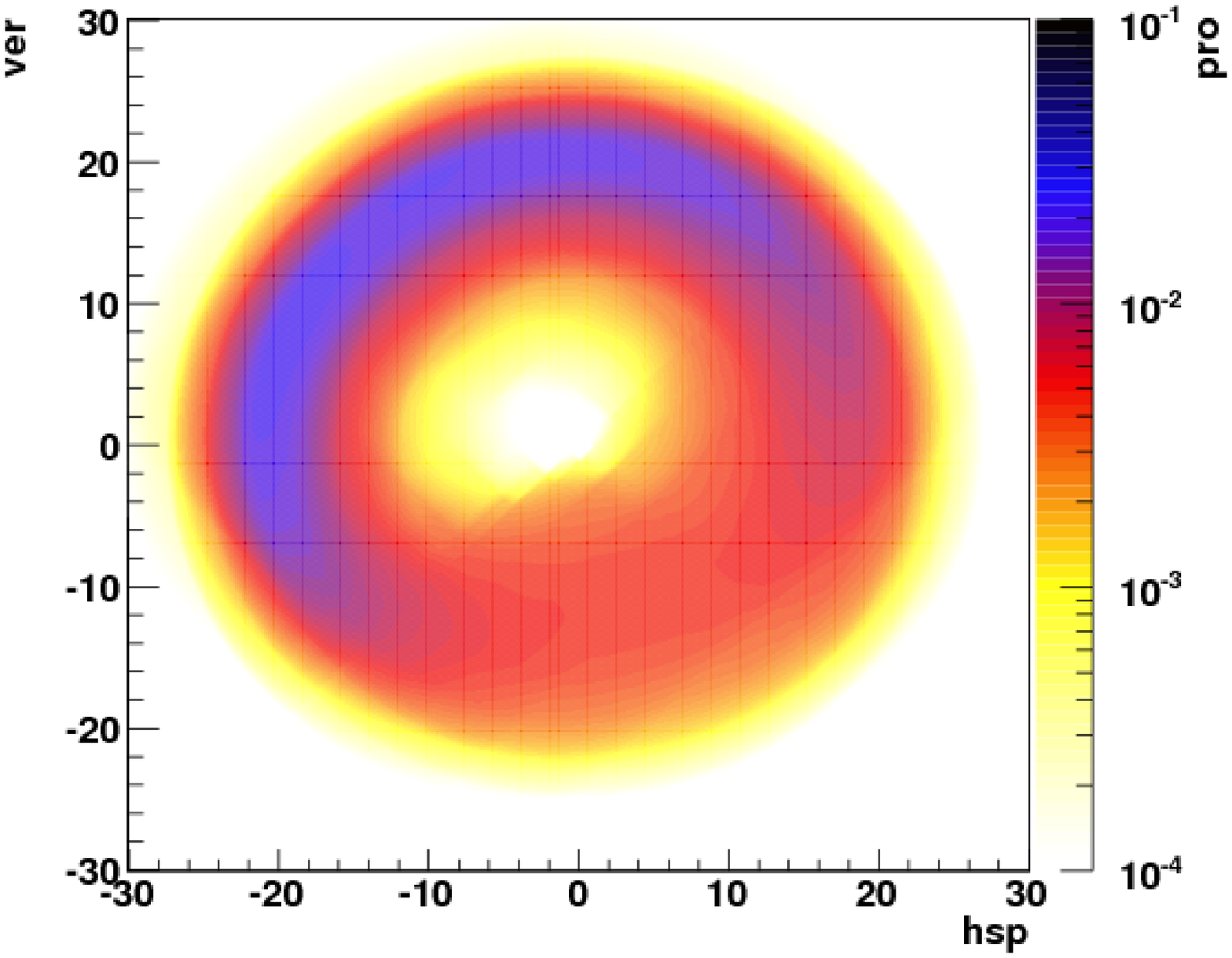}}}
\caption{
The figures show the simulated light flux, $\Phi(t)$ and the probability
distribution $f_{\ppdf}(t)$ of the light emitted from an idealized
shower placed 1730~m below the glacier surface.
The snapshot is taken $t=100$~ns after light emission at the
origin. The shower direction is $\Theta_{\src}=135$\degr, as for the
 muon in figure~\ref{infmuonpic}. 
\label{showerpic}}
\end{figure*}

A detailed study of the properties of the glacial ice at the South Pole
has been performed by the \amanda collaboration\cite{amandaice}. The
ice is very clear in the optical and near UV region with
absorption lengths of 20--120~m, depending on wavelength. The effective
scattering lengths are around 25~m, less for shorter wavelengths. Both
scattering and absorption are strongly depth dependent. The variations
at depths greater than 1450 m, where air bubbles no longer exist, are
explained by changes 
in climatic conditions which correlate with
concentrations of insoluble dust deposits.
At each 10~m depth interval, the effective scattering and absorptions
lengths, $\lambda_{\eff}$ and $\lambda_{\abb}$ as function of wavelength
were determined. As an example, the time distributions corresponding to
two different wavelengths and light source--receiver positions were
calculated and compared with experimental distributions,
see figure~\ref{iceplots}.

\label{neuastro} \label{glacialphysics} The photon propagation and
recording methods were applied to various idealized event types, such as
the light emission from minimum ionizing muons, and from electromagnetic
showers generated when
ultra-relativistic electrons interact with the detector medium. Using a
charged particle propagator such as {MMC}\cite{ICRCphot_mmc}, the photon tables
of idealized events types are dynamically combined to describe realistic
composite events.
The \px~photon simulation results are accessed directly through \textsc{root}
compliant \cpp~interfaces. The IceCube simulation programs query these
interfaces and apply detector specific details such as simulation of
electronics, data acquisition and triggers.  For event
reconstruction, \px{} provides individual photon probability density
functions (pdfs), and the expected number of detected photons. These are
used by track-fitting algorithms, for example maximum-likelihood
routines.
The interface also delivers photon arrival times randomly drawn
from the cumulative arrival time distributions.

Figure~\ref{infmuonpic} shows the
light distribution of a simulated minimum ionizing muon traveling
diagonally upwards, on its way through the point \mbox{$x=0,z=0$}. At the front of
the track, we observe a cross section of the unscattered \chv wavefront,
followed by a diffuse light cloud as the photons are scattered away from
the geometrical \chv cone. We also observe a weak deflection of the photons with higher
flux $\Phi(t)$ both above and below the dusty ice region near $z=-350$ m.

Figure~\ref{showerpic} shows a cross section of the light distribution
of a simulated shower at a depth of 1730~m, 100 ns after light emission.
Ultra-relativistic electrons deposit their energy much quicker than
muons, confining most of the light emission to the vicinity of the
interaction point, depending on energy. At the same time, light may
propagate hundreds of meters into glacial layers with very different
optical properties. Shower-like event are more dependent on a complete
implementation of variations in ice properties with depth since the
localized light emission makes it harder to reconstruct the lepton
direction.
The use of \px~with heterogeneous ice models makes it possible
for IceCube to adequately handle such events.

\section{Conclusion}
 
New photon propagation methods were implemented, and are in use in the
simulation and reconstruction of particle physics events for IceCube.
The \px~ program is used for calculating and tabulating light
distributions of calibration sources and ultra-relativistic charged
particles, as a function of time and space in the heterogeneous South
Pole glacier.
The full depth and wavelength dependent ice description of
\cite{amandaice} was implemented. Shower-like events (induced by
ultra-relativistic electrons) are more
sensitive to depthwise ice property variations than are muons. This is
increasingly important for higher energies, as light
propagates further into different glacial layers.
The IceCube simulation can fully take into account how depth and
wavelength dependent variations of the optical properties of
the South Pole glacier distort the footprints of elementary particle
interactions.

\appendix

%\end{document}

\setcounter{figure}{0}
\setcounter{table}{0}
%\documentclass{article}
%\usepackage{icrctc07}

% old stuff taken from icrc29
%\usepackage{graphicx,amssymb,amsmath,times}
%\setcounter{page}{1}

\newcommand{\myedit}[1]{\textit{Editorial: #1}}
\newcommand{\ramble}[1]{\textsc{#1}}
\newcommand{\us}{\ensuremath{\,\mu\mathrm{s}}}
\newcommand{\meter}{\ensuremath{\,\mathrm{m}}}
\newcommand{\PeV}{\ensuremath{\,\mathrm{PeV}}}
\newcommand{\EeV}{\ensuremath{\,\mathrm{EeV}}}
\newcommand{\eV}{\ensuremath{\,\mathrm{eV}}}
\newcommand{\NPE}{\ensuremath{N_\mathit{pe}}}
\newcommand{\Ldir}{\ensuremath{L_\mathit{dir}}}
\newcommand{\tres}{\ensuremath{t_\mathit{res}}}
\newcommand{\mutot}{\ensuremath{\mu_\mathit{tot}}}
\newcommand{\DOM}{DOM}
\newcommand{\DOMs}{DOMs}
%\newcommand{\etal}{\textit{et al.}}
%
%Title of paper
\title{Reconstruction of high energy muon events in IceCube using waveforms}
%Short title to print in the headers to the final publication (Not showed in this print).
\shorttitle{Reconstruction of high energy muon events in IceCube using waveforms}
%All paper authors
\authors{S.~Grullon$^1$, D.J.~Boersma$^1$, G.~Hill$^1$, K.~Hoshina$^1$, K.~Mase$^2$ for the IceCube Collaboration$^A$}
\shortauthors{S.~Grullon et al.}
%All the affiliations.
\afiliations{$^1$ The IceCube Project,
                  UW Madison,
                  222 West Washington Avenue,
                  Madison WI, USA \\
             $^2$ Chiba University, Yayoi-tyo 1-33, Inage-ku, 
		  Chiba-shi 263-8522, Japan \\
	     $^A$ See special section of these proceedings
}
\email{grullon@icecube.wisc.edu}

\abstract{

We present a method to reconstruct the geometry and energy of high
energy muon tracks in IceCube. Through a log-likelihood optimization
procedure, an event hypothesis is obtained by maximizing the agreement
of the expected amount of light (as function of time) in the optical
modules with the shapes of the pulses recorded in the optical modules.
This reconstruction method aims to use all information contained in
the waveforms recorded in the IceCube digital optical modules (\DOMs),
by comparing those waveforms directly with the expected arrival time
distribution of Cherenkov photons at the \DOM\ after emission from
a hypothetical track, taking into account the optical properties of
the South Pole ice. We expect that this method will be effective
in particular for highly energetic events in which a significant
fraction of the \DOMs\ records many photo-electrons.  Currently,
for simulated events within an energy range of $100\TeV$ to $32\PeV$
which were reconstructed as throughgoing,
we obtain an energy resolution of
$0.34$ in $\mathrm{Log}(E/\mathrm{GeV})$ and an angular resolution of
$0.62^\circ$.

}

%\begin{document}

\maketitle

\section{Introduction}\label{ICRCwf_intro}

%\ramble{Some nice paragraph about high energy cosmic rays and GZK neutrinos.}
%
%Extremely high energy cosmic rays (EHECRs) ($>10^{20}\eV$)
%are observed by several experiments such as AGASA\cite{agasa},
%HiRes\cite{hires} and Auger\cite{auger}.  There are several theories to
%explain the EHECRs.  They are usually classified into two models, namely
%``bottom up'' and ``top down''.  In the more conventional bottom up model,
%the EHECRs are explained by statistical acceleration\cite{fermiacc}
%in sources such as AGNs and GRBs. In the more exotic top down models,
%the EHECR are generated through topological defects or the decay
%of supermassive particles. In many of these models high energy neutrinos
%are produced.
%
%An EHECR loses its energy during its travel through interaction
%with the cosmic microwave background, well known as the GZK
%mechanism\cite{GZK1966}.  Through this mechanism, extremely high energy (EHE)
% neutrinos will
%be generated.  A measurement of (or an upper limit on) the flux of EHE
%neutrinos will help to constrain the models for EHECR generation.

%\ramble{Some words about how icecube works, provide essential information so
%that a new reader can have a rough idea of what we are talking about
%when we use words like "ATWD waveform". E.g. mention ATWD is relatively
%short and fine grained, but high dynamic range, while FADC is long with
%low dynamic range.}

The IceCube telescope is being deployed in the Antarctic ice with its main 
goal to detect high energy neutrinos arriving from astrophysical sources.
Nearly one third of the detector is installed and currently operational
\cite{ice3status}.  
When fully deployed, the instrumented volume will be approximately $1\km^3$.

When a neutrino interacts in the ice in or near the detector, it produces
a track or cascade signature. Some of the Cherenkov light emitted by the
charged lepton and secondary charged particles triggers the \DOMs. A \DOM\
digitizes the signal from a $10\,\mathrm{inch}$ photo-multiplier in two
ways: with an analog transient waveform digitizer (ATWD) and with a fast
analog to digital converter (fADC) \cite{firstyear}.

The main purpose of the ATWD is to record precise timing information
of photons arriving in \DOMs\ relatively close to the track or
cascade. Therefore, it reads the same signal in 3 channels operating
on different gains. Each channel has up to 128~bins with a bin witdh of
$3.6\ns$.  The main purpose of the fADC is to measure pulses with a wider
time distribution from a further away track or cascade. It has 256~bins
with a bin width of $25\ns$, giving a total time window of $6.4\us$.

Given that the IceCube neutrino observatory records the full waveform
information, a new likelihood reconstruction technique to exploit
the full waveform information is the goal of the research described
in this paper.  Conventional reconstruction techniques~\cite{nimreco}
ported to IceCube from its predecessor, AMANDA, do not use the complete
waveform. This is a reflection of the original AMANDA data acquisition
system which recorded only the leading edge time of the pulse, the total
charge of the pulse, and the total time over threshold of the pulse.
These conventional reconstruction techniques in IceCube utilize this
information by extracting pulse shapes from the ATWD or fADC waveforms
and reconstruct a cascade or a muon hypothesis based on this information.

In this paper, we focus on the likelihood reconstruction of high energy
muon tracks arising from extremely high energy (EHE) neutrinos with
energies up to $10^{11}\GeV$.
%In section~\ref{gzk} we show how this reconstruction can be
%for example be used in the search for so-called GZK neutrinos, which
%should be produced when EHE cosmic rays interact with cosmic microwave
%background\cite{GZK1966}.
EHE neutrinos should be produced when EHE cosmic rays interact with the
cosmic microwave background \cite{GZK1966}. The significant background
due to atmospheric muons presents a major challenge, however.  Since the
zenith and energy distributions are different for signal and background,
good geometry and energy reconstruction are vital for signal detection.

We hope that with the waveform-based event reconstruction method a
significant improvement in sensitivity can be achieved for events at a
wide energy range from $\sim10\TeV$ up to highest energies, $\sim\EeV$.
At energies above $1\PeV$ we expect to increase the sensitivity by
effectively reconstruction the energy of non-contained events.

%The waveform-based event reconstruction method is a powerful tool that
%can be applied to the GZK neutrino search.

\section{Method}

We define a function which gives the likelihood that the observed
waveforms in the DOMs are the result of a given muon track. Using a
standard minimizer algorithm, the track's position, direction and energy
are found for which the likelihood has a maximum.

\subsubsection{Expected photon arrival time distribution at a single DOM}
%\label{pdfsec}

A crucial element in the likelihood function is the description $\mu(t)$
of the expected number of photo-electrons as a function of time in a given
DOM for a given muon track.  This description consists of the expected
total number of photo-electrons $\mutot$ together with a probability
density function (PDF) $p(t)$ of the arrival time distribution of a
single photon: $\mu(t) = \mutot \cdot p(t)$.

The $\mutot$ and PDF depend on the energy, direction, and the distance of
the track to the \DOM, the relative orientation of the DOM with respect
to the track, and the optical properties of the ice between the track
and the \DOM.

%Typically, the PDF is a narrowly peaked distribution
%for \DOMs\ closer than one scattering length ($\sim20-40\meter$) to the track,
%while for \DOMs\ further away the distribution gets wider \cite{photonix}.

At energies of a few hundred TeV and higher, most of the Cherenkov light
is not emitted by the muon itself, but by its many secondaries and by
the stochastic showers.  For our reconstruction of a high energy muon
track, we assume that the muon track with stochastic showers can be
approximated by an "infinite cascade" which is a string of equidistant
average showers each with an energy deposit corresponding to the $dE/dX$
energy loss of the track in the ice.

% Next paragraph is going to be drastically shortened and detechnologized.
For the results in this paper, we took $\mutot$ and the PDF from a table
generated using the "photonics" light propagation code \cite{photonix}. An
alternative approach uses a parametrization of the average waveforms
obtained from the full IceCube simulation.

Fig.~\ref{photorec} shows the comparison between the expected
photo-electron distribution $\mu(t)$ as obtained with photonics and
individual waveforms as obtained in the full MC simulation.
\begin{figure}
\begin{center}
\includegraphics[width=0.48\textwidth]{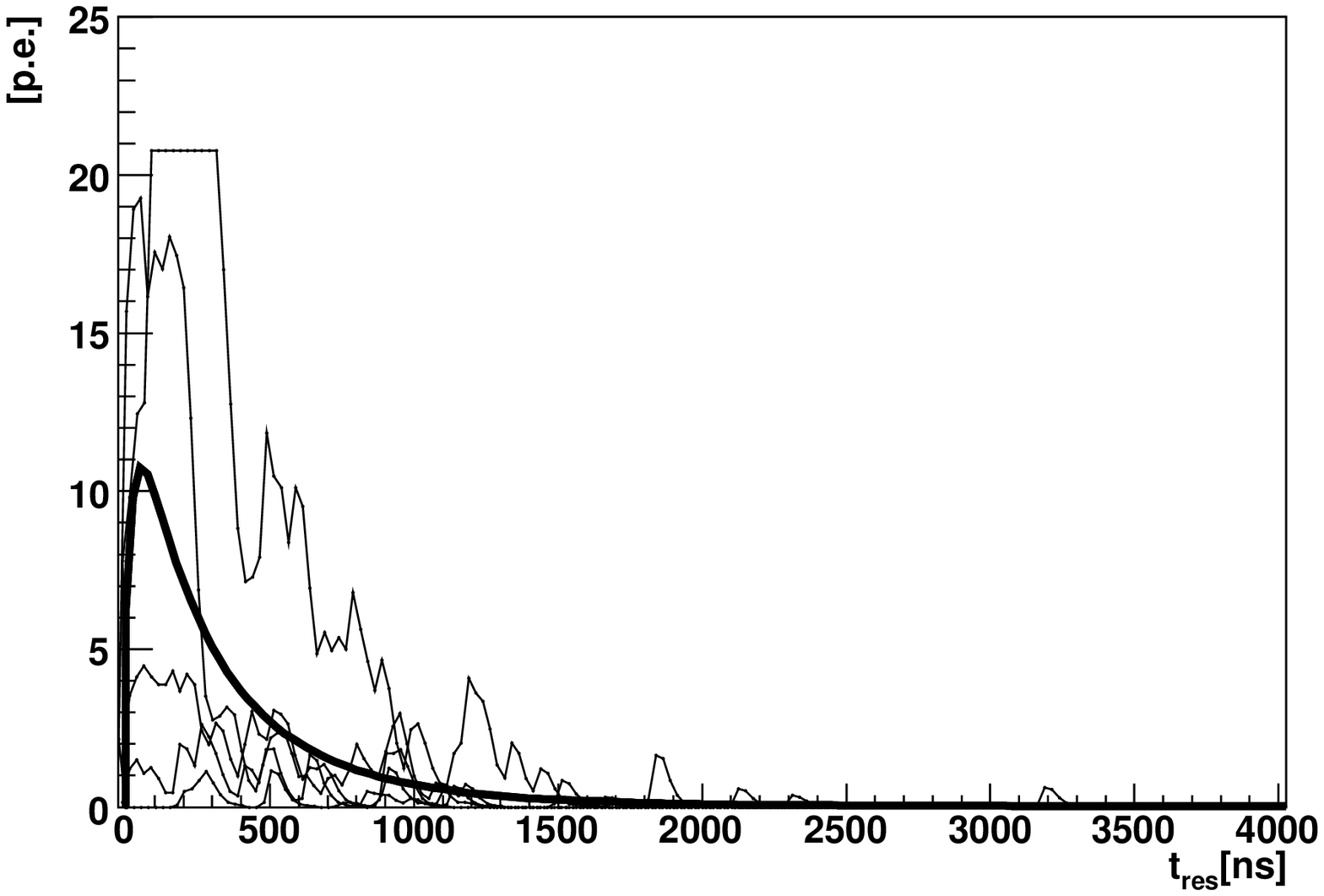}
\end{center}
\vspace{-10mm}
\begin{center}
\includegraphics[width=0.48\textwidth]{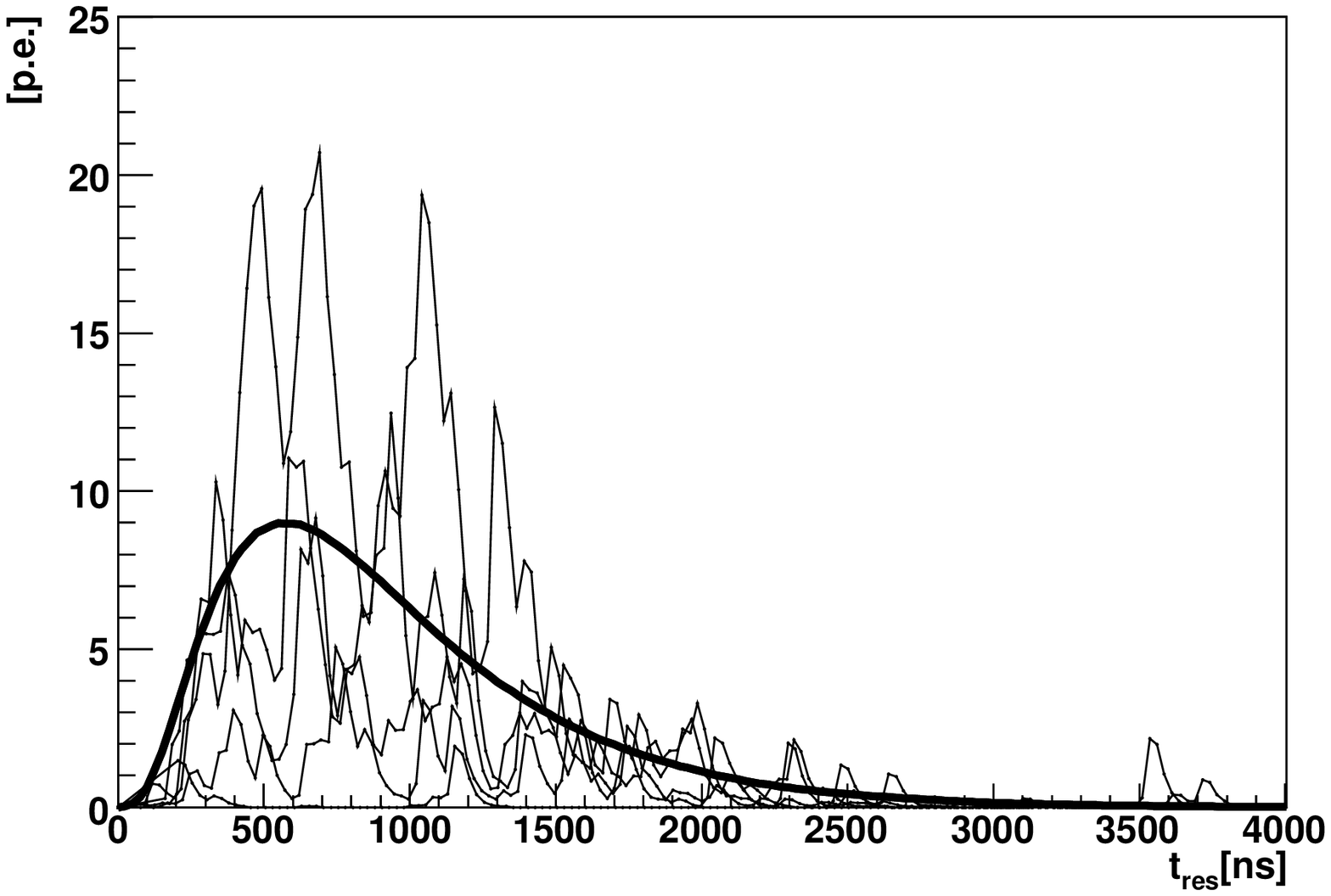}
\end{center}
\vspace{-5mm}
\caption{Comparison of the expected photo-electron distribution $\mu(t)$
(thick line) from photonics tables with some actual waveforms from the
full MC simulation of high energy muons (thin lines).
Upper figure: $1\PeV$ at $53\meter$,
lower figure: $100\PeV$ at $147\meter$.
}\label{photorec}
\end{figure}
%% KM: This is not true, I think.
%% KM: This is simply because we ignore the stocastic each shower,
%% KM: and take the ``continuous'' (averaged) cascade proximity.
%% KM: Howerver, this proximity should work on average.
%% KM: we can add this sentence, but I would rather put this sentences
%% KM: to the previous section (or at the caption of figure 1).
%%
%%It should be remarked that at lower energies,
%%the shape agreement between the PDF and the waveform is not perfect, see
%%figure~\ref{photorec}. Observed waveforms in this energy range have very
%%few photo-electrons showing up as narrow pulses, while the PDF can have
%%a fairly wide arrival time distribution.  At higher energies, however,
%%the photon flux is so high that the shapes of individual waveforms should
%resemble the shapes of the PDF more closely.

It should be remarked that the individual waveforms may resemble the
expected average waveform only at very high energies and in DOMs close
enough to the track.
In most events, the individual waveforms in various DOMs will look different,
as shown in Fig.~\ref{photorec}.
First, individual stochastics near the DOMs may produce fluctuations
beyond the statistical (Poissonian) fluctuations from the average as
modeled by the infinite cascade approximation.
Second, when the $\mu(t)$ times the width a of a single photoelectron
pulse is less than 1, then the of course the individual waveforms of
the occasional individual photoelectrons will not follow that low PDF.

%It should be remarked that only at very high energies and in DOMs
%close enough to the track, the individual waveforms should resemble the
%expected average waveform. In most events and many DOMs the individual
%waveforms will look different, as shown in Fig.~\ref{photorec}.  First,
%when the total $\mutot$ is low compared to the total width of the expected
%distribution of the arrival times of the photons, then the pulses of the
%individual photoelectrons will hardly overlap with each other. Second,
%our approximation of a very high energy muon as a string of equidistant
%showers works best at high energies, at intermediate energies the
%individual stochastics will be (much) more pronounced and produce larger
%fluctuations in \NPE\, the total number of measured photo-electrons,
%in individual DOMs.

\subsubsection{Poissonian likelihood for waveforms}

The conventional reconstruction strategy described in the introduction
works well for lower energy muon events in which the total charge
corresponds to only a few photo-electrons. High energy muon events on
the other hand are characterized by a large amount of deposited light and
therefore produce wide, complicated waveforms with many photo-electrons.
Reconstructing the geometry of a high energy muon track would benefit
from the complete waveform information, as the width of the observed
waveforms scales with the distance between the muon track and the DOM.
A likelihood reconstruction of the muon energy would also require the
complete waveform in order to measure the total amount of light deposited
in the IceCube detector since this correlates with the energy of the muon.

The likelihood function using the complete waveform is formulated as
follows. What is the probability of observing a waveform $f(t)$ given
an expected photo-electron distribution $\mu(t)$? The waveform $f(t)$
is measured from the ATWD or the fADC, and the expected photo-electron
arrival distribution is given by the PDF. The expected photo-electron
arrival distribution depends on the hypothesis parameters, namely the
geometry $\vec{x}$ (position of the muon at $t=t_0$ and its direction)
and the energy, $E$.  If you bin the waveform $f(t)$ into $K$ bins,
the probability of observing $n_{i}$ photons in the $i$th waveform bin
given an expectation of $\mu_{i}$ photons in the $i$th bin is given by
Poissonian statistics. The overall probability for a single OM is given
by the product over all waveform bins:

\begin{equation}
P(f(t)|\vec{x},E)=\prod_{i=1}^K\frac{e^{-\mu_{i}}}{n_{i}!} \mu_{i}^{n_{i}}
\label{prob}
\end{equation}

Taking the log of the Poissonian probability gives us:

\begin{eqnarray}
%\log P(f(t)|\vec{x},E)&=& \\
%\nonumber\sum_{i=1}^{K}\left(n_{i}\log\frac{\mu_{i}}{\mutot} 
%\right) \!\!\!\!&\!\!\!+\!\!\!&\!\!\! \NPE\log\mutot - \mutot
%\label{ICRCwf_llh}
\log P(f(t)|\vec{x},E) &=&  
\nonumber\sum_{i=1}^{K}\left(n_{i}\log\frac{\mu_{i}}{\mutot} \right)  \\
&\!\!\!\!\!\!\!\!\!\!\!\!\!\!\!\!\!\!\!\!\!\!\!\!\!\!\!\!\!\!\!+&
\!\!\!\!\!\!\!\!\!\!\!\!\!\!\!\!\!\!\!\! \NPE\log\mutot - \mutot 
\label{ICRCwf_llh}
\end{eqnarray}

% DJB: I replaced "photons" by "photo-electrons" in several places

%% KM: This part is a bit lengthy. I will make it short
%% DJB: I disagree; this formula is the heart of our method, we should
%% not economize on making its meaning clear.
The first term is a sum over all waveform bins. Each term in the sum
$n_{i}\log\frac{\mu_{i}}{\mutot}$ corresponds to the normalized timing
probability of observing a photo-electron in the $i$th waveform bin
weighted by the number of observed photo-electrons in the $i$th bin.

%Here, $\mutot$ is the total number of expected photo-electrons 
%(which depends on the geometry and the energy of the muon), and $N$
%is the total number of observed photo-electrons.

We evaluate Eq.~\ref{ICRCwf_llh} for all DOMs in the ice and sum these values
as our log-likelihood function which we then maximize with respect to
the free parameters of the track. This amounts to fitting the shape of
the PDF to the measured waveform. This allows the reconstruction of not
only the geometry of the muon, but also its energy.

One feature that needs to be addressed is the issue regarding the
saturation of the waveform which the likelihood formula does not take
into account. Currently, saturation is taken into account by simply
truncating both the PDF and the measured waveform at some level close
to the actual saturation level of the hardware, while sticking to the
formalism of Poissonian statistics.

% LAM
%Another correction may be necessary to account for the fact that
%our detector is currently operated in "hard local coincidence" mode,
%which means that DOMs for which the nearest neighbors see no signal are
%ignored. This has consequences for DOMs with low $\mutot$.
%This effect may be more significant at intermediate energies
%than at extremely high energies.

\subsubsection{Fitting strategy}

When reconstructing the muon track, there are in general six free
parameters to fit (the vertex, direction, and the energy). Fitting the
geometry and energy separately in three stages turns out to be more
efficient than fitting them all at once.
%This is because of $\mutot$ in Eq.~\ref{llh}
%which depends both on the geometry and the energy of the muon, so a decent
%estimate of the energy would lead to a better geometry reconstruction and
%vice-versa.  We therefore develop a three-stage reconstruction technique
%in reconstructing the geometry and the energy of the muon.
We seed the first stage of the reconstruction with a first guess of the
geometry and the energy and proceed to fit the geometry only (five free
parameters).  We then seed the second stage with this result, fitting
the energy only (one free parameter). Finally, we use this second stage
result to seed a third fit, which refits the geometry again (five free
parameters).

\section{Results}

\begin{figure}
\begin{center}
\vspace{-5mm}
\includegraphics[width=0.48\textwidth]{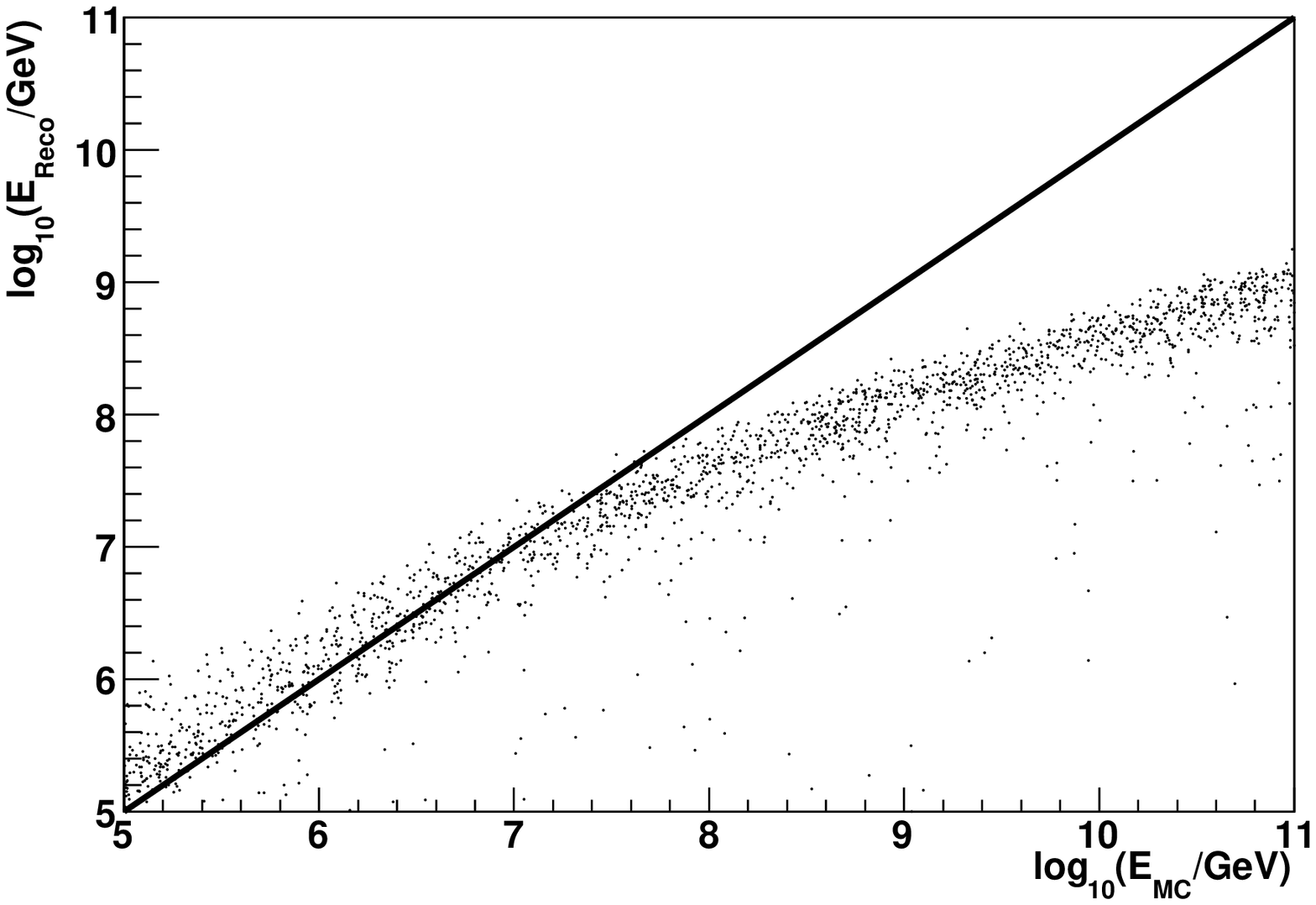}
\end{center}
\vspace{-5mm}
\caption{Reconstructed muon energy versus simulated energy for reconstructed tracks
that go through the IceCube detector (see text). The diagonal
line $E_{reco}=E_{true}$ is added to guide the eye.}
\label{energyres}
\end{figure}

% (DJB) I have rewritten this, please check
% (DJB) I have rewritten this even more, please check. I wanted to shorten
% the explanation of Ldir. Maybe we should omit this explanation altogether?

The energy reconstruction results are shown in Fig.~\ref{energyres} for a
MC event sample simulated with an $E^{-1}$ spectrum and an energy range
from $10\TeV$ to $100\EeV$ with $4\pi$ coverage in the full 80-string
IceCube geometry.  Only reconstructed throughgoing muon tracks are
selected, which are muon tracks whose point of closest approach to the
geometrical center of the IceCube detector is within the IceCube array.

At energies above $\sim 30\PeV$, the reconstructed energy is systematically low
due to saturation in the DOMs, which is currently not taken into account.
The slope of the distribution for energies below $30\PeV$ may be improved
by adjusting the "infinite cascade" model, in particular the relation
between the energy of the muon track and the energy in an average shower
of the infinite cascade. For energies below $30\PeV$, approximately $31 \%$ of the events are
reconstructed as throughgoing.

For throughgoing muon tracks and $E_{MC} < 30\PeV$ , the angular
resolution (defined as the median of the distribution of angular
differences of the reconstructed and simulated muon tracks) is
found to be $0.62^\circ$. Our obtained energy resolution is $0.34$
in $\mathrm{Log}(E/\mathrm{GeV})$. With the traditional AMANDA
style reconstruction about $35\%$ of the events are reconstructed as
throughgoing, with an angular resolution of $0.63^\circ$.

%\section{Discussion}
%\label{gzk}
%\myedit{To be written... See figures~\ref{gzk1} and \ref{gzk2} to understand why
%it is vital to have good direction and energy reconstruction also at extremely high
%energies. HOWEVER: with these figures, the paper will be too long...}
%
%
%\begin{figure}
%\begin{center}
%%\includegraphics[width=0.496\textwidth,height=0.472\textwidth]{EventRateVsZenithReco.gif}
%\end{center}
%\caption{Rate versus the cosine of the reconstructed zenith
%angle. Background in red: atmospheric muon events (probably with
%some strong cut on $N_\mathit{ch}$. Signal (GZK neutrinos) in
%black. \myedit{For the proceedings we should probably avoid relying
%on color. BTW: if Keiichi sent me EPS versions of these plots then I
%cannot find them back in my mail folder...}}\label{gzk1}
%\end{figure}
%
%\begin{figure}
%\begin{center}
%%\includegraphics[type=gif,width=0.496\textwidth,height=0.472\textwidth]{EvetRateRecoWeightCorrect.gif}
%\end{center}
%\caption{Rate versus energy. Background in red: atmospheric muon events
%(probably with some strong cut on $N_\mathit{ch}$. Signal (GZK neutrinos)
%in black. \myedit{For the proceedings we should probably avoid relying
%on color.}}\label{gzk2}
%\end{figure}

%\ramble{Show GZK log(E) plot.
%Show GZK cos(zenith) plot.
%Show GZK $A_\mathit{eff}$ plot.}

\section{Outlook}

The waveform based reconstruction as currently implemented performs
reasonably well. With a sample of simulated high energy events, we
obtain an angular resolution comparable or better than conventional
reconstruction methods.

We have identified several aspects of the algorithm and its implementation
which can still be improved, including a proper way to use the information
of saturated DOMs. This should further improve the energy resolution
(currently $0.34$ in $\mathrm{Log}(E/\mathrm{GeV})$) and extend the
energy range beyond 1EeV. The results of this paper are only for throughgoing muon 
tracks; we hope to present similar results for high energy non-contained events as well.

The method is in principle not limited to track-like events; it can be
applied to events of any signature, such as showers and possibly also
muon bundles.

% \myedit{I am still writing this, will be continued....}

\section{Acknowledgements}

The authors would like to acknowledge support from the Office of Polar
Programs of the National Science Foundation and the Japan Society for
the Promotion of Science.

%\end{document}

\setcounter{figure}{0}
\setcounter{table}{0}
%%
% International Cosmic Ray Conference 2007 Merida Yucatan Mexico
% In This file you will find detailed instructions to correctly
% typeset your document.
%
%
%

%Class Requeried
%\documentclass{article}
%The ICRC Style
%\usepackage{icrctc07}

%The paper title
\title{Radio Detection of GZK Neutrinos - AURA status and plans}
%Short title to print in the headers to the final publication (Not showed in this print).
\shorttitle{AURA - Radio GZK Detector }
%All paper authors
\authors{H. Landsman$^{1}$ For the IceCube Collaboration$^{2}$,  L. Ruckman$^{3}$, G. S. Varner$^{3}$}
%Short title to print in the headers to the final puplication (Not showed in this print).
\shortauthors{H.Landsman and et al}
%All the affiliations.
\afiliations{$^1$Department of Physics, University of Wisconsin, Madison, WI 53706, U.S.A \\ $^2$ See special section of these proceddings.\\ $^3$ Dept. of Physics and Astronomy, University of Hawaii, Manoa, HI 96822, U.S.A}

\email{hagar@icecube.wisc.edu}

%The abstract.
\abstract{
The excellent radiofrequency transparency of cold polar ice, combined with the coherent Cherenkov emission produced by neutrino-induced showers when viewed at wavelengths longer than a few centimeters, has spurred considerable interest in an ultimate, large-scale radiowave neutrino detector array. A statistically compelling GZK signal will require at least an order of magnitude improvement in the product of (livetime)x(Effective volume) over existing (RICE, ANITA, e.g.) neutrino detection experiments. Correspondingly, the AURA (Askaryan Underice Radio Array) experimental effort seeks to take advantage of the opportunity presented by IceCube drilling through 2010 to establish the radiofrequency technology needed to achieve $100-1000 \mit {km^3} $ effective volumes. We discuss three test strings co-deployed with IceCube in 2006-07 which combine fast in-ice digitization with an efficient, multi-tiered trigger scheme. Ultimately, augmentation of IceCube with large-scale $(1000 \mit{km^3sr})$ radio and acoustic arrays would extend the physics reach of IceCube into the EeV-ZeV regime and offer substantial technological redundancy.}

%\email{aastex-help@aas.org}

%%%%%%%%%%%%%%%%%%%% B E G I N   D O C U M E N T%%%%%%%%%%%%%%%%%%%%%%%
%\begin{document}
\maketitle
%Begin the section.

\section{Introduction and Detection Principle}

%In the rare event that a neutrino interacts in a medium, it creates or scatters charge particles that produce energetic cascades, emitting  multi-wavelength Cherenkov radiation. Neutrino detectors look for this radiation using an array of photo-detector in deep, dark, transparent media (like ice or water). Detecting this weak radiation signal may indicate a neutrino is passing through or near the detector.  
%Neutrinos from nuclear reactors and the Sun (MeV), and from the atmosphere (GeV) were detected and are used to study particle- and astro-physics. Higher energy neutrinos are believed to be created in astronomical objects like GRBs, AGNs, and SGRs). When cosmic Rays (CR) Protons (with energy more than $10^{19.5} eV$) interact with the background radiation, a cutoff in the flux of CR arriving to earth from large distances and the flux of the so-called GZK neutrinos results. Neutrinos may also be the signature of exotic and new physics like dark matter and monopoles.

The Astrophysical high energy neutrinos hold valuable information about their sources, either a point source like GRBs, AGNs, and SGRs, or high energy cosmic rays (through the GZK process). Consequently they can also teach us EHE particle physics in energies unreachable by earthbound accelerators. 

%Neutrinos can  be a powerful astronomy tool, and teach us about EHE particle physics in energies unreachable by earthbound accelerators.

As the energy of the neutrino increases the atmospheric neutrino background flux decreases and the interaction cross section of the neutrino increases, which favors the detection of HE neutrinos over low energy ones. On the other hand, the estimated fluxes of those high energy neutrinos exhibit an overall decrease with energy. The combination of a small flux, low neutrino interaction cross section, and limited life span of humans require the construction of large scale detectors to improve the detection probability.

%For the high flux of low energy neutrinos originating from nuclear reactors or from the sun, a detector with a volume of $\sim 10 m^3$ is sufficient. 

The $km^3$ scale detectors like IceCube, AMANDA, NEMO and Antares are (will be) made of thousands of photo-multiplier tubes, sensitive to optical photons.  They are sensitive to neutrinos with energies between $10^{2} GeV - 10^{10} GeV $. In order to survey the extreme high energy regime of more than $10^{10} GeV$, larger detectors are needed. % and a techniques more efficient, and more sensitive to the high energy cascades is needed. 

In 1968, G.A.Askaryan \cite{askaryan} suggested that cascades generated by high energy charged leptons moving through matter, produce an excess of negative charge moving at relativistic speed, thus emitting Cherenkov radiation. For radiation with shorter wavelength, like optical photons, the phase is random and the electric field is proportional to the square root of the net negative charge developed in the cascade. But for photons with wavelengths longer than the transverse dimensions of the cascade, like RF photons, the radiation is coherent and the electric field is proportional to the negative charge in the cascade. It is expected that neutrinos with energy of$\sim 10^{18} eV$ or more will produce cascades with transverse dimensions of order $\sim 0.1$ meters, thus emitting coherent RF radiation. Radio-frequency neutrino detectors are therefore more sensitive to such high energy events than optical detectors .

This effect was demonstrated in an accelerator measurement where coherent linearly polarized RF radiation was measured from the interaction of a beam dumped into RF transparent matter (sand, salt and ice)\cite{slac}.
The simpler installation of radio detectors, the long attenuation length of RF in ice and the sensitivity to EHE events makes the RF region a useful probe for EHE neutrino detection.

Several experiments are already using the Askaryan effect for neutrino detection in Antarctica: The RICE array was deployed with the AMANDA neutrino telescope near the South Pole at depths of 100-300 m. The array consists of 20 dipole antennas covering a volume of $200 \times 200 \times 200 m^3$, and is sensitive between 200 to 500 MHz. RICE established limits on high energy neutrino fluxes as well as investigated the radio-glacial properties of the deep ice \cite{rice}. The ANITA experiment, air borne at 40km, observed the Antarctic ice searching for RF emission. The high altitude makes the volume that ANITA covers large (1.5 million $km^3$), but the short flight time and the refraction of RF photons in the transition from ice to air limits the exposure time and the angular coverage of this experiment \cite{anita}. %Other experiments that use the RF technology are observing the Greenland ice (FORTE) or moon rocks (GLUE) as detection media.

\section{Detector design and 2006/2007 Deployment}
In the austral summer of 2006-2007, three Radio Clusters were co-deployed with the IceCube optical array as part of the AURA (Askaryan Under-ice Radio Array) experimental effort. Each cluster consists of up to four broadband dipole antennas, centered at 400MHz, and four metal tubes holding the front-end electronics including filters and amplifiers supporting these antennas: specifically, a $450$ MHz notch filter to reject constant noise from the South Pole communication channel, a $200$ MHz high pass filter and a $\sim 45dB$ amplifier. An additional $\sim 20dB$ amplification is done at later stage, for a total of $\sim 65dB$ amplification.  An additional antenna is used as a transmitter for calibration. 

The DRM (Digital Radio Module) within a 13 inch diameter glass sphere contains the triggering, digitization and communication electronics as well as a power converter. It holds the TRACR board(Trigger Reduction And Communication for RICE) that controls the calibration signal and the high triggering level, the SHORT board (SURF High Occupancy RF Trigger) that provides frequency banding of the trigger source,  the ROBUST card (Read Out Board UHF Sampling and Trigger) that provides band trigger development, high speed digitization and second level trigger discrimination, the LABRADOR (Large Analog Bandwidth Recorder And Digitizer with Ordered Readout)\cite{labrador} digitization chip, and a Motherboard that controls the power, communication and timing. 

%The sampling speed is 2 GSPS, 1.3 GHz bandwidth and 256 ns buffer depth. 

A 260-capacitor Switched Capacitor Array (SCA) continuously observes the input RF channels (two channels per antenna) and an additional timing channel. To reduce power consumption and dead times, the information is held and digitized only when a trigger is received. The sampling speed is two Giga-Samples Per Second, with a 256 ns buffer depth.
%In the offline processing a characteristic pedestal is subtracted from each SCA byte, and the waveform is time-ordered based on the trigger time. 
A 300 MHz on-board Advanced Transient Waveform Digitizer is used for precise trigger timing. 
A Wilkinson type ADC converts the measured voltage into a count value with a 12-bit dynamic range. 

Six cables are connected to the DRM. One for power and communication with the surface and five for the transmitter and receiver antennas. The spacing between the antennas is 13.3 meters, and the total length of the cluster is 40 meters. The AURA cluster is shown in figure \ref{ICRC1155_fig2}.

The fast and broadband nature of the Askaryan RF signal is exploited for background reduction. Once the voltage measured on an antenna crosses an adjustable threshold, the digitization is triggered and the signal is split into four frequency bands (200-400 MHz, 400-650 MHz, 650-880 MHz and 880-1200 MHz). If enough frequency bands are present in the signal, the channel associated with this antenna will trigger. In the current settings, at least two out of four bands are needed for triggering to happen. The cluster will trigger if enough channels trigger (current setting requires at least three out of four antennas).

The digitized data is sent to the surface using the IceCube in-ice and surface cables.

IceCube on-going construction activity made it possible to deploy clusters down to 1400 meters deep, a depth that is usually less favored by RF detectors due to warmer ice and high drilling cost. The clusters were deployed on the top of IceCube strings, at depths of 1400 or 400 meters.

Table \ref{Deploy_Table} summarizes the depth and location of the three units. Out of the 8 receivers deployed, 7 receivers are operational. One channel was tested fine before deployment, and most likely damaged during the freeze-in of the water surrounding the cluster after deployment. The data being taken consists of ambient and transient background studies, calibration runs using the AURA transmitter and the in-ice RICE transmitters.  

\begin{figure}
\begin{center}
\noindent
\includegraphics [width=0.45\textwidth]{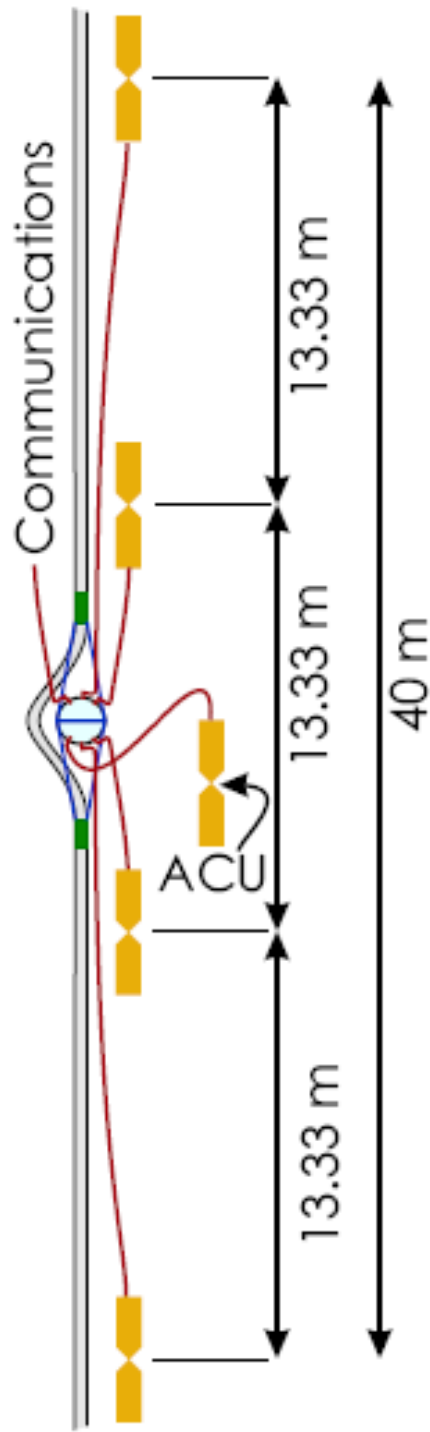}
\end{center}
\caption{
The radio cluster, made of a  DRM (Digital Radio Module), and 5 antennas (4 receivers and a transmitter).
}\label{ICRC1155_fig2}
\end{figure}

\begin{table*}
\begin{center}
\begin{tabular}{|c|c|c|c|c|l|}
\hline 
Cluster& num. Transmitters & num. Receivers & Location (x,y,z) in m & Front end amplifier brand\\
\hline
1 & 1 & 4 & $(50,500,-1400)$ & Miteq \\
2 & 1 & 4 & $(220,210,-250)$ & LNA-SSA \\
3 & 1 & 0 & $(195,120,-1400)$ & None\\
\hline
\end{tabular}
\caption{Locations of the deployed clusters. Coordinates are relative to IceCube center array at surface.} \label{Deploy_Table}
\end{center}
\end{table*}

\begin{figure*}
\begin{center}
\includegraphics [width=0.5\textwidth]{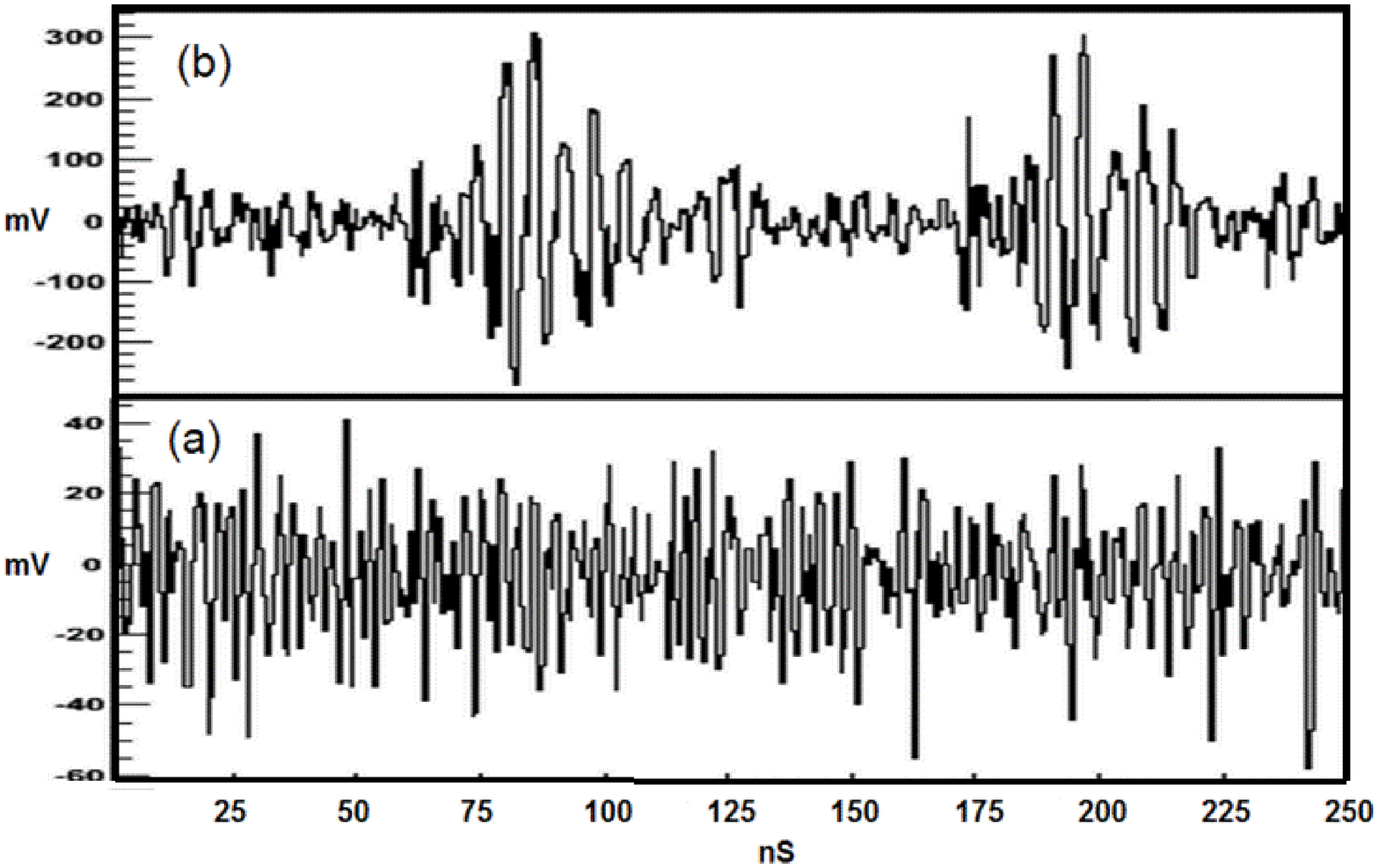}
\end{center}
\caption{
Wave form signals for a single antenna for background and calibration runs. (a) Background only (b) In ice transmitter pulse.
}\label{fig-wf}
\end{figure*}

The proximity of the South Pole station and especially the IceCube and AMANDA detectors may cause significant RF noise in the AURA sensitive band of $200-1200$ MHz. This noise pattern is being carefully studied and the amplified background noise frequency has a clear enhancement between $200-400$ MHz, with an amplitude of about $50mV$ corresponding to 7 ADC bits depth. The noise spectrum and intensity depends on the location of the antenna relative to the DRM and the type of front-end amplifier used. Background studies were also performed with the IceCube and AMANDA detectors turned off.  Figure \ref{fig-wf} shows sample waveforms taken for background studies with and without the transmitter antenna on for a single antenna.
%The pedestal subtraction is stable to within 1 bit.

\section{2008 Deployment and beyond}
The concept of a GZK radio frequency detector, deployed in shallow depths or in a surface array had been suggested more than 20 years ago \cite{array}.
A future large scale GZK $100 km^2$ scale detector will be a hybrid of different Cherenkov radiation detection techniques, allowing composite trigger and coincidence and can be built around IceCube. The long attenuation length of the ice (hundreds of meters), makes the South Pole ice a natural choice for deploying a RF detector.

In the next season (2007-2008) we plan to continue our efforts to design and build a shallow GZK neutrino detector. We will continue to use the IceCube deep holes and existing deployment and DAQ  infrastructure for  deploying additional clusters. We will investigate different depths ($1400, 200, <100$ meter and surface) and study the noise in lower frequencies ($<200$MHz) since the acceptance is expected to increase with wavelength, albeit at the expense of timing resolution.

A cluster will also be deployed $\sim 1km$ away from the IceCube array to study the ice and environment away from the IceCube array, and investigate possible solutions to communication and power distribution challenges that a large scale array presents.
 A surface array of radio detectors is relatively easy to deploy, but the refractive index difference between the ice, firn (soft ice layers on top of the glacier) and air decreases the angular acceptance of a surface detector due to total reflection of rays propagating between the layers. On the other hand, deeper deployments in depths of tens to hundreds of meters increases the technical difficulties and cost of such an array. 

 The design of the cluster will be similar to last year's clusters with possible minor changes to the antennas and electronics. By deploying at different depths and locations the RF properties of the ice, the suitability of ice for such of detector and studies of different cluster designs will be checked, while building a sub-GZK detector that will be able to detect HE events, reconstruct vertices, and look for events coincident with IceCube.

 Once completed, IceCube is expected to measure about 1 GZK event per year. A sucessful GZK detector deployed on surface or in shallow depth will have to measure at least $\sim 10 GZK$ events a year. A hybrid of the RF array and IceCube will give sub-samples of coincidences events with cross-calibration capabilities  and unique signal signatures.

\section{Summary}

Three radio clusters were deployed at the South Pole as an extension to the IceCube array. In the next year, we plan to deploy additional clusters to have a sufficient 3D array for vertices reconstruction, make radio-glaciological measurement at different depths and distances from the IceCube array, and check the suitability of the IceCube environment for RF detection.  
These are the first steps toward building a $100km^2$ GZK detector built around IceCube. Such a detector will be a powerful tool in investigating the EHE neutrino world.

\section{Acknowledgements}
This work is supported by the Office of Polar Programs of the National Science Foundation.

%This is the reference to .bib file (Whitout .bib!)
%\bibliography{libros}
%This in the bibtex style, is ok.
%\bibliographystyle{plain}

%\end{document}

%
\end{document}